\documentclass[11pt,a4paper, twosides]{book}

\usepackage{phdthesis}
\usepackage[latin1]{inputenc}
\usepackage{amssymb,amsmath,amsfonts,amsthm}
\usepackage{color}
\usepackage{graphicx} 
\usepackage{braket}
\usepackage[english] {babel}
\usepackage{slashed}
\usepackage{float} 
\usepackage[small,bf,hang]{caption2}
\usepackage{feynmf} 
\usepackage{subfigure}

\renewcommand{\d}{\ensuremath{\mathrm{d}}}
\newcommand{\ii}{\ensuremath{\mathrm{i}}}
\newcommand{\p}{\partial}
\newcommand{\Tr}{\ensuremath{\mathrm{Tr}}}
\newcommand{\e}{\ensuremath{\mathrm{e}}}
\newcommand{\GZ}{\ensuremath{\mathrm{GZ}}}
\newcommand{\RGZ}{\ensuremath{\mathrm{RGZ}}}
\newcommand{\AGZ}{\ensuremath{\mathrm{AGZ}}}
\newcommand{\cl}{\ensuremath{\mathrm{cl}}}
\newcommand{\s}{\ensuremath{\mathrm{s}}}
\newcommand{\nl}{\ensuremath{\mathrm{nl}}}
\newcommand{\h}{\ensuremath{\mathrm{h}}}
\newcommand{\sub}{\ensuremath{\mathrm{sub}}}
\newcommand{\modi}{\ensuremath{\mathrm{mod}}}
\newcommand{\intt}{\ensuremath{\mathrm{int}}}
\newcommand{\ext}{\ensuremath{\mathrm{ext}}}
\newcommand{\fin}{\ensuremath{\mathrm{fin}}}
\newcommand{\dive}{\ensuremath{\mathrm{div}}}
\newcommand{\bare}{\ensuremath{\mathrm{bare}}}
\newcommand{\subs}{\ensuremath{\mathrm{subs}}}
\newcommand{\phys}{\ensuremath{\mathrm{phys}}}
\newcommand{\R}{\ensuremath{\mathrm{R}}}
\newcommand{\YM}{\ensuremath{\mathrm{YM}}}
\newcommand{\FP}{\ensuremath{\mathrm{FP}}}
\newcommand{\gf}{\ensuremath{\mathrm{gf}}}
\newcommand{\m}{\ensuremath{\mathrm{m}}}
\newcommand{\lco}{\ensuremath{\overline{\varphi}\varphi-\overline{\omega}\omega}}
\newcommand{\quadr}{\ensuremath{\mathrm{quadr}}}

\newcommand{\MSbar}{\overline{\mbox{MS}}}
\newcommand{\en}{\ensuremath{\mathrm{en}}}
\newcommand{\lms}{\Lambda_{\overline{\mbox{\tiny{MS}}}}}

\newcommand{\glue}{\ensuremath{\mathrm{glue}}}

\newcommand{\Rglue}{\ensuremath{\mathrm{Rglue}}}
\newcommand{\sources}{\ensuremath{\mathrm{sources}}}
\newcommand{\fields}{\ensuremath{\mathrm{fields}}}
\newcommand{\Boltz}{\ensuremath{\mathrm{Boltz}}}
\newcommand{\new}{\ensuremath{\mathrm{new}}}
\newcommand{\mini}{\ensuremath{\mathrm{min}}}
\newcommand{\lat}{\ensuremath{\mathrm{lat}}}
\newcommand{\CGZ}{\ensuremath{\mathrm{CGZ}}}
\newcommand{\vac}{\ensuremath{\mathrm{vac}}}
\newcommand{\aux}{\ensuremath{\mathrm{aux}}}
\newcommand{\QCD}{\ensuremath{\mathrm{QCD}}}

\begin{document}

\thispagestyle{empty}

\hphantom{bla}
\vspace{-2cm}
\hspace*{-4.2cm}
\includegraphics[width= 22cm]{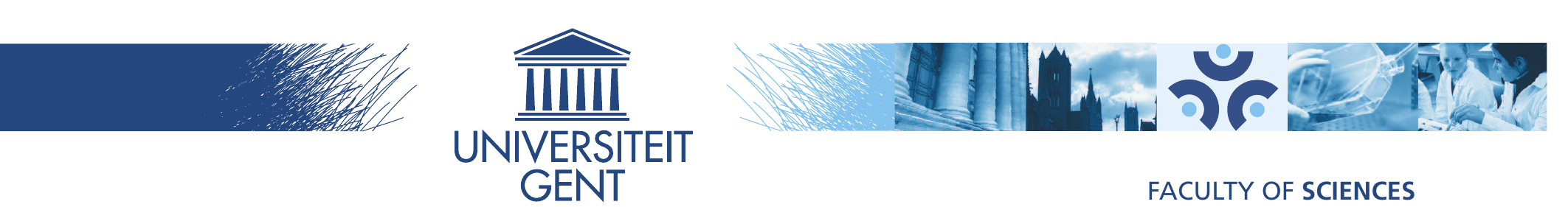} \\[2mm]
\begin{center}
Faculty of Sciences \\
Physics and Astronomy

\vspace*{3.5cm}

{\huge\bf A study of the Gribov-Zwanziger action:\\ \vspace{1cm}
 from propagators to glueballs } \\
\vspace{6cm}
{\LARGE Nele Vandersickel }

\vspace{2cm}

Thesis submitted in fulfillment of the requirements for the degree of \\ Doctor (Ph.D.) in Sciences: Physics
\\\vspace*{0.2cm}
Year 2010-2011 \\
\vspace*{0.2cm}Promotor : Prof. Dr. Henri Verschelde\\
\vspace*{0.2cm}Co-Promotor: Dr. David Dudal
\end{center}

\frontmatter
\newpage
\vphantom{lalala}\thispagestyle{empty}\newpage\thispagestyle{empty}
\newpage

\chapter*{Word of thanks}
\addcontentsline{toc}{chapter}{Word of thanks}

Writing a PhD thesis is something one cannot do alone, that is why I would like to thank a number a people who have helped me throughout this journey. \\
\\
Two people I have to thank in particular, as I could join the Belgium-Brazilian collaboration from the beginning and learn all the things about my research subject while also having the possibility to work on new research. First, I would like to thank David Dudal, for the guidance during my PhD and all the effort in bringing this thesis to a good end. Also Silvio P. Sorella, thank you for the nice collaboration during 3,5 years and the warm hospitality in Brazil (e também muito obrigada Simone!). Thank you also for reading chapter 2 in such great detail. I am also grateful to Henri Verschelde for giving me the opportunity to work in this interesting field and for the freedom he gave me during the PhD. \\
\\
Many other people have helped me as well. Dan Zwanziger, thank you for the many interesting discussions, for your patience in answering my questions and the effort for helping me to understand the GZ action (including the careful reading of chapter 3 of my thesis!). Orlando Oliviera, thank you for the nice collaboration, it was always a pleasure to discuss with you. Attilio Cucchieri, thank you for many insights in Lattice QCD and for giving clear answers to all my questions. Also thank you for a very detailed reading of this thesis. I would like to thank John Gracey, for the collaboration during the first two years of my PhD. \\
\\
Then I have a big list of people, who I met during my PhD, and in one way or another influenced my PhD. In alphabetic order, I wish to thank
Reinhard Alkofer,
Arlene Cristina Aguilar,
Laurent Baulieu,
Pedro Bicudo,
Daniele Binosi,
Maxim Chernodub,
Arturo Gomez,
Marcello Guimaraes,
Markus Huber,
Kei-Ichi Kondo,
Klaus Lichtenegger,
Axel Maas,
Vincent Mathieu,
Tereza Mendes,
Ana Mizher,
Stefan Olejnik,
Joannis Papavassiliou,
Andrea Quadri,
Lorenz Von Smekal. It was a pleasure to meet you all!\\
\\
I would also like to thank Arne Defauw who made an excellent master thesis in Gribov copies.\\
\\
To end the list, I want to thank my colleagues, Jutho, Nele, Benjamin, Dirk, Jos, Karel, Hans and David. A special word of thanks goes out to Jutho for helping me with numerous problems. I thank Inge Van der Vennet, for all the help with the administration.\\
\\
I also want to thank the FWO, for funding me during 4 years.\\
\\
I'm grateful to Annabelle Stevens, for all the help in the organization of our conference in Ghent.\\
\\
Finally, I wish to thank my my parents, for giving me the opportunity to study Physics, for their unconditional support, and a warm home which is always open for us. I'm grateful to my two fantastic sisters Els and Veerle, for always helping me in so many ways, and my friends for great times outside physics. Above all I want to thank Niels, for your enormous ability to always put a smile on my face. 
\chapter*{Preface}
\addcontentsline{toc}{chapter}{Preface}
This thesis presents a study of the Gribov-Zwanziger framework: from propagators to glueballs. The chapters \ref{algebraic} and \ref{chapgribovtoGZ} are meant as an introduction and only require a basic knowledge of quantum field theory. Chapter \ref{algebraic} explains the techniques behind algebraic renormalization, which shall be widely used throughout this thesis, while chapter \ref{chapgribovtoGZ} tries to give a pedagogic overview of the Gribov-Zwanziger framework as this is not available yet in the literature. The subsequent chapters contain own research. First in chapter \ref{scrutinizing}, we shall dig a bit deeper in the Gribov-Zwanziger framework, by exploring the BRST symmetry and the KO criterium. Next, in chapter \ref{refined} we shall elaborate on the ghost and the gluon propagator in the infrared and present a refined Gribov-Zwanziger action. Further, we present two chapters on the search for physical operators within the (refined) Gribov-Zwanziger framework, chapter \ref{chappart1} and  \ref{chappart2}. A small chapter \ref{last} is devoted to some values for different glueballs. We end this thesis with the conclusions, chapter \ref{conclusions}. 
\chapter*{Dutch Summary - Nederlandse Samenvatting}
\addcontentsline{toc}{chapter}{Dutch Summary}
We present here a Dutch Summary of this thesis.

\section*{Inleiding: QCD, een enorme lappendeken}
Quantum Chromodynamica (QCD) is de theorie van de sterke interacties, één van die vier fundamentele krachten van ons universum. Deze kracht beschrijft de interactie tussen quarks en gluonen, die fundamentele bouwstenen voorstellen van de gekende materie. Deze kracht is asymptotisch vrij bij hoge energieën, wat wil zeggen dat quarks en gluon zich als vrije deeltjes kunnen bewegen. Daarentegen bij lage energieën, wat overeenkomt met onze huidige wereld, vormen quarks en gluonen steeds gebonden toestanden, die we hadronen noemen. De best gekende voorbeelden hiervan zijn de protonen en neutronen, maar er bestaat een enorme verscheidenheid aan andere deeltjes die bijvoorbeeld in deeltjesversnellers kunnen gecreëerd worden. Quarks en gluonen dragen een soort lading die we kleur noemen, en hadronen zullen steeds manifestaties voorstellen van kleurloze objecten. Vreemd genoeg werd er nog nooit een vrije quark of gluon waargenomen. Dit fenomeen noemen we confinement en QCD is de enige fundamentele kracht met deze unieke eigenschap. Ondanks 40 jaar intensief onderzoek sinds de formulering van het standaardmodel, blijft het onduidelijk wat de precieze verklaring van confinement is. Zelf de formulering wat confinement eigenlijk is, staat ter discussie. \\
\\
Hoe komt dit nu, dat confinement zo een moeilijk te verklaren fenomeen is? De verkla\-ring ligt in het feit dat de standaardtechnieken die heel succesvol gebleken zijn in Quantum Elektrodynamica (QED), een andere fundamentele kracht, niet toepasbaar zijn in QCD. QED vertoont namelijk een compleet ander gedrag. Voor lage energieën heeft deze theorie een koppelingsconstante die almaar kleiner wordt, en dus kan je een techniek toepassen, perturbatietheorie genoemd, die een reeksontwikkeling in deze kleine parameter toelaat. Daarentegen bij QCD, is de koppelingsconstante veel te groot om nog perturbatietheorie te kunnen toepassen. Daarom moeten we onze toevlucht zoeken in niet perturbatieve technieken, die echter niet zo eenvoudig zijn. Er is ondertussen een enorme variëteit aan technieken ontwikkeld die elk op hun manier QCD proberen te benaderen. Alle methodes om QCD te beschrijven kan men beschouwen als een enorm lappendeken, elke methode belicht andere aspecten en heeft zo zijn eigen voordelen.\\
\\
Misschien is het interessant te vermelden dat ook zonder quarks, zogenaamde pure QCD, er ook confinement is. Uiteraard kan dit niet bevestigd worden door experimenten, maar rooster simulaties tonen aan dat dit inderdaad het geval is. Daarom is het reeds interessant om QCD zonder quarks te onderzoeken, omdat de sleutel tot de oplossing verborgen ligt in het gedrag van de gluonen. Daarom zullen we hier dan ook voornamelijk pure QCD onderzoeken.

\section*{De Gribov-Zwanziger actie}
In deze thesis zullen we een stukje van het lappendeken beschrijven. We zullen hiervoor starten van de $SU(N)$ Yang-Mills ijktheorie. Deze theorie is een veralgemening van pure QCD waarvoor je dan $N =3$ moet stellen. Als je deze theorie wil neerschrijven in een padintegraal, wordt het snel duidelijk dat je een ijk moet kiezen om een betekenisvolle padintegraal te kunnen neerschrijven. De ijk vastleggen betekent dat men per serie van ijkequivalente velden één veld uitpikt. Dit werd voor het eerst gedaan door Faddeev en Popov in het jaar 1967 \cite{Faddeev:1967fc}. Een heel populaire ijk is de Landau ijk, door zijn eenvoud, en daarom zullen we in deze thesis dan ook hoofdzakelijk in deze ijk werken. Ondanks het grote succes van de aanpak van Faddeev en Popov, o.a.~door het ontdekken van een resterende symmetrie, de BRST symmetrie, ontdekte Gribov dat er toch iets niet helemaal klopte, zie zijn beroemde artikel uit 1978 \cite{Gribov:1977wm}.  Elke normale ijk die men probeert te kiezen heeft namelijk het probleem dat er nog steeds Gribov kopieën aanwezig zijn. Dit zijn ijkequivalente velden die beide aan dezelfde ijkvoorwaarde voldoen. Dit wil zeggen dat de ijkfixing van Faddeev en Popov onvolledig zou zijn. Gribov beschreef in zijn artikel hoe dit gevolgen zou hebben op de gluon en op de ghost propagator. Door de Gribov kopieën uit te sluiten, vertoont de gluon propagator een ander gedrag, namelijk in het lage momentum regime wordt deze propagator onderdrukt. Ook de ghost propagator vertoont een ander gedrag: in het lage momentum gebied wordt deze propagator versterkt. Gribov zijn aanpak werd later veralgemeend door Zwanziger in 1989 \cite{Zwanziger:1989mf}. Hij construeerde een actie die een groot deel van de Gribov kopieën uitsluit, door de integratie van de gluonen te beperken tot een kleiner gebied, genaamd de Gribov regio: de Gribov-Zwanziger actie was geboren. Jammer genoeg breekt deze actie de BRST symmetrie, weliswaar enkel op een ``zachte manier'', dwz voor grote energieën herstelt deze symmetrie zich opnieuw.\\
\\
Elke theorie moet getest worden. Aangezien propagatoren sterk ijkafhankelijke grootheden zijn, kon deze enkel via roostersimulaties getest worden. Dit werd dan ook uitvoerig gedaan, en de Gribov-Zwanziger theorie werd steeds opnieuw bevestigd. De ghost propagator vertoonde versterkt gedrag en de gluon propagator was onderdrukt in het infrarood, schond de positiviteit en was nul bij momentum gelijk aan nul. Al deze voorspellingen werden inderdaad gedaan door de Gribov-Zwanziger actie.\\
\\
Dit tot er een nieuw geluid kwam in 2007. Toen werden er in \cite{Cucchieri:2007md} verrassende resultaten gepubliceerd. De roostersimulaties in deze paper waren uitgevoerd op enorme roosters zodat men meer data in het diep infrarode gebied kon verzamelen: de gluonpropagator leek te naderen tot een eindige waarde voor nul momentum en de spookpropagator was toch niet zo versterkt als vooralsnog steeds aangenomen werd. De hele gemeenschap stond voor een raadsel.

\section*{De verfijnde Gribov-Zwanziger actie}
De vraag die uiteraard rees was of het nog steeds mogelijk was deze resultaten te verklaren vanuit de Gribov-Zwanziger theorie. Daarom werd er op zoek gegaan naar andere niet perturbatieve effecten die in rekening moeten gebracht worden om dit te verklaren. Door het dynamisch in rekening brengen van condensaten, dit zijn vacuumverwachtingwaarden van een locaal samengestelde operator, in de Gribov-Zwanziger theorie, bleek het inderdaad mogelijk om opnieuw resultaten te vinden in overeenstemming met de roosterdata. Condensaten zijn typische niet-perturbatieve effecten die reeds lang gekend zijn. \\
\\
Na het verfijnen van de Gribov-Zwanziger actie, was er nu een nieuwe voorspelling voor de gluon propagator voorhanden. Deze voorspelling werd dan ook met succes gefit aan de reeds bestaande roosterdata, en gaf een bevestiging voor de verfijnde Gribov-Zwanziger theorie.

\section*{Glueballs}
Tot zover werden er enkel propagatoren onderzocht. Dit zijn sterk ijkafhankelijke grootheden en dus zou het interessant zijn om te onderzoeken of er we uit de (verfijnde) Gribov-Zwanziger actie ook voorspellingen voor fysische grootheden kunnen halen. Immers, we hebben nu een model gevonden dat de propagatoren goed beschrijft, maar als dit model wil overeind blijven, zou deze ook voorspellingen moeten kunnen doen over fysische deeltjes. Aangezien we nog steeds werken met pure Yang-Mills theorie, zouden deze deeltjes een kluwen van gluonen moeten zijn: glueballs. Hiervoor hebben we in een eerste poging de renormalizatie van $F^2_{\mu\nu}$ onderzocht, daar deze operator standaard geassocieerd wordt met de scalaire glueball. Door de gecompliceerde vorm van de Gribov-Zwanziger actie, was dit een niet triviale opdracht. We vonden dat deze operator zich mengt met andere $d=4$ operatoren. Daar de Gribov-Zwanziger actie de BRST symmetrie breekt, heeft dit belangrijke gevolgen voor de correlator. De correlator $\Braket{F^2_{\mu\nu}(x) F^2_{\mu\nu}(y)}$ is op zich geen welgedefinieerde grootheid, net door die menging. We vonden echter wel dat de volgende grootheid, $\Braket{\mathcal R(x) \mathcal R(y)}$, goed gedefinieerd is met $\mathcal R$ een renormalizatie groep invariante grootheid gegeven door
\begin{eqnarray}
\mathcal{R}= \frac{\beta(g^2)}{g^2} F^2_{\mu\nu} -2\gamma_c(g^2)  \mathcal E
\end{eqnarray}
waarbij $\mathcal E$ een niet BRST invariante combinatie van velden uit de Gribov-Zwanziger actie is.\\
\\
Nu we de renormalizatie van $F^2_{\mu\nu}$ achter de rug hebben, kunnen we de correlator $\Braket{\mathcal R(x) \mathcal R(y)}$ onderzoeken. Het idee is het volgende: onderstel een operator $\mathcal O$. Indien dit een fysische operator is, dan zouden we deze moeten kunnen schrijven in een K\"all\'{e}n-Lehmann representatie:
\begin{align}
 \Braket{ \mathcal O(k) \mathcal O(-k) }  =  \int_{\tau_{0}}^{\infty} \d\tau \; \rho({\tau}) \; \frac{1}{\tau+k^2} \;,
\end{align}
waarbij $\rho({\tau})>0$. M.a.w., de correlator $\Braket{ \mathcal O(k) \mathcal O(-k) } $ moet een vertakkingslijn vertonen op de negatieve\footnote{Dit omdat we steeds in de Euclidische ruimte werken.} $x$-as. Indien echter deze correlator (een) vertakkingslijn(en) vertoont die niet op de negatieve $x$-as ligt en/of we vinden een spectrale dichtheid $\rho({\tau})<0$ op de negatieve $x$-as, dan spreken we een niet fysische operator. Nu werd reeds gevonden dat $\Braket{F^2_{\mu\nu}(x) F^2_{\mu\nu}(y)}$ reeds een vertakkingslijn vertoont op de imaginaire as, wat erop wijst dat $F^2_{\mu\nu}$ en dus hoogstwaarschijnlijk ook $\mathcal R$ geen goede operator is.\\
\\
Daarom zijn we op zoek gegaan naar een andere operator, die wel fysisch is. Hiervoor hebben we een nieuw concept ingevoerd: $i$-particles. Op laagste orde zijn we erin geslaagd met behulp van de $i$-particles een operator te vinden die een goede K\"all\'{e}n-Lehmann representatie heeft. Jammer genoeg breekt deze operator de BRST op een harde manier, zodat deze operator er niet renormalizeerbaar uitziet.\\
\\
Tot besluit kunnen we stellen dat we enerzijds een renormalizeerbare operator hebben gevonden, maar met een slechte K\"all\'{e}n-Lehmann representatie en anderzijds een niet renormalizeerbare operator, maar met een goede K\"all\'{e}n-Lehmann representatie. Hoe we deze twee kunnen verenigen is voorlopig nog een vraagteken, maar dit wordt momenteel onderzocht.

\section*{Indeling van deze thesis}
Deze thesis start met een korte situering van het werk in hoofdstuk \ref{chapintro}. Daar we in deze thesis heel vaak gebruik zullen maken van techniek van algebraïsche renormalizatie, geven we hierover eerst een inleiding, zie hoofdstuk \ref{algebraic}. Net omdat dit geen standaard techniek is, gaan we enkel uit van een basiskennis Quantumveldentheorie. Hierna volgt hoofdstuk \ref{chapgribovtoGZ}: hierin geven we een overzicht van de ideeën van Gribov tot de constructie van de Gribov-Zwanziger actie. Daar in de literatuur dit overzicht ontbreekt, hoopt dit hoofdstuk deze leemte te vullen. Op het einde van dit hoofdstuk bewijzen we dat deze actie renormalizeerbaar is met de technieken uitgelegd in hoofdstuk  \ref{algebraic}. Vanaf dan start het eigenlijke onderzoek. In hoofdstuk \ref{scrutinizing} gaan we nog wat dieper in op de Gribov-Zwanziger actie. We leggen de betekenis uit van de breking van de BRST symmetrie en we tonen aan hoe we deze kunnen herstellen door het invoeren van nieuwe velden. Ook leggen we een verband tussen de Gribov-Zwanziger actie en het beroemde Kugo-Ojima theorema. Vervolgens verfijnen we de Gribov-Zwanziger actie om in overeenstemming te komen met de roosterdata in hoofdstuk \ref{refined}. De volgende twee hoofdstukken starten dan de zoektocht naar fysische operatoren. In hoofdstuk \ref{chappart1} onderzoeken we de algebraïsche renormalizatie van $F^2_{\mu\nu}$ in de Gribov-Zwanziger actie, terwijl we in hoofdstuk \ref{chappart2} de $i$-deeltjes invoeren. In hoofdstuk \ref{last} bespreken we nog enkele resultaten voor de massa's van enkele glueballs die berekend werden in het kader van de $i$-particles. We eindigen tot slot deze thesis met een besluit en vooruitzichten in hoofdstuk \ref{conclusions}.

\tableofcontents

\mainmatter

\chapter{Introduction\label{chapintro}}
\section{The patchwork quilt of QCD}
Quantum Chromodynamics (QCD) is the theory which describes the strong interaction, one of the four fundamental forces in our universe. This force describes the interactions between quarks and gluons, which are fundamental building blocks of our universe. At very high energies, QCD is asymptotically free, meaning that quarks and gluons should be detected as free particles\footnote{E.g.~in deep inelastic scattering (DIS) experiments, quarks can be treated as free particles.}. However at low energies, i.e~our daily world, due to the strong force, quarks and gluons interact and form bound states called hadrons. A well known example of these bound states are the proton and the neutron, but a whole zoo of hadrons has been observed in particle detectors. In fact, the only way to obtain information about the strong force is through bound states, as no free quark or gluon has even been detected. We call this phenomenon \textit{confinement}. Although since the formulation of the standard model (which includes QCD), 40 years of intensive research have past, no good answer has been found to probably one the most fundamental questions in QCD. Even the formulation of what confinement really is, is under discussion.\\
 \\
The difficulty for solving confinement lies in the fact that the standard techniques which have been so successful in QED, are not applicable in QCD. In QED the coupling constant is small enough\footnote{This depends of course on the energy range of interest.}, so one can apply perturbation theory, which amounts to writing down a series in the coupling constant. As in QCD, the coupling constant increases for decreasing energies, the coupling constant is too large and perturbation theory alone can never give a good description of the theory. Therefore, other techniques are required which we call non-perturbative methods. There exists a wide range of non-perturbative methods, which all try to approach QCD from one way or another. One should not see these different techniques as competing, but as patchwork trying to cover all aspects of QCD.\\
\\
Even if one omits quarks in QCD, one can still call the remaining theory confining. Although there is no real experimental evidence, lattice simulations have shown that gluons form bound states which we call glueballs and no free gluon shall be detected. Therefore, it is already interesting to investigate pure QCD without quarks, and try to find out what happens. One could say that confinement is hidden in the behavior of the gluons.

\section{The Yang-Mills theory: definitions and conventions \label{yangmillsintro}}
Now what is QCD? QCD is a gauge theory, as all theories for the fundamental forces. In fact, QCD is only a special case of the more general $SU(N)$ Yang-Mills theory, but with $N=3$.\\
\\
In order to know what we are talking about, let us already introduce the Yang-Mills theory here, as this theory shall be the main building block of this thesis. We shall try to be consistent and use the conventions outlined in this section. The derivation of the Yang-Mills action can be found in any standard textbook on quantum field theory \cite{Peskin}.\\
\\
We start with the compact group $SU(N)$ of $N \times N$ unitary matrices $U$ which have determinant one. We can write these matrices as
\begin{eqnarray}\label{U}
U &=& \e^{-\ii g \theta_a X_a} \;,
\end{eqnarray}
whereby $X^a$ represent the generators of the $SU(N)$ group. The index $a, b, c, \ldots$ is called the color index and runs from $\{1, \ldots, N^2 -1\}$. These generators obey the following commutation rule
\begin{eqnarray}
[X^a, X^b] &=& \ii f_{abc} X^c\;,
\end{eqnarray}
and thus the $SU(N)$ group corresponds to a simple Lie group. We can choose these generators to be hermitian, $X^\dagger = X$, and normalize them as follows
\begin{eqnarray}
\Tr[X_a X_b] &=& \frac{\delta_{ab}}{2}\;.
\end{eqnarray}
The generators $X_a$ belong to the adjoint representation of the group $SU(N)$, i.e.
\begin{eqnarray}
U X_a U^\dagger &=& (D^A)_{ab} X_b\;,
\end{eqnarray}
with $(D^A (X_a))_{bc}= - \ii f_{abc}$. Now we can construct a Lagrangian, which is symmetric under this group.
\\
\\
Firstly, we define the standard $SU(N)$ Yang-Mills action as
\begin{eqnarray}
S_{\YM} &=& \int \d^4 x \frac{1}{2} \Tr F_{\mu\nu}  F_{\mu\nu}\;,
\end{eqnarray}
whereby $F_{\mu\nu}$ is the field strength
\begin{eqnarray}\label{ff}
F_{\mu\nu}  &=& \p_\mu A_{\nu}  - \p_\nu A_\mu  - \ii g [A_{\mu}, A_{\nu}]\;,
\end{eqnarray}
and $A_\mu$ the gluon fields which belongs to the adjoint representation of the $SU(N)$ symmetry, i.e.
\begin{eqnarray}
A_\mu &=& A_\mu^a X^a \;.
\end{eqnarray}
The field strength can thus also be written as
\begin{eqnarray}
F_{\mu\nu} &=& F_{\mu\nu}^a X^a\;,
\end{eqnarray}
whereby
\begin{eqnarray}
F_{\mu\nu}^a \ =\ \p_\mu A_{\nu}^a  - \p_\nu A_\mu^a  +g  f_{akl} A_{\mu}^k A_{\nu}^l	\;.
\end{eqnarray}
Under the $SU(N)$ symmetry, we define $A_\mu$ to transform as
\begin{eqnarray}\label{notinf}
A_{\mu}' &=& U A_\mu U^{\dagger} - \frac{\ii}{g} (\p_\mu U) U^{\dagger}\;,
\end{eqnarray}
and one can check that this is compatible with the fact the $A_\mu$ belongs to the adjoint representation. Consequently, from \eqref{ff} we find
\begin{eqnarray}
F_{\mu\nu}' &=& U F_{\mu\nu} U^{\dagger}\;,
\end{eqnarray}
and therefore the Yang-Mills action is invariant under the $SU(N)$ symmetry. Infinitesimally, the transformation \eqref{notinf} becomes
\begin{eqnarray}\label{infinitesimal}
\delta A_{\mu}^a &=& -D_\mu^{ab} \theta^b \;,
\end{eqnarray}
with $D_\mu^{ab}$ the covariant derivative in the adjoint representation
\begin{eqnarray}\label{covariantderivativeadjoint}
D_\mu^{ab} &=& \p_\mu \delta^{ab} - g f^{abc} A_\mu^c\;.
\end{eqnarray}
Secondly, we can also include the following matter part in the action, when considering full QCD
\begin{eqnarray}
S_{\m} &=& \int \d^4 x \left( \overline \psi^i_\alpha (\gamma_\mu)_{\alpha \beta} D_\mu^{ij} \psi^j_\beta \right)\;,
\end{eqnarray}
which contains the matter fields $\overline \psi_i$ and $\psi_i$ belonging to the fundamental representation of the $SU(N)$ group, i.e.
\begin{eqnarray}
\overline \psi_i' &=& U_{ij} \overline \psi_j\;,
\end{eqnarray}
or infinitesimally
\begin{eqnarray}
\delta \overline \psi_i' &=& -\ii \theta^a X^a_{ij} \overline \psi_j\;.
\end{eqnarray}
The index $i$ runs from $\{1, \ldots, N\}$. Every $\overline \psi_i$ and $\psi_i$ is in fact a spinor, which is indicated with the indices $\{\alpha, \beta, \ldots$\}. $D_\mu^{ij}$ is the covariant derivative in the fundamental representation
\begin{eqnarray}
    D_\mu^{ij} &=& \p_\mu \delta^{ij} - \ii g A_\mu^a (X^a)^{ij} \;,
\end{eqnarray}
and $\gamma_\mu$ are the Dirac gamma matrices. One can again check that also the matter part is invariant under the $SU(N)$ symmetry. The matter field $\overline \psi$ represents the quarks of our model. As is known from the standard model, there is more than one type of quark, which we call flavors. For each flavor, we would need to add a term like $S_\m$, but keeping the notation simple, we shall not introduce a flavor index here. The starting point of the Yang-Mills theory including quarks is thus given by
\begin{eqnarray}\label{ym}
S &=& S_\YM + S_\m \;.
\end{eqnarray}
Most of the time, we shall however omit the matter part, and work with pure Yang-Mills theory.

\section{Outline of the thesis}
Now that we have introduced the Yang-Mills theory, let us explain the goal of this thesis. In this thesis, we shall try to cover one square of the patchwork quilt. Most of the time, we shall omit quarks and work with pure Yang-Mills theory, commonly known as gluodynamics. \\
\\
For this, we need techniques. In everything which we shall do, we believe that renormalizability is of great importance. To prove that a theory is renormalizable, we use a beautiful technique called algebraic renormalization. This is an extremely powerful tool, which is unfortunately not widely known. Therefore, we have given a general introduction of the formalism, starting only from basic knowledge of quantum field theory. This is the topic of chapter \ref{algebraic}.\\
\\
In chapter \ref{chapgribovtoGZ}, we shall introduce the model which we shall explore in great detail in this thesis, i.e.~the Gribov-Zwanziger (GZ) formalism. As no overview of this model has been presented so far, we give a detailed description of the origin of the GZ action. For this, we shall start from the Yang-Mills action, and show that one is obliged to choose a gauge when quantizing the theory. The way to implement a gauge is by following the Faddeev-Popov method and we shall mainly work in the Landau gauge. However, if one carefully does the derivation, one sees that the Faddeev-Popov method contains flaws, i.e.~the gauge is not fixed uniquely and Gribov copies arise. However, at the time Faddeev and Popov invented their method, they were only interested in the perturbative regime, where the flaws can be neglected. But as we explained, the non-perturbative regime is the interesting regime where confinement sets in. Gribov was the first to notice that these flaws could have serious consequences on the ghost and the gluon propagators at low momenta. Subsequently, Zwanziger found a way to construct an action which implemented the ideas of Gribov: the Gribov-Zwanziger action was born. With the help of chapter \ref{algebraic}, we shall then prove that this action is renormalizable.\\
\\
Once the Gribov-Zwanziger action is explained, we shall elaborate on several aspects concerning this action. This was done in chapter \ref{scrutinizing}. In particular, we shall show that the GZ action breaks the BRST symmetry, and we shall further elaborate  on this. Another important point shall be the relation of the GZ action to the Kugo-Ojima confinement criterium.\\
\\
A theory is only a model until it has been tested. As we are working with pure Yang-Mills theory, we cannot use experimental data, but we can compare our analytical results with lattice calculations. Two particular quantities have been tested in great detail. Namely the ghost and the gluon propagator, as they are believed to play an important role in confinement scenarios. Until 2007, the following results were reported: the gluon propagator was positivity violating and infrared suppressed in 2d, 3d and 4d. At zero momentum, in 2d the gluon propagator was vanishing \cite{Maas:2007uv}, in 3d the gluon propagator was not in contradiction with a vanishing gluon propagator \cite{Cucchieri:2003di}, while in 4d, not so much was known. The ghost propagator was believed to be enhanced. A vanishing gluon propagator and an enhanced ghost propagator exactly agrees with the predictions of the GZ framework, increasing at that time the success of the model. However, in 2007, the papers \cite{Cucchieri:2007md,Bogolubsky:2007ud} appeared, with simulations on huge lattices. What they found was striking: the gluon propagator did not seem to vanish at zero momentum and the ghost propagator did not display enhancement. Therefore, it seemed that something was missing in the GZ framework. Perhaps other non-perturbative effects should be included? This is the topic of chapter \ref{refined}. We shall show that within the GZ formalism, we can still obtain results which are in agreement with the lattice data, by including condensates into the theory, which we call the Refined GZ (RGZ) action. We shall corroborate this with some explicit fits with the lattice data. \\
\\
So far, we only investigated propagators. However, these are highly gauge dependent quantities. If the (R)GZ action really is the correct theory to describe QCD, we should be able to the find bound states, i.e.~glueballs. This is the topic of chapter \ref{chappart1} and \ref{chappart2}. In chapter \ref{chappart1} we shall investigate the operator $F^2_{\mu\nu}$, which is commonly believed to be related to the scalar glueball operator, and we shall show that we can renormalize this operator to all orders within the (R)GZ framework. We shall even be able to construct a renormalization group invariant. However, the correlator $\braket{F^2 (x) F^2(y)}$ displays unphysical cuts. Therefore, we shall look in chapter \ref{chappart2} for operators which have physical properties and could be really related to glueballs. In chapter \ref{last} we shall discuss some results for the masses of the lowest lying glueballs in the framework of  $i$-particles. We shall formulate our conclusions and outlook in chapter \ref{conclusions}.

\chapter{Algebraic renormalization \label{algebraic}}
In this chapter we shall provide the reader with an overview of the algebraic framework of renormalization, as a large part of this thesis is based on this technique. This overview is mostly based on the book \cite{Piguet1995}, with the focus on the practical implementation of the algebraic renormalization.

\section{The generating functionals}
Let us start with a general classical Lagrangian $\mathcal L$ involving a certain set of fields $\phi_i$, with $i$ denoting the kind of field, as well as the spin and internal degrees of freedom, and working in $d$-dimensional space-time. The Lagrangian can always be decomposed in a quadratic part and an interacting part
\begin{eqnarray}
\mathcal L(\phi_i (x)) &=&  \mathcal L_0(\phi (x)) + \mathcal L_{\intt}(\phi (x))\;,
\end{eqnarray}
whereby we can write the quadratic part as
\begin{eqnarray}
\mathcal L_0(\phi (x)) &=& \frac{1}{2} \phi_i (x) K^{ij}(\partial) \phi_j (x) \;.
\end{eqnarray}
The action is then defined by
\begin{eqnarray}
S  (\phi_i) &=& \int \d x \mathcal L(\phi_i(x)) \;.
\end{eqnarray}
Now that we have introduced the action, we can introduce the Feynman path integral
\begin{eqnarray}\label{si}
\int [\d \phi] \e^{\frac{-1}{\hbar}S(\phi_i) }\;,
\end{eqnarray}
whereby we shall work in Euclidean space-time. 
This object is the basis of quantum field theory. 
Expression \eqref{si} forms the basis for the calculation of Green's functions in Euclidean quantum field theory. Green's functions are vacuum expectation values of the fields,
\begin{eqnarray}\label{si2}
 \Braket{\phi_{i_1} (x_1) \phi_{i_2} (x_2) \ldots \phi_{i_n} (x_n)   } \equiv  \int [\d \phi]  \phi_{i_1} (x_1) \phi_{i_2} (x_2) \ldots \phi_{i_n} (x_n)  \e^{\frac{-1}{\hbar }S(\phi_i) }\;.
\end{eqnarray}
These Green's function are objects of interest in Quantum Field theory, as they can enter the formula of the cross section corresponding to the scattering processes of particles. The main question is thus how one can calculate these Green's functions. Now we shall construct three different kinds of generating functionals which generate different kinds of Green's functions \cite{Zinn-Justin2002,Peskin,Ryder,Zee:2003mt,Casalbuoni}.

\subsection{The generating functional $Z(J)$ \label{thegeneratingfunctional}}
Firstly, we start with the vacuum expectation value of a products of fields as given in equation \eqref{si2},\begin{eqnarray}\label{npoint}
\Braket{ \phi_{i_1} (x_1) \ldots \phi_{i_N} (x_N)} &=& \int [\d \phi]  \phi_{i_1} (x_1) \ldots \phi_{i_N} (x_N) \e^{-\frac{1}{\hbar}S(\phi_i) }\;.
\end{eqnarray}
Let us first elaborate on the meaning of these Green's functions. The Green's function \eqref{npoint} is called a $n$-point function. Diagrammatically we can think about $n$-legs, namely $\phi_{i_1}, \ldots, \phi_{i_N}$ starting from the space-time points $x_1, \ldots, x_N$, which have to be connected in some way. How one can connect these lines, is determined by the action $S$, and different possibilities are often possible also with a different number of internal loops. We order the diagrams according to the number of loops, which shall correspond to a certain order in $\hbar$: we call this perturbation theory. Purely written as in \eqref{npoint}, also diagrams shall be created which are completely disconnected from the space time points $x_i$. These diagrams are called vacuum diagrams, and by choosing an appropriate normalization factor $\mathcal N$, one can show that one can get rid of this kind of diagrams.\\
\\
The expectation values \eqref{npoint} can be derived from the following generating functional,
\begin{eqnarray}\label{genZ}
Z(J) &=& \mathcal N \int [\d \phi] \e^{-\frac{1}{\hbar}\left(  S(\phi_i) + \int \d x J^i (x) \phi_i (x) \right)}\;,
\end{eqnarray}
whereby $J^i (x)$ is an external source for field $\phi_i$. Here we have immediately introduced the normalization factor $\mathcal N$ to omit diagrams which contain pure vacuum terms. If we take $\mathcal N = Z^{-1} (J \equiv 0)$, one can easy appreciate that no pure vacuum terms will appear in the diagrams. In what follows, we shall not write explicitly $\mathcal N$, always assuming its presence. If we derive the functional $Z(J)$ with respect to the corresponding sources and set these sources equal to zero afterwards, we find indeed
\begin{eqnarray}
\left. \frac{\delta^n Z(J)}{\delta J_i(x_1) \ldots \delta J_i (x_N)} \right|_{J_i = 0} &=&\left(\frac{-1}{\hbar}\right)^N \int [\d\phi]  \phi_i (x_1) \ldots \phi_N (x_N) \e^{-\frac{1}{\hbar}S(\phi_i) }\;.
\end{eqnarray}
In this sense, we can write $Z(J)$ as a series in $J$:
\begin{equation}\label{expZ}
Z(J) = \sum_{N= 0}^{\infty} \frac{(-1/\hbar)^N}{N!} \int \d x_1  \int \d x_2 \ldots \int \d x_N J^{i_1}(x_1) \ldots J^{i_N}(x_N) \Braket{ \phi_{i_1}(x_1) \ldots \phi_{i_N}(x_N) }\;.
\end{equation}

\subsection{The generating functional $Z^c(J)$ of the connected Green's functions}
So far, connected as well as disconnected diagrams were taken into account. One can therefore also define the following Green's function,
\begin{eqnarray}
\Braket{\phi_{i_1} (x_1) \ldots \phi_{i_N} (x_N)}_c\;,
\end{eqnarray}
which only takes into account connected diagrams. \\
\\
The generating functional of the connected Green's functions is related to $Z(J)$ in the following way,
\begin{eqnarray}\label{exp}
Z(J) &=& \e^{-\frac{1}{\hbar} Z^c (J)}\;,
\end{eqnarray}
where again
\begin{equation}\label{Zc}
Z^c(J) = \sum_{N= 1}^{\infty} \frac{(-1/\hbar)^{N-1}}{N!} \int \d x_1  \int \d x_2 \ldots \int \d x_N J^{i_1}(x_1) \ldots J^{i_N}(x_N) \Braket{\phi_{i_1}(x_1) \ldots \phi_{i_N}(x_N)  }_c \;.
\end{equation}
This can be understood as follows.  A generic Green's function can always be described by a sum of connected Green's functions,
\begin{eqnarray}\label{greenconn}
\Braket{ \phi_{i_1} (x_1 )} &=&  \Braket{ \phi_{i_1} (x_1)}_c \nonumber\\
\Braket{ \phi_{i_1} (x_1) \phi_{i_2} (x_2)} &=&  \Braket{ \phi_{i_1} (x_1) \phi_{i_2} (x_2)}_c +  \Braket{  \phi_{i_1} (x_1)}_c  \Braket{\phi_{i_1} (x_2)}_c \nonumber\\
\Braket{ \phi_{i_1} (x_1) \phi_{i_2} (x_2) \phi_{i_2} (x_3)} &=&  \Braket{\phi_{i_1} (x_1) \phi_{i_2} (x_2) \phi_{i_3} (x_3)}_c + \Braket{  \phi_{i_1} (x_1)}_c  \Braket{ \phi_{i_2} (x_2) \phi_{i_3} (x_3)}_c \nonumber\\
&&+  \Braket{  \phi_{i_2} (x_2)}_c  \Braket{  \phi_{i_1} (x_1) \phi_{i_3} (x_3)}_c \nonumber\\
&& + \Braket{  \phi_{i_3} (x_3)}_c  \Braket{ \phi_{i_1} (x_1) \phi_{i_2} (x_2)}_c \nonumber\\
 &&+  \Braket{  \phi_{i_1} (x_1)}_c  \Braket{  \phi_{i_2} (x_2)}_c  \Braket{  \phi_{i_3} (x_3)}_c   \\
 &\vdots& \nonumber
\end{eqnarray}
as one can read from the following diagrams
\begin{eqnarray}
\parbox{35mm}{ \begin{fmffile}{a1}
  \begin{fmfgraph*}(20,10)
  \fmfleft{i}
  \fmfright{j}
  \fmf{plain}{i,j}
  \fmfblob{.30w}{j}
  \end{fmfgraph*}
  \end{fmffile} } &=& \parbox{35mm}{ \begin{fmffile}{a2}
  \begin{fmfgraph*}(20,10)
  \fmfleft{i}
  \fmfright{j}
  \fmf{plain}{i,j}
  \fmfv{decor.shape=circle,decor.filled=empty,decor.size=.30w,label=$c$,label.dist=-1.2mm}{j}
  \end{fmfgraph*}
  \end{fmffile} } \nonumber\\
  \parbox{35mm}{ \begin{fmffile}{a3}
  \begin{fmfgraph*}(20,10)
  \fmfleft{i}
  \fmfright{j}
  \fmf{plain}{i,k}
  \fmf{plain}{k,j}
  \fmfblob{.30w}{k}
  \end{fmfgraph*}
  \end{fmffile} } &=& \parbox{35mm}{ \begin{fmffile}{a4}
  \begin{fmfgraph*}(20,10)
  \fmfleft{i}
  \fmfright{j}
  \fmf{plain}{i,k}
  \fmf{plain}{k,j}
  \fmfv{decor.shape=circle,decor.filled=empty,decor.size=.30w,label=$c$,label.dist=-1.2mm}{k}
  \end{fmfgraph*}
  \end{fmffile} } +
  \parbox{35mm}{ \begin{fmffile}{a5}
  \begin{fmfgraph*}(20,7)
  \fmfleft{i,j}
  \fmfright{k,l}
  \fmf{plain}{i,v1}
  \fmf{phantom}{v1,k}
  \fmf{plain}{v2,l}
  \fmf{phantom}{j,v2}
  \fmfv{decor.shape=circle,decor.filled=empty,decor.size=.30w,label=$c$,label.dist=-1.2mm}{v1}
  \fmfv{decor.shape=circle,decor.filled=empty,decor.size=.30w,label=$c$,label.dist=-1.2mm}{v2}
  \end{fmfgraph*}
  \end{fmffile} } \nonumber\\
  &&\nonumber\\
  &&\nonumber\\
    \parbox{35mm}{ \begin{fmffile}{a6}
  \begin{fmfgraph*}(20,10)
  \fmfleft{i,j}
  \fmfright{k}
  \fmf{plain}{i,m}
  \fmf{plain}{j,m}
  \fmf{plain}{m,k}
  \fmfblob{.30w}{m}
  \end{fmfgraph*}
  \end{fmffile} } &=&  \parbox{25mm}{ \begin{fmffile}{a7}
  \begin{fmfgraph*}(20,10)
  \fmfleft{i,j}
  \fmfright{k}
  \fmf{plain}{i,m}
  \fmf{plain}{j,m}
  \fmf{plain}{m,k}
  \fmfv{decor.shape=circle,decor.filled=empty,decor.size=.30w,label=$c$,label.dist=-1.2mm}{m}
  \end{fmfgraph*}
  \end{fmffile} }
  +  3 \parbox{35mm}{ \begin{fmffile}{a8}
  \begin{fmfgraph*}(20,10)
  \fmfleft{i,j}
  \fmfright{k,l}
  \fmf{plain}{i,k}
  \fmf{plain}{j,m}
  \fmf{plain}{m,l}
  \fmfv{decor.shape=circle,decor.filled=empty,decor.size=.30w,label=$c$,label.dist=-1.2mm}{k}
  \fmfv{decor.shape=circle,decor.filled=empty,decor.size=.30w,label=$c$,label.dist=-1.2mm}{m}
  \end{fmfgraph*}
  \end{fmffile} }
  +  \parbox{35mm}{ \begin{fmffile}{a9}
  \begin{fmfgraph*}(20,15)
  \fmfleft{i,j,k}
  \fmfright{o,p,q}
  \fmf{plain}{i,x}
  \fmf{phantom}{x,o}
  \fmf{plain}{y,p}
  \fmf{phantom}{j,y}
  \fmf{plain}{k,z}
  \fmf{phantom}{z,q}
  \fmfv{decor.shape=circle,decor.filled=empty,decor.size=.30w,label=$c$,label.dist=-1.2mm}{x}
  \fmfv{decor.shape=circle,decor.filled=empty,decor.size=.30w,label=$c$,label.dist=-1.2mm}{y}
  \fmfv{decor.shape=circle,decor.filled=empty,decor.size=.30w,label=$c$,label.dist=-1.2mm}{z}
  \end{fmfgraph*}
  \end{fmffile} } \nonumber\\
   &\vdots&
\end{eqnarray}
whereby the $c$ stands for the connected diagrams. This expansion can be found back if we rewrite \eqref{exp}:
{\allowdisplaybreaks
\begin{eqnarray}
Z(J) &=& \e^{-\frac{1}{\hbar} \sum_{N= 1}^{\infty} \frac{(-1/\hbar)^{N-1}}{N!} \int \d x_1  \int \d x_2 \ldots \int \d x_N J^{i_1}(x_1) \ldots J^{i_N}(x_N) \Braket{\phi_{i_1}(x_1) \ldots \phi_{i_N}(x_N)  }_c} \nonumber\\
&=& 1 +  \left(\frac{-1}{\hbar}\right) \Biggl[ \int \d x_1   J^{i_1}(x_1)  \Braket{ \phi_{i_1}(x_1) }_c \nonumber\\
&&  +  \frac{1}{2!}\left(\frac{-1}{\hbar}\right) \int \d x_1 \d x_2  J^{i_1}(x_1) J^{i_2}(x_2)  \Braket{ \phi_{i_1}(x_1) \phi_{i_2}(x_2) }_c   \nonumber\\
 && +  \frac{1}{3!} \left(\frac{-1}{\hbar}\right)^2 \int \d x_1 \d x_2 x_3 J^{i_1}(x_1) J^{i_2}(x_2) J^{i_3}(x_3) \Braket{\phi_{i_1}(x_1) \phi_{i_2}(x_2) \phi_{i_3}(x_3) }_c  + \ldots \Biggr] \nonumber\\
 && + \frac{1}{2!} \left(\frac{-1}{\hbar}\right)^2 \Biggl[ \int \d x_1   J^{i_1}(x_1)  \Braket{ \phi_{i_1}(x_1) }_c \int \d x_2   J^{i_2}(x_2)  \Braket{ \phi_{i_2}(x_2) }_c  \nonumber\\
 && +    \frac{-1}{\hbar}  \int \d x_1   J^{i'_1}(x_1)  \Braket{ \phi_{i'_1}(x_1) }_c   \int \d x_1 \d x_2  J^{i_1}(x_1) J^{i_2}(x_2)  \Braket{ \phi_{i_1}(x_1) \phi_{i_2}(x_2) }_c  + \ldots \Biggr]  \nonumber\\
 && + \frac{1}{3!} \left(\frac{-1}{\hbar}\right)^3 \Biggl[ \int \d x_1   J^{i_1}(x_1)  \Braket{ \phi_{i_1}(x_1) }_c \int \d x_2   J^{i_2}(x_2)  \braket{ \phi_{i_2}(x_2) }_c \nonumber\\
 && \times \int \d x_3   J^{i_3}(x_3)  \braket{ \phi_{i_3}(x_3) }_c + \ldots \Biggr]\;.
\end{eqnarray}}

\noindent To guide the eye, we explicitly write the expansion \eqref{expZ} up to third order
\begin{eqnarray*}
Z(J) &=& 1 +  \frac{-1}{\hbar} \int \d x_1  J^{i_1}(x_1)  \Braket{ \phi_{i_1}(x_1)   } \nonumber\\
&&+   \frac{1}{2!} \left(\frac{-1}{\hbar}\right)^2  \int \d x_1  \d x_2 J^{i_1}(x_1) J^{i_2}(x_2) \Braket{ \phi_{i_1}(x_1)  \phi_{i_2}(x_2)  } \nonumber\\
&& + \frac{1}{3!} \left(\frac{-1}{\hbar}\right)^3  \int \d x_1  \d x_2 \d x_3  J^{i_1}(x_1) J^{i_2}(x_2)J^{i_3}(x_3)  \Braket{ \phi_{i_1}(x_1)  \phi_{i_2}(x_2) \phi_{i_3}(x_3) }\;.
\end{eqnarray*}
Comparing equal orders in $\hbar$, we indeed nicely recover the relations \eqref{greenconn}.\\
\\
Let us have a look at the meaning of two point functions. Starting from \eqref{Zc}, we obtain
\begin{eqnarray}\label{two point}
 \frac{  \delta^2 Z^c(J) }{ \delta J_j(x) \delta J_k(y)} &=& \braket{  \phi_j(x) \phi_k(y)}_c\;,
\end{eqnarray}
which represents the exact propagator of the fields $ \phi_j(x)$ and $\phi_k(y)$.

\subsection{The generating functional $\Gamma$ of the 1PI Green's functions}
Next, we shall discuss the generating functional $\Gamma[\phi^\cl]$, which is also called the effective action. The effective action is of uttermost importance for the concept of algebraic renormalization. Not only is $\Gamma[\phi^\cl]$ the generator of the 1PI (one particle irreducible) Green's functions, i.e.~Green's functions with amputated external legs, we shall also prove that $\Gamma[\phi^\cl]$ reduces to the classical action $S$ at lowest order. The effective action is also of paramount importance as it contains the information regarding the vacuum expectation values of the fields.\\
\\
Let us define the effective action. First, we need to introduce the quantity $\phi^{\cl}_{i}$, called the ``generalized'' classical field, by
\begin{eqnarray}\label{defphicl}
\phi^{\cl}_{i}(x)  = \frac{\delta Z^c(J)}{\delta J_i(x)} &=&  \frac{ \int [\d \phi] \phi_i(x) \e^{-\frac{1}{\hbar}\left( S(\phi_i) + \int \d x J^i (x) \phi_i (x) \right)}   }{Z (J)} \;,
\end{eqnarray}
which is equal to the vacuum expectation value of $\phi_i$ in the presence of external sources $J_i$ and therefore depends\footnote{In order to not overload the notation, we did not explicitly write this dependence: $\phi^{\cl}_{i}(x)$ should be written as $\phi^{\cl}_{i}(x,J)$.}  on $J_i$. The classical field should certainly not be confused with the dummy integration variable $\phi$, and can be interpreted as the real classical field in the presence of a source $J$. With this classical field we can define the effective action as the Legendre transform of $Z^c (J)$,
\begin{eqnarray}\label{legendre}
\Gamma (\phi^{\cl}) &=& Z^c(J) - \int \d x J^i (x) \phi^\cl_i (x) \;.
\end{eqnarray}
The definition is simple, however, one should be careful about its meaning. The effective action defines a functional of $\phi^\cl (x)$ through the implicit dependence of $J^i$ on $\phi^\cl_i$.\\
\\
A feature of the effective action is that the functional derivative w.r.t.~$\phi^\cl$ becomes nice and simple:
\begin{eqnarray}\label{eff}
\frac{\delta \Gamma (\phi^{\cl})}{\delta \phi^\cl_i (z)} &=& \int \d y \underbrace{\frac{\delta Z^c(J) }{\delta J^i(y)}}_{= \phi^{\cl}_{i}(y)} \frac{\delta J^i(y)}{ \delta \phi^\cl_i (z)} - \int \d x \frac{\delta J^i (x)}{\delta \phi^\cl_i (z) } \phi^\cl_i (x) - J_i(z) \nonumber\\
&=& - J_i(z) \;.
\end{eqnarray}
Therefore, if we set the external sources equal to zero, we find that
\begin{eqnarray}
 \left. \frac{\delta \Gamma (\phi^{\cl})}{\delta \phi^\cl_i (z)} \right|_{J=0}&=& 0\;,
\end{eqnarray}
and thus the solutions to this equation provides us the vacuum expectation values of the fields $\phi_i$, i.e.~$\braket{\phi_i (x)}$. This is a very valuable relation, the main difficulty relies of course in the evaluation of $\Gamma(\phi^\cl)$.\\
\\
Let us now investigate the relation between $\Gamma$ and the Green's functions. Firstly, we start again from relation \eqref{eff}. We immediately find that
\begin{eqnarray}
\frac{\delta}{\delta J_j(x)} \frac{\delta \Gamma (\phi^{\cl})}{\delta \phi^\cl_i (z)} &=& - \delta^{ij}\delta (z-x) \;.
\end{eqnarray}
On the other hand, by implementing the chain rule, we can also write the left hand side of this equation as
\begin{eqnarray}
\frac{\delta}{\delta J_j(x)} \frac{\delta \Gamma (\phi^{\cl})}{\delta \phi^\cl_i (z)} &=& \int \d y  \frac{\delta \phi_k^\cl (y)}{ \delta J_j(x)} \frac{\delta^2 \Gamma (\phi^\cl)}{ \delta \phi^\cl_k (y) \delta \phi^\cl_i (z)}\;.
\end{eqnarray}
Now inserting the definition \eqref{defphicl} we obtain
\begin{eqnarray}
\frac{\delta}{\delta J_j(x)} \frac{\delta \Gamma (\phi^{\cl})}{\delta \phi^\cl_i (z)} &=& \int \d y  \frac{  \delta^2 Z^c(J) }{ \delta J_j(x) \delta J_k(y)} \frac{\delta^2 \Gamma (\phi^\cl)}{\delta \phi^\cl_k (y) \delta \phi^\cl_i (z)} \;.
\end{eqnarray}
Invoking equation \eqref{two point}, we reveal the following interesting relation
\begin{eqnarray}\label{inverse}
 \int \d y   \braket{ \phi_j(x) \phi_k(y)}_c  \frac{\delta^2 \Gamma (\phi^\cl)}{\delta \phi^\cl_k (y) \delta \phi^\cl_i (z)}  &=& - \delta^{ij}\delta (z-x)\;,
\end{eqnarray}
which represents the fact that two infinite dimensional matrices are inverses of each other. This means that $\frac{\delta^2 \Gamma (\phi^\cl)}{\delta \phi^\cl_i (x) \delta \phi^\cl_j (y)}$ represents the inverse of the connected two point function or equivalently the exact propagator of the fields $\phi_i (x)$ and $\phi_j (y)$. This is an important result as this means that the masses of the particles are encoded in the second derivatives of the effective action: the roots of the 1PI propagator correspond to the poles of the connected propagator. Next, we can investigate higher order derivatives of the effective action. Let us rewrite \eqref{inverse} in a different notation,
\begin{eqnarray}
\frac{  \delta^2 Z^c(J) }{ \delta J_i(x) \delta J_j(y)} &=& -\left( \frac{\delta^2 \Gamma (\phi^\cl)}{\delta \phi^\cl_i (x) \delta \phi^\cl_j (y)} \right)^{-1}\;,
\end{eqnarray}
and differentiate w.r.t.~$\frac{\delta }{\delta J_k (z)}$
\begin{eqnarray}
\frac{  \delta^3 Z^c(J) }{ \delta J_i(x) \delta J_j(y) \delta J_k (z)} &=& -\frac{\delta }{ \delta J_k (z)} \left( \frac{\delta^2 \Gamma (\phi^\cl)}{\delta \phi^\cl_i (x) \delta \phi^\cl_j (y)} \right)^{-1}\;.
\end{eqnarray}
We apply again the chain rule,
\begin{eqnarray}
\frac{  \delta^3 Z^c(J) }{ \delta J_i(x) \delta J_j(y) \delta J_k (z)} &=& -\int \d v  \frac{\delta \phi^\cl_\ell (v) }{ \delta J_k (z)} \frac{\delta}{\delta \phi^\cl_\ell(v) } \left( \frac{\delta^2 \Gamma (\phi^\cl)}{\delta \phi^\cl_i (x) \delta \phi^\cl_j (y)} \right)^{-1}\;,
\end{eqnarray}
and the standard rule for differentiating the inverse of a matrix, $\frac{\p }{\p x} M^{-1}(x) = - M^{-1} \frac{\p M}{\p x} M^{-1}$,
\begin{align}\label{1pi}
&\frac{  \delta^3 Z^c(J) }{ \delta J_i(x) \delta J_j(y) \delta J_k (z)} \nonumber\\
&= \int \d w  \frac{\delta \phi^\cl_\ell (w) }{ \delta J_k (z)} \int \d u \d v \left( \frac{\delta^2 \Gamma (\phi^\cl)}{\delta \phi^\cl_i (x) \delta \phi^\cl_m (u)}  \right)^{-1} \left( \frac{\delta^3 \Gamma (\phi^\cl)}{ \delta \phi^\cl_\ell(w) \delta \phi^\cl_m (u) \delta \phi^\cl_n (v)}  \right)  \left( \frac{\delta^2 \Gamma (\phi^\cl)}{\delta \phi^\cl_n (v) \delta \phi^\cl_j (y)}  \right)^{-1} \nonumber\\
&= \int \d w  \d u \d v   \braket{ \phi_\ell (w) \phi_k (z) }_c \braket{  \phi_i (x) \phi_m (u)}_c  \braket{ \phi_j (y) \phi_n (v) }_c   \underline{\left( \frac{\delta^3 \Gamma (\phi^\cl)}{ \delta \phi^\cl_\ell(w) \delta \phi^\cl_m (u) \delta \phi^\cl_n (v)}  \right)} \;.\nonumber\\
\end{align}
This expression can be better understood  diagrammatically,\\
\begin{eqnarray}
 \parbox{35mm}{ \begin{fmffile}{b2}
  \begin{fmfgraph*}(40,40)
  \fmftop{i}
  \fmfbottom{j,k}
  \fmf{plain}{i,v1}
  \fmf{plain}{j,v2}
  \fmf{plain}{k,v3}
  \fmf{plain}{v,v1}
  \fmf{plain}{v,v2}
  \fmf{plain}{v,v3}
  \fmfdot{i,j,k}
  \fmflabel{$x_1$}{i}
  \fmflabel{$x_2$}{j}
  \fmflabel{$x_3$}{k}
  \fmfv{decor.shape=circle,decor.filled=empty,decor.size=.30w}{v}
  \fmfv{decor.shape=circle,decor.filled=10,decor.size=.10w}{v1}
    \fmfv{decor.shape=circle,decor.filled=10,decor.size=.10w}{v2}
      \fmfv{decor.shape=circle,decor.filled=10,decor.size=.10w}{v3}
  \end{fmfgraph*}
  \end{fmffile} }
 \quad &=& \quad \parbox{35mm}{ \begin{fmffile}{b1}
  \begin{fmfgraph*}(40,40)
  \fmftop{i}
  \fmfbottom{j,k}
  \fmf{plain}{i,v}
  \fmf{plain}{j,v}
  \fmf{plain}{k,v}
  \fmfdot{i,j,k}
  \fmflabel{$x_1$}{i}
  \fmflabel{$x_2$}{j}
  \fmflabel{$x_3$}{k}
  \fmfv{decor.shape=circle,decor.filled=10,decor.size=.40w}{v}
  \end{fmfgraph*}
  \end{fmffile} }
\end{eqnarray}
whereby the left hand side of this figure represents the 1PI three point Green's function, where at each external leg a fully dressed propagator is attached. This operation reconstructs the connected three point function, as expressed in fact by equation \eqref{1pi}. We observe that the connected three point Green's function is split in different parts: three legs which contain the full propagators and a remaining part. This remaining part shall be always connected in the sense that one cannot divide the diagram in two separate pieces just by cutting one line, i.e.~the white blob represents a 1PI diagram. Therefore the underlined part in expression \eqref{1pi} is the 1PI three point Green's function and the effective action has generated this Green's function, i.e.
\begin{eqnarray}
\frac{\delta^3 \Gamma (\phi^\cl)}{ \delta \phi^\cl_\ell(v) \delta \phi^\cl_m (u) \delta \phi^\cl_n (v)}  &=& \braket{ \phi_\ell(v) \phi_m (u) \phi_n (v)}_{1PI} \;.
\end{eqnarray}
One can also calculate higher order derivatives of $Z^c(J)$. These calculations become more complicated, though finally result in the simple expression
\begin{eqnarray}
\frac{\delta^n \Gamma (\phi^\cl)}{ \delta \phi^\cl_{i_1}(x_1) \delta \phi^\cl_{i_2} (x_2) \ldots \delta \phi^\cl_{i_n} (x_n)}  &=& \braket{ \phi_{i_1}(x_1) \phi_{i_2} (x_2)  \ldots \phi_{i_n} (x_n) }_{1PI}\;.
\end{eqnarray}
This is a very important result as it shows that the effective action contains the complete set of physical predictions of a quantum field theory. Indeed, as is well known, 1PI Green's functions are related to the $S$ matrix elements, which contain the information of scattering processes, as established by the LSZ reduction formulas \cite{Lehmann:1954rq}. \\
\\
As a final property, let us show that we can loop expand the effective action with the zeroth order equal to the classical action. We start with expression \eqref{exp} of the generating functional $Z^c$ and substitute the expression of $Z(J)$ from \eqref{genZ},
\begin{eqnarray}\label{exp2}
\e^{-\frac{1}{\hbar} Z^c(J)} &=& \int [\d \phi] \e^{-\frac{1}{\hbar}\left( S(\phi_i) + \int \d x J^i (x) \phi_i (x) \right)}\;.
\end{eqnarray}
The calculation of $Z^c(J)$ shall be inspired by the method of the saddle point approximation of the path integral. For an integral of the type,
\begin{eqnarray}
I&=& \int \d x \e^{-f(x)}\;,
\end{eqnarray}
whereby we suppose that $f(x)$ is stationary at some point $x_0$, i.e.~$f'(x_0)=0$, we can Taylor expand $f(x)$ in the region near $x_0$,
\begin{eqnarray}\label{Taylor}
f(x) &=& f(x_0) + \frac{1}{2} (x-x_0)^2 f''(x_0) + \ldots\;,
\end{eqnarray}
so that the integral $I$ becomes
\begin{eqnarray}
I\approx \e^{-f(x_0)} \int \d x \e^{-\frac{1}{2} (x-x_0)^2 f''(x_0)}\;.
\end{eqnarray}
In this way, the integral has become a Gaussian integral which we can evaluate.
Let us therefore try to do the same for the right hand side of \eqref{exp2}. For this, we need to find a stationary point for $S(\phi_i) + \int \d x J^i (x) \phi_i (x)$, which is the classical field equation:
\begin{eqnarray}
\left. \frac{\delta S (\phi_i)}{\delta \phi_i} \right|_{\phi = \phi_0} + J_i &=& 0\;,
\end{eqnarray}
whereby $\phi_0$ is the classical field. At leading order, we just substitute this solution into the right hand side of \eqref{exp2},
\begin{eqnarray}
\e^{-\frac{1}{\hbar} Z^c(J)} &=& \int [\d \phi] \e^{-\frac{1}{\hbar}\left( S(\phi_{i,0}) + \int \d x J^i (x) \phi_{i,0} (x)  + O(\hbar)\right)}\;,
\end{eqnarray}
and thus
\begin{eqnarray}
Z^c(J) &=& S(\phi_{i,0}) + \int \d x J^i (x) \phi_{i,0} (x)\;.
\end{eqnarray}
Let us now calculate the  classical field $\phi^\cl$, see expression \eqref{defphicl},
\begin{eqnarray}
\phi^{\cl}_{i}(x)  = \frac{\delta Z^c(J)}{\delta J_i(x)} &=&  \phi_{i,0}(x)\;,
\end{eqnarray}
so we find that in leading order $\phi^{\cl}_{i}(x) =  \phi_{i,0}(x)$ and thus
\begin{eqnarray}
Z^c(J) &=& S(\phi_{i}^\cl) + \int \d x J^i (x) \phi_{i}^\cl (x)\;.
\end{eqnarray}
After performing the Legendre transformation \eqref{legendre} we find,
\begin{eqnarray}\label{seriesgamma}
\Gamma (\phi^{\cl}) &=&  S(\phi_{i}^\cl) \;,
\end{eqnarray}
which is what we wanted to show. In fact, this is perfectly logical as the only 1PI zero-loop graphs are the ones corresponding to the classical action. 
These vertices correspond to a term in the Lagrangian. Higher order corrections of the effective action, e.g.~$\Gamma^{(n)}$ shall correspond to $n$-loop graphs.

\section{Composite operators\label{sico}}
In an algebraic renormalization analysis, composite operators shall play an important role. Therefore, we shall elucidate the concept here. \\
\\
Let us first explain what composite operators are. Classically, it is a local polynomial of fields and derivatives at the same space time point. One should treat composite operators as separate objects when investigating the renormalization of a theory, and not just as the product of fields, as they can induce new infinities. The problem lies in the fact that the limit
\begin{eqnarray}
 \lim_{x_1 \to x_2} \Braket{\phi_{i_1} (x_1) \phi_{i_2} (x_2)} \;,
\end{eqnarray}
can become singular. Therefore, if we want to consider a composite operator in our theory, we introduce them into the classical action coupled to a source:
\begin{eqnarray}
S (\phi_i, \rho_k) &=& S (\phi_i) +  \int \d x \rho_k Q^k \;,
\end{eqnarray}
with $Q^k$ the composite operators, and $\rho_k$ the appropriate sources.\\
\\
Now we can generalize the definitions of the generating functionals. Firstly, in order to construct Green's function of the type
\begin{multline}
 \Braket{ \phi_{i_1} (x_1) \ldots \phi_{i_N} (x_N) Q_{k_1} (y_1) \ldots  Q_{k_m} (y_m)  }\\
 = \int [\d \phi]  \phi_{i_1} (x_1) \ldots \phi_{i_N} (x_N) Q_{k_1} (y_1) \ldots  Q_{k_m} (y_m)  \e^{-\frac{1}{\hbar}S(\phi_i) }\;,
\end{multline}
the expression \eqref{genZ} becomes
\begin{eqnarray}
Z(J, \rho) &=& \mathcal N \int [\d \phi] \e^{-\frac{1}{\hbar}\left( S(\phi_i) + \int d x \rho_k Q^k + \int \d x J^i (x) \phi_i (x) \right)}\;,
\end{eqnarray}
so we can generate the Green's function with the insertion of composite operators,
\begin{multline}
\left. \frac{\delta^{n + m} Z(J)}{\delta J_{i_1}(x_1) \ldots \delta J_{i_n} (x_n) \delta \rho_{k_1} (y_1) \ldots \rho_{k_m} (y_m) } \right|_{J_i, \rho_i = 0} \\
= \Braket{T \phi_{i_1} (x_1) \ldots \phi_{i_N} (x_N) Q_{k_1} (y_1) \ldots  Q_{k_m} (y_m)  }\;.
\end{multline}
The related Feynman graphs shall now contain new vertices corresponding to the insertion of the field polynomials $Q^k$. Secondly, we generalize expression \eqref{Zc}
\begin{eqnarray}
Z(J,\rho) &=& \e^{-\frac{1}{\hbar} Z^c (J, \rho)}\;,
\end{eqnarray}
and finally, the effective action becomes
\begin{eqnarray}
\Gamma (\phi^{\cl}, \rho) &=& Z^c(J, \rho) - \int \d x J^i (x) \phi^\cl_i (x)\;,
\end{eqnarray}
with $\phi^\cl$ still defined as in \eqref{defphicl}. $\Gamma (\phi^{\cl}, \rho)$ is thus the generating functional of 1PI Green's functions,
\begin{multline}
\left. \frac{\delta^n \Gamma (\phi^\cl)}{ \delta \phi^\cl_{i_1}(x_1)  \ldots \delta \phi^\cl_{i_n} (x_n) \delta \rho_{k_1} (y_1) \ldots \delta \rho_{k_m} (y_m)} \right|_{\rho =0}\\ = \braket{ \phi_{i_1}(x_1) \ldots \phi_{i_n} (x_n) Q^{k_1}(y_1) \ldots Q^{k_m} (y_m) }_{1PI}\;.
\end{multline}
We can also formulate the previous generating functionals in a slightly different way. The following object
\begin{eqnarray}
\left. \frac{\delta Z(J, \rho)}{\delta \rho_k (y) } \right|_{ \rho = 0} &:=& Q^k(y) \cdot Z(J)\;,
\end{eqnarray}
is the generating functional of Green's functions with the insertion of the composite operator $Q^k (y)$. Analogously,
\begin{eqnarray}
\left. \frac{\delta Z^c(J, \rho)}{\delta \rho_k (y) } \right|_{ \rho = 0} &:=& Q^k(y) \cdot Z^c(J)\;,
\end{eqnarray}
generates the connected Green's functions and
\begin{eqnarray}\label{genZ2}
\left. \frac{\delta \Gamma(\phi^\cl, \rho)}{\delta \rho_k (y) } \right|_{ \rho = 0} &:=& Q^k(y) \cdot \Gamma(\phi^\cl)\;,
\end{eqnarray}
the connected 1PI Green's function, both with the insertion of the composite operator. As at the lowest order, $ \Gamma(\phi^\cl, \rho)$ is equal to the classical action $S (\phi_i, \rho_k)$, see expression \eqref{seriesgamma}, we can expand expression \eqref{genZ2},
\begin{eqnarray}\label{seriesinsertion}
\left. \frac{\delta \Gamma(\phi^\cl, \rho)}{\delta \rho_k (y) } \right|_{ \rho = 0} &=& Q^k(y) + O(\hbar)\;,
\end{eqnarray}
whereby $Q^k(y)$ is the classical composite operator.

\section{The QAP, the key to algebraic renormalization}
In this section, we shall explain the procedure of algebraic renormalization which is based on the quantum action principle (QAP). Basically, algebraic renormalization is based on the concept of symmetries. The whole idea is that a Lagrangian and its quantum corrections are completely determined by the underlying symmetry content. From this viewpoint, the symmetry content of a given quantum field theory model plays a fundamental role, as both the classical  Lagrangian and the corresponding Feynman rules are determined by its symmetries. One should therefore ask what happens with the classical symmetries at the quantum level, in order to construct a meaningful consistent perturbation theory at higher orders. This is the scope of this section.

\subsection{The idea of symmetries \label{The idea of symmetries}}
The first step in the process of algebraic renormalization is the detection of the symmetries of the classical action. Let us therefore start with the classical action again, $S(\phi_i)$. This action has a continuous symmetry if the following infinitesimal variation of the fields
\begin{eqnarray} \label{symmetry}
\delta \phi_i &=& P_i (\phi)\;,
\end{eqnarray}
whereby $P_i$ contains an infinitesimal parameter, leaves the action invariant,
\begin{eqnarray}
\delta S (\phi_i) &=& 0\;.
\end{eqnarray}
We assume the functions $P_i$ to be a local polynomial in the fields and their derivatives. All the symmetries we shall encounter shall belong to a representation of a Lie group $G$. If $X_a$ are the generators of the Lie algebra $G$, they obey the following commutation relation
\begin{eqnarray}\label{Lie}
[X_a ,X_b] &=& \ii f_{abc} X_c \;,
\end{eqnarray}
with $f_{abc}$ the structure constants of the Lie group. Next to discrete symmetries, we can distinguish between two different kinds of continuous symmetries, linear and non linear symmetries, which we shall discuss below. Linearly broken symmetries shall be discussed too, as they are also allowed according to the Quantum Action Principle.\\
\\
The symmetries can also be divided into local and global symmetries. In the case of a global symmetry, we can write
\begin{eqnarray}\label{sivar}
P_i(\phi) &=&\ii \varepsilon^a P_i^a (\phi)\;,
\end{eqnarray}
with $\varepsilon^a$ a constant infinitesimal parameter. In the case of a local symmetry, $\varepsilon^a$ becomes dependent on the space time coordinates.

\subsubsection{Linear symmetry}
The symmetry \eqref{symmetry} is linear if
\begin{eqnarray}\label{si8}
P_i (\phi) &=&a_{ij} \phi_j\;,
\end{eqnarray}
with $a_{ij}$ quantities independent of the fields. Now consider a Green's function
\begin{eqnarray}
\braket{ \phi^{i_1} \phi^{i_2} \ldots \phi^{i_n} } &=& \int [\d \phi]  \phi_{i_1} (x_1) \phi_{i_2} (x_2) \ldots \phi_{i_n} (x_n)  \e^{-\frac{1}{\hbar }S(\phi_i) } \;,
\end{eqnarray}
as the classical action $S(\phi_i)$ is invariant under the symmetry transformation, we easily find that
\begin{equation}\label{Wasi}
\delta \braket{ \phi^{i_1} \phi^{i_2} \ldots \phi^{i_n} } = \sum_{\ell =1}^n      \braket{\phi^{i_1} (x_1) \ldots \phi^{i_{\ell-1}} (x_{ \ell -1})  a_{i_\ell j} \phi^{j} (x_\ell) \phi^{i_{\ell + 1}} (x_{\ell + 1}) \ldots \phi^{i_n} (x_n) } \;,
\end{equation}
a sum of elementary Green's functions.\\
\\
The invariance of the action can be written in functional form as follows,
\begin{align}\label{W1}
\mathcal W S &= 0\;, &  \mathcal W &= \int \d x P_i (\phi(x)) \frac{\delta}{\delta \phi_i (x)} \;,
\end{align}
where $\mathcal W$ is called the Ward operator associated to the invariance \eqref{si8}. In the case of a global symmetry, the Ward identities is an integrated identity. However, when dealing with local symmetries, $\varepsilon$ is a function of the space time coordinate $x$, see equation \eqref{sivar}. Therefore, due to the linear independence, the Ward identity shall be a non integrated identity. Non integrated identities are usually more powerful then integrated identities.

\subsubsection{Non-linear symmetry }
More generally, we assume a symmetry which is non linear in the fields. As an example, let us assume a symmetry which is quadratic in the fields,
\begin{eqnarray}\label{symmetry2}
P_i(\phi) &=& b_{ijk} \phi_j (x) \phi_k(x) \;,
\end{eqnarray}
We shall encounter this kind of symmetry typically when dealing with the BRST symmetry of non abelian gauge theories. What we now see is that the variation of a Green's function under this symmetry does not result in a sum of elementary Green's functions,
\begin{equation}
\delta \braket{ \phi^{i_1} \phi^{i_2} \ldots \phi^{i_n} } =  \sum_{\ell =1}^n  \braket{\phi^{i_1} (x_1) \ldots \phi^{i_{\ell-1}} (x_{ \ell -1})  b_{i_\ell j k} \underline{\phi^{j} (x_\ell) \phi^{k} (x_\ell)} \phi^{i_{\ell + 1}} (x_{\ell + 1}) \ldots \phi^{i_n} (x_n) }\;.
\end{equation}
What we find is the insertion of the composite operators \eqref{symmetry2} into the Green's functions. It is therefore necessary, in the case of non linear symmetries, to explicitly add the corresponding composite operator \eqref{symmetry2} to the starting action as they will be needed for the renormalization process. The classical action shall therefore become
\begin{eqnarray}
\Sigma &=& S(\phi) + S_{\ext}\;,
\end{eqnarray}
with
\begin{eqnarray}
S_{\ext} &=& \int \d x \rho_i P_i (\phi)\;,
\end{eqnarray}
whereby $\rho_i$ is a source coupled to the composite operator, see section \ref{sico}. Of course, one wishes to retain the symmetry \eqref{symmetry2} for the new action $\Sigma$. We should then check if $S_{\ext}$ is left invariant\footnote{We define $\delta \rho^i =0$. }. If not, i.e.~ if $\delta P_i (\phi) = Q_i \not= 0$, we need to introduce another source coupled to $Q$.  In this case, the action becomes larger again
\begin{eqnarray}
\Sigma' &=& \Sigma + S_{\ext,2}\;,
\end{eqnarray}
with
\begin{eqnarray}
S_{\ext,2} &=& \int \d x R_i Q_i \;.
\end{eqnarray}
If $\delta Q^i$ is also nonzero, we can continue this process until we find a variation which is zero.
However, it is also possible that this process is infinite. In such cases, to avoid the infinite number of sources, the best way to proceed in through the introduction of a nilpotent BRST operator encoding the information of the Lie algebra of the system, see \cite{Piguet1995}. \\
\\
Now suppose $ \delta Q_i = 0$. We can write the symmetry \eqref{symmetry2} in functional form as follows
\begin{align}\label{W2}
\mathcal W (S) = \int \d x \frac{\delta S}{\delta \rho_i} \frac{\delta S}{\delta \phi_i} = 0\;.
\end{align}
This way of writing shall turn out useful later on.

\subsubsection{Linearly broken symmetries}
Other symmetries which shall play an important role in the process of algebraic renormalization are the linearly broken symmetries. Moreover, we can generalize expression \eqref{W1} and \eqref{W2} to write down the most general Ward identity compatible with the Quantum Action Principle. Assume we have a classical action $S$ which also depends -besides on the fields $\phi_i$ and the sources $\rho_i$- on a set of parameters\footnote{$\lambda_i$ can be e.g.~the coupling constant, the masses, \ldots.} $\lambda_i$. The most general Ward identity is then given by,
\begin{align}\label{mostgen}
\mathcal W S (\lambda_i, \phi_i, \rho_i) &= \Theta \;,
\end{align}
whereby $\mathcal W$ and the classical breaking $ \Theta$ are given by
\begin{eqnarray}
\mathcal W &=& \omega_i \frac{\delta }{\delta \lambda_i} + \int \d x \left( \sigma_i \frac{ \delta}{\delta \phi_i} + T_{ij} \phi^j \frac{\delta}{ \delta \phi_i} + \frac{\delta S}{\delta \rho_i }  \frac{\delta}{\delta \phi_i }  \right)\;,\nonumber\\
\Theta &=& \int \d x \left( \alpha_i \phi_i + \beta_{ij} \rho_i \phi_j \right)\;.
\end{eqnarray}
In the expression $\omega_j$ are constants, while $\sigma_i$, $T_{ij}$, $\alpha_i$ and $\beta_{ij}$  are quantities independent of the fields, but they can be functions of partial derivatives. In the case that $\mathcal W$ is a local non integrated identity, i.e.~$\mathcal W = \mathcal W (x)$, the corresponding breaking $\Theta$ shall be a non integrated expression. Notice that in the case of a local Ward identity the global part $\omega_i \frac{\delta }{\delta \lambda_i}$ cannot appear.

\subsection{The Quantum Action Principle}
After determining the symmetries of the classical action, the main question is what happens with these symmetries at the quantum level. Here the Quantum Action Principle (QAP) comes in the game, which was worked out in the early 70's in \cite{Lowenstein1971,Lam1972,Clark1976,Lowenstein:1971jk,Lam:1973qa}. The QAP gives information about the structure of the Ward identities at the quantum level. For the QAP to be applicable to a certain theory, we always assume that this theory is local, Lorentz invariant and power-counting renormalizable. In addition, we also assume that the propagators of the theory have the following behavior in the UV,
\begin{equation}
\lim_{k \to \infty }\braket{\phi_i (k) \phi_j(-k)} \sim \frac{R(k)}{k^2}\;,
\end{equation}
with $R(k)$ a certain polynomial of the momentum and $k^2 = k^2_1+ k^2_2+ k^2_3 +k^2_4 $ as we are working in the Euclidean space time.

\subsubsection{The QAP for linear symmetries}
Let us continue with the Ward identity \eqref{W1}. The Quantum Action Principles states that the classical Ward identity \eqref{W1} becomes at the quantum level\footnote{For a nice review on the QAP, see e.g.~\cite{Haussling}.}
\begin{eqnarray}\label{QAPlinear}
\int \d x P_i(x) \frac{\delta \Gamma}{\delta \phi_i (x)} &=& \Delta \cdot \Gamma\;,
\end{eqnarray}
whereby $\Delta$ is an insertion as explained in \eqref{genZ2}. This insertion has certain properties:
\begin{itemize}
\item $\Delta(x)$ is an integrated local polynomial in the sources and fields and in their partial derivatives, $\Delta = \int \d x \Delta(x)$.
\item   $\Delta$ has a dimension which is bounded by $(d - d_i + d_{P}) $, whereby $d$ is the space-time dimension, $d_i$ is the mass dimension of the field $\phi_i$ and $d_{P}$ is the dimension of $P$.
\item  $\Delta$ has the same quantum numbers as $\mathcal W$ ( e.g.~charge conjugation, parity, global indices, etc.)
\end{itemize}

\subsubsection{The QAP for non linear symmetries}
In the case of a non linear symmetry \eqref{W2}, the QAP states that
\begin{eqnarray}
\int \d x \frac{\delta \Gamma}{\delta \rho^a_i (x)} \frac{\delta \Gamma}{\delta \phi_i (x)} &=& \Delta^{a} \cdot \Gamma\;,
\end{eqnarray}
with $\Delta^{a}$ obeying the similar properties as in the linear case.

\subsubsection{The QAP for linearly broken symmetries}
Now the most general Ward identity compatible with the QAP is given in \eqref{mostgen}.
At the quantum level, the identity \eqref{mostgen} becomes
\begin{eqnarray}\label{QAP3}
\omega^{a}_j \frac{\delta \Gamma }{\delta \lambda_i} + \int \d x \left( \sigma^a_i \frac{ \delta  \Gamma}{\delta \phi_i} + T^a_{ij} \phi^j \frac{\delta  \Gamma}{ \delta \phi_i} + \frac{\delta  \Gamma}{\delta \rho_i }  \frac{\delta  \Gamma}{\delta \phi_i }  \right) &=& \Theta^a + \Delta^{a} \cdot \Gamma \;.
\end{eqnarray}
As we have seen in expression \eqref{seriesgamma}, the effective action $\Gamma$ can be expanded in a power series of $\hbar$, recovering at zeroth order the classical action again. Therefore, we loop expand $\Gamma$
\begin{eqnarray}
\Gamma &=& \sum_{n=0}^{\infty} \hbar^n \Gamma^{(n)}\;,
\end{eqnarray}
with $\Gamma^{(0)}$ equal to the classical action $S$. Let us therefore expand expression \eqref{QAP3} in series of $\hbar$. At zeroth order we need to recover the classical Ward identity,
\begin{eqnarray}\label{QAP3b}
\omega^{a}_j \frac{\delta S }{\delta \lambda_i} + \int \d x \left( \sigma^a_i \frac{ \delta  S}{\delta \phi_i} + T^a_{ij} \phi^j \frac{\delta  S}{ \delta \phi_i} + \frac{\delta  S}{\delta \rho_i }  \frac{\delta  S}{\delta \phi_i }  \right) &=& \Theta^a  \;,
\end{eqnarray}
from which we can see that $\Delta^{a} $ can only start from order $\hbar$.

\subsubsection{An easy example}
Let us give an easy example to fix the thoughts. We shall study the quantum extension of the $U(1)$ global symmetry in a scalar model. We start with the following classical action
\begin{eqnarray}
S &=& \int \d^4 x \left( \p_\mu \overline \varphi \p_\mu \varphi + m^2 \overline \varphi \varphi + \frac{g}{4} (\overline \varphi \varphi)^2 \right)\;,
\end{eqnarray}
whereby we work in 4 space time dimensions. The fields $\phi_i$ are thus given by the complex conjugate pair $(\overline \varphi, \varphi)$ and the parameters $\lambda_i$ are $(m, g)$, where $m$ stands for the mass and $g$ is the coupling constant. One easily sees that this classical action is invariant under the following linear Ward identity
\begin{eqnarray}
\mathcal W S &=&0 \;, \nonumber\\
\mathcal W &=&  \int \d^4 x \left( \varphi \frac{\delta }{\delta \varphi} - \overline \varphi \frac{\delta }{\delta \overline \varphi}   \right)\;,
\end{eqnarray}
expressing in functional form the $U(1)$ global invariance:
\begin{eqnarray}
\delta \varphi &=& \ii \alpha \varphi \;,\nonumber\\
\delta \overline \varphi &=& -\ii \alpha \overline \varphi \;,
\end{eqnarray}
whereby $\alpha$ is constant parameter. In addition, the action is invariant under the following discrete symmetry $\kappa$
\begin{align}\label{discrete}
\varphi &\to \overline \varphi  &  \overline\varphi &\to  \varphi \;,
\end{align}
while $\mathcal W$ is odd under this symmetry $\kappa$
\begin{eqnarray}
\kappa \mathcal W &=& - \mathcal W\;.
\end{eqnarray}
Now we can apply the QAP, implying that
\begin{eqnarray}
\mathcal W \Gamma &=& \Delta \cdot \Gamma \;.
\end{eqnarray}
We can prove that $\mathcal W \Gamma  = 0$ by a recursive argument. At one loop,
\begin{eqnarray}
\hbar \mathcal W \Gamma^{(1)} &=& \hbar \Delta  + O(\hbar^2) \;\;.
\end{eqnarray}
whereby we have made explicit that $\Delta$ can only start from order $\hbar$, see expression \eqref{QAP3b}. $\Delta$ is an integrated polynomial, with dimension bounded by four and is odd under the discrete symmetry \eqref{discrete}, as described below equation \eqref{QAPlinear}. We can therefore parameterize $ \Delta $ as
\begin{eqnarray}
\Delta &=&\int \d^4 x \Bigl[ a_1 \left( \varphi^2 - \overline \varphi^2 \right) + a_2  \left( \varphi^3 \overline \varphi - \overline \varphi^3 \varphi \right) + a_3   \left( \p \varphi \p \varphi - \p \overline \varphi \p \overline \varphi  \right) + a_4 \left( \varphi^4 - \overline \varphi^4 \right) \Bigr] \nonumber\\
&=& \mathcal W  \Bigl[ a_1 \left( \varphi^2 + \overline \varphi^2 \right) + a_2  \left( \varphi^3 \overline \varphi + \overline \varphi^3 \varphi \right) + a_3   \left( \p \varphi \p \varphi + \p \overline \varphi \p \overline \varphi  \right) + a_4 \left( \varphi^4 + \overline \varphi^4 \right) \Bigr]\nonumber\\
&=& \mathcal W  \sum_{i=1}^4 a_i \widetilde \Delta_i\;,
\end{eqnarray}
with $a_i$ arbitrary coefficients. Now we can replace $\Gamma^{(1)}$ by
\begin{eqnarray}\label{a2}
\overline \Gamma^{(1)} &=& \Gamma^{(1)} -  \sum_{i=1}^4 a_i \widetilde \Delta_i\;,
\end{eqnarray}
so we obtain
\begin{eqnarray}\label{QAPU1model2}
\hbar \mathcal W \overline \Gamma^{(1)} &=&    O(\hbar^2) \;\;.
\end{eqnarray}
We can repeat the procedure order by order. In this way, we have proven the effective action to obey the Ward identity also at the quantum level.


\subsection{Anomaly}
There are two different possibilities for the breaking of the Ward identity $\Delta$. Firstly, the breaking can be trivial as in the example shown previously. In this case, one can always absorb the breaking by introducing local non invariant counterterms in the effective action $\Gamma$ in order to restore the Ward identity at the quantum level. However, a second possibility is that the breaking is non trivial. In this case, there is no possibility of restoring the Ward identity by the introduction of local counterterms in the effective action. In this case, we call the Ward identity anomalous. The symmetry is thus broken by quantum corrections. Proving that the breaking $\Delta$ is non anomalous shall often rely on solving consistency conditions. Let us give an example. Suppose the linear Ward identity $\mathcal W$ in equation \eqref{W1} has a group index $a$ belonging to a Lie algebra, so we can write
\begin{eqnarray}
\mathcal W^a S = \int \d x P_i^a(\phi) \frac{\delta}{\delta \phi_i} S = 0\;.
\end{eqnarray}
If the theory is power counting renormalizable, this symmetry becomes at quantum level,
\begin{eqnarray}\label{288}
\mathcal W^a \Gamma = \Delta^a \cdot \Gamma = \hbar \Delta^a + O(\hbar^2)\;.
\end{eqnarray}
From the commutation relations we find,
\begin{eqnarray}\label{liew}
[\mathcal W^a, \mathcal W^b] &=& f_{abc} \mathcal W^c\;.
\end{eqnarray}
Applying this equation to $\Gamma$ and using equation \eqref{288}, we find at lowest order,
\begin{eqnarray}
\mathcal W^a \Delta^b - \mathcal W^b \Delta^a &=& f_{abc} \Delta^c \;,
\end{eqnarray}
which is called the Wess-Zumino consistency condition. Therefore we have proven that the commutations relations between the Ward operators $\mathcal W^a$, i.e.~expression \eqref{liew}, put a restriction on the breaking $\Delta^a$.

\subsection{Stability}
Let us now suppose that we are dealing with a Ward identity which is not anomalous. This does not yet mean that our theory is  renormalizable. For this, we have to prove the stability of the action. Let us explain this concept with the help of our easy example, the scalar $U(1)$ model. We have shown that the theory is non anomalous, therefore at first order, $\mathcal W \overline \Gamma^{(1)} =  0$. Now, we can split $\overline \Gamma^{(1)}$ into two independent parts: a finite part and a divergent part,
\begin{eqnarray}\label{finitedivergent}
\overline \Gamma^{(1)} &=& \overline \Gamma^{(1)}_\fin + \overline \Gamma^{(1)}_\dive \;,
\end{eqnarray}
where $\overline \Gamma^{(1)}_\dive$ is an integrated local functional of the fields. Due to the linearity of $\mathcal W$ we have that
\begin{eqnarray}\label{vraag}
\mathcal W \overline \Gamma^{(1)}_\dive &=&   0\;.
\end{eqnarray}
The classical action itself shall, when calculating loop diagrams, be responsible for divergences, while, after the renormalization process, the total effective action is finite, representing the physical content of the theory. Therefore, we can think of the effective action as follows
\begin{eqnarray}
\Gamma &=& S + \sum_{n=1}^{\infty}\overline \Gamma^{(n)} ~=~ S + \sum_{n=1}^{\infty}\hbar^n \overline \Gamma^{(n)}_{\fin} + \sum_{n=1}^{\infty}\hbar^n \overline \Gamma^{(n)}_\dive \;.
\end{eqnarray}
We now introduce the bare action in order to reabsorb  the divergent part of the effective action by means of a suitable redefinition of the fields and parameters of the action $S$,
\begin{eqnarray}\label{sibare}
 S(\varphi_i, \lambda_i, \rho_i) + \sum_{n=1}^{\infty} \overline \Gamma^{(n)}_{\dive}(\varphi_i, \lambda_i, \rho_i) = S (\phi_{i,0}, \lambda_{i,0}, \rho_{i, 0})  \;,
\end{eqnarray}
whereby $\phi_{i,0}$, $\lambda_{i,0}$, $\rho_{i, 0}$ are given by a suitable   redefinition:
\begin{align}
\lambda_{i,0} &= (1 + \hbar a) \lambda_i  & \rho_{i,0} &= (1 + \hbar b) \rho_i  & \phi_{i,0} &= (1 + \hbar c) \phi_i \;.
\end{align}
Equation \eqref{sibare} expresses the stability of the classical action $S$, meaning that the divergences occurring at the quantum level can be reabsorbed by the introduction of local counterterms obtained by redefining the fields and parameters of $S(\varphi_i, \lambda_i, \rho_i) $. The quantity $S (\phi_{i,0}, \lambda_{i,0}, \rho_{i, 0})$ is usually called the bare action in the literature. It is easy to see, that if the action is stable at order $\hbar$, one can prove by induction that it is stable at all orders.

\subsubsection{An easy example: continuation}
Let us now prove the action to be stable for the scalar $U(1)$ model. $\Gamma^{(1)}_\dive$ is given by
\begin{eqnarray}
\Gamma^{(1)}_\dive &=& x_1 \int \d^4 x \p\overline \varphi \p\varphi  + x_2 \int \d^4 x m^2 \overline \varphi \varphi  + x_3 \int \d^4 x \frac{g}{4} (\overline \varphi \varphi)^2\;,
\end{eqnarray}
with $x_1$, $x_2$ and $x_3$ arbitrary parameters. One can check that they are indeed the only invariant terms of dimension 4 which fulfil the Ward identity. We now need to prove that
\begin{eqnarray*}
\int \d^4 x \left( \p \overline \varphi_0 \p \varphi_0 + m_0^2 \overline \varphi_0 \varphi_0 + \frac{g_0}{4} (\overline \varphi_0 \varphi_0)^2 \right) = \int \d^4 x \left( \p\overline \varphi \p\varphi + m^2 \overline \varphi \varphi + \frac{g}{4} (\overline \varphi \varphi)^2 \right) + \hbar \Gamma^{(1)}_\dive \;,
\end{eqnarray*}
which can be done by a redefinition of the fields and parameters
\begin{align}
\overline \varphi_0 &= \left(1 + \hbar \frac{x_1}{2} \right)  \overline \varphi  &  \varphi_0 &= \left(1 + \hbar\frac{x_1}{2} \right)   \varphi  \nonumber\\
m_0^2 & = \left(1 + \hbar \left( \frac{x_2}{2} - x_1 \right)\right) m^2    &  g_0 & = \left(1 + \hbar \left(\frac{x_3}{2} - 2 x_1\right) \right) g \;.
\end{align}
We have thus proven that the $U(1)$ model is renormalizable.

\section{Intermezzo: cohomologies}\label{intermezzo:cohomolgy}
Before tackling the renormalization of the Yang-Mills action, we shall introduce some useful concepts here.

\subsection{Cohomology}
Suppose $\delta$ is a nilpotent operator, $\delta^2 = 0$. The cohomology of $\delta$ is given by the solutions of the equation
\begin{eqnarray}\label{298}
\delta \Delta &=& 0\;,
\end{eqnarray}
which cannot be written in the form
\begin{eqnarray}\label{coho}
 \Delta = \delta \Omega \;.
\end{eqnarray}
A quantity $\Delta$ obeying equation \eqref{298} is called \textit{closed}, while a quantity of the form \eqref{coho} is called \textit{exact}. The cohomology of $\Delta$ is thus identified by quantities which are closed but not exact. More precisely, a non trivial quantity $\Delta$ is always defined up to the addition of an arbitrary exact part, i.e.~one speaks of cohomology classes. In fact, take now two closed quantities $\Delta_1$ and $\Delta_2$. These quantities belong to the same cohomology class if
\begin{eqnarray}
\Delta_1 - \Delta_2 &=& \delta (\ldots) \;,
\end{eqnarray}
i.e.~when $\Delta_1$ and $\Delta_2$ differ by an exact part.\\
\\
In this way one can always write $\Delta$ obeying \eqref{298} as a sum of a trivial part and a non trivial part.
\begin{eqnarray}\label{siex}
\Delta &=& \Delta_{\mathrm{n.triv}} + \underbrace{ \delta (\ldots)}_{\Delta_{\mathrm{triv}}}\;,
\end{eqnarray}
whereby $ \Delta_{\mathrm{n.triv}}$ does not contain parts that can be written as $\delta (\ldots)$. In quantum field theory, these non trivial parts shall be the most interesting parts, as they will be related to the renormalization of the physical parameters of the theory.

\subsection{Doublet theorem}\label{doublettheorem}
Now there is a very important theorem which shall be very useful later on.  Suppose our theory contains a pair of fields, sources or parameters $(u_i,v_i)$ which form a doublet:
\begin{align}\label{uomzetten}
\delta u_i &= v_i & \delta v_i &= 0\;,
\end{align}
whereby the subscript $i$ is a certain index (e.g.~color). We assume $u_i$ to be commuting, while $v_i$ is an anticommuting quantity. Then we can prove that $u_i$ and $v_i$ shall never enter the non trivial part of the cohomology of $\delta$.\\
\\
The proof is as follows. We introduce two operators $\hat{P}$ and $\hat{A}$
\begin{eqnarray}
\hat{P} &=& \int \d x \left(  u_i \frac{\partial}{\partial u_i} + v_i \frac{\partial}{\partial v_i} \right )\nonumber \\
\hat{A} &=&  \int \d x\ \left( u_i \frac{\partial}{\partial v_i}   \right)\;,
\end{eqnarray}
Functionally, we write for the nilpotent operator $\delta$
\begin{eqnarray}
\delta &=& v_i \frac{\p }{\p u_i}\;,
\end{eqnarray}
so we obtain
\begin{align*}
\delta\  \hat{A}   = \int \d x\left( v_i \frac{\partial}{\partial v_i} + v_j u_i \frac{\partial}{\partial u_j}\frac{\partial}{\partial v_i} \right), \nonumber\\
 \hat{A}\ \delta \  = \int \d x \left( u_i \frac{\partial}{\partial u_i} -   u_i v_j \frac{\partial}{\partial v_i} \frac{\partial}{\partial u_j} \right)\;,
\end{align*}
and thus
\begin{equation} \label{antic2}
\{ \delta, \hat{A} \} = \hat{P}.
\end{equation}
Analogously we also have
\begin{eqnarray*}
 \hat{P}  \delta &=&  \int \d x \left(  u_i v_j \frac{\partial}{\partial u_i} \frac{\partial}{\partial u_j}  + v_i \frac{\partial}{\partial u_i} - v_i v_j \frac{\partial}{\partial v_i} \frac{\partial}{\partial u_j}  \right) \nonumber\\
\delta \ \hat{P}  &=&\int \d x \left(  v_j \frac{\partial}{\partial u_j} +  v_j u_i  \frac{\partial}{\partial u_j} \frac{\partial}{\partial u_i} + v_j v_i \frac{\partial}{\partial u_j} \frac{\partial}{\partial v_i} \right)\;,
\end{eqnarray*}
and thus
\begin{equation}\label{antic}
[\delta \ , \hat{P}] = 0.
\end{equation}
As $\widehat P$ is a counting operator for the total number of $u_i$ and $v_i$, we can expand\footnote{We assume $\Delta$ to be a polynomial in $u_i$ and $v_i$. } $\Delta$, see expression \eqref{siex}, in eigenvectors of $\widehat P$,
\begin{equation}
\Delta = \sum_{n \geq 0} \Delta_n\;,
\end{equation}
whereby $\widehat P  \Delta_n = n \Delta_n$ and $n$ represents the total number of $u_i$ and $v_i$ in $\Delta_n$. Now from the cohomology condition \eqref{298} and the commutation relation \eqref{antic}, we find that
\begin{equation}
0 = \sum_{n \geq 0} n \delta \Delta_n \quad \Rightarrow \quad \sum_{n \geq 1} n \delta \Delta_n = 0 \;.
\end{equation}
Looking at expression \eqref{uomzetten}, we easily obtain that 
\begin{equation}
 \delta \Delta_n = 0  \quad \forall n \geq 1 \;.
\end{equation}
Finally, using this property and invoking expression \eqref{antic2}, we obtain
\begin{eqnarray}
\Delta &=&  \Delta_0 + \sum_{n \geq 1} \frac{1}{n}  \hat{P} \Delta_n \nonumber\\
  &=&  \Delta_0  + \sum_{n \geq 1} \frac{1}{n} \delta \hat{A} \Delta_n \nonumber\\
  &=& \Delta_0 + \delta (\ldots) \;.
\end{eqnarray}
In conclusion, as $\delta^2 = 0$
\begin{equation}
\delta  \Delta = \delta \Delta_0 \;,
\end{equation}
whereby $ \Delta_0$ is independent of the doublet $(u_i,\ v_i)$. The quantities $u_i$ and $v_i$ shall thus never enter the non trivial part of the cohomology.


\section{The algebraic renormalization of the Yang-Mills action}
In this section, we shall explore the details of the algebraic renormalization of the Yang-Mills action including a matter part, see equation \eqref{ym}. We shall show that the gauge fixed action has a non linear symmetry, called the BRST symmetry, which is therefore a nice example for applying the QAP. Firstly, we shall search for all the possible Ward identities of the classical action. Next, we investigate these Ward identities at the quantum level by applying the QAP and scrutinize the possibility for an anomaly.  In the final part, we shall prove that the action is stable and therefore renormalizable.

\subsection{Gauge fixing the Yang-Mills action and looking for all the symmetries}
When trying to formulate a path integral for the Yang-Mills action \eqref{ym}, it was noticed that the path integral was ill-defined. Therefore, one needed to include a gauge fixing \cite{'tHooft:2005cq}. We shall elaborate on this in the next chapter, but here we just accept the following well defined action,
\begin{eqnarray}
S &=& S_\YM + S_\m +  S_\gf \;,
\end{eqnarray}
with $S_\gf$ the gauge fixing,
\begin{eqnarray}\label{BRs}
S_\gf &=& \int \d^4 x \left( b^a \p_\mu A_\mu^a + \alpha \frac{(b^a)^2}{2} + \overline{c}^{a}\partial _{\mu } D_{\mu}^{ab}c^b \right) \;,
\end{eqnarray}
whereby $\alpha$ is the gauge parameter, $c^a$ and $\overline c^a$ are anti-commuting fields and $b$ is a bosonic field. $b^a$ is also an auxiliary field (sometimes referred to as the Nakanishi-Lautrup field \cite{Nakishi:1966di}), as it has no interaction vertices. Therefore, one can easily integrate out this field by invoking the equations of motion of the $b^a$ field to obtain the following equivalent gauge fixing
\begin{eqnarray}\label{Yangmillsgaugefixalpha}
S_{\gf,2} &=&\int \d^4 x\left(  -\frac{1}{2\alpha}  \int \d^4 x (\p_\mu A_\mu)^2 +  \overline{c}^{a}\partial _{\mu } D_{\mu}^{ab}c^b\right)\;,
\end{eqnarray}
which is perhaps more familiar.  However, as we shall see, the form \eqref{BRs} is more suitable for proving the algebraic renormalizability of the Yang-Mills action. In the case that $\alpha = 0$, we are in the Landau gauge.\\
\\
Let us now have a look at all the symmetries and/or Ward identities of the classical action $S$. Now that we have introduced a gauge fixing, obviously, the $SU(N)$ gauge symmetry is broken. However, it was found by Becchi, Rouet and Stora \cite{Becchi:1974xu} and independently by Tyutin, that this action $S$ still enjoys a remaining symmetry, called the BRST symmetry \cite{Becchi:1996yh},
\begin{eqnarray}
s S &=& 0 \;,
\end{eqnarray}
with
\begin{align}\label{BRST}
sA_{\mu }^{a} &=-\left( D_{\mu }c\right) ^{a}\,, & sc^{a} &=\frac{1}{2}gf^{abc}c^{b}c^{c}\,,   \nonumber \\
s\overline{c}^{a} &=b^{a}\,,&   sb^{a}&=0\,, \nonumber\\
s \psi^i_\alpha &= - \ii g   c^a (X^a)^{ij} \psi^j_\alpha & s \overline \psi^i_\alpha &= - \ii g  \overline \psi^j_\alpha c^a (X^a)^{ji} \;.
\end{align}
One can check that $s$ is nilpotent,
\begin{eqnarray}
s^2 &=& 0\,.
\end{eqnarray}
a property which shall turn out to be very important\footnote{Notice that without the introduction of the $b$-fields, the BRST operator would be nilpotent only on-shell, i.e.~using the equation of motion.}. As we explained in section \ref{The idea of symmetries}, we need to couple a source to each non linear variation of the fields, i.e.~to $sA_{\mu }^{a}$, $sc^{a}$, $s \psi^i_\alpha$ and $s \overline \psi^i_\alpha$. Therefore, we add the following auxiliary part to the action,
\begin{eqnarray}
\Sigma &=& S_\YM + S_\m + S_\gf + S_\ext \\
S_\ext &=& \int  \d^4 x\left( K_{\mu }^{a} s A_\mu^a  + L^{a} s c^a + \overline Y_\alpha^i s \psi^i_\alpha + s \overline \psi^i_\alpha  Y_\alpha^i\right) \nonumber\\
 &=& \int  \d^4 x\left( - K_\mu^a D_\mu^{ab} c^b  + \frac{1}{2} g  L^{a}f^{abc}c^b c^c    - \ii g   \overline Y_\alpha^i  c^a (X^a)^{ij} \psi^j_\alpha  - \ii g  \overline \psi^j_\alpha c^a (X^a)^{ji}   Y_\alpha^i \right) \;,\nonumber
\end{eqnarray}
whereby $K_\mu^a$ is a Grassmann source, $L^{a}$ is a bosonic source and $\overline Y_\alpha^i$ and $Y_\alpha^i$ are spinor sources transforming in the same representation as $\psi$. The $s$ variation of all these sources is equal to zero. Notice that due to the nilpotency of the BRST symmetry $s$, we were able to add this part in an $s$ exact fashion. We have summarized all the quantum numbers of the fields in table \ref{tabel1} and \ref{tabel2}.
 Now that we have introduced this extra part, we summarize the Ward identities. \label{chap1wardidentities}
\begin{itemize}
\item Firstly, the Slavnov-Taylor identity is now given by
\begin{eqnarray}
\mathcal{S}(\Sigma ) &=&\int \d^{4}x\left( \frac{\delta \Sigma}{\delta K_{\mu }^{a}}\frac{\delta \Sigma }{\delta A_{\mu}^{a}}+\frac{\delta \Sigma }{\delta L^{a}}\frac{\delta \Sigma }{\delta c^{a}}+b^{a}\frac{\delta \Sigma}{\delta \overline{c}^{a}} + \frac{\delta \Sigma}{ \delta \overline Y_\alpha^i} \frac{\delta \Sigma}{ \delta \psi^i_\alpha} - \frac{\delta \Sigma}{\delta Y^i_\alpha} \frac{\delta \Sigma}{\delta \overline \psi^i_\alpha}  \right) = 0 \,. \nonumber\\
\end{eqnarray}
\item Secondly, if we derive the action $\Sigma$ w.r.t.~$b^a$, we find the gauge condition,
\begin{eqnarray}\label{gaugefixing}
\frac{\delta \Sigma}{ \delta b^a } &=& \p_\mu A_\mu^a + \alpha b^a\;.
\end{eqnarray}
Notice that this symmetry is linearly broken. However, this is allowed according to the QAP as explained in the previous section.
\item Thirdly, the action enjoys the antighost equation:
\begin{eqnarray}\label{antighost}
\left( \frac{\delta }{\delta \overline c^a} +  \p_\mu \frac{\delta }{\delta K_\mu^a}  \right) \Sigma &=& 0 \;.
\end{eqnarray}
\item Finally, from the table \ref{tabel1} and \ref{tabel2} we also notice that the action preserves the ghost number and the spinor number:
\begin{eqnarray}\label{gn}
\mathcal{G}n( \Sigma) &=& \int \d^4 x \left( c^a \frac{\delta }{\delta c^a} -  \overline c^a \frac{\delta }{\delta \overline c^a}  - K_\mu^a \frac{\delta}{\delta K_\mu^a} - 2 L^a \frac{\delta}{\delta L^a} - \overline Y^i_\alpha \frac{\delta}{\delta \overline Y^i_\alpha} -  Y^i_\alpha \frac{\delta}{\delta Y^i_\alpha} \right) \Sigma= 0 \nonumber\\
\mathcal{S}n (\Sigma) &=&\int \d^4 x \left( Y^i_\alpha \frac{\delta }{\delta Y^i_\alpha} -  \overline Y^i_\alpha \frac{\delta }{\delta \overline Y^i_\alpha}  +   \psi^i_\alpha \frac{\delta }{\delta \psi^i_\alpha} -  \overline \psi^i_\alpha \frac{\delta }{\delta \overline \psi^i_\alpha}   \right)\Sigma = 0\;.
\end{eqnarray}
However, we should notice that the spinor number is not a necessary Ward identity to prove the renormalizability of the Yang-Mills action. Therefore, we shall not further discuss this quantum number.
\end{itemize}

\begin{table}[H]
\begin{center}
        \begin{tabular}{|c|c|c|c|c|c|c|}
        \hline
        & $A_{\mu }^{a}$ & $c^{a}$ & $\overline{c}^{a}$ & $b^{a}$ & $\overline \psi^i_\alpha$ & $\psi^i_\alpha$  \\
        \hline
        \hline
        \textrm{dimension} & $1$ & $0$ &$2$ & $2$  & $3/2$ & $3/2$  \\
        \hline
        $\mathrm{ghost\; number}$ & $0$ & $1$ & $-1$ & $0$ & $0$ & $0$   \\
        \hline
        $\mathrm{spinor\; number}$  & $0$ & $0$ & $0$ & $0$ & $-1$ & $1$ \\
        \hline
        \end{tabular}
        \caption{Quantum numbers of the fields.}\label{tabel1}
      \end{center}  \end{table}
        \begin{table}[H]
        \begin{center}
    \begin{tabular}{|c|c|c|c|c|}
        \hline
        &$K_{\mu }^{a}$&$L^{a}$& $\overline Y^i_\alpha$& $Y^i_\alpha $   \\
        \hline
        \hline
        \textrm{dimension} & $3$ & $4$ & $5/2$ & $5/2$     \\
        \hline
        $\mathrm{ghost\; number}$ & $-1$ & $-2$ & $-1$ & $-1$  \\
        \hline
        $\mathrm{spinor\; number}$ & $0$ & $0$ & $-1$ & $1$  \\
        \hline
        \end{tabular}
        \caption{Quantum numbers of the sources.}\label{tabel2}
\end{center}
\end{table}

\subsection{The Ward identities at the quantum level}

We shall now try to prove that all the Ward identities can be transformed to the quantum level and that no anomaly is present.

\subsubsection{The gauge condition\label{chap1gaugecond}}
Let us start with the gauge condition \eqref{gaugefixing}, we would like to prove that
\begin{eqnarray}\label{bgamma}
\frac{\delta  \Gamma}{ \delta b^a } &=& \p_\mu A_\mu^a + \alpha b^a   \;.
\end{eqnarray}
The QAP translates the symmetry to the quantum level:
\begin{eqnarray}
\frac{\delta \Gamma}{ \delta b^a } &=& \p_\mu A_\mu^a + \alpha b^a + \Delta^a \cdot \Gamma\;.
\end{eqnarray}
We know that $\Delta$ can only start from order $\hbar$. Let us now assume that it starts at order $h^n$, with $n \geq$ 1:
\begin{eqnarray}
\frac{\delta \Gamma}{ \delta b^a } &=& \p_\mu A_\mu^a + \alpha b^a + \hbar^n \Delta^a + O(h^{n+1}) \;.
\end{eqnarray}
From the properties below equation \eqref{QAPlinear}, we can determine $\Delta^a$: $\Delta^a$ is a local polynomial of the sources and fields of dimension two with ghost number zero, therefore from the tables \ref{tabel1} and \ref{tabel2} we can deduce that
\begin{eqnarray}
\Delta^a(x) &=& F^a (A, c, \overline c)(x) + \omega^{ab} b^b(x) \;,
\end{eqnarray}
with $F^a$ a local polynomial in the fields $A$, $c$ and $\overline c$, and $\omega^{ab}$ certain constants. No other combinations are possible, as the dimensionality of the other fields and sources are too high for constructing a dimension 2 polynomial. Now we also know that
    \begin{eqnarray} \label{gaugefixingcond}
    \left[ \frac{\delta }{\delta b^a(x)} , \frac{\delta }{\delta b^b(y) }\right] &=& 0 \;,
    \end{eqnarray}
so that acting on $\Gamma$ gives,
\begin{eqnarray}\label{a1}
            && [ \frac{\delta }{\delta b^a(x)} , \frac{\delta }{\delta b^b (y)}]  \Gamma = 0 \nonumber\\
\Rightarrow &&  \frac{\delta }{\delta b^a(x)} ( \p_\mu A_\mu^b + \alpha b^b + \hbar^n \Delta^b  )(y) -  \frac{\delta }{\delta b^b (y)} ( \p_\mu A_\mu^a + \alpha b^a + \hbar^n \Delta^a  )(x) = 0 \nonumber\\
\Rightarrow && \frac{\delta }{\delta b^a(x)} \Delta^b(y) -   \frac{\delta }{\delta b^b(y)} \Delta^a(x) = 0 \;,
\end{eqnarray}
returns us a consistency condition. Filling in the expression for $\Delta^a $ we immediately obtain that $\omega^{ba} = \omega^{ab}$. Now we can rewrite  \eqref{a1}
\begin{eqnarray}
 \frac{\delta }{\delta b^b(y)} \Delta^a(x)  = \frac{\delta }{\delta b^a(x)} (F^b (A, c, \overline c) + \omega^{bc} b^c)(y) \;,
\end{eqnarray}
so we obtain after integration
\begin{eqnarray}
 \Delta^a(x)  = \frac{\delta }{\delta b^a(x)} \int \d^4 y \left( F^b (A, c, \overline c) b^b (y) + \frac{1}{2}\omega^{bc} b^b b^c (y) \right)\;.
\end{eqnarray}
Just as in equation \eqref{a2}, we can redefine the effective action $\Gamma$,
\begin{eqnarray}\label{redefine}
\overline \Gamma &=& \Gamma - \hbar^n \int \d^4 y \left( F^b (A, c, \overline c) b^b (y) + \frac{1}{2}\omega^{bc} b^b b^c (y) \right) \;,
\end{eqnarray}
so that
\begin{eqnarray}\label{sfe}
\frac{\delta \overline \Gamma}{ \delta b^a } &=& \p_\mu A_\mu^a + \alpha b^a  + O(h^{n+1}) \;.
\end{eqnarray}
We can repeat this argument at each consecutive order, so that we have indeed proven \eqref{bgamma}.

\subsubsection{Antighost equation}
 Let us now investigate the antighost equation \eqref{antighost}. We would like to prove that
 \begin{eqnarray}\label{antighostgamma}
\left( \frac{\delta }{\delta \overline c^a} +  \p_\mu \frac{\delta }{\delta K_\mu^a}  \right) \overline \Gamma &=& 0 \;,
\end{eqnarray}
whereby we continue to work with $\overline \Gamma$ defined in \eqref{redefine}. We can rewrite this equation into a more simple form by performing the following transformation
\begin{eqnarray}\label{Ktransform}
 \begin{cases}
\tilde K_\mu &= K_\mu + \p_\mu \overline c \;,\\
\tilde{\overline c}  &= \overline c \;,
\end{cases}
\end{eqnarray}
so that
\begin{eqnarray}\label{Ktransform2}
\begin{cases}
\frac{\delta}{\delta K_\mu(x)^a} &= \frac{\delta}{\delta \tilde K_\mu(x)^a} \;, \nonumber\\
\frac{\delta}{\delta{ \overline c}(x)^a} &= \frac{\delta}{\delta \tilde{ \overline c}(x)^a} - \p_\mu  \frac{\delta}{\delta \tilde K^a_\mu}\;.
\end{cases}
\end{eqnarray}
and thus the antighost equation becomes
\begin{eqnarray}\label{newantighost}
 \frac{\delta }{ \delta \tilde{\overline c}^a} \Sigma &=& 0 \;,
\end{eqnarray}
with $\Sigma$ now in the new variables ($\tilde K_\mu$, $\tilde{\overline c}$, \ldots). We can now repeat the proof of the gauge condition in the previous section \ref{chap1gaugecond}.  Applying the QAP yields\footnote{We omit the $\tilde{\phantom{c}}$ notation for the rest of the paragraph.}
\begin{eqnarray}
\frac{\delta \overline \Gamma}{ \delta \overline c^a } &=& \Delta^a \cdot \overline \Gamma\;.
\end{eqnarray}
We assume again that the breaking $\Delta$ starts at order $h^n$, with $n \geq$ 1,
\begin{eqnarray}
\frac{\delta \overline \Gamma}{ \delta \overline c^a } &=& \hbar^n \Delta^a  + O(\hbar^{n+1})\;,
\end{eqnarray}
with $\Delta$ again a local polynomial of the sources and fields of dimension two however with ghost number +1 and thus given by,
\begin{eqnarray}
\Delta^a(x) &=& G^a (A, c, b)(x) + \upsilon^{ab}(c) \overline c^b(x) \;,
\end{eqnarray}
whereby $ \upsilon^{ab}(c)$ is has to function of $c$ to combine to ghost number $+1$. However, we have to keep in mind that $\overline \Gamma$ obeys the identity \eqref{sfe}. Therefore, we know that $G^a$ cannot depend on $b$ and thus
\begin{eqnarray}
\Delta^a(x) &=& G^a (A, c)(x) + \upsilon^{ab}(c) \overline c^b(x) \;.
\end{eqnarray}
Here, we have that
    \begin{eqnarray}\label{a3}
    \left\{ \frac{\delta }{\delta \overline c^a(x)} , \frac{\delta }{\delta \overline c^b(y) } \right\} &=& 0 \;,
    \end{eqnarray}
as we are working with ghost fields here. Multiplying this equation with the effective action $\Gamma$ gives
\begin{eqnarray}
\frac{\delta }{\delta \overline c^a(x)} \Delta^b(y) +   \frac{\delta }{\delta \overline c^b(y)} \Delta^a(x) = 0 \;,
\end{eqnarray}
and thus $\upsilon^{ab}(c) = -\upsilon^{ba}(c)$. We can solve this equation again,
\begin{eqnarray}
 \Delta^a(x)  = \frac{\delta }{\delta \overline c^a(x)} \int \d^4 y \left(  \overline c^b G^b (A, c)  (y) + \frac{1}{2} \upsilon^{bc}(c) \overline c^b \overline c^c (y) \right)\;.
\end{eqnarray}
We can redefine the action analogously as in equation \eqref{redefine}, so the antighost as well as the gauge condition hold to order $\hbar^n$. By induction, we have thus proven that these identities hold to all orders\footnote{In order not to overload the notation, we call the redefined effective action again $\Gamma$.}.

\subsubsection{Ghost number} 
We can easily prove that the ghost number 
remains zero at the quantum level.
According to the QAP, the first equation of expression \eqref{gn} becomes
\begin{eqnarray}
\mathcal{G}n (\Gamma) &=& \Delta \cdot \Gamma\;,
\end{eqnarray}
whereby $\Delta$ is an integrated local polynomial of dimension 4 which has the same quantum numbers as the Ward identity $\mathcal{G}n$. From equation \eqref{gn}, we see that  the Ward identity $\mathcal{G}n$ itself has ghost number zero. Therefore, $\Delta$ also has ghost number zero. Assuming that the breaking $\Delta$ starts at order $\hbar^n$, we rewrite
\begin{eqnarray}\label{due}
\mathcal{G}n (\Gamma) &=& \hbar^n \Delta + O(\hbar^{n+1})\;.
\end{eqnarray}
We can parameterize $\Delta$ as
\begin{equation}\label{deltass}
\Delta =  \int \d^4 x a_1 ( \p_\mu \overline c^a + K_\mu^a) \p_\mu c^a + \int \d^4 x a_2 ( \p_\mu \overline c^a + K_\mu^a) A_\mu^b  c^c f^{abc} + a_3 \int \d^4 x L^a c^b c^c f^{abc} + \mathcal S (A)\;,
\end{equation}
with $ \mathcal S (A)$ given by
\begin{equation}
\int \d^4 x \left( b_1 A \p \p A   + b_2 A A \p A + b_3 A A A A  \right) \;.
\end{equation}
The parameters $a_1, \ldots, b_3$ are arbitrary and in the last expression, one can contract the different fields $A$ and partial derivatives with all possible Lorentz contractions and color contractions\footnote{Giving rise to more than three terms. For simplicity, we have not written down all possible contractions.}. For the derivation of expression \eqref{deltass}, we have also kept in mind the validity of the gauge condition and the antighost equation. \\
\\
Let us now prove that all the coefficients $a_1, \ldots, b_3$ vanish. We start with $a_1$. If we act on both sides of the equation \eqref{due} with the test operator $\frac{\delta}{\delta \overline c (x)} \frac{\delta}{\delta c (y)}$ and set all sources and fields equal to zero, we find
\begin{equation}
\left. \frac{\delta}{\delta \overline c (x)} \frac{\delta}{\delta c (y)} \mathcal{G}n (\Gamma) \right|_{\fields,\sources = 0} = \left. \frac{\delta}{\delta \overline c (x)} \frac{\delta}{\delta c (y)}  \hbar^n \Delta \right|_{\fields,\sources = 0}\;,
\end{equation}
or thus, with equation \eqref{gn}, we obtain
\begin{equation}
\left. \frac{\delta^2 \Gamma}{\delta \overline c(x) \delta c(y)}\right|_{\fields,\sources = 0} +  \left. \frac{\delta^2 \Gamma}{\delta \overline c(y) \delta \overline c(x)}\right|_{\fields,\sources = 0} = \hbar a_1 \p_y^2 \delta^4 (x-y)\;.
\end{equation}
As the l.h.s.~of this equation is clearly equal to zero, it follows that $a_1$ is also equal to zero. In an analogical fashion, one can prove that all other coefficients are zero. As an example, acting with $\frac{\delta^4}{\delta A \delta A \delta A \delta A}$ on equation \eqref{due} and setting all fields and sources equal to zero, immediately gives $b_3 = 0$. Therefore, we have proven $\Gamma$ to have ghost number zero to order $\hbar$. We can however prove this order by order, so in summary, we have
\begin{eqnarray}\label{ghostnumberidentity}
\int \d^4 x \left( c^a \frac{\delta }{\delta c^a} -  \overline c^a \frac{\delta }{\delta \overline c^a}  - K_\mu^a \frac{\delta}{\delta K_\mu^a} - 2 L^a \frac{\delta}{\delta L^a} - \overline Y^i_\alpha \frac{\delta}{\delta \overline Y^i_\alpha} -  Y^i_\alpha \frac{\delta}{\delta Y^i_\alpha} \right) \Gamma&=& 0 \;. 
\end{eqnarray}

\subsubsection{The Slavnov-Taylor identity}
 Finally, we need to show that the Slavnov-Taylor identity can be extended to the quantum level. From the QAP we learn that
\begin{eqnarray}\label{notlin}
\mathcal{S}(\Gamma ) &=&\int \d^{4}x\left( \frac{\delta \Gamma}{\delta K_{\mu }^{a}}\frac{\delta \Gamma }{\delta A_{\mu}^{a}}+\frac{\delta \Gamma }{\delta L^{a}}\frac{\delta \Gamma }{\delta c^{a}}+b^{a}\frac{\delta \Gamma}{\delta \overline{c}^{a}} + \frac{\delta \Gamma}{ \delta \overline Y_\alpha^i} \frac{\delta \Gamma}{ \delta \psi^i_\alpha} - \frac{\delta \Gamma}{\delta Y^i_\alpha} \frac{\delta \Gamma}{\delta \overline \psi^i_\alpha}  \right) = \Delta \cdot \Gamma \,. \nonumber\\
\end{eqnarray}
Again, let us assume that the breaking $\Delta$ starts at order $\hbar^n$,
\begin{eqnarray}\label{a4}
\mathcal{S}(\Gamma ) &=& \hbar^n \Delta  + O(h^{n+1})\;,
\end{eqnarray}
then we have to show that we are able to restore the Slavnov-Taylor identity by a redefinition of the effective action $\Gamma$.\\
\\
Firstly, we shall derive a condition for $\Delta$, called the consistency condition. We can exploit the nilpotency of the Slavnov-Taylor identity. For this we need to introduce the linearized version of \eqref{notlin},
\begin{eqnarray}
\mathcal{S}_\Gamma &=&\int \d^{4}x\left(  \frac{\delta \Gamma}{\delta K_{\mu }^{a}}\frac{\delta  }{\delta A_{\mu}^{a}}+  \frac{\delta  \Gamma}{\delta A_{\mu}^{a}}\frac{\delta }{\delta K_{\mu }^{a}}  +  \frac{\delta \Gamma }{\delta L^{a}}\frac{\delta  }{\delta c^{a}}+ \frac{\delta \Gamma }{\delta c^{a}} \frac{\delta }{\delta L^{a}} + b^{a}\frac{\delta }{\delta \overline{c}^{a}}  \right. \nonumber\\
 &&\left. \qquad \qquad + \frac{\delta \Gamma}{ \delta \overline Y_\alpha^i} \frac{\delta }{ \delta \psi^i_\alpha} +  \frac{\delta \Gamma}{ \delta \psi^i_\alpha} \frac{\delta }{ \delta \overline Y_\alpha^i}  - \frac{\delta \Gamma}{\delta Y^i_\alpha} \frac{\delta }{\delta \overline \psi^i_\alpha} -    \frac{\delta }{\delta \overline \psi^i_\alpha} \frac{\delta }{\delta Y^i_\alpha} \right)  \,.
\end{eqnarray}
We can now prove that
\begin{eqnarray}\label{theor1}
\mathcal{S}_\Gamma \mathcal{S}(\Gamma )  &=& 0 \;.
\end{eqnarray}
To prove this, we abbreviate the formula \eqref{notlin},
\begin{eqnarray}
\mathcal{S}(\Gamma) &=&  \int \d^{4}x\left( \frac{\delta \Gamma}{\delta \mathcal K_i }\frac{\delta \Gamma }{\delta \mathcal A_i } + b^a \frac{\delta \Gamma}{\delta \overline c^a}\right)\;,
\end{eqnarray}
whereby $\mathcal K_i$ symbolizes the ghost fields/sources and $\mathcal A_i$ symbolizes the bosonic fields/sources. For the linearized Slavnov-Taylor identity, we write in this notation
\begin{eqnarray}
\mathcal{S}_\Gamma &=&  \int \d^{4}x\left( \frac{\delta \Gamma}{\delta \mathcal K_i }\frac{\delta }{\delta \mathcal A_i }  + \frac{\delta \Gamma }{\delta \mathcal A_i }\frac{\delta }{\delta \mathcal K_i } + b^a \frac{\delta }{\delta \overline c^a}\right)\;.
\end{eqnarray}
Now we unleash $\mathcal{S}_\Gamma$ on $\mathcal{S}(\Gamma)$ to find
\begin{eqnarray}
\mathcal{S}_\Gamma \mathcal{S}(\Gamma)  &=&  \int \d^{4}x\left( \frac{\delta \Gamma}{\delta \mathcal K_i }\frac{\delta }{\delta \mathcal A_i }  + \frac{\delta \Gamma }{\delta \mathcal A_i }\frac{\delta }{\delta \mathcal K_i } + b^a \frac{\delta }{\delta \overline c^a}\right)   \int \d^{4}y\left( \frac{\delta \Gamma}{\delta \mathcal K_j }\frac{\delta \Gamma }{\delta \mathcal A_j } + b^b \frac{\delta \Gamma}{\delta \overline c^b}\right)\nonumber\\
&=& \int \d^{4}x \int \d^{4}y  \left( \frac{\delta \Gamma}{\delta \mathcal K_i }\frac{\delta^2 \Gamma }{\delta \mathcal A_i \delta \mathcal K_j } \frac{\delta \Gamma }{\delta \mathcal A_j }  + \frac{\delta \Gamma}{\delta \mathcal K_i }  \frac{\delta \Gamma}{\delta \mathcal K_j }\frac{\delta^2 \Gamma }{\delta \mathcal A_i \delta \mathcal A_j } +  \frac{\delta \Gamma }{\delta \mathcal A_i } \frac{\delta^2 \Gamma}{\delta \mathcal K_i \delta \mathcal K_j }\frac{\delta \Gamma }{\delta \mathcal A_j } \right. \nonumber\\
&&   - \frac{\delta \Gamma }{\delta \mathcal A_i } \frac{\delta \Gamma}{\delta \mathcal K_j }\frac{\delta^2 \Gamma }{\delta \mathcal K_i  \delta \mathcal A_j } +    b^a  \frac{\delta^2 \Gamma}{ \delta \overline c^a \delta \mathcal K_j }\frac{\delta \Gamma }{\delta \mathcal A_j } -
b^a  \frac{\delta \Gamma}{  \delta \mathcal K_j }\frac{\delta^2 \Gamma }{ \delta \overline c^a \delta \mathcal A_j } \nonumber\\
 && \left. + b^b\frac{\delta \Gamma}{\delta \mathcal K_i } \frac{\delta^2 \Gamma}{\delta \mathcal A_i \delta \overline c^b} + b^b \frac{\delta \Gamma }{\delta \mathcal A_i }  \frac{\delta^2 \Gamma}{\delta \mathcal K_i \delta  \overline c^b}   + b^a b^b   \frac{\delta^2 \Gamma}{ \delta \overline c^a \delta \overline c^b} \right) \;,
\end{eqnarray}
whereby the indices $i,a$ and $j,b$ belong to the $x$ resp.~$y$ dependent fields/sources. Now it is easy to see that all terms cancel pairwise or that they are a contraction of a symmetric and an antisymmetric part. Therefore we have proven \eqref{theor1}. Notice that $\Gamma$ plays no particular role in this proof and that it can be generalized to all functions\footnote{ This proof here hold only for functions of the fields and sources with even ghost number, however, also for odd ghost number, this identity holds, but the form of the linearized Slavnov-Taylor identity changes.} $\mathcal F$.
Unleashing $\mathcal{S}_\Gamma$ on equation \eqref{a4}, we find that
\begin{eqnarray}
 \mathcal{S}_\Gamma \Delta &=& 0 + O(\hbar)\;.
\end{eqnarray}
We can expand $\mathcal{S}_\Gamma$, the lowest order contribution, which gives
\begin{eqnarray}\label{consistency}
 \mathcal{S}_\Sigma \Delta &=& 0 \;.
\end{eqnarray}
This is called the consistency solution.\\
\\
Now we have two options. The consistency condition \eqref{consistency} can have a trivial solution, $\Delta = \mathcal S_\Sigma \Lambda$, or it has a non trivial solution and cannot be written as $\Delta = \mathcal S_\Sigma (\ldots)$. Only in the first case, we can prove that the Slavnov-Taylor identity holds at the quantum level. Indeed, if we define
\begin{eqnarray}\label{overlinegamma}
\overline \Gamma &=& \Gamma - \hbar^n \Lambda\;,
\end{eqnarray}
we have that
\begin{eqnarray}
\mathcal S (\overline \Gamma) &=&   \int \d^{4}x\left( \frac{\delta (\Gamma - \hbar^n \Lambda)}{\delta \mathcal K_i }\frac{\delta (\Gamma - \hbar^n \Lambda)}{\delta \mathcal A_i } + b^a \frac{\delta (\Gamma - \hbar^n \Lambda)}{\delta \overline c^a}\right)\nonumber\\
&=& \mathcal S (\Gamma) - \hbar^n \underbrace{\mathcal S_\Gamma}_{= \mathcal S_\Sigma + O(\hbar)} \Lambda + O(\hbar^{n+1}) \nonumber\\
&=&  \mathcal S (\Gamma) - \hbar^n \Delta + O(\hbar^{n+1}) \nonumber\\
&=& O(\hbar^{n+ 1})\;.
\end{eqnarray}
The only thing which we need to check is that $\overline \Gamma$ does not violate the other Ward identities. This is the subject of the next paragraphs.\\
\\
Firstly, we can prove that $\Delta$ obeys the antighost equation \eqref{antighost} and does not depend on the $b$-field:
\begin{align}\label{a6}
 \frac{\delta}{\delta b^a } \Delta &=  0 &  \left( \frac{\delta }{\delta \overline c^a} +  \p_\mu \frac{\delta }{\delta K_\mu^a}  \right) \Delta &= 0\;.
\end{align}
The first relation can be derived from the following expression:
\begin{eqnarray}\label{a5}
\frac{\delta }{\delta b^a(y)} \mathcal S (\Gamma) -  \mathcal S_\Gamma \left(\frac{\delta \Gamma}{ \delta b^a(y)} - \p_{y,\mu} A^a_\mu \right) &=& \left( \frac{\delta }{\delta \overline c^a(y)} +  \p_\mu \frac{\delta }{\delta K_\mu^a(y)}  \right) \Gamma\;.
\end{eqnarray}
Indeed, we observe that due to the antighost equation \eqref{antighostgamma} and the gauge equation \eqref{bgamma}, the above expression becomes
\begin{align}
\frac{\delta }{\delta b^a(y)} \mathcal S (\Gamma) &=0  &\Rightarrow  && \frac{\delta }{\delta b^a(y)} \Delta &=0 \;.
\end{align}
The proof of the expression \eqref{a5} is straightforward. We compute that
\begin{multline}
\frac{\delta}{\delta b^a(y)} \int \d^{4}x\left( \frac{\delta \Gamma}{\delta \mathcal K_i }\frac{\delta \Gamma }{\delta \mathcal A_i } + b^b \frac{\delta \Gamma}{\delta \overline c^b}\right) \nonumber\\
= \int \d^{4}x\left( \frac{\delta^2 \Gamma}{\delta \mathcal K_i \delta b^a(y)} \frac{\delta \Gamma}{\delta A_i} + \frac{\delta \Gamma}{\delta \mathcal K_i} \frac{\delta^2 \Gamma}{\delta b^a(y) \delta A_i} + b^b \frac{\delta^2 \Gamma}{\delta b^a(y)\delta \overline c^b} \right) + \frac{\delta \Gamma }{\delta \overline c^a(y)}\;,
\end{multline}
and
\begin{multline*}
\mathcal S_\Gamma \left(\frac{\delta \Gamma}{ \delta b^a(y)} - \p_{y,\mu} A^a_\mu(y) \right) \\
= \int \d^4 x \left( \frac{\delta \Gamma}{\delta \mathcal K_i} \frac{\delta^2 \Gamma}{\delta A_i \delta b^a(y)} + \frac{\delta \Gamma}{\delta A_i} \frac{\delta^2 \Gamma}{\delta \mathcal K_i \delta b^a(y)} + b^b \frac{\delta^2 \Gamma}{ \delta b^a(y) \delta \overline c^b} \right) - \p_{y,\mu} \frac{\delta \Gamma}{\delta K_\mu^a(y) }\;,
\end{multline*}
and thus the remainder of these two expressions gives exactly the right hand side of \eqref{a5}. Notice again that $\Gamma$ plays no specific role in this derivation. The second relation of \eqref{a6} can be derived from
\begin{eqnarray}
\left( \frac{\delta }{\delta \overline c^a} +  \p_\mu \frac{\delta }{\delta K_\mu^a}  \right) \mathcal S (\mathcal F) + \mathcal S_{\mathcal F} \left( \frac{\delta }{\delta \overline c^a} +  \p_\mu \frac{\delta }{\delta K_\mu^a}  \right) \mathcal F &=& 0\;,
\end{eqnarray}
for $\mathcal F$ a function with even ghost number. Replacing $\mathcal F$ with $\Gamma$ returns us the second relation of \eqref{a6}. The expression above can be proven in a similar fashion as \eqref{a5}. \\
\\
Secondly, we can also easily show that $\Delta$ has ghost number one. 
Unlashing the operator
\begin{eqnarray}\label{ghostnumberoperator}
\mathcal{G}n &=& \int \d^4 x \left( c^a \frac{\delta }{\delta c^a} -  \overline c^a \frac{\delta }{\delta \overline c^a}  - K_\mu^a \frac{\delta}{\delta K_\mu^a} - 2 L^a \frac{\delta}{\delta L^a} - \overline Y^i_\alpha \frac{\delta}{\delta \overline Y^i_\alpha} -  Y^i_\alpha \frac{\delta}{\delta Y^i_\alpha} \right)\;,
\end{eqnarray}
on equation \eqref{a4}, we see that
\begin{eqnarray}
\mathcal {G}n \left( \mathcal S[\Gamma]  \right)&=& \hbar^n \mathcal{G}n( \Delta)\;.
\end{eqnarray}
As $\Gamma$ has ghost number zero, and the operation $\mathcal S$ rises the ghost number with one, $\Delta$ also has ghost number one. 
\\
\\
So far, we have shown that $\Delta$ does not depend on $b$, obeys the antighost equation and has ghost number one. 
If we can prove that the same properties hold for $\Lambda$, $\overline \Gamma$ (see equation \eqref{overlinegamma}) obeys all the Ward identities. For this, we shall work in the transformed fields ($\tilde K_\mu$, $\tilde{\overline c}$, \ldots) as introduced in equation \eqref{Ktransform}. We recall that this transformation changes the functional derivatives, see expression \eqref{Ktransform2}
and thus the antighost equation has changed form, e.g.~for $\Delta$
\begin{eqnarray}
\frac{ \delta }{\delta \tilde{\overline c^a}} \Delta (\tilde K_\mu, \tilde{\overline c}, \ldots) &=& 0 \;.
\end{eqnarray}
We now define the following counting operator:
\begin{eqnarray}
\mathcal N &=& \int d^4 x \left( b^a \frac{\delta }{\delta b^a} + \tilde{\overline c}^a \frac{\delta }{\delta \tilde{\overline c}^a}\right)\;,
\end{eqnarray}
which counts the sum of the number of $b$ and $\overline c$ fields. Expanding $\Lambda$ according to the eigenvalues of this operator yields
\begin{eqnarray}
\Lambda &=& \sum_{n \geq 0} \Lambda_n \;,
\end{eqnarray}
whereby $\mathcal N \Lambda_n = n \Lambda_n$. $\Lambda_0$ is thus the part which does not contain any $b$ nor $\tilde{\overline c}$ fields. As $\Delta$ obeys the antighost equation and does not depend on $b$, unlashing the counting operator on $\Delta = \mathcal S_\Sigma \Lambda$ gives
\begin{eqnarray}\label{finally}
0 &=& \mathcal N \mathcal S_\Sigma \Lambda = \mathcal S_\Sigma \mathcal N  \Lambda =   \sum_{n \geq 0} n \mathcal S_\Sigma \Lambda_n \;,
\end{eqnarray}
whereby we have used the property that $\mathcal N$ and $\mathcal S_\Sigma$ commute. This can be checked by writing $\mathcal S_\Sigma$,
\begin{multline}
\mathcal{S}_\Sigma =\int \d^{4}x\left( -\left( D_{\mu }c\right) ^{a}    \frac{\delta  }{\delta A_{\mu}^{a}}+  \left( \frac{\delta S_{\YM}}{\delta A_\mu^a} - \p_\mu b^a   - g f_{bac} \p_\mu \overline c^b c^c - g f_{bac} K^b_\mu c^c  \right) \frac{\delta }{\delta K_{\mu }^{a}}  \right.\\
 \left. + \frac{1}{2}gf^{abc}c^{b}c^{c}  \frac{\delta  }{\delta c^{a}}+ \left( - D_\mu^{ab} \p_\mu \overline c^b - D_\mu^{ab} K_\mu^b +  g f_{dac} L^d c^c \right) \frac{\delta }{\delta L^{a}} + b^{a}\frac{\delta }{\delta \overline{c}^{a}}   + \ldots \right)  \,.
\end{multline}
in the new variables\footnote{The $\ldots$ stand for the parts related to the quarks, which are irrelevant here. },
\begin{multline}
\mathcal{S}_\Sigma =\int \d^{4}x\left( -\left( D_{\mu }c\right) ^{a}    \frac{\delta  }{\delta A_{\mu}^{a}}+  \left( \frac{\delta S_{\YM}}{\delta A_\mu^a}   - g f_{bac} \tilde K^a_\mu c^c  \right) \frac{\delta }{\delta \tilde K_{\mu }^{a}}  \right.  \\
 \left. + \frac{1}{2}gf^{abc}c^{b}c^{c}  \frac{\delta  }{\delta c^{a}}+ \left(   g f_{dac} L^d c^c \right) \frac{\delta }{\delta L^{a}} + b^{a}\frac{\delta }{\delta \tilde{\overline{c}}^{a}}   + \ldots \right)  \,.
\end{multline}
One now easily sees that $\mathcal{S}_\Sigma$ and $\mathcal N$ commute. Finally, from \eqref{finally}, we find that $\Lambda_n$ for $n \geq 1$ must be equal to zero and only $\Lambda_0$ survives. This means that $\Lambda$ is independent from $b$ and obeys the antighost equation.\\
\\
In conclusion, if the consistency condition \eqref{consistency} has a trivial solution, $\Delta = \mathcal S_\Sigma \Lambda$, the Slavnov-Taylor identity holds to all orders. Otherwise, we can have an anomaly, and the Slavnov-Taylor identity is broken. In this case, we shall not be able to prove that the action is renormalizable as the Slavnov-Taylor identity is a crucial identity. But when exactly do we have an anomaly? Firstly, one can prove \cite{Peskin} that when the spinors $\psi$ are a sum of a left handed spinor and a right handed spinor, no anomaly is present and the consistency condition always has a trivial solution. Secondly, in the case our matter field $\psi$ would be a left handed spinor, i.e.
\begin{eqnarray}
\psi = \psi_\mathrm{left}  \qquad \mathrm{whereby} \qquad \frac{1}{2} (1- \gamma_5) \psi_\mathrm{left} ~=~\psi_\mathrm{left}\;,
\end{eqnarray}
there is the possibility of an anomaly. This anomaly is called the Adler-Bardeen anomaly. However, this anomaly depends on the number of flavours, and if chosen correctly, this anomaly can vanish, and it can be proven to vanish to all orders. Such a construction plays an important role in the elektroweak theory. 

\subsection{Stability of the Yang-Mills action}
Let us consider the case without anomaly and prove that the Yang-Mills action is stable in this case. In the previous section, we have determined all the Ward identities which the effective action obeys, therefore we can characterize the first order counterterm $\overline \Gamma_\dive^{(1)}$ and try to reabsorb this into the action as described in equation \eqref{sibare}. If this works, we have proven the action to be stable.\\
\\
For all the Ward identities we have described previously, we can write down the identities for the first order counterterm $\overline \Gamma_\dive^{(1)}$. The gauge condition and the antighost equation read
\begin{subequations}
 \begin{eqnarray}
\frac{\delta   \overline \Gamma_\dive^{(1)} }{ \delta b^a } &=& 0 \label{gaugec1} \;,  \\
\left( \frac{\delta }{\delta \overline c^a} +  \p_\mu \frac{\delta }{\delta K_\mu^a}  \right) \overline \Gamma_\dive^{(1)} &=& 0 \label{gaugec2}\;,
\end{eqnarray}
\end{subequations}
and the Slavnov-Taylor identity to first order yields,
\begin{eqnarray}
\mathcal S_\Sigma \overline \Gamma_\dive^{(1)} &=& 0\;.
\end{eqnarray}
whereby we repeat that
\begin{eqnarray}
\mathcal{S}_\Sigma&=&\int \d^{4}x\left(  \frac{\delta \Sigma}{\delta K_{\mu }^{a}}\frac{\delta  }{\delta A_{\mu}^{a}}+  \frac{\delta \Sigma }{\delta A_{\mu}^{a}} \frac{\delta }{\delta K_{\mu }^{a}} + \frac{\delta \Sigma }{\delta L^{a}}\frac{\delta  }{\delta c^{a}}+ \frac{\delta  \Sigma}{\delta c^{a}} \frac{\delta  }{\delta L^{a}} + b^{a}\frac{\delta }{\delta \overline{c}^{a}}  \right. \nonumber\\
 &&\left. \qquad \qquad + \frac{\delta \Sigma}{ \delta \overline Y_\alpha^i} \frac{\delta }{ \delta \psi^i_\alpha} +  \frac{\delta \Sigma}{ \delta \psi^i_\alpha} \frac{\delta }{ \delta \overline Y_\alpha^i}  - \frac{\delta \Sigma}{\delta Y^i_\alpha} \frac{\delta }{\delta \overline \psi^i_\alpha} -   \frac{\delta \Sigma}{\delta \overline \psi^i_\alpha} \frac{\delta }{\delta Y^i_\alpha}  \right)  \;,
\end{eqnarray}
with $\mathcal S_\Sigma ^2 = 0$. Beside these identities, we have also shown that $\overline \Gamma_\dive^{(1)}$ has ghost number zero 
as one can read from \eqref{ghostnumberidentity}.\\
\\
Let us now construct this counterterm, which is the most general local polynomial of dimension four obeying the previous constraints. Here we can benefit from the nilpotency of the Slavnov-Taylor identity. As explained in section \ref{intermezzo:cohomolgy}, we can write the most general form of $\overline \Gamma_\dive^{(1)}$ as a sum of two parts
\begin{eqnarray}
\overline \Gamma_\dive^{(1)} &=& \Sigma_{\mathrm{n.triv.}}  + \Sigma_{\mathrm{triv.}}\;,
\end{eqnarray}
whereby the trivial part can be written as
\begin{align}
\Sigma_{\mathrm{triv.}} &= \mathcal S_\Sigma \Omega & \Rightarrow&& &\mathcal S_\Sigma \Sigma_{\mathrm{triv.}} = 0 \;.
\end{align}
Let us first try to construct the non trivial part, $\Sigma_{\mathrm{n.triv.}}$. First, we notice that there is one doublet in the game, $(\overline c^a, b^a)$. Due to the doublet theorem from section \ref{intermezzo:cohomolgy}, this implies that these fields cannot enter the non trivial part. After some combinatorial effort, one sees that also $K_\mu^a$, $L^a$ and $c^a$ can not enter the non trivial part.  Also the spinor sources $\overline Y^i_\alpha$ and $Y^i_\alpha $ are unable to enter the non trivial part, as it is impossible to construct a term with 
 ghost number zero including these sources. Only with the spinor fields $\overline \psi^i_\alpha$ and $\psi^i_\alpha$ we can construct a term with the right quantum numbers and invariant under $\mathcal S_\Sigma$:
\begin{eqnarray}\label{proofw}
 \overline \psi^i_\alpha (\gamma_\mu)_{\alpha \beta} D_\mu^{ij} \psi^j_\beta\;.
\end{eqnarray}
However, we shall prove later that this term can be written as a variation of $\mathcal S_\Sigma$, therefore, this term is trivial. In conclusion, only the gluon field $A_\mu$ shall enter the non trivial part. One can check that the only combination which is invariant under the BRST symmetry\footnote{As $ \Sigma_{\mathrm{n.triv.}}$ contains only gluon fields, the Slavnov-Taylor identity $\mathcal S_\Sigma$ reduces to the usual BRST.} is the Yang-Mills action itself. Therefore,
\begin{eqnarray}
 \Sigma_{\mathrm{n.triv.}}  &=& a_0 \frac{1}{4} \int \d^4 x F^2_{\mu\nu} \;,
\end{eqnarray}
whereby $a_0$ is an arbitrary parameter. Secondly, we have to scrutinize the trivial part. For this part, we need to construct the most general local polynomial $\Omega$ of all the fields and sources of ghost number $-1$ 
as $\mathcal S_\Sigma$ increases the ghost number with 1. With the help of the tables \ref{tabel1} and \ref{tabel2} we find
\begin{multline}\label{tegentermYM}
\Omega =\int \d^4 x \Bigl( a_1 A_\mu^a K_\mu^a + a_2 \p_\mu \overline c^a A_\mu^a + a_3 c^a L^a + a_4 \overline c^a b^a + a_5 \frac{g}{2} f^{abc} c^a \overline c^b \overline c^c \\
+ a_6 \overline \psi_\alpha^i Y_\alpha^i + a_7 \overline Y^i_\alpha \psi^i_\alpha  + a_8 \overline \psi_\alpha^i \overline Y_\alpha^i  + a_9  Y^i_\alpha \psi^i_\alpha \Bigr)\;,
\end{multline}
whereby $a_1, \ldots, a_9$ are arbitrary parameters. Unleashing $\mathcal S_\Sigma$ at each term, we find
{\allowdisplaybreaks
\begin{align}
 &a_1 \mathcal S_\Sigma \int \d^4 x A_\mu^a K_\mu^a &&=  a_1  \int \d^4 x\Bigl(A_\mu^a \frac{\delta S_\YM }{\delta A_\mu^a} + b^a \p_\mu A_\mu^a + \overline c^a \p_\mu (g f_{akb} A^k_\mu c^b ) - K_\mu^a g f_{akb} A^k_\mu c^b  \nonumber\\
 &&&\qquad - \ii \overline \psi^i_\alpha (\gamma_\mu)^{\alpha \beta} g A_\mu^a (X^a)^{ij} \psi^j_\beta - D_\mu c^a K_\mu^a \Bigr)\nonumber\\
 &a_2 \mathcal S_\Sigma \int \d^4 x   \p_\mu \overline c^a A_\mu^a &&= a_2 \int \d^4 x\Bigl(  \p_\mu b^a A_\mu^a +   \p_\mu \overline c^a  D_\mu c^a \Bigr)\nonumber\\
 &a_3 \mathcal S_\Sigma \int \d^4 x  c^a L^a &&= a_3 \int \d^4 x\Bigl(\frac{1}{2}f_{abc} c^b c^c L^a - \overline{c}^{a}\partial _{\mu } D_{\mu}^{ab}c^b + K_\mu^a D_\mu^{ab} c^b  -  g  L^{a}f^{abc}c^b c^c \nonumber\\
 &&& \qquad  + \ii g   \overline Y_\alpha^i  c^a (X^a)^{ij} \psi^j_\alpha  + \ii g  \overline \psi^j_\alpha c^a (X^a)^{ji}   Y_\alpha^i    \Bigr) \nonumber\\
 &a_4 \mathcal S_\Sigma \int \d^4 x \overline c^a b^a &&= -a_4 \int \d^4 x\Bigl( \overline c^a \overline c^a\Bigr) = 0 \nonumber\\
 &a_5 \mathcal S_\Sigma \int \d^4 x  \frac{g}{2} f^{abc} c^a \overline c^b \overline c^c &&= a_5  \frac{g}{2} f^{abc}  \int \d^4 x\Bigl( \frac{g}{2} f_{ade} c^d c^e \overline c^b \overline c^c - 2  c^a b^b \overline c^c  \Bigr) \nonumber\\
 &a_6 \mathcal S_\Sigma \int \d^4 x  \overline \psi_\alpha^i Y_\alpha^i  &&= a_6 \int \d^4 x\Bigl(-  \overline \psi_\alpha^i (\gamma_\mu)^{\alpha \beta} D^{ij}_\mu \psi^j_\beta \Bigr) \nonumber\\
 &a_7 \mathcal S_\Sigma \int \d^4 x  \overline Y^i_\alpha \psi^i_\alpha && \sim a_8 \mathcal S_\Sigma \int \d^4 x \overline \psi_\alpha^i \overline Y_\alpha^i \sim a_9 \mathcal S_\Sigma \int \d^4 x Y^i_\alpha \psi^i_\alpha \sim \mathcal S_\Sigma \int \d^4 x  \overline \psi_\alpha^i Y_\alpha^i  \;.\nonumber\\
\end{align}}

\noindent We have still have to impose the gauge condition and antighost equation. From the gauge condition \eqref{gaugec1} and the antighost equation \eqref{gaugec2} we learn that
\begin{align}
a_5 &= 0  & a_1 & =  a_2\;.
\end{align}
We also notice that the terms in $a_6$, $a_7$, $a_8$ and $a_9$ are equal, so we can take them together. Here we have also proven the term \eqref{proofw} to be an $\mathcal S_\Sigma$ variation and thus belonging to the trivial part.
Taking all previous results together, the trivial term is thus given by
\begin{eqnarray}\label{triviaal}
 \Sigma_{\mathrm{triv.}}  &=& a_1  \int \d^4 x \Bigl( A_\mu \frac{\delta S_\YM }{\delta A_\mu^a} +   \p_\mu \overline c^a  \p_\mu c^a   +  K_\mu^a \p_\mu c^a  -  \ii \overline \psi^i_\alpha (\gamma_\mu)^{\alpha \beta} g A_\mu^a (X^a)^{ij} \psi^j_\beta \Bigr)  \nonumber\\
 & & + a_2  \int \d^4 x\Bigl(-\frac{1}{2}f_{abc} c^b c^c L^a - \overline{c}^{a}\partial _{\mu } D_{\mu}^{ab}c^b + K_\mu^a D_\mu^{ab} c^b   + \ii g   \overline Y_\alpha^i  c^a (X^a)^{ij} \psi^j_\alpha \nonumber\\
 && + \ii g  \overline \psi^j_\alpha c^a (X^a)^{ji}   Y_\alpha^i )   \Bigr) + a_3 \int \d^4 x\Bigl(  \overline \psi_\alpha^i (\gamma_\mu)^{\alpha \beta} D^{ij}_\mu \psi^j_\beta \Bigr) \;,
\end{eqnarray}
whereby we have three independent parameters here\footnote{$a_2$ and $a_3$ are not equal to the ones in expression \eqref{tegentermYM}, but set again in a logical order. }.\\
\\
 For the action to be stable, we need show that (see equation \eqref{sibare})
\begin{eqnarray}\label{stable2}
\Sigma_{\bare} (\phi_{i,0}, \rho_{i, 0}, \lambda_{i,0})  &=&   \Sigma (\phi_i, \rho_i, \lambda_i)  + \hbar  \overline \Gamma^{(1)}_{\dive} + O(\hbar^2)\;,
\end{eqnarray}
with $\phi_i = \{ A, c, \overline c, b, \psi, \overline \psi \}$, $\rho_i =   \{ K, L, Y, \overline Y \} $, $\lambda_i =   \{ g, \alpha \} $ and the bare fields, sources and parameters defined as
\begin{align}
\phi_{i,0} &= Z_{\phi_i}^{1/2} \phi_i \;,& \rho_{i,0} &= Z_{\rho_i} \rho_i\;, & \lambda_{i,0} &= Z_{\lambda_i} \lambda_i\;.
\end{align}
Let us work out one example and try to determine $Z_A$. To determine this particular constant, \eqref{stable2} is
\begin{multline}
\int \d^4 x \frac{1}{4} (\p_\mu A^a_{\nu,0} - \p_\nu A^{a}_{\mu, 0})^2  = Z_A \int \d^4 x \frac{1}{4} (\p_\mu A^a_{\nu} - \p_\nu A^{a}_{\mu})^2 \\
= \int \d^4 x \frac{1}{4} (\p_\mu A^a_{\nu} - \p_\nu A^{a}_{\mu})^2  - \hbar\Bigl( a_0 \int \d^4 x \frac{1}{4} (\p_\mu A^a_{\nu} - \p_\nu A^{a}_{\mu})^2 + 2a_1 \int  \d^4 x\frac{1}{4} (\p_\mu A^a_{\nu} - \p_\nu A^{a}_{\mu})^2  \Bigr)\;,
\end{multline}
and thus
\begin{align}\label{Za}
Z_A &= 1 - \hbar\left( a_0 + 2 a_1\right) & & \Rightarrow & Z_A^{1/2} &= 1 - \hbar\left(\frac{a_0}{2} + a_1  \right)\;.
\end{align}
The other renormalization factors can be determined in a similar fashion,
\begin{align}\label{Zalles}
&Z_g = 1 + \hbar\left(\frac{a_0}{2}   \right) \;,  & &Z_K = Z_c \;, \nonumber\\
&Z_b^{1/2}  = Z_A^{1/2} \;,   & &Z_L = Z_A^{1/2} = Z_g Z_A^{1/2} Z_c^{1/2} Z_\psi^{-1/2} \;,\nonumber\\
&Z_\alpha = Z_A \;,  && Z_Y = Z_{\overline Y}\;. \nonumber\\
&Z_c^{1/2} = Z_{\overline c}^{1/2} = 1 + \frac{\hbar}{2} \left(a_1 + a_2 \right)\;, \nonumber\\
&Z_\psi^{1/2}= Z_{\overline \psi}^{1/2} = 1 - \hbar\left(\frac{a_3}{2}   \right)\;,
\end{align}
As all counterterms can be absorbed, we have proven the action to be renormalizable. Notice that the coupling constant $g$ is renormalized with the non trivial counterterm, while all the other (non physical) fields and parameters are renormalized with the trivial counterterm.

\subsection{Algebraic renormalization of the YM action in the Landau gauge\label{5.5}}
We can also prove the Yang-Mills action to be renormalizable when choosing immediately the Landau gauge, $\alpha = 0$. The action remains stable under quantum corrections as an extra identity comes into the game, namely the Ghost-Ward identity,
\begin{equation}\label{GW}
\mathcal{G}^{a}( S_\YM + S_\gf)=\Delta _{\mathrm{cl}}^{a}\,,
\end{equation}
with
\begin{eqnarray}
\mathcal{G}^{a} &=&\int \d^dx\left( \frac{\delta }{\delta c^{a}}+gf^{abc} \overline{c}^{b}\frac{\delta }{\delta b^{c}} \right) \,,
\end{eqnarray}
and
\begin{equation}
\Delta _{\mathrm{cl}}^{a}=g\int \d^{4}xf^{abc}\left( K_{\mu}^{b}A_{\mu }^{c}-L^{b}c^{c}\right) \;,
\end{equation}
a linear breaking. In an analogical fashion as before, one can show that this identity shall put a constraint on the counterterm, namely
\begin{eqnarray}
\mathcal{G}^{a} \Sigma_{\mathrm{n.triv.}} &=& 0 \;.
\end{eqnarray}
Therefore, in the counterterm \eqref{tegentermYM}, $a_3$ shall be equal to zero. This comes down to setting $a_2 = 0$ in \eqref{triviaal} and it is trivial so see that the counterterm can be absorbed with the renormalization factors equal to expression \eqref{Za}-\eqref{Zalles} whereby $a_2 = 0$.\\
\\
For other fixed values of $\alpha$, $\alpha = \alpha^*$, the action is not stable as we cannot reabsorb the divergences. Though, is it necessary to let $\alpha$ to be free, as we did previously. We needed to introduce the associated $Z_\alpha$-factor, thereby ensuring renormalizability for the general $\alpha$ case. Consequently, $\alpha$ itself has become a running quantity.
Since keeping $\alpha=0$ also gives a renormalizable action, we conclude that $\alpha=0$ must be a fixed point of the $\alpha$ renormalization group equation.

\subsection{Conclusion}
We have here shown in detail that the Yang-Mills action is renormalizable when no anomaly is present. From this example, one can see that algebraic renormalization provides a very systematic and powerful technique to prove that an action is renormalizable. As the Yang-Mills action is the basis of the standard model, we have here given a first peek view on how one could renormalize the standard model. However, this is far more elaborated and beyond the scope of this small introduction and we refer to the literature for an algebraic proof of the renormalization of the elektroweak standard model \cite{Kraus:1997bi}.


\chapter{From Gribov to the Gribov-Zwanziger action\label{chapgribovtoGZ}}
In this chapter we shall give an overview of the literature concerning the Gribov problem. First, we shall uncover the Gribov problem in detail by reviewing the Faddeev-Popov quantization \cite{Faddeev:1967fc}. Next, we shall treat the Gribov problem semi-classically, as done in \cite{Gribov:1977wm}. Then, we shall try to translate the ideas of Gribov into a quantum field theory by formulating the Gribov-Zwanziger action \cite{Zwanziger:1989mf,Zwanziger:1992qr}.

\section{The Faddeev-Popov quantization}
In this section we shall repeat the Faddeev-Popov quantization which solved the quantization problem of the Yang-Mills action\cite{DeWitt:1964yg,DeWitt:1967yk,Faddeev:1967fc}. Although this has now become a standard textbook item (see \cite{Weinberg:1996kr,Masujima,Ryder,Zinn-Justin2002,thooft2007} for some examples), we shall go into the details of the calculations to point out some subtleties which are rarely mentioned.

\subsection{Zero modes}
Let us start again with the Yang-Mills action as introduced in section \ref{yangmillsintro} of chapter \ref{algebraic}. For this part, we shall only consider the pure gluonic part, as the quantization problem is related this sector. We recall that
\begin{eqnarray}
S_{\YM} &=& \int \d^d x \frac{1}{4}  F_{\mu\nu}^a  F_{\mu\nu}^a \;.
\end{eqnarray}
Naively, we would assume the generating functional $Z(J)$ to be defined by equation \eqref{genZ},
\begin{eqnarray}\label{genym}
Z(J) &=& \int [\d A] \e^{-S_\YM + \int \d x J_\mu^a A_\mu^a}\;.
\end{eqnarray}
Unfortunately, this functional is not well defined. Indeed, taking only the quadratic part of the action,
\begin{eqnarray}
Z(J)_\quadr &=& \int [\d A] \e^{- \frac{1}{4} \int \d x (\p_\mu A_\nu(x) - \p_\nu A_\mu(x) )^2 + \int \d x J_\mu^a(x) A_\mu^a(x)} \nonumber\\
            &=& \int [\d A] \e^{ \frac{1}{2} \int \d x \d y  A_\nu^a(x)\left[ \delta^{ab} \delta (x -y)  ( \p^2 \delta_{\mu\nu} - \p_\mu \p_\nu )\right] A_\mu^b(y) + \int \d x J_\mu^a(x) A_\mu^a (x)}\;,
\end{eqnarray}
and performing a Gaussian integration \eqref{gauss1}
\begin{eqnarray}
Z(J)_\quadr            &=& (\det A)^{-1/2} \int [\d A] \e^{ -\frac{1}{2} \int \d x \d y  J_\nu^a(x)  A_{\mu\nu}(x,y)^{-1} J_\mu^a(y)} \;,
\end{eqnarray}
with $A_{\mu\nu}(x,y) =  \delta (x-y)  ( \p^2 \delta_{\mu\nu} - \p_\mu \p_\nu )$, we see that this expression is ill-defined as the matrix $A_{\mu\nu}(x,y)$ is not invertible. This matrix has vectors with zero eigenvalues, e.g.~the vector $ Y_\mu(x) = \p_\mu \chi (x)$,
\begin{eqnarray}
\int \d y A_{\mu\nu}(x,y) Y_\nu(y)~=~ \int \d y [\delta (x-y)  ( \p^2 \delta_{\mu\nu} - \p_\mu \p_\nu )] \p_\nu \chi (y) &=& 0\;.
\end{eqnarray}
Therefore, something is wrong with the expression of the generating function \eqref{genym}. Notice that this problem is present for $SU(N)$ Yang-Mills action as well as for QED, i.e.~the abelian version of the Yang-Mills action.\\
 \\
The question is now where do these zero modes come from? Let us consider a gauge transformation \eqref{notinf} of $A_\mu = 0$, whereby we take $U = \exp \ii g X^a \chi^a$,
\begin{eqnarray}
A_{\mu}' &=& - \frac{\ii}{g} (\p_\mu U) U^{\dagger} ~=~  X_a \p_\mu \chi^a \;,
\end{eqnarray}
or thus $A_\mu^{a\prime} = \p_\mu \chi^a$. This means that our examples of zero modes $Y_\mu$ are in fact gauge transformations of $A_\mu =0$. As we are integrating over the complete space of all possible gluon fields $A_\mu$, we are also integrating over gauge equivalent fields. As these give rise to zero modes, we are taking too many configurations into account.

\subsection{A two dimensional example}
\begin{figure}[H]
\begin{center}
\includegraphics[width=8cm]{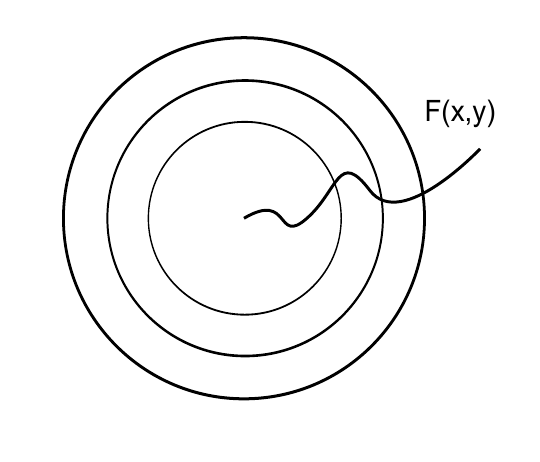}
\caption{Gauge orbits of a system with rotational symmetry in a plane and a function $\mathcal F$ which picks one representative from each gauge.}\label{2fig1}
\end{center}
\end{figure}

To fix the thoughts, let us consider a two dimensional example. Consider an action, $S(r)$, invariant under a rotation in a two-dimensional space,
\begin{eqnarray}\label{start}
W &=& \int \d \vec r \e^{ -S(r)} ~=~ \int_0^{2\pi} \d \theta \int_0^{\infty} r \d r \e^{-S(r)} \;.
\end{eqnarray}
The ``gauge orbits'' of this example are concentric circles in a plane, see Figure \ref{2fig1}. All the points on the same orbit, give rise to the same value of the action $S(\vec r)$. Therefore to calculate $W$, we could also pick from each circle exactly one point, i.e.~the representative of the ``gauge orbit'', and multiply with the number of points on the circle (see Figure \ref{2fig1}). Now how exactly can we implement this?  Mathematically, we know that for each real-valued function, we have that \cite{Riley}
\begin{eqnarray}
\delta ( f (x)) &=& \sum_i \frac{\delta ( x - x_i)}{ | f' (x_i)|} \;,
\end{eqnarray}
with $x_i$ the solutions of $f ( x) = 0$ and provided that $f$ is a continuously differentiable function with $f'$ nowhere zero.
Integrating over $x$ yields
\begin{eqnarray}
\int \d x  \delta (f ( x)) &=& \sum_i \frac{1}{| f' (x_i)|}  \;,
\end{eqnarray}
or thus we find the following unity
\begin{eqnarray}
\frac{1}{\sum_i \frac{1}{ |f' (x_i)|}} \int \d x  \delta ( f ( x))  &=&  1 \;.
\end{eqnarray}
Applying this formula in our 2 dimensional plane, we can write
\begin{eqnarray}\label{gribovc}
\frac{1}{\sum_i \frac{1}{ \left. \left| \frac{\p \mathcal F (r, \phi) }{\p \phi} \right| \right|_{ \mathcal F( r,\phi) =0} }}   \int \d \phi  \delta ( \mathcal F (r, \phi))  &=&  1 \;,
\end{eqnarray}
whereby $\mathcal F$ represents the line which intersects each orbit. Now assuming \textit{that our function $\mathcal F$ intersects each orbit only once}, we can write
\begin{eqnarray}
 \left. \left| \frac{\p \mathcal F (r, \phi) }{\p \phi} \right| \right|_{ \mathcal F( r,\phi) =0}    \int \d \phi  \delta ( \mathcal F (r, \phi))  &=&  1 \;.
\end{eqnarray}
However, to make to analogy with the Yang-Mills gauge theory in the next section, we rewrite this,
\begin{equation} \label{unity}
\left. \left| \frac{\p \mathcal F (r, \theta + \phi ) }{\p \phi} \right| \right|_{ \mathcal F( r,\theta + \phi) =0}    \int \d \phi  \delta ( \mathcal F (r, \theta + \phi))  =  \left. \left| \frac{\p \mathcal F (\vec r^\phi ) }{\p \phi} \right| \right|_{ \mathcal F( \vec r^\phi) =0}    \int \d \phi  \delta ( \mathcal F (\vec r^\phi))  = 1 \;,
\end{equation}
and thus in this notation $\phi$ represents the rotation angle of the vector $\vec r$. This unity shall allow us to pick on every orbit (a given $r$) only one representative, where $\mathcal F( \vec r^\phi) = 0$. Notice however that for every representative we have to multiply with a Jacobian, i.e.~the derivative of the function $\mathcal F$ in this point with respect to the symmetry parameter $\phi$. As this Jacobian shall only depend on the distance $r$, we denote this measure as follows
\begin{equation}\label{prefaddeev}
\Delta_{\mathcal F}(r) = \left. \left| \frac{\p \mathcal F (\vec r^\phi ) }{\p \phi} \right| \right|_{ \mathcal F( \vec r^\phi) =0} \;.
\end{equation}
Now inserting this unity into expression \eqref{start},
\begin{equation}
W = \int \d \theta \int r \d r  \Delta_{\mathcal F}(r)     \int \d \phi   \delta ( \mathcal F ( r, \theta + \phi )) \e^{\ii S(r)} \;,
\end{equation}
and transforming $\theta \to \theta -\phi$,
\begin{eqnarray} \label{2transf}
W  &= &\int \d \phi  \int \d \theta \int r \d r  \Delta_{\mathcal F}(r)     \delta ( \mathcal F ( r, \theta )) \e^{\ii S(r)}\;,
\end{eqnarray}
we are able to perform the integration over $\phi$, which gives a factor of $2 \pi$.
\begin{equation}
W = 2 \pi \int \d \theta \int r \d r  \Delta_{\mathcal F}(r)     \delta ( \mathcal F ( r, \theta )) \e^{\ii S(r)} \;.
\end{equation}
This factor represents the ``volume'' of each orbit. We shall see that we obtain something similar for the Yang-Mills action.\\
\\
Finally, let us remark that it is of uttermost importance that $\mathcal F$ intersects each orbit only once. Else, this derivation is not valid, and one should continue with formula \eqref{gribovc}.

\subsection{The Yang-Mills action}
We can repeat an analogous story for the Yang-Mills action \cite{Faddeev:1967fc,bach,Pokorski}, by keeping in mind the pictorial view of the previous section. Due to the gauge invariance of the Yang-Mills action, we can also divide the configuration space ${A_\mu (x)}$ into gauge orbits of equivalent classes. Two points of one equivalency class are always connected by a gauge transformation $U = \exp(-\ii g X^a \theta^a)$,
\begin{eqnarray}\label{AU}
A^U = U A_\mu U^{\dagger} - \frac{\ii}{g} (\p_\mu U) U^{\dagger} \;,
\end{eqnarray}
 see equations \eqref{notinf} and \eqref{U}. In analogy with the two dimensional example, we shall therefore also try to pick only one representative from each gauge orbit, which shall define a surface in gauge-field configuration space. As we are now working in a multivariable setting, i.e.~infinite dimensional space time coordinate system $x$, and the $N^2 - 1$ dimensional color coordinate system $a, b, \ldots$, the analogue of the unity \eqref{unity} becomes,
\begin{equation}\label{unity2}
\Delta_{\mathcal F}  \int  [\d U]  \delta ( \mathcal F (A^U ))  =  1 \;,
\end{equation}
whereby we have used a shorthand notation,
\begin{eqnarray}
\delta ( \mathcal F (A^U )) &=& \prod_{x } \prod_a \delta ( \mathcal F^a (A_\mu^U (x) ) \nonumber\\
\left[\d U\right] & \sim & \prod_x \prod_a \d \theta^a (x) \;.
\end{eqnarray}
Due to the multivariable system, the Jacobian \eqref{prefaddeev} needs to be replaced by the absolute value of the determinant,
\begin{equation}\label{absvalue}
\Delta_{\mathcal F} (A)=  |\det \mathcal  M_{ab} (x,y) | \qquad \text{with} \qquad  \mathcal M_{ab} (x,y) = \left. \frac{\delta \mathcal F^a (A_\mu^U (x))  }{\delta \theta^b(y)} \right|_{\mathcal F(A^U) =0} \;.
\end{equation}
This determinant is called the Faddeev-Popov determinant. Just as in the two dimensional example, this determinant is independent from the gauge parameter $\theta^a$. \\
\\
Inserting this unity into the generating function \eqref{genym} gives
\begin{eqnarray}\label{chap2Z2}
Z &=&  \int [ \d U] \int [\d A]  \Delta_{\mathcal F}(A)    \delta ( \mathcal F (A^U))   \e^{-S_\YM } \;,
\end{eqnarray}
whereby we omit for a moment the part $J A$. Analogous to the two dimensional example (see equation \eqref{2transf}), we perform a gauge transformation of the field $A \to A^{U^\dagger}$, so that $A_\mu^U$ transforms back to $A_\mu$:
\begin{equation}
A^U_\mu = U A_\mu U^{\dagger} - \frac{\ii}{g} (\p_\mu U) U^{\dagger} \qquad \to\qquad  U A_\mu^{U^\dagger} U^{\dagger} - \frac{\ii}{g} (\p_\mu U) U^{\dagger} = A_\mu \;.
\end{equation}
Expression \eqref{chap2Z2} becomes
\begin{eqnarray}
Z &=&  \int [ \d U] \int [\d A]  \Delta_{\mathcal F}(A)    \delta ( \mathcal F (A))   \e^{-S_\YM } \;,
\end{eqnarray}
as the action, the measure $ [\d A] $ and the Faddeev-Popov determinant are invariant under gauge transformations. Now we have isolated the integration over the gauge group $U$, so we find
\begin{eqnarray}\label{genZZZ}
Z &=& V \int [\d A]  \Delta_{\mathcal F}(A)    \delta ( \mathcal F (A))   \e^{-S_\YM } \;,
\end{eqnarray}
with $V$ an infinite constant. As explained in section \ref{thegeneratingfunctional} of chapter \ref{algebraic}, one can always omit constant factors. It is exactly this infinite constant which made the path integral \eqref{genym} ill-defined.\\
\\
Let us now work out the Faddeev-Popov determinant. This determinant is gauge invariant, and does not depend on $\theta^a$, therefore we can choose $A$ so that if satisfies the gauge condition  $\mathcal F (A) = 0$. In this case, we can set $\theta^a =0$,
\begin{equation}
  \mathcal M_{ab} (x,y) = \left. \frac{\delta \mathcal F^a (A_\mu^U (x))  }{\delta \theta^b(y)} \right|_{\theta =0 \& \mathcal F(A) =0 } \;.
\end{equation}
Applying the chain rule yields,
\begin{eqnarray}
 \mathcal M_{ab} (x,y) &=& \int \d z \left. \frac{\delta \mathcal F^a (A_\mu (x))   }{\delta A_\mu^c (z)} \frac{\delta A^{c,U}_\mu (z)} {\delta \theta^b(y)}    \right|_{\theta =0 \& \mathcal F(A) =0 }\;.
\end{eqnarray}
First working out expression \eqref{AU} for small $\theta$
\begin{eqnarray}
A_\mu^U &=& A_\mu - (D_\mu \theta)^a X^a + O(\theta^2) \;,
\end{eqnarray}
and thus
\begin{eqnarray}
 \mathcal M_{ab} (x,y) &=& \int \d z \left. \frac{\delta \mathcal F^a (A_\mu (x))   }{\delta A_\mu^c (z)} (- D_\mu^{bc} \delta (y-z) )   \right|_{ \mathcal F(A) =0 }\;.
\end{eqnarray}
This is the most general expression one can obtain without actually choosing the gauge condition, $\mathcal F$. Let us now continue to work out the Faddeev-Popov determinant for the linear covariant gauges. For this, we start from the Lorentz condition, i.e.
\begin{eqnarray}\label{notideal}
\mathcal F^a (A_\mu (x)) &=& \p_\mu A^{\mu a} (x) - B^a (x) \;,
\end{eqnarray}
with $B^a (x)$ an arbitrary scalar field.  With this condition, $ \mathcal M_{ab} (x,y) $ becomes,
\begin{eqnarray}\label{mab}
 \mathcal M_{ab} (x,y) &=&  \left.    - \p_\mu  D_\mu^{ab} \delta (y-x)    \right|_{ \mathcal F(A) =0 } \;.
\end{eqnarray}
Because in the delta function in expression \eqref{genZZZ}, the condition $\mathcal F(A) =0$ is automatically fulfilled, so we find,
\begin{eqnarray}
Z&=& \int [\d A] [\det [ - \p_\mu  D_\mu^{ab} \delta (y-x) ]]    \delta (  \p A - B)   \e^{-S_\YM } \;.
\end{eqnarray}
Still, we cannot calculate with this form. However, luckily, there is a way to lift this determinant into the action. For this, we need to introduce Grasmann variables. As described in the appendix in expression \eqref{ghostapp}, we have that (by setting $\eta = \overline \eta =0$)
\begin{eqnarray}\label{absvalue2}
\det \mathcal M_{ab} (x,y) &=& \int [\d c] [\d \overline c] \exp \int \d x \d y \overline c^a (x) \mathcal M_{ab} (x,y) c^b (y)\;,
\end{eqnarray}
and thus
\begin{eqnarray}\label{genz5}
Z&=& \int [\d A][\d c] [\d \overline c]      \delta (\p A - B )   \exp \left[- S_\YM  -   \int \d x  \overline c^a (x)  \p_\mu  D_\mu^{ab}  c^b (x)  \right] \;,
\end{eqnarray}
and we have been able to convert to determinant into a part in the action.\\
\\
Finally, we would like to get rid of the dirac delta function. For this, we can perform a little trick: since gauge-invariant quantities should not be sensitive to changes of auxiliary conditions, we average over the arbitrary field $B^a (x)$ by multiplying with a Gaussian factor,
\begin{eqnarray}
  \int [\d B] \delta (\p A - B ) \exp \left( \frac{1}{2 \alpha} \int \d x B^2 \right)  &=& \exp \left( \frac{1}{2 \alpha} \int \d x (\p_\mu A_\mu^a )^2 \right)\;,
\end{eqnarray}
whereby $\alpha$ corresponds to the width of the Gaussian distribution. Taking all the results together, we obtain the following gauge fixed action:
\begin{eqnarray}\label{actiongauge}
S &=& S_\YM + \underbrace{\int \d x \left(   \overline c^a   \p_\mu  D_\mu^{ab}  c^b -  \frac{1}{2 \alpha} (\p_\mu A_\mu^a )^2   \right)}_{S_\gf}\;,
\end{eqnarray}
which is exactly the action \eqref{Yangmillsgaugefixalpha} where we started from in chapter \ref{algebraic}. This gauge is called the \textit{linear covariant gauge}. Taking the limit $\alpha \to 0$ returns the \textit{Landau gauge}. In this case the width $\alpha$ vanishes and thus the Landau gauge is equivalent to the Lorentz gauge \eqref{notideal} with $B = 0$. Another widely know gauge is the \textit{Feynman gauge} whereby $\alpha =1$. The Landau gauge has the advantage of being a fixed point under renormalization, while in the Feynman gauge, the form of the gluon propagator has the most simple form.\\
\\
In conclusion, we have obtained the following well defined generating functional:
\begin{eqnarray}
Z(J) &=&   \int [\d A][\d c] [\d \overline c]         \exp \left[- S + \int \d x J_\mu^a A_\mu^a  \right]\;,
\end{eqnarray}
with the action $S$ given in equation \eqref{actiongauge}.

\subsubsection{Two important remarks concerning the Faddeev-Popov derivation}
We need to make two important remarks.
\begin{itemize}\label{twoimportantremarks}
\item First, notice that in fact, we need to take \textit{the absolute value} of the determinant, see expression \eqref{absvalue}. In most standard textbooks, this absolute value is immediately omitted, without mentioning that mathematically, it should be there. Subsequently, in equation \eqref{absvalue2}, we have neglected this absolute value in order to introduce the ghosts. It was thus implicitly assumed that this determinant is always positive. However, in the next section, we shall prove that this is not always the case. Only when considering infinitesimal fluctuations around $A_\mu =0$, i.e.~in perturbation theory, this determinant is a positive quantity (see section \ref{thegribovproblem}.)
\item Secondly, closely related to the first remark, this derivation is done is the assumption of having a gauge condition which intersects with each orbit \textit{only once}. We call this an \textit{ideal} condition. If this is not the case, one should in fact continue with an analogous formula as \eqref{gribovc}, namely,
    \begin{eqnarray}
    1 + N(A) &=& \Delta_{\mathcal F}  \int  [\d U]  \delta ( \mathcal F (A^U ))\;,
    \end{eqnarray}
    whereby $N(A)$ is the number of Gribov copies for a given orbit\footnote{For each copy, the Faddeev-Popov determinant shall be the same, therefore, the sum in equation \eqref{gribovc} can be replaced by $1 + N(A)$.}.  Again, in the next section, we shall show that the condition \eqref{notideal} is not ideal by demonstrating that the orbit can be intersected more than once. In fact, a mistake is made here.
\end{itemize}

\subsubsection{Other gauges}
Here we have worked out the Faddeev-Popov quantization for the \textit{Linear covariant gauges} which encloses the \textit{Landau} and \textit{Feynman gauge} as a special case. However, many other gauges are possible. We can divide the gauges in several classes. The first class are the \textit{covariant gauges}, which besides the linear covariant gauges also holds e.g.~the \textit{'t Hooft gauges} \cite{'tHooft:1971fh} and the \textit{background fields gauges} \cite{DeWitt:1967ub}. A second class of gauges are the noncovariant gauges, i.e.~gauges which break Lorentz invariance. The most famous example is probably the \textit{Coulomb gauge}, whereby $\mathcal F^a = \nabla_i A_i^a$ (see e.g.~\cite{Kleinert:2006} for a derivation). Some other examples are the \textit{axial gauge}, the \textit{planar gauge}, \textit{light-cone gauge} and the \textit{temporal gauge}. A nice overview on the second class can be found in \cite{Leibbrandt:1987qv}.
Finally, there are some other gauges like the \textit{Maximal Abelian gauge} \cite{Shinohara:2001cw}, which breaks color symmetry, and some more exotic gauges which break translation invariance. For a nice overview on different gauges, we refer to \cite{Klaus}.\\
\\
For this thesis, we shall mainly work in the Landau gauge.


\subsubsection{The BRST symmetry}\label{globalmark}
Now that we have fixed the gauge, the local gauge symmetry is obviously broken. Notice that on the other hand the global gauge symmetry is still present as one can check by performing a global gauge transformation on the action \eqref{actiongauge}. Fortunately, as discussed in chapter \ref{algebraic}, after fixing the gauge a remaining symmetry is still present, namely the BRST symmetry. This BRST symmetry is most easily seen when introducing the $b$-field, as done is equation \eqref{BRs}
\begin{eqnarray}
S &=& S_\YM +  \int \d^d x \left( b^a \p_\mu A_\mu^a + \alpha \frac{(b^a)^2}{2} + \overline{c}^{a}\partial _{\mu } D_{\mu}^{ab}c^b \right) \;,
\end{eqnarray}
whereby the nilpotent BRST symmetry $s$ is given by expression \eqref{BRST}.\\
\\
\label{physicalsubspace}This BRST symmetry is of uttermost importance, as it is useful for several properties. Firstly, as we have already proven, it was the key to the proof of the renormalizability of the Yang-Mills action. Secondly, the BRST symmetry shall also be the key to the proof that the Yang-Mills action is \textit{unitary}. Let us explain what unitarity means. We define the physical state space $\mathcal H_\subs$, which is a subspace of the total Hilbert space, as the set of all physical states $\ket{\psi}_\phys$. A physical state is defined by the cohomology of the free BRST symmetry\footnote{This means, switch off interactions or set $g= 0$.} \cite{Slavnov:1989jh,Frolov:1989az}
\begin{equation}
s_0 \ket{\psi}_\phys = 0 \qquad \text{and} \qquad  \ket{\psi}_\phys \not= s_0 (\ldots) + \ldots \;,
\end{equation}
with $s_0$ is the free BRST symmetry. Now a theory is unitary if
\begin{enumerate}
\item Starting from physical states belonging to $\mathcal H_\subs$, after these states have interacted, one should end up again with physical states $\in \mathcal H_\subs$.
\item All physical states should have a positive norm.
\end{enumerate}
It will exactly be the BRST symmetry which will allow to prove these properties\footnote{See \cite{Lehmann:1954rq,Henneaux:1992ig} for the original proofs, or \cite{Dudal:2007ch} for a more recent version of the proof.}. Also notice that by fixing the gauge, we have introduced extra particles, the ghost particles $c$ and $\overline c$. As these particles are scalar and anticommuting, they violate the spin statistics theorem. If the theory wants to be valid, these ghost particles have to be excluded from the physical spectrum. This is of course related to issue of unitary and one can show that the ghosts are indeed excluded from $\mathcal H_\subs$ by invoking the BRST symmetry.\\
\\
Finally, let us remark that in some approaches, BRST symmetry is regarded as a first principle of a gauge fixed Lagrangian \cite{Baulieu:1983tg,Nakanishi:1990qm}, rather than instead using Faddeev Popov quantization.

\section{The Gribov problem\label{thegribovproblem}}
Let us now explicitly show that in the Landau gauge\footnote{From now on, we shall work in the Landau gauge, unless explicitly mentioned.}, the gauge condition is not ideal. Gribov demonstrated this first in his famous article \cite{Gribov:1977wm} in 1977, which has been reworked pedagogically in \cite{Sobreiro:2004us}. For the gauge condition \eqref{notideal}, he explained that one can have three possibilities. A gauge orbit can intersect with the gauge condition only once ($L$), more than once ($L'$) or it can have no intersection ($L''$). In \cite{Gribov:1977wm} Gribov explains that no examples of the type $L''$ are known, however that many examples of the type $L'$ are possible.

\begin{figure}[H]
\begin{center}
\includegraphics[width=8cm]{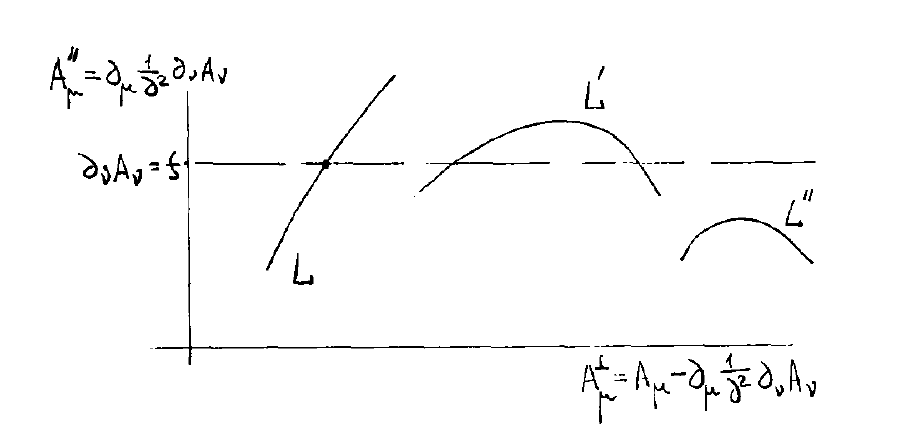}
\caption{The three possibilities for a gauge orbit w.r.t.~a gauge condition. Original figure from \cite{Gribov:1977wm}.}\label{2figgribov}
\end{center}
\end{figure}

\noindent Let us quantify this. Take two equivalent fields, $A_\mu$ and $A_\mu'$ which are connected by a gauge transformation \eqref{notinf}. If they both satisfy the same gauge condition, e.g.~the Landau gauge, we call $A_\mu$ and $A_\mu'$ \textit{Gribov copies}. We can work out this condition a bit further,
\begin{eqnarray}
&A_\mu' = U A_\mu U^{\dagger} - \frac{\ii}{g} (\p_\mu U) U^{\dagger}\;,  \qquad  \p_\mu A_\mu =0 \quad \& \quad  \p_\mu A_\mu' = 0 \;,&\nonumber\\
&\Downarrow& \nonumber\\
& \p_\mu U A_\mu U^{\dagger} +  U A_\mu \p_\mu U^{\dagger} - \frac{\ii}{g} (\p^2_\mu U) U^{\dagger} - \frac{\ii}{g} (\p_\mu U) (\p_\mu U^{\dagger}) =0 \;.&
\end{eqnarray}
Taking an infinitesimal transformation, $U = 1 + \alpha$, $U^\dagger = 1- \alpha$, with $\alpha = \alpha^a X^a$, this expression can be expanded to first order,
\begin{eqnarray}\label{zeromode}
-\p_\mu (\p_\mu \alpha + \ii g  [\alpha, A_\mu]) &=& 0\;,
\end{eqnarray}
or from equation \eqref{covariantderivativeadjoint} we see that this is equivalent with  
\begin{eqnarray}
- \p_\mu D_\mu \alpha &=& 0\;.
\end{eqnarray}
In conclusion, the existence of (infinitesimal) Gribov copies is connected to the existence of zero eigenvalues of the Faddeev-Popov operator. This is a very important insight as now we can understand the two remarks made in the previous section. \\
\\
Firstly, for small $A_\mu$, this equation reduces to $-\p_\mu^2 \alpha = 0$. However, it is obvious that the eigenvalue equation
\begin{equation}
-\p_\mu^2 \psi = \epsilon \psi \;,
\end{equation}
only has positive eigenvalues $\epsilon = p^2 > 0$. This means that also for small values of $A_\mu$ we can expect the eigenvalues $\epsilon (A)$ to be larger than zero. However, for larger $A_\mu$, this cannot be guaranteed anymore. Therefore, negative eigenvalues can appear (and will appear) and thus the Faddeev-Popov operator shall also have zero eigenvalues. This means that our gauge condition is not ideal. Secondly, if the Faddeev-Popov operator has negative eigenvalues, the determinant of this operator can switch sign and the positivity of this determinant is no longer ensured. An explicit construction of a zero mode of the Faddeev Popov operator has been worked out in \cite{Henyey:1978qd,vanBaal:1991zw,Sobreiro:2004us}.\\
\\
Finally, let us also mention that in QED no Gribov copies are present. We can show this with a simple argument. In QED, the gauge transformations are given by
\begin{eqnarray}
A_\mu' &=& A_\mu - \p_\mu \chi \;,
\end{eqnarray}
and thus, for the Landau gauge, $\p_\mu A_\mu =0$, the condition for $A_\mu'$ to be a gauge copy of $A_\mu$ becomes
\begin{eqnarray}
 \p_\mu A_\mu' = 0 &\Rightarrow& \p_\mu^2 \chi = 0 \;,
\end{eqnarray}
which does not have any solutions besides plane waves. As a plane wave does not vanish at infinity, they cannot be used for constructing a gauge copy $A_\mu'$.

\subsection{The Gribov region: a possible solution to the Gribov problem?}
\subsubsection{Definition of the Gribov region}
\begin{figure}[h]
\begin{center}
\includegraphics[width=8cm]{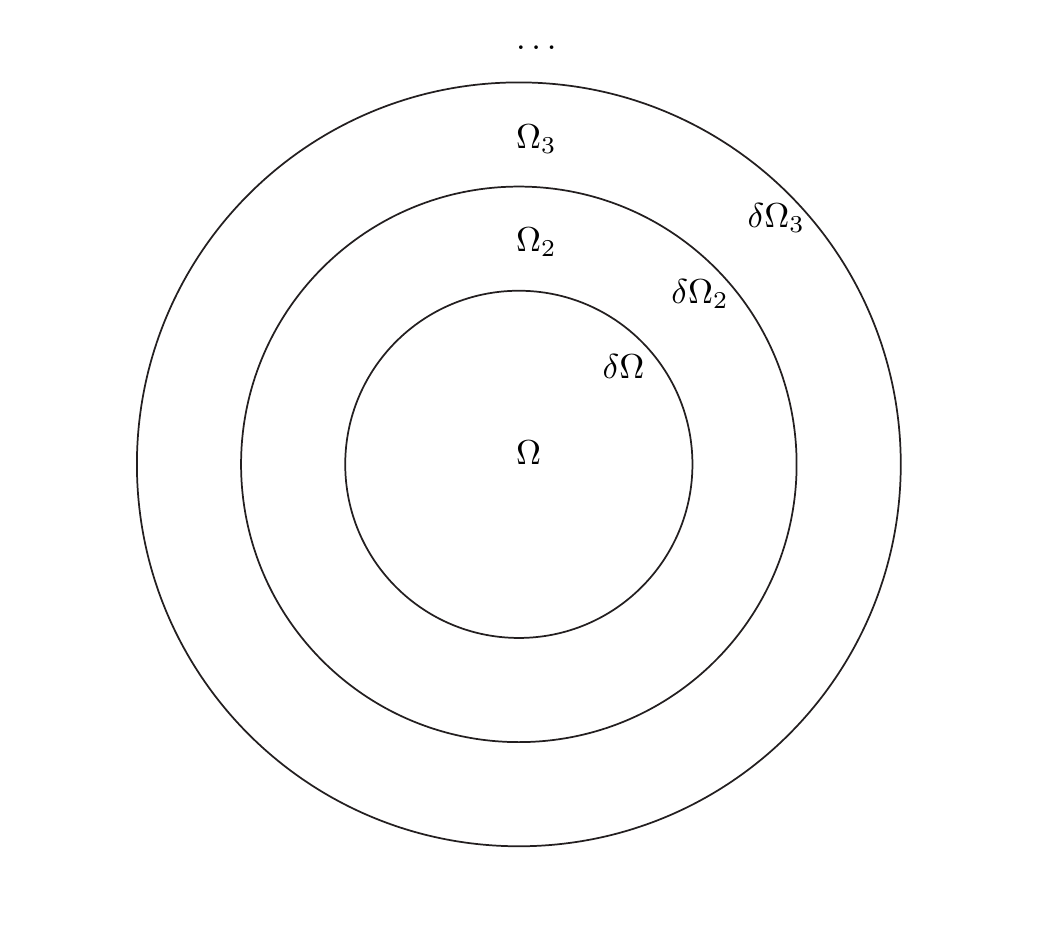}
\caption{The different regions in the hyperspace $\p A = 0$.}\label{2fighorizon}
\end{center}
\end{figure}

Now that we have shown that the Faddeev-Popov quantization is incomplete, we need to improve the gauge fixing. Gribov was the first to propose in 1977 \cite{Gribov:1977wm} to further restrict to a region of integration, the so-called Gribov region $\Omega$, which is defined as follows:
\begin{eqnarray}\label{defgribovregion}
\Omega &\equiv &\{ A^a_{\mu}, \, \p_{\mu} A^a_{\mu}=0, \, \mathcal{M}^{ab}  >0  \} \,,
\end{eqnarray}
whereby the Faddeev-Popov operator $\mathcal{M}^{ab}$ is given by equation \eqref{mab}
\begin{eqnarray}\label{mab2}
\mathcal M^{ab}(x,y) &=& - \p_\mu  D_\mu^{ab} \delta(x-y)  = \left( - \p_\mu^2 \delta^{ab} +  \p_\mu f_{abc} A_\mu^c \right) \delta(x-y)\;.
\end{eqnarray}
This is the region of gauge fields obeying the Landau gauge and for which the Faddeev-Popov operator is positive definite. We recall that a matrix is positive definite if for all vectors $\omega$,
\begin{eqnarray}\label{positivedef}
\int \d x \d y \omega^a(x)  \mathcal M^{ab} (x,y) \omega^b(y) > 0 \;.
\end{eqnarray}
In this way the problem of the absolute value of the determinant would already be solved (first remark on p.\pageref{twoimportantremarks}). The border of this region $\delta \Omega$ is called the first Gribov horizon and at this border the first eigenvalue of the Faddeev-Popov operator becomes zero. Crossing this horizon, this eigenvalue becomes negative. This is depicted in Figure \ref{2fighorizon}. Consecutively, one can define the other horizons similarly, as drawn on the picture, where the second ($\delta \Omega_2$), the third ($\delta \Omega_3$), \ldots eigenvalue becomes zero. However, keep in mind that this picture is a very simplified pictorial view. In reality, the space of gauge fields is much more complicated.

\subsubsection{An alternative formulation of the Gribov region}
We can also define the Gribov region as the set of \textit{relative} minima of the following functional\footnote{The original derivation can be found in \cite{Semenov1986,Maskawa:1978rf,Zwanziger1982a}, a more recent version in the appendix of \cite{Capri:2005dy}. }
\begin{eqnarray}\label{functional}
||A^U ||^2 &=&   \Tr \int \d x A^U_\mu (x) A^U_\mu (x)  = \frac{1}{2} \int \d x A^{a U}_\mu (x) A^{a U}_\mu (x) \;,
\end{eqnarray}
which corresponds to selecting on each gauge orbit the gauge configurations which minimizes $A^2$. Notice that there can be more than one minimum. It can be seen relatively easy  that this definition agrees with the Gribov region.
Assume we have a gluon field $A_\mu$ which minimizes the functional \eqref{functional}. Firstly, in order to have an extremum, varying $||A||^2$ w.r.t.~an infinitesimal gauge transformation \eqref{infinitesimal} must be zero,
\begin{eqnarray}
\delta ||A||^2 &=& \delta  \left( \frac{1}{2} \int \d x A^{a }_\mu (x) A^{a }_\mu (x) \right) = \int \d x \delta A^{a }_\mu (x) A^{a }_\mu (x) = -\int \d x D_\mu^{ab} \theta^b(x) A^{a }_\mu (x) \nonumber\\
&=& - \int \d x \p_\mu \theta^a(x) A^{a }_\mu (x)  =  \int \d x \theta^a(x) \p_\mu  A^{a }_\mu (x) = 0\;.
\end{eqnarray}
As this equation must be zero for all $\theta^a$, we must have that $\p_\mu  A^{a }_\mu (x) = 0$. Secondly, this extremum must be a minimum, therefore, differentiating again,
\begin{eqnarray}
\delta^2 ||A||^2 &=&  - \int \d x \p_\mu  \theta^a (x) \delta A^{a }_\mu (x)  = \int \d x  \theta^a (x) (-\p_\mu  D_\mu^{ab}) \theta^b (x) > 0 \quad \forall\ \theta \;.
\end{eqnarray}
This implies that the operator $-\p_\mu  D_\mu^{ab} = \mathcal M^{ab}$ must be positive definite, see equation \eqref{positivedef}.

\subsubsection{Properties of the Gribov region}
How exactly one can implement this restriction, is the topic of the next sections. First, we shall answer some profound questions and discuss some properties of the Gribov region.
\begin{itemize}
\item \label{propertiesofgribovregion} Firstly, does each orbit of gauge equivalent fields intersect with the Gribov region? This property is of course of paramount importance as it would not make any sense to integrate over an incomplete region of gauge fields. A first step towards establishing this property was made in \cite{Gribov:1977wm} where is was proved that for every field infinitesimally close to the horizon $\delta \Omega$, there exists an gauge copy at the other side of the horizon, infinitesimally close again. Later, it was then actually proven that every gauge orbit indeed intersects with the Gribov region \cite{Zwanziger1982a,Dell'Antonio:1991xt}. In \cite{Dell'Antonio:1991xt} is was mathematically rigorously proven that for every gauge orbit, the functional \eqref{functional} achieves its absolute minimum. Moreover, since every minimum belongs to the Gribov region, every gauge orbit intersects with the Gribov region.
\item Does $A_\mu =0$ belong to the Gribov region? This is important as this means that the perturbative region is also incorporated in the Gribov region. In fact, we can prove this very easily (see the appendix of \cite{Zwanziger:2003cf}).  Taking $A_\mu =0$, the Faddeev-Popov operator becomes $\mathcal M^{ab} = -\p^2 \delta^{ab}$, which is positive definite.
\item We can also prove that the Gribov region is convex \cite{Zwanziger:2003cf}. This means that for two gluon fields $A_\mu^1$ and $A_\mu^2$ belonging to the Gribov region, also the gluon field $A_\mu = \alpha A_\mu^1 + \beta A_\mu^2$ with $\alpha, \beta \geq 0$ and $\alpha + \beta  =1$, is inside the Gribov region. To demonstrate this, we need to show that $\mathcal M^{ab} ( \alpha A_\mu^1 + \beta A_\mu^2)$ is positive definite. However, from expression \eqref{mab2}, we immediately see that
    \begin{eqnarray*}
    \mathcal M^{ab} ( \alpha A_\mu^1 + \beta A_\mu^2) &=& \alpha \mathcal M^{ab} (A_\mu^1) + \beta \mathcal M^{ab} (A_\mu^2) \;.
    \end{eqnarray*}
    As $\alpha, \beta \geq 0$, this sum of two positive definite matrices is again a positive definite matrix.
\item Finally, we can also show rather easily that the Gribov region is bounded in every direction \cite{Zwanziger:2003cf}. Assume we have a gluon field $A_\mu$ part of the Gribov region $\Omega$, then we can show that the gluon field $\lambda A_\mu$ with $\lambda$ large enough, shall be located outside of $\Omega$. Firstly, as the matrix $ \mathcal M^{ab}_2(A_\mu) = \p_\mu f_{abc} A_\mu^c$ is traceless, the sum of all the eigenvalues of $\mathcal M^{ab}_2$ is zero. Therefore, for $A_\mu \not= 0$, there should     exist at least one eigenvector\footnote{We assume this eigenvector to have norm 1.} $\omega$ with negative eigenvalue $\kappa$, i.e.
     \begin{eqnarray}
      \int \d x \d y \omega^a(x)  \mathcal M_2^{ab} (x,y) \omega^b(y) = \kappa  < 0 \;.
     \end{eqnarray}
     Secondly, as $\mathcal M^{ab}_2(A_\mu)$ is linear in $A_\mu$,  $\mathcal M^{ab}_2( \lambda A_\mu) = \lambda \mathcal M^{ab}_2( A_\mu)$ has the same eigenvector $\omega$ with eigenvalue $\lambda \kappa$. Therefore,
     \begin{eqnarray}
       \int \d x \d y \ \omega^a(x) \mathcal M^{ab}( \lambda A_\mu) (x,y) \omega^b(y)  &=& \int \d x \  \omega^a(x) ( -  \p_\mu^2 )  \omega^a(x) +  \lambda \kappa \;,
     \end{eqnarray}
     which shall become negative for large enough $\lambda$. Consequently, $\mathcal M^{ab}( \lambda A_\mu)$ is no longer positive definite and $\lambda A_\mu$ is located outside the horizon. Therefore, $\Omega$ is bounded in every direction. Moreover, in \cite{Dell'Antonio:1989jn} it has been proven that the Gribov region is contained within a certain ellipsoid.
\end{itemize}
With all these properties, restricting the integration of gluon fields to the Gribov region looks like a very attractive option to improve the gauge fixing. Unfortunately, the Gribov region still contains Gribov copies. This was first discussed in \cite{Semenov1986}. Let us repeat their reasoning. Assume a gluon field $A_\mu$ belonging to the boundary of the Gribov region, then we have that
\begin{eqnarray}
\delta ||A||^2 &=& 0 \;,\nonumber\\
\delta^2 ||A||^2 &\not>& 0 \quad\Rightarrow \quad \exists \theta, \quad  \int \d x  \theta^a (x) \mathcal M^{ab}(x) \theta^b (x) = 0\;.
\end{eqnarray}
As the Faddeev-Popov determinant has zero modes, this means that it is inconclusive whether $||A||^2$ is a minimum\footnote{This is a consequence of the second derivative test as can be found in any textbook on basic mathematics.}. We have to consider the third variation $\delta^3 ||A||^2$,
\begin{eqnarray}
\delta^3 ||A||^2 &=& g f_{abc} \int \d x \p_\mu \theta^a(x) \theta^b(x) D_\mu^{cd}(x) \theta^d(x) \;,
\end{eqnarray}
which is, generally speaking, not zero\footnote{One can compare this with $x^3$ which has a saddlepoint at $x=0$. The first and second derivative are zero, while the third derivative is positive. }. Therefore, a gluon field on the boundary of the Gribov region is not a relative minimum of the functional \eqref{functional} and thus there must exist a transformation $\tilde U$:
\begin{equation}\label{AUA}
||A||^2 > ||A^{\tilde U}||^2 \;.
\end{equation}
Now, suppose we have a function $f$ from $X$ to $Y$, where $X, Y$ are topological spaces. We say a function $f$ is continuous at $x$ for some $x \in X$ if for any neighborhood $V$ of $f(x)$, there is a neighborhood $U$ of $x$ such that $f(U) \subseteq V$. In this view, let us define the following function, $f(A, U) = ||A^U||$, which we can consider a continuous function $\forall A, U$.  Moreover, the function $g(A) = f(A,I) - f(A, \tilde U) $ is also a continuous function. Therefore, for certain $A$ located on the boundary of the Gribov region, we know that $g(A) > 0$, see equation \eqref{AUA}. Due to continuity, we can take a region $V$ around $g(A)$ which is still positive, such that there is a region $U$ around $A$ for which $g(U) \subseteq V$. We consider the field $(1-\epsilon)A_\mu$, with $\epsilon > 0$ and as small as needed to lie inside $U$. From the convexity of the Gribov region, it is easy to understand that $(1-\epsilon)A_\mu \in \Omega$. Therefore, we can conclude that $||[(1-\epsilon) A]^{\tilde U}||^2 < ||(1-\epsilon)A||^2$ and we have that the relative minimum $||(1-\epsilon)A||^2$ is \textit{not} the absolute minimum.\\
\\
Also numerical results have confirmed this, see e.g.~\cite{Cucchieri:1997dx}. In fact, it is not surprising that the Gribov region still contains copies. By looking at the functional \eqref{functional}, it seems obvious that on a gauge orbit, the functional \eqref{functional} can have more than one relative minima. Two relative minima of \eqref{functional} on the same orbit are Gribov copies which both belong to $\Omega$. Also Gribov was already aware of this possibility \cite{Gribov:1977wm}. 


\subsection{The fundamental modular region (FMR)}
\subsubsection{Definition of the FMR}
What is then the configuration space free from Gribov copies? It is obvious that from the functional \eqref{functional} which defines the Gribov region, we can also define a more strict region, i.e.~the set of \textit{absolute} minima of the functional \eqref{functional}. As we take for each gauge orbit, the absolute minimum, we shall select, on a given orbit, only one gluon field, namely the gauge configuration closest to the origin. This region is then called \textit{the fundamental modular region} $\Lambda$. Restricting to this region of integration is also called \textit{the minimal Landau gauge}\footnote{Sometimes this is also called the \textit{absolute Landau gauge}, while the minimal Landau gauge can refer to taking one arbitrary minimum of the functional \eqref{functional}, depending on the author or article. }. $\Lambda$ is then a proper subset of $\Omega$, $\Lambda \subset \Omega$.
Notice that the absolute minimum of the functional \eqref{functional} can only determine the minimum up to a global gauge transformation. Indeed, as mentioned on p.\pageref{globalmark}, fixing the gauge does not break the global gauge symmetry and by performing a global gauge transformation $H$ independent from the space time coordinate $x$, expression \eqref{functional} does not change,
\begin{eqnarray}
||A^U ||_H^2 &=&   \Tr \int \d x H A^U_\mu (x) H^\dagger  H A^U_\mu (x) H^\dagger  =  \Tr \int \d x  A^U_\mu (x) A^U_\mu (x) = ||A^U ||^2  \;.
\end{eqnarray}
Therefore, saying that we picked out from a gauge orbit exactly one configuration always means modulo global gauge transformations.\\
\\
In fact, the FMR $\Lambda$ would be the exact gauge fixing if the global minima of the functional \eqref{functional} are non-degenerate. However, it is proven that degenerate minima can only occur on the boundary of the FMR, $\delta \Lambda$ \cite{Zwanziger:1993dh}. Therefore, if one would integrate over
\begin{eqnarray}
Z &=& \int_{\Lambda} [\d A] \e^{-S_\YM} \;,
\end{eqnarray}
these degenerate minima do not play any role, as they have zero measure. This agrees with endpoints of a function which do not play a role when integrating over a function.

\subsubsection{Properties of the FMR}
Let is discuss again some properties of the FMR, many similar as the Gribov region.
\begin{itemize}
\item Firstly, all gauge orbits intersect with the FMR. This is in fact already demonstrated in the first bullet point on p.\pageref{propertiesofgribovregion}.
\item $A_\mu = 0$ belongs to the FMR as 0 is the smallest possible norm.
\item $\Lambda$ is convex \cite{Semenov1986}. This is a bit more involved to prove than for the case of the Gribov region $\Omega$. We have to show that if $A^1_\mu, A^2_\mu \in \Lambda$, also $B_\mu = t A_\mu^1 + (1-t) A_\mu^2$, with $t \in [0,1]$. For this we work out the functional \eqref{functional},
\begin{eqnarray*}
||A^U ||^2 &=& \Tr \int \d x A^U_\mu (x) A^U_\mu (x)  \nonumber\\
&=& \Tr \int \d x \left( U A_{\mu} U^\dagger - \frac{\ii}{g} (\p_\mu U ) U^\dagger  \right) \left( U A_{\mu} U^\dagger - \frac{\ii}{g} (\p_\mu U ) U^\dagger  \right).	 \nonumber\\
&=& ||A||^2 - 2 \frac{\ii}{g} \Tr \int \d x \left( A_{\mu} U^\dagger \p_\mu U  \right) - \frac{1}{g^2} \Tr \int \d x \left( (\p_\mu U) U^\dagger (\p_\mu U ) U^\dagger  \right) \;.
\end{eqnarray*}
As $A_\mu^1$ and $A_\mu^2$ both belong to the FMR, we have that
\begin{eqnarray*}
||A^{1,U} ||^2 - ||A^1||^2 \geq 0 &\Leftrightarrow& - 2 \frac{\ii}{g} \Tr \int \d x \left( A^1_{\mu} U^\dagger \p_\mu U  \right) - \frac{1}{g^2} \Tr \int \d x \left( (\p_\mu U) U^\dagger (\p_\mu U ) U^\dagger  \right) \geq 0 \nonumber\\
||A^{2,U} ||^2 - ||A^2||^2 \geq 0 &\Leftrightarrow& - 2 \frac{\ii}{g} \Tr \int \d x \left( A^2_{\mu} U^\dagger \p_\mu U  \right) - \frac{1}{g^2} \Tr \int \d x \left( (\p_\mu U) U^\dagger (\p_\mu U ) U^\dagger  \right) \geq 0 \;.
\end{eqnarray*}
Or thus, combining these two inequalities,
\begin{eqnarray}
- 2 \frac{\ii}{g} \Tr \int \d x \left( (t A_\mu^1 + (1-t) A_\mu^2) U^\dagger \p_\mu U  \right) - \frac{1}{g^2} \Tr \int \d x \left( (\p_\mu U) U^\dagger (\p_\mu U ) U^\dagger  \right) \geq 0 \;,
\end{eqnarray}
from which follows
\begin{eqnarray}
||B^{U} ||^2 - ||B||^2 \geq 0 \;.
\end{eqnarray}
Therefore, $B$ belongs to the FMR and the FMR is convex.

\item $\Lambda$ is bounded in every direction. This is obvious as $\Lambda \subset \Omega$ with $\Omega$ bounded in every direction.
\item The border of $\Lambda$, $\delta \Lambda$ has some points in common with the Gribov horizon \cite{vanBaal:1991zw}.
\end{itemize}

\subsection{Other solutions to the Gribov problem}
Firstly, a very important result has been proven by Singer in \cite{Singer:1978dk}, whereby it was shown that all Lorentz invariant gauges suffer from Gribov copies. Therefore, a gauge free a Gribov copies shall thus manifestly break Lorenz invariance, and are therefore very difficult to handle in calculations. These gauges do exist, e.g.~the space like planar gauge \cite{Bassetto:1983rq} which has no Gribov copies. 
\\
\\
Many other attempts have been done, like improving the Faddeev-Popov gauge fixing in \cite{Ghiotti:2005ih} whereby they have managed to lift the absolute value of the Faddeev-Popov determinant into the action. However, they were not able to take into account the number of copies into the action, and therefore, as far as we know, no further calculations have been done in their framework. Also in \cite{Slavnov:2008xz,Quadri:2010vt} an attempt to improve the gauge fixing has been done. However, as far as we know, the meaning of their model remains unclear in the infrared. 
\\
\\
In \cite{Zwanziger:1981kg}, another method is proposed which avoids the Gribov ambiguity. The method is based on stochastic quantization and is an improved implementation of the method proposed in \cite{Parisi:1980ys} where the gauge was not fixed. However, this method is not so convenient to use. \\
\\
Finally, to be complete, we should also mention that the following paper \cite{Hirschfeld:1978yq} claimed that the Faddeev-Popov quantization is correct although Gribov copies are present. Also in \cite{Friedberg:1995ty}, an attempt is done to try to take into account all Gribov copies, by working in a Coulomb-like gauge. In this way, they have avoided the use of a boundary condition like restricting to the Gribov region.

\subsection{Summary}
In conclusion, to be absolutely sure that one has a correct quantization of the Yang-Mills theory, one should really restrict to the FMR in order to have a completely correct gauge fixing whereby only one gauge configuration is chosen per orbit. However, no practical implementation of this region has been found so far in the continuum. Some other attempts of improving gauge fixing are interesting, but not very convenient or too difficult to handle. However, if we restrict ourself to the Gribov region, it is possible to perform practical calculations. Gribov has done this semi-classically, and Zwanziger has managed building an action which automatically restricts to the Gribov region. One can still object as the Gribov region still contains Gribov copies, but there has been a conjecture by Zwanziger \cite{Zwanziger:2003cf,Greensite:2004ke}, that the important configurations lie on the common boundary $\delta \Lambda \cap \delta \Omega$ of the Gribov region $\Omega$ and the FMR $\Lambda$. Therefore, the extra copies inside the Gribov region would not play a significant role, and it would be sufficient to restrict to the Gribov region.

\subsubsection{Important configurations lie on the common boundary $\delta \Lambda \cap \delta \Omega$}
Let us go a bit more into the details of the statement that the important configurations lie on the common boundary $\delta \Lambda \cap \delta \Omega$, as a clear reasoning is not available in literature \cite{private}. \\
\\
We shall start by proving the following statement: {\it If a configuration $A$ satisfies the two conditions (i) it is an absolute minimum of the minimizing functional $||A^U ||^2$, see equation \eqref{functional} and (ii) there exists an $x$-dependent solution $\omega$, that is with $\p_\mu \omega \neq 0$, to the equation,
\begin{equation}\label{degenerate}
D_\mu(A)\omega = 0\;,
\end{equation}
then configuration $A$ lies on the common boundary $\p\Lambda \cap \p\Omega$ of the Gribov region $\Omega$ and the FMR $\Lambda$.}  Here $\omega$ is a common eigenvector of each of the $D_\mu(A)$ for $\mu = 1,\ldots, d$.  \\
\\
To prove the statement we first observe that the equation,
\begin{equation}\label{2zhorizon}
\p_\mu D_\mu(A) \omega = 0\;,
\end{equation}
follows from \eqref{degenerate}. The existence of an $x$-dependent solution $\omega$ to \eqref{2zhorizon} is the defining condition for configurations $A$ to lie on the Gribov horizon $\p\Omega$.\footnote{The condition that $\omega$ be $x$-dependent, $\p_\mu \omega \neq 0,$ is necessary because the minimal Landau gauge condition does not fix global gauge transformations $U = \exp\omega$, and global gauge transformations have as infinitesimal generator $x$-independent $\omega$, with $\p_\mu \omega = 0$.  These satisfy \eqref{2zhorizon} for every transverse configuration $A$ (including those in the interior of $\Lambda$), when $\omega$ is $x$-independent, $\p_\mu\omega = 0$, as is easily verified.  Indeed when $A$ is transverse, we have $\p_\mu D_\mu(A) = D_\mu(A) \p_\mu$, and so $\p_\mu D_\mu(A)\omega = D_\mu(A) \p_\mu \omega = 0$ for any $x$-independent $\omega$.}  Moreover, by condition (i), configuration $A$ is an absolute minimum of $||A^U||$ so it satisfies the defining condition of the fundamental modular region $\Lambda$, and we have $A \in \Lambda$. Furthermore $\Lambda$ is included in $\Omega$, $\Lambda \subset \Omega$, and so, since $A$ lies on the boundary of $\Omega$, $A \in \p\Omega$, it necessarily lies on the boundary of $\Lambda$, $A \in \p\Lambda$. QED.\\
\\
Now we give some argument that part of the configurations on this common boundary $\delta \Lambda \cap \delta \Omega$, shall dominate. The argument goes as follows. In the maximal Abelian gauge, it is stated that Abelian configurations (or more precisely those nearby) dominate the functional integral \cite{Ezawa:1982bf}, and would be responsible for confinement. In $SU(2)$, any Abelian configuration can be written as\footnote{This can be generalized to other $SU(N)$.}
\begin{equation}\label{abelian}
{A'}_\mu^a(x) = \delta^{a3} a_\mu(x)\;,
\end{equation}
whereby $a_\mu(x)$ is an arbitrary function. This configuration possesses the degeneracy property \eqref{degenerate}.  Indeed for ${\omega'}^a = c \delta^{a3}$, where $c$ is a constant, we have
\begin{equation}
D_\mu(A')\omega' = \p_\mu \omega' - \ii g [A'_\mu, \omega] = 0\;.
\end{equation}
When this configuration is gauge transformed to the absolute minimum, $A = A^{\prime U}$ on its gauge orbit, we have $A \in \Lambda$ and $D_\mu(A) \omega = 0$, as $D_\mu(A) \omega = 0$ is a gauge invariant equation, which is proven in appendix \ref{sigma}\ref{appgaugeinv}. When the gauge transform of $\omega'$, $ \omega = U \omega' U^{\dagger}$, is no longer a constant function, conditions (i) and (ii) are satisfied and the Abelian configuration \eqref{abelian} are equivalent with configurations on the common boundary $\delta \Lambda \cap \delta \Omega$ of the Gribov region $\Omega$ and the FMR $\Lambda$.\\
\\
For completeness, we should stress that certainly not all Abelian configurations shall have an equivalent configuration on the boundary $\delta \Lambda \cap \delta \Omega$. It is always possible that an Abelian configuration already lies inside the FMR $\Lambda$. Indeed, we may make an Abelian gauge transformation so the Abelian configuration is transverse, $\p_\mu a_\mu = 0$, and is thus simultaneously in Landau gauge and maximal Abelian gauge.  Each such configuration corresponds to a unique distinct Abelian field tensor $f_{\mu \nu}(x) = \p_\mu a_\nu - \p_\nu a_\mu$, , with inversion $a_\nu = (\p^2)^{-1} \p_\mu f_{\mu \nu}$ automatically satisfying $\p_\nu a_\nu = 0$, so different transverse Abelian configurations are gauge inequivalent.  Thus the set of gauge inequivalent Abelian configurations consists of unbounded rays $\lambda a_\mu$, where $\lambda > 0$ is any positive number.  On the other hand the FMR, $\Lambda$, in minimal Landau gauge is bounded in every direction.  So some transverse Abelian configurations lie inside $\Lambda$ and some lie outside.  When the transverse Abelian configurations are gauge transformed to the absolute minimum (by non-Abelian gauge transformations), those Abelian configurations that were originally inside $\Lambda$ remain there.  (For them $\omega$ is $x$-independent, $\p_\mu \omega = 0$.)  Only those transverse Abelian configurations that were originally outside $\Lambda$ get mapped onto $\p\Lambda \cap \p\Omega$ by the minimizing non-Abelian gauge transformation.\\
\\
We also note that the corresponding statements hold in lattice gauge theory, as discussed in \cite{Greensite:2004ke}.  
\\
\\
As a side remark, notice that if we have one gauge field $A$, satisfying equation \eqref{degenerate}, with $\omega$ different from zero, the whole gauge orbit shall satisfy this equation due to its gauge invariance. Such a gauge orbit has a peculiar feature. In one direction, namely the direction defined by $\omega$, it is degenerated because the infinitesimal gauge transformation,
\begin{equation}
A_\mu' = A_\mu + D_\mu(A) \varepsilon \omega = A_\mu\;,
\end{equation}
with $\varepsilon$ infinitesimal, leaves $A$ invariant. Therefore, the gauge orbit through $A$ has one dimension less that a generic gauge orbit.  We call \eqref{degenerate} the ``degeneracy property''. Since this property characterizes a gauge orbit it is of geometrical significance, whereas the Gribov horizon is more an artifact of gauge fixing to the minimal Landau gauge.

\section{Semi classical solution of Gribov}
\subsection{The no-pole condition}
Gribov was the first one to try to restrict the region of integration to the Gribov region \cite{Gribov:1977wm,Sobreiro:2004us}, which was done in a semi-classical way. He restricted the generating functional to the Gribov region by introducing a factor $V(\Omega)$ in expression \eqref{genz5},
\begin{eqnarray}\label{ZJ}
Z(J) &=& \int_\Omega [\d A]   \exp \left[- S_\YM   \right] \nonumber\\
&=& \int [\d A][\d c] [\d \overline c]    V(\Omega)  \delta (\p A )   \exp \left[- S_\YM  -   \int \d x  \overline c^a (x)  \p_\mu  D_\mu^{ab}  c^b (x)  \right] \;,
\end{eqnarray}
whereby we are working in the Landau gauge,  $\delta (\p A )$. Now the question is how to determine this factor $V(\Omega)$. One can see that there is a close relationship between the ghost sector and the Faddeev-Popov determinant, which is clear from calculating the exact ghost propagator. For this, we start from expression \eqref{ghostapp}
\begin{eqnarray}
	I &=& \int [\d c][\d \overline{c}] \exp\left[ \int \d^d x \d^d y \ \overline{c}_a(x) A_{ab} (x,y)c_b(y) + \int \d^d x \ (J^{a}_c(x) c_a(x) + \overline{c}_a(x) J_{\overline{c}}^a (x)) \right]  \nonumber \\
	&=&C \det A \exp  -\int \d^d x \d^d y \   J_c^a(x) A^{-1}_{ab}(x,y) J_{\overline{c}}^b(y)\;,
\end{eqnarray}
whereby in our case:
\begin{eqnarray}
A_{ab}(x,y) &=& -\p_\mu D_\mu^{ab} \delta(x-y)\;.
\end{eqnarray}


\noindent From this we can calculate the ghost propagator,
\begin{eqnarray}
\braket{ \overline c_a(x)  c_b (y)}_c  &=&  \frac{\delta }{\delta \tilde J_{ c }^b (y)} \frac{\delta }{\delta \tilde J_{\overline c}^a(x)} Z\nonumber\\
&=&  \int [\d A]  V(\Omega) \delta(\partial_{\mu}A_{\mu}^{a})\det(-\partial_{\mu}D_{\mu}^{ab})  A_{ab}^{-1}(x,y) \e^{-S_{YM}}\;.
\end{eqnarray}
Taking the Fourier transform and keeping in mind that we have conservation of momentum
\begin{equation}\label{ghosthorbla}
\braket{ \overline c_a(p)  c_b (-p)}_c   =  \int [\d A]  V(\Omega) \delta(\partial_{\mu}A_{\mu}^{a})\det(-\partial_{\mu}D_{\mu}^{ab})\left( \int \d (x-y) \e^{\ii p (x-y)} A_{ab}^{-1}(x,y) \right)\e^{-S_{YM}}\;,
\end{equation}
we can compare this expression with the one loop renormalization improved ghost propagator starting from the Faddeev-Popov action,
\begin{eqnarray}\label{ghosthor}
\braket{\overline c_a(p)  c_b (k)}_c  &=&   \delta(p + k) \delta^{ab} \mathcal G(k^2) \nonumber\\
\mathcal G(k^2) &=& \underbrace{\frac{1}{k^2}}_{\mathcal P_1} \underbrace{ \frac{1}{ \left( 1 - \frac{11 g^2 N}{48 \pi^2} \ln \frac{\Lambda^2}{k^2} \right)^{\frac{9}{44}}}}_{\mathcal P_2}\;.
\end{eqnarray}
From this expression, we can make some interesting observation. Firstly, for large momentum $k^2$ we are within the Gribov region $\Omega$, as perturbation theory should work there. Indeed, for large $k^2$, $\mathcal G(k^2) \approx 1/ (k^2 \ln \frac{\Lambda}{k^2} )$, which is the perturbative result. Secondly, we notice this expression to have two poles: one pole at $k^2 = 0$ and one pole at $k^2 = \Lambda^2 \exp \left(- \frac{1}{g^2} \frac{48 \pi^2}{11 N} \right)$. The first pole indicates that for $k^2 \approx 0$, we are approaching a horizon, see expression \eqref{ghosthor}. As for all $k^2$, $\mathcal P_1$ is always positive, we stay inside the Gribov region. The second part of the ghost propagator $\mathcal P_2$ is not always positive for all $k^2$. For $k^2 < \Lambda^2 \exp \left(- \frac{1}{g^2} \frac{48 \pi^2}{11 N} \right)$, $\mathcal P_2$ shall become complex, indicating that we have left the Gribov region. Therefore, $V(\Omega)$ makes it impossible for a singularity to exist except at $k^2 = 0$.\\
\\
From these observations, we can construct the no-pole condition. For this, we shall calculate  $\mathcal G (k^2, A)_{ab}$ whereby the gluon field is considered as an external field. This comes down to calculating $\det(-\partial_{\mu}D_{\mu}^{ab})  A^{-1}(x,y)$ from expression \eqref{ghosthorbla}, i.e. we shall calculate the following diagrams:

\begin{figure}[H]
\begin{center}
\includegraphics[width=16cm]{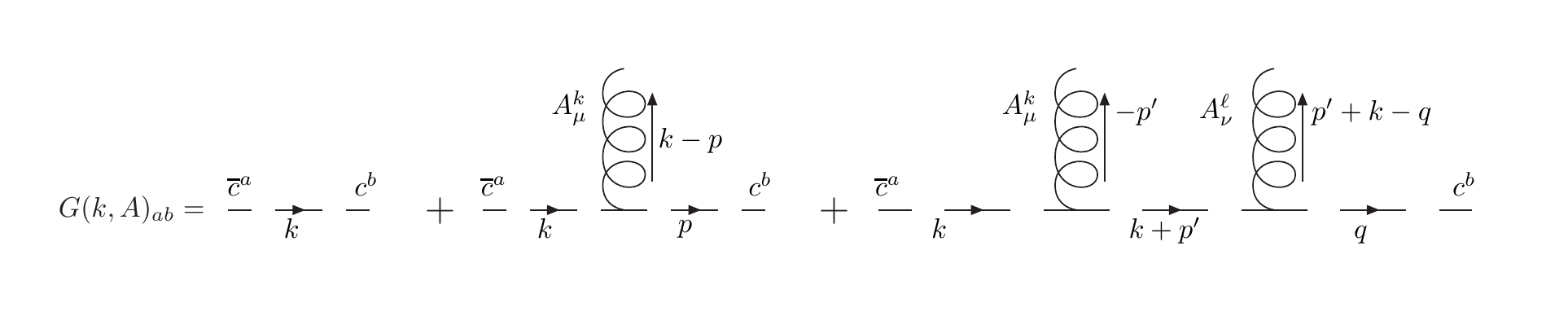}
\caption{The ghost propagator with external field to second order.}\label{ghost}
\end{center}
\end{figure}

\noindent In momentum space, these three diagrams are given by\footnote{The Feynman rule for the ghost-ghost gluon vertex is given by $\ii k_\mu f_{akb}$ with $akb$ resp. from $\overline c$, $A$, and $c$ whereby the outgoing momentum $k_\mu$ stems from $\overline c$.}
\begin{eqnarray}\label{2.66}
I_1 &=& \delta^{ab} (2\pi)^d \delta(k-q) \frac{1}{k^2}  \nonumber\\
I_2 &=& g\frac{1}{k^2}  \frac{1}{p^2} f_{a k b}\  \ii p_\mu    A_\mu^k (k -p) \nonumber\\
I_3 &=& g^2\int \frac{\d^d p'}{(2\pi)^d} \frac{1}{k^2} \frac{1}{(p' + k)^2} \frac{1}{q^2} f_{a k c}\ \ii ( p' + k)_\mu    A_\mu^k ( -p')  f_{c \ell b}\ \ii q_\nu A_\nu^\ell (p' + k - q)\;.
\end{eqnarray}
In fact, this is all we can say about these diagrams, unless we take into account that after determining $V(\Omega)$, which shall be a function of the external gluon field, we shall always need to integrate over $A$. This means that the gluon lines are connected, rendering the second diagram to be equal to zero. For the third diagram, the incoming momentum $k$ shall equal the outcoming momentum $q$. Formally, we can therefore rewrite the third diagram as
\begin{eqnarray}
I_3 &=& - g^2\frac{\delta(k-q) (2\pi)^d}{V}  \frac{1}{k^4} f_{a k c}  f_{c \ell b} \int \frac{\d^d p'}{(2\pi)^d} \frac{ k_\nu (p'+ k)_\mu }{(p' + k)^2}      A_\mu^k ( -p')  A_\nu^\ell (p')\;,
\end{eqnarray}
whereby we have introduced the infinite volume factor $V$, to maintain the right dimensionality. Moreover, we also know that the color indices  $k = \ell$. In order to calculate the correct prefactor, we take the sum over the color factors, see formula \eqref{liestructure}
\begin{eqnarray}\label{coloraway}
\mathcal G (k^2, A) &=& \frac{1}{N^2 -1}\delta_{ab} \mathcal G (k^2, A)_{ab}=  \frac{1}{k^2} + \frac{1}{V}\frac{1}{k^4} \frac{N g^2}{N^2 - 1} \int\frac{ \d^d q}{(2 \pi)^d} A_\mu^\ell(-q) A_\nu^{\ell} (q)  \frac{(k-q)_\mu  q_\nu}{(k-q)^2} \nonumber\\
&=& \frac{1}{k^2}\left( 1 + \sigma(k, A) \right)\;,
\end{eqnarray}
whereby
\begin{eqnarray}
\sigma(k,A) &=& \frac{1}{V}\frac{1}{k^2} \frac{Ng^2}{N^2 - 1} \int\frac{ \d^d q}{(2 \pi)^d} A_\mu^\ell(-q) A_\nu^{\ell} (q) \frac{ (k-q)_\mu  k_\nu }{(k-q)^2}\;.
\end{eqnarray}
Now we can rewrite this
\begin{eqnarray}
\mathcal G (k^2, A) &\approx&  \frac{1}{k^2}\frac{1}{ 1 - \sigma(k, A) }\;,
\end{eqnarray}
as in this way we are considering the inverse, or the 1PI diagram, see expression \eqref{inverse}. This inverse contains more information as we are in fact resumming an infinite tower of Feynmandiagrams. The condition that the Faddeev-Popov operator has no zero modes, reduces to the requirement that
\begin{eqnarray}\label{condition}
\sigma(k,A) < 1\;.
\end{eqnarray}
We can work out this requirement a bit more. As we are working in the Landau gauge, $q_\mu A_\mu = 0$, and thus $A_\mu^\ell A_\nu^\ell$ is transversal,
\begin{eqnarray}
A_\mu^\ell(-q) A_\nu^{\ell}(q) &=& \omega(A) \left( \delta_{\mu\nu} - \frac{q_\mu q_\nu}{q^2} \right) = \omega (A) P_{\mu\nu}\;,
\end{eqnarray}
moreover, multiplying with $\delta_{\mu\nu}$, we find that $\omega(A) = \frac{1}{d-1}A_\mu^\ell A_\mu^\ell $, with $d$ the number of dimensions. Therefore, we can simplify $\sigma$,
\begin{eqnarray}\label{2.67}
\sigma(k,A) &=& \frac{1}{V}\frac{1}{d-1} \frac{Ng^2}{N^2 - 1} \frac{k_\mu  k_\nu}{k^2} \int\frac{ \d^d q}{(2 \pi)^d} A_\alpha^\ell(-q) A_\alpha^{\ell} (q) \frac{ 1 }{(k-q)^2} P_{\mu\nu} \;.
\end{eqnarray}
As it is possible to prove that $\sigma(k,A)$ decreases with increasing $k^2$, see appendix \ref{sigma}, whereby we have assumed that $A_\alpha^\ell(-q) A_\alpha^{\ell}(q)$ is positive, the condition \eqref{condition} becomes,
\begin{eqnarray}\label{nopolecondition}
\sigma(0, A) < 1\;.
\end{eqnarray}
Taking the limit $k^2 \to 0$ in $\sigma(k,A)$ yields,
\begin{eqnarray}\label{nopole}
\sigma(0,A) &=& \frac{1}{V}\frac{1}{d-1} \frac{Ng^2}{N^2 - 1} \lim_{k^2 \to 0} \frac{k_\mu  k_\nu}{k^2} \frac{d-1}{d} \delta_{\mu\nu}  \int\frac{ \d^d q}{(2 \pi)^d} A_\alpha^\ell(-q) A_\alpha^{\ell} (q) \frac{ 1 }{q^2} \nonumber\\
&=& \frac{1}{V} \frac{1}{d} \frac{Ng^2}{N^2 - 1} \int\frac{ \d^d q}{(2 \pi)^4} A_\alpha^\ell(-q) A_\alpha^{\ell} (q) \frac{ 1 }{q^2}\;,
\end{eqnarray}
whereby we used the fact that $\int \d^dq f(q^2) q_\mu q_\nu/q^2 = 1/d\  \delta_{\mu\nu} \int \d^d q f(q^2)$. \\
\\
In summary, the no-pole condition is given by
\begin{eqnarray}
V(\Omega) &=& \theta (1 - \sigma(0,A))\;,
\end{eqnarray}
with $\sigma(0,A)$ given by expression \eqref{nopole}, or thus, using the Heaviside function,
\begin{eqnarray}\label{thetareplace}
V(\Omega) &=&  \int_{- \ii \infty + \epsilon}^{+ \ii \infty + \epsilon}\frac{\d \beta}{2\pi \ii \beta} \e^{\beta (1 - \sigma(0,A))} \;,
\end{eqnarray}
we can insert this into the path integral \eqref{ZJ}.

\subsection{The gluon and the ghost propagator\label{sectiongribov}}
\subsubsection{The gluon propagator}
Our goal is calculate the gluon propagator in Fourier space
\begin{eqnarray}
\Braket{A_\mu^a (k) A_\nu^b(p)}\;,
\end{eqnarray}
at lowest order including the restricting to the Gribov region. We start from the path integral \eqref{ZJ}, while introducing appropriate sources for the gluons,
\begin{equation*}
Z(J) =  \mathcal N \int \frac{\d \beta}{2\pi \ii \beta} \int [\d A]    \e^{\beta (1 - \sigma(0,A))}    \exp -\left[ S_\YM^\quadr   + \int \d^d x \frac{1}{2 \alpha} (\p_\mu A_\mu)^2 + \int \d^d x A_\mu^a(x) J_\mu^a (x) \right]\;,
\end{equation*}
with $N= Z^{-1}(J=0)$. We do not need to take into account the integration over $[\d c][\d \overline c]$ as we are only calculating the free gluon propagator, also we only need the free part of $S_\YM$.  Translating this in Fourier space, we have that
\begin{multline*}
 S_\YM^\quadr   + \int \d^d x \frac{1}{2 \alpha} (\p_\mu A_\mu)^2 + \int \d^d x A_\mu^a(x) J_\mu^a (x) \\
 =  \int \frac{\d^d k}{(2\pi)^d} \left( \frac{1}{2} A_\mu^a(k)  \left(\delta_{\mu\nu} k^2 + \left(\frac{1}{\alpha} - 1 \right)k_\mu k_\nu  \right) A_\nu^a(-k)   -  A_\mu^a(k) J_\mu^a (-k) \right) \;,
\end{multline*}
or thus,
\begin{multline}
\Braket{A_\mu^a (k) A_\nu^b(p)} \\ \left. = \frac{\delta^2 }{\delta J_\mu^a(-k) \delta J_\nu^b(-p)}   \int \frac{\d \beta\e^{\beta}}{2\pi \ii \beta}   \int [\d A]  \e^{ -\int \frac{\d^d k}{(2\pi)^d} \frac{1}{2} A_\mu^a(k)   K_{\mu \nu}^{ab} (k) A_\nu^b(-k) +    \int \frac{\d^d k}{(2\pi)^d}  A_\mu^a(k) J_\mu^a (-k) } \right|_{J=0} \;,
\end{multline}
whereby
\begin{eqnarray}
K_{\mu \nu}^{ab} (k) &=&\delta^{ab} \left( \beta\frac{1}{V} \frac{2}{d} \frac{N g^2}{N^2 - 1} \delta_{\mu\nu} \frac{1}{k^2} + \delta_{\mu\nu} k^2 + \left(\frac{1}{\alpha} - 1 \right)k_\mu k_\nu \right)\;,
\end{eqnarray}
also includes the part stemming from $\sigma(0,A)$ in expression \eqref{nopole}. Now invoking the Fourier transform of \eqref{gauss1}, we find
\begin{equation}
\Braket{A_\mu^a (k) A_\nu^b(p)} = \delta(k+p) \mathcal N  \int \frac{\d \beta\e^{\beta}}{2\pi \ii \beta}  (\det  K_{\mu \nu}^{ab} )^{-1/2}  (K_{\mu \nu}^{ab})^{-1} (k)\;.
\end{equation}
This determinant has been worked out in appendix \ref{sigma}, resulting in
\begin{eqnarray}
 (\det  K_{\mu \nu}^{ab} )^{-1/2}  &=&  \exp\left[ -\frac{d-1}{2} (N^2 - 1) V \int \frac{\d^d q}{(2\pi)^d} \ln \left( q^2 + \frac{\beta N g^2}{N^2 - 1} \frac{2}{d V}\frac{1}{q^2} \right)  \right]\;,
\end{eqnarray}
and thus
\begin{align}
\Braket{A_\mu^a (k) A_\nu^b(p)} =& \delta(k+p)\mathcal N \int \frac{\d \beta}{2\pi \ii }  \e^{f(\beta)} (K_{\mu \nu}^{ab})^{-1} (k)\;, \nonumber\\
f(\beta) =& \beta - \ln \beta  -\frac{d-1}{2} (N^2 - 1) V \int \frac{\d^d q}{(2\pi)^d} \ln \left( q^2 + \frac{\beta N g^2 }{N^2 - 1} \frac{2}{d V}\frac{1}{q^2} \right)\;.
\end{align}
As we assume $(K_{\mu \nu}^{ab})^{-1} (k)$ not to be oscillating too much, we apply the method of steepest descent\footnote{The infinite parameter to apply the method of steepest descent is $1/\hbar$, which is not written anymore.} to evaluate the integral over $\beta$,
\begin{eqnarray}
\Braket{A_\mu^a (k) A_\nu^b(p)} &=& \delta(k+p)\mathcal N' \e^{f(\beta_0)} \left. (K_{\mu \nu}^{ab})^{-1} (k) \right|_{\beta = \beta_0}\;,
\end{eqnarray}
whereby we have absorbed $2\pi \ii$ into $\mathcal N$. $\beta_0$ is the minimum of $f(\beta)$, i.e.
\begin{eqnarray}\label{betaverwaar}
&&f'(\beta_0) =  0 \nonumber\\
&\Rightarrow& 1 =\frac{1}{\beta_0}  +\frac{d-1}{d}N g^2 \int \frac{\d^d q}{(2\pi)^d} \frac{1}{ \left( q^4 + \frac{\beta_0 N g^2}{N^2 - 1}  \frac{2}{d V} \right) } \;.
\end{eqnarray}
We define the Gribov mass,
\begin{eqnarray}
\gamma^4 &=& \frac{\beta_0 N }{N^2 - 1} \frac{2}{d V} g^2\;,
\end{eqnarray}
which serves as an infrared regulating parameter in the integral. As in fact, $V$ is equal to infinity, in order to have a finite $\gamma$, $\beta_0 \sim V$. Therefore, $1/\beta_0$ can be neglected and we obtain the following gap equation,
\begin{eqnarray}\label{gapequation}
1 = \frac{d-1}{d}N g^2 \int \frac{\d^d q}{(2\pi)^d} \frac{1}{ \left( q^4 + \gamma^4 \right) } \;,
\end{eqnarray}
which shall determine $\gamma^4$. Now, we only have to calculate the inverse of
\begin{eqnarray}
(K_{\mu \nu}^{ab})(k) &=&\delta^{ab} \left( \gamma^4 \delta_{\mu\nu} \frac{1}{k^2} + \delta_{\mu\nu} k^2 + \left(\frac{1}{\alpha} - 1 \right)k_\mu k_\nu \right)\;,
\end{eqnarray}
whereby we have set $\beta = \beta_0$ which yields,
\begin{eqnarray}
(K_{\mu \nu}^{ab})(k)^{-1} &=&\delta^{ab} \left( \frac{k^2}{k^4 + \gamma^4} P_{\mu\nu}(k) + \alpha \frac{k^2}{\alpha \gamma^4 + k^4} \frac{k_\mu k_\mu}{k^2} \right) \;,
\end{eqnarray}
as one can check by calculating $(K_{\mu \nu}^{ab})^{-1} (k) (K_{\nu\kappa}^{bc}) (k) = \delta^{ac} \delta^{\mu\kappa}$. For $\alpha = 0$, the inverse becomes tranverse and the gluon propagator is given by
\begin{eqnarray}\label{gluonprop}
\Braket{A_\mu^a (k) A_\nu^b(p)} &=& \delta(k+p) \delta^{ab} \frac{k^2}{k^4 + \gamma^4} P_{\mu\nu}(k) \;,
\end{eqnarray}
as $\mathcal N'$ shall cancel $\e^{f(\beta_0)}$ due to normalization.

\subsubsection{The ghost propagator}
Now that we have found the gluon propagator, we can calculate the ghost propagator. In fact, this comes down to connecting the gluon legs in expression \eqref{2.66}. We easily find (analogous as expression \eqref{2.67}),
\begin{eqnarray}\label{ghostpropagator}
\mathcal G^{ab} (k^2) &=& \delta^{ab} \frac{1}{k^2}\frac{1}{ 1 - \sigma(k) }\;, \nonumber\\
\sigma(k) &=& Ng^2  \frac{k_\mu  k_\nu}{k^2} \int\frac{ \d^d q}{(2 \pi)^d} \frac{q^2}{q^4 + \gamma^4} \frac{ 1 }{(k-q)^2} \left( \delta_{\mu\nu} - \frac{q_\mu q_\nu}{q^2}\right)\;.
\end{eqnarray}
To calculate $1-\sigma(k)$, we rewrite the gap equation \eqref{gapequation} as
\begin{eqnarray}
1 = \frac{k_\mu k_\nu}{k^2} N g^2 \int \frac{\d^d q}{(2\pi)^d} \frac{1}{ \left( q^4 + \gamma^4 \right) } \left( \delta_{\mu\nu} - \frac{q_\mu q_\nu}{q^2}\right)\;,
\end{eqnarray}
and thus, we have written the unity in a complex way
\begin{equation*}
1-\sigma(k) = \frac{k_\mu k_\nu}{k^2} N g^2 \int \frac{\d^d q}{(2\pi)^d} \frac{1}{ \left( q^4 + \gamma^4 \right) } \left( \delta_{\mu\nu} - \frac{q_\mu q_\nu}{q^2}\right) \left( 1- \frac{q^2}{(k-q)^2}\right) = \frac{k_\mu k_\nu}{k^2} N g^2 R_{\mu\nu} (k)\;.
\end{equation*}
To investigate the infrared behavior, we expand this integral for small $k^2$, whereby up to order $k^2$
\begin{eqnarray}
\left( 1- \frac{q^2}{(k-q)^2}\right) &=& 1 - \frac{1}{\frac{k^2}{q^2} - \frac{2 k_\mu q_\mu}{q^2} + 1} = \frac{k^2}{q^2} - \frac{2 k_\mu q_\mu}{q^2} - 4 \left( \frac{k_\mu q_\mu}{q^2}\right)^2\;,
\end{eqnarray}
and thus we can split $R_{\mu\nu}$ in three parts. The first part is given by
\begin{equation}
R_{\mu\nu}^1 (k) = k^2 \int \frac{\d^d q}{(2\pi)^d} \frac{1}{q^2 \left( q^4 + \gamma^4 \right) } \left( \delta_{\mu\nu} - \frac{q_\mu q_\nu}{q^2}\right) = \frac{d-1}{d}\delta_{\mu\nu} k^2   I_\gamma \;,
\end{equation}
whereby $I_\gamma = \int \frac{\d^d q}{(2\pi)^d} \frac{1}{q^2 \left( q^4 + \gamma^4 \right) } $ is a number depending on $\gamma$.  The second part is zero, at is it odd in $q$, and the third part is given by
\begin{equation}
R_{\mu\nu}^3 (k) = -4 \delta_{\mu\nu} k_\alpha k_\beta \int \frac{\d^d q}{(2\pi)^d} \frac{q_\alpha q_\beta}{q^4 \left( q^4 + \gamma^4 \right) } + 4 k_\alpha k_\beta \int \frac{\d^d q}{(2\pi)^d} \frac{q_\alpha q_\beta q_\mu q_\nu}{q^6 \left( q^4 + \gamma^4 \right) } \;.
\end{equation}
The first term of this expression is given by,
\begin{equation*}
 - 4 \delta_{\mu\nu} \frac{\delta^{\alpha \beta}}{d} k_\alpha k_\beta \int \frac{\d^d q}{(2\pi)^d} \frac{1}{q^2 \left( q^4 + \gamma^4 \right) } = -\frac{4}{d}  \delta_{\mu\nu} k^2 I_\gamma \;,
\end{equation*}
while the second term is given by
\begin{multline*}
 4 k_\alpha k_\beta (\delta_{\alpha \beta} \delta_{\mu\nu} +  \delta_{\alpha \mu} \delta_{\beta\nu} + \delta_{\alpha \nu} \delta_{\beta\mu}) \frac{1}{d^2 + 2d} \int \frac{\d^d q}{(2\pi)^d} \frac{1}{q^2 \left( q^4 + \gamma^4 \right) } \\ = 4 (k^2 \delta_{\mu\nu} + 2 k_\mu k_\nu)  \frac{1}{d^2 + 2d} I_\gamma\;.
\end{multline*}
Therefore,
\begin{eqnarray}
R_{\mu\nu}^3 (k) &=&  4\left( - \frac{1}{d}  \delta_{\mu\nu} k^2  + (k^2 \delta_{\mu\nu} + 2 k_\mu k_\nu)  \frac{1}{d^2 + 2d} \right)I_\gamma\;.
\end{eqnarray}
Taking all results together, we obtain,
\begin{eqnarray}
1-\sigma(k) &=& N g^2\frac{k_\mu k_\nu}{k^2}\left[ \frac{d-1}{d}\delta_{\mu\nu} k^2     +  4\left( - \frac{1}{d}  \delta_{\mu\nu} k^2  + (k^2 \delta_{\mu\nu} + 2 k_\mu k_\nu)  \frac{1}{d^2 + 2d} \right) \right]I_\gamma \nonumber\\
&=&  N g^2k^2 \left[ \frac{d-1}{d}    +  4\left( - \frac{1}{d}   +   \frac{3}{d^2 + 2d} \right) \right]I_\gamma \nonumber\\
&=& N g^2 k^2 \frac{d^2 - 3 d + 2}{d^2 +2d} I_\gamma \;,
\end{eqnarray}
or thus, the ghost propagator is enhanced,
\begin{eqnarray}
\mathcal G^{ab} (k^2) &=& \delta^{ab} \frac{1}{k^4}   \frac{d^2 +2d}{d^2 - 3 d + 2} \frac{1}{N g^2 I_\gamma}\;.
\end{eqnarray}
As an example, for $d = 4$, we find easily that $I_\gamma^{d=4} = 1/(32\pi^2 \gamma^2)$ and thus
\begin{eqnarray}
\left.\mathcal G^{ab} (k^2)\right|_{d=4} &=& \delta^{ab} \frac{1}{k^4}    \frac{128\pi^2 \gamma^2}{Ng^2 }\;.
\end{eqnarray}
Also in three dimension we find enhancement of the ghost. In two dimensions, the calculations are not so straightforward as there is a problem with switching the limit and the integration. We refer to section \ref{chap42dform} of chapter \ref{refined} for more details. The conclusion shall however remain the same, also in  $2$ dimensions the ghost propagator is enhanced.\\
\\
\label{latticeI}For a long time, these results for the ghost and gluon propagator were confirmed by lattice calculations and therefore considered important, see \cite{Cucchieri:1997fy,Cucchieri:1999sz,Oliveira:2004gy,Bloch:2003sk,Furui:2003jr} for some examples and \cite{Cucchieri:2010xr} for a nice recent overview. In fact, looking at the calculations, $\sigma = 1$ means that the $\theta$-function has become a $\delta$-function. This is due to the fact that we could neglect $1/\beta_0$ in expression \eqref{betaverwaar} as we are working in an infinite volume $V \to \infty$. In other words, by limiting to the Gribov region, the ghost propagator has an extra pole, which indicates that the region close to the boundary has an important effect on the ghost propagator.

\section{The Gribov-Zwanziger action}
After the publication of Gribov, Zwanziger tried to generalize his results to all orders by constructing an action which implements the restriction to the Gribov region, we shall call this action the Gribov-Zwanziger (GZ) action. In this section we shall first analyze a toy model to demonstrate how the GZ action was obtained \cite{Zwanziger1989}.


\subsection{A toy model}
We start with the following simple quadratic action in one dimension
\begin{eqnarray}
S &=& \int \d x\frac{1}{2} (\p A(x))^2\;,
\end{eqnarray}
whereby we omit color and Lorentz indices. We assume the Gribov region is contained within the following ellipsoid,
\begin{eqnarray}\label{ellips}
\underbrace{\frac{1}{2}\int \frac{\d k}{(2\pi)} \frac{A(k) A(-k)}{k^2}}_{f(A)} = c  \;.
\end{eqnarray}
Therefore, to restrict to the Gribov region, we need to consider the following generating functional
\begin{eqnarray}\label{gentheta}
Z &=& \int [\d A] \theta(c - f(A)) \e^{-S}\;,
\end{eqnarray}
as the $\theta$-function assures that $f(A) < c$.\\
\\
Next, if we replace $A(k)/ \sqrt{k^2}$ by $y(k)$, and we work in a finite volume, $f(A)< c$ can be written as
\begin{eqnarray}
\frac{1}{V} \sum_k y_k y_{-k} &<& c \;,
\end{eqnarray}
and thus expression \eqref{ellips} can be seen as an hypersurface in an infinite dimensional space. As is known for hyper spheres, the volume gets more and more concentrated on the surface as the dimension grows.  Therefore, we can replace the $\theta$-function with a $\delta$-function and the generating functional \eqref{gentheta} becomes
\begin{eqnarray}\label{gendelta}
Z &=& \int [\d A] \delta(c - f(A)) \e^{-S}\;.
\end{eqnarray}
Let us remark that also Gribov already noticed this, see the end of the previous section.\\
\\
We use the formula
\begin{eqnarray}
\delta (x - y) &=& \int_{-\infty}^{\infty} \frac{\d t}{2\pi} \ \e^{\ii (x-y) t} \;,
\end{eqnarray}
which we Wick rotate in order to get rid of the imaginary numbers, so we find
\begin{eqnarray}\label{wick}
Z &=&\int_{-\ii \infty + \epsilon}^{\ii \infty + \epsilon} \frac{\d \beta}{2\pi \ii} \int [\d A]  \e^{-S} \e^{\beta (c - f(A))} \nonumber\\
&=&  \int_{-\ii \infty + \epsilon}^{\ii \infty + \epsilon} \d \beta \e^{-g(\beta)} \;,
\end{eqnarray}
whereby $g(\beta) = -\ln \int [\d A] \e^{\beta(c-f(A))} \e^{-S}$. Now we shall perform a saddle point approximation to integrate over $\beta$,
\begin{eqnarray}
Z &\approx&  \e^{-g(\beta_0)}\;,
\end{eqnarray}
whereby $\beta_0$ is determined by
\begin{eqnarray}\label{determiningbeta}
g'(\beta_0) &=& 0 \nonumber\\
c&=& \frac{\int [\d A] f(A) \e^{\beta(c-f(A))} \e^{-S} }{\int [\d A] \e^{\beta(c-f(A))} \e^{-S}}  \nonumber\\
c&=& \frac{\int [\d A] f(A) \e^{V\beta^*(-f(A))} \e^{-S} }{\int [\d A] \e^{V\beta^*(-f(A))} \e^{-S}} \nonumber\\
c &=& \Braket{f(A)}_{\beta^*} \;,
\end{eqnarray}
whereby we have set $\beta = V \beta^*$. Therefore, we find the following generating functional
\begin{eqnarray}\label{partZ}
Z &=& \int [\d A] \e^{- (V \beta^*f(A) + S))} \;,
\end{eqnarray}
with $\beta^*$ determined by \eqref{determiningbeta}. In fact, the saddle point approximation becomes exact in the equivalence between microcanonical and canonical ensemble \cite{munster}. This can be proven by showing that $f(A)$ has zero variance. Let us investigate the distribution of $f(A)$,
\begin{eqnarray}\label{pr1}
\braket{f(A)} &=&  \left.\int \frac{\d k}{(2\pi)} \frac{1}{2 k^2} \int [\d A] A(k) A(-k)  \e^{- \frac{1}{2}\int \frac{\d k}{2\pi} A(k) \left[ k^2 +  \frac{\beta}{k^2} \right] A(-k) } \right|_{J =0}\nonumber\\
&=& \left. \int \frac{\d k}{(2\pi)} \frac{1}{2 k^2} \frac{\delta^2}{\delta J(k) \delta J(-k) }  \int [\d A]   \e^{-\frac{1}{2} \int \frac{\d k}{2\pi} A(k) \left[ k^2 +  \frac{\beta}{k^2} \right] A(-k)  +  \int \frac{\d k}{(2\pi)} A(k) J(-k)} \right|_{J =0}\nonumber\\
&=& \left. \int \frac{\d k}{(2\pi)} \frac{1}{2 k^2} \frac{\delta^2}{\delta J(k) \delta J(-k) }    \e^{ \frac{1}{2}\int \frac{\d k}{2\pi} J(k) \left[ k^2 +  \frac{\beta}{k^2} \right]^{-1} J(-k)} \right|_{J =0}\nonumber\\
&=& \int \frac{\d k}{(2\pi)} \frac{1}{2}\frac{1}{k^4 + \beta} \;,
\end{eqnarray}
whereby we used the Fourier transform of expression \eqref{gauss1} in the appendix.  Also,
\begin{align}\label{pr2}
\braket{f^2(A)} &=\frac{1}{4}\left. \int \frac{\d k}{(2\pi)}\frac{\d p}{(2\pi)} \frac{1}{k^2}\frac{1}{p^2} \frac{\delta^4}{\delta J(k) \delta J(-k) \delta J(p) \delta J(-p) }  \int [\d A]   \e^{ \frac{1}{2}\int \frac{\d k}{2\pi} J (k) \left[ k^2 +  \frac{\beta}{k^2} \right]^{-1} J(-k)} \right|_{J =0}\nonumber\\
&= \frac{1}{2}\int \frac{\d k}{(2\pi)} \frac{1}{k^4 + \beta} \frac{1}{2}\int \frac{\d p}{(2\pi)} \frac{1}{p^4 + \beta} \nonumber\\
  & \hspace{3cm}+ \frac{1}{2}\int \frac{\d k}{(2\pi)}\frac{\d p}{(2\pi)} \frac{1}{k^2}  \frac{1}{p^2}  \frac{\delta (k + p)}{V} \frac{\delta (k + p)}{V} \left[ p^2 +  \frac{\beta}{p^2} \right]^{-2} \nonumber\\
&= \braket{f(A)}^2 + \frac{1}{2}\frac{ \delta(0)}{V} \int \frac{\d p}{(2\pi)} \frac{1}{(p^4 + \beta)^2} \;,
\end{align}
whereby we have formally written $\frac{\delta J(-k)}{\delta J(p)} = \delta(k+p)/V$. In the thermodynamic limit, $V\to \infty$, we find that $\braket{f^2(A)} = \braket{f(A)}^2$, thus $f(A)$ has no variance and behaves like a $\delta$-function.

\subsection{The non-local Gribov-Zwanziger action}
To restrict the region of integration of the Yang-Mills action to the Gribov region, we need to consider the following path integral,
\begin{eqnarray}\label{startder}
Z &=& \int [\d A] \e^{S_{\YM} + S_\gf} \theta( \lambda (A))\;,
\end{eqnarray}
whereby $\lambda (A)$ is the lowest eigenvalue of the Faddeev-Popov operator,
\begin{equation}
\mathcal M^{ab} = \mathcal M_0^{ab} + \mathcal M_1^{ab} =  - \p^2 \delta^{ab} + g f_{abc} A^c_\mu \p_\mu\;,
\end{equation}
whereby we are working on-shell, $\p_\mu A_\mu = 0$. In this way, the lowest eigenvalue is always greater than zero. Firstly, we notice that all constant vectors are eigenvectors of the Faddeev-Popov operator, with zero eigenvalue. As these eigenvalues never become negative, we shall not consider these vectors and work in the space orthogonal to this space.

\subsubsection{Degenerate perturbation theory}
In order to find the lowest lying (non-trivial) eigenvalue $\lambda (A)$, we shall apply degenerate perturbation theory whereby $\mathcal M_0^{ab}$ is the unperturbed operator. For the moment, we work in a finite box with length $L$, which shall approach infinity in the infinite volume limit. The vectors which form a basis of eigenvectors of the operator $\mathcal M_0^{ab}$ and span the entire Hilbert space are called $\ket{\Psi^{(0)}_{\vec n s}}$: $\vec n \in \mathbb Z^d / \{ 0\}$, while $s$ runs over all the colors, $s = 1 , \ldots, N^2 -1$. $\vec n_0$ represents the lowest momenta, i.e.~$\vec n_0 = (0,\ldots,0, \pm 1, 0, \ldots, 0)$. Therefore, $\ket{\Psi^{(0)}_{\vec n_0 s}}$ represent the vectors belonging to the lowest eigenvalue\footnote{The lowest lying eigenvalue is of course zero, belonging to the constant vectors, but we are not considering these constant vectors anymore.}  $\lambda_{\vec n_0 s}^{(0)}$ which is given by
\begin{equation}
\lambda_{\vec n_0}^{(0)} = \left( \frac{2 \pi}{L} \right)^2\;.
\end{equation}
Notice that the space spanned by the vectors $\ket{\Psi^{(0)}_{\vec n_0 s}}$ is $2d(N^2 - 1)=T$ dimensional. We call this space $\mathcal H_0$
\begin{equation}\label{hho}
\mathcal H_0 = \text{span}(\ket{\Psi^{(0)}_{\vec n_0 s}}, s=1, \ldots, T)\;.
\end{equation}
The projector onto this space $\mathcal H_0$ is easily identified as
\begin{eqnarray}
P_0 = \sum_s  \ket{\Psi^{(0)}_{\vec n_0s}} \bra{\Psi^{(0)}_{\vec n_0s}}\;,
\end{eqnarray}
which is of course an operator working in the entire Hilbert space. The other eigenvectors $\ket{\Psi^{(0)}_{\vec n s}}$ have corresponding eigenvalue $\lambda_{\vec n s}^{(0)}$ given by
\begin{equation}
\lambda_{\vec n s}^{(0)} = \left( \frac{2 \pi}{L} \right)^2 || \vec n ||^2 \;.
\end{equation}
The entire space can be decomposed into $\mathcal H_0 + \mathcal H_1 + \mathcal H_2 + \ldots$, whereby the $\mathcal H_n$ are defined as the spaces spanned by the vectors belonging to the same eigenvalue\footnote{We can order the eigenvalues by size, $\mathcal H_n$ belongs to the $(n+1)$th eigenvalue.}.\\
\\
In configuration space, the eigenvectors are given by
\begin{equation}\label{conf1}
\Braket{x, a|\Psi^{(0)}_{\vec n s}} = \delta^{as} \left( \frac{1}{L} \right)^\frac{d}{2} \e^{\ii \frac{2\pi}{L} \vec n \cdot \vec x}\;,
\end{equation}
which shall lead to
\begin{equation}\label{conf2}
\Braket{x, a| \widetilde \Psi^{(0)}_{\vec k s}} = \delta^{as} \left( \frac{1}{2\pi} \right)^\frac{d}{2} \e^{\ii \vec k \cdot \vec x}\;,
\end{equation}
in the infinite volume limit.\\
\\
Let us now switch on the perturbation $\mathcal M_1$. The degenerate eigenvalues shall split up in different eigenvalues, while the eigenvectors $\ket{\Psi^{(0)}_{\vec n s}}$ shall evolve into $\ket{\Psi_{\vec n s}}$. In order to solve the eigenspace problem, we shall try to find a matrix $\kappa$ and $S$, so that
\begin{equation}\label{ms}
\mathcal M S = S \kappa \;,
\end{equation}
whereby $S$ is a matrix whereby the columns are given by the exact vectors $\ket{\Psi_{\vec n_0 s}}$ and $\kappa$ is a diagonal matrix with the diagonal elements given by $\lambda_{\vec n_0 s}$, with $\vec n_0 = (0,\ldots,0, \pm 1, 0, \ldots, 0)$ and $s = 1,\ldots, N^2 -1$. By introducing $S$, we can restrict ourselves to the space of interest, namely the space of eigenvectors  $\ket{\Psi_{\vec n_0 s}}$ with the lowest lying eigenvalues $\lambda_{\vec n_0 s}$.\\
\\
The next step is to determine $S$ and $\kappa$. We shall write them as a perturbation series:
\begin{align}
S & = \sum_{n = 0}^\infty S_n & \kappa &= \sum_{n = 0}^\infty \kappa_n \;.
\end{align}
By substituting them into \eqref{ms}, and identifying equal orders, we find
\begin{subequations}
\begin{eqnarray}\label{uitgebreid}
\mathcal M_0 S_0 &=& S_0 \kappa_0 \;,\\
\mathcal M_0 S_1 + \mathcal M_1 S_0 &=& S_1 \kappa_0 + S_0 \kappa_1 \label{rxs} \;,\\
\mathcal M_0 S_2 + \mathcal M_1 S_1 &=& S_2 \kappa_0 + S_1 \kappa_1 + S_0 \kappa_2 \label{qrz} \;, \\
\mathcal M_0 S_3 + \mathcal M_1 S_2 &=&  S_3 \kappa_0 + S_2 \kappa_0 + S_1 \kappa_2 + S_0 \kappa_3 \;,\\
&\vdots& \nonumber
\end{eqnarray}
\end{subequations}
The first equation is the free equation, thus $S_0$ is the matrix whereby the columns are given by $\ket{\Psi_{\vec n_0 s}^{(0)}}$, and $\kappa_0$ is the $T\times T$ dimensional matrix with diagonal elements $\lambda_{\vec n_0}^{(0)}$. Notice that we can write $P_0 = S_0 S_0^\dagger$.\\
\\
In order to solve the higher order equations, we shall use the normalization condition, i.e.~$S_n$ for $n \geq 1$ does not contain any vectors from the space $\mathcal H_0$:
\begin{equation}
P_0 S_n  = 0 \quad \forall n \geq 1\;,
\end{equation}
which we can also write as
\begin{equation}
S_0^\dagger S_n = 0 \quad \forall n \geq 1 \;.
\end{equation}
We now multiply the remaining equations with $S_0^\dagger$. Firstly, we find
\begin{eqnarray}
S_0^\dagger \mathcal M_0 S_1 + S_0^\dagger \mathcal M_1 S_0 &=&  S_0^\dagger S_0 \kappa_1 = \kappa_1
\end{eqnarray}
As the columns of $S_1$ contain vectors which live in the orthogonal complement of $\mathcal H_0$, also the columns of $\mathcal M_0 S_1$ shall live in this space. Therefore, $S_0^\dagger \mathcal M_0 S_1 = 0$ and we find
\begin{equation}\label{kappa1}
\kappa_1 = S_0^\dagger \mathcal M_1 S_0 \;.
\end{equation}
In an analogous fashion, we find for the other equations
\begin{align}\label{kappa2}
\kappa_2 &= S_0^\dagger \mathcal M_1 S_1  & \kappa_3 &= S_0^\dagger \mathcal M_1 S_2 & \ldots \;.
\end{align}
To find $S_1$, we start from equation \eqref{rxs},
\begin{eqnarray}
\mathcal M_0 S_1 - S_1\kappa_0 &=&S_0 \kappa_1     - \mathcal M_1 S_0 \;.
\end{eqnarray}
As $\kappa_0 = \lambda_{\vec n_0}^{(0)} I^T$, with $I^T$ the $T\times T$ unity matrix, we find upon using equation \eqref{kappa1}
\begin{equation}\label{S12}
\mathcal M_0 S_1 -  \lambda_{\vec n_0}^{(0)} S_1  =S_0 \kappa_1     - \mathcal M_1 S_0 \Rightarrow S_1 =  \left( \mathcal M_0 -  \lambda_{\vec n_0}^{(0)}  I^\infty   \right)^{-1}  (P_0 - I^\infty) \mathcal M_1 S_0 \;.
\end{equation}
In an analogous fashion, we can deduce $S_2$ from equation \eqref{qrz},
\begin{eqnarray}
 \mathcal M_0 S_2 - S_2 \kappa_0 &=& - \mathcal M_1 S_1 + S_1 \kappa_1 + S_0 \kappa_2 \nonumber\\
\Rightarrow (\mathcal M_0 - \lambda_{\vec n_0}^{(0)}  I^\infty  ) S_2 &=&   (P_0 - I^\infty) \mathcal M_1 \left( \mathcal M_0 -  \lambda_{\vec n_0}^{(0)}  I^\infty   \right)^{-1} (P_0 - I^\infty) \mathcal M_1 S_0 \nonumber\\
&&+ \left( \mathcal M_0  -  \lambda_{\vec n_0}^{(0)}  I^\infty   \right)^{-1} (P_0 - I^\infty) \mathcal M_1 P_0 \mathcal M_1 S_0\;,
\end{eqnarray}
or thus
\begin{equation}
S_2 = \left[ \left( \mathcal M_0 -  \lambda_{\vec n_0}^{(0)}  I^\infty   \right)^{-1} (P_0 - I^\infty) \mathcal M_1  \right]^2 S_0 + \left( \mathcal M_0  -  \lambda_{\vec n_0}^{(0)}  I^\infty   \right)^{-2} (P_0 - I^\infty) \mathcal M_1 P_0 \mathcal M_1 S_0 \;,
\end{equation}
whereby we made use of equations \eqref{kappa1}, \eqref{kappa2} and \eqref{S12}. With the expressions for $S_1$ and $S_2$, we write
\begin{eqnarray}
\kappa_0 &=& \lambda_{\vec n_0}^{(0)} I^T \;,\nonumber\\
\kappa_1 &=& S_0^\dagger \mathcal M_1 S_0 \;, \nonumber\\
\kappa_2 &=& S_0^\dagger \mathcal M_1 \left( \mathcal M_0 -  \lambda_{\vec n_0}^{(0)}  I^\infty   \right)^{-1}  (P_0 - I^\infty) \mathcal M_1 S_0 \;, \nonumber\\
\kappa_3 &=& S_0^\dagger \mathcal M_1 \left[ \left( \mathcal M_0 -  \lambda_{\vec n_0}^{(0)}  I^\infty   \right)^{-1} (P_0 - I^\infty) \mathcal M_1  \right]^2 S_0 \;,\nonumber\\
 &&+ S_0^\dagger \mathcal M_1 \left( \mathcal M_0  -  \lambda_{\vec n_0}^{(0)}  I^\infty   \right)^{-2} (P_0 - I^\infty) \mathcal M_1 P_0 \mathcal M_1 S_0 \;,
\end{eqnarray}
or as the column $s$ of $S_0$ is given by the vector $\ket{\Psi_{\vec n_0s}^{(0)}}$, we can write the previous expressions in terms of its matrix elements,
\begin{eqnarray}
\kappa_0^{\vec n_0' u, \vec n_0 t} &=& \lambda_{\vec n_0}^{(0)} \delta^{ut} \delta_{\vec n_0, \vec n_0'} \;,\nonumber\\
\kappa_1^{\vec n_0' u, \vec n_0 t} &=& \bra{\Psi_{\vec n_0' u}^{(0)}} \mathcal M_1 \ket{\Psi_{\vec n_0 t}^{(0)}} \;, \nonumber\\
\kappa_2^{\vec n_0' u, \vec n_0 t} &=& - \bra{\Psi_{\vec n_0'  u}^{(0)}} \mathcal M_1 \left( \mathcal M_0  - \lambda_{\vec n_0}^{(0)}  I^\infty      \right)^{-1}  (I^\infty - P_0 ) \mathcal M_1 \ket{\Psi_{\vec n_0  t}^{(0)}} \;, \nonumber\\
\kappa_3^{\vec n_0' u, \vec n_0 t} &=& \bra{\Psi_{\vec n_0' u}^{(0)}} \mathcal M_1 \left[ \left(\mathcal M_0  - \lambda_{\vec n_0}^{(0)}  I^\infty     \right)^{-1} (I^\infty - P_0  ) \mathcal M_1  \right]^2 \ket{\Psi_{\vec n_0 t}^{(0)}} \nonumber\\
 &&- \bra{\Psi_{\vec n_0' u}^{(0)}} \mathcal M_1 \left( \mathcal M_0  -  \lambda_{\vec n_0}^{(0)}  I^\infty   \right)^{-2} (I^\infty - P_0 ) \mathcal M_1 P_0 \mathcal M_1 \ket{\Psi_{\vec n_0 t}^{(0)}} \;.
\end{eqnarray}

\subsubsection{The infinite volume limit}
We shall now show that in the large volume limit, we can do some simplifications.\\
\\
We start with the expression of $\kappa_2$. We can rewrite $ (I^\infty - P_0) $ in terms of a bra-ket notation,
\begin{equation}
 I^\infty - P_0 = \sum_{||\vec n ||>0} \sum_{s=1}^{N^2-1} \ket{\Psi_{\vec n s}^{(0)}} \bra{\Psi_{\vec ns}^{(0)}} - \sum_{||\vec n || = 1} \sum_{s=1}^{N^2-1} \ket{\Psi_{\vec n_0 s}^{(0)}} \bra{\Psi_{\vec n_0s}^{(0)}} =  \sum_{||n||>1}^\infty \sum_{s=1}^{N^2-1} \ket{\Psi_{ns}^{(0)}} \bra{\Psi_{ns}^{(0)}} \;.
\end{equation}
If we now let the operator $\left( \mathcal M_0  -  \lambda_{\vec n_0}^{(0)}  I^\infty   \right)^{-1}$ act on this term, we obtain
\begin{align}
 \left( \mathcal M_0  -  \lambda_{\vec n_0}^{(0)}  I^\infty   \right)^{-1} (I^\infty - P_0) &=  \sum_{||\vec n||>1} \sum_{s=1}^{N^2-1}  (\lambda_{\vec n}^{(0)} - \lambda_{\vec n_0}^{(0)})^{-1}\ket{\Psi_{\vec n s}^{(0)}} \bra{\Psi_{\vec ns}^{(0)}} \nonumber\\
&= \sum_{||\vec n||>1} \sum_{s=1}^{N^2-1}  \left( \frac{1}{ \left( \frac{2\pi}{L} \right)^2 ||\vec n||^2 - \left( \frac{2\pi}{L} \right)^2 } \right)\ket{\Psi_{\vec n s}^{(0)}} \bra{\Psi_{\vec n s}^{(0)}} \;.
\end{align}
Inserting this in $\kappa_2^{\vec n_0' u, \vec n_0 t}$ yields,
\begin{equation}
\kappa_2^{\vec n_0' u, \vec n_0 t} = - \sum_{||\vec n||>1} \sum_{s=1}^{N^2-1}   \left( \frac{1}{ \left( \frac{2\pi}{L} \right)^2 ||\vec n||^2 - \left( \frac{2\pi}{L} \right)^2 } \right)  \bra{\Psi_{\vec n_0'  u}^{(0)}} \mathcal M_1 \ket{\Psi_{\vec n s}^{(0)}} \bra{\Psi_{\vec n s}^{(0)}}      \mathcal M_1 \ket{\Psi_{\vec n_0  t}^{(0)}} \;.
\end{equation}
We can work out the matrix elements with the help of equation \eqref{conf1}
\begin{eqnarray}\label{lalala}
\bra{\Psi_{\vec n s}^{(0)}}      \mathcal M_1 \ket{\Psi_{\vec n_0  t}^{(0)}} &=& \ii n_{0,\mu}\frac{2\pi}{L}  \frac{1}{L^d}\int \d^d x \ \e^{-\ii \frac{2\pi}{L} \vec n \cdot \vec x} g f_{stc} A_\mu^c(x)\e^{\ii \frac{2\pi}{L} \vec n_0 \cdot \vec x} \nonumber\\
\bra{\Psi_{\vec n_0'  u}^{(0)}} \mathcal M_1 \ket{\Psi_{\vec n s}^{(0)}}  &=& \ii n_{0,\mu}' \frac{2\pi}{L} \frac{1}{L^d} \int \d^d x \ \e^{-\ii \frac{2\pi}{L} \vec n_0' \cdot \vec x} g f_{usc} A_\mu^c(x)\e^{\ii \frac{2\pi}{L} \vec n \cdot \vec x} \;,
\end{eqnarray}
whereby we made use of partial integration and the Landau gauge condition $\p_\mu A_\mu = 0$ in the second matrix element. If we take the infinite volume limit, see expression \eqref{conf2}, we can write
\begin{eqnarray}
\bra{\Psi_{\vec n s}^{(0)}}      \mathcal M_1 \ket{\Psi_{\vec n_0  t}^{(0)}} &=& \ii\frac{2\pi}{L}  \left( \frac{2\pi}{L} \right)^d \int \d^d x \ \frac{1}{(2\pi)^{d/2}} \e^{-\ii \vec k \cdot \vec x} g f_{stc} A_\mu^c(x)  n_{0,\mu}  \frac{1}{(2\pi)^{d/2}}  \e^{\ii \vec 0 \cdot \vec x} \nonumber\\
&=& \ii\frac{2\pi}{L}  \left( \frac{2\pi}{L} \right)^d \Braket{ \widetilde \Psi_{\vec k  s}^{(0)} |\vec A \cdot \vec n_0| \widetilde \Psi_{\vec 0  t}^{(0)} } \;, \nonumber\\
\bra{\Psi_{\vec n_0'  u}^{(0)}} \mathcal M_1 \ket{\Psi_{\vec n s}^{(0)}}  &=& \ii\frac{2\pi}{L}  \left( \frac{2\pi}{L} \right)^d \Braket{ \widetilde \Psi_{\vec 0  u}^{(0)} |\vec A \cdot \vec n_0'| \widetilde \Psi_{\vec k  s}^{(0)} }\;,
\end{eqnarray}
with $(\vec A^{ab})_\mu= g f_{abc} A_\mu^c$. We can write $\kappa_2^{\vec n_0' u, \vec n_0 t}$ as
\begin{multline}
\kappa_2^{\vec n_0' u, \vec n_0 t} =\left( \frac{2\pi}{L}\right)^2 \left( \frac{2\pi}{L}\right)^d  \sum_{||\vec n||>1} \left( \frac{2\pi}{L}\right)^d  \left( \frac{1}{ \left( \frac{2\pi}{L} \right)^2 ||\vec n||^2 - \left( \frac{2\pi}{L} \right)^2 } \right)  C(\vec k)^{\vec n_0' u, \vec n_0 t}  \;,
\end{multline}
whereby we have parameterized
\begin{equation}
C(\vec k)^{\vec n_0' u, \vec n_0 t} = \sum_{s=1}^{N^2-1} \Braket{ \widetilde \Psi_{\vec 0  u}^{(0)} |\vec A \cdot \vec n_0'| \widetilde \Psi_{\vec k  s}^{(0)} }  \Braket{ \widetilde \Psi_{\vec k  s}^{(0)} |\vec A \cdot \vec n_0| \widetilde \Psi_{\vec 0  t}^{(0)} }\;.
\end{equation}
In the infinite volume limit, we can write
\begin{eqnarray}
\kappa_2^{\vec n_0' u, \vec n_0 t} &=&\left( \frac{2\pi}{L}\right)^2 \left( \frac{2\pi}{L}\right)^d  \int_{||\vec k||> \frac{2\pi}{L}} \d^d k  \left( \frac{1}{ k^2 - \left( \frac{2\pi}{L} \right)^2 } \right)  C(\vec k)^{\vec n_0' u, \vec n_0 t}  \nonumber\\
&=& \left( \frac{2\pi}{L}\right)^2 \left( \frac{2\pi}{L}\right)^d  \int_{||\vec k||> \frac{2\pi}{L}} \d^d k  \frac{1}{k^2} \left[\sum_{n=0}^{+\infty} \frac{(2\pi/L)^{2n}}{k^{2n}} \right]  C(\vec k)^{\vec n_0' u, \vec n_0 t} \;.
\end{eqnarray}
Now we can see that the higher order terms for $n\geq 1$ will vanish relatively in the infinite volume limit when comparing to the term for $n=0$. One can check this for all $d\geq 2$. Therefore, we obtain
\begin{eqnarray}
\kappa_2^{\vec n_0' u, \vec n_0 t} &=& \left( \frac{2\pi}{L}\right)^2 \left( \frac{2\pi}{L}\right)^d  \int_{||\vec k||> \frac{2\pi}{L}} \d^d k  \frac{1}{k^2}  C(\vec k)^{\vec n_0' u, \vec n_0 t} \nonumber\\
&=&\left( \frac{2\pi}{L}\right)^2 \left( \frac{2\pi}{L}\right)^d  \int_{||\vec k||\geq \frac{2\pi}{L}} \d^d k  \frac{1}{k^2}  C(\vec k)^{\vec n_0' u, \vec n_0 t} \;,
\end{eqnarray}
whereby we have included the endpoint $||\vec k||=\frac{2\pi}{L} $ in the region of integration, which is a valid operation. Going back to the original formulation, we thus find a very simple form for $\kappa_2^{\vec n_0' u, \vec n_0 t}$
\begin{eqnarray}
\kappa_2^{\vec n_0' u, \vec n_0 t} &=&  - \bra{\Psi_{\vec n_0' u}^{(0)}} \mathcal M_1 \mathcal M_0^{-1} \mathcal M_1 \ket{\Psi_{\vec n_0t}^{(0)}} \;.
\end{eqnarray}
In an analogous fashion, we can simplify $\kappa_3$ as well, leading to
\begin{equation}
\kappa_3^{\vec n_0' u, \vec n_0 t} = \bra{\Psi_{\vec n_0' u}^{(0)}} \mathcal M_1   \mathcal M_0^{-1}  \mathcal M_1  \mathcal M_0^{-1}  \mathcal M_1   \ket{\Psi_{\vec n_0 t}^{(0)}} - \bra{\Psi_{\vec n_0' u}^{(0)}} \mathcal M_1  \mathcal M_0^{-2}  \mathcal M_1 P_0 \mathcal M_1 \ket{\Psi_{\vec n_0 t}^{(0)}} \;.
\end{equation}
Moreover, we observe that the second term in the expression of $\kappa_3$ can be neglected in comparison with the first term due to the presence of $P_0$, an argument which can be generalized to $\kappa_n$. This can be checked by working these term out in a similar fashion, as done in equation \eqref{lalala}.
\noindent In conclusion, in the large volume limit, we obtain the following matrices
\begin{eqnarray}
\kappa_0^{\vec n_0' u, \vec n_0 t} &=& \lambda_{\vec n_0}^{(0)} \delta^{ut} \delta_{\vec n_0, \vec n_0'}  \;,\nonumber\\
\kappa_1^{\vec n_0' u, \vec n_0 t} &=& \bra{\Psi_{\vec n_0' u}^{(0)}} \mathcal M_1 \ket{\Psi_{\vec n_0 t}^{(0)}} \;, \nonumber\\
\kappa_2^{\vec n_0' u, \vec n_0 t} &=& - \bra{\Psi_{\vec n_0' u}^{(0)}} \mathcal M_1 \mathcal M_0^{-1} \mathcal M_1 \ket{\Psi_{\vec n_0t}^{(0)}} \;, \nonumber\\
\kappa_3^{\vec n_0' u, \vec n_0 t}&=& \bra{\Psi_{\vec n_0' u}^{(0)}} \mathcal M_1   \mathcal M_0^{-1}  \mathcal M_1  \mathcal M_0^{-1}  \mathcal M_1   \ket{\Psi_{\vec n_0 t}^{(0)}} \;.
\end{eqnarray}
We notice that
\begin{equation}
\mathcal M^{-1} = \frac{1}{1 + \mathcal M_0^{-1} \mathcal M_1} \mathcal M_0^{-1} =   \left( 1-\mathcal M_0^{-1} \mathcal M_1 + \mathcal M_0^{-1} \mathcal M_1  \mathcal M_0^{-1} \mathcal M_1  \right)\mathcal M_0^{-1} \;,
\end{equation}
or thus
\begin{equation}
\mathcal M^{-1} =    \mathcal M_0^{-1}  -\mathcal M_0^{-1} \mathcal M_1  \mathcal M_0^{-1}  + \mathcal M_0^{-1} \mathcal M_1  \mathcal M_0^{-1} \mathcal M_1  \mathcal M_0^{-1} \;.
\end{equation}
Therefore, we can now sum the whole series $\kappa_n$ starting from $n =2$,
\begin{equation}
\kappa_r = \sum_{n\geq2} \kappa_n =- \bra{\Psi_{\vec{n}_0' u}^{(0)}} \mathcal M_1 \mathcal M^{-1} \mathcal M_1  \ket{\Psi_{ \vec{n}_0't}^{(0)}} \;.
\end{equation}

\subsubsection{Taking the trace of $\kappa$}
Next, we replace the condition of all eigenvalues of $\kappa = \kappa_0 + \kappa_1 + \kappa_r$ to be positive with the weaker condition that the trace of $\kappa$ has to be positive, we refer to \cite{Zwanziger1989} for some remarks on this. Firstly, the trace of $\kappa_0$ is given by
\begin{equation}
\Tr \kappa_0 =\sum_{||\vec n|| =1} \sum_{u=1}^{N^2-1} \kappa_0^{\vec n u, \vec n u} =    \lambda_{\vec n_0}^{(0)} T \;.
\end{equation}
Secondly, the trace of $\kappa_1$ is zero,
\begin{equation}
\Tr \kappa_1 = \sum_{||\vec n|| =1} \sum_{u=1}^{N^2-1}  \kappa_1^{\vec n u, \vec n u}   = \sum_{||\vec n|| =1} \sum_{u=1}^{N^2-1}  \bra{\Psi_{\vec n u}^{(0)}} \mathcal M_1 \ket{\Psi_{\vec n u}^{(0)}} = 0\;,
\end{equation}
as $ M_1^{uu} = g f_{uuc} A^c_\mu \p_\mu = 0$. Finally, the trace of $\kappa_r$ is given by
\begin{eqnarray*}
\Tr \kappa_r &=& - \sum_{||\vec n|| =1} \sum_{u=1}^{N^2-1}  \bra{\Psi_{\vec n u}^{(0)}} \mathcal M_1 \mathcal M^{-1} \mathcal M_1  \ket{\Psi_{\vec n u}^{(0)}} \nonumber\\
&=& - \sum_{||\vec n|| =1} \sum_{u,a,b=1}^{N^2-1} \int \d^d x \int \d^d y  \Braket{\Psi_{\vec n u}^{(0)} |\mathcal M_1 | x,a}  \Braket{x,a|\mathcal M^{-1}|y,b} \Braket{y,b| \mathcal M_1  |\Psi_{\vec n u}^{(0)} } \;.
\end{eqnarray*}
Working out the matrix elements in the infinite volume limit gives:
\begin{eqnarray}
\Braket{y,b| \mathcal M_1  |\Psi_{\vec n u}^{(0)} } &=& g f_{buc} A_\mu^c(y) \p_\mu \left(\frac{1}{L^{d/2}} \e^{\ii \frac{2\pi}{L} \vec n \cdot \vec y} \right) \nonumber\\
&=& \ii \frac{2\pi}{L}  g f_{buc} n_\mu A_\mu^c(y)  \left(\frac{1}{L^{d/2}} \right) \nonumber\\
\Braket{\Psi_{\vec n u}^{(0)} |\mathcal M_1 | x,a} &=& \ii \frac{2\pi}{L}  g f_{uac} n_\mu A_\mu^c(x)  \left(\frac{1}{L^{d/2}} \right)\;,
\end{eqnarray}
and thus\footnote{The summation over the color indices is implicitly assumed.}
\begin{eqnarray}
\Tr \kappa_r &=& \int \d^d x \int \d^d y \sum_{||\vec n|| =1} \frac{2\pi}{L}  g f_{uac} n_\mu A_\mu^c(x)  \left(\frac{1}{L^{d/2}} \right)  (\mathcal M^{-1})^{ab} \delta(x-y) \nonumber\\
&& \times \frac{2\pi}{L}  g f_{buc} n_\nu A_\nu^c(y)  \left(\frac{1}{L^{d/2}} \right)\nonumber\\
&=& 2 \left(\frac{1}{L^d} \right)  \left(\frac{2\pi}{L} \right)^2  \int \d^d x \int \d^d y g f_{uac} A_\mu^c(x)   (\mathcal M^{-1})^{ab}(x,y)    g f_{buc} A_\mu^c(y)  \nonumber\\
&=& - 2 \left( \frac{2\pi}{L}  \right)^2 \frac{1}{V} \int \d^d x \int \d^d y g f_{ba \ell} A_\mu^a (x) (\mathcal M^{-1})^{\ell m}(x,y) g f_{bkm} A^k_\mu (y) \;.
\end{eqnarray}
In conclusion, the following quantity should be positive
\begin{multline*}
\Tr \kappa = 2 \left( \frac{2\pi}{L}  \right)^2  \left( d (N^2 -1) - \frac{1}{V} \int \d^d x \int \d^d y g f_{ba\ell} A_\mu^a (x) (\mathcal M^{-1})^{\ell m} \delta(x-y) g f_{bkm} A^k_\mu (y)\right) \\
> 0 \;.
\end{multline*}
Therefore, we can set
\begin{eqnarray}
Z &=& \int [\d A] \e^{S_{\YM} + S_\gf} \theta( d(N^2 -1) - h(A)) \;,
\end{eqnarray}
whereby
\begin{eqnarray}
h(A)&=& \frac{1}{V} \int \d^d x \int \d^d y g f_{ba\ell} A_\mu^a (x) (\mathcal M^{-1})^{\ell m} \delta(x-y) g f_{bkm} A^k_\mu (y) \;.
\end{eqnarray}

\subsubsection{The non-local GZ action}\label{sectnonlocal}
We can now follow the analysis of the toy model, to lift $h(A)$ into the action. The first step was to replace the $\theta$-function with a $\delta$-function. The argument here is that at lowest order, $h(A)$ reduces to
\begin{eqnarray}
h(A)&=& \frac{1}{V} \int \d^d x \int \d^d y g f_{ba\ell} A_\mu^a (x) (\delta^{\ell m} \frac{1}{-\p^2} \delta(x-y)) g f_{bkm} A^k_\mu (y) \nonumber\\
&=& \frac{N}{V} \int \d^d x A_\mu^a (x) \frac{1}{\p^2} A^a_\mu (x) =  \frac{N}{V} \int \frac{\d^d p}{(2\pi)^d} A_\mu^a(p) \frac{1}{p^2} A_\mu^a (-p) \;,
\end{eqnarray}
which is exactly the toy model bound \eqref{ellips} (up to a factor $N$, which can be removed by redefinition of $h(A)$). Therefore, at least at lowest order it is justified to replace $\theta$ with $\delta$. It is assumed that higher order corrections would not affect this statement. The second step was to write the $\theta$-function as an exponential, by applying equation \eqref{wick},
\begin{eqnarray}\label{theta}
Z &=& \int [\d A] \int \frac{\d \gamma}{2 \pi \ii}\e^{\gamma(d (N^2 -1 )-h(A))} \e^{-S_\YM - S_\gf} \nonumber\\
&=&   \int \frac{\d \gamma}{2\pi \ii} \e^{-v(\gamma)} \;,
\end{eqnarray}
whereby $v(\gamma) = - \ln \int [\d A] \e^{\gamma(d (N^2 -1 )-h(A))} \e^{-S_\YM - S_\gf}$. The next step is to apply the saddle point approximation,
\begin{eqnarray}
Z &\approx&    \e^{-v(\gamma_0)} =  \int [\d A]  \e^{-[S_\YM + S_\gf + V \gamma^* h(A) - \gamma^* \int \d^d x d (N^2 -1 )]} \;,
\end{eqnarray}
whereby $\gamma^*$ is determined by
\begin{eqnarray}
v'(\gamma_0) &=& 0 \nonumber\\
d (N^2 - 1)&=& \frac{\int [\d A] h(A) \e^{V\gamma^*(-h(A))} \e^{-S} }{\int [\d A] \e^{V\gamma^*(-h(A))} \e^{-S}} \nonumber\\
d (N^2 - 1) &=& \Braket{h(A)}_{\gamma^*} \;,
\end{eqnarray}
whereby $\gamma = V \gamma^*$, in order to show explicitly  that the action is an extensive quantity. The final step is to prove that the saddle point approximation becomes exact, which can be done analogously as in equations \eqref{pr1} and \eqref{pr2}.\\
\\
In conclusion, the Gribov-Zwanziger path integral becomes
\begin{eqnarray}
Z  =  \int [\d A][\d c][\d \overline c]  \e^{-[S_\YM + S_\gf +   \int \d^d x h_1 (x) - \gamma^4 \int \d^d x d (N^2 -1 )]} \;,
\end{eqnarray}
whereby we have replaced $\gamma^*$ with $\gamma^4$ for further convenience and $h_1(x)$ is given by
\begin{eqnarray}\label{2h1}
h_1(x) &=&  \gamma^4 \int \d^d y g f_{ba\ell} A_\mu^a (x) (\mathcal M^{-1})^{\ell m} \delta(x-y) g f_{bkm} A^k_\mu (y) \;.
\end{eqnarray}
The parameter $\gamma$ is fixed by the following gap equation
\begin{eqnarray}
\Braket{h_1(x)} &=& \gamma^4 d (N^2 - 1) \;.
\end{eqnarray}

\subsubsection{Remarks}
One thing which needs to be pointed out, is what happens at the boundary with the path integral. At the boundary, one of the eigenvalues $\lambda$ approaches zero. Because of $h_1$, which contains the inverse of the Faddeev-Popov determinant, the probability in the path integral reduces very quickly, roughly speaking, a factor $\e^{-\frac{1}{\lambda}}$ enters the path integral. On the other hand, we have argued that only the boundary of the Gribov region gives contributions for $V \to \infty$, by replacing the $\theta$ function with a $\delta$ function in equation \eqref{theta}. This could sound contradictory. However, we can give an example which demonstrates what is going on. The path integral shall be a result of two competing functions. Firstly, we have a factor $r^{N-1}$ stemming from the integration, whereby $N$ approaches infinity in the thermodynamic limit and we simple take $r$ to represents the fields, while secondly, a factor $\e^{-\frac{1}{R-r}}$, with $R$ the size of the boundary, represents the horizon function. The following function shows us what is going on in the path integral
\begin{equation}\label{expressionqq}
\lim_{N \to \infty} r^{N-1} \e^{-\frac{1}{R-r}} \;.
\end{equation}
In the figures below, one can see how for larger $N$, this function evolves into a delta function
\begin{figure}[H]
  \centering
     \includegraphics[width=5cm]{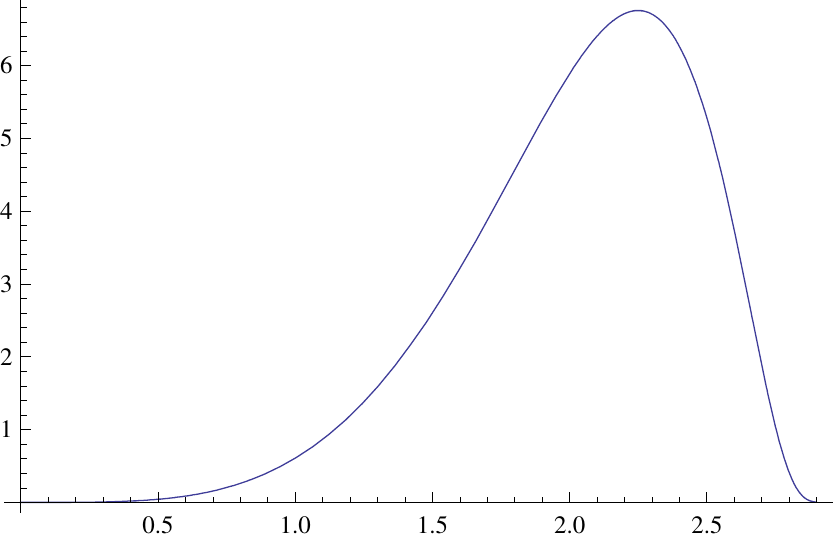}
     \includegraphics[width=5cm]{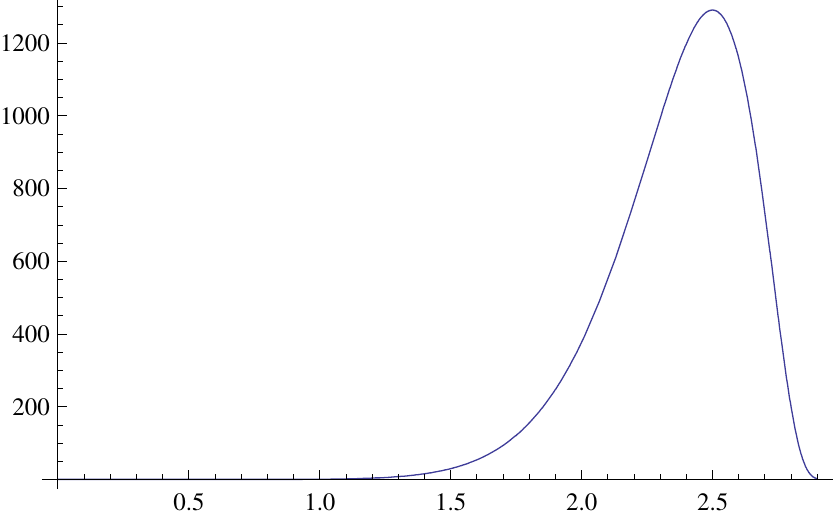}
     \includegraphics[width=5cm]{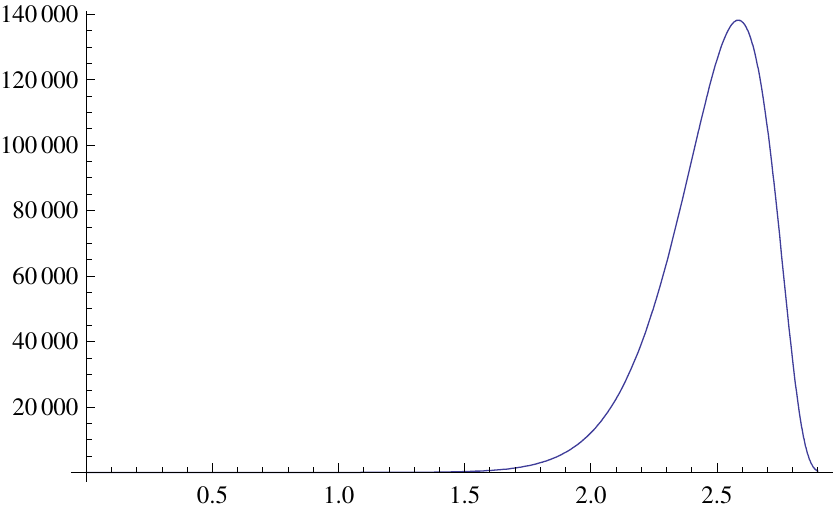}
     \includegraphics[width=5cm]{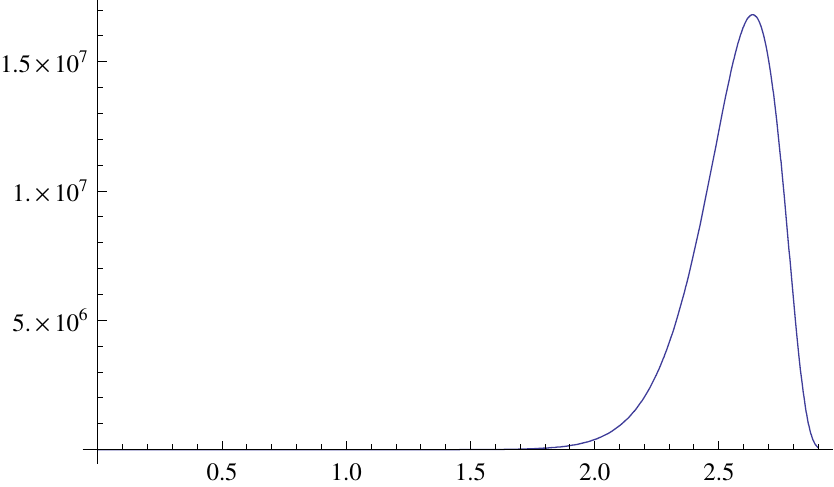}
  \caption{Evolution of the expression \eqref{expressionqq} for growing $N$ whereby we chose $R = 3$. }
\end{figure}

\noindent Another element which looks contradictory is the following \cite{private2}. In perturbation theory, or for large momenta, only the small area perturbing around $A =0$ is important, while we have just shown that the configurations get concentrated on the boundary of the Gribov region. Perhaps this can be explained by noticing that the Gribov parameter $\gamma^2$ cannot be accessed in perturbation theory. Indeed, as $\gamma^2 \propto \Lambda_\QCD^2 \propto  \e^{ -\frac{1}{g^2}}$, perturbatively, $\gamma = 0$.


\subsubsection{The correct horizon function}
In subsequent articles \cite{Zwanziger:1992qr,Zwanziger:1993dh}, Zwanziger refined the horizon function \eqref{2h1} into the following function
\begin{equation}\label{danhorizon}
S_\h = \lim_{\theta \to 0}  \int \d^d x h_2 (x) = \lim_{\theta \to 0}  \int \d^d x \int \d^d y   \left( D_\mu^{ac}(x) \gamma^2(x) \right)(\mathcal M^{-1})^{ab}(x,y) \left( D_\mu^{bc} (y) \gamma^2(y) \right) \;,
\end{equation}
whereby $\gamma(z)$ is defined through
\begin{eqnarray}\label{thetadep}
\gamma^2 (z) &=& \e^{\ii \theta z} \gamma^2 \;.
\end{eqnarray}
The $\lim_{\theta \to 0}$ operation corresponds to replacing the space time dependent $\gamma^2(z)$ with the constant Gribov parameter $\gamma^2$.  We observe that this horizon function shares a great resemblance with $h_1(x)$ from expression \eqref{2h1}. It is worth to point out here that the limit, $\lim_{\theta \to 0}$, in expression \eqref{danhorizon} is meant to be taken after an appropriate localization of the horizon function, a point which we shall outline in detail in what follows. \\
\\
In conclusion, the non-local action is given by
\begin{equation}\label{nonlocalGZ}
S_{\nl} = S_\YM + S_\gf + S_\h \;,
\end{equation}
with $S_\YM$ the Yang-Mills action and $S_\gf$ the gauge fixing and
\begin{eqnarray}\label{horizoncondition}
\braket{h_2(x)} &=& \gamma^4 d (N^2 -1) \,.
\end{eqnarray}

\subsection{The local Gribov-Zwanziger action\label{localization}}
In this section, we shall localize the action $S_\nl$ by introducing some extra fields. Looking at the following standard formula for Gaussian integration for bosonic fields, see expression \eqref{gauss8}
\begin{multline}\label{alg2}
 C \det A^{-1} \exp  \int \d^d x \d^d y \   J^a_\varphi(x) (A^{-1})^{ab}(x,y) J^b_{\overline{\varphi}}(y) \\
 = \int [\d \varphi][\d \overline{\varphi}] \exp\Bigl[ \int \d^d x \d^d y - \overline{\varphi}^a(x) A^{ab} (x,y)\varphi^b(y)
 + \int \d^d x \ (\varphi^a J^a_\varphi(x)  + \overline{\varphi}^a(x) J^a_{\overline{\varphi}} (x)) \Bigr] \;,
\end{multline}
we observe that we can get rid of the inverse of the Faddeev-Popov operator in $h_2(x)$ by introducing new fields. For every index $i$, defined by $\ldots_i^a = \ldots^{ac}_\mu$, we can write for $h_2(x)$
\begin{eqnarray*}
&&\exp\left( - \int \d^d x h_2(x) \right) = \prod_{i = 1}^{d (N^2 +1)} \det (-\mathcal M)\int [\d \varphi][\d \overline{\varphi}]   \exp \Biggl( \lim_{\theta\to 0} \Bigl[   \int \d^d x \int \d^d y  \nonumber\\
&&\overline \varphi_i^a (x) \mathcal M^{ab}(x,y)   \varphi^b_i (y) + \int \d^d x \left( D_i^{a} (x) \gamma^2(x) \right) \varphi_i^a(x) +  \left( D_i^a (x) \gamma^2(x) \right) \overline \varphi_i^a(x) \Bigr] \Biggr) \;,
\end{eqnarray*}
whereby we have introduced a pair of complex conjugate bosonic fields $\left(\overline \varphi_\mu^{ac}, \varphi_\mu^{ac}\right)$ $=$ $\left(\overline \varphi_i^{a}, \varphi_i^{a}\right)$. We can then also lift the determinants $ \det (-\mathcal M)$ into the exponential by introducing a pair of Grassmann fields $\left( \overline \omega_\mu^{ac},\omega_\mu^{ac} \right)$ $=$ $\left( \overline \omega_i^{a},\omega_i^{a} \right)$. Making use of the standard Gaussian formula for Grassmann variables
\begin{multline}\label{alg1}
	C \left(\det A\right) \exp\left(  -\int \d^d x \d^d y \   J_\omega^a(x) (A^{-1})^{ab}(x,y) J_{\overline{\omega}}^b(y) \right) \\
= \int [\d\omega][\d\overline{\omega}] \exp\left[ \int \d^d x \d^d y \ \overline{\omega}^a(x) A^{ab} \omega^b(y) + \int \d^d x \ (J^a_\omega(x) \omega^a(x) + \overline{\omega}^a(x) J_{\overline{\omega}}^a (x)) \right] \;,
\end{multline}
whereby we set the sources $J_\omega^a $ and $J_{\overline{\omega}}^b$ equal to zero, we obtain
\begin{eqnarray*}
\exp\left( - \int \d^d x h_2(x) \right) = \prod_{i = 1}^{d (N^2 +1)}  \int [\d\omega][\d\overline{\omega}][\d \varphi][\d \overline{\varphi}]
\exp \Bigl[  \int \d^d x \int \d^d y  \left( \overline \varphi_i^a (x) \mathcal M^{ab}(x,y)   \varphi^b_i (y)\right.\nonumber\\
 \left.- \overline \omega_i^a (x) \mathcal M^{ab}(x,y)   \omega^b_i (y)\right) +  \lim_{\theta\to 0} \int \d^d x \left( D_i^{a} (x) \gamma^2(x) \right) \varphi_i^a(x)  +  \left( D_i^a (x) \gamma^2(x) \right) \overline \varphi_i^a(x) \Bigr] \;.
\end{eqnarray*}
The new localized action thus becomes
\begin{eqnarray}\label{SGZphys}
S_\GZ &=& S_0' +  S_\gamma' \;,
\end{eqnarray}
with
\begin{equation}\label{SGZ}
S_0' = S_\YM + S_\gf   + \int \d^d x\left( \overline \varphi_\mu^{ac} \p_\nu D_\nu^{ab} \varphi_\mu^{bc}  - \overline \omega_\mu^{ac} \p_\nu D_\nu^{ab} \omega_\mu^{bc}   \right) \,,
\end{equation}
and with
\begin{eqnarray}\label{xhor1a}
S_\gamma' &=& -  \lim_{\theta \to 0} \int \d^d x\left[ \left( D_\mu^{ac} (x) \gamma^2(x) \right) \varphi_\mu^{ac}(x)  + \left( D_\mu^{ac} (x) \gamma^2(x) \right) \overline \varphi_\mu^{ac}(x)  \right]\nonumber   \\
&=&  \lim_{\theta\to 0} \int \d^d x \;  \gamma^2(x) D_\mu^{ca} (\varphi_\mu^{ac}(x) + \overline \varphi_\mu^{ac}(x))  \nonumber\\
&=&  \gamma^2 \int \d^d x  \; D_\mu^{ca} (\varphi_\mu^{ac}(x) + \overline \varphi_\mu^{ac}(x))\;. \label{bb1}
\end{eqnarray}
Notice that, as already remarked, the limit $\theta\to 0 $,  in equation \eqref{bb1} has been performed after localization. As one can see from \eqref{thetadep}, taking this limit is equivalent with setting $\gamma^2(x)$ equal to the constant $\gamma^2$. As at the level of the action, total derivatives are always neglected, $S_\gamma'$ becomes
\begin{eqnarray}\label{h1bis}
S_\gamma '&=&   \gamma ^{2}  \int \d^d x g  f^{abc}A_\mu^a \left( \varphi_\mu^{bc} +  \overline \varphi_\mu^{bc} \right) \,.
\end{eqnarray}
Notice here that starting from the first horizon function $h_1(x)$ given in \eqref{2h1} and undertaking the same procedure, we would end up with exactly the same action $S_\gamma'$. This can be understood as we have neglected the total derivatives. Although the local actions derived from $h_1(x)$ and $h_2(x)$ are the same, at the nonlocal level they are clearly different. We shall come back to this subtle point in chapter \ref{scrutinizing}.\\
\\
Let us now translate the nonlocal horizon condition \eqref{horizoncondition} into a local version \cite{Zwanziger:1992qr}. The local action $S_\GZ$ and the nonlocal action $S_\YM + S_\gf + S_\h$ are related as follows,
\begin{equation}
    \int [\d A] [\d b] [\d c][ \d \overline{c}] \e^{- (S_\YM + S_\gf + S_\h)} = \int [\d A ][\d b][ \d c ][\d\overline{c}][ \d \varphi][ \d\overline{\varphi}][ \d \omega][ \d \overline{\omega}]\e^{-S_\GZ} \;.
\end{equation}
Next, we take the partial derivative of both sides with respect to $\gamma^2$ so we obtain,
\begin{eqnarray}
-2 \gamma^2 \frac{\braket{h}}{\gamma^4} &=& \braket{g f^{abc} A^a_{\mu} ( \varphi^{bc}_{\mu} +  \overline{\varphi}^{bc}_{\mu})}  \;,
\end{eqnarray}
for both horizon functions $h_1(x)$ and $h_2(x)$. We recall that $\braket{\p_\mu \varphi^{aa}} = 0$ and $\braket{\p_\mu \overline \varphi^{aa}} = 0$, meaning that both horizon functions $h_1$ and $h_2$ give rise to the same local horizon condition.  Using these expressions and assuming that $\gamma \not= 0$, we can rewrite the horizon condition \eqref{horizoncondition}
\begin{eqnarray}\label{horizonconditionlocal}
\braket{g f^{abc} A^a_{\mu} ( \varphi^{bc}_{\mu} +  \overline{\varphi}^{bc}_{\mu})}   + 2 \gamma^2 d (N^2 -1)   = 0  \;.
\end{eqnarray}
By adding the vacuum term
\begin{eqnarray}\label{vac}
\int \d^d x \;  \gamma^4 d (N^2 - 1 ) \;,
\end{eqnarray}
to $S_\gamma'$, we can write the horizon condition as
\begin{eqnarray}\label{gapgamma}
\frac{\p \Gamma}{\p \gamma^2} &=& 0\;,
\end{eqnarray}
with $\Gamma$ the quantum action defined as
\begin{eqnarray}
\e^{-\Gamma} &=& \int [\d\Phi] \e^{-S_\GZ} \;,
\end{eqnarray}
where $\int [\d\Phi]$ stands for the integration over all the fields.\\
\\
For the Gribov-Zwanziger action to be renormalizable, it necessary to perform a shift over the field $\omega^a_i$, see \cite{Zwanziger:1992qr},
\begin{equation}\label{shift}
\omega^a_i (x) \to  \omega^a_i (x) + \int \d^d z (\mathcal M^{-1})^{ad} (x,z) g f_{dk\ell} \p_\mu [ D_\mu^{ke} c^e (z) \varphi^\ell (z)]\;,
\end{equation}
so that the action becomes
\begin{eqnarray}\label{SGZecht}
S_\GZ &=& S_0 +  S_{\gamma} \,,
\end{eqnarray}
whereby $S_0'$ has been replaced by $S_0$
\begin{eqnarray}\label{S0}
S_0 &=& S_0' + \int \d^d x \left( - g f^{abc} \p_\mu \overline \omega_i^a    D_\mu^{bd} c^d  \varphi_i^c \right)\,,
\end{eqnarray}
and the vacuum term is now included in $S_\gamma$
\begin{eqnarray}
S_{\gamma}&=& S_\gamma' + \int \d^d x  \; \gamma^4 d (N^2 - 1 ) \;.
\end{eqnarray}
In section \ref{algebraicrenormGZ} we shall prove the action $S_\GZ$ to be renormalizable. We would like to stress that this is far from being trivial, especially since no new parameter is needed to take into account vacuum divergences, which would lead to a modification of the vacuum term we introduced by hand in equation \eqref{vac}. In addition, the algebraic formalism employed in the next section also gives a more clean argument why the extra term appearing in equation \eqref{S0} is necessary, without the need of performing the nonlocal shift \eqref{shift}.

\subsection{The gluon and the ghost propagator\label{gluonghostchap2}}
Now that we have the local Gribov-Zwanziger action at our disposal, we can easily calculate the gluon and ghost propagator, at lowest order. We shall show that we obtain the same results as Gribov obtained, see section \ref{sectiongribov}.

\subsubsection{The gluon propagator}
To calculate the tree level gluon propagator, we only need the free part of the Gribov-Zwanziger action $S_\GZ$,
\begin{eqnarray}
    S_\GZ^0 &=& \int \d^d x \Bigl[ \frac{1}{4} \left( \p_{\mu} A_{\nu}^a - \p_{\nu}A_{\mu}^a\right)^2 + \frac{1}{2\alpha} \left( \p_{\mu} A^a_{\mu} \right)^2 +\overline{\varphi}^{ab}_{\mu} \p^2 \varphi^{ab}_{\mu} \nonumber\\
 &&  \hspace{1cm}- \gamma^2 g(f^{abc}A^a_{\mu} \varphi_{\mu}^{bc} + f^{abc}A^a_{\mu}\overline{\varphi}^{bc}_{\mu} )  +\ldots \Bigr]\;,
\end{eqnarray}
where the limit $\alpha \to 0$ is understood in order to recover the Landau gauge. The $\ldots$ stands for the constant term $-d (N^2 -1) \gamma^4$ and other terms in the ghost- and $\omega, \overline{\omega}$-fields irrelevant for the calculation of the gluon propagator. Next, we integrate out the $\varphi$- and $\overline{\varphi}$-fields. As we are only interested in the gluon propagator, we simply use the equations of motion, $\frac{\p S_\GZ^0 }{\p \overline{\varphi}_{\mu}^{bc}} = 0$ and $\frac{\p S_\GZ^0 }{\p\varphi_{\mu}^{bc}} = 0$, which give
\begin{eqnarray}
    \varphi^{bc}_{\mu} = \overline{\varphi}^{bc}_{\mu} =  \frac{1}{\p^2}\gamma^2 g f^{abc} A^a_{\mu} \;.
\end{eqnarray}
We use this result to rewrite $ S_\GZ^0$,
\begin{eqnarray}
 S_\GZ^0 &=& \; \int \d^d x \; \left[ \frac{1}{4} \left(\p_{\mu} A_{\nu}^a - \p_{\nu}A_{\mu}^a \right)^2+ \frac{1}{2\alpha} \left(\p_{\mu} A^a_{\mu} \right)^2  + \gamma^4 g^2 f^{abc} A_{\mu}^a \frac{1}{\p^2 } f^{dbc} A^d_{\mu} \right. \nonumber\\
&& \left. \hspace{1cm}- 2 \gamma^4 g(f^{abc}A^a_{\mu} \frac{1}{\p^2} g f^{dbc} A_{\mu}^d ) + \ldots \right] \nonumber\\
&=&  \; \int \d^dx \; \left[ \frac{1}{4} \left( \p_{\mu} A_{\nu}^a - \p_{\nu}A_{\mu}^a \right)^2+ \frac{1}{2\alpha} \left( \p_{\mu} A^a_{\mu} \right)^2  - N \gamma^4 g^2 A_{\mu}^a \frac{1}{\p^2 } A^a_{\mu} + \ldots \right]\;,
\end{eqnarray}
whereby the last step is explained with the relation \eqref{liestructure}. We continue rewriting $ S_\GZ^0$ so we can easily read off the gluon propagator
\begin{eqnarray}
  S_\GZ^0&=&  \; \int \d^d x \; \left[ \frac{1}{2} A^a_{\mu} \Delta^{ab}_{\mu\nu} A^b_{\nu} + \ldots \right] \;, \nonumber\\
\Delta^{ab}_{\mu\nu} &=&\left[ \left(-\p^2 - \frac{2 g^2 N \gamma^4}{\p^2} \right) \delta_{\mu\nu} - \p_{\mu}\p_{\nu} \left(\frac{1}{\alpha} - 1\right) \right]\delta^{ab} \;.
\end{eqnarray}
The gluon propagator can be determined by taking the inverse of $\Delta^{ab}_{\mu\nu}$ and converting it to momentum space. Doing so, we find the following expression
\begin{eqnarray}
  \; \Braket{ A^a_{\mu}(p) A^b_{\nu}(k)} &=& \delta(p+k) (2\pi)^d \underbrace{ \frac{p^2}{p^4 + 2g^2 N \gamma^4}}_{\mathcal{D}(p^2)}\left[\delta_{\mu\nu} - \frac{p_{\mu}p_{\nu}}{p^2} \right]\delta^{ab} \;,\nonumber\\
\end{eqnarray}
which is exactly the same expression as Gribov found, see equation \eqref{gluonprop}. We can already observe that this expression is suppressed in the infrared region,
while displaying complex poles at $k^{2}=\pm i{\hat \gamma}^{2}$. This structure does not allow us to attach the usual particle meaning to the gluon propagator, invalidating the
interpretation of gluons as excitations of the physical spectrum. In other words, gluons cannot be considered as part of the physical spectrum. In this sense, they are confined by the Gribov horizon, whose presence is encoded in the explicit dependence the propagator on the Gribov parameter $\gamma$.

\subsubsection{The ghost propagator}
\begin{figure}[H]
\begin{center}
\includegraphics[width=8cm]{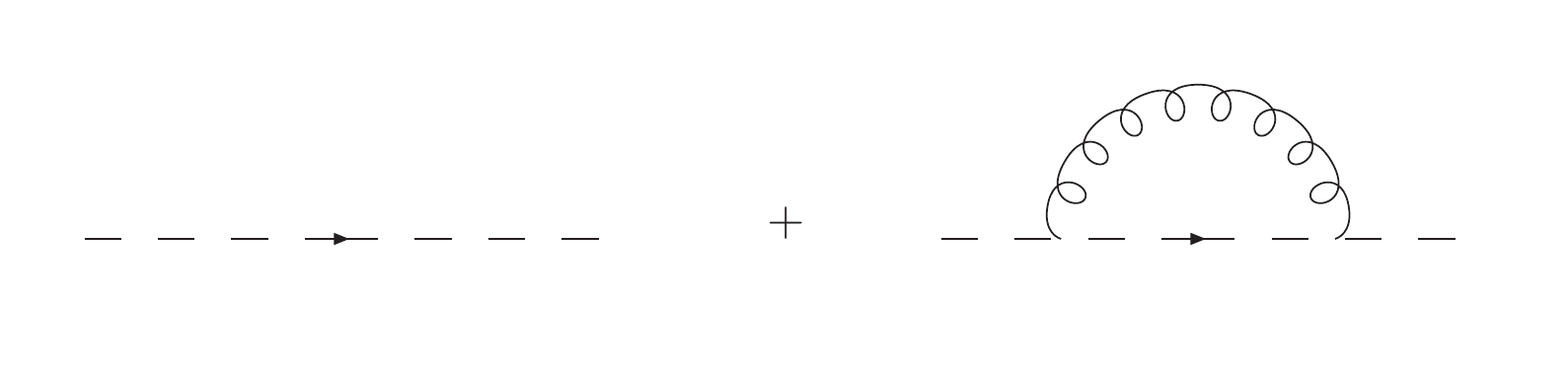}
\caption{The one loop corrected ghost propagator.}\label{2ghostprop}
\end{center}
\end{figure}

In order to find the ghost propagator, we need to calculate the one-loop corrected ghost propagator, see figure \ref{2ghostprop}

\noindent In momentum space, the ghost propagator is given by
\begin{eqnarray}
\Braket{c^a(p) \overline{c}^b(k)} &=&  \delta^{ab} (2\pi)^d \delta(k-p) \mathcal{G}(k^2)\;,
\end{eqnarray}
whereby
\begin{eqnarray}\label{ghostlabel}
\mathcal{G}(k^2) &=& \frac{1}{k^2} + \frac{1}{k^2} \left[g^2 \frac{N}{N^2 - 1} \int \frac{\d^d
q}{(2\pi)^4} \frac{(k-q)_{\mu} k_{\nu}}{(k-q)^2}
\frac{q^2}{q^4 + 2g^2 N \gamma^4} \right] P_{\mu \nu}(q) \frac{1}{k^2} \nonumber\\
&=& \frac{1}{k^2} (1+ \sigma(k^2)) +
\mathcal{O}(g^4)\;,
\end{eqnarray}
with
\begin{eqnarray}\label{ghi}
\sigma(k^2) &=& Ng^2 \frac{1}{k^2}\int \frac{\d^d q}{(2\pi)^d} \frac{(k-q)_{\mu} k_{\nu}}{(k-q)^2} \frac{q^2}{q^4 + 2g^2 N \gamma^4} \left(\delta_{\mu\nu} - \frac{q_\mu q_\nu}{q^2} \right)\nonumber\\
&=&  Ng^2\frac{k_\mu k_\nu}{k^2}\int \frac{\d^d q}{(2\pi)^d} \frac{1}{(k-q)^2} \frac{q^2}{q^4 + 2g^2 N \gamma^4} \left(\delta_{\mu\nu} - \frac{q_\mu q_\nu}{q^2} \right)\;,
\end{eqnarray}
which is of course similar to expression \eqref{ghostpropagator}. We are again interested in the low momentum behavior and therefore calculate $\sigma(0)$,
\begin{eqnarray}\label{sigmanul}
\sigma(0) &=&  \frac{Ng^2}{N^2 - 1} \frac{k_\mu k_\nu}{k^2} \delta_{\mu\nu} \frac{d-1}{d}\int \frac{\d^d q}{(2\pi)^d} \frac{1}{q^2} \frac{q^2}{q^4 + 2g^2 N \gamma^4}\nonumber\\
&=&  \frac{Ng^2}{N^2 - 1}  \frac{d-1}{d}\int \frac{\d^d q}{(2\pi)^d}  \frac{1}{q^4 + 2g^2 N \gamma^4} \;.
\end{eqnarray}
Notice that this integral diverges. \\
\\
To calculate this integral, we shall invoke the gap equation \eqref{gapgamma}. Firstly, we calculate the effective action. The one loop effective action $\Gamma_\gamma^{(1)}$ is obtained from the quadratic part of our action $S_\GZ^0$
\begin{equation}
\e^{-\Gamma_\gamma ^{(1)} }=\int [\d\Phi] \e^{-S_\GZ^0}\;,
\end{equation}
with
\begin{eqnarray*}
 S_\GZ^0 &=&   \; \int \d^dx \; \left[ \frac{1}{4} \left( \p_{\mu} A_{\nu}^a - \p_{\nu}A_{\mu}^a \right)^2+ \frac{1}{2\alpha} \left( \p_{\mu} A^a_{\mu} \right)^2  - N \gamma^4 g^2 A_{\mu}^a \frac{1}{\p^2 } A^a_{\mu}  -d(N^2 - 1) \gamma^4 + \ldots \right]\;.
\end{eqnarray*}
Notice that this time, we need to maintain the constant term $-d(N^2 - 1) \gamma^4$ as it will enter the horizon condition. After a straightforward calculation the one loop effective action in $d$ dimensions yields,
\begin{eqnarray}
\Gamma_\gamma^{(1)} &=& -d(N^{2}-1)\gamma^{4} +\frac{(N^{2}-1)}{2}\left( d-1\right) \int \frac{\d^{d}q}{\left(
2\pi \right) ^{d}} \ln \frac{q^4  + 2 g^2 N \gamma^4}{ q^2} \;.
\end{eqnarray}
Now we can apply the gap equation \eqref{gapgamma},
\begin{equation}
\frac{\p  \Gamma_\gamma^{(1)}  }  {\p \gamma^2} = - 2 \gamma^2 d(N^{2}-1) + 2g^2 N (N^2 -1)\gamma^2 (d-1)\int \frac{\d^{d}q}{\left(2\pi \right) ^d}  \frac{1}{q^4  + 2 g^2 N \gamma^4} ~=~ 0\;,
\end{equation}
or thus\footnote{The solution $\gamma = 0$ has to be disregarded. This is an artefact of the reformulation of the horizon condition.}
\begin{eqnarray}
1 &=&   g^2 N \frac{d-1}{d}\int \frac{\d^{d}q}{\left(2\pi \right) ^{d}}  \frac{1}{q^4  + 2 g^2 N \gamma^4}\;,
\end{eqnarray}
which exactly expresses
\begin{eqnarray}\label{sigmagelijkaan1}
\sigma(0) = 1\;,
\end{eqnarray}
see expression \eqref{sigmanul}. This means that the ghost propagator is enhanced, just as we expected from the semi-classical calculation of Gribov. Moreover, this result has been explicitly checked up to two loops, see \cite{Ford:2009ar,Gracey:2005cx}.

\section{Algebraic renormalization of the Gribov-Zwanziger action\label{algebraicrenormGZ}}
We shall now prove that the Gribov-Zwanziger action is renormalizable to all orders by using \textit{algebraic renormalization} as explained in chapter \ref{algebraic}. In \cite{Zwanziger:1989mf}, it was first shown that the Gribov-Zwanziger action was renormalizable. A first algebraic proof was given in \cite{Sobreiro:2004us}, and made complete in \cite{Dudal:2010fq}. A recent alternative proof can be found in \cite{Capri:2010hb}.

\subsection{The starting action and the BRST}
We start with the Gribov-Zwanziger action
\begin{eqnarray}\label{GZstart}
S_\GZ &=& S_\YM + S_\gf + S_{0} +  S_{\gamma} \,,
\end{eqnarray}
with
\begin{eqnarray}
S_{0}&=& \int \d^d x \left( \overline \varphi_i^a \p_\mu \left( D_\mu^{ab} \varphi^b_i \right)  - \overline \omega_i^a \p_\mu \left( D_\mu^{ab} \omega_i^b \right) - g f^{abc} \p_\mu \overline \omega_i^a    D_\mu^{bd} c^d  \varphi_i^c \right) \nonumber \;, \\
S_{\gamma}&=& -\gamma ^{2}g\int\d^{d}x\left( f^{abc}A_{\mu }^{a}\varphi _{\mu }^{bc} +f^{abc}A_{\mu}^{a}\overline{\varphi }_{\mu }^{bc} + \frac{d}{g}\left(N^{2}-1\right) \gamma^{2} \right) \;.
\end{eqnarray}
We recall that we have simplified the notation of the additional fields $\left( \overline \varphi_\mu^{ac},\varphi_\mu^{ac},\overline \omega_\mu^{ac},\omega_\mu^{ac}\right) $ in $S_0$ as $S_0$ displays a symmetry with respect to the composite index $i=\left( \mu,c\right)$. Therefore, we have set
\begin{equation}
\left( \overline \varphi_\mu^{ac},\varphi_\mu^{ac},\overline \omega_\mu^{ac},\omega_\mu^{ac}\right) =\left( \overline \varphi_i^a,\varphi_i^a,\overline \omega_i^a,\omega_i^a \right)\,.
\end{equation}
The BRST variations \eqref{BRST} can be logically extended for all the fields,
\begin{align}\label{BRST1}
sA_{\mu }^{a} &=-\left( D_{\mu }c\right) ^{a}\,, & sc^{a} &=\frac{1}{2}gf^{abc}c^{b}c^{c}\,,   \nonumber \\
s\overline{c}^{a} &=b^{a}\,,&   sb^{a}&=0\,,  \nonumber \\
s\varphi _{i}^{a} &=\omega _{i}^{a}\,,&s\omega _{i}^{a}&=0\,,\nonumber \\
s\overline{\omega}_{i}^{a} &=\overline{\varphi }_{i}^{a}\,,& s \overline{\varphi }_{i}^{a}&=0\,.
\end{align}
However, due to the $\gamma$ dependent term, $S_\gamma$, one can easily check that Gribov-Zwanziger action breaks this BRST symmetry softly\cite{Zwanziger:1989mf,Dudal:2008sp},
\begin{equation}\label{breaking}
s S_\GZ = s (S_{0} +  S_{\gamma}) ~=~ s (  S_{\gamma}) ~=~- g \gamma^2 \int \d^d x f^{abc} \left( A^a_{\mu} \omega^{bc}_\mu -
 \left(D_{\mu}^{am} c^m\right)\left( \overline{\varphi}^{bc}_\mu + \varphi^{bc}_{\mu}\right)  \right)\,.
\end{equation}
In chapter \ref{scrutinizing}, we shall elaborate on the meaning of this BRST breaking. \\
\\
In order to discuss the renormalizability of $S_\GZ$, we should treat the breaking as a composite operator to be introduced into the action by means of a suitable set of external sources. This procedure can be done in a BRST invariant way, by embedding $S_\GZ$ into a larger action, namely
\begin{eqnarray}\label{brstinvariant}
\Sigma_\GZ &=& S_{\YM} + S_{\gf} + S_0 + S_\s \,,
\end{eqnarray}
whereby
\begin{eqnarray}\label{previous}
S_\s &=& s\int \d^d x \left( -U_\mu^{ai} D_\mu^{ab} \varphi_i^b - V_\mu^{ai} D_{\mu}^{ab} \overline \omega_i^{b} - U_\mu^{ai} V_\mu^{ai}  + T_\mu^{a i} g f_{abc} D^{bd}_\mu c^d \overline \omega^c_i \right)\nonumber\\
&=& \int \d^d x \left( -M_\mu^{ai}  D_\mu^{ab} \varphi_i^b - gf^{abc} U_\mu^{ai}   D^{bd}_\mu c^d  \varphi_i^c   + U_\mu^{ai}  D_\mu^{ab} \omega_i^b - N_\mu^{ai}  D_\mu^{ab} \overline \omega_i^b - V_\mu^{ai}  D_\mu^{ab} \overline \varphi_i^b \right. \nonumber\\
&&\left.+ gf^{abc} V_\mu^{ai} D_\mu^{bd} c^d \overline \omega_i^c - M_\mu^{ai} V_\mu^{ai}+U_\mu^{ai} N_\mu^{ai}  + R_\mu^{ai} g f^{abc} D_\mu^{bd} c^d \overline \omega^c_i  + T_\mu^{ai} g f_{abc} D^{bd}_\mu c^d \overline \varphi^c_i\right) \,. \nonumber\\
\end{eqnarray}
We have introduced 3 new doublets ($U_\mu^{ai}$, $M_\mu^{ai}$), ($V_\mu^{ai}$, $N_\mu^{ai}$) and ($T_\mu^{ai}$, $R_\mu^{ai}$) with the following BRST transformations,
and
\begin{align}\label{BRST2}
sU_{\mu }^{ai} &= M_{\mu }^{ai}\,, & sM_{\mu }^{ai}&=0\,,  \nonumber \\
sV_{\mu }^{ai} &= N_{\mu }^{ai}\,, & sN_{\mu }^{ai}&=0\,,\nonumber \\
sT_{\mu }^{ai} &= R_{\mu }^{ai}\,, & sR_{\mu }^{ai}&=0\;.
\end{align}
We have therefore restored the broken BRST at the expense of introducing new sources. However, we do not want to alter our original theory \eqref{SGZ}. Therefore, at the end, we have to set the sources equal to the following values:
\begin{eqnarray}\label{physlimit}
&& \left. U_\mu^{ai}\right|_{\phys} = \left. N_\mu^{ai}\right|_{\phys} = \left. T_\mu^{ai}\right|_{\phys} = 0 \,, \nonumber\\
&& \left. M_{\mu \nu }^{ab}\right|_{\phys}= \left.V_{\mu \nu}^{ab}\right|_{\phys}=  -\left.R_{\mu \nu}^{ab}\right|_{\phys} = \gamma^2 \delta ^{ab}\delta _{\mu \nu } \,.
\end{eqnarray}
We emphasize that only in \cite{Dudal:2010fq} this doublet ($T_\mu^{ai}$, $R_\mu^{ai}$) was introduced, and therefore, the only correct algebraic renormalization of the Gribov-Zwanziger action can be found there. Otherwise, the intended physical limit does not exactly reproduce the original action \eqref{GZstart}. In the original article \cite{Zwanziger:1992qr}, these two sources were also not introduced. When taking the physical limit, an extra term was generated, which was then removed by doing a (nonlocal) shift in the $\omega$ field. Here, we have circumvented this unnecessary shift by introducing the doublet ($T_\mu^{ai}$, $R_\mu^{ai}$). Notice that the terms $gf^{abc} V_\mu^{ai} D_\mu^{bd} c^d \overline \omega_i^c$ and $ R_\mu^{ai} g f^{abc} D_\mu^{bd} c^d \overline \omega^c_i $ cancel in the physical limit.

\subsection{The Ward identities}
Following the procedure of algebraic renormalization outlined in the chapter \ref{algebraic}, we should try to find all possible Ward identities. Before doing this, in order to be able to write the Slavnov-Taylor identity, we first have to couple all nonlinear BRST transformations to a new source. Looking at \eqref{BRST1}, we see that only $A_\mu^a$ and $c^a$ transform nonlinearly under the BRST $s$. Therefore, we add the following term to the action $\Sigma_\GZ $,
\begin{eqnarray}\label{ext}
S_{\mathrm{ext}}&=&\int \d^d x\left( -K_{\mu }^{a}\left( D_{\mu }c\right) ^{a}+\frac{1}{2}gL^{a}f^{abc}c^{b}c^{c}\right) \;,
\end{eqnarray}
with $K_{\mu }^{a}$ and $L^a$ two new sources which shall be put to zero at the end,
\begin{eqnarray}\label{physlimit2}
\left. K_{\mu }^{a}\right|_{\phys} =\left. L^{a}\right|_{\phys}  = 0\;.&
\end{eqnarray}
These sources are invariant under the BRST transformation,
\begin{align}
s K_{\mu }^{a} &=0\;, & s L^{a} &= 0\;.
\end{align}
The new action is therefore given by
\begin{eqnarray}\label{enlarged}
\Sigma'_\GZ &=& \Sigma_\GZ + S_{\mathrm{ext}} \;.
\end{eqnarray}
The next step is now to find all the Ward identities obeyed by the action $\Sigma'_\GZ$. We have enlisted all the identities below:

\begin{table}
\caption{Quantum numbers of the fields.}
\label{2tabel1}
\begin{center}
\begin{tabular}{|c|c|c|c|c|c|c|c|c|}
\hline
& $A_{\mu }^{a}$ & $c^{a}$ & $\overline{c}^{a}$ & $b^{a}$ & $\varphi_{i}^{a} $ & $\overline{\varphi }_{i}^{a}$ &                $\omega _{i}^{a}$ & $\overline{\omega }_{i}^{a}$   \\
\hline
\textrm{dimension} & $1$ & $0$ &$2$ & $2$ & $1$ & $1$ & $1$ & $1$ \\
$\mathrm{ghost\; number}$ & $0$ & $1$ & $-1$ & $0$ & $0$ & $0$ & $1$ & $-1$ \\
$Q_{f}\textrm{-charge}$ & $0$ & $0$ & $0$ & $0$ & $1$ & $-1$& $1$ & $-1$ \\
\hline
\end{tabular}
\end{center}
\end{table}

\begin{table}
\caption{Quantum numbers of the sources.}
\begin{center}
\label{2tabel2}
\begin{tabular}{|c|c|c|c|c|c|c|c|c|}\hline
        $U_{\mu}^{ai}$&$M_{\mu }^{ai}$&$N_{\mu }^{ai}$&$V_{\mu }^{ai}$& $R_{\mu }^{ai}$  &  $T_{\mu }^{ai}$ &$K_{\mu }^{a}$&$L^{a}$  \\
\hline
         $2$ & $2$ & $2$ &$2$ & 2&2  & $3$ & $4$  \\
         $-1$& $0$ & $1$ & $0$ & 0& -1 & $-1$ & $-2$  \\
         $-1$ & $-1$ & $1$ & $1$ &1&1& $0$ & $0$  \\
\hline
\end{tabular}
\end{center}
\end{table}
\label{pagewardidentities}
\begin{enumerate}
\item The Slavnov-Taylor identity is given by
\begin{equation}\label{slavnov}
\mathcal{S}(\Sigma'_\GZ )=0\;,
\end{equation}
with
\begin{multline*}
\mathcal{S}(\Sigma'_\GZ ) =\int \d^d x\left( \frac{\delta \Sigma'_\GZ
}{\delta K_{\mu }^{a}}\frac{\delta \Sigma'_\GZ }{\delta A_{\mu
}^{a}}+\frac{\delta \Sigma'_\GZ }{\delta L^{a}}\frac{\delta \Sigma'_\GZ
}{\delta c^{a}} \right. \nonumber\\
\left. +b^{a}\frac{\delta \Sigma'_\GZ }{\delta \overline{c}^{a}}+\overline{\varphi }_{i}^{a}\frac{\delta \Sigma'_\GZ }{\delta \overline{\omega }_{i}^{a}}+\omega _{i}^{a}\frac{\delta \Sigma'_\GZ }{\delta \varphi _{i}^{a}} +M_{\mu }^{ai}\frac{\delta \Sigma'_\GZ}{\delta U_{\mu}^{ai}}+N_{\mu }^{ai}\frac{\delta \Sigma'_\GZ }{\delta V_{\mu }^{ai}} + R_{\mu }^{ai}\frac{\delta \Sigma'_\GZ }{\delta T_{\mu }^{ai}}\right) \;.
\end{multline*}

\item The $U(f)$ invariance is given by
\begin{equation}\label{ward1}
U_{ij} \Sigma'_\GZ =0\;,
\end{equation}
\begin{multline}
U_{ij}=\int \d^dx\Bigl( \varphi_{i}^{a}\frac{\delta }{\delta \varphi _{j}^{a}}-\overline{\varphi}_{j}^{a}\frac{\delta }{\delta \overline{\varphi}_{i}^{a}}+\omega _{i}^{a}\frac{\delta }{\delta \omega _{j}^{a}}-\overline{\omega }_{j}^{a}\frac{\delta }{\delta \overline{\omega }_{i}^{a}} \\
-  M^{aj}_{\mu} \frac{\delta}{\delta M^{ai}_{\mu}} -U^{aj}_{\mu}\frac{\delta}{\delta U^{ai}_{\mu}} + N^{ai}_{\mu}\frac{\delta}{\delta N^{aj}_{\mu}}
  +V^{ai}_{\mu}\frac{\delta}{\delta V^{aj}_{\mu}}    +  R^{aj}_{\mu}\frac{\delta}{\delta R^{ai}_{\mu}} + T^{aj}_{\mu}\frac{\delta}{\delta T^{ai}_{\mu}} \Bigr)  \;. \nonumber
\end{multline}
By means of the diagonal operator $Q_{f}=U_{ii}$, the
$i$-valued fields and sources turn out to possess an additional quantum number.
One can find all quantum numbers in Table \ref{2tabel1} and Table \ref{2tabel2}.

\item The Landau gauge condition reads
\begin{eqnarray}\label{gaugeward}
\frac{\delta \Sigma'_\GZ }{\delta b^{a}}&=&\partial_\mu A_\mu^{a}\;.
\end{eqnarray}

\item The antighost equation yields
\begin{eqnarray}
\frac{\delta \Sigma'_\GZ }{\delta \overline{c}^{a}}+\partial _{\mu}\frac{\delta \Sigma'_\GZ }{\delta K_{\mu }^{a}}&=&0\;.
\end{eqnarray}

\item The linearly broken local constraints yield
\begin{eqnarray}
\frac{\delta \Sigma'_\GZ }{\delta \overline{\varphi }^{a}_i}+\partial _{\mu }\frac{\delta \Sigma'_\GZ }{\delta M_{\mu }^{ai}} + g f_{dba}    T^{d i}_\mu \frac{\delta \Sigma'_\GZ }{\delta K_{\mu }^{b i}} &=&gf^{abc}A_{\mu }^{b}V_{\mu}^{ci} \;, \nonumber\\
\frac{\delta \Sigma'_\GZ }{\delta \omega ^{a}_i}+\partial _{\mu}\frac{\delta \Sigma'_\GZ }{\delta N_{\mu}^{ai}}-gf^{abc}\overline{\omega }^{b}_i\frac{\delta \Sigma'_\GZ }{\delta b^{c}}&=&gf^{abc}A_{\mu }^{b}U_{\mu }^{ci} \;.
\end{eqnarray}

\item The exact $\mathcal{R}_{ij}$ symmetry reads
\begin{equation}
\mathcal{R}_{ij}\Sigma'_\GZ =0\;,
\end{equation}
with
\begin{multline}\label{rij}
\mathcal{R}_{ij} = \int \d^dx\left( \varphi _{i}^{a}\frac{\delta}{\delta\omega _{j}^{a}}-\overline{\omega }_{j}^{a}\frac{\delta }{\delta \overline{\varphi }_{i}^{a}}+V_{\mu }^{ai}\frac{\delta }{\delta N_{\mu}^{aj}}-U_{\mu }^{aj}\frac{\delta }{\delta M_{\mu }^{ai}} + T^{a i}_\mu \frac{\delta }{\delta R_{\mu }^{aj}}  \right) \;.
\end{multline}

\item The integrated Ward identity is given by
\begin{equation}
\int \d^d x \left( c^a \frac{ \delta \Sigma'_\GZ }{ \delta \omega^{ a}_i} + \overline \omega^{a}_i \frac{ \delta \Sigma'_\GZ }{ \delta  \overline c^a } + U^{a i}_\mu \frac{ \delta \Sigma'_\GZ }{ \delta  K^a_\mu }  \right) = 0\;.
\end{equation}
\end{enumerate}
Here we should add that due to the presence of the sources $T_{\mu }^{ai}$ and $R_{\mu }^{ai}$, the Ghost-Ward identity is broken, see section \ref{5.5} of chapter \ref{algebraic}. However, it shall turn out that this is not a problem for the renormalization procedure being undertaken.

\subsection{The counterterm}
The next step in the algebraic renormalization is to translate all these symmetries into constraints on the counterterm\footnote{In the previous chapter, this counterterm was denoted by $\overline \Gamma^{(1)}$,e.g.~see expression \eqref{finitedivergent}.} $\Sigma_\GZ^c$, which is an integrated polynomial in the fields and sources of dimension four and with ghost number zero. The classical action $\Sigma_\GZ'$ changes under quantum corrections according to
\begin{eqnarray}\label{deformed}
    \Sigma_\GZ' \rightarrow \Sigma_\GZ' + h \Sigma_\GZ^c\,,
\end{eqnarray}
whereby $h$ is the perturbation parameter. Demanding that the perturbed action $(\Sigma_\GZ' + h \Sigma_\GZ^c)$ fulfills the same set of Ward identities obeyed by $\Sigma_\GZ'$, see chapter \ref{algebraic}, it follows that the counterterm $\Sigma_\GZ^c$ is constrained by the following identities.
\begin{enumerate}
\item The linearized Slavnov-Taylor identity yields
\begin{equation}
\mathcal{B}\Sigma_\GZ^{c}=0\;,
\end{equation}
with $\mathcal{B}$ the nilpotent linearized Slavnov-Taylor operator,
\begin{multline}
\mathcal{B}=\int \d^{4}x\Bigl( \frac{\delta \Sigma_\GZ'}{\delta K_{\mu }^{a}}\frac{\delta }{\delta A_{\mu }^{a}}+\frac{\delta \Sigma_\GZ' }{\delta A_{\mu }^{a}}\frac{\delta }{\delta K_{\mu }^{a}}+\frac{\delta \Sigma_\GZ' }{\delta L^{a}}\frac{\delta }{\delta c^{a}}+\frac{\delta\Sigma_\GZ' }{\delta c^{a}}\frac{\delta }{\delta L^{a}}+b^{a}\frac{\delta }{\delta \overline{c}^{a}}\\
+\overline{\varphi}_{i}^{a}\frac{\delta }{\delta \overline{\omega }_{i}^{a}}+\omega_{i}^{a}\frac{\delta }{\delta \varphi_{i}^{a}}
+M_{\mu }^{ai}\frac{\delta }{\delta U_{\mu }^{ai}} + N_{\mu }^{ai}\frac{\delta }{\delta V_{\mu }^{ai}} + R_{\mu }^{ai}\frac{\delta }{\delta T_{\mu }^{ai}}  \Bigr) \,,
\end{multline}
and
\begin{equation}
\mathcal{B}^2=0\;.
\end{equation}

\item The $U(f)$ invariance reads
\begin{eqnarray}
U_{ij} \Sigma_\GZ^{c} &=&0 \;.
\end{eqnarray}

\item The Landau gauge condition
\begin{eqnarray}\label{2lgc}
\frac{\delta \Sigma_\GZ^{c}}{\delta b^{a}}&=&0\,.
\end{eqnarray}

\item The antighost equation
\begin{eqnarray}\label{2age}
\frac{\delta \Sigma_\GZ^{c}}{\delta \overline c^{a}}+\p_\mu\frac{\delta \Sigma_\GZ^{c}}{\delta K_{\mu}^a} &=&0 \,.
\end{eqnarray}

\item The linearly broken local constraints yield
\begin{eqnarray}
\left( \frac{\delta  }{\delta \overline{\varphi }^{a}_i}+\partial _{\mu }\frac{\delta}{\delta M_{\mu }^{ai}} +\partial _{\mu }\frac{\delta }{\delta M_{\mu }^{ai}} + g f_{abc}    T^{b i}_\mu \frac{\delta }{\delta K_{\mu }^{c i}}\right) \Sigma_\GZ^{c} &=&0 \;,\nonumber\\
\left( \frac{\delta }{\delta \omega ^{a}_i}+\partial _{\mu}\frac{\delta }{\delta N_{\mu}^{ai}}-gf^{abc}\overline{\omega }^{b}_i \frac{\delta }{\delta b^{c}} \right) \Sigma_\GZ^{c}&=&0 \;.
\end{eqnarray}

\item The exact $\mathcal{R}_{ij}$ symmetry reads
\begin{equation}
\mathcal{R}_{ij}\Sigma_\GZ^{c}=0\;,
\end{equation}
with  $\mathcal{R}_{ij}$ given in \eqref{rij}.

\item Finally, the integrated Ward identity becomes
\begin{equation}
\int \d^d x \left( c^a \frac{ \delta \Sigma_\GZ^{c}}{ \delta \omega^{ a}_i} + \overline \omega^{a}_i \frac{ \delta \Sigma_\GZ^{c}}{ \delta  \overline c^a } + U^{a i}_\mu \frac{ \delta \Sigma_\GZ^{c}}{ \delta  K^a_\mu }  \right) = 0 \;.
\end{equation}
\end{enumerate}
Now we can write down the most general counterterm $\Sigma_\GZ^{c}$ of $d=4$, which obeys the linearized Slavnov-Taylor identity, has ghost number zero, and vanishing $Q_f$ number,
\begin{eqnarray}\label{counterterm}
\Sigma^c_\GZ &=& a_0 S_{\YM} + \mathcal{B} \int \d^d \!x\,   \biggl\{  a_{1} K_{\mu}^{a} A_{\mu}^{a} + a_2 \partial _{\mu} \overline{c}^{a} A_{\mu}^{a}+a_3 \,L^{a}c^{a}
+a_4 U_{\mu}^{ai}\,\partial _{\mu }\varphi _{i}^{a} +a_5 \,V_{\mu}^{ai}\,\partial _{\mu }\overline{\omega }_{i}^{a}\nonumber\\
&& + a_6 \overline{\omega}_{i}^{a} \partial^{2} \varphi_{i}^{a} + a_7 U_{\mu}^{ai}V_{\mu}^{ai} + a_8 gf^{abc}U_{\mu}^{ai}\varphi_{i}^{b}A_{\mu}^{c}+a_9 gf^{abc}V_{\mu}^{ai}\overline{\omega }_{i}^{b}A_{\mu }^{c}\nonumber \\
&& +a_{10}gf^{abc}\overline{\omega }_{i}^{a}A_{\mu }^{c}\,\partial _{\mu }\varphi _{i}^{b}+a_{11}gf^{abc}\overline{\omega }_{i}^{a}(\partial _{\mu }A_{\mu}^{c})\varphi _{i}^{b} +b_1 R_{\mu}^{ai} U_{\mu}^{ai} +b_2 T_{\mu }^{ai} M_{\mu }^{ai} \nonumber\\
&&+ b_3 g f_{abc} R_{\mu }^{ai} \overline{\omega }_{i}^{b} A_{\mu}^{c} + b_4 g f_{abc} T_{\mu}^{ai} \overline{\varphi }_{i}^{b} A_{\mu}^{c} + b_5 R_{\mu}^{ai} \p_\mu \overline{\omega }_{i}^{a}  + b_6 T_{\mu}^{ai} \p_\mu \overline{\varphi }_{i}^{a} \biggr\} \;,
\end{eqnarray}
with $a_0, \ldots, a_{11}$ arbitrary parameters. Now we can unleash the constraints on the counterterm. Firstly, although the ghost Ward identity \eqref{GW} is broken, we know that this is not so in the standard Yang-Mills case. Therefore, we can already set $a_3=0$ as this term is not allowed in the counterterm of the standard Yang-Mills action, which is a special case of the action we are studying\footnote{In particular, since we will always assume the use of a mass independent renormalization scheme, we may compute $a_3$ with all external mass scales (= sources) equal to zero. Said otherwise, $a_3$ is completely determined by the dynamics of the original Yang-Mills action, in which case it is known to vanish to all orders (see section \ref{5.5} of chapter \ref{algebraic}).}. Secondly, due to the Landau gauge condition \eqref{2lgc} and the antighost equation \eqref{2age} we find,
\begin{eqnarray}
a_1 &=& a_2\;.
\end{eqnarray}
Next, the linearly broken constraints (5.) give the following relations
\begin{align}
 a_1 &= -a_8  = - a_{9} = a_{10} = a_{11} = -b_3 = b_4\;, \nonumber\\
  a_4 &= a_5 = -a_6 = a_7\;, \quad b_1 =b_2 = b_5 = b_6 = 0 \;.
\end{align}
The $R_{ij}$ symmetry does not give any new information, while the integrated Ward identity relates the two previous strings of parameters:
\begin{multline}
 a_1 = -a_8  = - a_{9} = a_{10} = a_{11} = -b_3 = b_4  \equiv     a_4 = a_5 = -a_6 = a_7 \;.
\end{multline}
Taking all this information together, we obtain the following counterterm
\begin{multline}\label{final}
\Sigma^c= a_{0}S_{YM}  + a_{1}\int \d^dx\Biggl(  A_{\mu}^{a}\frac{ \delta S_{YM}}{\delta A_{\mu }^{a}}  + \p_\mu \overline{c}^a \p_\mu c^a  + K_{\mu }^{a}\partial _{\mu }c^{a}  + M_\mu^{a i} \p_\mu \varphi^{a}_i -  U_\mu^{a i} \p_\mu \omega^{a}_i \\
+ N_\mu^{a i} \p_\mu \overline{\omega}_i^{a} +  V_\mu^{a i}\p_\mu \overline{\varphi}^{a}_i   +  \p_\mu \overline{\varphi}^{a}_i \p_\mu \varphi^{a}_i +  \p_\mu \omega^{a}_i \p_\mu \overline{\omega}^{a}_i + V_\mu^{a i} M_\mu^{a i} - U_\mu^{a i}N_\mu^{a i}  - g f_{abc} U_\mu^{ai} \varphi^{b}_i \p_\mu c^c \\
- g f_{abc} V_\mu^{ai} \overline{\omega}^{b}_i \p_\mu c^c - g f_{abc} \p_{\mu} \overline{\omega}^a_i \varphi^{b}_i  \p_\mu c^c   - g f_{abc} R_\mu^{ai} \p_\mu c^b \overline \omega_i^c + g f_{abc} T_\mu^{ai} \p_\mu c^b \overline \varphi_i^c \Biggr) \;.
\end{multline}

\subsection{The renormalization factors}
As a final step, we have to show that the counterterm \eqref{final} can be reabsorbed by means of a multiplicative renormalization of the fields and sources.
If we try to absorb the counterterm into the original action, we easily find,
\begin{eqnarray}\label{Z1}
Z_{g} &=&1-h \frac{a_0}{2}\,,  \nonumber \\
Z_{A}^{1/2} &=&1+h \left( \frac{a_0}{2}+a_{1}\right) \,,
\end{eqnarray}
and
\begin{eqnarray}\label{Z2}
Z_{\overline{c}}^{1/2} &=& Z_{c}^{1/2} = Z_A^{-1/4} Z_g^{-1/2} = 1-h \frac{a_{1}}{2}\,, \nonumber \\
Z_{b}&=&Z_{A}^{-1}\,, \nonumber\\
Z_{K }&=&Z_{c}^{1/2}\,,  \nonumber\\
Z_{L} &=&Z_{A}^{1/2}\,.
\end{eqnarray}
The results \eqref{Z1} are already known from the renormalization of the original Yang-Mills action in the Landau gauge. Further, we also obtain
{\allowdisplaybreaks \begin{eqnarray}\label{Z3}
Z_{\varphi}^{1/2} &=& Z_{\overline \varphi}^{1/2} = Z_g^{-1/2} Z_A^{-1/4} = 1 - h \frac{a_1}{2}\,, \nonumber\\
Z_\omega^{1/2} &=& Z_A^{-1/2} \,,\nonumber\\
Z_{\overline \omega}^{1/2} &=& Z_g^{-1} \,,\nonumber\\
Z_M &=& 1- h\frac{a_1}{2} = Z_g^{-1/2} Z_A^{-1/4}\,, \nonumber\\
Z_N &=& Z_A^{-1/2} \,, \nonumber\\
Z_U &=& 1 + h \frac{a_0}{2} = Z_g^{-1} \,, \nonumber\\
Z_V &=& 1- h \frac{a_1}{2} = Z_g^{-1/2}Z_A^{-1/4} \,,  \nonumber\\
Z_T &=& 1+h \frac{a_0}{2} = Z_g^{-1}  \,,  \nonumber\\
Z_R &=& 1- h \frac{a_1}{2} = Z_g^{-1/2}Z_A^{-1/4}\;.
\end{eqnarray}}

\noindent This concludes the proof of the renormalizability of the action \eqref{GZstart} which is the physical limit of $\Sigma_\GZ' $. Notice that in the physical limit \eqref{physlimit}, we have that
\begin{eqnarray}\label{Zgamma}
Z_{\gamma^2} &=& Z_g^{-1/2} Z_A^{-1/4}\;.
\end{eqnarray}

\section{Relation between Gribov no-pole condition and the GZ action}
To end this chapter, we would like to point out that the equivalence between the no-pole condition and Zwanziger's horizon condition has been checked up to third order in the gauge fields. All the details of the calculation can be found in \cite{Gomez:2009tj}.

\chapter{Scrutinizing the Gribov-Zwanziger action \label{scrutinizing} }
In this chapter we shall elaborate on the Gribov-Zwanziger action by discussing a variety of topics. Firstly, we shall calculate all the propagators of this action and discuss the transversality of the gluon propagator. Secondly, we devote some effort in scrutinizing the BRST breaking of the GZ action and its consequences. We shall even demonstrate how one can restore this broken BRST by the introduction of additional fields. Next, we shall briefly touch the hermiticity of the GZ action and we shall also devote some words on the form of the horizon function in relation to the renormalizability. Finally, we shall dwell upon the Kugo-Ojima criterium in relation to the GZ action.

\section{The propagators of the GZ action\label{propGZ1}}
As it shall be useful later on in this thesis, we shall calculate all the propagators of the GZ action. In order to make no mistakes with minus signs, we shall perform the calculations with great care. We start by taking only the quadratic part of the action $S_{\RGZ}$ into account
\begin{multline}
S_{\GZ} ~=~ \int \d^4 x \Bigl[ \frac{1}{4} (\p_\mu A_\nu^a - \p_\nu A_\mu^a)^2 + b^a \p_\mu A_\mu^a + \overline c^a \p_\mu^2 c^a  + \overline \varphi^a_i \p_\mu^2 \varphi^a_i - \overline \omega^a_i \p_\mu^2 \omega^a_i -\gamma^2 g f^{abc} A_\mu^a \varphi^{bc}_\mu\\ -\gamma^2 g f^{abc} A_\mu^a \overline \varphi^{bc}_\mu \Bigr]\;.
\end{multline}
We see three different parts appear:
\begin{multline}\label{3parts}
S_{\GZ} = \int \d^4 x \Bigl[  \frac{A_\mu^a}{2} ( - \p^2 \delta^{\mu\nu} + \p_\mu \p_\nu) \delta^{ab} A_\nu^b   + \frac{1}{2} b^a \p_\mu A_\mu^a - \frac{1}{2} \p_\mu b^a A_\mu^a  +   \overline \varphi^{ab}_\mu \p^2  \varphi^{ab}_\mu  \\
 - \gamma^2 g f^{abc} A_\mu^a  (\varphi^{bc}_\mu  + \overline \varphi^{bc}_\mu) \Bigr]
+ \int \d^4 x \Bigl[  \overline \omega^a_i (-\p^2) \omega^a_i \Bigr] + \int \d^4 x \Bigl[ \overline c^a \p_\mu^2 c^a \Bigr]\;.
\end{multline}

\subsection*{The $\overline \omega \omega$ propagator}
The goal is to calculate the propagator
\begin{eqnarray}
\Braket{\widetilde{ \overline \omega}^a_\mu(p) \widetilde \omega^b_\nu(k)} \;.
\end{eqnarray}
As we are working in Euclidean space, the path integral is given by
\begin{eqnarray}
P &=& \int [\d \Phi] \e^{-S_\GZ}\;,
\end{eqnarray}
with $[\d \Phi]$ the integration over all the fields. In order to calculate the propagator in momentum space we can now employ formula \eqref{ghostapp}
\begin{align}
	I &= \int [\d\omega][\d\overline{\omega}] \exp\left[ \int \d^d x \d^d y \ \overline{\omega}_\mu(x) A_{\mu\nu} (x,y)\omega_\nu(y) + \int \d^d x \ (J^{\mu}_\omega(x) \omega_\mu(x) + \overline{\omega}_\nu(x) J_{\overline{\omega}}^\nu (x)) \right]  \nonumber \\
	&=C \det A \exp  -\int \d^d x \d^d y \   J_\omega^\mu(x) A^{-1}_{\mu\nu}(x,y) J_{\overline{\omega}}^\nu(y) \;,
\end{align}
whereby in our case:
\begin{eqnarray}
A(x,y) = \delta(x-y) (\p^2 ) \delta^{\mu\nu} \;, \nonumber\\
A^{-1}(x,y) = \delta(x-y)  \frac{1}{\p^2 }\delta^{\mu\nu} \;.
\end{eqnarray}
Now going to Fourierspace,
\begin{align}
	I &= \int [\d\omega][\d\overline{\omega}] \exp\left[ \int \frac{\d^d p}{(2\pi)^d}   \widetilde{\overline{\omega}}(-p)(-p^2 ) \widetilde{\omega}(p) + \int \frac{\d^d p}{(2\pi)^d}\ ( \widetilde {J}_\omega(-p) \widetilde{\omega}(p) + \widetilde{\overline{\omega}}(p) \widetilde{J}_{\overline{\omega}} (-p)) \right]  \nonumber \\
	&=C \det A \exp  -\int \frac{\d^d p}{(2\pi)^d}    \widetilde{J}_\omega(-p)  \frac{1}{p^2} \widetilde{J}_{\overline{\omega}}(p) \;,
\end{align}
we can calculate the propagator in Fourierspace:
\begin{equation}
\left. - (2\pi)^8 \frac{\delta }{\delta \widetilde J_{\overline{\omega}}(-p)} \frac{\delta }{\delta \widetilde J_\omega(-k)}I \right|_{J = 0} = \Braket{\widetilde{ \overline \omega}^{ab}_\mu(p) \widetilde \omega^{cd}_\nu(k)} ~=~ \delta^{ac}\delta^{bd} \delta^{\mu\nu} \frac{-1}{p^2 }\delta(p+ k) (2\pi)^4 \;.
\end{equation}


\subsection*{The ghost propagator}
Completely analogously, we find
\begin{eqnarray}
  \Braket{ \widetilde{ \overline c}^a(k) \widetilde c^b(p) } ~=~ \delta^{ab} \frac{1}{p^2} (2\pi)^4 \delta(p+ k) \;.
\end{eqnarray}

\subsection*{The mixed operators}
For the final part, we shall use the following general formula,
\begin{eqnarray}\label{general}
	I(A, J) &=& \int [\d \varphi] \exp \left[ - \frac{1}{2} \int \d^d x \d^d y\  \varphi(x) A(x,y) \varphi(y) +   \int \d^d x \ \varphi (x) J(x) \right] \nonumber \\
	&=& C (\det A)^{-1/2} \exp \frac{1}{2} \int \d^dx \d^dy\ J(x) A^{-1}(x,y) J(y) \;.
\end{eqnarray}
However, as we see in the action \eqref{3parts} we need to rewrite the complex conjugate Bose fields $\overline \varphi$ and $\varphi$ in terms of real fields. Therefore, we shall introduce the real fields $U$ and $V$
\begin{eqnarray}
\varphi_\mu^{ab} &=& \frac{V_\mu^{ab} + i U_\mu^{ab} }{2}  \;, \nonumber\\
\overline \varphi_\mu^{ab} &=& \frac{V_\mu^{ab} - i U_\mu^{ab} }{2} \;.
\end{eqnarray}
We can thus rewrite the relevant part of the action as
\begin{multline}\label{boz}
S_{\RGZ}' = \int \d^4 x \Bigl[  \frac{A_\mu^a}{2} ( - \p^2 \delta^{\mu\nu} + \p_\mu \p_\nu) \delta^{ab} A_\nu^b   + \frac{1}{2} b^a \p_\mu A_\mu^a - \frac{1}{2} \p_\mu b^a A_\mu^a  + \\
 \frac{1}{4} \left( V^{ab}_\mu \p^2 V^{ab}_\mu + U^{ab}_\mu \p^2  U^{ab}_\mu \right) - \frac{1}{2}\gamma^2 g f^{abc} A_\mu^a  V^{bc}_\mu  - \frac{1}{2}\gamma^2 g f^{abc} A_\mu^a  V^{bc}_\mu \Bigr]\;.
\end{multline}
We  rewrite this in matrixform as
\begin{multline*}
\exp[-S_\RGZ'] =\exp [- \frac{1}{2} \int \d^4 x \underbrace{ \begin{bmatrix}
 A^a_{\mu}(x) & b^m(x) &  V^{k \ell }_{\alpha}(x) &  U^{k \ell }_{\alpha}(x)
\end{bmatrix}}_{X} \nonumber\\
\times
\underbrace{\begin{bmatrix} - \p^2  P_{\mu\nu}  \delta^{ab}  &  - \p_\mu \delta^{an}  &  - \gamma^2 g   f^{aij}  \delta_{\mu\kappa} & 0 \\ \\
  \p_\mu \delta^{bm} &0&0&0 \\ \\
   - \gamma^2 g   f^{bk\ell }  \delta_{\alpha\nu} &0& \frac{1}{2} \p^2   & 0 \\
   && \delta^{\alpha \kappa} \delta^{ki} \delta^{\ell j} & \\
   0 & 0&0 &  \frac{1}{2} \p^2   \\
   &&& \delta^{\beta \lambda }\delta^{sp} \delta^{tq}
 \end{bmatrix}}_{A}
\begin{bmatrix}
A^b_{\nu}(x) \\ \\ b^n(x) \\ \\ V^{ij}_{\kappa}(x) \\ \\ U^{pq}_{\lambda}(x)
\end{bmatrix}] \;,
\end{multline*}
where
\begin{eqnarray}
P_{\mu\nu} = \delta_{\mu\nu} - \frac{\p_{\mu}\p_{\nu}}{\p^2}, \hspace{1cm}   L_{\mu\nu} =  \frac{\p_{\mu}\p_{\nu}}{\p^2}\;,
\end{eqnarray}
are the respective transverse and longitudinal projectors. Notice that we have rewritten \eqref{boz} in a symmetric way. Now we can apply the general formula \eqref{general}, meaning that we have to find the inverse of $A$.
\begin{multline*}
\begin{bmatrix} - \p^2  P_{\mu\nu}  \delta^{ab}  &  - \p_\mu \delta^{an}  &  - \gamma^2 g   f^{aij}  \delta_{\mu\kappa} & 0 \\ \\
  \p_\nu\delta^{mb} &0&0&0 \\ \\
   - \gamma^2 g   f^{bk\ell }  \delta_{\alpha\nu} &0& \frac{1}{2} \p^2  & 0 \\
   && \delta^{\alpha \kappa} \delta^{ki} \delta^{\ell j} & \\
   0 & 0&0 &  \frac{1}{2} \p^2   \\
   &&& \delta^{\beta \lambda }\delta^{sp} \delta^{tq}
 \end{bmatrix}
 \underbrace{
 \begin{bmatrix}
 A^{b\ c}_{\nu\ \tau} &  B^{b\ o}_{\nu } & C_{\nu\ \omega}^{b\ xy} & D^{b\ gh}_{\nu\ \chi} \\ \\
 E^{n\ c}_{\hphantom{f\ } \tau}   &  F^{n \ o} & G_{\hphantom{f\ } \omega}^{ n\ xy} & H^{n \ gh}_{\hphantom{f\ } \chi} \\ \\
 I^{ij\ c}_{\kappa\ \tau}  &J_{\kappa }^{ij\ o} & K_{\kappa \ \omega}^{ij \ xy} & L^{ij \ gh}_{\kappa \ \chi} \\ \\
 M^{pq \ c}_{\lambda \ \tau} & N^{pq\ o}_{\lambda} & O^{pq \ xy}_{\lambda \ \omega} & P^{pq \ gh}_{\lambda \ \chi}
 \end{bmatrix}}_{A^{-1}}
 \\ =
  \begin{bmatrix}
 \delta^{ac} \delta_{\mu \tau} & 0 & 0 & 0\\ \\0 & \delta^{mo}& 0&0 \\ \\
 0&0&  \delta^{kx} \delta^{\ell y} \delta_{\alpha \omega} & 0 \\ \\ 0&0& 0& \delta^{sg} \delta^{th} \delta_{\beta \chi}
 \end{bmatrix} \;.
\end{multline*}
\normalsize After some calculation we find for $A^{-1}(x) $:
\begin{eqnarray*}
 \begin{bmatrix}
 \delta^{bc} \left[ \frac{-\p^2}{\p^4  + 2 N g^2 \gamma^4}  P_{\nu \tau} \right] &   \frac{\p_\nu}{\p^2} \delta^{bo}  & f^{bxy} P_{\nu \omega} \frac{ -2 g \gamma^2}{ \p^4  + 2 g^2 N \gamma^4}    &0  \\  \\
  \p_\tau \delta^{nc} \frac{-1}{\p^2} & \delta^{no} \frac{ 2 g^2 N \gamma^4}{ \p^4} &  f^{nxy} \p_\omega \frac{- 2 g \gamma^2}{ \p^4} &  0 \\ \\
  P_{\kappa \tau} f^{ijc} \frac{ -2 g \gamma^2}{ \p^4  + 2 g^2 N \gamma^4}  & \p_\kappa f^{ijo} \frac{- 2 g \gamma^2}{ \p^4}  & f^{ijr} f^{xyr} P_{\kappa \omega} \frac{4 g^2 \gamma^4}{ (- \p^2)( \p^4   + 2 g^2 N \gamma^4 )} &  0 \\
  & & + \frac{-2}{-\p^2 } \delta^{ix} \delta^{jy} \delta_{\kappa \omega} &\\ \\
  0 & 0& 0& \frac{-2}{-\p^2 } \delta^{pg} \delta^{qh} \delta_{\lambda \chi}
 \end{bmatrix}\;.
\end{eqnarray*} \normalsize
For the propagators, we need to go to Fourierspace,
\begin{multline}\label{generalfourierspace}
\int [\d \varphi] \exp \left[ - \frac{1}{2} \int \d^d x \  X(x) A(x) X^T(x) +   \int \frac{\d^d p}{(2\pi)^d}  \ \widetilde{X} (-p) \widetilde{J}(p) \right] \\
	= C (\det A)^{-1/2} \exp \frac{1}{2} \int \frac{\d^d p}{(2\pi)^d}  \widetilde J^T( -p ) A^{-1}(p)  \widetilde J(p)\;,
\end{multline}
with
\begin{eqnarray}
\widetilde{J}^T &=& \begin{bmatrix}J_A & J_b& J_V &J_U  \end{bmatrix} \;,
\end{eqnarray}
and $A^{-1}(p)$ given by
\begin{eqnarray*}
 \begin{bmatrix}
 \delta^{bc} \left[ \frac{p^2 }{p^4  + 2 N g^2 \gamma^4}  P_{\nu \tau} \right] &   \frac{ - \ii p_\nu }{p^2} \delta^{bo}  & f^{bxy} P_{\nu \omega} \frac{ -2 g \gamma^2}{ p^4  + 2 g^2 N \gamma^4}    &0  \\  \\
   \ii p_\tau \delta^{nc} \frac{1}{p^2} & \delta^{no} \frac{ 2 g^2 N \gamma^4}{ p^4} &  f^{nxy}\ii p_\omega \frac{- 2 g \gamma^2}{ p^4} &  0 \\ \\
  P_{\kappa \tau} f^{ijc} \frac{ -2 g \gamma^2}{ p^4  + 2 g^2 N \gamma^4}  & \p_\kappa f^{ijo} \frac{- 2 g \gamma^2}{ p^4}  & f^{ijr} f^{xyr} P_{\kappa \omega} \frac{4 g^2 \gamma^4}{  p^2( p^4 + 2 g^2 N \gamma^4 )} &  0 \\
  & & + \frac{-2}{ p^2 } \delta^{ix} \delta^{jy} \delta_{\kappa \omega} &\\ \\
  0 & 0& 0& \frac{-2}{p^2} \delta^{pg} \delta^{qh} \delta_{\lambda \chi}
 \end{bmatrix}\;.
\end{eqnarray*}
We now have all the ingredients to calculate the propagators.\\
\\
\textbf{$AA$-propagator}\\
We have for example,
\begin{align*}
 \frac{\delta}{\delta \widetilde J_{A_\mu^a }(-p)} \frac{\delta}{\delta \widetilde  J_{A_\nu^b }(-k)  } I &= \frac{1}{(2\pi)^{8}} \Braket{ \widetilde A_\mu^a (p) \widetilde A_\nu^b (k)} \\
 &=  \frac{1}{(2\pi)^4}\delta^{ab} \delta(k+ p) \left[ \frac{ p^2 }{p^4 + 2 N g^2 \gamma^4}  P_{\mu \nu} \right]\;,
\end{align*}
or equivalently
\begin{eqnarray}
\Braket{ \widetilde A_\mu^a (p)  \widetilde A_\nu^b (k)}&=&   \frac{ p^2 }{p^4 + \lambda^4}  P_{\mu \nu} \delta^{ab} \delta(k+ p) (2\pi)^4 \;,
\end{eqnarray}
whereby we have defined
\begin{eqnarray}\label{deflambda}
\lambda^4  &=& 2 N g^2 \gamma^4 \;.
\end{eqnarray}
\textbf{$Ab$-propagator }\\
Next,
\begin{eqnarray}
 \frac{\delta}{\delta \widetilde J_{A_{\mu}^a}(-p) } \frac{\delta}{\delta \widetilde J_{b^b}(-k) } I &=& \frac{1}{(2\pi)^{8}} \Braket{  \widetilde A_\mu^{a}  (p) \widetilde b^c (k)} ~=~ - \ii \frac{p_\mu}{p^2} \delta^{ab}  \frac{ \delta(p+k)}{(2\pi)^4} \;,
\end{eqnarray}
or thus
\begin{eqnarray}
\Braket{\widetilde A_\mu^a(p) \widetilde b^b(k)} &=& - \ii \frac{p_\mu}{ p^2} \delta^{ab} \delta(p+k) (2\pi)^4 \;.
\end{eqnarray}
\textbf{$bb$-propagator }\\
\begin{eqnarray}
\Braket{b^a(p) b^b(k)} &=&   \delta^{ab} \frac{ \lambda^4}{p^4}\delta(p+k)(2\pi)^4 \;.
\end{eqnarray}
\textbf{The propagators with $U$ and $V$}\\
In an analogue fashion, we find
\begin{eqnarray}
\Braket{ \widetilde A^a_{\mu}(p) \widetilde{ V}^{bc}_{\nu}(k)} &=&  f^{abc}  \frac{- 2g \gamma^2}{p^4   + \lambda^4}  P_{\mu \nu}(p) (2\pi)^4 \delta(p+k) \;, \nonumber\\
 \Braket{ \widetilde b^a(p) \widetilde{ V}^{bc}_{\nu}(k)} &=& f^{abc} \ii p_\nu \frac{- 2 g \gamma^2}{ p^2 (p^2}  (2\pi)^4 \delta(p+k)  \;,\nonumber\\
  \Braket{\widetilde{ V}^{ab}_{\mu}(p) \widetilde{ V}^{cd}_{\nu}(k)} &=& \left(  f^{abr} f^{cdr} P_{\mu \nu} \frac{4 g^2 \gamma^4}{  p^2( p^4  + 2 g^2 N \gamma^4 )} +  \frac{-2}{ p^2} \delta^{ac} \delta^{bd} \delta_{\mu \nu}  \right)    (2\pi)^4 \delta(p+k) \;, \nonumber\\
   \Braket{\widetilde{ U}^{ab}_{\mu}(p) \widetilde{ U}^{cd}_{\nu}(k)} &=&  \frac{-2}{p^2 } \delta^{ac} \delta^{bd} \delta_{\mu \nu}    (2\pi)^4 \delta(p+k) \;, \nonumber\\
\Braket{ \widetilde A^a_{\mu}(p) \widetilde{ U}^{bc}_{\nu}(k)} &=&  \Braket{ \widetilde b^a(p) \widetilde{ U}^{bc}_{\nu}(k)} ~=~ \Braket{ \widetilde{ U}^{ab}_{\nu} (p) \widetilde{ U}^{cd}_{\nu}(k)} ~=~0 \;,
\end{eqnarray}
which can be rewritten in terms of $\varphi$ and $\overline \varphi$ again,
\begin{eqnarray}
\Braket{ \widetilde A^a_{\mu}(p) \widetilde{ \varphi}^{bc}_{\nu}(k)} &=& \Braket{ \widetilde A^a_{\mu}(p) \widetilde{\overline \varphi}^{bc}_{\nu}(k)} ~=~ f^{abc}  \frac{- g \gamma^2}{p^4  + \lambda^4}  P_{\mu \nu}(p) (2\pi)^4 \delta(p+k) \;, \nonumber\\
 \Braket{ \widetilde b^a(p) \widetilde{ \varphi}^{bc}_{\nu}(k)} &=&  \Braket{ \widetilde b^a(p) \widetilde{ \overline \varphi}^{bc}_{\nu}(k)} ~=~  f^{abc} \ii p_\nu \frac{-  g \gamma^2}{ p^4}  (2\pi)^4 \delta(p+k) \;, \nonumber\\
  \Braket{\widetilde{ \varphi}^{ab}_{\mu}(p) \widetilde{ \overline \varphi }^{cd}_{\nu}(k)} &=& \left(  f^{abr} f^{cdr} P_{\mu \nu} \frac{ g^2 \gamma^4}{  p^2( p^4 + 2 g^2 N \gamma^4 )} +  \frac{-1}{ p^2 } \delta^{ac} \delta^{bd} \delta_{\mu \nu}  \right)    (2\pi)^4 \delta(p+k) \;, \nonumber\\
   \Braket{\widetilde{ \varphi}^{ab}_{\mu}(p) \widetilde{ \varphi }^{cd}_{\nu}(k)} &=&  \Braket{\widetilde{\overline \varphi}^{ab}_{\mu}(p) \widetilde{\overline \varphi }^{cd}_{\nu}(k)} ~=~ f^{abr} f^{cdr} P_{\mu \nu} \frac{ g^2 \gamma^4}{ p^2( p^4 + 2 g^2 N \gamma^4 )}   (2\pi)^4 \delta(p+k) \;. \nonumber\\
\end{eqnarray}

\noindent In summary, we have the following large set of propagators in the theory:
\begin{eqnarray}\label{summarypropGZ}
\Braket{\widetilde{ \overline \omega}^{ab}_\mu(k) \widetilde \omega^{cd}_\nu(p)} &=& \delta^{ac}\delta^{bd} \delta^{\mu\nu} \frac{-1}{p^2 } \delta(p+k) (2\pi)^4 \;, \nonumber\\
\Braket{ \widetilde{ \overline c}^a(k) \widetilde c^b(p) } &=& \delta^{ab} \frac{1}{p^2} \delta(p+k) (2\pi)^4 \;,\nonumber\\
\Braket{ \widetilde A_\mu^a (p)  \widetilde A_\nu^b (k)}&=&   \frac{ p^2 }{p^4  + \lambda^4}  P_{\mu \nu} \delta^{ab} \delta(k+ p) (2\pi)^4\;, \nonumber\\
\Braket{\widetilde A_\mu^a(p) \widetilde b^b(k)} &=& - \ii \frac{p_\mu}{ p^2} \delta^{ab} \delta(p+k) (2\pi)^4 \;, \nonumber\\
\Braket{b^a(p) b^b(k)} &=&   \delta^{ab} \frac{ \lambda^4}{p^4 }\delta(p+k)(2\pi)^4 \;, \nonumber\\
\Braket{ \widetilde A^a_{\mu}(p) \widetilde{ \varphi}^{bc}_{\nu}(k)} &=& \Braket{ \widetilde A^a_{\mu}(p) \widetilde{\overline \varphi}^{bc}_{\nu}(k)} ~=~ f^{abc}  \frac{- g \gamma^2}{p^4   + \lambda^4}  P_{\mu \nu}(p) (2\pi)^4 \delta(p+k) \;, \nonumber\\
 \Braket{ \widetilde b^a(p) \widetilde{ \varphi}^{bc}_{\nu}(k)} &=&  \Braket{ \widetilde b^a(p) \widetilde{ \overline \varphi}^{bc}_{\nu}(k)} ~=~  f^{abc} \ii p_\nu \frac{-  g \gamma^2}{ p^4}  (2\pi)^4 \delta(p+k) \;, \nonumber\\
  \Braket{\widetilde{ \varphi}^{ab}_{\mu}(p) \widetilde{ \overline \varphi }^{cd}_{\nu}(k)} &=& \left(  f^{abr} f^{cdr} P_{\mu \nu} \frac{ g^2 \gamma^4}{ p^2( p^4  + 2 g^2 N \gamma^4 )} +  \frac{-1}{ p^2 } \delta^{ac} \delta^{bd} \delta_{\mu \nu}  \right)    (2\pi)^4 \delta(p+k)  \;,\nonumber\\
   \Braket{\widetilde{ \varphi}^{ab}_{\mu}(p) \widetilde{ \varphi }^{cd}_{\nu}(k)} &=&  \Braket{\widetilde{\overline \varphi}^{ab}_{\mu}(p) \widetilde{\overline \varphi }^{cd}_{\nu}(k)} ~=~ f^{abr} f^{cdr} P_{\mu \nu} \frac{ g^2 \gamma^4}{ p^2( p^4  + 2 g^2 N \gamma^4 )}   (2\pi)^4 \delta(p+k) \;.\nonumber\\
\end{eqnarray}

\section{The transversality of the gluon propagator}
One might wonder whether the gluon propagator still remains transverse in the presence of the Gribov horizon. As the gluon propagator is the connected two-point function, we ought to consider the generator $Z^c$ of connected Green functions, which can be constructed from the quantum effective action $\Gamma$ by means of a Legendre transformation. Due to the QAP, $\Gamma$ obeys the renormalized version of the Ward identity \eqref{gaugeward}, or
\begin{equation}\label{gp1}
    \frac{\delta\Gamma}{\delta b^a}=\p_\mu A_\mu^a\;.
\end{equation}
Introducing sources $I^a(J_\mu^a)$ for the fields $b^a (A_\mu^a)$ and performing the Legendre transformation, see expression \eqref{legendre}, the identity \eqref{gp1} translates into
\begin{equation}\label{gp2}
    I^a=\p_\mu\frac{\delta Z^c}{\delta    J_\mu^a}\;.
\end{equation}
Acting with $\frac{\delta}{\delta J_\mu^b}$ on this expression, and by setting all sources equal to zero, we retrieve
\begin{equation}\label{gp3}
    0=\p_\mu^x\left.\frac{\delta^2 Z^c}{\delta J_\mu^a(x) \delta    J_\mu^b(y)}\right|_{I,J=0}=\p_\mu^x\braket{A_\mu^a(x)A_\nu^b(y)}\;,
\end{equation}
which expresses nothing else but the transversality of the gluon propagator.

\section{The soft breaking of the BRST symmetry\label{sectbreakingBRST}}
\subsection{The breaking}
We recall here that the Gribov-Zwanziger action \eqref{GZstart} is not invariant under the BRST transformation \eqref{BRST1}. Indeed, in equation \eqref{breaking} we have found that
\begin{eqnarray}\label{deltabrst}
   \Delta_{\gamma} &\equiv& sS = sS_\gamma= -g \gamma^2 \int \d^4 x f^{abc} \left( A^a_{\mu} \omega^{bc}_\mu -
 \left(D_{\mu}^{am} c^m\right)\left( \overline{\varphi}^{bc}_\mu + \varphi^{bc}_{\mu}\right)  \right) \;.
\end{eqnarray}
We see that the presence of the Gribov parameter $\gamma$ prevents the action from being invariant under the BRST symmetry. Notice that if $\gamma = 0$, we can integrate out the fields $\overline \varphi, \varphi, \overline \omega, \omega$ and we left with the original Yang-Mills theory. Therefore, this breaking is clearly due to the introduction of the horizon into the Yang-Mills action. Nevertheless, this fact does not prevent the use of the Slavnov-Taylor identity to prove the renormalizability of the theory, which is very remarkable. Since the breaking $\Delta_{\gamma}$ is soft, i.e.~it is of dimension two in the fields, it can be neglected in the deep ultraviolet, where we recover the usual notion of exact BRST invariance as well as of BRST cohomology for defining the physical subspace, see p.\pageref{physicalsubspace}. However, in the nonperturbative infrared region, the breaking term cannot be neglected and the BRST invariance is lost.

\subsection{The BRST breaking as a tool to prove that the Gribov parameter is a physical parameter}
The breaking term \eqref{deltabrst} has an interesting consequence as it allows us to give a simple algebraic proof of the fact that the Gribov parameter $\gamma$ is a physical parameter of the theory, and that as such it can enter the explicit expression of gauge invariant correlation functions like for instance $\braket{F^{2}(x)F^{2}(y)} $ or the vacuum condensate $\braket{F^2}$.\\
\\
We shall demonstrate this as follows. Taking the derivative of both sides of equation \eqref{deltabrst} with respect to $\gamma ^{2}$ one gets,
\begin{eqnarray}
s\frac{\partial S}{\partial \gamma ^{2}}&=& \frac{1}{\gamma^2} \Delta _{\gamma } = -g  \int \d^4 x f^{abc} \left( A^a_{\mu}
\omega^{bc}_\mu - \left(D_{\mu}^{am} c^m\right)\left( \overline{\varphi}^{bc}_\mu + \varphi^{bc}_{\mu}\right)  \right)\;,
\end{eqnarray}
from which, keeping in mind that the BRST operator $s$ as defined in equation \eqref{BRST1} is nilpotent, it immediately follows that $\frac{\partial S}{\partial \gamma ^{2}}$ cannot be cast in the form of a BRST exact variation, namely
\begin{eqnarray}
\frac{\partial S}{\partial \gamma ^{2}}\neq s\widehat{\Delta}_{\gamma }\;,
\end{eqnarray}
for some local integrated dimension two quantity $\widehat{\Delta}_{\gamma }$. From this, we can conclude that the Gribov parameter $\gamma ^{2}$ is a physical parameter and that it can enter into the expectation values of gauge invariant quantities. Let us demonstrate this more explicitly by assuming for a moment the contrary
\begin{eqnarray}\label{d22}
sS_{\gamma }&=&0\;,
\end{eqnarray}
instead of inducing the breaking term $\Delta _{\gamma }$. Since $S_{\gamma }$ depends on the auxiliary fields $\bigl(\overline{\varphi }_{\mu }^{ac}$, $\varphi _{\mu} ^{ac}$, $\overline{\omega }_{\mu}^{ac}$, $\omega _{\mu }^{ac}\bigr)$ which constitute a set of BRST doublets, it would follow from equation \eqref{d22} that a local integrated polynomial $\widehat{S}_{\gamma }$ would exist such that
\begin{equation}  \label{d23}
S_{\gamma }=s\widehat{S}_{\gamma }\;.
\end{equation}
Subsequently, taking the derivative of both sides of expression \eqref{d23} with respect to $\gamma ^{2}$, one would obtain
\begin{equation}\label{d24}
\frac{\partial S_{\gamma }}{\partial \gamma ^{2}}=s\frac{\partial \widehat{S}_{\gamma }}{\partial \gamma ^{2}}\;,
\end{equation}
a relation implying that $\gamma ^{2}$ is an unphysical parameter. Indeed, for a gauge invariant quantity $\mathcal G$, we would have that
\begin{equation}
\frac{\p \braket{\mathcal{G}}}{\p\gamma^2} = \frac{\delta}{ \delta \gamma^2} \int [\d \phi] \mathcal G\ \e^{-S_\GZ} \sim \int [\d \phi] \mathcal G\ s \widehat{S}_{\gamma }\e^{-S_\GZ} \sim \int [\d \phi]s\left( \mathcal G\  \widehat{S}_{\gamma } \right) \e^{-S_\GZ} \sim  \Braket{s(\ldots)} = 0
\end{equation}
and correlation functions of gauge invariant operators would be completely independent from $\gamma^2$. We see thus that the presence of the soft breaking term $\Delta _{\gamma }$ plays an important role, ensuring that $\gamma ^{2}$ is a relevant parameter of the theory and that $\frac{\p \braket{\mathcal{G}}}{\p\gamma^2} \not= 0$. The existence of the breaking $\Delta _{\gamma }$ thus seems to be an important ingredient to introduce a nonperturbative mass gap in a local and renormalizable way.

\subsection{The Maggiore-Schaden construction revisited}
The authors of the paper \cite{Maggiore:1993wq} attempted to interpret the BRST breaking as a kind of \textit{spontaneous} symmetry breaking. We shall now re-examine this proposal and conclude that, instead, the BRST breaking has to be considered as an \textit{explicit} symmetry breaking, where we shall present a few arguments which have not been considered in \cite{Maggiore:1993wq}. Although this discussion might seem to be only of a rather academic interest, there is nevertheless a big difference between a \textit{spontaneously} or \textit{explicitly} broken continuous symmetry, since only in the former case a Goldstone mode would emerge. For the benefit of the reader, we shall  first explain in detail the approach of \cite{Maggiore:1993wq}.  One starts by adding the following BRST exact term to the Yang-Mills action:
\begin{eqnarray}
    S_1 &=& s \int \d^4 x \left( \overline{c}^{a} \p_{\mu}A^a_{\mu} + \overline{\omega}_{\mu}^{ac} \p_{\nu} D_{\nu}^{ab} \varphi^{bc}_{\mu} \right)
    \;, \label{mg1}
\end{eqnarray}
with $s$, the same nilpotent BRST  operator as defined in \eqref{BRST1}. The first term represents the Landau gauge fixing, while the second term is a BRST exact piece in the
fields $(\varphi,\omega, \overline{\varphi},\overline{\omega})$. Of course, from expression \eqref{mg1}, it follows that $s$ defines a symmetry of the action $S_{YM} + S_1 $.
 As a consequence,  the nilpotent operator $s$ allows us to define two doublets $(\varphi,\omega)$ and $(\overline{\varphi},\overline{\omega})$. This  doublet structure implies that we can exclude these fields from the physical subspace, see p.\pageref{doublettheorem}, which makes $S_{YM} + S_1 $ equivalent to the ordinary Yang Mills gauge theory. Next, Maggiore and Schaden introduced a set of shifted fields, which -translated to our conventions- are given by:
\begin{eqnarray}
\varphi^{ab}_{\mu} &=& \varphi^{\prime ab}_{\mu} + \gamma^2 \delta^{ab} x_{\mu} \;, \nonumber\\
\overline{\varphi}^{ab}_{\mu} &=& \overline{ \varphi}^{\prime ab}_{\mu} + \gamma^2 \delta^{ab} x_{\mu}\;, \nonumber\\
\overline{c}^{ a} &=& \overline{c}^{\prime a} + g \gamma^2 f^{abc} \overline{\omega}^{bc}_{\mu} x_{\mu} \;,\nonumber\\
b^{a} &=& b^{ \prime a} + g \gamma^2 f^{abc} \overline{\varphi}^{bc}_{\mu} x_{\mu}\;.
\end{eqnarray}
 All fields $(\varphi^{\prime ab}_{\mu}, \overline{\varphi}^{\prime ab}_{\mu}, \overline{c}^{\prime a}, b^{ \prime a}) $ have vanishing vacuum expectation value (VEV), namely
\begin{equation}
\langle \varphi^{\prime ab}_{\mu} \rangle = \langle \overline{ \varphi}^{\prime ab}_{\mu} \rangle = \langle \overline{c}^{\prime
a} \rangle = \langle b^{ \prime a} \rangle = 0 \;. \label{mg2}
\end{equation}
Along with these new fields ($\varphi^{\prime ab}_{\mu}$, $\overline{\varphi}^{\prime ab}_{\mu}$, $\overline{c}^{\prime a}$, $b^{\prime a}$),  one introduces a modified nilpotent
BRST operator $\widetilde{s}$ given by:
\begin{align}
\widetilde{s} \;\overline{c}^{\prime a} &=b^{\prime a}\;,&   \widetilde{s} b^{\prime a}&=0\;,  \nonumber \\
\widetilde{s}\varphi _{\mu}^{\prime ab} &=\omega _{\mu}^{ab}\;, & \widetilde{s} \overline{\varphi }^{\prime ab}_\mu &=0\;,  \nonumber \\
\widetilde{s} A_{\mu}^a &= -D_{\mu}^{ab} c^b \;,&  \widetilde{s} \omega_{\mu}^{ab} &= 0\;,
\end{align}
which looks exactly like \eqref{BRST1}. However, we emphasize that by introducing these new fields, the BRST operator $\widetilde{s}$ will give rise to an explicit $x$-dependence when acting on the field $\overline{\omega}_{\mu}^{ab}$:
\begin{eqnarray}
\widetilde{s} \overline{\omega}_{\mu}^{ab} &=&
\overline{\varphi}^{\prime ab}_{\mu} + \gamma^2 \delta^{ab} x_\mu
\;. \label{mg3}
\end{eqnarray}
Furthermore, by taking the vacuum expectation value of both sides of equation \eqref{mg3}, one gets
\begin{equation}
\langle \widetilde{s}\; \overline{\omega}_{\mu}^{ab} \rangle =
\gamma^2 \delta^{ab} x_\mu \;, \label{mg4}
\end{equation}
from which the authors of \cite{Maggiore:1993wq} infer that the BRST operator $\widetilde{s}$ suffers from spontaneous symmetry breaking.  Notice also that \eqref{mg4} gives a VEV to a quantity with a free Lorentz index.\\
\\
With the introduction of the shifted fields, we can rewrite the action $S_1$ as:
\begin{equation}
    S_1 = \widetilde{s} \int \d^4 x \left( \overline{c}^{\prime a} \p_{\mu}A^a_{\mu} + \overline{\omega}_{\mu}^{ac} \p_{\nu} D_{\nu}^{ab} \varphi^{\prime bc}_{\mu} + g \gamma^2 f^{abc} \overline{\omega}^{bc}_{\nu} x_{\nu} \p_{\mu} A^a_{\mu} + \gamma^2 \overline{\omega}^{ac}_{\mu} \p_{\nu} D_{\nu}^{ab} \delta^{bc} x_{\mu}  \right) \;.
\end{equation}
The last two terms can be simplified, leading to
\begin{equation}\label{actionbrst}
    S_1 = \widetilde{s} \int \d^4 x \left( \overline{c}^{\prime a} \p_{\mu}A^a_{\mu} + \overline{\omega}_{\mu}^{ac} \p_{\nu} D_{\nu}^{ab} \varphi^{\prime bc}_{\mu}
   - g \gamma^2 \overline{\omega}^{ab}_{\mu} f_{abc} A^c_{\mu}  \right) \;.
\end{equation}
If we calculate this action explicitly, we recover the original Gribov-Zwanziger action, without the constant part $4\gamma^4 (N^2 -1)$. For this reason  one adds $ - \gamma^2 \widetilde{s} \int \d^4 x \p_{\mu} \overline{\omega}^{aa}_{\mu}$ to the action $S_1$. Doing so,  one finds
\begin{align}
    S_1 =& \widetilde{s} \int \d^4 x \left( \overline{c}^{\prime a} \p_{\mu}A^a_{\mu} + \overline{\omega}_{\mu}^{ac} \p_{\nu} D_{\nu}^{ab} \varphi^{\prime bc}_{\mu}
   - g \gamma^2 \overline{\omega}^{ab}_{\mu} f_{abc} A^c_{\mu}  - \gamma^2 \p_{\mu} \overline{\omega}^{aa}_{\mu} \right) \nonumber\\
   =& \int \d^4 x \left[ b^{\prime a} \p_{\mu} A^a_{\mu} + \overline{c}^{\prime a} \p_{\mu} \left(D_{\mu}^{ab} c^b \right) \right] + \int \d^4 x \left[ \overline{\varphi}^{\prime ac}_{\mu} \p_\nu D_{\nu}^{ab} \varphi^{ \prime bc}_{\mu} + \gamma^2  x_{\mu}   \p_\nu D_{\nu}^{ab} \varphi^{\prime ba}_{\mu} \right.\nonumber\\
   & \left. +  \overline{\omega}^{ac}_{\mu} \p_{\nu} \left(g  f^{akb} D^{kd}_{\nu} c^d \varphi^{\prime bc}_{\mu}  \right) - \overline{\omega}^{ac}_{\mu} \p_{\nu} D^{ab}_{\nu} \omega^{bc}_{\mu} \right]  + \int \d^4 x\left[ -g \gamma^2 \overline{\varphi}^{\prime ab}_{\mu} f_{abc} A^c_{\mu} \right. \nonumber\\
   &\left.- g \gamma^2 \overline{\omega}^{ab}_{\mu} f_{abc} D^{cd}_{\mu} c^d - 4 \gamma^4 (N^2 - 1)\right]\;.
\end{align}
If we naively assume that we can perform a partial integration, we find after dropping the surface terms,
\begin{eqnarray}\label{sec}
S_{YM}+ S_1&=& S_\GZ -  \; g \gamma^2 f^{abc}
\int \d^4x \; \overline{\omega}_{\mu}^{ab} D_{\mu}^{cd} c^d \;,
\end{eqnarray}
whereby $S_\GZ$ is given in expression \eqref{GZstart}. The last expression reveals that one has recovered the Gribov-Zwanziger action from an exact $\widetilde{s}$-variation with the addition of an extra term $\left( \;- g \gamma^2 f^{abc} \int \d^4x \; \overline{\omega}_{\mu}^{ab} D_{\mu}^{cd} c^d \right)$. However, this term is irrelevant as we shall explain now. Assume that we want to compose an arbitrary Feynman diagram without any external $\overline{\omega}$ leg and thereby using the action \eqref{sec}. The second term from this action can never contribute to this Feynman diagram as it contains an external $\overline{\omega}$. Indeed, this leg requires an $\omega$-leg, which in its turn is always accompanied by an $\overline{\omega}$ leg. Hence, the action \eqref{sec} is equivalent to the standard Gribov-Zwanziger action \eqref{GZstart} when we exclude the diagrams containing external $\overline{\omega}$ legs\footnote{For our
purposes these diagrams are irrelevant, e.g.~the vacuum energy, the gluon and ghost propagator,\ldots .}. \\
\\
Although at first sight this construction might seem useful, it turns out that a few points have been overlooked. Let us investigate this in more detail. Firstly, we  point out that rather delicate assumptions have been made concerning the partial integration. To reveal the obstacle, we perform once more the partial integration explicitly,
\begin{eqnarray}\label{surface}
\int \d^4 x \gamma^2 x_{\mu} \p_{\nu} D_{\nu}^{ab} \varphi^{\prime ba}_{\mu} &=& \text{surface term} - \int \d^4 x \gamma^2 \delta_{\mu\nu} D^{ab}_{\nu} \varphi^{\prime ba}_{\mu} \;.
\end{eqnarray}
Normally, one drops the surface terms, as the fields vanish at infinity. However in this case, as $x_{\mu}$ does not vanish at infinity, it is not sure if the surface terms $ \propto x_\mu$ will be zero. One would have to impose extra conditions on the fields to justify the dropping of the surface terms. On the other hand,  when we do not perform the partial integration to avoid the surface terms, we are facing an explicit, unwanted $x$-dependence in the action, resulting in an explicit breaking of translation invariance. \\
\\
Another way of looking at the problem consists of performing a partial integration on the second term of the action \eqref{actionbrst} \textit{before} applying the BRST variation
$\widetilde{s}$. Doing so, we find,
\begin{eqnarray}\label{actionbrst2}
    S_1 &=& \widetilde{s} \int \d^4 x \left( \overline{c}^{\prime a} \p_{\mu}A^a_{\mu} -  \p_{\nu} \overline{\omega}_{\mu}^{ac}  D_{\nu}^{ab} \varphi^{\prime bc}_{\mu}
   - g \gamma^2 \overline{\omega}^{ab}_{\mu} f_{abc} A^c_{\mu} - \gamma^2 \p_{\mu} \overline{\omega}^{aa}_{\mu} \right) \;.
\end{eqnarray}
Subsequently, applying the BRST variation gives,
\begin{align}
    S_1 =& \int \d^4 x \left[ b^{\prime a} \p_{\mu}A^a_{\mu} + \overline{c}^{\prime a} \p_{\mu} \left(D_{\mu}^{ab} c^b \right) \right] + \int \d^4 x \left[- \p_\nu \overline{\varphi}^{\prime ac}_{\mu}  D_{\nu}^{ab} \varphi^{\prime bc}_{\mu} - \gamma^2 \delta^{\mu\nu}   D_{\nu}^{ab} \varphi^{\prime ba}_{\mu} \right. \nonumber\\
     & \left. - \left( \p_{\nu} \overline{\omega}^{ac}_{\mu} \right)  g  f^{akb} D^{kd}_{\nu} c^d \varphi^{\prime bc}_{\mu}  + \left( \p_{\nu} \overline{\omega}^{ac}_{\mu} \right)  D^{ab}_{\nu} \omega^{bc}_{\mu} \right] + \int \d^4 x\left[ -g \gamma^2 \overline{\varphi}^{\prime ab}_{\mu} f_{abc} A^c_{\mu}  \right.\nonumber\\
   & \left.- g \gamma^2 \overline{\omega}^{ab}_{\mu} f_{abc} D^{cd}_{\mu} c^d - 4 \gamma^4 (N^2 - 1)\right] \nonumber\\
   =&~ S_\GZ - g \gamma^2 f^{abc} \int \d^4x\overline{\omega}_{\mu}^{ab} D_{\mu}^{cd} c^d \;.
\end{align}
In this case, we do not encounter the problem of nonvanishing surface terms. To recapitulate, if we first let the BRST variation act on the action \eqref{actionbrst}, and then perform a partial integration, we find a different result than performing these two operations the other way around. This difference is exactly given by the surface term from equation \eqref{surface}. This discrepancy arises of course from the explicit $x$-dependence introduced in the BRST transformation $\widetilde{s}$, giving nontrivial contributions. For example, we introduced a term $ - \gamma^2 \widetilde{s} \int \d^4 x \p_{\mu} \overline{\omega}^{aa}_{\mu}$ which might seem to be zero since we are looking at the integral of
a complete derivative (thus usually taken to be a vanishing surface term), but when the BRST variation is taken first, a nontrivial integrated piece remains.\\
\\
Apparently, to find the correct Gribov-Zwanziger action with the Maggiore-Schaden argument, there is some kind of a ``hidden working hypothesis'' that \eqref{actionbrst2} is the correct action to start with, and that partial integration is not always allowed\footnote{If it would be allowed, one would be able to cross from the second action \eqref{actionbrst2} to the first one \eqref{actionbrst}, but as we have just shown, these two starting actions are inequivalent. }. The fact that there seems to be a kind of ``preferred'' action to start with, is just a signal that there is a problem with the boundary conditions for some of the fields and hence surface terms when integrating.\\
\\
In addition, a second problem arises. In the Gribov-Zwanziger approach, we recall that the parameter $\gamma$ is not free and is determined by the horizon condition \eqref{gapgamma}. Obviously, the solution $\gamma =0$ is excluded, as else, we would be back in the ordinary Yang-Mills theory. In \cite{Dudal:2005na}, a solution for $\gamma \not= 0$ is found, which gave rise to a positive vacuum energy $E_{\mathrm{vac}} > 0$. However, according to Maggiore-Schaden argument, at one loop  order the stable solution should be that corresponding to $\gamma=0$ \cite{Maggiore:1993wq}, as, if $\gamma =0$, the vacuum energy would be vanishing, i.e. $E_{\mathrm{vac}} = 0$, which is energetically favored over a positive vacuum energy. This delivers a contradiction with the Gribov-Zwanziger approach,  as the restriction to the Gribov region requires that $\gamma \not= 0$, thus giving a positive energy $E_{\mathrm{vac}} > 0$ at one loop.\\
\\
In conclusion, we believe that interpreting the BRST breaking as a kind of spontaneous symmetry breaking cannot be supported by calculations as some mathematical details were overlooked.

\section{Restoring the BRST}
\subsection{Adapting the BRST symmetry $s$ is not possible}
A question which arises almost naturally is whether it might be possible to modify the BRST operator, i.e. $s\rightarrow s_m$, in such a way that the new operator $s_m$ would be still nilpotent, while defining an exact symmetry of the action, $s_mS=0$. Although we are not going to give a formal proof, we can present a simple argument discarding such a possibility. We have already observed that the BRST transformation \eqref{BRST1} defines an exact symmetry of the action when $\gamma =0$, which corresponds to the physical
situation in which the restriction to the Gribov region has not been implemented. Hence, it appears that one should search for possible modifications of the BRST operator which depends on $\gamma $, namely
\begin{eqnarray}
s_m&=&s+s_\gamma  \;,
\end{eqnarray}
whereby
\begin{eqnarray}
s_\gamma&=&\gamma\textrm{-dependent}\;\textrm{terms}\;,
\label{mstwee}
\end{eqnarray}
so as to guarantee a smooth limit when $\gamma$ is set to zero. However, taking into account the fact that $\gamma $ has mass dimension one, that all auxiliary fields $\left( \overline{\varphi }_{\mu }^{ac},\varphi _{\mu }^{ac},\overline{\omega }_{\mu}^{ac},\omega _{\mu }^{ac}\right)$ have dimension one too, and that the BRST operator $s$ does not alter the dimension of the fields\footnote{It is understood that the usual canonical dimensions are assigned to the fields $A^{a}_{\mu}$, $b^a$, $c^a$, ${\bar c}^a$. It is apparent that the BRST operator $s$ does not alter the dimension of the fields.}, it does not seem possible to introduce extra $\gamma$-dependent terms in the BRST transformation of the fields $\left( \overline{\varphi }_{\mu }^{ac},\varphi _{\mu }^{ac},\overline{\omega }_{\mu }^{ac},\omega _{\mu }^{ac}\right) $ while preserving locality, Lorentz covariance as well as color group structure.

\subsection{Restoring the BRST by the introduction of new fields}
However, there is another way to restore the BRST \cite{Dudal:2010hj}. By introducing new fields, we shall demonstrate that one can in fact restore the BRST. However, this BRST symmetry shall turn out not to be nilpotent. The idea is the following. In \cite{Sorella:2009vt} it was shown that the broken BRST symmetry $s$ can be rewritten as a non-local symmetry, which is not nilpotent\footnote{At about the same time, the paper \cite{Kondo:2009qz} appeared, which showed that one could write a non-local and nilpotent BRST symmetry for the GZ action, however, this non-local symmetry is not so transparant.}. By introducing extra fields, we shall show that it is possible to localize this non-local symmetry.

\subsubsection{Constructing the non-local BRST symmetry $s'$}
We shall start from the standard GZ action \eqref{GZstart}, set $g\gamma^2=\theta^2$ and drop the vacuum term for notational shortness as this vacuum term will not influence any variation of the action,
\begin{multline}\label{GZb1}
 S_{\GZ} = S_\YM +  \int \d^d x\,\left( b^a \p_\mu A_\mu^a +\overline c^a \p_\mu D_\mu^{ab} c^b \right) + \int \d^d x\left( \overline \varphi_\mu^{ac} \p_\nu D_\nu^{ab} \varphi_\mu^{bc} \right. \\ \left. - \overline \omega_\mu^{ac} \p_\nu D_\nu^{ab} \omega_\mu^{bc}  +\theta^{2}  f^{abc}A_\mu^a \left( \varphi_\mu^{bc} +  \overline \varphi_\mu^{bc}\right) \underline{- g f^{abc} \p_\mu \overline \omega_\nu^{ae}    D_\mu^{bd} c^d  \varphi_\nu^{ce}} \right)\,.
\end{multline}
Let us first reconstruct the non-local BRST symmetry as proposed in \cite{Sorella:2009vt}. Firstly, following \cite{Sorella:2009vt}, we shall for the moment also drop the underlined term. Dropping this term temporarily only leads to a breaking in the BRST $s$ which is itself the $s$-variation of something, thus it is rather harmless. Later on, we shall take this term into account anyhow by the nature of the construction itself. For now, we shall thus study the following GZ action,
\begin{eqnarray}\label{GZb4}
\hat S_{\GZ} &=& S_\YM +  \int \d^d x\,\left( b^a \p_\mu A_\mu^a +\overline c^a \p_\mu D_\mu^{ab} c^b \right)  +\nonumber\\&&+ \int \d^d x\left( \overline \varphi_\mu^{ac} \p_\nu D_\nu^{ab} \varphi_\mu^{bc}  - \overline \omega_\mu^{ac} \p_\nu D_\nu^{ab} \omega_\mu^{bc}  +\theta^{2}  f^{abc}A_\mu^a \left( \varphi_\mu^{bc} +  \overline \varphi_\mu^{bc}\right)\right)\,.
\end{eqnarray}
Applying \eqref{BRST} yields
\begin{multline}
    s \hat S_{\GZ}=\int \d^d x\Bigl(\theta^2c^k D_\mu^{ka}f^{abc}\left(\varphi_\mu^{bc} +  \overline \varphi_\mu^{bc}\right)+\theta^2f^{abc}A_\mu^a\omega_\mu^{bc}\\
    +\underbrace{gf^{abc}(D_\nu^{bp}c^p)\left(\p_\nu\overline\varphi_\mu^{ae}\varphi_\mu^{ce}-\p_\nu\overline\omega_\mu^{ae}\omega_\mu^{ce}\right)}_{s(g f^{abc} \p_\mu \overline \omega_\nu^{ae}    D_\mu^{bd} c^d  \varphi_\nu^{ce})}\Bigr)\,.
\end{multline}
According to \cite{Sorella:2009vt}, the positivity of the Faddeev-Popov operator inside the Gribov region allows to rewrite \eqref{GZb4} as
\begin{align}\label{GZb6}
s\hat S_{\GZ}  &= \int \d^d x\left( c^a D_\nu^{ab}\Lambda_\nu^b + \theta^2 f^{abc} A_\mu^a \omega_\mu^{bc}\right)\nonumber\\
&=\int \d^d x\left((D_\nu^{ma}\Lambda_\nu^a)[(\p_\nu D_\nu)^{-1}]^{mc}\frac{\delta}{\delta \overline c^c} \hat S_{\GZ}-\theta^2f^{abc}A_\mu^a[(\p_\nu D_\nu)^{-1}]^{bm}\frac{\delta}{\delta \overline\omega_\mu^{mc}} \hat S_{\GZ}\right)\,,
\end{align}
with
\begin{equation}
    \Lambda_\nu^a= \theta^2 f^{abc}(\varphi_\nu^{bc}+\overline\varphi_\nu^{bc})-gf^{bap}\left(\p_\nu\overline\varphi_\mu^{bc}\varphi_\mu^{pc}-\p_\nu\overline\omega_\mu^{bc}\omega_\mu^{pc}\right)\,.
\end{equation}
From \eqref{GZb6}, we can read off a new, albeit nonlocal, BRST symmetry, $s'\hat S_{\GZ}=0$, generated by
\begin{eqnarray}\label{GZb7}
s'A_{\mu }^{a} &=&-D_{\mu }^{ab} c^b\,, \quad s'c^{a} ~=~\frac{1}{2}gf^{abc}c^{b}c^{c}\,, \quad s'\overline{c}^{a} ~=~b^{a}-(D_\nu^{kc}\Lambda_\nu^c)[(\p_\nu D_\nu)^{-1}]^{ka}\,, \quad   s'b^{a}~=~0\,,  \nonumber \\
s'\varphi _{\mu}^{ac} &=&\omega_{\mu}^{ac}\,,\quad s'\omega_{\mu}^{ac}~=~0\,, \quad s'\overline{\omega}_{\mu}^{ac} ~=~\overline{\varphi }_{\mu}^{ac}+\theta^2f^{qpc}A_\mu^q[(\p_\nu D_\nu)^{-1}]^{pa}\,,\quad s' \overline{\varphi }_{\mu}^{ac}~=~0\,.
\end{eqnarray}
One can now see that $s'$ is not nilpotent, $s'^2\neq0$, \cite{Sorella:2009vt}.

\subsubsection{Localization of the BRST variations}
We want to explore the possibility to localize the nonlocal expressions appearing in the BRST variations \eqref{GZb7}. We have in mind to introduce extra fields into the GZ action, in such a way that their equation of motions reproduce the nonlocal BRST expressions. As such, we can hope to establish a (at least on-shell) local version of the BRST symmetry $s'$. As it shall soon become clear, our localization procedure starts from the local GZ action itself and, at the end, we shall naturally come to the non-local BRST just described upon using some equations of motion\footnote{The non-local symmetry of \cite{Kondo:2009qz} seems to fall outside this construction.}. \\
\\
We shall treat the breaking proportional to $\Lambda_\nu^a$ in two parts and we introduce the notation
\begin{equation}
    \overline\Lambda_\nu^a= f^{abc}(\varphi_\nu^{bc}+\overline\varphi_\nu^{bc})\,,\qquad \hat{\Lambda}_\nu^a=-gf^{bap}\left(\p_\nu\overline\varphi_\mu^{bc}\varphi_\mu^{pc}-\p_\nu\overline\omega_\mu^{bc}\omega_\mu^{pc}\right)\;,
\end{equation}
for the true, resp.~fake BRST breaking content of $\Lambda_\nu^a\equiv\theta^2 \overline\Lambda_\nu^a+\hat \Lambda_\nu^a$ . The reason for this is that it will naturally lead to a modification of the \emph{complete} GZ action \eqref{GZb1} rather than of the reduced version \eqref{GZb4}.\\
\\
We start with the original BRST $s$, and introduce the following doublets
\begin{eqnarray}\label{GZb9}
s \alpha^a &=&\Omega^a\,,\quad   s\Omega^a~=~0\,,\quad s \overline\Omega^a ~=~\overline\alpha^a\,,\quad   s\overline\alpha^a~=~0\,,\nonumber\\
s \beta_\mu^{ab} &=&\Psi_\mu^{ab}\,, \quad  s\Psi_\mu^{ab}~=~0\,,\quad s \overline\Psi_\mu^{ab} ~=~\overline\beta_\mu^{ab}\,,\quad  s\overline\beta_\mu^{ab}~=0~\,.
\end{eqnarray}
The $\alpha^a$, $\overline{\alpha}^a$, $\beta_\mu^{ab}$ and $\overline\beta_\mu^{ab}$ are commuting, while $\Omega^a$, $\overline{\Omega}^a$, $\Psi_\mu^{ab}$ and $\overline\Psi_\mu^{ab}$ are anti-commuting fields. We also introduce the auxiliary action
\begin{align}
S_{\aux}&=s \int \d^4x\left(\alpha^a \p_\mu D_\mu^{ab}\overline\Omega^b -\overline\Omega^a D_\nu^{ab} \overline\Lambda_\nu^b+ \beta_\nu^{ac}\p_\mu D_\mu^{ab}\overline{\Psi}_\nu^{bc}-f^{abc}A_\mu^a\overline{\Psi}_\mu^{bc}\right)\nonumber\\
&=\int \d^4x\left(\Omega^a \p_\mu D_\mu^{ab}\overline \Omega^b + \alpha^a \p_\mu D_\mu^{ab}\overline \alpha^b + gf^{abc}(\p_\mu \alpha^a)(D_\mu^{bd}c^d)\overline \Omega^c-\overline\alpha^a D_\nu^{ab} \overline\Lambda_\nu^b+\overline\Omega^a s(D_\nu^{ab} \overline\Lambda_\nu^b)\right.\nonumber\\
&\left.+ \Psi_\nu^{ac}\p_\mu D_\mu^{ab}\overline \Psi_\nu^{bc}+\beta_\nu^{ac}\p_\mu D_\mu^{ab}\overline \beta_\nu^{bc} +gf^{abc} (\p_\mu\beta_{\nu}^{ae})(D_\mu^{bd}c^d)\overline \Psi_\nu^{ce}-f^{abc}A_\mu^a\overline{\beta}_\mu^{bc}-f^{abc}\overline{\Psi}_\mu^{bc}D_\mu^{ad}c^d\right)\,.
\end{align}
It is clear at sight that the equations of motions for $\alpha^a$ and $\beta_\mu^{ab}$ are closely related to the $\theta$-dependent part of the nonlocal expressions in the r.h.s.~of \eqref{GZb7}.\\
\\
For the moment, let us just change the GZ action \eqref{GZb4} by hand and consider
\begin{eqnarray}
\hat S_{\GZ}^{\modi}=\hat S_{\GZ}+S_{\aux}\,,
\end{eqnarray}
and define the transformation\footnote{$\delta$ itself will not correspond to a symmetry.} $\delta$ by means of
\begin{align}\label{GZb14}
    \delta \overline\alpha^a &=\theta^2 c^a\,, &  \delta \overline{c}^a&=-\theta^2 \alpha^a\,, & \delta \overline\beta_\mu^{bc} &=\theta^2 \omega_\mu^{bc}\,, &    \delta \overline{\omega}_\mu^{bc}&=\theta^2\beta_\mu^{bc}\,, &\delta(\text{rest})&=0\,.
\end{align}
Then, we find
\begin{eqnarray*}
    (s+\delta)(\hat S_{\GZ}^{\modi})&=&  \underbrace{s\hat S_{\GZ}}_{*} + \underbrace{\delta \hat S_{\GZ}}_{**} + \underbrace{\delta S_{\aux}}_{***}\nonumber\\
    &=&\underbrace{\int \d^d x\left(\theta^2 c^a D_\nu^{ab}\overline\Lambda_\nu^b + \theta^2 f^{abc} A_\mu^a \omega_\mu^{bc}+s(g f^{abc} \p_\mu \overline \omega_\nu^{ae}    D_\mu^{bd} c^d  \varphi_\nu^{ce})\right)}_{*}\nonumber\\
    &&+\underbrace{\int \d^dx \left(-\theta^2 \alpha^a \p_\mu D_\mu^{ab}c^b-\theta^2\beta_\mu^{ac}\p_\nu D_\nu^{ab}\omega_\mu^{bc}\right)}_{**}\nonumber\\
    &&+\underbrace{\int \d^dx \left(\theta^2 \alpha^a \p_\mu D_\mu^{ab}c^b -\theta^2c^a D_\nu^{ab}\overline\Lambda_\nu^b+\theta^2 \beta_\mu^{ac}\p_\nu D_\nu^{ab}\omega_\nu^{bc}-\theta^2 f^{abc}A_\mu^a \omega_\mu^{bc}\right)}_{***}\nonumber\\
    &=& (s+\delta)\int \d^dx\left(g f^{abc} \p_\mu \overline \omega_\nu^{ae}    D_\mu^{bd} c^d  \varphi_\nu^{ce}\right)-\delta\int \d^dx\left(g f^{abc} \p_\mu \overline \omega_\nu^{ae}    D_\mu^{bd} c^d  \varphi_\nu^{ce}\right)\,.
\end{eqnarray*}
We can rewrite this as
\begin{equation}\label{GZb16}
(s+\delta) \tilde S_{\GZ}^{\modi}= -\delta\int \d^dx\left(g f^{abc} \p_\mu \overline \omega_\nu^{ae}    D_\mu^{bd} c^d  \varphi_\nu^{ce}\right)=-\theta^2\int \d^dx\left(g f^{abc} \p_\mu \beta_\nu^{ae}    D_\mu^{bd} c^d  \varphi_\nu^{ce}\right)\,,
\end{equation}
whereby we introduced a modified version of the original GZ action \eqref{GZb1}, given by $\tilde S_{\GZ}^{\modi}=S_{\GZ}+ S_{\aux}$. Looking at \eqref{GZb16}, we have found that $s+\delta$ ``almost'' generates a symmetry of the foregoing action $\tilde S_{\GZ}^{\modi}$. In order to get an actual symmetry, we rewrite, using partial integration,
\begin{eqnarray*}
  -\int \d^dx\left(g f^{abc} \p_\mu \beta_\nu^{ae}    D_\mu^{bd} c^d  \varphi_\nu^{ce}\right)&=&  \int \d^dxD_\mu^{bd}\left(g f^{abc} \p_\mu \beta_\nu^{ae}\varphi_\nu^{ce}  \right)[(\p_\nu D_\nu)^{-1}]^{dq}\frac{\delta S_{\GZ}^{\modi}}{\delta \overline c^q}\,.
\end{eqnarray*}
We can localize the latter term again analogously as before, by extending the auxiliary action by a novel quartet of fields,
\begin{align}
s Q^a &=  R^a\,, & s R^a &=0 \,, & s \overline R^a &= \overline Q^a\,,  &       s \overline Q^a &=0 \;.
\end{align}
with $R$, $\overline R$ anti-commutating fields, while $Q$ and $\overline Q$ are bosonic fields. We introduce a second auxiliary action
\begin{eqnarray*}
S_{\aux, 2}&=& s\int \d^4x\left(Q^a \p_\mu D^{ab}_\mu \overline R^b - \overline R^d \underbrace{D_\mu^{bd} (g f^{abc} \p_\mu \beta_\nu^{ae} \varphi^{ce}_\nu}_{\kappa^d })\right) \nonumber\\
&=& \int \d^4x\left( R^a \p_\mu D^{ab}_\mu \overline R^b +  Q^a \p_\mu D_\mu^{ab}\overline Q^b + g f^{abc}\p_\mu Q^a  D_\mu^{bd} c^d \overline R^c - \overline Q^d \kappa^d + \overline R^d s (\kappa^d) \right)\;,
\end{eqnarray*}
and extend the $\delta$-transformation \eqref{GZb14} to
\begin{align}
    \delta \overline Q^a &= \theta^2 c^a\,, &  \delta \overline{c}^a&=-\theta^2 Q^a - \theta^2 \alpha^a\,.
\end{align}
In this way we have that
\begin{eqnarray}
\delta S_{\GZ} &=& -   \theta^2\int \d^4 x Q^a \p_\mu D_\mu^{ab} c^b\,,\quad \delta S_{\aux, 2} ~=~  \int \d^4 x (- \theta^2 c^d \kappa^d + \theta^2 Q^b \p_\mu D_\mu^{ab} c^b)\,,
\end{eqnarray}
which nicely cancels out in combination with \eqref{GZb16}.\\
\\
In summary, the following action
\begin{align}\label{GZmod}
S_{\GZ}^{\modi}&=S_{\GZ}+ S_{\aux} + S_{\aux, 2}\nonumber\\
&=S_\YM +  \int \d^d x\,\left( b^a \p_\mu A_\mu^a +\overline c^a \p_\mu D_\mu^{ab} c^b \right)+ \int \d^d x\left( \overline \varphi_\mu^{ac} \p_\nu D_\nu^{ab} \varphi_\mu^{bc}  - \overline \omega_\mu^{ac} \p_\nu D_\nu^{ab} \omega_\mu^{bc} \right. \nonumber  \\
& \left. +\theta^{2}  f^{abc}A_\mu^a \left( \varphi_\mu^{bc} +  \overline \varphi_\mu^{bc}\right)- g f^{abc} \p_\mu \overline \omega_\nu^{ae}    D_\mu^{bd} c^d  \varphi_\nu^{ce}\right) +\int \d^dx\left(\Omega^a \p_\mu D_\mu^{ab}\overline \Omega^b  \right. \nonumber\\
& \left.+ \alpha^a \p_\mu D_\mu^{ab}\overline \alpha^b + gf^{abc}(\p_\mu \alpha^a)(D_\mu^{bd}c^d)\overline \Omega^c-\overline\alpha^a D_\nu^{ab} \overline\Lambda_\nu^b+\overline\Omega^a s(D_\nu^{ab} \overline\Lambda_\nu^b)+ \Psi_\nu^{ac}\p_\mu D_\mu^{ab}\overline \Psi_\nu^{bc}\right.\nonumber\\
&\left.+\beta_\nu^{ac}\p_\mu D_\mu^{ab}\overline \beta_\nu^{bc} +gf^{abc} (\p_\mu\beta_{\nu}^{ae})(D_\mu^{bd}c^d)\overline \Psi_\nu^{ce}-f^{abc}A_\mu^a\overline{\beta}_\mu^{bc}-f^{abc}\overline{\Psi}_\mu^{bc}D_\mu^{ad}c^d\right)\nonumber\\
 &+\int \d^4x\left( R^a \p_\mu D^{ab}_\mu \overline R^b +  Q^a \p_\mu D_\mu^{ab}\overline Q^b + g f^{abc}\p_\mu Q^a  D_\mu^{bd} c^d \overline R^c - \overline Q^d \kappa^d + \overline R^d s (\kappa^d) \right)\;,
 \end{align}
enjoys the following modified BRST invariance $s_\theta \equiv s + \delta$,
\begin{align}\label{GZb18}
s_\theta A_{\mu }^{a} &=-\left( D_{\mu }c\right) ^{a}\,, &  s_\theta c^{a} &=\frac{1}{2}gf^{abc}c^{b}c^{c}\,,& s_\theta \overline{c}^{a} &=b^{a}-\theta^2\alpha^a -\theta^2 Q^a  \,,&  s_\theta b^{a}&=0\,,  \nonumber \\
s_\theta \varphi _{\mu}^{ac} &=\omega_{\mu}^{ac}\,,& s_\theta\omega_{\mu}^{ac}&=0\,,& s_\theta \overline{\omega}_{\mu}^{ac} &=\overline{\varphi }_{\mu}^{ac}+\theta^2\beta_\mu^{bc}\,,& s_\theta \overline{\varphi }_{\mu}^{ac}&=0\,,\nonumber\\
 s_\theta \alpha^a &=\Omega^a\,, &  s_\theta\Omega^a &=0\,,& s_\theta \overline\Omega^a &=\overline\alpha^a\,, &  s_\theta\overline\alpha^a &=\theta^2 c^a\,,\nonumber\\
s_\theta \beta_\mu^{ab} &=\Psi_\mu^{ab}\,,&   s_\theta\Psi_\mu^{ab}&=0\,,& s_\theta \overline\Psi_\mu^{ab} &=\overline\beta_\mu^{ab}\,,&  s_\theta\overline\beta_\mu^{ab}&=\theta^2 \omega_\mu^{ab}\,,\nonumber\\
s_\theta Q^a &=  R^a \,, & s_\theta R^a &=0 \,,& s_\theta \overline R^a &= \overline Q^a\,,& s_\theta \overline Q^a &= \theta^2 c^a \,.
\end{align}
We notice that we did not have to use the equations of motion to identify the $s_\theta$-symmetry of the modified GZ action $S_{\GZ}^{\modi}$. The newly constructed BRST is not nilpotent, $ s_\theta^2\neq0$, in analogy with its nonlocal version written down in \eqref{GZb7}. We have that $s_\theta^4 = 0$. Upon using the equations of motion of the new fields, we can derive from \eqref{GZb18} the on-shell (but nonlocal) BRST invariance of the original GZ action. Notice that this BRST will be slightly different from the one constructed in \cite{Sorella:2009vt}, the difference being caused by the fact that we constructed a BRST invariance of the \emph{complete} GZ action \eqref{GZb1} rather than of \eqref{GZb4}.

\subsection{A few properties of the modified GZ action}
\subsubsection{Other symmetries}
We shall identify some extra invariances of the modified GZ action $S_{\GZ}^{\modi}$. Firstly, as
\begin{equation}
 \int \d^dx \left(\overline\Omega^a \frac{\delta}{\delta \alpha^a} -\overline\alpha^a \frac{\delta}{\delta \Omega^a}\right) S_{\GZ}^{\modi} =\int \d^dx\left( \overline \Omega^a \p D^{ab} \overline \alpha^b - \overline \alpha^a \p D^{ab} \overline \Omega^b +  gf^{abc}(\p_\mu \overline\Omega^a)(D_\mu^{bd}c^d)\overline\Omega^c\right)
\end{equation}
and by rewriting the r.h.s.~of it by means of
\begin{eqnarray}
\int \d^dx\left( \overline \Omega^a \p D^{ab} \overline \alpha^b - \overline \alpha^a \p D^{ab} \overline \Omega^b \right) &=& \int \d^dx\left( g f^{akb} \p A^k \overline \alpha^b \right) = \int \d^dx\left( g f^{akb}  \overline \alpha^b \frac{\delta}{\delta\overline b^k}S_{\GZ}^{\modi}\right)\,,  \nonumber\\
\int \d^dx\left(-gf^{abc}(\p_\mu \overline\Omega^a)(D_\mu^{bd}c^d)\overline\Omega^c\right) &=& \int \d^dx\left(-\frac{1}{2}gf^{abc}(D_\mu^{bd}c^d)\p_\mu (\overline\Omega^a\overline\Omega^c)\right)\nonumber\\
       &=&\int \d^dx\left(\frac{1}{2}gf^{abc}\overline\Omega^a\overline\Omega^c\frac{\delta}{\delta\overline c^b}S_{\GZ}^{\modi}\right)\,,
\end{eqnarray}
we conclude that
\begin{equation}\label{GZb20}
    \Delta_1= \int \d^dx\left(\overline\Omega^a \frac{\delta}{\delta \alpha^a}-\overline\alpha^a \frac{\delta}{\delta \Omega^a}-\frac{1}{2}gf^{abc}\overline\Omega^a\overline\Omega^c\frac{\delta}{\delta\overline c^b} - g f^{akb} \overline \Omega^a \overline \alpha^b \frac{\delta }{\delta b^k}\right)\,,\qquad \text{with }\Delta_1^2=0\,,
\end{equation}
establishes a (nilpotent) symmetry of $S_{\GZ}^{\modi}$. Similarly, we also find the following symmetries of the action
\begin{align}\label{GZb21}
\Delta_2 &= \int \d^dx\left(\overline\Psi_{\nu}^{bc} \frac{\delta}{\delta \beta_\nu^{bc}}-\overline\beta_\nu^{bc} \frac{\delta}{\delta \Psi_\nu^{bc}} - \frac{1}{2}gf^{abc}\overline\Psi_\nu^{ae}\overline\Psi_\nu^{ce}\frac{\delta}{\delta\overline c^b} - g f^{akb} \overline \Psi^{ae}_\nu \overline \beta^{be}_\nu \frac{\delta }{\delta b^k} \right) \,, &&\text{with }\Delta_2^2&=0\,, \nonumber\\
\Delta_3&= \int \d^dx\left(\overline R^a \frac{\delta}{\delta Q^a}-\overline Q^a \frac{\delta}{\delta R^a}-\frac{1}{2}gf^{abc}\overline R^a \overline R^c\frac{\delta}{\delta\overline c^b} - g f^{akb} \overline R^a \overline Q^b \frac{\delta }{\delta b^k}\right)\,,&& \text{with }\Delta_3^2&=0\,.
\end{align}
We can also link some of the original fields with the newly introduced one through the symmetries
\begin{equation}
\Delta_4= \int \d^dx\left(\overline \Omega^a \frac{\delta}{\delta \overline c^a}+c^a \frac{\delta}{\delta \Omega^a} - g f^{akb} c^a \overline \Omega ^b \frac{\delta }{\delta b^k}\right)\,,\qquad \text{with }\Delta_4^2=0\,,
\end{equation}
and
\begin{equation}
\Delta_5= \int \d^dx\left(\overline R^a \frac{\delta}{\delta \overline c^a}+c^a \frac{\delta}{\delta R^a} - g f^{akb} c^a \overline R ^b \frac{\delta }{\delta b^k}\right)\,,\qquad \text{with }\Delta_5^2=0\,.
\end{equation}
Clearly, $\left\{\Delta_i,\Delta_j\right\}=0$, but $\left\{\Delta_i,s_\theta\right\}\neq0$ generate further symmetries. In addition, there might be more symmetries not related to this algebra, but we did not attempt to find such here.\\
\\
We notice that we can rewrite $S_{\GZ}^{\modi}$ as
\begin{align}\label{GZb22}
&S_{\GZ}^{\modi}=S_\YM +  \int \d^d x\,\left( b^a \p_\mu A_\mu^a +\overline c^a \p_\mu D_\mu^{ab} c^b \right) + \int \d^d x\left( \overline \varphi_\mu^{ac} \p_\nu D_\nu^{ab} \varphi_\mu^{bc}  - \overline \omega_\mu^{ac} \p_\nu D_\nu^{ab} \omega_\mu^{bc} \right.\nonumber\\
 &\left. +\theta^{2}  f^{abc}A_\mu^a \left( \varphi_\mu^{bc} +  \overline \varphi_\mu^{bc}\right)\right) +\int \d^4x\left(-gf^{abc}\p_\mu \overline\omega_\nu^{ae}D_\mu^{bd}c^d\varphi_\nu^{ce}+\Omega^a \p_\mu D_\mu^{ab}\overline \Omega^b + \alpha^a \p_\mu D_\mu^{ab}\overline \alpha^b \right.\nonumber\\
& + gf^{abc}(\p_\mu \alpha^a)(D_\mu^{bd}c^d)\overline \Omega^c + \Psi_\nu^{ac}\p_\mu D_\mu^{ab}\overline \Psi_\nu^{bc}+\beta_\nu^{ac}\p_\mu D_\mu^{ab}\overline \beta_\nu^{bc} +gf^{abc} (\p_\mu\beta_{\nu}^{ae})(D_\mu^{bd}c^d)\overline \Psi_\nu^{ce}  \nonumber\\
& \left.+ R^a \p_\mu D^{ab}_\mu \overline R^b +  Q^a \p_\mu D_\mu^{ab}\overline Q^b + g f^{abc}\p_\mu Q^a  D_\mu^{bd} c^d \overline R^c \right)+\Delta_1\int \d^4x \left(\Omega^a D_\nu^{ab} \Lambda_\nu^b+\alpha^a s(D_\nu^{ab} \Lambda_\nu^b)\right)\nonumber\\
& +\Delta_2\int \d^dx\left(f^{abc}A_\mu^a\Psi_\mu^{bc}-f^{abc}\beta_\mu^{bc}D_\mu^{ad}c^d\right) + \Delta_3 \int \d^4x \left(R^d \kappa^d +  Q^d s (\kappa^d))\right)\,.
\end{align}
Finally, we also observe that the action is left invariant under constant shifts of the fields $\Psi_\mu^{ac}$, $\overline c^a$, $\overline\omega_\mu^{ac}$, $\beta_\mu^{ac}$, $\alpha^a$, $\Omega^a$, $Q^a$ and $R^a$, expressed through the following identities
\begin{equation}\label{GZb22b}
    \int \d^4x \frac{\delta S_{\GZ}^{\modi}}{\delta \chi}=   0\,,\qquad \chi\in \left\{\Psi_\mu^{ac},\overline c^a,\overline\omega_\mu^{ac},\beta_\mu^{ac},\alpha^a,\Omega^a,Q^a,R^a\right\}\,.
\end{equation}

\subsubsection{Connection between the original Yang-Mills and modified GZ action in the case of vanishing Gribov mass}
An important property of the modified GZ action is to investigate what happens when we set $\theta^2=0$. If our $S_{\GZ}^{\modi}$ is meaningful, we expect to find back the original Yang-Mills theory in the Landau gauge (modulo trivial, unity-related, terms in the action). Setting $\theta^2=0$ yields
\begin{multline*}
\left.S_{\GZ}^{\modi}\right|_{\theta^2=0}   = S_\YM +  \int \d^d x\,\left( b^a \p_\mu A_\mu^a +\overline c^a \p_\mu D_\mu^{ab} c^b \right)  + \int \d^d x\left( \overline \varphi_\mu^{ac} \p_\nu D_\nu^{ab} \varphi_\mu^{bc}  - \overline \omega_\mu^{ac} \p_\nu D_\nu^{ab} \omega_\mu^{bc} \right.\nonumber\\
\left. -  \underline{g f^{abc} \p_\mu \overline \omega_\nu^{ae}    D_\mu^{bd} c^d  \varphi_\nu^{ce}}\right) +\int \d^dx\left(\Omega^a \p_\mu D_\mu^{ab}\overline \Omega^b + \alpha^a \p_\mu D_\mu^{ab}\overline \alpha^b +  \underline{gf^{abc}(\p_\mu \alpha^a)(D_\mu^{bd}c^d)\overline \Omega^c } \right.\nonumber\\
\left.+ \Psi_\nu^{ac}\p_\mu D_\mu^{ab}\overline \Psi_\nu^{bc}+\beta_\nu^{ac}\p_\mu D_\mu^{ab}\overline \beta_\nu^{bc} +  \underline{gf^{abc} (\p_\mu\beta_{\nu}^{ae})(D_\mu^{bd}c^d)\overline \Psi_\nu^{ce} } \right)\nonumber\\
+\int \d^4x\left( R^a \p_\mu D^{ab}_\mu \overline R^b +  Q^a \p_\mu D_\mu^{ab}\overline Q^b +  \underline{ g f^{abc}\p_\mu Q^a  D_\mu^{bd} c^d \overline R^c} -  \underline{\overline Q^d \kappa^d }+  \underline{\overline R^d s (\kappa^d)} \right)\,,
\end{multline*}
with
\begin{equation}
    \kappa^d = D_\mu^{bd} (g f^{abc} \p_\mu \beta_\nu^{ae} \varphi^{ce}_\nu )\,.
\end{equation}
Considering (ghost neutral) Green functions of the original Faddeev-Popov fields, it is then easily checked that the underlined terms will never contribute, as there are no propagators of the desired kind to attach them to the Green functions of interest. This argument is completely similar to the one given originally in \cite{Zwanziger:1992qr} which was related to the presence of the term $g f^{abc} \p_\mu \overline \omega_\nu^{ae}    D_\mu^{bd} c^d  \varphi_\nu^{ce}$ in the GZ action, which can also not couple. The residual terms are all forming unities, and upon integrating out these, we recover the Yang-Mills action.  Glancing at \eqref{GZb18}, we notice that $s_{\theta=0}=s$, with $s$ the original BRST, upon generalization to the new fields which become pairwise BRST $s$-doublets. In fact, since $s$ is now a symmetry and seeing that
\begin{equation}\label{Gzb27b}
    \left.S_{\GZ}^{\modi}\right|_{\theta^2=0}   = S_\YM +  \int \d^d x\,\left( b^a \p_\mu A_\mu^a +\overline c^a \p_\mu D_\mu^{ab} c^b \right)+s\,\int \d^dx \left(\ldots\right)\,,
\end{equation}
we did nothing more than writing down a somewhat more complicated version of the FP Landau gauge fixing. We shall thence also recover the original Yang-Mills BRST cohomology. All GZ-related fields, old or new, are then physically trivial as they appear as doublets of a nilpotent symmetry, see p.\pageref{doublettheorem}.

\subsubsection{Connection between the modified and original GZ action}
In this section we wish to verify that $S_{\GZ}^{\modi}$ and $S_{\GZ}$ are in fact equivalent in the sense that for any Green function built from the original GZ fields, i.e.~with $\phi\in\bigl\{A_\mu^a$, $b^a$, $\overline{c}^a$, $c^a$, $\varphi_\mu^{ab}$, $\overline\varphi_\mu^{ab}$, $\omega_\mu^{ab}$, $\overline\omega_\mu^{ab}\bigr\}$, we have the following identification
\begin{eqnarray}\label{GZb32}
  \Braket{\phi(x_1)\ldots\phi(x_n)}_{\modi}=  \int [\d\Phi]_{\modi} \phi(x_1)\ldots\phi(x_n) \e^{-S_{\GZ}^{\modi}} =   \Braket{\phi(x_1)\ldots\phi(x_n)}_{\GZ}\,.
\end{eqnarray}
Let us show this in two ways. Firstly, the nonlocal substitutions
\begin{eqnarray}\label{GZb33}
\Omega^a &=& \Omega'^a - g f^{\ell b q} (\p_\mu \alpha^\ell) (D_\mu^{bd} c^d) [\p D^{-1}]^{qa} + s (D^{qk}_\nu \overline \Lambda^k_\nu ) [(\p D^{-1})]^{qa}\,,  \nonumber\\
\Psi_\nu^{ae} &=& {\Psi'}_\nu^{ae} -g f^{\ell b q} (\p_\mu \beta^{ae}_\nu) (D_\mu^{bd} c^d) [\p D^{-1}]^{qa} - f^{\ell q e} (D^{\ell d}_\nu c^d ) [(\p D^{-1})]^{qa}\,,\nonumber\\
R^a &=& R'^a - g f^{\ell b q} (\p_\mu Q^\ell) (D_\mu^{bd} c^d) [\p D^{-1}]^{qa} +s (\kappa^q  ) [(\p D^{-1})]^{qa} \,,\nonumber\\
 \alpha^a&=&\alpha'^a + D_\nu ^{qd} \overline \Lambda^d_\nu [\p D^{-1} ]^{qa}\,, \nonumber\\
Q^a &=& Q'^a +  \kappa^q   [\p D^{-1} ]^{qa}\,,\quad \beta_\nu^{ac}~=~{\beta'}_\nu^{ac}+ f^{dqc} A^d_\nu [ (\p D)^{-1}]^{qa}\,,
\end{eqnarray}
which come with a trivial Jacobian, lead to
\begin{eqnarray}\label{GZb34}
  \Braket{\phi(x_1)\ldots\phi(x_n)}_{\modi}=  \int [\d\Phi]_{\modi} \phi(x_1)\ldots\phi(x_n) \e^{-{S'}_{\GZ}^{\modi}} \;,
\end{eqnarray}
after dropping the prime-notation again, with the shifted ${S'}_{\GZ}^{\modi}$ given by
\begin{align}
{S'}_{\GZ}^{\modi}&=S_\YM +  \int \d^d x\,\left( b^a \p_\mu A_\mu^a +\overline c^a \p_\mu D_\mu^{ab} c^b \right) + \int \d^d x\left( \overline \varphi_\mu^{ac} \p_\nu D_\nu^{ab} \varphi_\mu^{bc}  - \overline \omega_\mu^{ac} \p_\nu D_\nu^{ab} \omega_\mu^{bc} \right. \nonumber\\
 & \left.+\theta^{2}  f^{abc}A_\mu^a \left( \varphi_\mu^{bc} +  \overline \varphi_\mu^{bc}\right) - g f^{abc} \p_\mu \overline \omega^{ae}_\nu D^{bd}_\mu c^d \varphi^{ce}_\nu\right) +\int \d^4x\left( \Omega^a \p_\mu D_\mu^{ab}\overline\Omega^b + \alpha^a \p_\mu D_\mu^{ab} \overline \alpha^b \right. \nonumber\\
&\left.  + \Psi^{ac}_\nu \p_\mu D_\mu^{ab} \overline \Psi^{bc}_\nu + \overline{\beta}_\nu^{ac}\p_\mu D_\mu^{ab}\beta_\nu^{bc} + R^{a} \p_\mu D_\mu^{ab}R^{b} +  Q^{a} \p_\mu D_\mu^{ab} Q^{b}\right)\nonumber\,.
\end{align}
Consequently, we can perform the path integration over the new fields, which are pairwise unities, to discover that
\begin{eqnarray}\label{GZb36}
\Braket{\phi(x_1)\ldots\phi(x_n)}_{\modi}=  \int [\d\Phi]_{\GZ} \phi(x_1)\ldots\phi(x_n) \e^{-S_{\GZ}}=\Braket{\phi(x_1)\ldots\phi(x_n)}_{\GZ}\,.
\end{eqnarray}
This important formula means that the original GZ correlation functions can be evaluated with either the original or the modified GZ action. We may thus replace the original GZ action with nonlocal BRST with its modified version, enjoying a local version of the BRST. As a first corollary, we can reestablish that $\left.S_{\GZ}^{\modi}\right|_{\theta^2=0}$ is equivalent with normal Yang-Mills gauge theories, as we know that GZ is for $\theta^2=0$.\\
\\
Secondly, to avoid the use of nonlocal shifts to prove the important result \eqref{GZb36}, let us also present an alternative derivation. We reconsider the generic correlation function \eqref{GZb32}
and consequently recall the following Gaussian integration formula in case of two cc fields (see equation \eqref{gauss8}),
\begin{equation}
    \int \d\sigma \d\overline\sigma \e^{-\int \d^4x~ \left(\overline\sigma ~\Delta~ \sigma + J~\overline\sigma+\overline J~ \sigma\right)}\propto \det \Delta^{-1} \e^{-\overline J~ \Delta~ J}\,,
\end{equation}
which means that if $J=0$ or $\overline J=0$, we shall just pick up a determinant upon integration. A similar formula holds for anti-commuting fields, yielding the  inverse power of the determinant. If we then apply this to \eqref{GZb32} and adopt the integration
order $(\overline R, R)$, $(\overline Q, Q)$, $(\overline \Psi, \Psi)$, $(\overline \beta, \beta)$, $(\overline \Omega, \Omega)$, $(\overline \alpha, \alpha)$,  it is easily seen that everything neatly cancels, including the determinants. This once again leads to the result \eqref{GZb36}. As a second corollary, we can mention that the gap formulation of the horizon condition, $\frac{\p \Gamma}{\p \gamma^2}=0$, will also remain unchanged.

\subsection{Renormalization of the modified GZ action}
As a final step, we should try to prove the renormalizability of the modified GZ action \eqref{GZmod}. However, this would require a lot of calculational efforts and shall be far from straightforward. Therefore, we consider this as future work.

\section{The hermiticity of the GZ action}
Let us now shortly comment on the hermiticity of the GZ action. We recall that the action is given by equation \eqref{GZstart}
\begin{eqnarray}
S_\GZ &=& S_{0} +  S_{\gamma} \,,
\end{eqnarray}
with
\begin{eqnarray}
S_{0}&=&S_\YM + S_\gf + \int \d^d x \left( \overline \varphi_i^a \p_\mu \left( D_\mu^{ab} \varphi^b_i \right)  - \overline \omega_i^a \p_\mu \left( D_\mu^{ab} \omega_i^b \right) - g f^{abc} \p_\mu \overline \omega_i^a    D_\mu^{bd} c^d  \varphi_i^c \right) \nonumber \;, \\
S_{\gamma}&=& -\gamma ^{2}g\int\d^{d}x\left( f^{abc}A_{\mu }^{a}\varphi _{\mu }^{bc} +f^{abc}A_{\mu}^{a}\overline{\varphi }_{\mu }^{bc} + \frac{d}{g}\left(N^{2}-1\right) \gamma^{2} \right) \;.
\end{eqnarray}
If we define
\begin{align}
\varphi^\dagger &= \overline \varphi \;,   & \overline \varphi^\dagger &=  \varphi \;,  & \omega^\dagger &=  \omega \;, & \overline \omega^\dagger &= -\overline \omega \;,
\end{align}
we see that the GZ action is almost Hermitian up to the term $- g f^{abc} \p_\mu \overline \omega_i^a D_\mu^{bd} c^d  \varphi_i^c$. However, we recall that this term was introduced for renormalization reasons by the shift \eqref{shift}. Therefore, returning to the action before the shift, the GZ action is Hermitian. Moreover, for practical purposes, the term $- g f^{abc} \p_\mu \overline \omega_i^a D_\mu^{bd} c^d  \varphi_i^c$ is almost redundant as it cannot couple to any Feynman diagram without external $c$ and $\omega$ legs. Therefore, we can conclude that in practice, the GZ action is Hermitian.\\
\\
One comment which one can make, is that the kinetic term $\overline \varphi \p^2 \varphi$ in the GZ action, seems to have to wrong sign. For a path integral to be well defined, one would naively expect a minus sign in front of all the kinetic terms. We could rectify this by changing $\varphi \to - \varphi$, but this would spoil the hermiticity of the GZ action. In addition, we can also argue that this kinetic term  $\overline \varphi \p^2 \varphi$ is always accompanied by the term $-\overline \omega \p^2 \omega$, forming a unity which can be integrated out. Also, it is not so clear anymore how to interpret the sign of $\overline \varphi \p^2 \varphi$, as in fact, there is mixing with the gluon field $A$, for $\gamma^2 \not=0$ and one should consider the path integral as a whole. Moreover, as we are not integrating over the entire region of gluon fields, but only over the gluon fields belonging to the Gribov region, it is very hard to make any conclusion whether the path integral $\int [\d \phi] \e^{-S_\GZ}$ is well defined, a usual problem in quantum field theory.

\section{The horizon condition\label{secthorizon}}
In this section, we would like to elaborate on the form of the horizon function. We recall that the non-local horizon function, which is added to the Faddeev-Popov action in order to restrict to the Gribov region is given by
\begin{equation}\label{onlycorrect}
S_\h = \lim_{\theta \to 0}  \int \d^d x h (x) = \lim_{\theta \to 0}  \int \d^d x \int \d^d y   \left( D_\mu^{ac}(x) \gamma^2(x) \right)(\mathcal M^{-1})^{ab}(x,y) \left( D_\mu^{bc} (y) \gamma^2(y) \right) \;,
\end{equation}
whereby $\gamma(z)$ was defined as follows
\begin{eqnarray}\label{gammadefi}
\gamma^2 (z) &=& \e^{\ii \theta z} \gamma^2 \;.
\end{eqnarray}
see equation \eqref{danhorizon}. As it shall be useful later, we could also define $h'(x)$,
\begin{equation}\label{onlycorrect2}
S_\h = \gamma^4 \lim_{\theta \to 0}  \int \d^d x h' (x) = \gamma^4\lim_{\theta \to 0}  \int \d^d x \int \d^d y   \left( D_\mu^{ac}(x) \e^{\ii \theta x} \right)(\mathcal M^{-1})^{ab}(x,y) \left( D_\mu^{bc} (y) \e^{\ii \theta y} \right) \;,
\end{equation}
where $\gamma$ is pulled out of the integral.\\
\\
In \cite{Kondo:2009wk}, it has been argued that there would be another possible horizon function, namely
\begin{equation}\label{hkondo}
h_K(x) =  \gamma^4  \int \d^d y  g^2 f^{akc} A^k_\mu(x) (\mathcal M^{-1})^{ab} (x,y) f^{b\ell c} A^\ell_\mu (y)\,,
\end{equation}
In fact, one can understand that certain doubt has arisen on which horizon function is the correct one, as one can easily see that setting $\gamma(x)$ in $h(x)$ immediately equal to a constant parameter, $\gamma(x)\equiv\gamma$, which agrees with switching the limit $\lim_{\theta \to 0}$ and the integration signs, we obtain the horizon function $h_K(x)$. One can therefore appreciate that the difference between $h(x)$ and $h_K(x)$ is very subtle. In addition, when localizing the horizon function $h_K (x)$, done in similar way as in section \ref{localization} of chapter \ref{chapgribovtoGZ}, one finds exactly the same local function. This can be understood as we have neglected the total derivatives. In conclusion, although the local actions derived from $h(x)$ and $h_K(x)$ are the same, at the nonlocal level they are clearly different, which also follows from \cite{Kondo:2009wk}. This is important when one is doing manipulations at the level of the nonlocal action as has been done in \cite{Kondo:2009wk}.\\
\\
In this section, we shall therefore shown that there is only \textit{one well defined horizon function}, namely $h(x)$ as defined in \eqref{onlycorrect}. We shall prove this in two different ways. This part is based on \cite{Dudal:2010fq}.

\subsection{The composite operators, part I}
We shall provide a strong argument which illustrates that only the horizon function $h(x)$ as defined in \eqref{onlycorrect} possesses a clear meaning at the quantum level. For this, we first need to demonstrate the following equality
\begin{eqnarray} \label{head}
\Braket{ f(A) (\mathcal M^{-1})^{ab}(x,y)  } &=& - \Braket{ f(A) c^a (x) \overline c^b(y)  } =  \frac{1}{N_c}\Braket{ f(A) \omega^a_i (x) \overline \omega^b_i(y)  }\;,
\end{eqnarray}
where $f(A)$ stands for an arbitrary quantity depending on the gauge fields $A^a_\mu$, and $N_c$ is the number of colors.
Let us start with
\begin{eqnarray}
 \Braket{ f(A) c^a (x) \overline c^b(y)  } &=& \int [\d \Phi] f(A) c^a (x) \overline c^b(y) \e^{-S_\GZ}\;.
\end{eqnarray}
We can rewrite this expression by adding corresponding sources for the ghost and the gluon fields,
\begin{equation} \label{eq}
 \Braket{ f(A) c^a (x) \overline c^b(y)  }
 = \left.  f\left(\frac{\delta}{\delta J_A} \right)\frac{ \delta}{\delta J_{\overline c}^b(y)} \frac{ \delta}{\delta J_c^a(x)}  \int [\d \Phi]   \e^{-S_\GZ + S_\sources} \right|_{\sources = 0}\;,
\end{equation}
with
\begin{equation}
S_\sources = \int \d^d x \left\{ \begin{bmatrix}
J_{\omega_i}^a & J_c^a
\end{bmatrix}
\begin{bmatrix}
\omega^a_i \\ c^a
\end{bmatrix}  + \begin{bmatrix}
\overline \omega^a_i & \overline c^a
\end{bmatrix}
\begin{bmatrix}
\overline J_{\overline \omega_i}^a \\ \overline J_{\overline c}^a
\end{bmatrix}   + J_A A\right\}\;,
\end{equation}
and with the full expression for $S_\GZ$ given in \eqref{SGZecht}. We can perform the integration over the ghosts $c$, $\overline c$, $\omega$, $\overline \omega$ as this is just a Gaussian integration. The relevant piece of the action is given by
\begin{equation}
S_\GZ = \int \d^d x \begin{bmatrix}
\overline \omega_i^a & \overline c^a
\end{bmatrix}\underbrace{
\begin{bmatrix}
 \mathcal M^{ab} & g f_{ak\ell} \p_\mu ( \varphi_i^\ell D_\mu^{kb} ) \\
0 &  -\mathcal M^{ab}
\end{bmatrix}}_{K^{ab}}
\begin{bmatrix}
\omega_i^c  \\ c^d
\end{bmatrix}  + \ldots \;.
\end{equation}
Making use of the formula \eqref{alg1} gives
\begin{multline}
 \Braket{ f(A) c^a (x) \overline c^b(y)  } \\=   f\left(\frac{\delta}{\delta J_A} \right)\frac{ \delta}{\delta J_{\overline c}^b(y)} \frac{ \delta}{\delta J_c^a(x)}
  \left.\int [\d \Phi]   \exp \left\{  \begin{bmatrix}
J_{\omega_i}^a & J_c^a
\end{bmatrix}  (K^{-1})^{ab}   \begin{bmatrix}
\overline J_{\overline \omega_i}^b \\ \overline J_{\overline c}^b
\end{bmatrix}  + \ldots \right\}  \right|_{\sources = 0} \;.
\end{multline}
The matrix $(K^{-1})^{ab}$ can be computed as
\begin{eqnarray}
(K^{-1})^{ab} &=&
 \begin{bmatrix}
(\mathcal M^{-1} )^{ab} & \chi\\
0 & -(\mathcal M^{-1} )^{ab}
\end{bmatrix}\;,
\end{eqnarray}
whereby $\chi$ is some function of the relevant fields. From \eqref{eq}, we now see that
\begin{eqnarray}
 \Braket{ f(A) c^a (x) \overline c^b(y)  } &=&   - \int [\d \Phi] f(A) (\mathcal M^{-1} )^{ab} \e^{-S_\GZ }\;,
\end{eqnarray}
while
\begin{multline}
\Braket{ f(A) c^a (x) \overline c^b(y)  } = - \frac{1}{N_c} f\left(\frac{\delta}{\delta J_A} \right)\frac{ \delta}{\delta J_{\overline \omega_i}^b(y) } \frac{ \delta}{\delta J_{\omega_i}^a(x)} \\\left.    \int [\d \Phi]    \exp \left\{  \begin{bmatrix} J_{\omega_i}^a & J_c^a \end{bmatrix}  (K^{-1})^{ab}   \begin{bmatrix}
\overline J_{\overline \omega_i}^b \\ \overline J_{\overline c}^b
\end{bmatrix}  + \ldots \right\}  \right|_{\sources = 0}  = -\frac{1}{N_c}\Braket{ f(A) \omega_i^a (x) \overline \omega_i^b(y)  }\;,
\end{multline}
which proves relation \eqref{head}.

\subsubsection{The horizon function $h(x)$}
We shall now prove that the expectation value $\Braket{h(x)}$, see equation \eqref{onlycorrect}, is renormalizable. We can apply the formula \eqref{head} to the horizon condition $\Braket{h(x) }$, yielding
\begin{eqnarray}\label{ident}
\int \d^d x \Braket{h(x)} &=& \lim_{\theta \to 0} \int \d^d x \int \d^d y  \Braket{ \left( D_\mu^{ac}(x) \gamma^2(x) \right)  (\mathcal M^{-1})^{ab}(x,y) \left( D_\mu^{bc} (y) \gamma^2(y) \right) }\nonumber  \\
&=& \frac{1}{N_c}\Braket{ \left( D_\mu^{ac}(x) \gamma^2(x) \right) \omega_i^a (x) \overline \omega_i^b (y)  \left( D_\mu^{bc} (y) \gamma^2(y)\right) } \nonumber  \\
&=& \frac{1}{N_c}\gamma^4 \int \d^d x \int \d^d y \braket{ ( D_\mu  \omega_i)^a (x) ( D_\mu \overline \omega_i)^a (y) } \;.
\end{eqnarray}
In order to check whether this horizon term $h(x)$ is well defined, the following correlator $\braket{ ( D_\mu  \omega_i)^a (x) ( D_\mu \overline \omega_i)^a (y) }$ including the composite operators $( D_\mu  \omega_i)^a(x)$ and $( D_\mu \overline \omega_i)^a(y)$ should be renormalizable. In fact, this turns out to be the case. We can prove this easily from the  algebraic renormalization of the Gribov-Zwanziger action given in section \ref{algebraicrenormGZ} of chapter \ref{chapgribovtoGZ}. Indeed, one can immediately obtain the correlator $\braket{ ( D_\mu  \omega_i)^a (x) ( D_\mu \overline \omega_i)^a (y) }$ by deriving the action $\Sigma_\GZ$ \eqref{brstinvariant} with respect to the sources\footnote{We recall here that in order to consistently discuss composite operators at the quantum level, they need to be introduced into the theory by means of suitable sources.} $N^{a}_{i}(x)$ and $U^{a}_{i}(y)$,
\begin{eqnarray}\label{aa}
&&\int [\d \Phi] \left. \frac{\delta }{\delta N^{a}_{i}(x) }  \frac{\delta }{\delta U^{a}_{i}(y) } \e^{- \Sigma_\GZ} \right|_{\text{all sources = 0}} \nonumber\\
&& \hspace{2cm} =  \int [\d \Phi] \left[ - g f^{abc} (D_\mu c)^b (x) \varphi_i^c (x) +  ( D \omega_i)^a(x) \right] ( D \overline \omega_i)^a(y)      \e^{- \Sigma_\GZ} \nonumber\\
&& \hspace{2cm} =  \Braket{ - g f^{abc} (D_\mu c)^b (x) \varphi_i^c (x)  ( D \overline \omega_i)^a(y)  }  +   \Braket{( D \omega_i)^a(x)  ( D \overline \omega_i)^a(y)  } \;.
\end{eqnarray}
We shall now show  that the first correlator in expression \eqref{aa} vanishes,
\begin{eqnarray}\label{green}
\Braket{  (g f^{abc} (D_\mu c)^b  \varphi_i^c) (x)  ( D \overline \omega_i)^a(y)  } &=& 0\;.
\end{eqnarray}
In fact, equation \eqref{green} belongs to a more general class of Green functions which are all zero, namely
\begin{eqnarray}\label{green2}
\Braket{  \Theta (x)  \Lambda(y)  } &=& 0\;,
\end{eqnarray}
with $ \Theta (x) $ a function of fields \textit{not} containing the field $\omega$, while $\Lambda(y)$ is a function containing $\overline \omega$. We can show that all these Green functions are zero by using an elementary  diagrammatical argument. It is impossible to construct any diagram which has an $\overline \omega$ leg starting from a space time point $y$ which has to be connected in some way to a space time point $x$, where no $\omega$ leg is present. Indeed, every $\overline \omega$ requires an $\omega$ leg to propagate, $\omega$ in his turn shall always produce another $\overline \omega$ leg in all vertices as can be seen from the action \eqref{SGZecht}.  Moreover,  the field $\overline \omega$ needs again another $\omega$ leg to propagate. Therefore, an $\omega$ leg is required in the space time point $y$ to close the diagram.  As a consequence, all Green functions of the type of equation \eqref{green2} are zero.\\
\\
Hence, we conclude that the Green function $\Braket{( D \omega_i)^a(x)  ( D \overline \omega_i)^a(y)  } $ is multiplicatively renormalizable as follows from
\begin{equation}\label{reshor}
  \Braket{( D \omega_i)^a(x)  ( D \overline \omega_i)^a(y)  }_0 = Z_U^{-1} Z_N^{-1}   \Braket{( D \omega_i)^a(x)  ( D \overline \omega_i)^a(y)  } \;,
\end{equation}
whereby
\begin{eqnarray}
 Z_U^{-1} Z_N^{-1}  &=& Z_A^{1/2}  Z_g=Z_c^{-1} \;,
\end{eqnarray}
see expression \eqref{Z3}.

\subsubsection{The horizon function $h_K$}
We shall now prove that the horizon function \eqref{hkondo} implies a horizon condition which is not multiplicatively renormalizable. In an analogous fashion as in the previous subsection, we can write
\begin{eqnarray}
\int \d^d x \Braket{h_K(x)} &=&   \gamma^4  \int \d^d x \int \d^d y \Braket{ g^2 f^{akc} A^k_\mu(x) (\mathcal M^{-1})^{ab} (x,y) f^{b\ell c} A^\ell_\mu (y) } \nonumber\\
&=&  - \frac{1}{N_c}\gamma^4  \int \d^d x \int \d^d y  \Braket{ (g f^{akc} A^k_\mu \omega^a_i )(x)  (g  f^{b\ell c} A^\ell_\mu \overline \omega^b_i)  (y) } \;.
\end{eqnarray}
We can demonstrate that the composite operators $g f^{akc} A^k_\mu \omega^a_i $ and $g  f^{b\ell c} A^\ell_\mu \overline \omega^b_i$ are not renormalizable. This is due to the fact that, at the quantum level, those composite operators will unavoidably mix with the operators $\p_\mu \omega^{a}_i$, $\p_\mu \overline\omega^{a}_i$, which have the same quantum numbers. For this, we need to consider the operators $\p_\mu \omega^{a}_i$ and $g f_{akb} A^k_\mu \omega^{b}_i$  as separate operators, each coupled to their own source, instead of $D_\mu^{ab} \omega_i^b$ being coupled to the single source $U_\mu^{a i}$, and similarly for $\p_\mu\overline \omega^{a}_i$ and $g f_{akb} A^k \overline \omega^{b}_i$. In Appendix \ref{appendixalternative}, we have given an alternative proof of the renormalizability of the Gribov-Zwanziger action, where we have coupled the following sources to the following operators, see equation \eqref{s2},
\begin{eqnarray}
   \p_\mu \omega_i^a &\rightarrow& U_\mu^{ai} \;, \nonumber\\
     g f^{akb} A_\mu^k \omega^{b}_i-g f_{abc} D_\mu^{bd }c^d \varphi_i^c & \rightarrow&  U_\mu^{\prime ai} \;,\nonumber\\
     \p_\mu \overline \omega_i^a  & \rightarrow& -N_\mu^{ai} \;,  \nonumber\\
      g f^{akb} A_\mu^k \overline \omega^{b}_i  & \rightarrow&   - N^{\prime a i}_{\mu} \;.
\end{eqnarray}
In \eqref{zmatrix1}, we have found that the sources $U$ and $U'$ mix, as well as  $N$ mixes with $N'$. Taking the inverse of this matrix yields
\begin{align}
 \left[
  \begin{array}{c}
    U \\
    U' \\
  \end{array}
\right]&=\left[
          \begin{array}{cc}
               Z_A^{-1/2} & a_1 \\
            0 & Z_g^{1}   \\
          \end{array}
        \right]
\left[
  \begin{array}{c}
    U_0\\
    U'_0
  \end{array}
\right]\,, \nonumber\\
 \left[
  \begin{array}{c}
    N \\
    N' \\
  \end{array}
\right]&=\left[
          \begin{array}{cc}
              Z_g^{-1}  & a_1 \\
            0 &  Z_A^{1/2}  \\
          \end{array}
        \right]
\left[
  \begin{array}{c}
    N_0\\
    N'_0
  \end{array}
\right]\,.
\end{align}
We recall that insertions of an operator can be obtained by taking derivatives of the generating functional $Z^c(U,U',N,N')$ w.r.t.~to the appropriate source. For example,
\begin{equation*}
(g f^{akb} A_\mu^k \overline \omega^{b}_i)_0 \sim \frac{\delta Z^c((U,U',N,N')}{\delta N'_0} = \frac{\delta N}{\delta N'_0}\frac{\delta Z^c(U,U',N,N') }{\delta N}+\frac{\delta N'}{\delta N'_0}\frac{\delta Z^c(U,U',N,N')}{\delta N }\;,
\end{equation*}
so that
\begin{eqnarray}\label{vgl1}
    (g f^{akb} A_\mu^k \overline \omega^{b}_i)_0  &=& a_1 ( \p_\mu \overline \omega^{a}_i ) + Z_A^{1/2} (g f^{akb} A_\mu^k \overline \omega^{b}_i)\;.
\end{eqnarray}
We can do the same for the other operator
\begin{equation}\label{vgl2}
     (g f^{akb} A_\mu^k \omega^{b}_i-g f_{abc} D_\mu^{bd }c^d \varphi_i^c)_0 = a_1 ( \p_\mu  \omega^{a}_i ) + Z_g (g f^{akb} A_\mu^k \omega^{b}_i - g f_{abc} D_\mu^{bd }c^d \varphi^c_i)\;.
\end{equation}
Taking the partial derivatives of $\Sigma^2_\GZ$ given in equation \eqref{B1} w.r.t.~$ N^{\prime a}_{i}(x)$ and $U^{\prime a}_{i}(y)$, we find
\begin{eqnarray*}\label{2terms}
&&\int [\d \Phi] \left. \frac{\delta }{\delta N^{\prime a}_{i}(x) }  \frac{\delta }{\delta U^{\prime a}_{i}(y) } \e^{- \Sigma_\GZ} \right|_{\text{all sources = 0}} \nonumber\\
&&\hspace{2cm} =  \int [\d \Phi] \left(g f^{akb} A_\mu^k \omega^{b}_i-g f_{abc} D_\mu^{bd }c^d \varphi_i^c \right)(x) (g f^{a\ell c} A_\mu^\ell \overline \omega^{ c }_i)(y)      \e^{- \Sigma_\GZ} \nonumber\\
&&\hspace{2cm} =  \Braket{ (g f^{akb} A_\mu^k \omega^{b}_i) (x)  (g f^{a\ell c} A_\mu^\ell \overline \omega^{c }_i)(y) } -   \Braket{ (g f_{abc} D_\mu^{bd }c^d \varphi_i^c) (x)  (g f^{a\ell c} A_\mu^\ell \overline \omega^{c}_i)(y) }\;.
\end{eqnarray*}
The correlator $\Braket{ (g f_{abc} D_\mu^{bd }c^d \varphi_i^c) (x)  (g f^{akb} A_\mu^k \overline \omega^{b}_i)(y) }$ also belongs to the class \eqref{green2} and is therefore equal to zero. However, the remaining Green function \\$\Braket{ (g f^{akb} A_\mu^k \omega^{b}_i) (x)  (g f^{a\ell c} A_\mu^\ell \overline \omega^{c}_i)(y) }$ is no longer multiplicatively renormalizable. Indeed, due to the mixing we have found, the bare correlator can be written as follows
\begin{multline}\label{renorm}
\Braket{ (g f^{akb} A_\mu^k \omega^{b}_i) (x)  (g f^{a\ell c} A_\mu^\ell \overline \omega^{c}_i)(y) }_0 = a_1 Z_A^{1/2} \Braket{ (\p_\mu \omega^{a}_i) (x)  (g f^{a\ell c} A_\mu^\ell \overline \omega^{ c}_i)(y) } \\
+ a_1 Z_g \Braket{ (g f^{akb} A_\mu^k \omega^{b}_i) (x)  (\p_\mu \overline \omega^{a}_i)(y) }
+ Z_g Z_A^{1/2} \Braket{ (g f^{akb} A_\mu^k \omega^{b}_i) (x)  (g f^{a\ell c} A_\mu^\ell \overline \omega^{ c}_i)(y) }\;.
\end{multline}
This is a strong argument why one should not rely on this horizon function $h_K$. When renormalizing this Green function, the ``missing" terms stemming from the covariant derivative re-enter again.

\subsection{The composite operators, part II}
\subsubsection{The horizon function $h (x)$}
Let us now look at the composite operators which are involved in the horizon term $h(x)$ \eqref{onlycorrect} after having performed the necessary localization, see section \ref{localization} of chapter \ref{chapgribovtoGZ}. In particular, we are interested in the composite operators appearing in expression \eqref{xhor1a} before setting $\gamma(x)$ equal to a constant, namely $D_\mu^{bc} \varphi^{cb}_{\mu}$ and $D_\mu^{bc} \overline{\varphi}^{bc}_{\mu}$. If the horizon function is well defined, we expect these composite operators to be renormalizable. We can check this again from the algebraic renormalization in section \ref{algebraicrenormGZ} of the previous chapter. Firstly, from \eqref{brstinvariant} we can deduce
\begin{eqnarray}
(D_\mu^{ab} \varphi^{b}_i)_0 &=& Z_M^{-1}  (D_\mu^{ab} \varphi^{b}_i ) = Z_g^{1/2} Z_A^{1/4} (D_\mu^{ab} \varphi^{b}_i ) \;.
\end{eqnarray}
Secondly, we can do something analogous for $D_\mu^{ab} \overline \varphi^{b}_i$. We see that this operator is a linear combination of two renormalizable composite operators, namely the composite operator coupled to $V_\mu^{ai}$, i.e.~$-D_\mu^{ab} \overline \varphi^b_i + g f_{abc} D^{bd}_\mu c^d \overline \omega^c_i$ and to $R_\mu^{ai}$, i.e.~$- g f_{abc} D^{bd}_\mu c^d \overline \omega^c_i$. Luckily, the two composite operators have the same renormalization constant, as without this property, the linear combination would not be renormalizable. Therefore,
\begin{equation*}
(D_\mu^{ab} \overline \varphi^b_i)_0   = - (-D_\mu^{ab} \overline \varphi^b_i + g f_{abc} D^{bd}_\mu c^d \overline \omega^c_i)_0 - (- g f_{abc} D^{bd}_\mu c^d \overline \omega^c_i)_0 = Z_V^{-1} (D_\mu^{ab} \overline \varphi^b_i)\;,
\end{equation*}
whereby $Z_V^{-1} = Z_g^{1/2}Z_A^{1/4}$ as can be found in \eqref{Z3}. In conclusion, both composite operators $D_\mu^{bc} \varphi^{cb}_{\mu}$ and $D_\mu^{bc} \overline{\varphi}^{bc}_{\mu}$ are renormalizable.

\subsubsection{The horizon function $h_K(x)$}
When taking the limit $\theta(x) \to 0$, we can drop the total derivatives. However, the remaining composite operators $g f_{akb} A^k \varphi^{b}_i$ and $g f_{akb} A^k \overline \varphi^{bi}$ are not multiplicative renormalizable. We can prove this in an analogous fashion as in the previous paragraph. In Appendix \ref{appendixalternative}, we have considered the operators $\p_\mu \varphi^{a}_i$ and $g f_{akb} A^k_\mu \varphi^{b}_i$  as separate operators instead of being coupled only to one source, and similarly for $\p_\mu\overline \varphi^{a}_i$ and $g f_{akb} A^k \overline \varphi^{b}_i$.  To that purpose, in the Appendix \ref{appendixalternative}, we have coupled the following sources to the following operators, see equation \eqref{s2}:
\begin{eqnarray}
   \p_\mu  \varphi^{a}_i &\rightarrow& M_{\mu}^{ai} \;,\nonumber\\
    g f^{akb} A_\mu^k \varphi^{b}_i & \rightarrow& M^{\prime ai}_{\mu} \;,\nonumber\\
     \p_\mu \overline \varphi^{a}_i  & \rightarrow& V_{\mu}^{ai} \;,\nonumber\\
       (g f_{akb} A_\mu^{k} \overline \varphi^{b}_i - g f_{abc} D^{bd}_\mu c^d \overline \omega^c_i)  & \rightarrow&  V_{\mu}^{\prime ai}\;,
\end{eqnarray}
and we have reworked out the complete algebraic renormalization in this alternative setting. In \eqref{zmatrix1}, we have found that the sources $M$ and $M'$ mix, and also $V$ mixes with $V'$. Taking the inverse of this matrix yields
\begin{align}
 \left[
  \begin{array}{c}
    M \\
    M' \\
  \end{array}
\right]&=\left[
          \begin{array}{cc}
              Z_g^{-1/2} Z_A^{-1/4} & a_1 \\
            0 & Z_g^{1/2} Z_A^{1/4}  \\
          \end{array}
        \right]
\left[
  \begin{array}{c}
    M_0\\
    M'_0
  \end{array}
\right]\,,  \nonumber\\
 \left[
  \begin{array}{c}
    V\\
    V' \\
  \end{array}
\right]&=\left[
         \begin{array}{cc}
              Z_g^{-1/2} Z_A^{-1/4} & a_1 \\
            0 & Z_g^{1/2} Z_A^{1/4}  \\
          \end{array}
        \right]
\left[
  \begin{array}{c}
    V_0\\
    V'_0
  \end{array}
\right]\,.
\end{align}
As in the previous section we have that
\begin{equation}
    (g f^{akb} A_\mu^k \varphi^{b}_i)_0 = a_1 ( \p_\mu  \varphi^{a}_i ) + Z_g^{-1/2} Z_A^{-1/4} (g f^{akb} A_\mu^k \varphi^{b}_i)\;.
\end{equation}
We can do the same for the other operator. Only some care has to be taken as the source $V_\mu^{\prime ai}$ couples to a sum of two operators $ (g f_{akb} A_\mu^{k} \overline \varphi^{b}_i - g f_{abc} D^{bd}_\mu c^d \overline \omega^c_i) $. We therefore have to subtract the operator coupled to the source $R_\mu^{ai}$, namely
\begin{eqnarray}
(  g f_{akb} A_\mu^{k} \overline \varphi^{b}_i)_0 &=&(g f_{akb} A_\mu^{k} \overline \varphi^{b}_i - g f_{abc} D^{bd}_\mu c^d \overline \omega^c_i)_0 - (- g f_{abc} D^{bd}_\mu c^d \overline \omega^c_i)_0 \nonumber\\
  &=& a_1 ( \p_\mu  \overline \varphi_i^{a} ) + Z_g^{-1/2} Z_A^{-1/4} (g f^{akb} A_\mu^k \overline \varphi^{b}_i  - g f_{abc} D^{bd}_\mu c^d \overline \omega^c_i)\nonumber\\
   &&\hspace{2cm}-  Z_g^{1/2} Z_A^{1/4}(- g f_{abc} D^{bd}_\mu c^d \overline \omega^c_i) \nonumber\\
  &=& a_1 ( \p_\mu  \overline \varphi_i^{a} ) + Z_g^{-1/2} Z_A^{-1/4} (g f^{akb} A_\mu^k \overline \varphi^{b}_i  ) \;.
\end{eqnarray}
We can thus conclude that the operators $g f_{akb} A^k \varphi^{b}_i$ and $g f_{akb} A^k \overline \varphi^{b}_i$ are not multiplicatively renormalizable and that they mix with the operators $\p_\mu \varphi^{a}_i$ and $\p_\mu\overline \varphi^{a}_i$, respectively. Therefore, one should keep in mind that the limit $\theta \to 0$ has to be taken as the final step, and that it can only be taken at the local level. Furthermore, the mixing we have found, tells us that one should always leave the covariant derivative of a field ``in one piece''. As a consequence, much care has to be taken at the nonlocal level when deriving all kinds of results.

\section{The GZ action and its relation to the Kugo-Ojima confinement criterium}
\subsection{Introduction: the Kugo-Ojima criterium}
Some important results concerning a possible origin of confinement, were given in \cite{Kugo:1979gm,Kugo:1995km}. In these papers, confinement was related to the enhancement of the ghost propagator.\\
\\
Let us explain this a bit more in detail. The whole starting point of the analysis by Kugo and Ojima is a well defined nilpotent (BRST) symmetry $Q_B$ and a ghost charge. In their analysis, they present two results. Firstly, they have shown that when having a well defined nilpotent symmetry, it is possible to show that all unphysical states, see p.\pageref{globalmark} shall form so-called quartets \cite{Kugo:1979gm} and decouple from the physical spectrum. In this way, only physical states, which are closed under the symmetry $Q_B$ but not exact, survive. In this way they have proven that the longitudinal and temporal gauge polarization, the ghost and the antighost fields can be excluded from the physical spectrum. In fact, this idea is very general, and can be applied when you have a system with a nilpotent symmetry $s$.\\
\\
Secondly, for the Faddeev-Popov action, they also showed the following. Using the equation of motion for the gluon field, the conserved global color current can be written as
\begin{equation}
J_\mu^a = \p_\mu F^a_{\mu\nu} + \{ Q_B , D_\mu^{ab} \overline c^b \} \;,
\end{equation}
and thus the charge is given by the integrated zero component of $J_\mu^a$, i.e.
\begin{equation}
Q^a = \int \d^3 x \left( \p_i F^a_{0i} + \{ Q_B, D_0^{ab} \overline c^b \} \right) \;.
\end{equation}
Now there are two criteria which need to be satisfied in order to have color confinement. The first criterium states that the gluon propagator cannot have massless poles, so the  first term of the previous expression vanishes as it is integral over a total derivative\footnote{If the gluon propagator has massless poles, the first term is ill-defined.}. The second criterium is that $\{ Q_B, D_0^{ab} \overline c^b \} $ is well defined, which is the case when
\begin{equation}\label{KO0}
 u(0)=-1\;,
\end{equation}
with $u(p^2)$ defined through the following Green function\footnote{Strictly speaking, the KO analysis is done in Minkowski space. We shall however, as any functional or lattice approach, consider the corresponding operator in Euclidean space.},
\begin{equation}\label{KO1}
    \int \d^dx \e^{\ii px}\Braket{D_\mu^{ad}c^d(x) D_\nu^{be}\overline c^e(0)}_{\FP}= \left(\left(\delta_{\mu\nu}-\frac{p_\mu p_\nu}{p^2}\right)u(p^2)-\frac{p_\mu p_\nu}{p^2}\right)\delta^{ab}\;.
\end{equation}
$\braket{O}_{\FP}$ stands for the expectation value taken with the Faddeev-Popov action. If the two criteria are met, $Q^a$ is well defined so we have that $Q^a \ket{\psi}_\phys = 0$ and color confinement is guaranteed.\\
\\
The second criterium can be connected to the ghost propagator. Indeed, in \cite{Kugo:1995km}, it was shown that one can parameterize the ghost propagator, defined as follows
\begin{eqnarray}\label{KO5}
\Braket{c^a(-p) \overline{c}^b(p)}&=\delta^{ab} G (p^2)\;,
\end{eqnarray}
in terms of\footnote{Actually, in \cite{Kugo:1995km}, another notation $v(p^2)$ has been used instead of $w(p^2)$, the relation being  $ w(p^2) = p^2v(p^2)$.}
\begin{eqnarray}\label{KO6}
G (p^2) &=& \frac{1}{p^2(1 + u(p^2) + w (p^2))}\;.
\end{eqnarray}
This relation was also discussed in \cite{Zwanziger:1992qr,Kondo:2009wk,Kondo:2009ug,Aguilar:2009nf,Boucaud:2009sd}. It is usually assumed that\footnote{$w (p^2) = 0$ has been checked up to two loops, see \cite{Gracey:2005cx}.} $w (p^2) = 0$, so that $u = -1$ implies an enhanced ghost propagator. Notice that this scenario is exactly predicted by the GZ framework.

\subsection{Important criticisms towards the KO criteria}
Although the argument of Kugo and Ojima sounds very attractive, two comments are in order. First of all, in the KO framework \cite{Kugo:1979gm,Kugo:1995km}, the existence of a globally well-defined BRST charge is assumed. Thus, the issue of the (non)existence of a nonperturbatively valid BRST symmetry is not explicitly faced. Secondly, the derivation of the two criteria was done by employing the usual Faddeev-Popov gauge fixed action. As such, the Gribov problem is  simply not addressed.\\
\\
Taking these two criticisms into account, the KO cannot really hold for the GZ action, as (1) the GZ breaks the BRST symmetry, see equation \eqref{deltabrst} and (2) the GZ action does take into account Gribov copies.\\
\\
Another point which should be mentioned is that Kugo and Ojima did not impose the criterium \eqref{KO0}, but they derived it as a condition to be checked/calculated. Though, nowadays, in functional formalisms as in \cite{Fischer:2008uz}, the criterion is used as an input. The ghost enhancement is often imposed as a boundary condition in order to favor the so-called scaling type solution of the Schwinger-Dyson and/or Functional Renormalization Group equations \cite{Fischer:2008uz}.

\subsection{Imposing $u(0)=-1$ as a boundary condition in the Faddeev-Popov action}
Now let us try the following \cite{Dudal:2009xh}. What if we impose the constraint $u(0)=-1$  directly into the Faddeev-Popov theory, by appropriately modifying the measure one starts from? We shall see in fact that the resulting action will be exactly the same as the GZ action.

\subsubsection{Microcanonical ensemble and equivalence with the canonical Boltzmann ensemble in the thermodynamic limit}
We shall first give an overview of some results from thermodynamics we intend to employ, see also section \ref{secttheGZaction} of chapter \ref{chapgribovtoGZ}.\\
\\
We consider a discrete system, whose Hamiltonian is $H(q,p)$, with $3N$ degrees of freedom. The averages in the microcanonical ensemble
are constructed out of
\begin{equation*}
\Sigma (E)=\int_{H=E}\d\mu \ =\int d\mu \;\delta (E-H)\;,
\end{equation*}
where $\d\mu =\d^{3N} q \d^{3N} p$ represents the classical phase space where $E$ stands for the constant energy of the system. Averages in the microcanonical ensemble are defined by $\braket{O}_{\mathrm{Micr}}=\frac{\int_{H=E}d\mu \ O}{\int_{H=E}d\mu }$. In order to establish the equivalence between the microcanonical and the (Boltzmann) canonical ensemble we rewrite the quantity $\Sigma (E)$ in the
following form
\begin{equation*}
\Sigma (E) =\int \d\mu \ \delta (E-H)=\int \d\mu \int_{-i\infty +\varepsilon }^{i\infty +\varepsilon }\frac{\d\beta }{2\pi i}\e^{\beta (E-H)}   =\int \frac{d\beta }{2\pi i}f(\beta )\ =\int \frac{d\beta }{2\pi i} \e^{-\omega (\beta )}\;,
\end{equation*}
whereby
\begin{equation}\label{dd5}
f(\beta )=\int \d\mu \e^{\left( \beta (E-H)\right) }\;, \qquad \omega (\beta)=-\ln f(\beta )\;.
\end{equation}
It can be shown  that, in the thermodynamic limit, $N,V\rightarrow \infty $, with $N/V$ fixed, the saddle point approximation becomes exact. We refer to \cite{munster} for an overview of the proof. So,
\begin{equation}\label{d8}
\Sigma (E)=\frac{1}{2\pi i}f(\beta ^{\star })\;,\text{with}\;\omega ^{\prime}(\beta ^{\star })=\frac{f^{\prime }(\beta ^{\star })}{f(\beta ^{\star })}~=~0\;.
\end{equation}
From equation \eqref{d8} it follows that
\begin{equation}\label{d10}
E=\braket{H}_{\mathrm{Boltz}}=\frac{\int d\mu ~H \e^{-\beta ^{\star }H}}{\int d\mu ~\e^{-\beta ^{\star }H}}\;.
\end{equation}
This is the gap equation determining the critical parameter $\beta ^{\star }$. Analogously, it can also be shown that \cite{munster} $\braket{O}_{\mathrm{Micr}}=\braket{O}_{\mathrm{Boltz}}=\frac{\int d\mu ~O\ \e^{-\beta ^{\star }H}}{\int d\mu~\e^{-\beta ^{\star }H}}$ for the average of any quantity $O(q,p)$.

\subsubsection{Imposing the KO criterion yields the GZ framework}
Starting from \eqref{KO1} and performing Lorentz and color contractions and taking the $p\rightarrow 0$ limit, we can write
\begin{equation}
(V)^{-1}\int d^d y \int d{^{d}}x\braket{ (D_\mu c)^{a}(x) (D_\mu \overline c)^a (y)}_{\FP} = (N^{2}-1)((d-1)u(0)-1)\;,
\end{equation}
after passing to Euclidean space, as any functional or lattice approach. $V$ denotes the spacetime volume.  Now using the result from equation \eqref{ident} and \eqref{head} from the previous section, we can write the previous expression as
\begin{multline}\label{KO2}
-\frac{1}{V} \frac{1}{\gamma^4} \lim_{\theta \to 0} \int \d^d x \int \d^d y  \Braket{ \left( D_\mu^{ac}(x) \gamma^2(x) \right)  (\mathcal M^{-1})^{ab}(x,y) \left( D_\mu^{bc} (y) \gamma^2(y) \right) }
 \\= (N^{2}-1)((d-1)u(0)-1)\;,
\end{multline}
whereby $\gamma^2(x)$ is defined in \eqref{gammadefi}. In order to get rid of the parameter $\gamma$, we rewrite \eqref{KO2}
\begin{multline}
-\frac{1}{V}  \lim_{\theta \to 0} \int \d^d x \int \d^d y  \Braket{ \left( D_\mu^{ac}(x) \e^{\ii \theta x} \right)  (\mathcal M^{-1})^{ab}(x,y) \left( D_\mu^{bc} (y) \e^{\ii \theta y} \right) } \\= (N^{2}-1)((d-1)u(0)-1)\;,
\end{multline}
In fact, the l.h.s.~of this expression contains the Zwanziger horizon function $h'(x)$, see equation \eqref{onlycorrect2}, so we can write
\begin{equation}
-\Braket{h'(x)} = (N^{2}-1)((d-1)u(0)-1)\;,
\end{equation}
We observe that the KO condition cannot be realized with the standard Faddeev-Popov measure $\d \mu _{\FP}$, otherwise we would have
\begin{equation}\label{ex1}
\braket{h'(x)}_{\FP}=d(N^{2}-1) \Leftrightarrow  \braket{h(x)}_{\FP}= \gamma^4 d(N^{2}-1) \;,
\end{equation}
which would contradict Zwanziger's result \eqref{horizoncondition}. We now implement the KO criterion $u(0)=-1$ as a boundary condition, amounting to start from the modified measure
\begin{equation}\label{ex2}
\d\mu _{\FP}   \to  \d\mu^\prime\equiv \d\mu _{\FP}\ \delta \left(Vd(N^{2}-1)-\int \d^d x h'(x)\right) \ ,
\end{equation}
which clearly implements $u(0)=-1$, yielding in fact
\begin{equation}\label{ex3}
\braket{h'(x)} =d(N^{2}-1)\ .
\end{equation}
We are thus led to consider the partition function
\begin{eqnarray}\label{d20}
\int \d \mu^{\,\prime} &=& \int \d\mu _{\FP}\ \delta \left( Vd(N^2 - 1 )-\int \d^d x h'(x)\right) \nonumber \\
&=& \int \d A\ \delta (\partial A)\ \det \mathcal{M}\ \e^{-S_{\mathrm{YM}}}\ \delta \left( Vd(N^2 - 1 )-\int \d^d x h'(x)\right) \nonumber \\
&=&\int \d\Phi \delta \left( V d(N^2-1)-\int \d^d x h'(x)\right) \e^{-S_{\FP}}\;.
\end{eqnarray}
Expression \eqref{d20} defines a microcanonical ensemble. Since we are working in a continuum field theory, we are working in
the thermodynamic limit, hence we have an equivalence with a Boltzmann canonical ensemble as outlined in the previous section. Using analogous
arguments as there, we arrive at
\begin{equation}\label{d21}
\int \d\mu^{\,\prime} =\int \d\mu _{\FP} \e^{\gamma^{4}\int \d^d x h'(x)}  \equiv\int \d\mu _{\FP}\ \e^{-S_\mathrm{H}} \;,
\end{equation}
where the mass parameter $\gamma$ follows from the gap equation
\begin{equation}
d \left( N^{2}-1\right) =\braket{h'(x)}_{\Boltz} = \frac{\int \d\mu _{\FP}\ \e^{-S_{\mathrm{H}}}\ h'(x)}{\int \d\mu _{\FP} \ \e^{-S_{\mathrm{H}}}}\;,
\end{equation}
which is the analogue of \eqref{d10}. We conclude that we can consistently encode the boundary condition \eqref{KO0} at the level of the action, which turns out to be identical to the GZ action, equation \eqref{nonlocalGZ}. Of course, we can localize it into the form \eqref{GZstart}.

\subsection{Conclusion}
We can thus conclude the following points
\begin{itemize}
\item Imposing a boundary condition into a theory can have serious consequences and one should really impose them from the beginning, to fully grasp all nontrivial aspects of the boundary. Here we have seen that imposing the boundary condition $u(0) = -1$ in the Faddeev-Popov measure with BRST symmetry, leads us to the GZ action without BRST symmetry. Therefore, imposing a boundary condition can change the symmetry content of a theory.
\item In the GZ formalism, the meaning of the KO criterium becomes unclear. This is due to the fact that the KO analysis was done in the Faddeev-Popov formalism, and not in the GZ formalism. Due to the breaking of the BRST symmetry in the GZ action, one cannot simply redo the KO analysis.
\end{itemize}
Perhaps another point which indicates that that KO confinement scenario is not the answer to confinement, is that the lattice data do not support an enhanced ghost anymore. This shall be the topic of the next chapter.


\chapter{The Refined Gribov-Zwanziger action \label{refined} }

\section{Introduction}
\subsection{The lattice results on the ghost and gluon propagator\label{chap2lattice}}
During the last 2 decades, there has been an intensive discussion about two particular quantities, i.e.~the gluon and the ghost propagator, mainly in the Landau gauge. We recall the following conventions for these propagators
\begin{align}
\Braket{A_\mu^a(k) A_\nu^b (p)} &=\delta^{ab} \delta(k-p) (2\pi)^d \mathcal{D}(p^2)\left(\delta_{\mu\nu}-\frac{p_\mu{p}_\nu}{p^2}\right)\;, \nonumber\\
 \Braket{c^a(k) \overline{c}^b(p)}&=\delta^{ab}\delta(k-p) (2\pi)^d  \mathcal{G}(p^2)\;.
\end{align}
For a long time, lattice results in 4, 3 and 2 dimensions have shown an infrared suppressed, positivity violating gluon propagator which seemed to tend towards zero for zero momentum, i.e.~$\mathcal{D}(0) =0$, and a ghost propagator which was believed to be enhanced in the infrared \cite{Cucchieri:2004mf,Sternbeck:2004xr}, $\mathcal{G}(k^2 \approx 0) \sim 1/k^{2 + \kappa}$ with $\kappa > 0$. Besides the Gribov-Zwanziger framework, also other analytical approaches were in agreement with these results.  For instance, several works based on the Schwinger-Dyson or Exact Renormalization Group equations reported an infrared enhanced ghost propagator and an infrared suppressed, vanishing gluon propagator, obeying a power law behavior characterized by a unique infrared exponent, as stated by a sum rule discussed in \cite{Alkofer:2000wg,Lerche:2002ep,Pawlowski:2003hq,Alkofer:2003jj}.\\
\\
However, in 2007 lattice data in 3d and 4d at larger volumes displayed an infrared suppressed, positivity violating gluon propagator, which is \textit{non-vanishing} at zero momentum, i.e.~$\mathcal{D}(0) \not=0$, and a ghost propagator which is \textit{no longer enhanced}, $\mathcal{G}(k^2 \approx 0) \sim 1/k^{2}$ \cite{Cucchieri:2007rg,Cucchieri:2007md}. This implied that the previous mentioned analytical approaches were not conclusive. Surprisingly, in 2d, the ghost propagator still displays an enhanced behavior while the gluon propagator does vanish at the origin \cite{Cucchieri:2007rg,Cucchieri:2008fc,Maas:2007uv}.\\
\\
After this paper, \cite{Cucchieri:2007md}, an avalanche of papers followed concerning the behavior of the propagators.  In the lattice community, similar results were obtained in \cite{Sternbeck:2007ug,Bogolubsky:2007ud,Bogolubsky:2009dc,Bornyakov:2008yx,Gong:2008td}, and see \cite{Cucchieri:2010xr} for a recent overview. The question which obviously arose, is there also a possible way to explain these results in the Gribov-Zwanziger framework?  This is the topic of this chapter, which is based on the following papers: \cite{Dudal:2007cw,Dudal:2008sp,Dudal:2008rm,Dudal:2008xd,Dudal:2010tf}.\\
\\
For completeness, let us mention that also within Schwinger-Dyson methods, solutions of the ghost and gluon propagator were obtained which were in agreement with the lattice data, see \cite{Binosi:2009qm,Fischer:2008uz,Boucaud:2008ky} and their references within.\\

\subsection{Condensates}
How could we explain the behavior of the gluon and the ghost propagator in the Gribov-Zwanziger framework in 4d and 3d? It seems logical to take into account other nonperturbative effects. A well-known important source of nonperturbative effects in gauge theories are condensates, viz.~vacuum expectation values of certain local operators. Next to the famous gauge invariant condensate $\braket{F_{\mu\nu}^2}$, of paramount importance for phenomenological applications \cite{Shifman:1978bx}, recent years\footnote{The $d=2$ gluon condensate was already considered in \cite{Greensite:1985vq,Lavelle:1988eg}.} have also witnessed an increased interest in the dimension two condensate $\braket{A^2}$ in the Landau gauge \cite{Gubarev:2000eu,Gubarev:2000nz}, and related to it the issue of $1/Q^2$ power corrections \cite{Narison:2005hb}. The latter corrections would correspond to an extension of the usual SVZ sum rule study of physical correlators. Some important early contributions to this field of research can be found in, for example, \cite{Grunberg:1997ud,Akhoury:1997ys,Akhoury:1997by,Gubarev:1998ew,Chetyrkin:1998yr,Zakharov:1999jj,Burgio:1997hc,Bali:1999ai,Chernodub:2000bk}. These works were based on renormalon analysis, lattice considerations of the interquark potential and condensates, nonperturbative short distance physics, $\ldots$.  Also at the propagator level such power corrections were identified in \cite{Boucaud:2000ey}. If we call the minimum of the functional \eqref{functional} $A^2_{\min}$,  it is clear that $\braket{A^2_{\min}}$ is a gauge invariant quantity by construction. This leads very naturally to the introduction of $\braket{A^2}$ in the Landau gauge since we can write \cite{Lavelle:1995ty}
\begin{eqnarray*}
A_{\min }^{2} &=&\frac{1}{2}\int \d^{d}x\left[ A_{\mu }^{a}\left( \delta _{\mu \nu }-\frac{\partial _{\mu }\partial _{\nu }}{\partial ^{2}}\right) A_{\nu
}^{a}-gf^{abc}\left( \frac{\partial _{\nu }}{\partial ^{2}}\partial
A^{a}\right) \left( \frac{1}{\partial ^{2}}\partial {A}^{b}\right)
A_{\nu }^{c}\right] \;+O(A^{4})\,,  \label{min1}
\end{eqnarray*}
from which it easily follows that $\braket{A^2_{\min}}=\braket{A^2}$ in the Landau gauge. This condensate then made its appearance in a variety of works, see e.g.~\cite{Dudal:2005na,Boucaud:2001st,Boucaud:2002nc,Boucaud:2008gn,Verschelde:2001ia,Dudal:2002pq,Dudal:2003vv,Browne:2003uv,Vercauteren:2007gx,Chernodub:2008kf,Dudal:2009tq,Kondo:2001nq,Li:2004te,Gubarev:2005it,Kekez:2005ie,Andreev:2006vy,RuizArriola:2004en,RuizArriola:2006gq,Megias:2009mp}. In the works \cite{Gubarev:2000eu,Gubarev:2000nz}, the relation was explored between this condensate and magnetic degrees of freedom, which are generally believed to play an important role for confinement.  Recently, this was further investigated by looking at the electric and magnetic components of $\braket{A^2}$ at finite temperature, hinting towards an interesting connection with the phase diagram \cite{Chernodub:2008kf,Vercauteren:2010rk,Dudal:2009tq}.\\
\\
Measurements of $\braket{A^2}$ at $T=0$ have been obtained using the lattice gluon propagator and the Operator Product Expansion (OPE) in \cite{Boucaud:2008gn}, based on earlier work \cite{Boucaud:2001st,Boucaud:2002nc}, giving the following estimate
\begin{equation}\label{A3}
\braket{g^2A^2} =5.1^{+0.7}_{-1.1}~\text{GeV}^2 \;,
\end{equation}
at the renormalization scale $\mu=10~\text{GeV}$. $\braket{A^2}$ also appeared as a source of power corrections in e.g.~\cite{RuizArriola:2006gq,Megias:2009mp}. An independent estimate using the OPE and the quark propagator in a quenched lattice simulation gave \cite{RuizArriola:2004en}
\begin{equation}\label{A3bis}
\braket{g^2A^2} =4.4\pm0.4~\text{GeV}^2\,.
\end{equation}
An ab initio calculation of $\braket{A^2}$ was presented in \cite{Verschelde:2001ia,Browne:2003uv}. It was shown that it is possible to construct an effective potential for $\braket{A^2}$ which is consistent with the renormalization (group) \cite{Verschelde:2001ia,Dudal:2002pq}. A non-vanishing condensate due to dimensional transmutation was favored as it lowered the vacuum energy. Using a resummation of Feynman diagrams, more evidence for $\braket{A^2}\neq0$ was given in  \cite{Dudal:2003vv}.

\subsection{The condensate $\Braket{A^2}$ in the GZ action\label{condensateAkwadraat}}
The extension of the effective potential formalism to the Gribov-Zwanziger case was first tackled in \cite{Dudal:2005na}. In this paper, the presence of the condensate $\braket{A^2}$ was investigated and it was shown that this condensate does not spoil the renormalizability of the GZ action. The GZ action with inclusion of the local composite operator $A_\mu^a A_\mu^a$ is given by
\begin{eqnarray}
S_\AGZ &=& S_\GZ + S_{A^2}\;,
\end{eqnarray}
whereby
\begin{eqnarray}
S_{A^2} &=& \int \d^d x \left( \frac{\tau }{2}A_{\mu }^{a}A_{\mu }^{a}-\frac{\zeta }{2}\tau ^{2}\right)\;,
\end{eqnarray}
with $\tau$ a new source invariant under the BRST transformation $s$ and $\zeta$ a new parameter. However, it is easily checked that taking into account this condensate does not alter the qualitative features of the gluon and the ghost propagator, the former one will vanish at zero momentum, while the latter one will still be enhanced.\\
\\
As it will be useful later, let us repeat the algebraic renormalization of the action with the inclusion of the operator $A^2$. This can be done similarly as in section \ref{algebraicrenormGZ} of chapter \ref{chapgribovtoGZ}.

\subsubsection{The starting action and the BRST}
Again, we shall make $S_\AGZ$ BRST invariant. We define
\begin{eqnarray}\label{AGZ}
\Sigma_\AGZ &=& \Sigma'_\GZ + \Sigma_{A^2} \;,
\end{eqnarray}
whereby $\Sigma'_\GZ$ is given in expression \eqref{enlarged} and
\begin{eqnarray}\label{sigmaakwadraat}
\Sigma_{A^2} &=& \int \d^{4}x s\left( \frac{\eta }{2}A_{\mu }^{a}A_{\mu}^{a}-\frac{\zeta }{2}\tau ^{2}\right)~=~ \int \d^4 x \left[\frac{1}{2} \tau A_{\mu }^{a}A_{\mu }^{a}+\eta A_{\mu }^{a}\partial _{\mu }c^{a}-\frac{1}{2}\zeta \tau ^{2} \right] \;,
\end{eqnarray}
with $\eta$ a new source and $s \eta = \tau$, so that $(\eta, \tau)$ forms a doublet. In the end, we replace all the sources with their physical values, see expression \eqref{physlimit} and \eqref{physlimit2}, and in addition
\begin{eqnarray}
 &\left. \eta \right|_{\phys} = 0\;,&
\end{eqnarray}
so one recovers $S_\AGZ$ again.

\subsubsection{The Ward identities}
It is now easily checked that the Ward identities 1-7 on p.\pageref{pagewardidentities} remain preserved. Obviously, the Slavnov-Taylor identity receives an extra term,
\begin{equation}
\mathcal{S}(\Sigma_\AGZ )=0\;,
\end{equation}
whereby
\begin{multline*}
\mathcal{S}(\Sigma_\AGZ ) =\int \d^d x\left( \frac{\delta \Sigma_\AGZ }{\delta K_{\mu }^{a}}\frac{\delta \Sigma_\AGZ}{\delta A_{\mu}^{a}}+\frac{\delta \Sigma_\AGZ  }{\delta L^{a}}\frac{\delta \Sigma_\AGZ}{\delta c^{a}} \right. \nonumber\\
\left. +b^{a}\frac{\delta \Sigma_\AGZ }{\delta \overline{c}^{a}}+\overline{\varphi }_{i}^{a}\frac{\delta \Sigma_\AGZ }{\delta \overline{\omega }_{i}^{a}}+\omega _{i}^{a}\frac{\delta \Sigma_\AGZ }{\delta \varphi _{i}^{a}} +M_{\mu }^{ai}\frac{\delta \Sigma_\AGZ }{\delta U_{\mu}^{ai}}+N_{\mu }^{ai}\frac{\delta \Sigma_\AGZ}{\delta V_{\mu }^{ai}} + R_{\mu }^{ai}\frac{\delta \Sigma_\AGZ }{\delta T_{\mu }^{ai}} + \tau  \frac{\delta \Sigma_\AGZ }{\delta \eta} \right) \;.
\end{multline*}

\subsubsection{The counterterm}
As all the Ward identities remain the same, it is easy to check that the counterterm is given by
\begin{eqnarray}\label{countertermAGZ}
\Sigma^c_\AGZ &=& \Sigma^c_\GZ + \int \d^{4}x\left( \frac{a_{2}}{2}\tau A_{\mu }^{a}A_{\mu }^{a}+ \frac{a_{3}}{2}\zeta \tau ^{2}+\left( a_{2}-a_{1}\right) \eta A_{\mu
}^{a}\partial _{\mu }c^{a}\right) \;,
\end{eqnarray}
whereby $\Sigma^c_\GZ$ is the counterterm \eqref{final}. This counterterm can be absorbed in the original action, $\Sigma_\AGZ$ leading to the same renormalization factors as in equations \eqref{Z1}-\eqref{Z3}.  \\
\\
In addition $Z_{\tau}$ is related to $Z_{g}$ and $Z_{A}^{1/2}$ \cite{Dudal:2005na}:
\begin{eqnarray}
    Z_\tau=Z_gZ_A^{-1/2}\;,
\end{eqnarray}
and $Z_\zeta$ and $Z_\eta$ are given by
\begin{eqnarray}
Z_{\zeta }&=&1+ h (-a_{3}-2a_{2}+4a_{1}-2a_{0}) \;,\nonumber\\
Z_\eta &=& 1 + h (\frac{a_0}{2} - \frac{3}{2}a_1 + a_2 ) \;.
\end{eqnarray}

\section{Refinement of the GZ action in 4 dimensions}
\subsection{Taking into account the dynamics of the action}
Now with these new lattice data \cite{Cucchieri:2007rg,Cucchieri:2007md}, the leap was quickly taken to investigate other $d=2$ condensates. We can provide two reasons for this. Firstly, we can imagine the fields $\left(\overline{\varphi }_{\mu}^{ac},\varphi _{\mu }^{ac}, \overline{\omega }_{\mu}^{ac},\omega_{\mu }^{ac}\right)$ introduced to localize the horizon function appearing in section \ref{localization} of chapter \ref{chapgribovtoGZ}, to correspond to the nonlocal dynamics of the GZ action. Once these fields are present, they will quite evidently develop their own dynamics at the quantum level, which might include further nonperturbative effects, not yet accounted for. These effects can induce important additional changes in the infrared region. More precisely, looking at the $A\varphi$-coupling present at tree level in the action \eqref{GZstart}, one can suspect that a non-trivial effect in the $\varphi$-sector might be able to modify the gluon propagator in the desired way.  Secondly, notice that the horizon condition \eqref{horizonconditionlocal} is in fact equivalent with giving a particular value to a dimension 2 $A\varphi$-condensate, more precisely $\braket{gf^{abc}A_\mu^a(\varphi_{\mu}^{bc}+\overline{\varphi}_{\mu}^{bc})}=-2d\left(N^2-1\right)\gamma^2$. Therefore, it seems reasonably fair to consider a possible $\overline{\varphi}\varphi$-condensation. Moreover, we shall show in section \ref{sectionperturbative} that the condensate of interest is already non-vanishing at the perturbative level. \\
\\
We shall now explore the effects of a dynamical mass generation for the $\varphi$-fields. This can be done by introducing the local composite operator $\overline{\varphi}\varphi$ into the action $S_\GZ$, (expression \eqref{GZstart}). It shall turn out to be possible without spoiling the renormalizability of the theory. We shall add a massive term of the form $J\overline{\varphi}^{a}_i \varphi^{a}_i$, with $J$ a new source. First of all, for renormalization purposes, we have to add this term in a BRST invariant way. Secondly, in analogy with \cite{Dudal:2005na}, we will also need a term $\propto J^2$, indispensable to kill potential novel divergences $\propto J^2$ in
the generating functional. Let us also immediately include the local composite operator $A_\mu^a A_\mu^a$ and consider the following extended action:
\begin{eqnarray}\label{nact}
    S_\RGZ &=& S_\GZ + S_{A^2} +  S_{\overline{\varphi} \varphi}, \\
    S_{\overline{\varphi} \varphi} &=& \int \d^4 x \left[s(-J \overline{\omega}^a_i \varphi^a_{i}) + \rho J \tau \right]  \nonumber\\
    &=&\int \d^4 x \left[-J\left( \overline{\varphi}^a_i \varphi^a_{i} - \overline{\omega}^a_i \omega^a_i \right) + \rho J \tau \right]   \;,
\end{eqnarray}
with $\rho$ a new dimensionless quantity and $J$ a new source invariant under the BRST transformation, $sJ = 0$. We underline that the final mass operator, $\overline{\varphi}\varphi-\overline{\omega}\omega$, is BRST invariant.\\
\\
For the rest of the text, we shall interchange the dimension two source  $J$ often with the mass $M^2$, and the source $\tau$ coupled to $A^2_\mu$, see expression \eqref{sigmaakwadraat}, shall be interchanged with $m^2$.

\subsection{The condensate at the perturbative level\label{sectionperturbative}}
To prove that it is useful to include the condensate $\braket{(\overline{\varphi}\varphi-\overline{\omega}\omega)}$, let us show that already at the perturbative level, this condensate is non-vanishing. Hence, one is almost obliged to incorporate the effects related to the operator $\overline{\varphi}\varphi-\overline{\omega}\omega$. We start from
\begin{equation}\label{pertsol}
    \braket{\overline{\varphi}\varphi-\overline{\omega}\omega}_{\mathrm{pert}}=-\left.\frac{\p W(J)}{\p
    J}\right\vert_{J=0} \;,
\end{equation}
with $W(J)$ the generating functional defined in our case as\footnote{$W(J)$ is commonly used, but is equal to the generator $Z_c(J)$ of connected diagrams, see equation \eqref{exp}. However, here only a source connected to the operator $\overline{\varphi}\varphi-\overline{\omega}\omega$ was considered, not to all the fields, and therefore can only generate a subclass of connected diagram related to the operator.}
\begin{eqnarray}
\e^{-W(J)} &=& \int [\d \Psi] \e^{-S_\RGZ}\;,
\end{eqnarray}
and with $S_\RGZ$ the extended Gribov-Zwanziger action given in \eqref{nact}. The one loop generating functional $W_{0}(J)$, can be obtained from the quadratic part of the action,
so we find for $W_{0}(J)$ in 4d,
\begin{equation}\label{wnull}
W_{0}(J) = - \frac{4 (N^2 - 1)}{2 g^2 N} \lambda^4  +  \frac{3(N^2 - 1)}{64 \pi^2} \left( \frac{8}{3} \lambda^4 + m_1^4 \ln \frac{m_1^2}{\overline{\mu}^2} + m_2^4 \ln \frac{m_2^2}{\overline{\mu}^2} - J^2 \ln \frac{J}{\overline{\mu}^2}\right) \;.
\end{equation}
whereby we have worked in the $\MSbar$ scheme, and used the following notational shorthand
\begin{align}\label{mm}
m_1^2 &= \frac{  (m^2 + J) - \sqrt{(J + m^2)^2  - 4( m^2 J + \lambda^4)  }}{2} \;, \nonumber\\
 m_2^2 &= \frac{ (m^2 + J) + \sqrt{(J + m^2)^2  - 4( m^2 J + \lambda^4)  }}{2} \;,
\end{align}
with $\lambda$ defined in equation \eqref{deflambda}. This calculation is explained in detail in appendix \ref{sigma}\ref{appendix3}. Now using the explicit expression \eqref{wnull}, we can obtain the perturbative value of the condensate \eqref{pertsol}, reading
\begin{multline}
\braket{\overline{\varphi}\varphi-\overline{\omega}\omega}_{\mathrm{pert}}= \frac{3(N^2 - 1)}{64 \pi^2} \left[\left(2 m_1^2 \ln \frac{m_1^2}{\overline \mu^2} + m_1^2\right) \frac{\p m_1^2}{\p J} + \left(2 m_2^2 \ln \frac{m_2^2}{\overline \mu^2} + m_2^2\right) \frac{\p m_2^2}{\p J} \right] \\
= \frac{3(N^2 - 1)}{64 \pi^2}  \left[ m^2 \ln \frac{4\lambda^4}{\overline \mu^2} + \frac{1}{2} \left( \frac{m^4}{ \sqrt{m^4 - 4 \lambda^4}} + \sqrt{m^4 - 4 \lambda^4}\right) \ln \frac{m^2 + \sqrt{m^4 - 4 \lambda^4}}{m^2 - \sqrt{m^4 - 4 \lambda^4}} \right]\;,
\end{multline}
where $\lambda$ is the nonzero solution of $\frac{\p\Gamma(\lambda)}{\p \lambda}=0$. This condensate is clearly different from zero. Even for $m^2 = 0$, we obtain
\begin{equation}\label{condpert}
\braket{\overline{\varphi}\varphi-\overline{\omega}\omega}_{\mathrm{pert}} = - \frac{3(N^2 - 1)}{64 \pi} \lambda^2\;,
\end{equation}
which is always different from zero, $\forall \lambda \not= 0$.

\subsection{Renormalization of the refined action\label{chap2renrom}}
Again, the investigation of the Refined GZ action can be done similarly as in section \ref{algebraicrenormGZ} of chapter \ref{chapgribovtoGZ}.

\subsubsection{The starting action and the BRST}
As $S_{\overline{\varphi} \varphi}$ is BRST invariant, the BRST invariant action is given by
\begin{eqnarray}\label{RGZ}
\Sigma_\RGZ &=& \Sigma_\AGZ  + S_{\overline{\varphi} \varphi}\;,
\end{eqnarray}
whereby $\Sigma_\AGZ$ is given in expression \eqref{AGZ}.

\subsubsection{The Ward identities}
It is now easily checked that the Ward identities 1-6 on p.\ref{pagewardidentities} remain preserved up to potential harmless linear breaking terms.
Only the integrated Ward identity 7. is broken due to the introduction of the mass operator $\overline{\varphi}\varphi-\overline{\omega}\omega$. However, as we are using mass independent renormalization schemes, the value of the new mass dimension 2 source $J$ cannot influence the explicit expression of the counterterm $\Sigma^c $ of expression \eqref{countertermAGZ}. This is quite logical, as a mass term can only add its own renormalization factor, it does not affect the renormalization factors of other quantities, which can be equally well computed with $J=0$. This implies that the counterterm  $\Sigma^{c}_\RGZ$
corresponding to the action $\Sigma_\RGZ$ is now given by
\begin{eqnarray} \label{tegenterm}
\Sigma^{c}_\RGZ &=& \Sigma^c_\AGZ + \Sigma^c_{\varphi\overline{\varphi}} \;, \nonumber\\
 \Sigma^c_{\varphi\overline{\varphi}} &=& \int \d^d x a_4 J \tau \;,
\end{eqnarray}
with $a_4$ an arbitrary parameter. This counterterm can be absorbed into the original action $\Sigma_\RGZ$. If we define
\begin{align}
   J_{0} &= Z_{J} J \;,  &  \rho_{0} &= Z_{\rho} \rho  \;,
\end{align}
we find
\begin{align}
Z_J &= Z_{\varphi}^{-1} = Z_g Z_A^{1/2}\;, &    Z_{\rho} &= 1 + \eta (a_4 - \frac{a_0}{2} - a_2)\;.
\end{align}
Hence we have proven the renormalizability of our extended action. \\
\\
As one can notice, symmetries do also not prevent a term $\kappa J^2$ to occur, with $\kappa$ a new parameter, but we can  argue that $\kappa$ is in fact a redundant parameter, as no divergences in $J^2$ will occur. A term of this form is independent of the fields, hence  it would only be necessary to get rid of the infinities in the functional
energy, which we calculate by integrating the action over all the fields \label{redundant}
\begin{equation}
\int [\d \Phi] \e^{-S_\RGZ} = \e^{-W(J)}\;.
\end{equation}
Seen from another perspective, we need a counterterm $\propto J^2$ to remove possible divergences in the vacuum correlators $\Braket{\bigl(\overline{\varphi}\varphi-\overline{\omega}\omega\bigr)_x\bigl(\overline{\varphi}\varphi-\overline{\omega}\omega\bigr)_y}$
for $x\to y$. Such new divergences are typical when a local composite operator (LCO) of dimension 2 is added to the theory in 4d. An a priori arbitrary new coupling $\kappa$ is then needed to reabsorb these divergences. In general, it can be made a unique function of $g^2$ such that $W(J)$ obeys a standard homogeneous linear renormalization group equation \cite{Dudal:2005na}. This is a good sign, as we do not want new independent couplings entering our action or results. A nice feature of the LCO under study, i.e. $(\lco)$, is that divergences $\propto J^2$ are in fact absent in the correlators, so there is even no need for  the coupling $\kappa$ here. The argument goes as follows. The Ward identities prohibit terms in $J\gamma^2$ from occurring. Notice that this is not a trivial point, as naively we expect it to occur from the dimensional point of view. It is only by making use of the extended action and its larger symmetry content that we can exclude a term $\propto J\gamma^2$ from the game. Hence, we can set $\gamma^2=0$ to find the vacuum divergence structure $\propto J^2$, as we will employ as usual mass independent renormalization schemes like the $\MSbar$ scheme. Now, there are two ways to understand that no divergences in $J$ will occur. Firstly, at the level of the action is easily recognized that the term $g\left( \partial _{\nu } \overline{\omega }_{i}^{a}\right) f^{abm}\left(D_{\nu }c\right)^{b}\varphi _{i}^{m}$ in the action is irrelevant for the computation of the generating functional as the associated vertices cannot couple to anything without external $\omega$- and $c$-legs. Thus forgetting about this term, the $(\overline{\varphi},\varphi)$- and $(\overline{\omega},\omega)$-integrations can be done exactly, and they neatly cancel due to the opposite statistics of both sets of fields. Hence, all $J$-dependence is in fact lost, and a fortiori no divergences arise. Secondly, for $\gamma^2=0$, the action $S_\RGZ^{\gamma^2 = 0}$ is BRST invariant, $sS_\RGZ^{\gamma^2 = 0}=0$. Consequently, the vacuum correlators $\Braket{\bigl(\overline{\varphi}\varphi-\overline{\omega}\omega\bigr)_x\bigl(\overline{\varphi}\varphi-\overline{\omega}\omega\bigr)_y}=
\Braket{s\left[(\varphi\overline{\omega})_x\bigl(\overline{\varphi}\varphi-\overline{\omega}\omega\bigr)_y\right]}=0$. Therefore, we have again proven that no divergences in $J$ appear. For $\gamma^2\neq0$, the BRST transformation $s$ no longer generates a symmetry (see section V), hence a non-vanishing result for the correlator
$\Braket{\bigl(\overline{\varphi}\varphi-\overline{\omega}\omega\bigr)_x\bigl(\overline{\varphi}\varphi-\overline{\omega}\omega\bigr)_y}$
or the condensate $\Braket{\overline{\varphi}\varphi-\overline{\omega}\omega}$ is allowed. A non-vanishing VEV for our new mass operator is thus exactly allowed since the BRST is already broken by the restriction to the horizon. From the first viewpoint, the $(\overline{\varphi},\varphi)$- and $(\overline{\omega},\omega)$-integrations will no longer cancel against each other, giving room for $J$-dependent contributions in the generating functional, albeit without generating any new
divergences.

\subsection{Modifying the effective action in order to stay within the horizon\label{chap3modifying}}
\subsubsection{Extended action}
A very important fact is to check if it is still possible to stay within the Gribov region $\Omega$, after adding this new mass term.  This can be investigated with the help of the ghost propagator $\mathcal{G}(k^2)$, which is given by
\begin{eqnarray}
\mathcal{G}^{ab}(k^2) &=&   \delta^{ab} \frac{1}{k^2} (1+ \sigma(k^2)) +
\mathcal{O}(g^4)\;,
\end{eqnarray}
with
\begin{eqnarray}\label{4ghi}
\sigma(k^2) &=& \frac{N}{N^2 - 1} \frac{g^2}{k^2}\int \frac{\d^d q}{(2\pi)^4} \frac{(k-q)_{\mu} k_{\nu}}{(k-q)^2} \Braket{A^a_{\mu}A^a_{\nu}} \;,
\end{eqnarray}
see expression\footnote{Of course, in expression \eqref{ghostlabel}, the gluon propagator is already filled in.} \eqref{ghostlabel}. We recall that being inside the region $\Omega$, is equivalent to state that
\begin{eqnarray}
\sigma(k^2) \leq 1 \;,
\end{eqnarray}
see equation \eqref{nopolecondition}. In this case, the ghost propagator could be rewritten in the following form,
\begin{eqnarray}\label{ghostom}
\mathcal{G}(k^2) &=&  \frac{1}{k^2} \frac{1}{1 - \sigma(k^2)} + \mathcal{O}(g^4)\;,
\end{eqnarray}
which represents the fact that we are working at the level of the inverse propagator or equivalently, at the level of the 1PI $n$-point functions, which are generated by the effective action $\Gamma$. This form is more natural, as we can now impose the gap equation \eqref{gapgamma}, which is also formulated at the level of the effective action. However, in the next section, it shall become clear that the current action $S_\RGZ$ does not guarantee us that we are located within  the region $\Omega$ as
$\sigma(0) \geq 1$. Therefore, we add a second term to the action,
$S_{\en}$, given by
\begin{eqnarray}\label{nieuweterm}
S_{\en} &=& 2 \frac{d (N^2 -1)}{\sqrt{2 g^2 N}}  \int \d^d x\ \varsigma \ \gamma^2 J \;,
\end{eqnarray}
with $\varsigma$ a new parameter. We have introduced the particular prefactor of $ 2 \frac{d (N^2 -1)}{\sqrt{2 g^2 N}}$ for later convenience. As it is a constant term, is it comparable with the term $\left( - \int \d^d x 4 (N^2 - 1) \gamma^4 \right)$ in the original Gribov-Zwanziger formulation \eqref{GZstart}. Therefore, it can be responsible for allowing us to stay inside the Gribov horizon by  enabling $\sigma$ to be smaller than 1. The explicit calculation of $\sigma$ will be done in the next section, but we can already intuitively sketch the reasoning why $\sigma$ will be altered. As this new term is independent of the fields, it will only enter the expression for the vacuum energy. However, due
to the gap equation \eqref{gapgamma}, it will also enter in the expression of the ghost propagator (and analogously any other quantity which contains $\gamma^2$). Recapitulating, the complete action now reads,
\begin{eqnarray}\label{RGZprime}
S_\RGZ' &=& S_\RGZ + S_{\en}\;,
\end{eqnarray}
with $S_\RGZ$ given in equation \eqref{RGZ}.

\subsubsection{Renormalizability}
The renormalizability of $S'_\RGZ$ can be easily verified. Therefore, we replace $S_{\en}$ with
\begin{eqnarray}\label{4sigma_en}
\Sigma_{\en} &=&  \int \d^d x \varsigma \Theta J \;,
\end{eqnarray}
with $\Theta$ a color singlet and BRST invariant source, $s \Theta =0$. In the end, we give $\Theta$ the physical value of
\begin{eqnarray}\label{4sigma_en2}
\left. \Theta \right|_{\phys} &=&  2 \frac{d (N^2 -1)}{\sqrt{2 g^2 N}} \gamma^2 \;,
\end{eqnarray}
to return to the original action $S'_\RGZ$. Again, we embed the action $S'_\RGZ$ into a larger action $\Sigma'_\RGZ$,
\begin{eqnarray}
\Sigma'_\RGZ &=& \Sigma_\RGZ + \Sigma_{\en} \;,
\end{eqnarray}
with $\Sigma_\RGZ$ given by \eqref{RGZ}. Firstly, as it is easily checked, the term $\Sigma_{\en}$ can only give rise to an additional harmless classical breaking in the Ward identities. Therefore, all the previous Ward identities will remain valid. Secondly, we have the following additional Ward identity,
\begin{eqnarray}
\frac{\delta \Sigma'_\RGZ }{\delta \Theta} &=&  \varsigma  J\;,
\end{eqnarray}
which implies that the counterterm is independent from $\Theta$. Taking these two argument together, we can conclude that the counterterm will be exactly the same as before, given by \eqref{tegenterm}. Therefore,
\begin{eqnarray}
\frac{\varsigma_0 \gamma_0^2 J_0}{ g_0}   &=& \frac{\varsigma \gamma^2 J}{ g}\;,
\end{eqnarray}
and consequently, no new renormalization factor is necessary,
\begin{eqnarray}
Z_{\varsigma} &=& Z_g Z_{\gamma^2}^{-1} Z_J^{-1} \;.
\end{eqnarray}

\subsubsection{Boundary condition}
Introducing a new parameter $\varsigma$, requires a second gap
equation  in order to determine this new parameter. We
recall that, in the case  in which $M^2 = 0$ or
equivalently in the original Gribov-Zwanziger formulation,
 we have
\begin{eqnarray}
\sigma(k^2 \approx 0) &=& 1 - C k^2 \;,
\end{eqnarray}
with $C$ a certain positive constant, which causes the enhancement of the ghost propagator $\mathcal{G}(k^2)$ at zero momentum,
\begin{eqnarray}
\mathcal{G}(k^2 \approx 0)&\sim& \frac{1}{C k^4} \;.
\end{eqnarray}
Therefore, we know that at zero momentum, slowly switching off $M^2$, will cause $\sigma (k^2 =0)$ going to 1. It is therefore very natural to demand that this transition has to occur smoothly by imposing the following boundary condition,
\begin{eqnarray} \label{bound}
\left.\frac{\partial \sigma(0)}{\partial M^2} \right|_{M^2=0}&=& 0\;.
\end{eqnarray}
In summary, we have now two gap equations. Firstly, the gap equation $\frac{\p \Gamma}{\p \gamma^2} =0$ fixes $\gamma^2$ as a function of $M^2$ and secondly, demanding that $\left.\frac{\partial \sigma(0)}{\partial M^2} \right|_{M^2=0}= 0$ will uniquely fix $\varsigma$. This leaves us with one free parameter, $M^2$, the fixation of which shall be discussed in section \ref{sectie4}.

\subsection{The modified gluon and ghost propagator}
Now that we have constructed the action $S'_\RGZ$, by adding two additional terms $S_{\overline{\varphi} \varphi}$ and $S_{\en}$ to the original Gribov-Zwanziger action, we investigate the gluon and the ghost propagator in detail.

\subsubsection{The gluon propagator}
We shall first examine the tree level gluon propagator. This propagator can be calculated in a completely analogous fashion as in section \ref{gluonghostchap2} of chapter \ref{chapgribovtoGZ}. Here, the free action reads
\begin{multline}
     S_\RGZ^{\prime 0} = \int \d^dx  \Bigl[ \frac{1}{4} \left( \p_{\mu} A_{\nu}^a - \p_{\nu}A_{\mu}^a \right)^2 + \frac{1}{2\alpha} \left( \p_{\mu} A^a_{\mu} \right)^2 +
\overline{\varphi}^{ab}_{\mu} \p^2 \varphi^{ab}_{\mu} \\
- \gamma^2 g(f^{abc}A^a_{\mu} \varphi_{\mu}^{bc} + f^{abc}A^a_{\mu} \overline{\varphi}^{bc}_{\mu} ) - M^2
\overline{\varphi}_{\mu}^{ab}\varphi_{\mu}^{ab} + \frac{m^2}{2} A^2_{\mu} +\ldots \Bigr]\;,
\end{multline}
where we already integrated out the $b$ field. The equations of motion for the $\varphi$ and $\overline \varphi$ fields are now given by
\begin{eqnarray}
    \varphi^{bc}_{\mu} = \overline{\varphi}^{bc}_{\mu} =  \frac{1}{\p^2 - M^2} \gamma^2 g f^{abc} A^a_{\mu} \;,
\end{eqnarray}
and with this result, we can easily read the gluon propagator
\begin{eqnarray}
  S_\RGZ^{\prime 0} &=&  \; \int \d^4 x \; \left[ \frac{1}{2} A^a_{\mu} \Delta^{ab}_{\mu\nu} A^b_{\nu} + \ldots \right] \;, \nonumber\\
\Delta^{ab}_{\mu\nu} &=&\left[ \left(-\p^2 + m^2 - \frac{2 g^2 N \gamma^4}{\p^2 - M^2} \right) \delta_{\mu\nu} - \p_{\mu}\p_{\nu}
\left(\frac{1}{\alpha} - 1\right) \right] \delta^{ab} \;.
\end{eqnarray}
Taking the inverse of $\Delta^{ab}_{\mu\nu}$ and converting it to momentum space gives
\begin{eqnarray}\label{gluonprop2}
  \; \Braket{ A^a_{\mu}(p) A^b_{\nu}(-p)} &=& \frac{1}{p^2+ m^2 + \frac{2g^2 N \gamma^4}{p^2 + M^2}}\left[\delta_{\mu\nu} - \frac{p_{\mu}p_{\nu}}{p^2} \right]\delta^{ab} \nonumber\\
 &=& \underbrace{\frac{p^2 + M^2}{p^4 + (M^2+m^2)p^2 + 2 g^2 N \gamma^4 + M^2 m^2 }}_{\mathcal{D}(p^2)}\left[\delta_{\mu\nu} - \frac{p_{\mu}p_{\nu}}{p^2} \right]\delta^{ab} \;.
\end{eqnarray}
From this expression we can already make two observations:
\begin{itemize}
\item $\mathcal{D}(p^2)$ enjoys infrared suppression.
\item $\mathcal{D}(0)\propto M^2$, so the gluon propagator does not vanish at the origin. Even if we set $m^2=0$ we still find a non-vanishing
gluon propagator, so we want to stress that this different result is clearly due to the novel mass term proportional to $\overline{\varphi}\varphi-\overline{\omega}\omega$.
\end{itemize}
In section \ref{positivityviolation} we shall uncover a third property, namely that $\mathcal{D}(p^2)$ displays a positivity violation. Also this observation is in accordance with the lattice results \cite{Bowman:2007du}.

\subsubsection{The ghost propagator}
The observation that $m^2 =0$ does not qualitatively alter the gluon propagator, will be repeated for the ghost propagator. Henceforth, we set $m^2 =0$, which also improves the readability. However, all calculations could in principle be repeated with the inclusion of the mass $m^2$.  \\
\\
We start with the expression for the ghost propagator. Substituting the expression of the gluon propagator \eqref{gluonprop2} in the expression of $\sigma$ \eqref{4ghi}
\begin{eqnarray}\label{exact}
\sigma(k^2) &=&  Ng^2 \frac{k_{\mu} k_{\nu}}{k^2} \int \frac{\d^d q}{(2\pi)^d} \frac{1}{(k-q)^2} \frac{q^2 + M^2}{q^4 + M^2 q^2 + \lambda^4 }\left[ \delta_{\mu\nu} - \frac{q_{\mu}q_{\nu}}{q^2} \right]\;.
\end{eqnarray}
We have also defined
\begin{eqnarray}\label{lambda4}
\lambda^4 &=& 2 g^2 N \gamma^4\;.
\end{eqnarray}
As we are again interested in the infrared behavior of this propagator, we expand the previous expression for small $k^2$,
\begin{eqnarray}
\sigma (k^2\approx 0) &=&    Ng^2 \frac{d-1}{d}  \int \frac{\d^d q}{(2\pi)^d} \frac{1}{q^2} \frac{q^2 + M^2}{q^4 + M^2 q^2 + \lambda^4 } + O(k^2) \;.
\end{eqnarray}
For later use, let us rewrite $\sigma(0)$ as
\begin{equation}\label{3}
\sigma (0) =   Ng^2 \frac{d-1}{d}  \int \frac{\d^d q}{(2\pi)^d} \frac{1}{q^4 + M^2 q^2 + \lambda^4 } +  Ng^2 M^2 \frac{d-1}{d}\int \frac{\d^d q}{(2\pi)^d} \frac{1}{q^2} \frac{ 1}{q^4 + M^2 q^2 + \lambda^4 } \;.
\end{equation}
Notice that the first integral in the right hand side of equation \eqref{3} diverges while the second integral is UV finite in 4d.\\
\\
We continue with the derivation of the gap equations as we would like to write $\lambda^2$ as a function of $M^2$, i.e.~$\lambda^2(M^2)$, in expression \eqref{3}. Firstly, we calculate the horizon condition \eqref{gapgamma} explicitly starting from the effective action. The one loop effective action $\Gamma_\gamma^{(1)}$ is obtained from the quadratic part of our action $S_\RGZ'$
\begin{equation}
\e^{-\Gamma_\gamma ^{(1)} }=\int [\d\Phi] \e^{-S^{\prime 0}_\RGZ}\;.
\end{equation}
This time, the terms $-d(N^2 - 1) \gamma^4$ and $ 2\frac{d (N^2 -1)}{\sqrt{2 g^2 N}}  \varsigma \ \gamma^2 M^2 $ have to be maintained, as they will enter the horizon condition. After a straightforward calculation the one loop effective action in $d$ dimensions yields,
\begin{equation}
\Gamma_\gamma^{(1)} = -d(N^{2}-1)\gamma^{4} + 2\frac{d (N^2 -1)}{\sqrt{2 g^2 N}}  \varsigma \ \gamma^2 M^2 +\frac{(N^{2}-1)}{2}\left( d-1\right) \int \frac{\d^{d}q}{\left(
2\pi \right) ^{d}} \ln \frac{q^4 + M^2 q^2 + \lambda^4}{ q^2 + M^2} \;.
\end{equation}
We rewrite the previous expression,
\begin{eqnarray}
\mathcal{E}^{(1)} = \frac{\Gamma_\gamma^{(1)}}{N^2 - 1} \frac{2 g^2 N}{d} ~=~ - \lambda^4  + 2 \varsigma \lambda^2 M^2 + g^2 N \frac{ d-1}{d} \int \frac{\d^{d}q}{\left(
2\pi \right) ^{d}} \ln \frac{q^4 + M^2 q^2 + \lambda^4}{ q^2 + M^2} \;,
\end{eqnarray}
and apply the gap equation \eqref{gapgamma},
\begin{eqnarray}\label{gapeq}
\frac{\p \mathcal{E}^{(1)}  }{\p \lambda^2} &=& 2 \lambda^2\left( -1 + \varsigma \frac{M^2}{\lambda^2} + g^2 N \frac{d-1}{d} \int \frac{\d^{d}q}{\left(
2\pi \right) ^{d}}  \frac{1}{q^4 + M^2 q^2 + \lambda^4} \right) ~=~ 0\;.
\end{eqnarray}
Secondly, we impose the boundary condition \eqref{bound} in order to obtain an explicit value for $\varsigma$. Instead of explicitly starting from expression \eqref{exact} to fix $\varsigma$, there is a much simpler way to find the corresponding $\varsigma$. Therefore, we act with $\frac{\p}{\p M^2}$ on the gap equation \eqref{gapeq}. Subsequently setting $M^2 =0$, gives
\begin{equation}\label{simpel}
 \varsigma\frac{1}{\lambda^2(0)}    -\frac{d-1}{d}g^2N\int    \frac{\d^dq}{(2\pi)^d}\frac{1}{q^2}\frac{1}{q^4+\lambda^4(0)}~=~0\;,
\end{equation}
where we imposed \eqref{bound}.  Proceeding, we find
\begin{eqnarray}\label{laatste}
    &&-\frac{d-1}{d}g^2N\int  \frac{\d^dq}{(2\pi)^d}\frac{1}{q^2}\frac{1}{q^4+\lambda^4(0)}+\varsigma\frac{1}{\lambda^2(0)}~=~0 \nonumber\\
    \Rightarrow&&  \varsigma ~=~ \lambda^2(0) \frac{3}{4}g^2N\int  \frac{\d^4q}{(2\pi)^4}\frac{1}{q^2}\frac{1}{q^4+\lambda^4(0)}  \nonumber\\
        \Rightarrow&& \varsigma ~=~  \frac{3g^2 N}{128 \pi}\;,
\end{eqnarray}
which determines $\varsigma$ at the current order.\\
\\
With the help of the latter two gap equations \eqref{gapeq} and \eqref{laatste}, we can rephrase the correction to the self energy of the ghost. Combining equation \eqref{3} and \eqref{gapeq} we can write
\begin{eqnarray}\label{sigmapre}
\sigma(0) &=& 1 + M^2 g^2 N \frac{d-1}{d}\int \frac{\d^d q}{(2\pi)^d} \frac{1}{q^2} \frac{ 1}{q^4 + M^2 q^2 + \lambda^4(M^2) } - \varsigma \frac{M^2}{\lambda^2(M^2)}\;.
\end{eqnarray}
From this expression, we can make several observations. Firstly, when $M^2 =0$, from the previous expression it immediately follows that
\begin{eqnarray}
\sigma(0) &=& 1\;,
\end{eqnarray}
which  gives back the ordinary Gribov-Zwanziger result, see \eqref{sigmagelijkaan1}. Secondly, when $M^2 \not= 0$, we notice that the ghost propagator is no longer enhanced and behaves like $1/k^2$, which is in qualitative agreement with the lattice results from \cite{Cucchieri:2007md}. This behavior is clearly due to the novel mass
term $ M^2\int \d^4 x\;\left( \overline{\varphi}^a_i \varphi^a_{i}- \overline{\omega}^a_i \omega^a_i \right)$. Thirdly, we see that the term in $\varsigma$ is crucial in
order to obtain a $\sigma(0)$ which is smaller than 1. Omitting this term would result in $\sigma(0) > 1$ in the case that $M^2 \not= 0$. However, including this term, we can easily prove that $\sigma \leq 1$. Indeed, taking expression \eqref{sigmapre} and replacing $\varsigma$ with the integral in \eqref{laatste}, we find
\begin{eqnarray}
\sigma(0) &=& 1 + M^2 g^2 N \frac{3}{4}\int \frac{\d^4 q}{(2\pi)^4} \frac{1}{q^2} \frac{ 1}{q^4 + M^2 q^2 + \lambda^4(M^2) }\nonumber\\
 && \hspace{2cm}-  \frac{M^2}{\lambda^2(M^2)} \lambda^2(0) \frac{3}{4}g^2N\int  \frac{\d^4q}{(2\pi)^4}\frac{1}{q^2}\frac{1}{q^4+\lambda^4(0)} \nonumber\\
&=& 1 + \frac{3}{4} \frac{M^2}{\lambda^2(M^2)} g^2 N \int \frac{\d^4 p}{(2\pi)^4} \frac{1}{p^2} \frac{ 1}{p^4 + \frac{M^2}{\lambda^2(M^2)} p^2 + 1 } \nonumber\\
&& \hspace{2cm}-   \frac{3}{4} \frac{M^2}{\lambda^2(M^2)} g^2 N \frac{3}{4}g^2N\int  \frac{\d^4 p}{(2\pi)^4}\frac{1}{p^2}\frac{1}{p^4+1} \nonumber\\
&=& 1 - \frac{3 x^2}{4}  g^2 N \int \frac{\d^4 p}{(2\pi)^4} \frac{1}{p^2}  \frac{ 1}{(p^4 + x p^2 + 1)(p^4 + 1) }  \;,
\end{eqnarray}
with $x = \frac{M^2}{\lambda^2(M^2)} \geq 0$, hence  $\sigma(0) \leq 1 $. At this point, we can really appreciate the role of the novel vacuum term \eqref{nieuweterm}. It serves as a stabilizing term for the horizon condition.  Indeed, without the term \eqref{nieuweterm}, we would end up outside of the Gribov region for some $k^2>0$, even for an infinitesimal\footnote{Notice that we must take $M^2 \geq 0$ to avoid unwanted tachyonic instabilities. } $M^2>0$. In this sense, the action $S'_\RGZ$ constitutes a refinement of the original Gribov-Zwanziger action, which is a smooth limiting case of $S'_\RGZ$.\\
\\
For later use, we can evaluate the integral in expression \eqref{sigmapre} as it is finite. The explicit one loop value for $\sigma(0)$ yields
\begin{multline}\label{sigmafinal}
\sigma(0) = 1 + M^2 \frac{3 g^2 N}{64 \pi^2}  \frac{1}{ \sqrt{M^4 - 4 \lambda^4}} \left[  \ln \left(  M^2  +  \sqrt{M^4 - 4 \lambda^4} \right) - \ln \left(M^2 -  \sqrt{M^4 - 4 \lambda^4} \right) \right]\\
 - \left( \frac{3g^2 N}{128 \pi} \right) \frac{M^2}{\lambda^2(M^2)}\;,
\end{multline}
where we have substituted the value \eqref{laatste} for $\varsigma$.\\
\\
In summary, we have found a ghost propagator which is no longer enhanced. So far, we have fixed $\lambda^2$ in function of $M^2$ and we have found a constant value for $\varsigma$. However, we have not  yet fixed $M^2$. This will be the task of the next section.

\subsection{A dynamical value for $M^2$ \label{sectie4}}
Up to this point, we have only introduced the mass $M^2$ by hand, however this value is in fact a dynamical value. In this thesis, we shall present two methods to find such a value. Firstly, in the appendix, we explain how to obtain a dynamical value for $M^2$ with the help of the effective action. However, the calculations become too involved, and therefore, we are not able to go beyond perturbation theory. Henceforth, in this section we have investigated a second method, the variational principle, and applied this to the ghost and gluon propagator, with more success.  \\
\\
Let us explain the variational method which we shall use. Along the lines of \cite{Jackiw:1995nf}, we introduce a formal loop counting parameter $\ell\equiv 1$ by replacing the action $S$ with $\frac{1}{\ell}S$. At the same time, we replace all the fields $\Phi$ by $\sqrt{\ell}\Phi$. Symbolically,
\begin{equation}\label{vpt1}
    S(\Phi,g)\to \frac{1}{\ell}S(\sqrt{\ell}\Phi,g)\;.
\end{equation}
It is readily derived that multiplying each field with a factor of $\sqrt{\ell}$ and performing an overall $1/\ell$ rescaling is the same as replacing the coupling $g$ with
$\sqrt{\ell}g$, so we can replace \eqref{vpt1} with
\begin{equation}\label{vpt2}
    S(\Phi,g)\to S(\Phi,\sqrt{\ell}g)\;.
\end{equation}
In this fashion, the free (quadratic) part of the action is $\ell$-invariant, while every interaction terms contains powers\footnote{We recall that the perturbative expansion is one in powers of $g^2$, and thus in integer powers of $\ell$.} of $\sqrt{\ell}$. The first order in the $\ell$-expansion, obtained by setting $\ell=0$, then corresponds to the free theory. More generally, the $\ell$-expansion is equivalent with the loop expansion, where it is understood that we put the formal bookkeeping parameter $\ell=1$ at
the end.\\
\\
The next step is to introduce the variational parameter $M^2$ into the theory. This is done in  a specific way: we add the quadratic mass term $S_M\equiv  M^2 \int  \d^4x\left[\left(\overline{\varphi}\varphi-\overline{\omega}\omega\right) + \frac{2 (N^2 -1)}{ g^2 N}   \varsigma  \lambda^2  \right]$ to the action, but subtract it again at higher order in $\ell$, i.e. we consider the action
\begin{equation}\label{vpt3}
    S(\Phi,g)\to  S(\Phi,\sqrt{\ell}g)+ S_M -\ell^k S_M\;,
\end{equation}
with $k > 0$. Since $\ell\equiv1$, we did not change the actual starting action at all.\\
\\
However, we maintain the strategy of performing an expansion in powers of $\ell$. Since the mass term is split up into 2 parts $\sim (1-\ell^k)M^2$, both parts will enter the $\ell$-expansion in a different way. At the end, we must set $\ell=1$ again. If we could compute an arbitrary quantity $\cal Q$ exactly, the $M^2$-independence would of course be apparent since the theory is not altered. However, at any finite order in $\ell$, a residual $M^2$-dependence will enter the result for $\mathcal{Q}$ due to the re-expanded powers series in $\ell$. Said otherwise, we have partially resummed the perturbative series for $\cal Q$ by making use of the parameter $\ell$. The hope is that some nontrivial information, encoded by the operator coupled to $1-\ell^k$, will emerge in the final expression for $\mathcal{Q}$. One query remains: how to handle the $M^2$ which appears in the approximate $\mathcal{Q}$? Therefore we can rely on the lore of minimal sensitivity \cite{Stevenson:1981vj}: we know that the exact $\mathcal{Q}$ cannot depend on $M^2$, hence it is very natural to demand that also at a finite order $\frac{\p \cal {Q}}{\p M^2}=0$, leading to a dynamical optimal value for the yet free parameter $M^2$. \\
\\
The described method of variationally introducing extra parameters into a quantum field theory provides us with a powerful tool to study nontrivial dynamical effects in an approximate fashion, yet the calculational efforts do not exceed those of conventional perturbation theory.\\
\\
We still  have to choose a value for $k$. We recall that the constant term, $S_{\en}$, was introduced in order to stay within the horizon. Therefore, we want to retain this term when we are applying the variational principle. However, we are working up to first order, meaning that we shall expand the quantity $\mathcal{Q}$ up to first order in $\ell$ and subsequently set $\ell = 1$. Hence, taking $k =1$ is  not a good  option as the constant term would vanish and have no influence. Therefore, a better option is to take e.g.~$k =2$, to assure the consistency of the variational setup with the restriction to the Gribov region. In this way, we are simply coupling the variational parameter $M^2$ directly to the theory.

\subsubsection{The ghost propagator}
\begin{figure}[H]
   \centering
       \includegraphics[width=8cm]{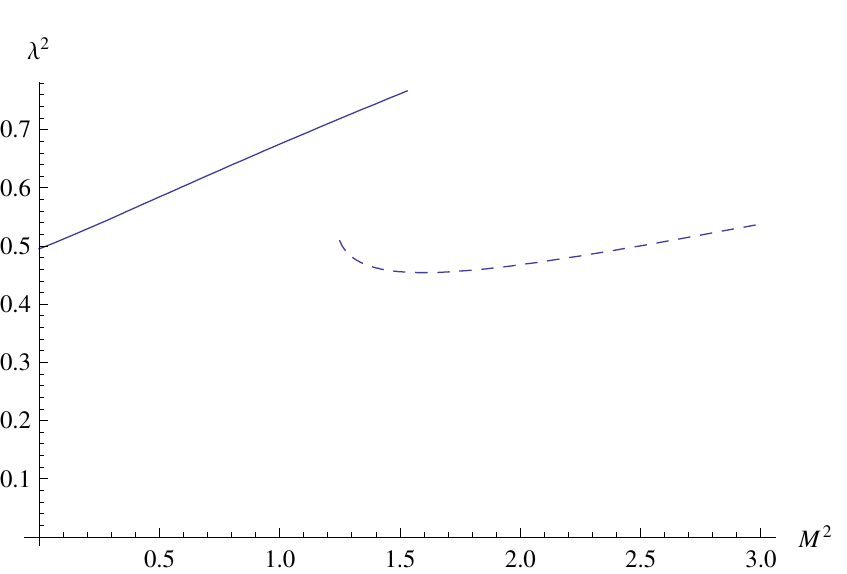}
   \caption{$\lambda^2$ in function of $M^2$ in units $\lms = 1$.}
   \label{gapdep}
\end{figure}

We start from the expression \eqref{ghostom} of the ghost propagator
\begin{eqnarray}
\mathcal{G}(k^2) &=&  \frac{1}{k^2} \frac{1}{1 - \sigma(k^2)}\;,
\end{eqnarray}
and apply the variational principle on the ghost propagator near zero momentum. We have,
\begin{eqnarray} \label{same}
\sigma (k^2\approx 0) &=&  Ng^2 \frac{d-1}{d}  \int \frac{\d^d q}{(2\pi)^d} \frac{1}{q^2} \frac{q^2 + M^2}{q^4 + M^2 q^2 + \lambda^4 } + O(k^2)\;.
\end{eqnarray}
As explained above, we replace $g^2 \rightarrow \ell g^2$ and $M^2 \rightarrow (1-\ell^2)M^2$.  Subsequently, we expand $\mathcal{G}(k^2)_{k^2 \approx 0}$ in powers of $\ell$ corresponding to a re-ordered loop expansion. As we have calculated the ghost propagator up to one loop, we only need to expand the above expression to the first power of $\ell$,
\begin{eqnarray}
\sigma (0) &=&  Ng^2 \ell \frac{d-1}{d}  \int \frac{\d^d q}{(2\pi)^d} \frac{1}{q^2} \frac{q^2 + M^2}{q^4 + M^2 q^2 + \lambda^4 } \;.
\end{eqnarray}
As indicated earlier, setting $\ell ~=~ 1$ gives
\begin{eqnarray}
\sigma (0) &=&  Ng^2  \frac{d-1}{d}  \int \frac{\d^d q}{(2\pi)^d} \frac{1}{q^2} \frac{q^2 + M^2}{q^4 + M^2 q^2 + \lambda^4 } \;,
\end{eqnarray}
which is exactly the same as \eqref{same}. This expression not only depends on $M^2$, but also on $\lambda^2$. However, we already know that $\lambda^2$ and $M^2$ are not independent variables, as they are related through the gap equation \eqref{gapeq},
\begin{eqnarray} \label{gapp}
  &&  -1 + \varsigma \frac{M^2}{\lambda^2} + g^2 N \frac{d-1}{d} \int \frac{\d^{d}q}{\left(
2\pi \right) ^{d}}  \frac{1}{q^4 + M^2 q^2 + \lambda^4}  ~=~ 0 \;.
\end{eqnarray}
Following the variational principle, we replace $M^2$ with $(1-\ell^2)M^2$ and $g^2$ with $\ell g^2$, expand the equation up to order $\ell^1$, and set $\ell =1$ in the end. Doing so, we recover again expression \eqref{gapp}. At this point, it can be clearly seen that $k = 1$ in equation \eqref{vpt3} would cancel the effect of the constant term $\frac{\varsigma M^2}{\lambda^2}$, while $k = 2$ is a better choice\footnote{Actually, every value for $k$, with $k \geq 2$ is allowed.}. Evaluating the integral in expression \eqref{gapp}, we find
\begin{eqnarray}
0&=&-1 + \frac{N g^2}{64 \pi^2}\left( \frac{5}{2} + 3\frac{m_1^2}{\sqrt{M^4-4\lambda^4}} \ln \frac{m_1^2}{\overline{\mu}^2} - 3\frac{m_2^2}{\sqrt{M^4-4\lambda^4}} \ln \frac{m_2^2}{\overline{\mu}^2}\right) + \varsigma \frac{M^2}{\lambda^2}\;.
\end{eqnarray}
This integral could be similarly calculated as done in appendix \ref{sigma}\ref{appendix3}. We recall that from the boundary condition \eqref{laatste}, we have already determined $\varsigma ~=~  \frac{3g^2 N}{128 \pi}$. \\
\\
We still require an appropriate value for $\overline{\mu}$. Therefore, we fix $\overline{\mu}^2 = \frac{3}{2}\left|M^2 + \sqrt{M^4 - 4 \lambda^4}\right|$ which was chosen as in \cite{Dudal:2005na}. We have opted for this specific renormalization scale $\overline{\mu}^2$ which shall result in an acceptably small effective expansion parameter $\frac{g^2 N}{16 \pi^2}$. Consequently, from the following equation,
\begin{eqnarray}\label{gkwadraat2}
g^2(\overline{\mu}^2)&=& \frac{1}{\beta_0\ln\frac{\overline{\mu}^2}{\lms^2} }\;,\;\;\;\;\;\textrm{with }\;\;\;\;\beta_0=\frac{11}{3}\frac{N}{16\pi^2}\;,
\end{eqnarray}
we find
\begin{eqnarray}
\frac{g^2 N}{16 \pi^2}&=& \frac{3}{11 \ln \left(\frac{3}{2}\left| M^2 + \sqrt{M^4 - 4 \lambda^4}\right| \right)}\;,
\end{eqnarray}
in units of $\lms = 1$.\\
\\
In summary, as $\sigma(0)$ remained the same after applying the variational principle, we can take the expression \eqref{sigmafinal} for $\sigma(0)$,
\begin{multline}
\sigma(0) =  1 + M^2 \frac{3 g^2 N}{64 \pi^2}  \frac{1}{ \sqrt{M^4 - 4 \left(\lambda^2(M^2)\right)^2}} \Bigl[  \ln \left(  M^2  +  \sqrt{M^4 - 4 \left(\lambda^2(M^2)\right)^2} \right) \\
- \ln \left(M^2 -  \sqrt{M^4 - 4 \left(\lambda^2(M^2)\right)^2} \right) \Bigr]  - \left( \frac{3g^2 N}{128 \pi} \right) \frac{M^2}{\lambda^2(M^2)}\;,
\end{multline}
where $\lambda^2(M^2)$ is determined by the gap equation,
\begin{equation}\label{gapecht}
0 = -1 + \frac{N g^2}{64 \pi^2}\left( \frac{5}{2} + 3\frac{m_1^2}{\sqrt{M^4-4\lambda^4}} \ln \frac{m_1^2}{\overline{\mu}^2} - 3\frac{m_2^2}{\sqrt{M^4-4\lambda^4}} \ln \frac{m_2^2}{\overline{\mu}^2}\right) + \frac{3g^2 N}{128 \pi} \frac{M^2}{\lambda^2}\;.
\end{equation}
Before continuing the analysis, let us first have a look at the gap equation. The gap equation solved for $\lambda^2$  as a function of $M^2$ is depicted in Figure \ref{gapdep}. We find two emerging branches, displayed by a continuous and a dashed line. The former solution exists in the interval $[0, 1.53]$, while the latter one only exists in $[1.25, \infty[$. As the latter branch does not exist around $M^2 =0$, we shall not consider this solution because the boundary condition \eqref{bound} demands a smooth
transition for the $M^2 \to 0$ limit.

\begin{figure}[H]
   \centering
          \includegraphics[width=8cm]{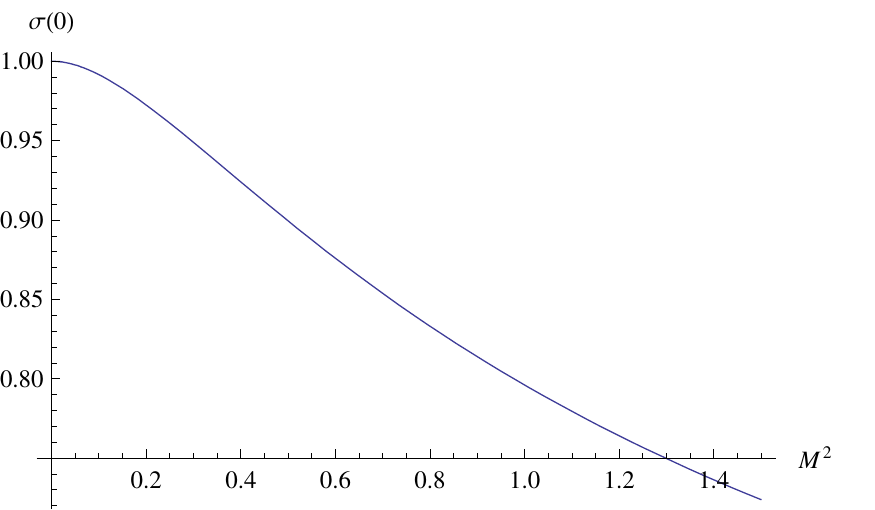}
   \caption{$\sigma(0)$ drawn in function of $M^2$ in units $\lms = 1$.}
   \label{ghost1}
\end{figure}

\noindent We can now have a closer look at the ghost propagator or equivalently $\sigma(0)$. We have graphically depicted $\sigma(0)$ in Figure \ref{ghost1}. Firstly, from the figure, we see that $\sigma(0)$ is nicely smaller than 1 for all $M^2$ in the interval $[0, 1.53]$. This is a remarkable fact as it implies that we have managed to stay within the horizon.  Secondly, we notice that the boundary condition $\left. \frac{\p \sigma (0)}{\p M^2} \right|_{M^2 =0} = 0$ is indeed fulfilled, which is a nice check on our result. We can now apply the minimal sensitivity approach on the quantity $\sigma(0)$. From Figure \ref{ghost1} we immediately see that there is no extremum. However, looking at the derivative of $\sigma(0)$ with respect to $M^2$, we do find a point of inflection at $M^2 = 0.37 \lms^2$. Demanding $\frac{\p^2 \sigma(0) }{ (\p M^2) ^2 } = 0$ is an alternative option when no extremum is found \cite{Stevenson:1981vj}. Taking this value for $M^2$, we find
\begin{eqnarray}
\sigma(0) &=&  0.93 \;.
\end{eqnarray}
The effective coupling is given by
\begin{eqnarray}\label{koppel1}
\frac{g^2 N }{ 16 \pi^2} &=& 0.53 \;,
\end{eqnarray}
which is smaller than 1.

\begin{figure}[H]
   \centering
          \includegraphics[width=8cm]{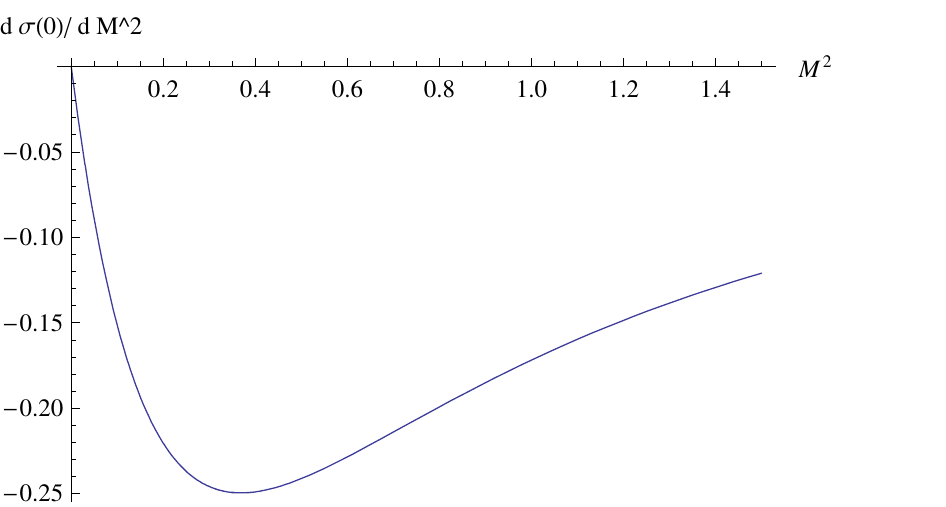}
   \caption{$ \frac{ \d \sigma(0)}{\d M^2}$ drawn in function of $M^2$ in units $\lms = 1$.}
   \label{ghost2}
\end{figure}

\subsubsection{The gluon propagator}
In order to apply the variational principle to the gluon propagator, we require its one loop correction. However, due to the rather complicated form of the propagator, obtaining the full exact expression for its one loop correction is not possible. Indeed to appreciate how cumbersome such an expression could be one has only to examine the $M^2$~$=$~$m^2$~$=$~$0$ case, \cite{Gracey:2006dr}, where all the one loop corrections to the propagators are given explicitly. However, despite this, it is possible to calculate the one loop gluon propagator directly in the zero momentum limit without knowledge of the full correction. Details of how to calculate this one loop correction with {\sc Form} can be found in \cite{Dudal:2008sp}\footnote{Afterwards, it has appeared that not all propagators were included in the calculations. Therefore, the results obtained here can only be considered in a qualitative nature. }

\begin{figure}[H]
   \centering
       \includegraphics[width=0.45\textwidth]{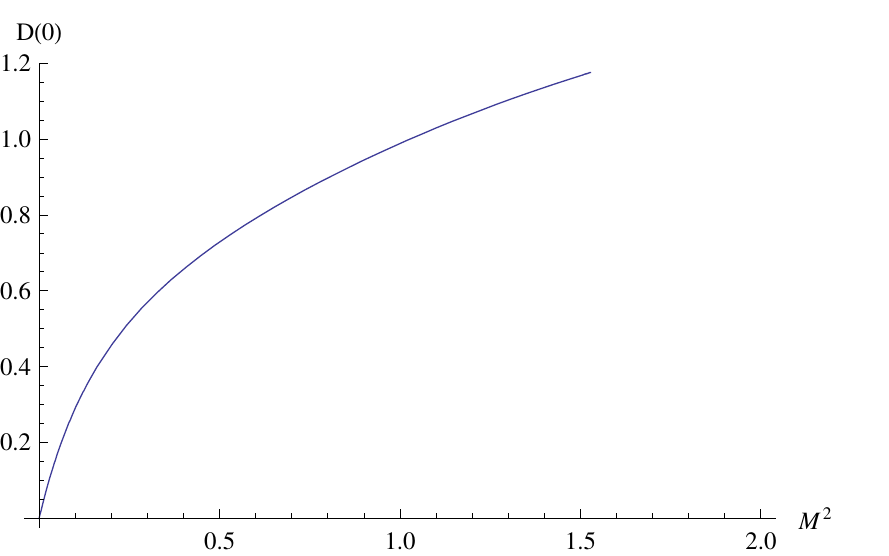}
   \caption{The gluon propagator $\mathcal{D}^{(1)}(0)$ drawn in function of $M^2$ in units $\lms = 1$.}
   \label{gluon1}
\end{figure}

\noindent The following expression was obtained
\begin{multline}\label{oneloop}
\mathcal{D}^{(1)}(0) = \frac{M^2}{\lambda^4}-\frac{g^2 N}{16\pi^2} \frac{M^4}{\lambda^8} \Biggl[  \frac{M^4}{\lambda^4}\frac{9}{16}\sqrt{M^4 - 4 \lambda^4} \ln \frac{m_2^2}{m_1^2}
+  \frac{M^6}{\lambda^4} \left( \frac{9}{16} \ln \frac{\lambda^4}{M^4} \right) \\
- \frac{15}{16}M^2 \lambda^4 \frac{1}{M^4 - 4 \lambda^4} + \frac{3}{2}\lambda^4 \frac{1}{\sqrt{M^4 - 4 \lambda^4}} \ln \frac{m_2^2}{m_1^2} + \frac{15}{8} \lambda^8 \frac{1}{(\!\sqrt{M^4 - 4 \lambda^4})^3} \ln \frac{m_2^2}{m_1^2} \\
+ M^2 \left( \frac{9}{8} - \frac{21}{16} \ln \frac{\lambda^4}{M^4} \right) - \frac{3}{16}   \sqrt{M^4 - 4 \lambda^4}\ln \frac{m_2^2}{m_1^2} \Biggr]\;,
\end{multline}
for the one loop correction at zero momentum where all mass variables correspond to renormalized ones.
\\
\\
We apply the variational principle to the gluon propagator in a completely similar manner as in the case of the ghost propagator. Therefore, we replace $M^2$ with $(1- \ell^2)M^2$ and $g^2$ with $\ell g^2$ in the expression \eqref{oneloop}, expand up to order $\ell^1$, and set $\ell =1$. Doing so, we find the original expression \eqref{oneloop} for the gluon propagator back. Firstly, we try to apply the principle of minimal sensitivity. Therefore, we have depicted the gluon propagator in Figure \ref{gluon1}. First, we notice that $ D^{(1)}(0)$ is positive for all $M^2 \in [0, 1.53]$. Unfortunately, we do find neither a minimum nor a point of inflection in this interval. Therefore, we shall take the value of $M^2$, which was obtained in the study of the ghost propagator (see previous section). Hence, setting $M^2 = 0.37 \lms^2$, gives
\begin{eqnarray}
\mathcal{D}^{(1)}(0) = \frac{0.63}{\lms^2} = \frac{11.65}{ \mathrm{GeV^2}}\;.
\end{eqnarray}
Evidently, the effective coupling is still smaller than 1, cfr.~\eqref{koppel1}.

\noindent In summary, the infrared value of the ghost propagator and the zero momentum gluon propagator seem to be reasonable. We find a non-enhanced ghost propagator and a gluon propagator which is non-zero at zero momentum. Our results for the gluon and ghost propagator are of a qualitative nature as we are only working in a first order approximation. In order to improve these numerical results, higher order calculations are recommendable. This is however far beyond the scope of this thesis.

\subsection{The temporal correlator: violation of positivity\label{positivityviolation}}
With the help of the variational technique, we can also show that the gluon propagator displays a violation of positivity. If we rewrite the gluon propagator in the K\"{a}ll\'{e}n-Lehmann spectral representation,
\begin{eqnarray} \label{pos}
\mathcal{D}(p^2) &=& \int_{0}^{+\infty} \d M_p^2 \frac{\rho(M_p^2)}{p^2 + M_p^2}\;,
\end{eqnarray}
$\rho(M_p^2)$ should be a positive function in order to interpret the fields in terms of stable particles. If $\rho(M_p^2) < 0$ for certain $M_p^2$, $\mathcal{D}(p^2)$ is positivity violating. As a practical way to uncover this property, one defines the temporal correlator \cite{Cucchieri:2004mf}
\begin{eqnarray}\label{tempcorrelator}
\mathcal{C}(t)&=& \int_0^{+\infty} \d M_p \rho(M_p^2) \e^{-M_pt}  = \frac{1}{2\pi} \int_{-\infty}^{+ \infty} \e^{-ipt}\mathcal{D}(p^2) \d p\;.
\end{eqnarray}
Consequently, if we can show that $\mathcal{C}(t)$ becomes negative for certain $t$, $\rho(M_p^2)$ cannot be positive for all $M_p^2$, resulting in a positivity violating gluon propagator. If the gluon propagator vanishes at zero momentum, $\mathcal{D}(0) = 0$, one can immediately verify from \eqref{pos} that $\rho(M_p^2)$ cannot be a positive quantity. However, having $\mathcal{D}(0) \not= 0$, does not exclude a positivity violation as we shall soon find out.\\
\\
We can now apply the variational technique on the temporal correlator. At tree level, this $\mathcal{C}(t)$ is given by
\begin{eqnarray}
\mathcal{C}(t, M^2)&=& \frac{1}{2\pi} \int_{-\infty}^{+ \infty}
\e^{-ipt}\frac{p^2 + M^2}{p^4 + M^2 p^2 +
\left(\lambda^2(M^2)\right)^2} \d p\;,
\end{eqnarray}
where $\lambda^2(M^2)$ is still determined by the gap equation \eqref{gapecht}. Replacing $M^2 \rightarrow (1-\ell^2)M^2$ and $g^2 \rightarrow \ell g^2$ is redundant in this case, as we only have the tree level gluon propagator $\mathcal{D}(p^2)$ at our disposal. We shall now implement the minimal sensitivity principle as follows: for each different value of $t$, we minimize the temporal correlator with respect to $M^2$. $C(t)$ displays a minimum at $M_{\mathrm{min}}^2 \neq 0$, for $t \gtrsim 6 /\lms $. In Table \ref{tabelM}, some values for $M_{\mathrm{min}}^2(t)$ for different $t$ are presented. For $t \lesssim 6$, we have taken $M^2 = 0$; it is clearly visible from the table below that $M^2_{\min} \to 0$ for decreasing $t$.

\begin{table}[H]
\renewcommand{\arraystretch}{2}
\begin{center}
\begin{tabular}{|c||ccccc|}
        \hline
        $t $  & 6 & 7 & 8 & 9 & 10 \\
        \hline
        $M_{\mathrm{min}}^2$  & 0 & 0.16 & 0.35 & 0.51 & 0.65  \\
        \hline
\end{tabular}
\end{center}
\caption{Some $M^2_{\min}$ for different $t$ in units $\lms = 1$.}\label{tabelM}
\end{table}

\begin{figure}[H]
   \centering
       \includegraphics[width=8cm]{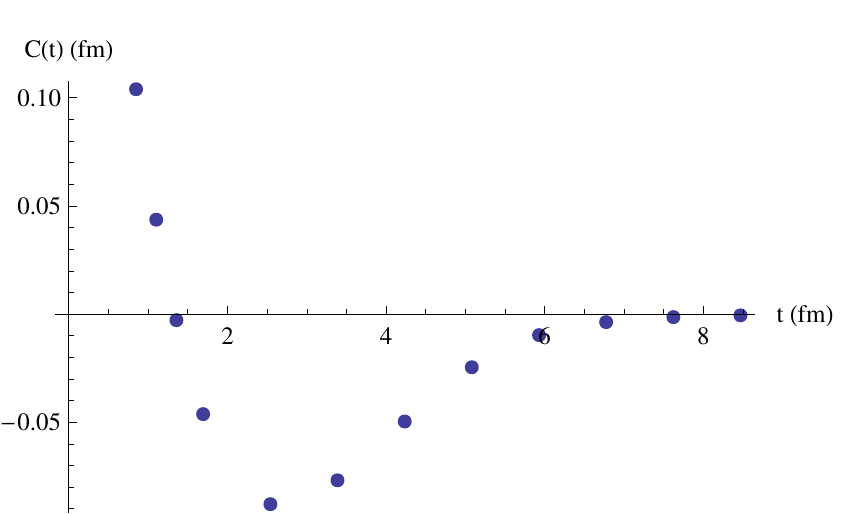}
   \caption{$\mathcal{C}(t)$ (fm) in function of $t$ (fm).}
   \label{ct}
\end{figure}

\begin{figure}[H]
   \centering
       \includegraphics[width=8cm]{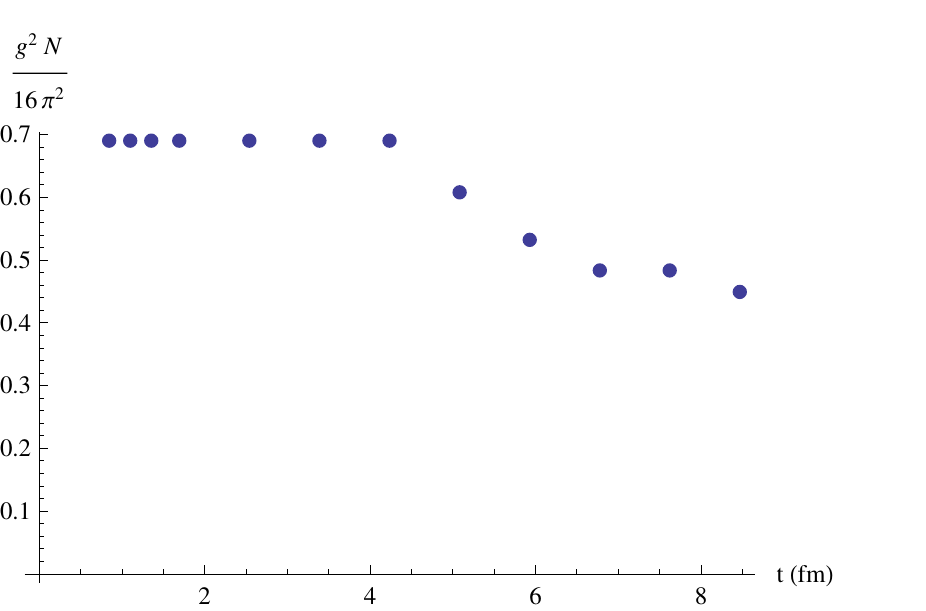}
   \caption{$g^2N/(16 \pi^2)$ in function of $t$ (fm).}
   \label{ctg}
\end{figure}

\noindent The corresponding $\mathcal{C}(t,M_{\mathrm{min}}^2)$ is depicted in \mbox{Figure \ref{ct}}. Both the $x$-axis and $y$-axis are shown in units fm ($1/\lms = 0.847$ fm), in order to compare our results with \cite{Bowman:2007du, Silva:2006bs}.  Not only do we find a positivity violating gluon propagator as $\mathcal{C}(t)$ becomes negative, but even the shape of this function is consistent with the lattice results\footnote{\cite{Bowman:2007du} included quarks, while \cite{Silva:2006bs} considered gluodynamics as we are studying in this thesis.} \cite{Bowman:2007du, Silva:2006bs}. Moreover, in \cite{Bowman:2007du, Silva:2006bs}, the positivity violation starts from $t \sim 1.5$ fm, in good agreement with our results. Finally, Figure \ref{ctg} displays the corresponding values of $g^2 N/16 \pi^2$. We can conclude that the previous  results are reliable for $t \lesssim 8$ as $g^2N/16 \pi^2$ is smaller than one.

\subsection{A remark about the strong coupling constant}
A renormalization group invariant definition of an effective strong coupling constant $g^2_{\mathrm{eff}}$ can be written down from the knowledge of the gluon and ghost propagators as
\begin{equation}
g^2_{\textrm{eff}}(p^2)=g^2 \left(\overline \mu^2\right) \widetilde{\mathcal{D}}\left(p^2,\overline \mu^2\right) \widetilde{\mathcal{G}}^2 \left(p^2,\overline \mu^2\right)\;,\label{coup}
\end{equation}
see e.g. \cite{Alkofer:2000wg}. $\widetilde{\mathcal{D}}$ and $\widetilde{\mathcal{G}}$ stand for the gluon and ghost form factor, defined by
\begin{eqnarray}
  \widetilde{\mathcal{D}}(p^2) &=& p^2\mathcal{D}(p^2)\;, \nonumber\\
  \widetilde{\mathcal{G}}(p^2) &=& p^2\mathcal{G}(p^2)\;.
\end{eqnarray}
The definition \eqref{coup} represents a kind of nonperturbative extension of the nonrenormalization of the ghost-gluon vertex. At the perturbative level, this is assured by the identity\footnote{This identity is also valid for the Yang-Mills action is the Landau gauge and is independent from the GZ action.} $Z_g=Z_c^{-1}Z_A^{-1/2}$, see \eqref{Z2}. Usually, this is assumed to remain valid at the nonperturbative level. Although this cannot be proven, this hypothesis has been corroborated by lattice studies like \cite{Cucchieri:2006tf,Cucchieri:2008qm}.\\
\\
In recent years, there was accumulating evidence that $g^2_{\textrm{eff}}(p^2)$ would reach an infrared fixed point different from zero: see e.g. \cite{Alkofer:2000wg, Lerche:2002ep,Zwanziger:2001kw,Pawlowski:2003hq} for a Schwinger-Dyson analysis, \cite{Gracey:2006dr,Gracey:2007bj} in the ordinary Gribov-Zwanziger approach and \cite{Bloch:2003sk,Furui:2004cx} for lattice results. These studies are mostly done in a MOM renormalization scheme, with the exception of \cite{Gracey:2006dr} where the $\MSbar$ scheme was employed. The manifestation of this infrared fixed point was motivated in Schwinger-Dyson studies and the ordinary Gribov-Zwanziger case by means of the power law behavior of the form factors,
\begin{eqnarray}
\widetilde{\mathcal{D}}(p^2)_{p^2\approx0}&\propto&\left(p^2\right)^{2\alpha}\;,
\nonumber\\
\widetilde{\mathcal{G}}(p^2)_{p^2\approx0}&\propto&\left(p^2\right)^{-\alpha}\;,\label{form2}
\end{eqnarray}
being expressible in terms of a single exponent $\alpha$. The Schwinger-Dyson community heralded in a variety of studies the value $\alpha\approx0.595$, whereas the Gribov-Zwanziger scenario gives $\alpha=1$. Anyhow, substituting a behavior like \eqref{form2} into the definition \eqref{coup} leads to $g^2_{\textrm{eff}}(p^2)_{p^2\approx 0}\propto (p^2)^0$, opening the door for a finite value.\\
\\
However, once again quoting the more recent large volume lattice data of \cite{Cucchieri:2007md,Cucchieri:2007rg,Cucchieri:2008qm,Cucchieri:2008fc}, the power law behavior \eqref{form2} seem to be excluded in favor of
\begin{eqnarray}
\widetilde{\mathcal{D}}(p^2)_{p^2\approx0}&\propto&p^2\;,
\nonumber\\
\widetilde{\mathcal{G}}(p^2)_{p^2\approx0}&\propto&\left(p^2\right)^{0}\;,\label{form3}
\end{eqnarray}
leading to a vanishing infrared effective strong coupling constant at zero momentum since $g^2_{\textrm{eff}}(p^2)_{p^2\approx 0}\propto p^2$. The refined analysis in this paper of the extended Gribov-Zwanziger action, including an additional dynamical effect, allows us to draw a similar conclusion up to the one loop level, i.e. an infrared vanishing $g^2_{\textrm{eff}}$. Certain lattice studies also pointed towards this particular scenario \cite{Ilgenfritz:2006gp}.

\subsection{Conclusion in 4d}
We can clearly conclude that by taking into account a condensation of the operator $\left(\overline{\varphi}\varphi-\overline{\omega}\omega\right)$ (which is already present perturbatively), it is possible to obtain a gluon propagator which is non-vanishing at zero momentum and a ghost propagator which is no longer enhanced. However, our values obtained are not very satisfactory and can only be seen in a qualitative nature. We shall namely show later that the condensate $\braket{A^2}$ plays an important role, while in all calculations, we have neglected this condensate as the calculations became too involved.\\
\\
For completeness, we have also put all the refined propagators in the appendix \ref{propagatorsRGZ}.

\section{Refinement of the GZ action in 3 dimensions}
\subsection{Introduction}
We are of course curious to see what happens in 3 dimensions. As described in section \ref{chap2lattice}, in 3 dimensions, the lattice data have found the same results, i.e.~a non enhanced ghost propagator and an infrared suppressed gluon propagator which is non-vanishing at zero momentum. Therefore, we shall again consider the action $S_\RGZ$,
\begin{eqnarray}\label{RGZ3d}
S_\RGZ &=& S_\GZ + S_{\overline{\varphi} \varphi}, \nonumber\\
S_{\overline{\varphi} \varphi} &=& \int \d^d x \left[s(-J \overline{\omega}^a_i \varphi^a_{i}) \right]  = \int \d^d x \left[-J\left( \overline{\varphi}^a_i \varphi^a_{i} - \overline{\omega}^a_i \omega^a_i \right) \right]\;,
\end{eqnarray}
as given in expression \eqref{nact}, whereby we have set $m^2$ immediately equal to zero in this section. $d$ is now equal to 3. \\
\\
Besides the fact that the dimensionality of many quantities change, e.g.~$\dim(g^2)=1$ and $\dim(\gamma^2)=3/2$, the proof of the renormalization is completely the same as in the previous section \eqref{chap2renrom}. Therefore, we shall not repeat this proof in this section.

\subsection{The condensate at the perturbative level}
We can again show that the local composite operator (LCO) $(\overline{\varphi}\varphi-\overline{\omega}\omega)$ has already a non-vanishing perturbative expectation value.  We recall that
\begin{equation}\label{apot4}
    \braket{\overline{\varphi}\varphi-\overline{\omega}\omega}_{\mathrm{pert}}=-\left.\frac{\p W(J)}{\p
    J}\right\vert_{J=0} \;,
\end{equation}
with $W(J)$ the generating functional,
\begin{eqnarray}\label{genfunc}
\e^{-W(J)} &=& \int [\d \Psi] \e^{-S_\RGZ}\;,
\end{eqnarray}
and with $S_\RGZ$ the extended Gribov-Zwanziger action given in \eqref{RGZ3d}. The lowest order expression for $W(J)$ has been calculated in the appendix \ref{sigma}\ref{appendix3}, where we have found in expression \eqref{Wnull3d},
\begin{equation}
    W^{(0)}(J)=-3(N^2-1)\frac{\lambda^4}{2g^2N}+\frac{N^2-1}{6\pi}\left(-m_1^3-m_2^3+J^{3/2}\right) \;.
\end{equation}
For $m^2 = 0$, $m_1$ and $m_2$ are given by
\begin{align}
m_1^2 &= \frac{  J - \sqrt{J^2  - 4 \lambda^4  }}{2} \;, & m_2^2 &= \frac{ J + \sqrt{J^2  - 4 \lambda^4  }}{2} \;.
\end{align}
Using this explicit expression, we can easily obtain the perturbative value of the condensate \eqref{apot4}, reading
\begin{equation}\label{condpert2}
\braket{\overline{\varphi}\varphi-\overline{\omega}\omega}_{\mathrm{pert}}=\sqrt{2}\frac{N^2-1}{8\pi}\lambda\approx 0.056(N^2-1)\lambda\;,
\end{equation}
where $\lambda$ is the nonzero solution of $\frac{\p\Gamma(\lambda)}{\p \lambda}=0$. Since at one loop
\begin{eqnarray}\label{gribovpuur}
 \Gamma(\lambda)&=&-d(N^2-1)\frac{\lambda^4}{2Ng^2}+\frac{N^2-1}{2}(d-1)\int \frac{\d^d q}{(2\pi)^d}\ln\left(q^4+\lambda^4 \right)\\
 &=&-3(N^2-1)\frac{\lambda^4}{2g^2N}+\frac{\sqrt{2}}{6\pi}(N^2-1)\lambda^3\;,
\end{eqnarray}
we find that
\begin{equation}\label{gribovpuuropl}
    \lambda=\frac{\sqrt{2}}{12\pi}g^2N \;,
\end{equation}
with
\begin{equation}\label{gribovpuurvac}
    E_{\mathrm{vac}}=g^6\frac{N^3(N^2-1)}{10368\pi^4}>0 \;.
\end{equation}
We notice that the one loop vacuum energy corresponding to the Gribov-Zwanziger action is \emph{positive}\footnote{The same feature was also observed in the one loop 4d case, see \eqref{positiveenergy}.}.

\subsection{Infrared problems in 3d}
Notice that in the case of $\gamma^2=0$, the 3d theory will not be well defined. In the absence of an infrared regulator, the perturbation theory of a super-renormalizable 3d gauge theory is ill-defined due to severe infrared instabilities \cite{Jackiw:1980kv}. This can be intuitively understood as the coupling constant $g^2$ carries the dimension of a mass. In the absence of an infrared regulator, the effective expansion parameter will look like $g^2/p$ with $p$ a certain (combination of) external momentum/-a. For $p\gg g^2$, very good ultraviolet behavior is apparent, but for $p\ll g^2$, infrared problems emerge. The presence of (a) dynamical mass scale(s) $m\propto g^2$ could ensure a sensible perturbation series, even for small $p$, as a natural expansion parameter is then provided by $g^2/m$. From this perspective, a nonvanishing Gribov mass $\gamma^2$ could also serve as infrared cut-off. This feature is also explicitly seen from equation \eqref{gribovpuuropl}, from which an effective dimensionless expansion parameter can be derived as
\begin{eqnarray}\label{exppar}
  \frac{g^2N}{(4\pi)^{3/2}\lambda} &=&\frac{3}{2\sqrt{2\pi}}\approx 0.6\;,
\end{eqnarray}
a quantity which is at least smaller than $1$. The inverse factor $(4\pi)^{3/2}$ is the generic loop integration factor generated in 3d.

\subsection{Modifying the effective action in order to stay within the horizon}
Similar as in section \ref{chap3modifying}, we need to add an extra term to the action in order assure that we are still inside the horizon. We define the new action again as
\begin{eqnarray}\label{RGZprime3d}
S_\RGZ' &=& S_\RGZ + S_{\en}\;,
\end{eqnarray}
with
\begin{eqnarray}\label{nieuweterm3d}
S_{\en} &=& 2 \frac{d (N^2 -1)}{\sqrt{2 g^2 N}}  \int \d^d x\ \varsigma \ \gamma^2 J \;.
\end{eqnarray}
We recall the following boundary condition which assures a smooth limit
\begin{eqnarray} \label{bound3d}
\left.\frac{\partial \sigma(0)}{\partial M^2} \right|_{M^2=0}&=& 0\;.
\end{eqnarray}

\subsection{The gluon and the ghost propagator}
\subsubsection{The gluon propagator}
The tree level gluon propagator corresponding to the action \eqref{RGZprime3d} is again given by
\begin{equation}
    \mathcal{D}(p^2)=\frac{p^2 + M^2}{p^4 + M^2p^2 + \lambda^4}\;,
\end{equation}
see expression \eqref{gluonprop2} and enjoys the same properties as described in the 4d case on page \pageref{gluonprop2}, namely
\begin{itemize}
\item $D(p^2)$ is infrared suppressed due to the presence of the mass scales $M^2$ and $\lambda^4$.
\item $D(0)=\frac{M^2}{\lambda^4}$, i.e. the gluon propagator does not vanish at zero momentum if $M^2$ is different from zero.
\end{itemize}
These properties seem to be in qualitative accordance with the lattice data \cite{Cucchieri:2007md,Bogolubsky:2007ud,Cucchieri:2007rg}. We also want to stress that the mass term related to $\overline{\varphi}\varphi-\overline{\omega}\omega$ plays a crucial role in having $ D(0)\neq0$, since in the standard Gribov-Zwanziger scenario, the gluon propagator necessarily goes to zero.

\subsubsection{The ghost propagator}
In 3d, the one loop corrected ghost propagator can again be written as
\begin{eqnarray}
\mathcal{G}(k^2) &=&  \frac{1}{k^2} \frac{1}{1- \sigma(k^2)}\;,
\end{eqnarray}
where $\sigma(k^2)$ is the following momentum dependent function
\begin{eqnarray}\label{sigmak}
\sigma(k^2) &=& g^2N \frac{k_{\mu} k_{\nu}}{k^2} \int \frac{\d^3 q}{(2\pi)^3} \frac{1}{(k-q)^2} \frac{q^2 + M^2}{q^4 + M^2q^2 + \lambda^4}\mathcal{P}_{\mu\nu}(q) \;.
\end{eqnarray}
Calculating this integral explicitly, we find \small
\begin{multline}\label{klo}
\sigma(k^2)= \frac{N g^2}{32 k^3 \pi} \Biggl\{ \frac{1}{M^2- \sqrt{M^4 - 4 \lambda^4}} \left( 1 + \frac{M^2}{\sqrt{M^4 - 4 \lambda^4}} \right) \Biggl[ \sqrt{2} k^3 \sqrt{M^2 - \sqrt{M^4 - 4 \lambda^4}}\\
 - \frac{k}{\sqrt{2}} \left( M^2 - \sqrt{M^4 - 4 \lambda^4}\right)^{3/2} - k^4 \pi + \frac{1}{2} \left(2 k^2 + M^2 - \sqrt{M^4 - 4 \lambda^4} \right)^2 \arctan \frac{\sqrt{2} k}{ \sqrt{M^2 - \sqrt{M^4 - 4 \lambda^4}}} \Biggr]\\
+ \frac{1}{M^2 + \sqrt{M^4 - 4 \lambda^4}}  \left( 1 - \frac{M^2}{\sqrt{M^4 - 4 \lambda^4}} \right) \Biggl[ \sqrt{2} k^3 \sqrt{M^2 + \sqrt{M^4 - 4 \lambda^4}} - \frac{k}{\sqrt{2}} \left( M^2 + \sqrt{M^4 - 4 \lambda^4}\right)^{3/2} \\
- k^4 \pi + \frac{1}{2} \left(2 k^2 + M^2 + \sqrt{M^4 - 4 \lambda^4} \right)^2 \arctan \frac{\sqrt{2} k}{\sqrt{M^2 + \sqrt{M^4 - 4 \lambda^4}}} \Biggr]\Biggr\} \;.
\end{multline} \normalsize
In order to find the behavior of the ghost propagator near zero momentum we take the limit $k^2 \to 0$ in equation \eqref{sigmak},
\begin{eqnarray}\label{gg}
\sigma(0) &=& g^2N \frac{2}{3}\int\frac{\d^3q}{(2\pi)^3}\frac{1}{q^2}\frac{q^2 + M^2}{q^4+ M^2 q^2 + \lambda^4} ~=~  \frac{ g^2N}{6\pi} \frac{M^2 +\lambda^2}{\lambda^2 \sqrt{M^2 + 2 \lambda^2}}\;,
\end{eqnarray}
which can of course be obtained by taking the limit $k^2 \to 0$ in expression \eqref{klo}. Similarly, one can check that $\sigma\to0$ for $k^2\to\infty$ and/or $M^2\to\infty$.\\
\\
Before drawing any conclusions, we still need to have a look at the gap equations, which shall fix $\lambda^2$ as a function of $M^2$.

\paragraph{The gap equations} \ \\
We begin with the first gap equation \eqref{gapgamma} in order to express $\lambda$ as a function of $M^2$. The effective action at one loop order is given by
\begin{equation*}
\Gamma_\gamma^{(1)} = -d(N^{2}-1)\gamma^{4} + 2\frac{d (N^2-1)}{\sqrt{2 g^2 N}}  \varsigma \ \gamma^2 M^2+\frac{(N^{2}-1)}{2}\left( d-1\right) \int \frac{\d^{d}q}{\left(
2\pi \right) ^{d}} \ln \frac{q^4 + M^2 q^2 + 2 g^2 N \gamma^2}{ q^2+ M^2} \;.
\end{equation*}
With $\lambda^4 = 2 g^2 N \gamma^4$, we rewrite the previous expression,
\begin{equation*}
\mathcal{E}^{(1)} =  \frac{\Gamma_\gamma^{(1)}}{N^2 - 1} \frac{2 g^2 N}{d} ~=~ - \lambda^4  + 2 \varsigma \lambda^2 M^2 + g^2 N \frac{ d-1}{d} \int \frac{\d^{d}q}{\left( 2\pi \right) ^{d}} \ln \frac{q^4 + M^2 q^2 + \lambda^4}{ q^2 + M^2} \;,
\end{equation*}
and apply the gap equation \eqref{gapgamma},
\begin{eqnarray}\label{gapintegraal}
0&=&  -1 + \varsigma \frac{M^2}{\lambda^2} + g^2 N \frac{d-1}{d} \int \frac{\d^{d}q}{\left(2\pi \right) ^{d}}  \frac{1}{q^4 + M^2 q^2+ \lambda^4} \;.
\end{eqnarray}
In 3 dimensions, the integral in this gap equation is finite, resulting in,
\begin{eqnarray}\label{gapuitgerekend}
0&=&  -1 + \varsigma \frac{M^2}{\lambda^2} + \frac{g^2 N}{6\pi}
\frac{1}{\sqrt{M^2 + 2 \lambda^2}} \;.
\end{eqnarray}
This expression will fix $\lambda^2$ as a function of $M^2$, i.e.~$\lambda^2(M^2)$ once we have found an explicit value for $\varsigma$.\\
\\
This explicit value for $\varsigma$ will be provided by the second gap equation \eqref{bound}. From expression \eqref{sigmak}, one finds
\begin{align*}
\sigma(0) &= g^2N \frac{d- 1}{d}\int\frac{\d^dq}{(2\pi)^d}\frac{1}{q^2}\frac{q^2 + M^2}{q^4+ M^2 q^2 + \lambda^4} \nonumber\\
   &= g^2N  \frac{d-1}{d} \int\frac{\d^dq}{(2\pi)^d}\frac{1}{q^4+ M^2 q^2 + \lambda^4} + M^2 g^2N \frac{d-1}{d}\int\frac{\d^dq}{(2\pi)^d}\frac{1}{q^2}\frac{1}{q^4+ M^2 q^2 + \lambda^4}\;.
\end{align*}
Therefore, we can rewrite the first gap equation
\eqref{gapintegraal} as
\begin{equation}\label{sigmazero}
    0~=~\sigma(0) - 1-M^2\frac{d-1}{d}g^2N\int \frac{\d^dq}{(2\pi)^d}\frac{1}{q^2}\frac{1}{q^4+M^2q^2+\lambda^4}+\varsigma\frac{M^2}{\lambda^2} \;.
\end{equation}
The second gap equation can then subsequently be obtained by acting with $\frac{\p}{\p M^2}$ on the previous expression and setting $M^2=0$. Doing so, we find
\begin{equation}\label{4simpel}
    -\frac{d-1}{d}g^2N\int \frac{\d^dq}{(2\pi)^d}\frac{1}{q^2}\frac{1}{q^4+\lambda^4(0)}+\varsigma\frac{1}{\lambda^2(0)}~=~0\;,
\end{equation}
by keeping \eqref{bound} in mind. Setting $M^2 = 0$ in \eqref{gapuitgerekend} yields,
\begin{eqnarray}
\sqrt{2\lambda^2(0)} &=&   \frac{g^2 N}{6 \pi}\;.
\end{eqnarray}
Proceeding with equation \eqref{4simpel}, we find the following simple solution for $\varsigma$,
\begin{equation}\label{varsigma}
    \varsigma~=~\frac{1}{12\pi}\frac{2}{\sqrt{2}}\frac{g^2N}{\lambda(0)}=1\;.
\end{equation}
In summary, the following expression,
\begin{eqnarray}\label{gapfinal}
 \frac{g^2 N}{6\pi} \frac{1}{\sqrt{M^2 + 2 \lambda^2}} &=&   1 -  \frac{M^2}{\lambda^2} \;,
\end{eqnarray}
fixes $\lambda^2(M^2)$.

\paragraph{The ghost propagator at zero momentum}\ \\
At this point, we have all the information we need to take a closer look at the ghost propagator at zero momentum. From equation \eqref{gg} and \eqref{gapfinal}, we find
\begin{eqnarray}
    \sigma(0) &=& \left( \frac{M^2}{\lambda^2} + 1 \right)  \left(  1 -  \frac{M^2}{\lambda^2} \right) ~=~ 1 -  \frac{M^4}{\lambda^4}\;.
\end{eqnarray}
From this expression we can make several observations. Firstly, when $M^2 =0$, we find that $\sigma(0) = 1$, which is exactly the result obtained in the original Gribov-Zwanziger action \cite{Gribov:1977wm,Zwanziger:1992qr}. Consequently, the ghost propagator \eqref{ghost} is enhanced and behaves like $1/k^4$ in the low momentum region. Secondly, for any $M^2 >0$, $\sigma(0)$ is smaller than 1. By contrast, without the inclusion of the extra vacuum term, $\sigma(0)$ would always be bigger than 1, which can be observed from expression \eqref{sigmazero}. Therefore, it is absolutely necessary to include this term. With $\sigma(0)$ smaller than 1, the ghost propagator is not enhanced and behaves as $1/k^2$.

\subsection{A dynamical value for $M^2$}
We shall follow section \ref{sectie4} in order to obtain a dynamical value for $M^2$ in 3d. Therefore, we introduce $M^2$ as a variational parameter into the theory by replacing the action $S_\RGZ$ by
\begin{equation}\label{var}
S_{GZ} +   (1-\ell^k)M^2 \int \d^d x \left[ - \left(\overline{\varphi}^a_i \varphi^a_{i} - \overline{\omega}^a_i \omega^a_i\right)   + 2 \frac{d (N^2 -1)}{\sqrt{2 g^2 N}} \varsigma \ \gamma^2 \right]\;,
\end{equation}
where $\ell$ serves as the loop counting parameter, and formally $\ell=1$ at the end. In this fashion, it is clear that the original starting action $S_{\GZ}$ has not been
changed. We have in fact added the terms in $M^2$, and subtracted them again at $k$ orders higher in the loop expansion. We shall set $k = 2$ as we are working up to one loop. Taking $k=1$ would destroy the effect of the vacuum term $S_{\en}$, which would be inconsistent as explained before.

\subsubsection{The ghost propagator}
\begin{figure}[H]
  \centering
     \includegraphics[width=7cm]{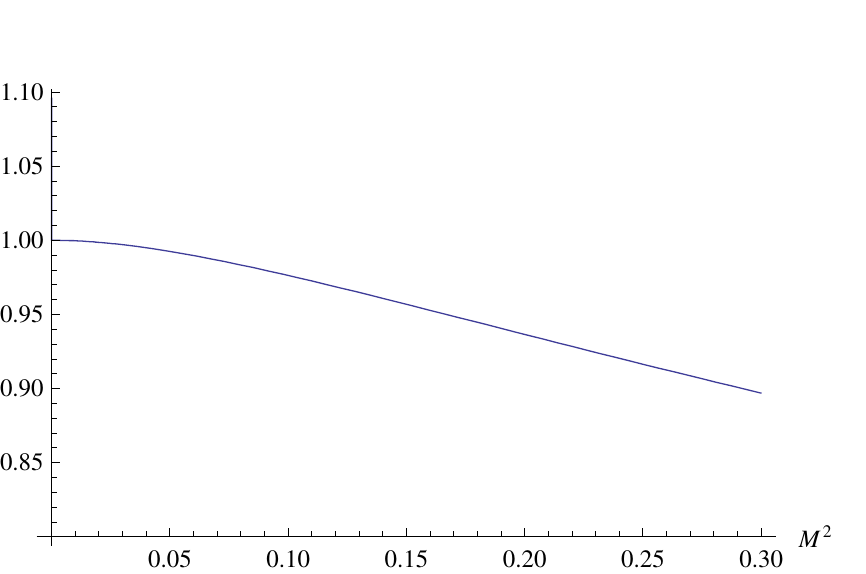}
     \includegraphics[width=7cm]{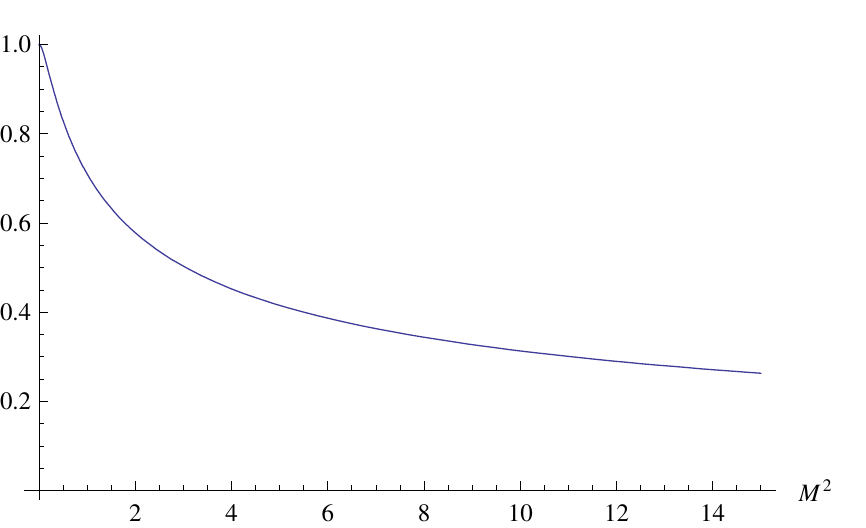}
  \caption{$\sigma(0)$ in function of $M^2$ in units $\frac{g^2 N}{6 \pi} =1$.}  \label{3fig1}
\end{figure}

In order to obtain a dynamical value for the ghost propagator, we start with the expression \eqref{sigmak} for $\sigma(k^2)$ and replace $g^2$ with $\ell g^2$ and $M^2$ with $(1-\ell^2)M^2$, expand up to order $\ell$ and set $\ell=1$. Doing so, we recover again the same expression \eqref{sigmak}. Following an analogous procedure for the gap equations, also results in the same expression \eqref{gapfinal}. This latter equation determines $\lambda^2(M^2)$, which can be plugged in the expression of $\sigma(k^2)$, thereby making
$\sigma(k^2)$ only a function of $M^2$, next to the momentum dependence. \\
\\
Firstly, let us investigate $\sigma(k^2)$ at zero momentum, which is the key-point of this paper. Figure \ref{3fig1} displays $\sigma(0)$ as a function of $M^2$. We observe that $\sigma(0)$ is indeed smaller than 1 for all $M^2 > 0$ as already shown analytically in the previous section. We also find a smooth limit of $\sigma(0)$ for $M^2 \to 0$ required by the second gap equation \eqref{bound}, as can be seen from the left figure. According to the principle of minimal sensitivity, we have to search for an extremum, i.e.~$\frac{\p  \sigma(0)}{\p M^2 }=0$. Unfortunately, there is no such an extremum present.  Nevertheless, we do find a point of inflection, $\frac{\p^2 \sigma(0)}{ (\p M^2)^2 }=0$ at $M^2 = 0.185 \left(\frac{g^2N}{6\pi}\right)^2$. Taking this value for $M^2$, we find
\begin{eqnarray}
\sigma(0) &=&  0.94\;,
\end{eqnarray}
which is indeed smaller than one and results in a non-enhanced behavior of the ghost propagator. This value needs to be compared with the lattice value of $\sigma(0) = 0.79$, which can be extracted from the data in \cite{Cucchieri:2008fc}.

\begin{figure}[H]
  \centering
      \includegraphics[width=10cm]{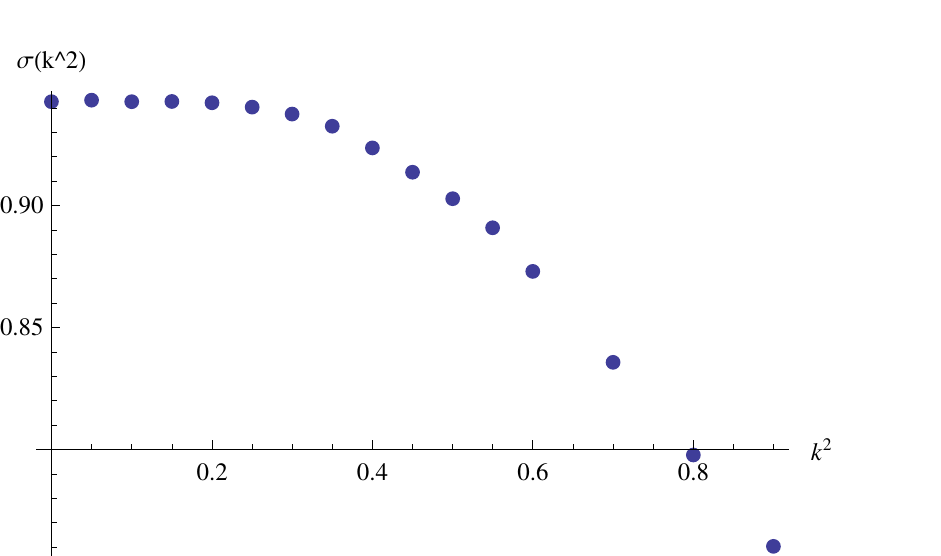}
  \caption{The optimal $\sigma(k^2)$ in function of $k^2$ in units of $\frac{g^2 N}{6 \pi }=1$.}  \label{fig2}
\end{figure}

\noindent Secondly, let us have a look at this point of inflection, when ``turning on'' the momentum $k^2$. As shown in Table \ref{table1}, we observe that $M^2$ will decrease, until it will vanish at $k^2 \approx 0.55$. The corresponding $\sigma(k^2)$ is displayed in Figure \ref{fig2}. We thus see that we find some kind of a momentum dependent effective mass $M^2(k^2)$, which disappears when $k$ grows. This could have been anticipated, as we naturally expect that the deep ultraviolet sector should be hardly affected. Let us also notice here that $\sigma(k^2)$ is a decreasing function, as can be explicitly checked from Figure \ref{fig2}. This of course means that we are staying within the horizon for any value of the momentum.

\begin{table}[H]
\renewcommand{\arraystretch}{2}
\begin{center}
\begin{tabular}{|c||cccccccccccc|}
        \hline
        $k^2 $& 0 & 0.05 & 0.1 & 0.15 & 0.2 & 0.25 & 0.30 & 0.35 & 0.40 & 0.45 &0.50 & 0.55  \\
        \hline
        $M_{\mathrm{min}}^2$& 0.19 & 0.16 & 0.14 & 0.12 & 0.10 & 0.08 & 0.06 & 0.04 & 0.03 & 0.02 & 0.01 & 0 \\
        \hline
\end{tabular}
\end{center}
\caption{Some $M^2_{\min}$ for different $k^2$ in units $\frac{g^2 N}{6 \pi}=1$.}\label{table1}
\end{table}

\begin{figure}[H]
  \centering
      \includegraphics[width=10cm]{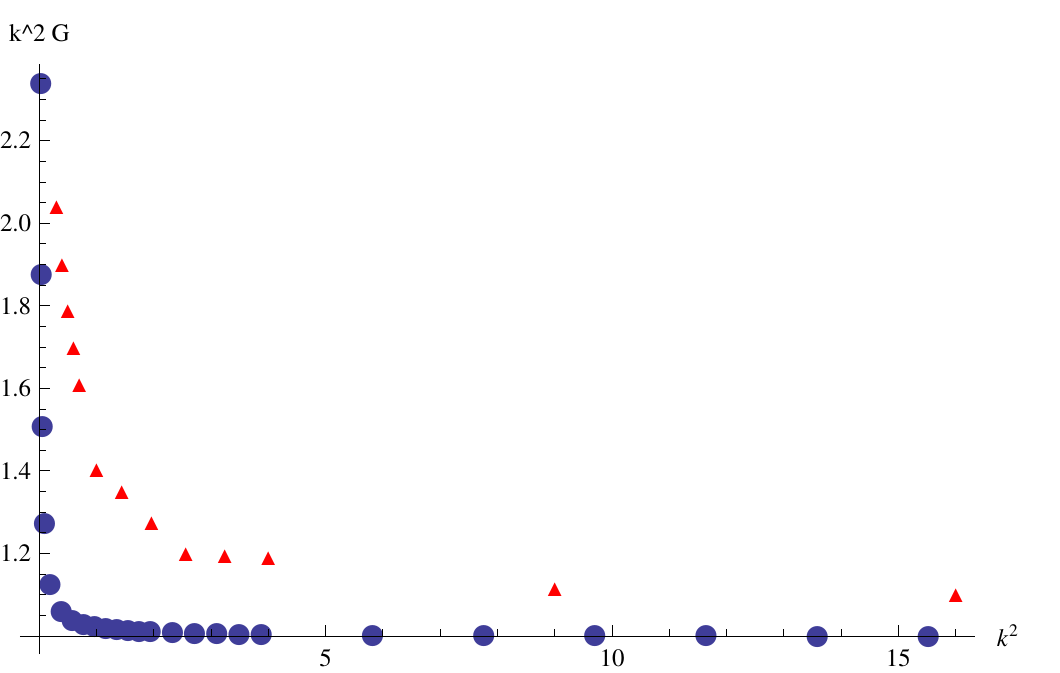}
  \caption{The optimal $k^2 \mathcal{G}(k^2)$ in function of $k^2$ in units of GeV. The lattice (our analytical) results are indicated with triangles (dots). The error bars on the
  lattice data are roughly of the size of the triangles. }  \label{fig3}
\end{figure}

\begin{figure}[H]
  \centering
      \includegraphics[width=10cm]{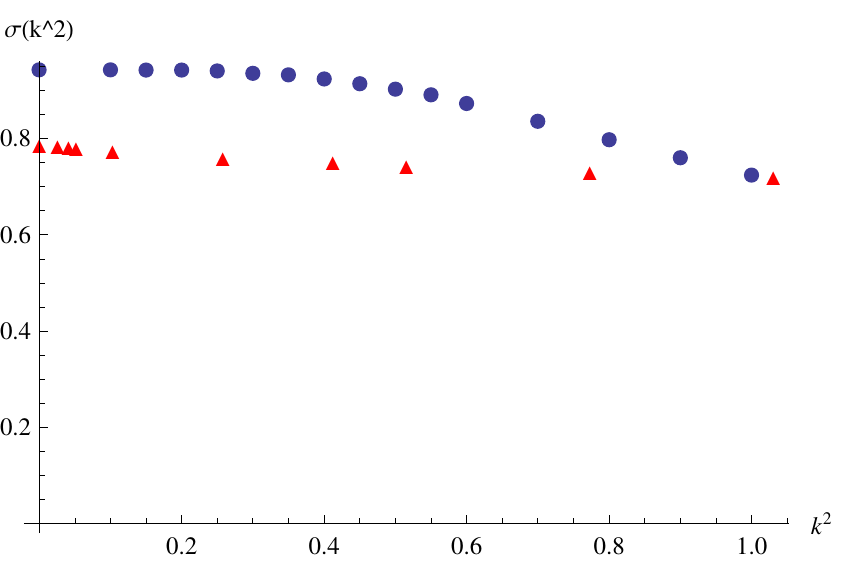}
  \caption{The optimal $\sigma(k^2)$ in function of $k^2$ in units of $\frac{g^2 N}{6 \pi }=1$. The lattice (our analytical) results are indicated with triangles (dots).}  \label{fig3accent}
\end{figure}

\noindent To compare our results with available lattice data, we must make the conversion to physical units of GeV. In \cite{Lucini:2002wg} a continuum extrapolated value for the ratio $\sqrt{\sigma}/g^2$ was given for several gauge groups; in particular, $\sqrt{\sigma}/g^2 \approx0.3351$ for $SU(2)$. Further, $\sqrt{\sigma}$ stands for the square root of the string tension. For this quantity, we used the input value of $\sqrt{\sigma} = 0.44$ GeV as in \cite{Cucchieri:2004mf}. Therefore, for $SU(2)$, we find,
\begin{eqnarray}
\left(\frac{g^2 N}{6 \pi}\right)^2 & \approx & 0.0194 \mathrm{GeV}^2\;.
\end{eqnarray}
In Figure \ref{fig3}, we have plotted the lattice as well our analytical result for the ghost dressing function, $k^2\mathcal{G}(k^2)$, in units of GeV. We used the numerical data
of \cite{Cucchieri:2008fc,Cucchieri:2008qm}, adapted to our needs. We observe that for sufficiently large $k^2$, the lattice data and our analytical results converge. In this case, the novel mass $M^2$ becomes zero as advocated earlier, meaning that we are back in the usual Gribov-Zwanziger scenario. For smaller $k^2$, we found it more instructive to compare the lattice estimate of $\sigma(k^2)$ with our value as the errors on $k^2\mathcal{G}(k^2) = \frac{1}{1- \sigma(k^2)}$ are becoming large when looking at  $\sigma(k^2)$ close to 1. Figure \ref{fig3accent} displays $\sigma(k^2)$ in units of $\frac{g^2 N}{6 \pi }=1$ up to $k^2 = 1 \times \left(\frac{g^2 N}{6 \pi }\right)^2 = 0.0194\ \mathrm{GeV}^2$. We see that both results are in reasonable agreement, especially if we keep in mind that we have only calculated $\sigma(k^2)$ in a first order approximation.
\vspace{-0.5cm} \enlargethispage{\baselineskip}

\subsubsection{The gluon propagator}
\begin{figure}[H]
  \centering
      \includegraphics[width=10cm]{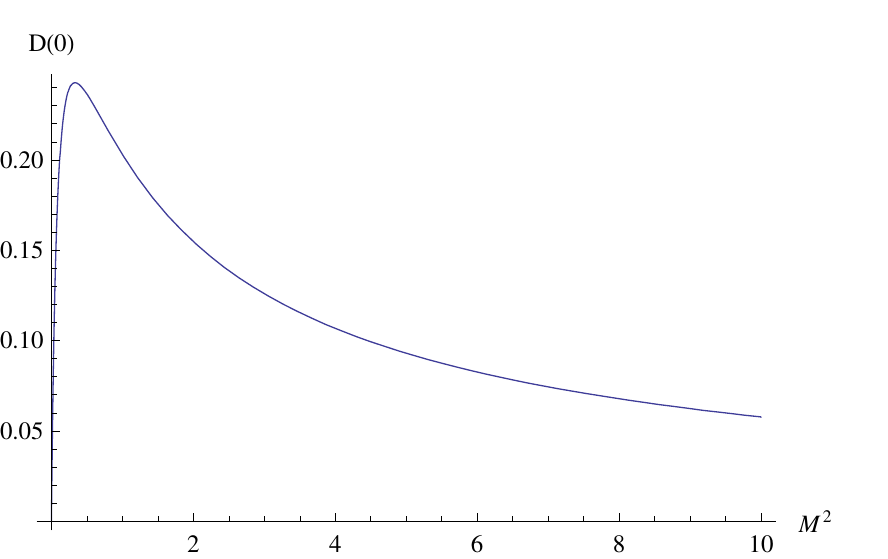}
  \caption{$\mathcal D(0)$ in units $\frac{g^2N}{6\pi}=1$.} \label{fig4}
\end{figure}

In order to apply an analogous procedure for the gluon propagator, we require its one loop correction. This correction can be found in \cite{Dudal:2008rm}\footnote{Afterwards, it has appeared that not all propagators were included in the calculations. Therefore, the results obtained here can only be considered in a qualitative nature. },
\begin{multline}\label{dnul}
\mathcal{D}(0) = \frac{M^2}{\lambda^4}+ \frac{g^2 N}{4 \pi} \frac{M^4}{\lambda^8} \left[\frac{ M^4}{\lambda^4}  \sqrt{M^2 + 2 \lambda^2}  -  \frac{M^5}{\lambda^4}  -  \frac{M^2}{\lambda^2} \sqrt{M^2 + 2 \lambda^2} -\frac{17}{12} M^2 \lambda^2 \frac{\sqrt{M^2 + 2 \lambda^2}}{M^4 - 4 \lambda^4}  \right. \\
\left. + \frac{13}{4} \lambda^4 \frac{\sqrt{M^2 + 2 \lambda^2}}{M^4 - 4 \lambda^4} -\frac{5}{3}\lambda^6 M^2 \frac{\sqrt{M^2 + 2 \lambda^2}}{(M^4-4 \lambda^4)^2}+ \frac{10}{3}\lambda^8 \frac{\sqrt{M^2 + 2 \lambda^2}}{(M^4 - 4 \lambda^4)^2} + \frac{7}{4} M - \frac{1}{4}\sqrt{M^2 + 2 \lambda^2} \right] \;.
\end{multline}
This propagator at zero momentum, $\mathcal{D}(0)$ is displayed in Figure \ref{fig4}. We immediately see that there is an extremum at $M^2 = 0.33\ \left(\frac{g^2 N}{6\pi}\right)^2$, resulting in
\begin{eqnarray}
\mathcal{D}(0) &=& \frac{0.24}{  \left(\frac{g^2 N}{6
\pi}\right)^2}.
\end{eqnarray}
Doing the conversion to physical units again, we find for $SU(2)$,
\begin{eqnarray} \label{eindres}
\mathcal{D}(0) &\approx&  \frac{12}{  \mathrm{GeV}^2}\,.
\end{eqnarray}
This value can be compared with the bounds derived from a partially numerical and partially analytical derivation \cite{Cucchieri:2007rg}:
\begin{eqnarray}
  \frac{1.2}{  \mathrm{GeV}^2} < \mathcal{D}(0) <  \frac{12}{
  \mathrm{GeV}^2}\;.
\end{eqnarray}
We recall that our value $\mathcal{D}(0) = 12 /\mathrm{GeV}^2$ is a first order approximation and is only of qualitative nature. Nevertheless, this value is still consistent with the boundaries of the lattice data set in \cite{Cucchieri:2007rg}.

\subsection{The temporal correlator: violation of positivity}
\begin{figure}[H]
  \begin{center}
    \subfigure[\;$t=1$]{\includegraphics[width=5cm]{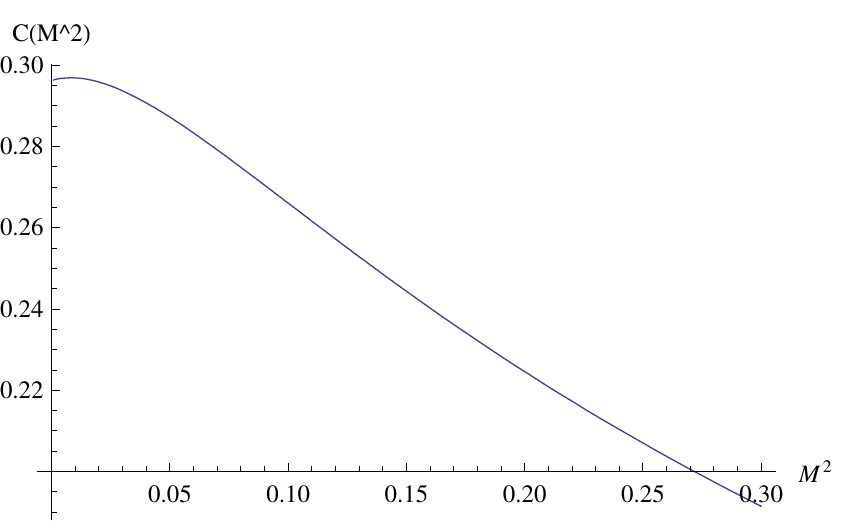} \label{fig1a}} \hspace{0.5cm}
    \subfigure[\;$t=1.5$]{\includegraphics[width=5cm]{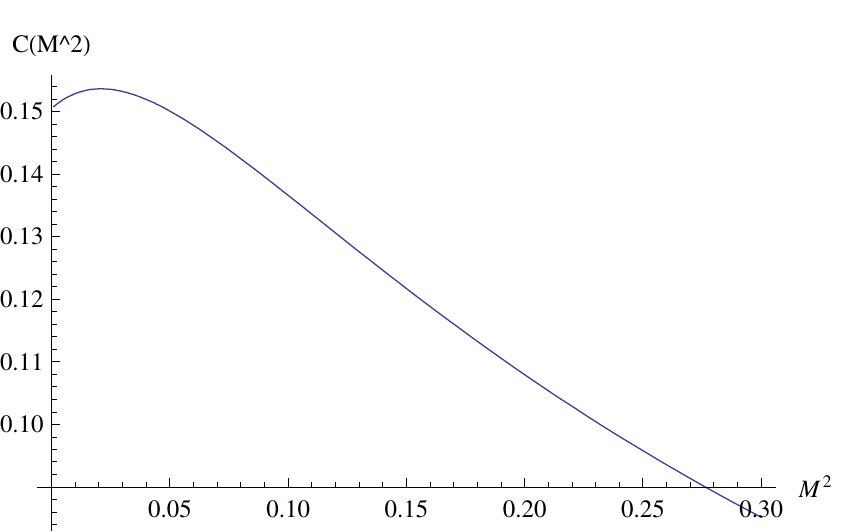}\label{fig1b}} \hspace{0.5cm}
    \subfigure[\;$t=2$]{\includegraphics[width=5cm]{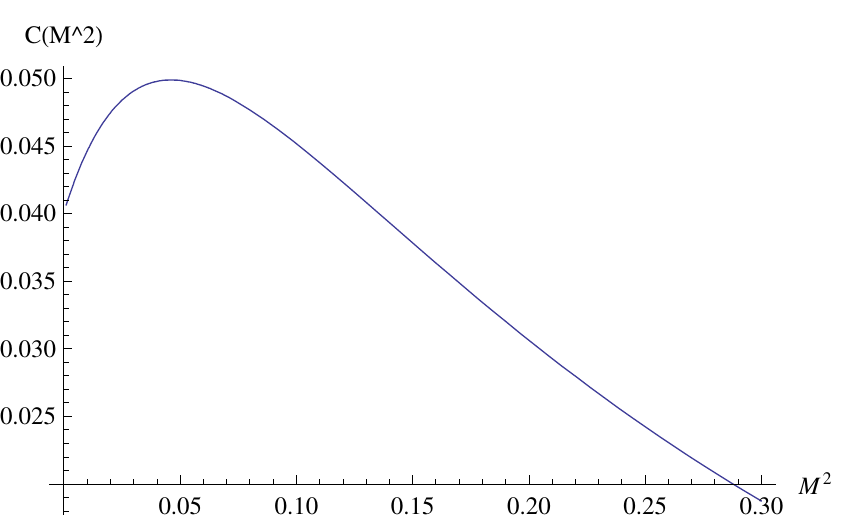}\label{fig1c}}
    \\
    \subfigure[\;$t=2.5$]{\includegraphics[width=5cm]{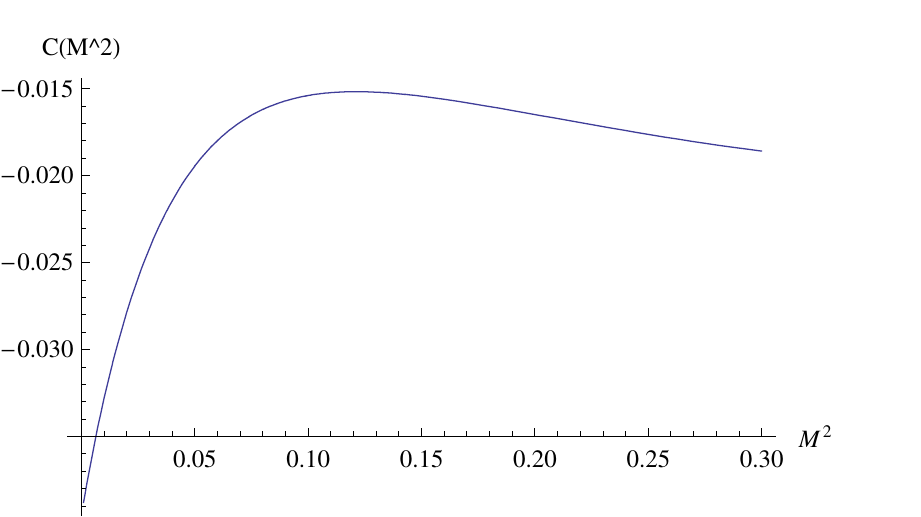}\label{fig1d}}
    \hspace{0.5cm}
    \subfigure[\;$t=3$]{\includegraphics[width=5cm]{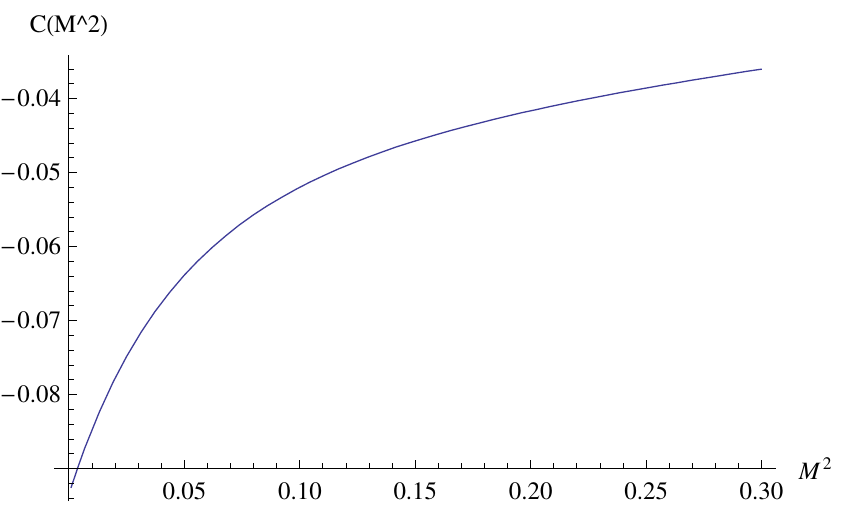}\label{fig1e}}
  \end{center} \vspace{-0.5cm}
  \caption{$\mathcal{C}(t,M^2)$ for a few values of $t$ in function of $M^2$, in units of fm.} \label{fig5}
\end{figure}

We shall investigate if a gluon propagator of the type \eqref{gluonprop2} displays a violation of positivity, another fact which is reported by the lattice data \cite{Cucchieri:2004mf}.  We shall again calculate the $1d$ Fourier transformation of $D(p^2)$, see expression \eqref{tempcorrelator}. In Figure \ref{fig5} the Fourier transforms, $\mathcal C (t, M^2)$ are shown for different $t$ in units of fm.

\begin{figure}[H]
  \begin{center}
    \subfigure[\;Refined case]{\includegraphics[width=5.5cm]{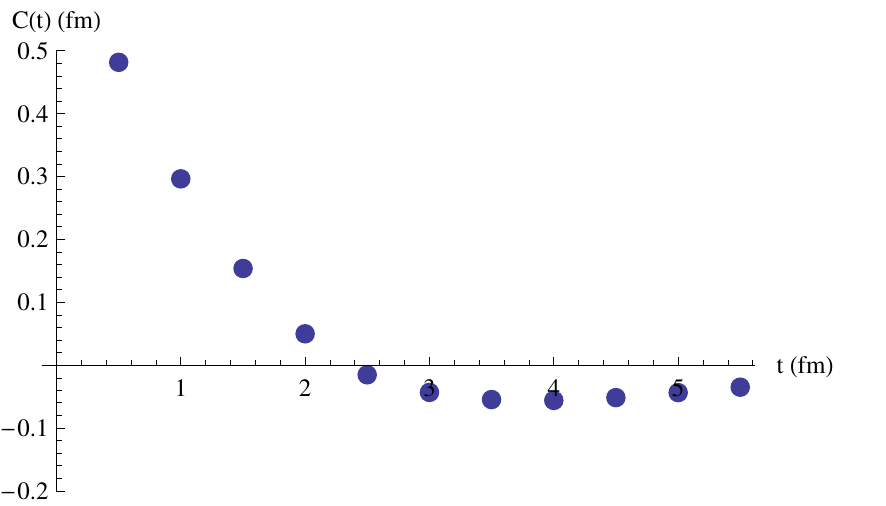} \label{fig6}} \hspace{0.5cm}
    \subfigure[\;Pure GZ case]{\includegraphics[width=5.5cm]{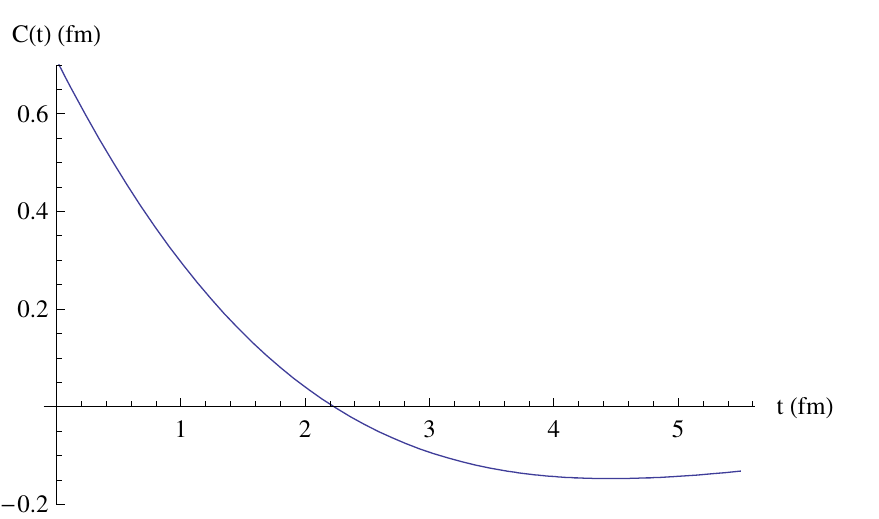} \label{fig7}} \hspace{0.5cm}
    \end{center} \vspace{-0.5cm}
\caption{$\mathcal{C}(t)$ in terms of $t$ in units of fm in the refined GZ case and in the pure GZ case. }
\end{figure}

\noindent To determine $M^2$, we again rely on the variational setup. We observe that for small $t$, there is no real extremum. However, for a certain $t\sim 1$, an extremum emerges at $M^2\sim 0$. This extremum starts to grow for increasing $t$, but at the same time, the curve flattens out. Therefore, starting from $t\sim 2.6 $, the
extremum disappears again. Hence, we have taken\footnote{$M^2\sim0.18$ is the maximal extremal value, corresponding to $t\sim2.6$.} $M^2=0$ for $t<1$, and $M^2=0.18$ for
$t>2.6$. The resulting temporal correlator is displayed in Figure \ref{fig6}.

\noindent We clearly observe a violation of positivity, which is in agreement with the lattice data (see Figure4 of \cite{Cucchieri:2004mf}). Although this agreement is only at a qualitative level, the shape of $\mathcal{C}(t)$ is very similar. \\
\\
It would be interesting to have a look at the temporal correlator in the pure Gribov-Zwanziger case ($M^2=0$). Therefore, $\mathcal{C}(t)$ is displayed in Figure \ref{fig7}. We can conclude that this plot is grosso modo the same as in the refined Gribov-Zwanziger setup.

\enlargethispage{\baselineskip}

\vspace{-0.8cm}

\subsection{Conclusion in 3d}
Again, we conclude that by taking into account the condensation of the operator $\left(\overline{\varphi}\varphi-\overline{\omega}\omega\right)$, we have obtain a positivity violating gluon propagator which is non-vanishing at zero momentum and a ghost propagator which is no longer enhanced.

\section{Refinement of the GZ action in 2 dimensions is impossible}
\subsection{Introduction}
Although 2d gauge theories share some similarities with their also confining 3d or 4d counterparts, there are nevertheless some notable differences. Firstly, at the classical level, as the gauge field $A_{\mu}$ contains only two degrees of freedom in 2d, imposing e.g. the Landau gauge condition, $\partial_{\mu} A_{\mu} = 0$, already removes these two degrees of freedom from the physical spectrum. Therefore, as no physical degrees of freedom remain, confinement seems to be a rather ``trivial'' phenomenon, if one sees confinement as the absence of the elementary gluon degrees of freedom. In contrast, in 3d and 4d, one respectively two degrees of freedom are maintained, hence confinement seems to be more than ``trivial''. Secondly, also at the quantum level, the 2d situation is different from the 4d case. In 2d, the coupling $g$ acquires the dimension of a mass and thus the theory becomes highly superrenormalizable. However, a drawback of the superrenormalizability is the appearance  of severe infrared instabilities. Therefore an infrared regulator, usually put in by hand, is necessary. We emphasize that caution is anyhow at place when performing calculations in 2d gauge theories as discussed in \cite{Bassetto:1999ah}.\\
\\
Therefore, it might not sound so surprising that the behavior of the ghost and the gluon propagator at low momentum is different in 2d than in 3d and 4d. Indeed, as we already mentioned in the introduction of this chapter, the ghost propagator is enhanced in 2d, while the gluon propagator is zero at zero momentum \cite{Cucchieri:2007rg,Cucchieri:2008fc,Maas:2007uv}.\\
\\
By analogy with the 4d and 3d case, we shall also add a mass term of the following form, $M^2 \int \d^2 x\;\Bigl( \overline{\varphi}_\mu^{ab}\varphi_\mu^{ab} -\overline{\omega}_\mu^{ab} \omega_\mu^{ab} \Bigr)$, to the localized Gribov-Zwanziger action $S_\GZ$ in 2d. Though, in the next sections, we shall demonstrate that including this mass term will give rise to infrared instabilities. However, purely from the algebraic and dimensional viewpoint, this mass term cannot be excluded in 2d just as in 3d or 4d. We shall thus start from the refined action
\begin{eqnarray}\label{actielocal2}
S_\RGZ' &=&S_{\GZ}+S_{\overline{\varphi} \varphi} + \int \d^2 x\left(d\frac{N^2-1}{g^2N}\varsigma M^2\lambda^2\right)\;, \nonumber\\
S_{\overline{\varphi} \varphi} &=& \int \d^2 x \left[-M^2\left( \overline{\varphi}^a_i \varphi^a_{i} - \overline{\omega}^a_i \omega^a_i \right) \right]\;.
\end{eqnarray}
The role of vacuum term proportional to the dimensionless parameter $\varsigma$ is a bit redundant in the 2d case, as the problems we shall encounter are neither related to nor curable by this quantity $\varsigma$, which played a pivotal role in 3d and 4d. For completeness and comparability with the 3d or 4d case, we have included it nevertheless.\\
\\
Subsequently, we compute the one loop quantum effective action $\Gamma$ as
\begin{equation}
\Gamma = -d(N^{2}-1)\frac{\lambda^{4}}{2g^2N}+\frac{(N^{2}-1)}{2}\left(d-1\right) \int \frac{\d^{d}p}{\left(2\pi \right) ^{d}}\ln  \left[p^2 \left( p^{2} + \frac{\lambda^4}{p^2+M^2}\right)\right]+d\frac{N^2-1}{g^2N}\varsigma M^2\lambda^2\;.
\end{equation}
The gap equation \eqref{gapgamma} is then determined by
\begin{eqnarray}\label{gappie0}
\frac{2}{g^2N}&=&\int\frac{\d^2p}{(2\pi)^2}\frac{1}{p^4+M^2p^2+\lambda^4}+\frac{2}{g^2N}\varsigma \frac{M^2}{\lambda^2}\;,
\end{eqnarray}
for $d=2$.

\subsection{Two reasons why the refined Gribov-Zwanziger action is excluded in 2d}
In this section, we shall provide two reasons why it is not possible to add the novel mass $\propto \overline{\varphi}\varphi-\overline{\omega}\omega$ to the standard Gribov-Zwanziger action. It shall become clear that it is exactly the fact that we are working in 2d which does signal us that the theory with $\overline{\varphi}\varphi-\overline{\omega}\omega$ coupled to it is not well defined.

\subsubsection{The first reason why $\overline{\varphi}\varphi-\overline{\omega}\omega$ is problematic in 2d}
We originally started the study of the dynamical effects associated to the operator $\overline{\varphi}\varphi-\overline{\omega}\omega$ in 3d and 4d because we found a non-vanishing vacuum expectation value for the operator $\overline{\varphi}\varphi-\overline{\omega}\omega$ already at the perturbative level, namely $\braket{\overline{\varphi}\varphi-\overline{\omega}\omega}\propto\gamma^2$, see equations \eqref{condpert} and \eqref{condpert2}. \\
\\
We shall now verify that our original rationale behind the study of $\overline{\varphi}\varphi-\overline{\omega}\omega$ no longer applies in 2d, showing that this operator cannot be consistently introduced in 2d. It should not come as a too big surprise that the difficulties related to the operator $\overline{\varphi}\varphi-\overline{\omega}\omega$ rely on the appearance of infrared instabilities, typical of 2d, which prevents the analogue phenomenon as in 3d or 4d to occur in 2d.\\
\\
Let us take a look at the condensate $\braket{\overline{\varphi}\varphi-\overline{\omega}\omega}$. We define the energy functional as
\begin{equation}\label{W}
  \e^{-W(J,\gamma^2)}=\int \d\Psi \e^{-S_{\GZ}+\int \d^2x
    J(\overline{\varphi}\varphi-\overline{\omega}\omega)+\varsigma'J\lambda^2}\;.
\end{equation}
Here, we suitably rescaled $\varsigma$ into $\varsigma'$ for notational convenience, $\varsigma' = d \frac{N^2 -1}{g^2 N} \varsigma $. We have also replaced the mass $M^2$ by the more conventional notation for a source, i.e.~$J$.\\
\\
Nextly, let us consider the perturbative value of the condensate, which is explicitly given by
\begin{equation}
    \braket{\overline{\varphi}\varphi-\overline{\omega}\omega}_{\mathrm{pert}}=-\left.\frac{\p W}{\p  J}\right\vert_{J=0}-\varsigma'\lambda^2\,.
\end{equation}
To calculate this quantity we evaluate the one loop energy functional,
\begin{equation}\label{wj}
W(J) = -d(N^{2}-1)\gamma^{4}+\frac{(N^{2}-1)}{2}\left( d-1\right) \int \frac{\d^{d}p}{\left(2\pi \right) ^{d}}\ln  \left[ p^2 \left( p^{2} + \frac{\lambda^4}{p^2+ J}\right)\right]-\varsigma'\lambda^2\;.
\end{equation}
With the help of dimensional regularization we find the following finite result,
\begin{equation}
W(J) =  - \frac{\lambda^4}{g^2 N}(N^2 -1) -\frac{N^2 - 1}{16 \pi}\left[ J \ln \frac{4 \lambda^4}{J^2} - \sqrt{J^2 - 4\lambda^4}\ln\frac{J- \sqrt{J^2 - 4 \lambda^4}}{J+\sqrt{J^2 - 4\lambda^4}} \right]-\varsigma'\lambda^2\,.
\end{equation}
This expression is well-defined when taking the limit $J\to0$. This corresponds to the pure Gribov-Zwanziger case, where $M^2 = J =0$. However, the derivative w.r.t.~$J$ is singular for $J = 0$. Indeed, we find
\begin{equation}
\frac{ \partial W(J)}{ \partial J} =  -\frac{N^2 - 1}{16 \pi} \left[ \frac{-J}{\sqrt{J^2 - 4 \lambda^4}} \ln\frac{J- \sqrt{J^2 - 4 \lambda^4}}{J+\sqrt{J^2 - 4 \lambda^4}}  + \ln \frac{4 \lambda^4}{J^2} \right]-\varsigma'\lambda^2\;,
\end{equation}
in which the second term diverges for $J\to0$. This would imply that
\begin{eqnarray}
 \Braket{\overline{\varphi}\varphi-\overline{\omega}\omega }= \infty\;.
\end{eqnarray}
This strongly suggests that is it impossible to couple the operator to the theory without even causing pathologies already in perturbation theory. A way to appreciate that this divergence is stemming from the infrared region is to derive first expression \eqref{wj} w.r.t. $J$ (assuming this is allowed) and then set $J=0$, in which case
\begin{eqnarray}
\left.\frac{ \partial W(J)}{ \partial J}\right\vert_{J=0}
&=&\frac{N^2-1}{2}(d-1)\left(\int\frac{\d^dp}{(2\pi)^d}\frac{p^2}{p^4+\lambda^4}-\int\frac{\d^dp}{(2\pi)^d}\frac{1}{p^2}\right)-\varsigma'\lambda^2\;.
\end{eqnarray}
The second term in the previous expression is typically zero in dimensional regularization, \emph{except} when $d=2$ as it then develops an infrared pole.\\
\\
Having revealed a first counterargument against the introduction of the mass operator $M^2(\overline{\varphi}\varphi-\overline{\omega}\omega)$ in 2d, let us give an even stronger objection in the following subsection.

\subsubsection{The second (main) reason why $\overline{\varphi}\varphi-\overline{\omega}\omega$ is problematic in 2d: the ghost propagator\label{chap42dform}}
\paragraph{The case $M^2 \not= 0$}\ \\
Let us consider the one loop ghost propagator. Explicitly, the one loop correction to the ghost self energy reads
\begin{eqnarray}\label{sigmadef}
\sigma(k) &=& g^2N \frac{k_{\mu} k_{\nu}}{k^2} \int \frac{\d^2
q}{(2\pi)^2} \frac{1}{(k-q)^2} \frac{q^2 + M^2}{q^4 + M^2q^2 +
\lambda^4}\left(\delta_{\mu\nu}-\frac{q_\mu q_\nu}{q^2}\right)\;.
\end{eqnarray}
Looking at the integral \eqref{sigmadef}, the term $\sim\frac{1}{(q-k)^2}$ which could potentially lead to an infrared singularity upon integration, is partially ``protected'' by the external momentum $k$. One might expect that the infrared divergence will only reveal itself in the limit $k\to0$.\\
\\
Bearing this in mind, let us determine $\sigma(k)_{k^2\sim0}$ by performing the $\vec{q}$-integration in \eqref{sigmadef} exactly for an arbitrary momentum $\vec{k}$. We shall invoke polar coordinates. Without loss of generality, we can put the $q_x$-axis along $\vec{k}$ to write
\begin{equation}
\sigma(k) = \frac{g^2N}{4\pi^2}\int_0^{\infty}q \d q \frac{q^2+M^2}{q^4+M^2q^2+\lambda^4}\int_{0}^{2\pi}\d \theta \frac{1}{k^2+q^2-2qk\cos\theta}(1-\cos^2\theta)\;,
\end{equation}
where we made use of $\vec{k}\cdot\vec{q}=kq\cos\theta$. The Poisson-like $\theta$-integral can be easily calculated using a contour integration,
\begin{equation}\label{contourint}
    \int_{0}^{2\pi}\d\theta
\frac{1-\cos^2\theta}{k^2+q^2-2qk\cos\theta}=\left\{\begin{array}{c}
                                                    \frac{\pi}{q^ 2}\qquad\mbox{if\;} k^2\leq q^2 \\
                                                    \frac{\pi}{k^ 2}\qquad\mbox{if\;} q^2\leq k^2
                                                  \end{array}\right.\;,
\end{equation}
so we obtain
\begin{eqnarray}\label{sigma2}
\sigma(k) &=&
\frac{g^2N}{4\pi}\left(\frac{1}{k^2}\int_0^{k}\frac{q(q^2+M^2)}{q^4+M^2q^2+\lambda^4}\d
q+\int_k^{\infty}\frac{q^2+M^2}{q(q^4+M^2q^2+\lambda^4)}\d
q\right)\;.
\end{eqnarray}
It appears that both integrals are well-behaved in the infrared and ultraviolet for $k>0$.\\
\\
Notice that we did not invoke the gap equation \eqref{gappie0} yet. This is possible, but neither necessary nor instructive at this point. In order to have a better understanding of the $k\to0$ behavior, we can calculate the integrals in \eqref{sigma2}, and extract the small momentum behavior. Doing so, one finds
\begin{equation}\label{sigma3}
    \left.\sigma(k)\right|_{k^2\sim 0}\sim     -\frac{g^2N}{8\pi}\frac{M^2}{\lambda^4}\ln(k^2)\;,
\end{equation}
in the case that $M^2\neq0$, which is a well-defined result, in
contrast with \eqref{sigmang}.\\
\\
However, there is still an infrared instability in the theory due to the final $\ln(k^2)$-factor appearing in $\sigma(k)$ for small $k$. This is our second main argument why coupling the mass operator $(\overline{\varphi}\varphi-\overline{\omega}\omega)$ to the theory causes problems:
\begin{itemize}
\item The quantum correction to the self energy explodes for small $k$, completely invalidating the loop expansion. This problem does not occur in 3d or 4d, since there
      $\sigma\leq 1$. It is not difficult to imagine that the infrared $\ln(k^2)$-singularity will spread itself through the theory, making everything ill-defined for small $k$.
\item Moreover, we also encounter a problem of a more fundamental nature. The starting point of the whole construction was to always stay within the Gribov horizon $\Omega$. This can be assured by the so called no-pole condition, i.e.~$\sigma(k^2) \leq 1$ as stated in the original article by Gribov \cite{Gribov:1977wm}. Since $M^2$ must be
positive\footnote{A negative $M^2$ would lead to tachyonic instabilities in the theory, see e.g. the vacuum functional as an example.}, we clearly see from \eqref{sigma3} that
\begin{equation}
\left.\sigma(k)\right|_{k^2\sim0}\gg1\;,
\end{equation}
hence $\mathcal G (k) = 1/k^2 1/ (1-\sigma(k))$ is signalling us that we have crossed the horizon.
\end{itemize}
This confirms again that $M^2=0$ is the only viable option, i.e.~we cannot go beyond the standard Gribov-Zwanziger action if we want to avoid the appearance of destructive infrared issues, which unavoidably force the theory to leave the Gribov region.\\
\\
\textbf{Remark.} In the previous paragraph, in order to calculate \eqref{sigmadef}, we have first determined the integral in expression \eqref{sigmadef} exactly and then we have taken the limit $k^2 \to 0$. However, one usually \cite{Gribov:1977wm,Sobreiro:2005ec} first expands the integrand for small $k^2$ and then performs the loop integration, as this
considerably reduces the calculational effort.  In the current case, this course of action unfortunately leads to incorrect results. Indeed, doing so, we would reexpress ``1'' as
\begin{eqnarray}
  1 &=& g^2N\frac{k_\mu k_\nu}{k^2}\int \frac{\d^2q}{(2\pi)^2} \frac{1}{q^4+M^2q^2+\lambda^4}\left(\delta_{\mu\nu}-\frac{k_\mu k_\nu}{k^2}\right)+\varsigma\frac{M^2}{\lambda^2}\;,
\end{eqnarray}
an operation which is based on the gap equation \eqref{gappie0}. Subsequently we rewrite $1-\sigma(k)$,
\begin{eqnarray}
  1-\sigma(k) &=& g^2N\frac{k_\mu k_\nu}{k^2}\int \frac{\d^2q}{(2\pi)^2}
  \frac{1}{q^4+\lambda^4}\left(1-\frac{q^2}{(k-q)^2}\right)\left(\delta_{\mu\nu}-\frac{k_\mu
  k_\nu}{k^2}\right)+\varsigma \frac{M^2}{\lambda^2}\nonumber\\
&+&g^2N \frac{k_{\mu} k_{\nu}}{k^2} \int \frac{\d^2 q}{(2\pi)^2}
\frac{1}{(k-q)^2} \frac{M^2}{q^4 + M^2q^2 +
\lambda^4}\left(\delta_{\mu\nu}-\frac{q_\mu q_\nu}{q^2}\right)
  \;,
\end{eqnarray}
and then we expand the integrand\footnote{We notice that there will be no terms of odd order in $k$. This would correspond to an odd power of $q$, which will vanish upon integration due to reflection symmetry.} around $k^2\sim0$ to find at lowest order,
\begin{eqnarray}\label{sigmang}
  \left.(1-\sigma(k))\right\vert_{k^2\sim0} &=&\frac{g^2N}{2}\int \frac{\d^2q}{(2\pi)^2}\frac{M^2}{q^2(q^4+M^2q^2+\lambda^4)} +\varsigma\frac{M^2}{\lambda^2}+\mathcal{O}(k^2)\;.
\end{eqnarray}
From this expression, we are led to believe that $1-\sigma(k)$, hence $\sigma(k)$, is ill-defined at small $k^2$, due to an infrared singularity which makes the integral in the r.h.s. of \eqref{sigmang} to explode. However, this is not true, as in this case, the limit and the integration cannot be exchanged. The only correct way is to first calculate the integral and then take the limit as was done in the previous paragraph. Further on this section, we shall explicitly explain why expression \eqref{sigmang} is wrong by exploring the $M^2 =0$ case in more detail.

\paragraph{The case $M^2 =0$}\ \\
It is instructive to take a closer look at the usual Gribov-Zwanziger scenario.  One finds for $M^2=0$ that
\begin{equation}\label{4sigma3}
    \left.\sigma(k)\right|_{k^2\sim 0}\sim \frac{g^2N}{4\pi}\left(\frac{\pi}{4\lambda^2}-\frac{k^2}{4\lambda^4}\right)\;,
\end{equation}
a result which is indeed free of infrared instabilities. We also point out that ordinary (perturbative) Yang-Mills theory is recovered when $\lambda=0$. It is hence nice to observe that this again causes troubles in the infrared since the $\lambda\to0$ limit diverges. This is just a manifestation of the fact that 2d gauge theories are infrared sick at the perturbative level, and need some (dynamical) regularization. Apparently, at least at the level of the ghost propagator at one loop, the Gribov mass acts a natural regulator in the infrared sector.\\
\\
We should still use the gap equation in \eqref{4sigma3} to find the correct ghost propagator. The gap equation \eqref{gapgamma} for $M^2=0$ is readily computed as
\begin{eqnarray}\label{gappie}
\frac{2}{g^2 N} &=& \int \frac{\d^2 p}{(2\pi)^2} \frac{1}{p^4 +
\lambda^4} = \frac{1}{8 \lambda^2}\;.
\end{eqnarray}
Evoking this gap equation, we find
\begin{eqnarray}\label{zinvol}
1-\sigma(k) &=& 1
-\frac{g^2N}{4\pi}\left(\frac{\pi}{4\lambda^2}-\frac{k^2}{4\lambda^4}\right)=
\frac{g^2N}{4\pi}\frac{k^2}{4\lambda^4}=\frac{16 k^2}{\pi g^2N}\;.
\end{eqnarray}
Henceforth, we obtain
\begin{eqnarray}\label{gh2}
\left.{\cal G}^{ab}(k)\right|_{k^2\sim0}  &=& \left.\delta^{ab}
\frac{1}{k^2} \frac{1}{1-\sigma(k)}\right|_{k^2\sim0}=\frac{\pi
g^2N}{16 k^4}\;.
\end{eqnarray}
We conclude that the ghost propagator is clearly enhanced and displays the typical behavior $\sim 1/k^4$ in the deep infrared, in accordance with the usual Gribov-Zwanziger scenario.\\
\\
\textbf{Remark.} As we already announced earlier in this section, let us have a closer look at the $M^2=0$ case. In a way completely similar to the $M^2\neq0$ case, we find, around $k^2\sim0$,
\begin{equation}
  \left.(1-\sigma(k))\right\vert_{k^2\sim0} = g^2N\frac{k_\mu k_\nu}{k^2}\int \frac{\d^2q}{(2\pi)^2}
  \frac{1}{q^4+\lambda^4}\left(\frac{k^2}{q^2}-4\frac{(k\cdot q)^2}{q^2}\right)\left(\delta_{\mu\nu}-\frac{k_\mu
  k_\nu}{k^2}\right)+\mathcal{O}(k^4)\;,
\end{equation}
where we have expanded the integrand w.r.t. $q$ before integrating. Exploiting polar coordinates once more, we are now brought to
\begin{equation}
  \left.(1-\sigma(k))\right\vert_{k^2\sim0} =  \frac{g^2N}{4\pi^2}k^2\int_{0}^{+\infty}\frac{q\d
  q}{q^2}\frac{1}{q^4+\lambda^4}\int_0^{2\pi}(1-4\cos^2\theta)(1-\cos^2\theta)\d\theta+\mathcal{O}(k^4)\;.
\end{equation}
Surprisingly, the $\theta$-integral vanishes, as it can be easily checked. In fact, one can extend this observation to all orders in
$k$. To do so, we write
\begin{eqnarray}
\frac{q^2}{(q-k)^2}=\frac{q^2}{q^2+k^2-2qk\cos\theta}=\frac{1}{1+\frac{k^2}{q^2}-2\frac{k}{q}\cos\theta}=\sum_{n=0}^{\infty}\left(\frac{k}{q}\right)^n{\cal
U}_n(\cos\theta)\;,
\end{eqnarray}
where we introduced the Chebyshev polynomials of the second kind, ${\cal U}_n(x)$. It holds that \cite{wolfram}
\begin{eqnarray}\label{cheb}
  {\cal U}_n(\cos\theta) &=& \frac{\sin((n+1)\theta)}{\sin\theta}\;.
\end{eqnarray}
Subsequently, we can rewrite
\begin{eqnarray}
1-\sigma(k)&=& g^2N \int
\frac{\d^2q}{(2\pi)^2}\sum_{n=1}^{\infty}(1-\cos^2\theta){\cal U}_n(\cos\theta)\left(\frac{k}{q}\right)^n\frac{1}{q^4+\lambda^4}\;,
\end{eqnarray}
where use has been made of ${\cal U}_0(x)=1$. Assuming that the integral and the infinite sum can be interchanged, we are led to
\begin{eqnarray}\label{slecht}
1-\sigma(k)&=& \frac{g^2N}{4\pi^2}
\sum_{n=1}^{\infty}k^n\int_{0}^{+\infty} \frac{\d q}{q^{n-1}}
\frac{1}{q^4+\lambda^4}\int_0^{2\pi}(1-\cos^2\theta){\cal U}_n(\cos\theta)\d\theta\;.
\end{eqnarray}
Since $n\geq 1$ and making use of \eqref{cheb}, for the $\theta$-integration we find
\begin{eqnarray}
\int_0^{2\pi} (1-\cos^2\theta){\cal U}_n(\cos\theta)\d\theta &=& \int_0^{2\pi}\sin\theta\sin((n+1)\theta) \d\theta\nonumber\\&=&\int_0^{2\pi}\frac{\cos(n\theta)-\cos((n+2)\theta)}{2} \d\theta=0\;.
\end{eqnarray}
However, this does not make the integral in \eqref{slecht} well defined, as the remaining $q$-integral is infrared singular for any occurring value of $n$! In fact, exactly these infrared divergences forbid the interchange of integral and of the infinite sum. This is a nice example of the fact that the integral of a infinite sum can be well defined, whereas the (sum of the) individual integrals are not.\\
\\
When we first integrate exactly for any $k$ and then  expand in powers of $k^2$, we do recover the meaningful result \eqref{zinvol} at $k^2\sim0$.

\subsection{Conclusion in 2d}
From the previous subsection, we can conclude that it is not possible to ``refine'' the Gribov-Zwanziger action in 2d, in contrast with the 3d or 4d case. We are thus back in the  usual Gribov-Zwanziger scenario, which predicts a $1/k^4$ singularity for the ghost propagator, and a vanishing gluon propagator at zero momentum, $D(0)=0$.

\section{Lattice study of the RGZ propagators}
\subsection{Introduction}
In this section, we shall discuss the agreement of the gluon propagator \eqref{gluonprop2} in 4d, i.e.
\begin{equation}\label{RZGtreelevel}
 D(p^2) = \frac{p^2 + M^2}{p^4 + \left(M^2 + m^2\right) p^2 + \underbrace{ 2 g^2 N \gamma^4 + M^2 m^2}_{\varpi}  }\;,
\end{equation}
with the lattice data. We shall base ourselves on \cite{Dudal:2010tf}. We are curious whether the propagator \eqref{RZGtreelevel} can reproduce not only qualitatively the gluon propagator, but that it also works out well at the \emph{quantitative} level. We shall therefore analyze the lattice gluon propagator in pure SU(3) Yang-Mills gauge theories in 4d and investigate to what extent the propagator \eqref{RZGtreelevel} can match the data, by treating the mass scales $m^2$, $M^2$ and $\gamma^4$ as fitting parameters. It shall turn out that an important role is played by the presence of the dimension two gluon condensate $\braket{A^2}$, represented by $m^2$.\\
\\
The lattice calculations were done by our collaborator O.~Oliveira, whereby the data of Table \ref{LatticeSetup} were used. Of the three $\beta$ values, $\beta = 6.0$ was used to perform an extrapolation to the infinite volume limit, whilst the Berlin-Moscow-Adelaide at $\beta = 5.7$ and $\beta = 6.2$ was be used to cross-check the final results. We refer to \cite{Dudal:2010tf} for the details concerning the lattice calculations. Let us now glance trough the results.

\begin{table}[H]
   \begin{center}
   \begin{tabular}{l@{\hspace{0.5cm}}c@{\hspace{0.3cm}}c@{\hspace{0.3cm}}c@{\hspace{0.3cm}}c@{\hspace{0.3cm}}c}
      \hline
       \multicolumn{6}{c}{$\beta = 5.7$ \hspace{2cm} $a = 0.1838$ fm} \\
       \hline
      $L$                 &    64    &  72    &  80    &  88    &  96 \\
      $aL$ (fm)   &  11.8  & 13.2 &  14.7 & 16.2 & 17.6 \\
      \# Conf       &   14     & 20    &  25    &  68   &  67   \\
      \hline
       \multicolumn{6}{c}{$\beta = 6.0$ \hspace{2cm} $a = 0.1016$ fm} \\
       \hline
      $L$                 &    32    &  48    &  64    &  80    &   \\
      $aL$ (fm)   &  3.25  & 4.88 &  6.50 & 8.13 &  \\
      \# Conf       &   126   & 104  &  120  &  47   &    \\
      \hline
       \multicolumn{6}{c}{$\beta = 6.2$ \hspace{2cm} $a = 0.07261$ fm} \\
       \hline
      $L$                 &   48    &  64    &   &      &   \\
      $aL$ (fm)   &  3.49 & 4.65 &    &     &  \\
      \# Conf       &   88   & 99     &    &    &    \\
      \hline
   \end{tabular}
   \caption{The lattice setup. For the conversion to physical units we took the lattice spacing measure from the string tension \cite{Bali:1992ru}. The first set of configurations, i.e.~those with $\beta = 5.7$, were generated by the Berlin-Moscow-Adelaide group and the results published in \cite{Bogolubsky:2009dc}. Note that in their paper, the lattice spacing was taken from $r_0$. The  Berlin-Moscow-Adelaide data was rescaled appropriately to follow our conventions.} \label{LatticeSetup}
   \end{center}
\end{table}

\subsection{Gluon propagator and evidence for the $d=2$ gluon condensate $\braket{A^2}$}
\begin{figure}[H]
   \centering
   \includegraphics[scale=0.5]{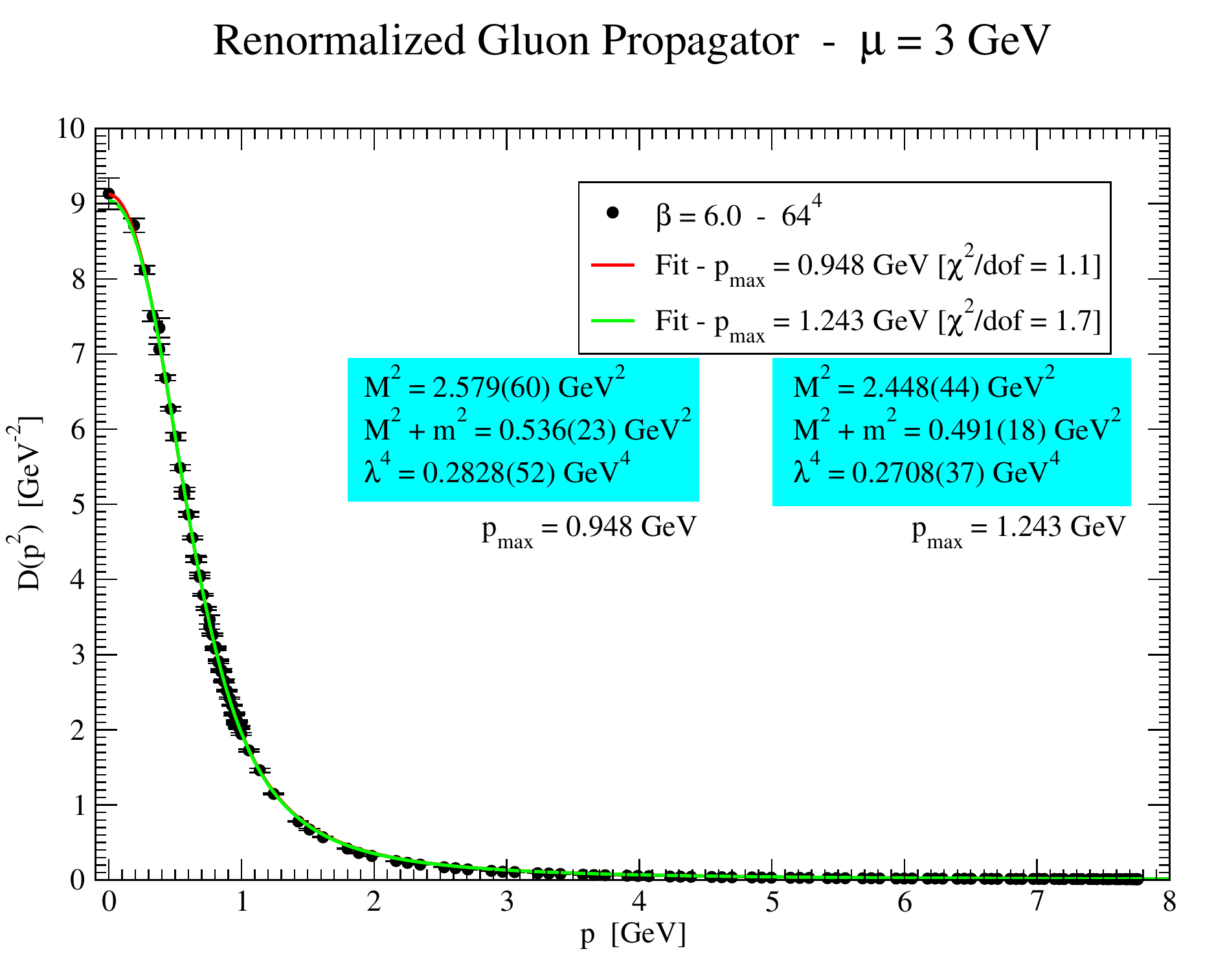}
   \caption{Gluon propagator and fit to \eqref{RZGtreelevel} using the momentum range $[0 , p_{max} ]$.
                   $p_{max} = 1.243$ GeV is the largest fitting range which has a $\chi^2/d.o.f. < 2$. The figure
                   includes the outcome of the fits for the two fitting ranges considered.}
   \label{FullPropFit}
\end{figure}

\begin{figure}[H]
   \centering
   \includegraphics[scale=0.4]{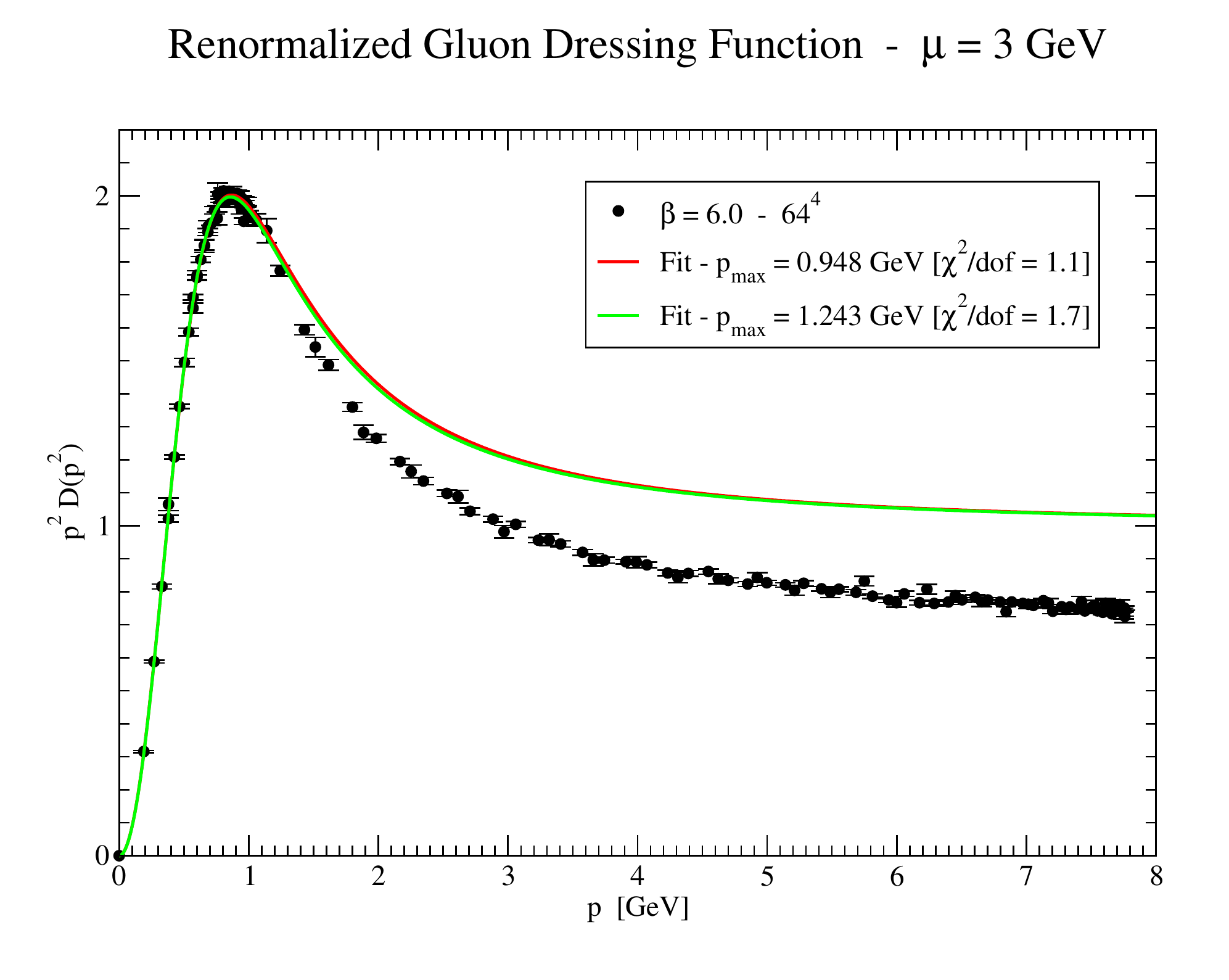}
      \caption{The same as in Figure \ref{FullPropFit} but for the gluon dressing function $p^2 D(p^2)$. The dressing function
                    provides a clear picture of the differences between \eqref{RZGtreelevel} and the lattice data in the
                    ultraviolet region.}
   \label{FullDressFit}
\end{figure}

In this section, we shall show that $m^2$ is of paramount importance for the fit of the gluon propagator \eqref{RZGtreelevel}. In principle, it is sufficient to consider the condensation of the operator  $\overline{\varphi}^a_i \varphi^a_{i} - \overline{\omega}^a_i \omega^a_i$ to have a gluon propagator that does not vanish at zero momentum. Therefore, we can test whether the lattice data can be fitted by the gluon propagator \eqref{RZGtreelevel} with $m^2 = 0$, i.e.
\begin{equation}\label{NoA2}
   D(p^2) = \frac{p^2 + M^2}{p^4 + M^2 p^2 + 2 g^2 N \gamma^4 } \,.
\end{equation}
Although expressions \eqref{RZGtreelevel} and \eqref{NoA2} have a very similar structure, the lattice data distinguishes quite clearly between the two functional forms. Indeed, while \eqref{RZGtreelevel} is able to reproduce the lattice propagator on a wide range of momentum starting at 0 GeV and going up to $1 - 1.5$ GeV, in the sense that the
corresponding fit have $\chi^2 /d.o.f. < 2$,  the fits corresponding to \eqref{NoA2} always have a $\chi^2 /d.o.f.$ larger than three, and should as such be rejected.\\
\\
To give an example of a fit with $m^2\neq0$, we have depicted the renormalized gluon propagator computed using the $\beta = 6.0$ and $64^4$ lattice and the fits corresponding to \eqref{RZGtreelevel}, see Figure \ref{FullPropFit}.  Although the fits use only the momentum in $[0 , p_{max} ]$, in Figure \ref{FullPropFit} we show the propagator if one uses \eqref{RZGtreelevel} over the entire momentum region. There is a small difference between the lattice data and the prediction of \eqref{RZGtreelevel} in the ultraviolet region which is clearly seen in the gluon dressing function, see Figure \ref{FullDressFit}. These small observed differences\footnote{For the highest lattice momenta $p = 7.76$ GeV, the measured propagator is 0.01205(32) GeV$^{-2}$, while \eqref{RZGtreelevel} predicts 0.0172 GeV$^{-2}$.} are expected as \eqref{RZGtreelevel} does not take into account the perturbative logarithmic corrections \cite{Dudal:2010tf}.\\
\\
In conclusion, the fits to the propagators \eqref{RZGtreelevel} and \eqref{NoA2} seem to point towards a non-vanishing gluon condensate $\braket{A^2}$. Later, we shall discuss this in more detail and extract an estimate for $\braket{A^2}$.

\subsection{Measuring the scales in the Refined Gribov-Zwanziger gluon propagator using the lattice data \label{orlandof}}
\begin{figure}[H]
   \centering
   \includegraphics[scale=0.4]{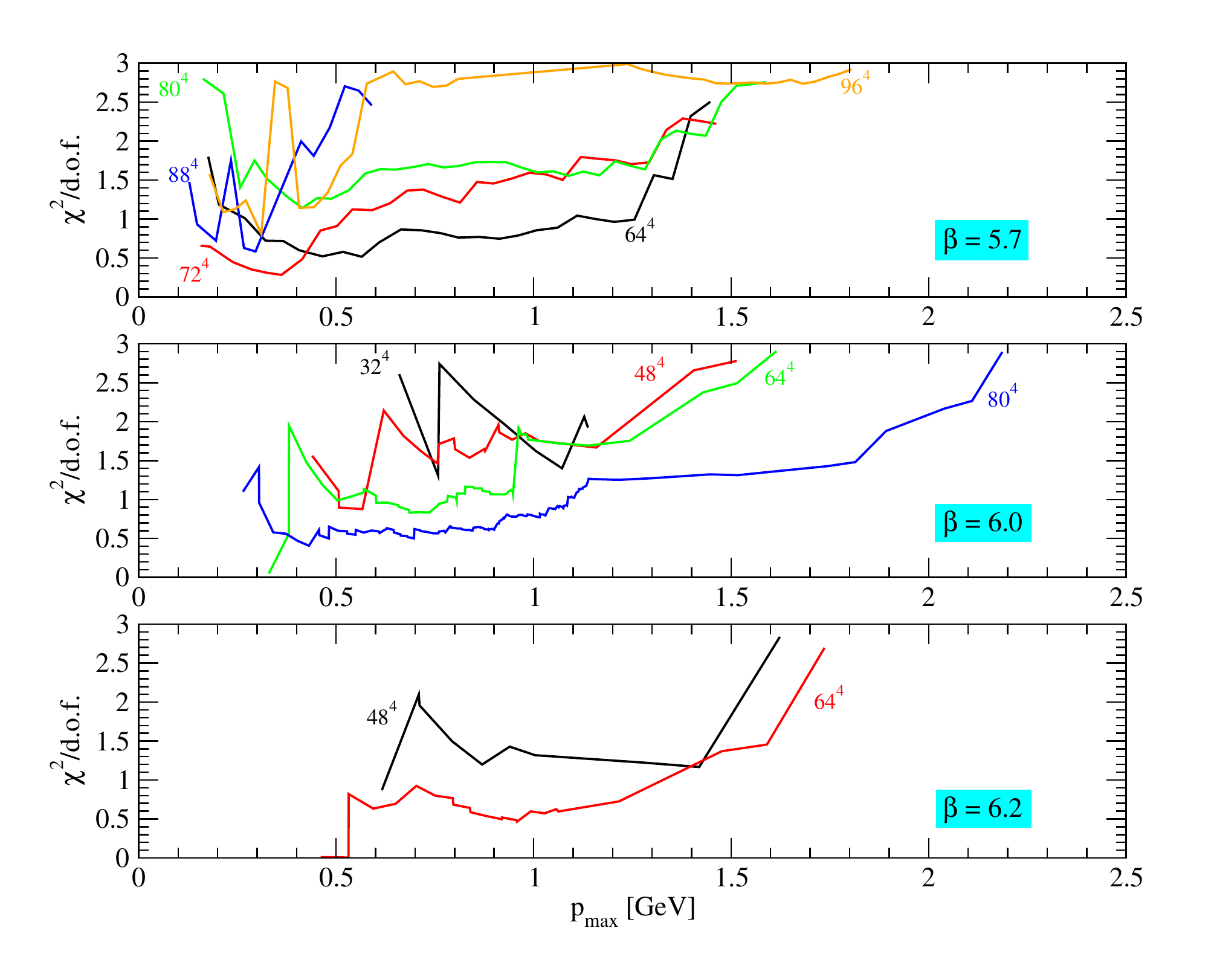}
\vspace{-0.4cm}
   \caption{Fitting the propagator to \eqref{RZGtreelevel}: $\chi^2/d.o.f.$ as a function of the maximum fitting
                  momenta $p_{max}$ for each lattice.}
   \label{Figchi2Pmax}
\end{figure}

Let us now give some estimates for the different mass parameters in \eqref{RZGtreelevel}. It is not expected that \eqref{RZGtreelevel} is able to describe the lattice propagator for the full range of momenta, due to logarithmic corrections.  Therefore, a sliding window analysis was performed, i.e.~the propagator was fitted using momenta in $[0 , p]$, with increasing values for $p$. Then, the $\chi^2/d.o.f.$ was used to establish a maximum range of momenta described by \eqref{RZGtreelevel}, see Figure \ref{Figchi2Pmax}. For the largest two $\beta$ values and for the largest lattices, the Refined Gribov-Zwanziger tree level propagator is able to describe the lattice data well above 1 GeV. In particular, for the largest volume, being the $\beta = 6.0$ and $80^4$ case, the lattice gluon propagator can be fitted by \eqref{RZGtreelevel} beyond 1.5 GeV.

\vspace{-0.4cm}
\vspace{-0.2cm}

\begin{figure}[H]
   \begin{center}
   \includegraphics[scale=0.35]{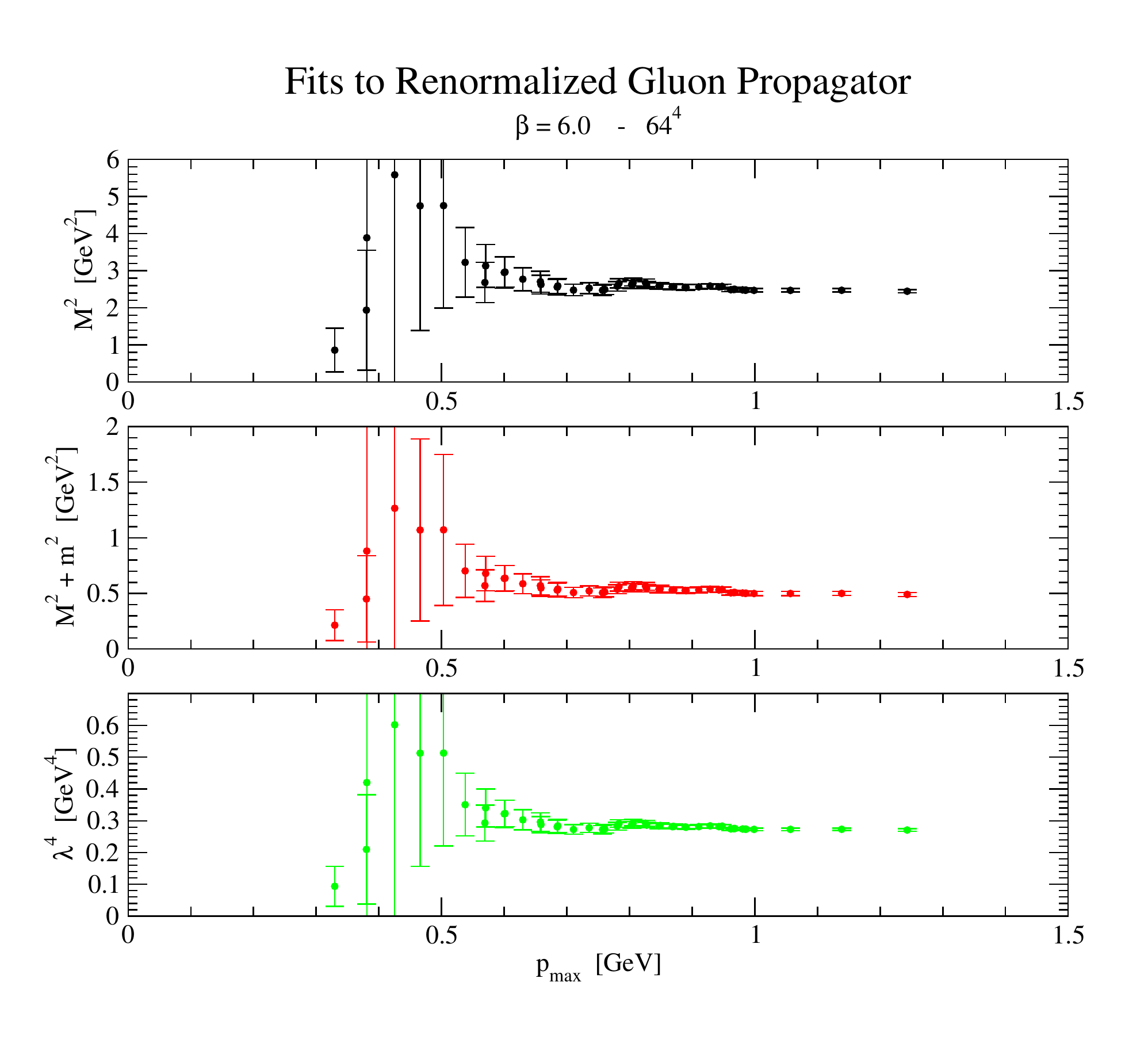}
   \end{center}
\vspace{-0.4cm}
\caption{Evolution of the fitting parameters with $p_{max}$ for $\beta = 6.0$ and $64^4$ data.}
   \label{Fitspmax64}
\end{figure}

\enlargethispage{\baselineskip}
\noindent In Figure \ref{Fitspmax64} the result of fitting \eqref{RZGtreelevel} is reported to the renormalized gluon propagator computed from the $\beta = 6.0$ and $64^4$ lattice data as a function of the fitting range $[ 0 , p_{max}]$. Similar plots can be shown for the remaining fits. As Figure \ref{Fitspmax64} shows, the estimated values for $M^2$, $M^2 + m^2$ and $\varpi^4 = 2 g^2 N \gamma^4 + M^2m^2$ are stable against a change on $p_{max}$. For each simulation, as a set of values, we choose those which correspond to the largest fitting range with a $\chi^2/d.o.f. \sim 1$. For example, for the $\beta = 6.0$ and $64^4$ data, we take $p_{max} = 0.929$ GeV and $M^2 =  2.589 \pm 0.068$ GeV$^2$, $M^2 + m^2 =  0.539 \pm 0.025$ GeV$^2$,  $\varpi^4 =   0.2837   \pm 0.0059$ for a $\chi^2 /d.o.f. =  1.07$. When the $\chi^2/d.o.f.$ never crosses or becomes to close to 1, such as happens in the smallest fitting lattice volume, we choose the set of values which minimizes $\chi^2/d.o.f.$ for the largest possible fitting range.

\vspace{-0.4cm}

\begin{figure}[H]
   \centering
   \includegraphics[scale=0.4]{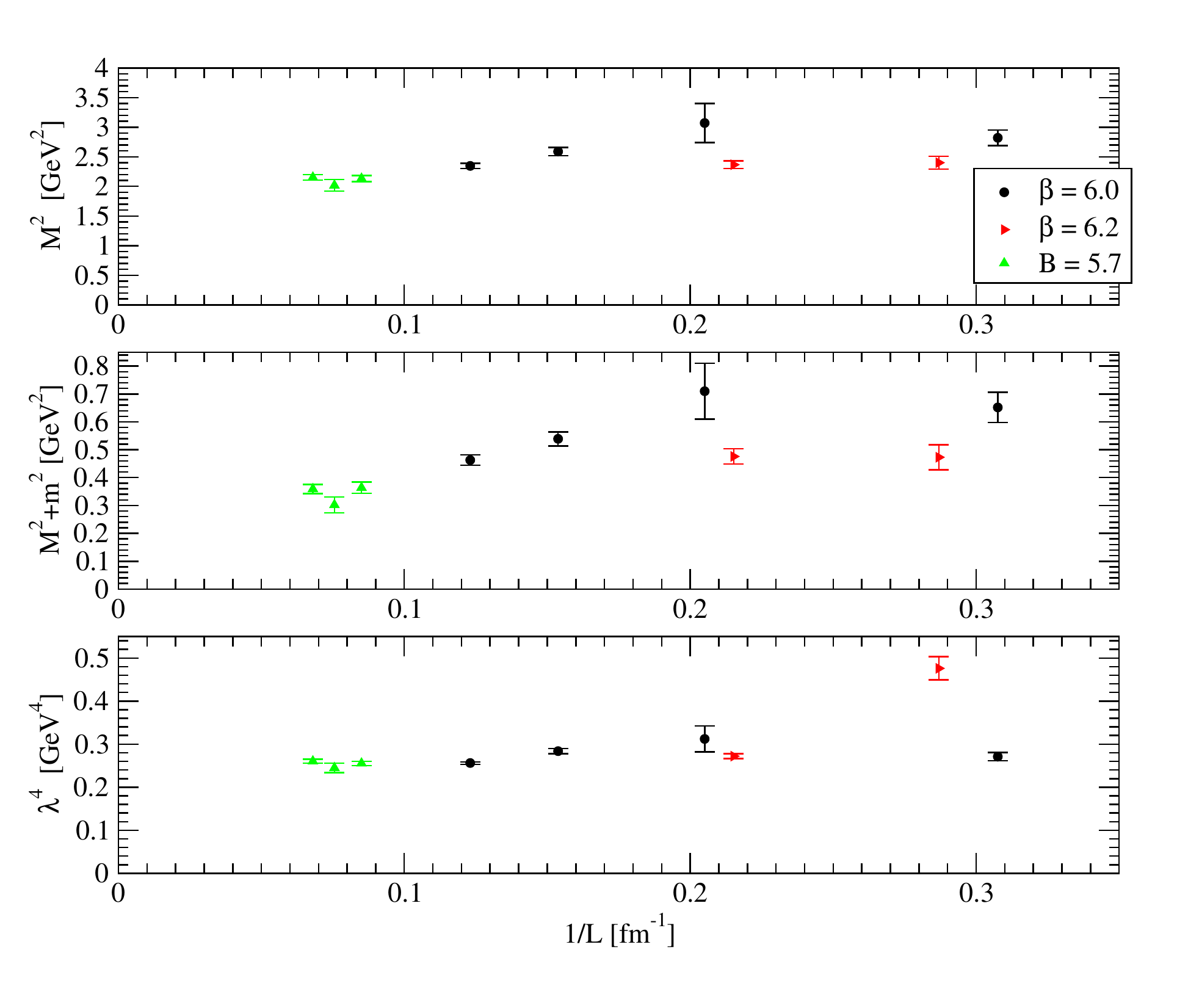}
   \caption{Parameters for the tree level gluon propagator of the Refined Gribov-Zwanziger action, computed fitting
                  the renormalized gluon propagator, as a function of the inverse of the lattice length $L$. The observed fluctuations in the $\beta = 5.7$ results are explained in \cite{Dudal:2010tf}.}
   \label{FigFitsfinais}
\end{figure}

\noindent In Table \ref{Fitsfinais} we report the estimates of the different parameters defining the Refined Gribov-Zwanziger tree level gluon propagator for each lattice simulation. The values are plotted in Figure \ref{FigFitsfinais} as a function of the inverse of the lattice length $L$. The data shows a small dependence on $1/L$, especially for $M^2+m^2$, and on the lattice spacing, i.e.~on $\beta$. Nevertheless, for $\beta = 6.0$, the four volumes can be combined to perform a linear
extrapolation to the infinite volume limit.

\begin{table}[H]
   \centering
   \begin{tabular}{l@{\hspace{0.7cm}} l@{\hspace{0.5cm}}l@{\hspace{0.5cm}}l@{\hspace{0.5cm}}l@{\hspace{0.5cm}}l@{\hspace{0.5cm}}l}
      \hline
      $L$    &  $p_{max}$  &  $M^2$                        & $M^2+m^2$                 & $\varpi^4$               &  $\chi^2 / d .o.f.$ \\
      \hline
      \multicolumn{6}{c}{$\beta = 5.7$} \\
     64   &  1.255          &  $2.132  \pm  0.052$  &   $0.364 \pm  0.020$  &  $0.2553 \pm 0.0051$  &  0.99  \\
     72   &   0.814          &  $2.017 \pm  0.097$  &   $0.302 \pm  0.028$  &  $0.245    \pm  0.011$   & 1.21  \\
     80   &   1.089          &  $2.151  \pm  0.047$ &   $0.359 \pm  0.016$  &  $0.2604 \pm 0.0049$  & 1.55  \\
      \hline
      \multicolumn{6}{c}{$\beta = 6.0$} \\
      32  &    1.072         &  $2.82 \pm 0.13$       &    $0.652 \pm 0.054$  &  $0.2708  \pm 0.0096$ &   1.40  \\
      48  &    0.757         &   $3.07  \pm 0.33$     &    $0.71   \pm 0.10$     &  $0.312 \pm 0.030$        &  1.46  \\
      64  &    0.929         &   $2.589 \pm 0.068$  &   $0.539 \pm 0.025$   &  $0.2837  \pm 0.0059$  &  1.07  \\
      80  &    1.103         &   $2.346 \pm 0.043$  &   $0.463 \pm 0.019$   &  $0.2561  \pm 0.0030$  &  1.03  \\
      \hline
      \multicolumn{6}{c}{$\beta = 6.2$} \\
     48   &    1.419         &   $2.40  \pm 0.11$      & $0.473 \pm 0.045$     &  $0.2677 \pm 0.0095 $  &  1.17 \\
     64   &    1.476         &   $2.366 \pm 0.066$   & $0.476 \pm 0.027$    &   $0.2721 \pm- 0.0057$  &  1.37 \\
         \hline
   \end{tabular}
   \caption{Tree level gluon propagator parameters from fitting the Refined Gribov-Zwanziger propagator \eqref{RZGtreelevel} to the renormalized lattice gluon propagator.
                 The errors reported are statistical and computed assuming Gaussian error propagation. }
   \label{Fitsfinais}
\end{table}

\begin{figure}[H]
   \centering
   \includegraphics[scale=0.4]{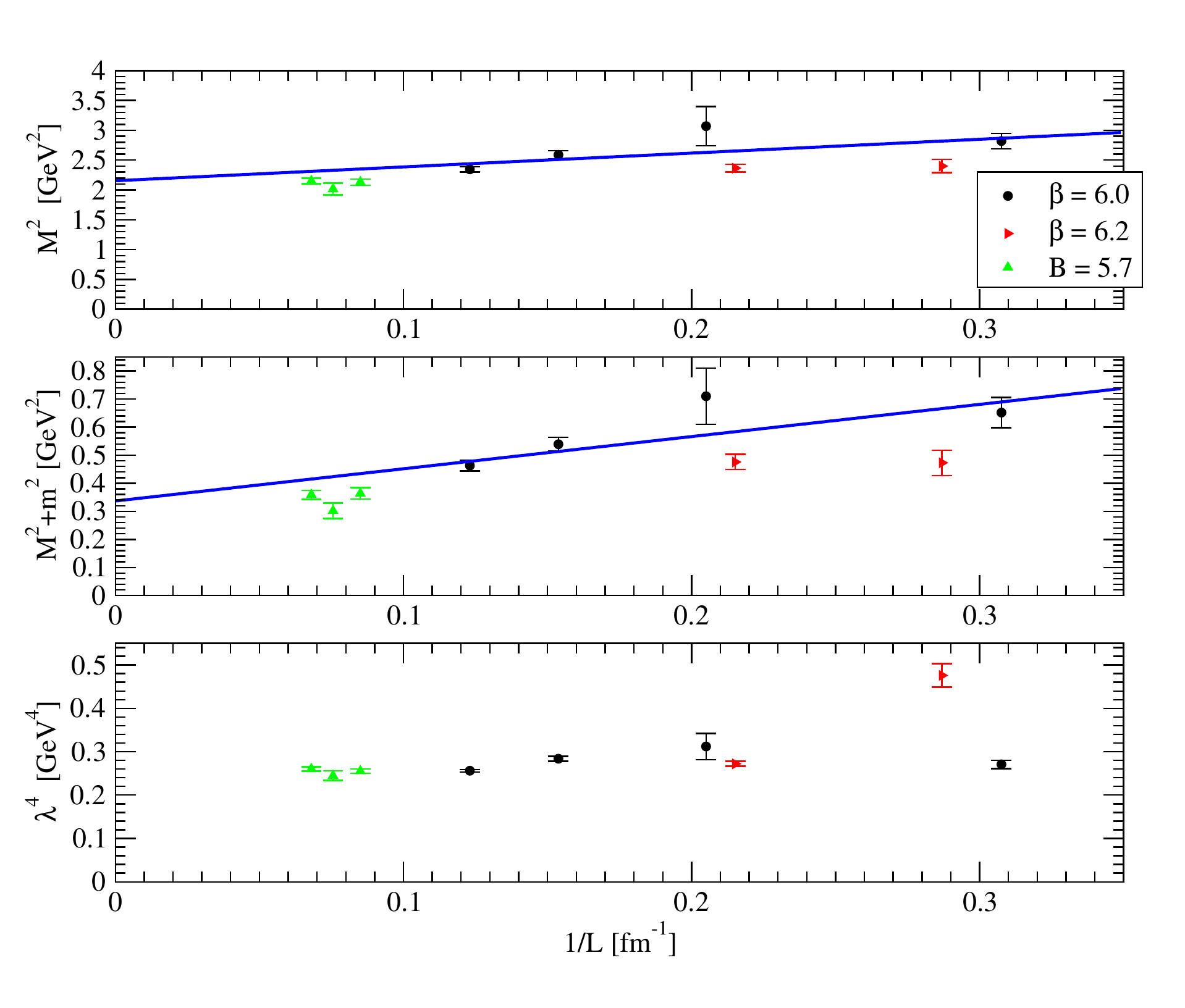}
   \caption{The same as Figure \ref{FigFitsfinais} but including the linear extrapolations for $M^2$ and
                   $M^2 + m^2$, which are obtained using the $\beta=6.0$ data. The large volume $\beta=5.7$ data serves as a consistency check, as explained before in the text.}
   \label{FigFitsfinaisFits}
\end{figure}

\noindent Now let us give the final values. Firstly, $M^2$ is reasonably well described by a linear function as a function of $1/L$. Indeed, the $\chi^2/d.o.f.$ of the fit is 2.13, giving
\begin{equation}\label{scale1}
M^2 = 2.15 \pm 0.13~\text{GeV}^2\,,
\end{equation}
which is in good agreement with the value computed from the largest $\beta = 5.7$ volume. \\
\\
Secondly, for $M^2 + m^2$, the linear fit gives an infinite volume value of
\begin{equation}\label{scale2}
M^2+m^2 = 0.337 \pm   0.047~\text{GeV}^2\,,
\end{equation}
for a $\chi^2/d.o.f. = 2.04$.\\
\\
Thirdly, for $\varpi^4$, the linear extrapolation has a $\chi^2/d.o.f.$ larger than 3. Fortunately, it seems that $\varpi^4$ shows the smallest dependence on $1/L$ and the lattice spacing, with the largest volumes providing numbers which are compatible, within one standard deviation. Therefore, given the results reported in Table \ref{Fitsfinais}
for the largest volumes, one can claim that
\begin{equation}\label{scale3}
\varpi^4 = 0.26~\text{GeV}^4\,,
\end{equation}
which are the reliable digits from the largest two lattices, see the Table \ref{Fitsfinais}. The linear extrapolations can be seen in Figure \ref{FigFitsfinaisFits}. We observe that the figures for the $\beta = 5.7$ data and the linearly extrapolated results are pretty close, giving us further confidence in the extrapolation.

\noindent From these three numbers, we can extract
\begin{equation}\label{scale4}
m^2 = - 1.81 \pm 0.14~\text{GeV}^2\,,
\end{equation}
and simultaneously
\begin{equation}
2 g^2 N \gamma^4 = 4.16 \pm 0.38~\text{GeV}^4\,.
\end{equation}
Therefore, assuming that \eqref{RZGtreelevel} describes the infrared gluon propagator, we have that
\begin{equation}
   D(0) = \frac{M^2}{\varpi^4} = 8.3 \pm 0.5 ~\text{GeV}^{-2}\,.
\end{equation}
The zero momentum gluon propagator computed using the extrapolated values for $M^2$ and $\varpi^4$ is in excellent agreement, within one standard deviation, with the lattice $D(0)$ computed from lattice QCD for $\beta = 5.7$ where $D(0) \sim 7 - 8.5$ GeV$^{-2}$, $\beta = 6.0$ and $80^4$ data where $D(0) = 8.93 \pm 0.47$ GeV$^{-2}$ and for $\beta = 6.2$ and $64^4$ data which has a $D(0) = 8.95 \pm 0.22$ GeV$^{-2}$.

\subsection{Extracting a value for the dimension two gluon condensate $\braket{g^2A^2}$}
In order to obtain an estimate that can be compared with other values available on the market, we shall rely on the renormalization group. In particular, we wish to compare with the values \eqref{A3} and \eqref{A3bis}, being
\begin{equation}\label{A3nog}
\braket{g^2A^2} =5.1^{+0.7}_{-1.1}~\text{GeV}^2\,,
\end{equation}
and
\begin{equation}\label{A3bisnog}
\braket{g^2A^2} =4.4\pm0.4~\text{GeV}^2\,.
\end{equation}
For this, we shall need the following correspondence between the tree level gluon mass $m^2$ and the condensate $\braket{A^2}$ \cite{Verschelde:2001ia,Dudal:2005na}
\begin{equation}\label{corr}
    \braket{g^2A^2}= -\zeta_0 m^2\,,\qquad \zeta_0=\frac{9}{13}\frac{N^2-1}{N}\,,
\end{equation}
Hence, our estimate \eqref{scale4} corresponds to a positive gluon condensate, as using \eqref{corr} yields for $N=3$
\begin{equation}\label{A2}
\braket{g^2A^2} = 3.35 \pm 0.26~\text{GeV}^2\,.
\end{equation}
As here, we have renormalized at a scale $\mu=3$ GeV, while the value \eqref{A3nog} was obtained at a renormalization scale $\mu=10$ GeV, this value needs to be rescaled. For this, we need the following one loop renormalization group equations,
\begin{eqnarray}\label{corr2}
    \mu\frac{\p}{\p \mu} g^2&=& -2\beta_0g^4\,,\qquad \beta_0=\frac{11}{3}\frac{N}{16\pi^2}\,,\nonumber\\
        \mu\frac{\p}{\p \mu} m^2&=& \gamma_0 g^2m^2\,,\qquad \gamma_0=-\frac{3}{2}\frac{N}{16\pi^2}\,.
\end{eqnarray}
which are valid in any (massless) renormalization scheme\footnote{We recall that the lowest order anomalous dimensions are universal quantities.} \cite{Verschelde:2001ia,Dudal:2005na}. We can solve the differential equation for $g^2$ at one loop to find
\begin{equation}
g^2 = \frac{1}{2 \beta_0} \ln \frac{\mu}{\Lambda_T}\;,
\end{equation}
whereby $\Lambda_T$ is the scale at which the corresponding coupling constant $g^2$ blows up. Therefore, we have at one loop
\begin{equation}\label{A5}
    \mu\frac{\p}{\p \mu} m^2 = \frac{\gamma_0}{2\beta_0} \frac{1}{\ln\frac{\mu}{\Lambda_T}}m^2\,,
\end{equation}
which by introducing the auxiliary variable $\xi=\ln\frac{\mu}{\Lambda_T}$, can be easily integrated to
\begin{equation}
    m^2=m_0^2 \left(\frac{\xi}{\xi_0}\right)^{\frac{\gamma_0}{2\beta_0}}=m_0^2\left(\frac{\ln\frac{\mu}{\Lambda_T}}{\ln\frac{\mu_0}{\Lambda_T}}\right)^{-9/44}\,.
\end{equation}
The fundamental scale $\Lambda_T$ of this $T$-scheme is related to the  conventional $\MSbar$ one through the conversion formula \cite{Boucaud:2008gn}
\begin{equation}\label{A4}
\Lambda_T=\lms e^{507/792}\,.
\end{equation}
Thus using the estimate $\lms=0.224$ GeV$^2$ determined in \cite{Boucaud:2008gn} consequently leads to
\begin{equation}
    \braket{g^2A^2}^{\mu=10\,\mathrm{GeV}}= 3.03 \pm 0.24~\text{GeV}^2\,,
\end{equation}
whereby we have employed \eqref{corr} and $\braket{g^2A^2}= 3.35$ GeV$^2$ at $\mu_0=3$ GeV$^2$ as input values. A remarkable result is that our estimate is at least in the same ballpark as the ones of \eqref{A3nog} and \eqref{A3bisnog}, which were obtained in a completely independent way.

\subsection{Comparing the lattice estimate of $M^2$ with the analytic obtained value}
Finally, although the lattice data seem to support the gluon propagator \eqref{RZGtreelevel}, the value obtained for $M^2$, namely $M^2 = 2.15 \pm 0.13$ GeV$^2$, see equation \eqref{scale1}, does not agree with the values obtained in section \ref{sectie4}, i.e. $M^2 = 0.37 \lms^2 = 0.08288$ GeV$^2$ , whereby we used $\lms=0.224$ GeV$^2$ determined in \cite{Boucaud:2008gn}. This latter value was obtained at a different renormalization scale, but is far too much off to be in the same ballpark as the value obtained by the lattice \eqref{scale1}. However, notice that in section \ref{sectie4}, we did not take into account the condensate $\Braket{A^2}$, as we have always set $m^2 = 0$. Therefore, it is possible that this has a great influence on the results in section \ref{sectie4}. In any way, we can conclude that the results of section \ref{sectie4} are not very satisfactory and more research should be done on this topic.

\section{But there is more}
So far, we have refined the GZ action by including a dimension 2 condensate for which the corresponding operator is BRST invariant, namely $\overline{\varphi}^a_i \varphi^a_{i} - \overline{\omega}^a_i \omega^a_i = s(\overline \omega^a_i \varphi^a_i)$. However, as we have discussed in great extend in chapter \ref{scrutinizing}, the BRST symmetry is softly broken in the GZ action. Therefore, one could ask why we have investigated a BRST invariant $d=2$ operator. In fact, we could split the operator $\overline{\varphi}^a_i \varphi^a_{i} - \overline{\omega}^a_i \omega^a_i$ into two separate operators, i.e.~coupled to different sources. Moreover, there are other $d=2$ operators. In fact, all possible renormalizable $d=2$ operators $\mathcal O_i$ in the GZ action, which have ghost number zero, are given by
\begin{equation}\label{operators}
\mathcal O_i = \{  A_\mu A_\mu,  \varphi_i^a  \varphi_i^a, \varphi_i^a \overline \varphi_i^a,   \overline{\varphi}^a_i \overline \varphi^a_i , \overline \omega^a_i \omega^a_i \}\;.
\end{equation}
We shall only investigate condensates which are fully contracted over the indices, e.g.~like $\varphi_i^a \overline \varphi_i^a = \varphi_\mu^{ac} \overline \varphi_\mu^{ac}$, in order to have a lorentz invariant object. However, as one can find in \cite{Gracey:2010cg}, there are other possibilities to combine the color indices. If one wants to be absolutely complete, one would have to take into account all possible color contractions. Unfortunately, taking all possible color contractions into account would be hopelessly complicated and we hope that we have captured the physics by taking only one color combination. However, in principle, different color combinations are possible.\\
\\
As an advantage of the new insight that we can split the operator $\overline{\varphi}^a_i \varphi^a_{i} - \overline{\omega}^a_i \omega^a_i$, we shall be able to calculate the effective potential. This was impossible for the RGZ action as explained in section \ref{sectie4}. In this way, we shall be able to provide profound indications that the condensates are indeed present.\\
\\
Therefore, the purpose of this section is twofold. (1) We want to show that there are more condensates than taken in consideration so far. (2) We want to show explicitly that the minimum of the effective potential including the condensates is a non trivial minimum, i.e.~in this minimum the condensates are present.

\subsection{A further refining of the Gribov-Zwanziger action}
We propose to study the following extended action,
\begin{eqnarray}\label{CGZ}
\Sigma_\CGZ &=& \Sigma_\GZ'  + \Sigma_{A^2} + S_{\varphi \overline \varphi}  + S_{\overline \omega \omega} + S_{\overline{\varphi} \overline \varphi, \overline \omega \overline \varphi }  + S_{\varphi \varphi, \omega \varphi } + S_\vac\;,
\end{eqnarray}
whereby $\Sigma_\GZ'$ is given by equation \eqref{enlarged}, $\Sigma_{A^2}$ by \eqref{sigmaakwadraat} and
\begin{eqnarray}
S_{\varphi \overline \varphi } & =&  \int \d^4 x s(  P \overline \varphi^a_i \varphi^a_i ) ~=~  \int \d^4 x \left[  Q  \overline \varphi^a_i \varphi^a_i- P  \overline \varphi^a_i \omega^a_i\right]  \;, \nonumber\\
S_{\overline \omega \omega } & =&  \int \d^4 x s(  V \overline \omega^a_i \omega^a_i) ~=~  \int \d^4 x \left[ W\overline \omega^a_i \omega^a_i - V \overline \varphi^a_i \omega^a_i \right]  \;, \nonumber\\
S_{\overline{\varphi} \overline \varphi, \overline \omega \overline \varphi } & =&  \frac{1}{2} \int \d^4 x s(  \overline G^{ij} \overline \omega^a_i \overline \varphi^a_j) ~=~  \int \d^4 x \left[  \overline H^{ij} \overline \omega^a_i \overline \varphi^a_j + \frac{1}{2} \overline G^{ij} \overline \varphi^a_i \overline \varphi^a_j \right]  \;, \nonumber\\
S_{\varphi \varphi, \omega \varphi } &=& \frac{1}{2} \int \d^4 x s(  H^{ij} \varphi^a_i  \varphi^a_j) ~=~   \int \d^4 x  \left[ \frac{1}{2} G^{ij} \varphi^a_i \varphi^a_j - H^{ij} \omega^a_i \varphi^a_j \right] \;,\nonumber\\
S_\vac &=&  \int \d^4 x \left[ \kappa (G^{ij} \overline G^{ij} - 2 H^{ij} \overline H^{ij}) +  \lambda (G^{ii} \overline G^{jj} - 2 H^{ii} \overline H^{jj}) \right] \nonumber\\
 && - \int \d^4 x \left[ \alpha   (Q Q +  Q W) + \beta ( Q W + W W)  + \chi Q \tau + \delta W \tau \right]\;.
\end{eqnarray}
We have introduced 4 new doublets of sources, i.e.
\begin{eqnarray}
s P &=& Q \;, \nonumber\\
s V &=& W \;, \nonumber\\
s  \overline G^{ij} &=& 2 \overline H^{ij} \;, \nonumber\\
s H^{ij} &=& G^{ij}\;,
\end{eqnarray}
whereby $P$, $V$, $H^{ij}$ and $\overline H^{ij}$  behave like Grassmann quantities. For consistency, the sources with double index $^{ij}$ sources are symmetric in these indices. In this light, we use the following definition for the derivative w.r.t.~a symmetric source $\Lambda_{\mu\nu}$:
\begin{equation}\label{der}
    \frac{\delta \Lambda_{ij}}{\delta    \Lambda_{k\ell}}=\frac{1}{2}\left(\delta_{ik}\delta_{j\ell}+\delta_{i\ell}\delta_{jk}\right)\;.
\end{equation}
Notice that some sources have double indices, e.g.~$H^{ij}$, while other sources have no indices, e.g.~$P$. The reason for this is only for the of the algebraic renormalization in order to keep certain symmetries, and they have no further meaning. \\
\\
We have also introduced a vacuum term which shall be important for the renormalization of the vacuum energy. As shown in \cite{Verschelde:2001ia}, the dimensionless local composite operator (LCO) parameters $\alpha $, $\beta$, $\chi$ and $\delta$ of the quadratic terms in the sources are needed to account for the divergences present in the correlation functions like
$\braket{\mathcal O_i(k) \mathcal O_j(-k) }$, with $\mathcal O_i$ one of the operators given in expression \eqref{operators}.\\
\\
Now we can prove that the action \eqref{CGZ} is renormalizable to all orders. The proof is very similar to the proof of the renormalizability of the RGZ action, the only difference is that algebraically, mixing is allowed between different sources and parameters. We refer to the appendix \ref{renvery} for all the details.\\
\\
For the rest of the story, we are only interested in a number of condensates.  Therefore, we first set the source $W =0$, which is coupled to $\overline \omega \omega$, as this is not our current interest\footnote{There is no lowest order coupling of $\omega$ and $\overline \omega$ to the gluon sector. } and we also set $P = V = \eta = 0$, as we have introduced these only to preserve the BRST. Secondly, we also set $H^{ij} = \overline H^{ij} = 0$ and we set $G^{ij} = \delta^{ij} G$ and $\overline G^{ij} = \delta^{ij} \overline G$. The action \eqref{CGZ} becomes,
\begin{eqnarray}\label{startxx}
\Sigma_\CGZ &=& S_\GZ + \int \d^4 x  \left[ Q  \overline \varphi^a_i \varphi^a_i   + \frac{1}{2} \tau A_{\mu }^{a}A_{\mu }^{a}  - \frac{1}{2}\zeta \tau ^{2}  - \alpha   Q Q    - \chi Q \tau\right] \nonumber\\
&& + \int \d^4 x \left[ \frac{1}{2} \overline G \overline \varphi^a_i \overline \varphi^a_i + \frac{1}{2} G \varphi^a_i \varphi^a_i  + \varrho G \overline G \right]\;,
\end{eqnarray}
whereby $(\kappa d (N^2 -1) + \lambda d^2(N^2 -1)^2)$ was replaced by one parameter $\varrho$.

\subsection{A diagrammatical look at the mixing and vacuum divergences\label{diagrammatical}}
\begin{figure}[H]
   \centering
       \includegraphics[width=5cm]{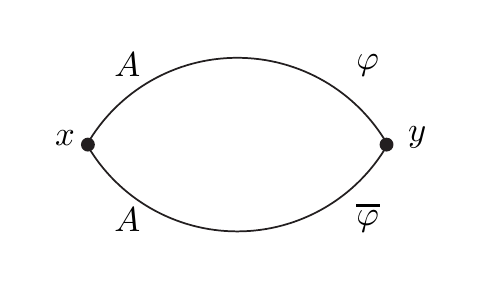}
   \caption{1-loop diagram.}
   \label{1loopd}
\end{figure}

Before starting the calculation of the effective action, we can provide some simplification of the action with the help of a diagrammatical argument. Firstly, looking at the action \eqref{startxx}, we see that a term $\chi Q \tau$ is present. This term is responsible for killing the divergences in the vacuum correlators $\Braket{A^2 (x) \overline \varphi \varphi (y)}$ for $x\to y$. However, we can prove that there are no divergences of this kind in diagrams at one loop. Let us start with these one loop diagrams. There is only one possible type of diagram for $\Braket{A^2 (x) \overline \varphi \varphi (y)}$, as can be found in Figure \ref{1loopd}.

\noindent The UV behavior of this diagram is finite, as can be extracted from the list of propagators \eqref{summarypropGZ}. Indeed, for large momenta, the corresponding integral of the diagram \eqref{1loopd} behaves like $\sim \int \d^4 p \frac{1}{p^4}\frac{1}{p^4}$, which is perfectly finite. Therefore, $ \lim_{x \to y}\Braket{A^2 (x) \overline \varphi \varphi (y)}$ is not divergent at one loop. In the next section, we shall explicitly prove this.\\
\\
At two loops, it is not possible to present the same argument as there exists a diagram which can be logarithmic divergent:

\begin{figure}[H]
   \centering
       \includegraphics[width=4.5cm]{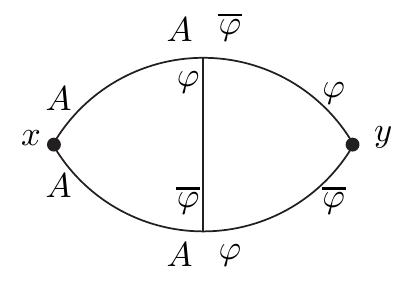}
   \caption{A possible divergent 2-loop diagram.}
   \label{1loopd2}
\end{figure}

\noindent as can be checked from the list of propagators \eqref{summarypropGZ}.\\
\\
Secondly, we can also have a look at the mixing of the operators $A^2$ and $\overline \varphi \varphi$. In the algebraic analysis, see appendix \ref{renvery}, we have found that a mixing is possible between the different operators, see equation \eqref{mixing}. This means that algebraically, a counterterm in $Q A_\mu A_\mu$ is allowed. This counterterm is needed to cancel the infinities of the following type of diagrams:

\begin{figure}[H]
   \centering
       \includegraphics[width=4.5cm]{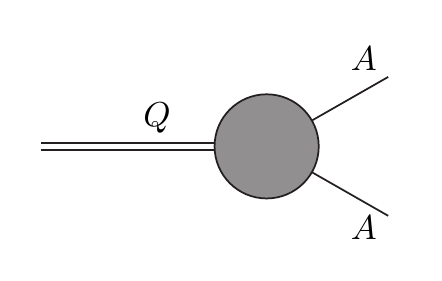}
   \label{mixxx}
\end{figure}

\noindent However, we can prove that there are no infinities at one loop, as the only possible diagram is given by,

\begin{figure}[H]
   \centering
       \includegraphics[width=4.5cm]{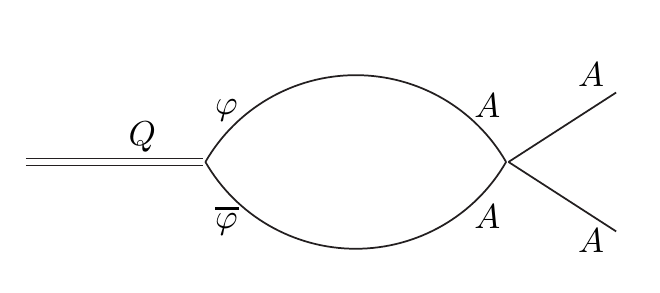}
   \label{mixx2}
\end{figure}

\noindent which is similar to the diagram in Figure \ref{1loopd}. We can thus conclude that the mixing can only start at two loops. Again, we cannot exclude divergences at two loops, due to a similar diagram as in Figure \ref{1loopd2}.

\subsection{The effective action}
In this section, we shall try to calculate the effective action. The calculation is quite technical and  shall therefore be split in different steps, although the result is reasonably compact and can be immediately found in expression \eqref{finaleffective}.\\
\\
The energy functional can be written as
\begin{eqnarray}\label{enfunc}
\e^{- W(Q, \tau, G, \overline G)} &=& \int [\d A_\mu][\d c][\d\overline c][\d b] [\d \varphi] [\d\overline \varphi] [\d \omega] [\d \overline \omega] \e^{-\Sigma_\CGZ}\;,
\end{eqnarray}
with $\Sigma_\CGZ$ given by equation \eqref{startxx}. We recall that in $d = 4 -\epsilon$ dimensions, we have the following dimensionalities,
\begin{eqnarray}
\left[A_\mu\right] &=& \left[\varphi \right] =  \frac{d-2}{2} = 1- \frac{\epsilon}{2}\;, \nonumber\\
\left[g\right] &=& \frac{4-d}{2} = \frac{\epsilon}{2}\;, \nonumber\\
\left[\tau \right] &=& \left[Q\right] = \left[G\right] = \left[\overline G\right] = 2   \;, \nonumber\\
\left[\zeta\right] &=& \left[\alpha\right]  =  \left[\chi\right] = \left[\varrho\right] = d-4 = -\epsilon\;.
\end{eqnarray}

\subsubsection{The LCO formalism}
In order to calculate the effective action, we shall follow the local composite operator (LCO) formalism developed  in \cite{Verschelde:2001ia,Verschelde:1995jj}. Let us outline the main idea. We start from a LCO $\mathcal O$, in our case a local dimension two operator within a dimension four theory. As is already done multiple times, we couple the operator(s) of interest to an appropriate source(s) $J$, and add this term $J \mathcal O$ to the Lagrangian. This gives rise to a functional $W (J)$, which we need to Legendre transform to find the effective potential, see equation \eqref{legendre}. However, as already made clear, novel infinities shall arise, which are proportional to $J^2$. This infinities are due to the divergences in the correlator $ \lim_{ x \to y}\braket{\mathcal O(x) \mathcal O(y)}$, as already explained in section \ref{diagrammatical}. Therefore, in general, a term proportional to $J^2$ is always needed in the counterterm, and the starting action needs a term\footnote{For an example, see the action \eqref{startxx}, where we the term $ - \frac{1}{2}\zeta \tau ^{2}  - \alpha   Q Q    - \chi Q \tau$ is needed in the starting action. The sources $Q$ and $\tau$ are coupled to the LCO operators $\mathcal O_1 = \overline \varphi_i \varphi_i $ and $\mathcal O_2 = A_\mu A_\mu$. Note that here, there is also a mixing term $\chi Q \tau$ for the divergences in $\lim_{x \to y}\braket{\mathcal O_1(x) \mathcal O_2 (y)}$.  } $\zeta J^2$. This $\zeta$ is called the LCO parameter, and is needed to absorb the divergences in $J^2$, i.e.~$\delta \zeta J^2$. The functional $W(J)$ obeys the following homogeneous RGE
\begin{equation}\label{RGEplus}
\left( \mu\frac{ \p}{\p \mu} + \beta(g^2) \frac{\p}{\p g^2} - \gamma_J(g^2) \int \d^4 x J \frac{\delta}{\delta J} + \eta(g^2 , \zeta) \frac{\p}{\p \zeta}  \right)W(J) = 0\;.
\end{equation}
with $\eta(g^2, \zeta)$ the running of $\zeta$,
\begin{equation}
\mu \frac{\p}{\p \mu} \zeta = \eta (g^2 , \zeta)
\end{equation}
Notice that it is necessary to include the running of $\zeta$ at this point.\\
\\
Now the question is, how can we determine this seemingly arbitrary parameter $\zeta$? This is possible by employing the renormalization group equations. We can write
\begin{equation}
\zeta_0 J_0^2 = \mu^{-\epsilon} (\zeta J^2 + \delta \zeta J^2)\;,
\end{equation}
which is the translation of expression \eqref{sibare}. As the l.h.s.~is independent from $\mu$, we can derive both sides w.r.t.~$\mu$ to find:
\begin{equation}\label{diffeq}
- \epsilon  (\zeta + \delta \zeta) + \left( \mu \frac{\p}{\p \mu} \zeta + \mu \frac{\p}{\p \mu} (\delta \zeta) \right) - 2 \gamma_J(g^2) (\zeta + \delta \zeta) = 0\;,
\end{equation}
whereby $\gamma_J(g^2)$ is the anomalous dimension of $J$. As we can consider $\zeta$ to be a function of $g^2$, and by evoking the $\beta$ function,
\begin{equation}
\beta (g^2) = \mu \frac{\p}{\p \mu} g^2
\end{equation}
the equation \eqref{diffeq} becomes,
\begin{equation}\label{diffeqx}
\beta(g^2)  \frac{\p}{\p g^2} \zeta (g^2) =   2 \gamma_J(g^2) \zeta +  f(g^2)\;.
\end{equation}
with $f(g^2) =  \epsilon  \delta \zeta  - \beta(g^2)  \frac{\p}{\p g^2} (\delta \zeta)  + 2 \gamma_G(g^2)  \delta \zeta $. The general solution of this differential equation reads
\begin{equation}
\zeta(g^2) = \zeta_p(g^2) + \alpha \exp \left( 2\int_1^{g^2}  \frac{\gamma_J(z)}{\beta (z)} \d z\right)\;,
\end{equation}
with $\zeta_p(g^2)$ a particular solution of \eqref{diffeqx}. A possible particular solution is given by
\begin{equation}
\zeta_p(g^2) = \frac{c_0}{g^2} + c_1 \hbar + c_2 g^2 \hbar^2 + \ldots \;.
\end{equation}
whereby we have temporarily introduced the dependence on $\hbar$. Notice therefore that the $n$-loop result for $\zeta(p^2)$ will require the $(n+ 1)$ loop results of $\beta(g^2)$, $\gamma_J(g^2)$ and $f(g^2)$. As we would like $\zeta$ to be multiplicatively renormalizable, we set $\alpha = 0$. In this case we have that
\begin{equation}
\zeta(g^2) + \delta \zeta(g^2) = \zeta_0 = Z_\zeta \zeta (g^2)\;.
\end{equation}
and we loose the independent parameter $\alpha.$ Also, now that $\zeta$ is a function of $g^2$, the RGE \eqref{RGEplus} becomes
\begin{equation}
\left( \mu\frac{ \p}{\p \mu} + \beta(g^2) \frac{\p}{\p g^2} - \gamma_J(g^2) \int \d^4 x J \frac{\delta}{\delta J}  \right)W(J) = 0\;,
\end{equation}
as deriving w.r.t.~$\zeta$ is now incorporated in deriving w.r.t.~$g^2$.\\
\\
After determining the LCO parameter $\zeta$, the next step is to calculate the effective action by doing a Legendre transformation. However, it shall be easier to perform a Hubbard-Stratonovich transformation on $W(J)$, whereby we introduce an auxiliary field $\sigma$ describing the composite operator $\mathcal O$. In this way, we immediately loose the quadratic term in $J^2$, and a clear relation with the effective action emerges. How this is done, shall be demonstrated in this section. We only need to mention that the case we are handling here is a bit more complicated due to the mixing of the operators $\mathcal O_1 = \overline \varphi_i \varphi_i $ and $\mathcal O_2 = A_\mu A_\mu$. However, the principles stay the same.

\subsubsection{Differential equation for the LCO parameters $\zeta$, $ \alpha$, $ \chi$ and $\varrho$}
We shall try to determine the four LCO parameters $\zeta$, $\alpha$, $\chi$ and $\varrho$. We shall first derive a differential equation for these parameters, in an analogous way as in \cite{Verschelde:2001ia,Dudal:2009tq}. As there can be mixing, we shall define $\delta \zeta$, $\delta \omega$ and $\delta \chi$ as follows
\begin{equation}\label{notatie}
-\frac{1}{2} \zeta_0 \tau_0^2 - \alpha_0 Q_0^2 - \chi_0 Q_0 \tau_0  = -\mu ^{-\epsilon} \left( \frac{1}{2} \zeta \tau^2 + \alpha Q^2 + \chi Q \tau + \frac{1}{2} \delta \zeta \tau^2 + \delta \alpha Q^2 +\delta \chi Q \tau \right)\;,
\end{equation}
while $\delta \varrho$ can be defined independently:
\begin{equation}\label{nata2}
\varrho_0 G_0 \overline G_0 =  \mu ^{-\epsilon}  Z_\varrho Z_G Z_{\overline G} \varrho G \overline G =  \mu ^{-\epsilon}  \left(1 + \frac{\delta \varrho}{\varrho} \right) \varrho G \overline G \;.
\end{equation}
We further define the $\beta$ function,
\begin{equation}
\mu \frac{\p}{\p \mu} g^2 = \beta (g^2)\;,
\end{equation}
and the anomalous dimension of $G$,
\begin{equation}\label{anoG}
\mu \frac{\p}{\p \mu} \ln Z_G   = \gamma_G(g^2) \quad \Rightarrow \quad \mu \frac{\p}{\p \mu} G = - \gamma_G(g^2)G \;,
\end{equation}
which is exactly the same as the anomalous dimension of $\overline G$ as $Z_G = Z_{\overline G}$. To define the anomalous dimensions of $Q$ and $\tau$, we start from equation \eqref{mixing}:
\begin{eqnarray}
  \underbrace{\begin{bmatrix}
     Q_0 \\
     \tau_0
  \end{bmatrix}}_{X_0}
 &= &
          \underbrace{\begin{bmatrix}
            Z_{QQ} &0 \\
           Z_{\tau Q} &  Z_{\tau\tau}          \end{bmatrix}}_{Z}
\underbrace{\begin{bmatrix}
    Q\\
    \tau
  \end{bmatrix}}_{X}\,,
\end{eqnarray}
a relation stemming from the algebraic renormalization. With the matrix $Z$, we can associate the anomalous dimension matrix $\Gamma$:
\begin{equation}
\mu \frac{\p}{\p \mu} Z = Z \Gamma\;,
\end{equation}
and thus
\begin{equation}\label{gammas}
\Gamma = Z^{-1} \mu \frac{\p}{\p \mu} Z = \begin{bmatrix}
              Z_{QQ}^{-1} \mu \frac{\p}{\p \mu} Z_{QQ} & 0 \\
              -Z_{\tau Q} \mu \frac{\p}{\p \mu} Z_{QQ} + Z_{\tau \tau}^{-1} \mu \frac{\p}{\p \mu} Z_{\tau Q} &  Z_{\tau\tau}^{-1} \mu \frac{\p}{\p \mu} Z_{\tau \tau}          \end{bmatrix} =  \begin{bmatrix}
              \gamma_{QQ} & 0 \\
              \Gamma_{21} &  \gamma_{\tau \tau}          \end{bmatrix}\;.
\end{equation}
This matrix is then related to the anomalous dimension of the operators:
\begin{align}
X_0 &= Z X &\Rightarrow 0 &= \mu \frac{\p Z}{ \p \mu}  X + Z \mu \frac{\p X}{\p \mu} &\Rightarrow \mu \frac{\p X}{\p \mu} &= - \Gamma X \,,
\end{align}
so the anomalous dimensions of the sources $Q$ and $\tau$ is given by
\begin{equation}
\mu \frac{\p}{\p \mu} \begin{bmatrix} Q \\ \tau \end{bmatrix} = \begin{bmatrix} -\gamma_{QQ} & 0  \\ - \Gamma_{21} & - \gamma_{\tau\tau} \end{bmatrix} \begin{bmatrix} Q \\ \tau \end{bmatrix}\;.
\end{equation}
With these definitions in mind, we can derive a differential equation for $\delta \zeta$, $\delta \omega$, $\delta \chi$ and $\delta \varrho$. We start with that of $\delta \varrho$. Starting from expression \eqref{nata2} and deriving w.r.t.~$\mu$, we find
\begin{equation}
- \epsilon  (\varrho + \delta \varrho) + \left( \mu \frac{\p}{\p \mu} \varrho + \mu \frac{\p}{\p \mu} (\delta \varrho) \right) - 2 \gamma_G(g^2) (\varrho + \delta \varrho) = 0\;.
\end{equation}
As we can consider $\varrho$ to be a function of $g^2$, according to the standard LCO formalism, we can rewrite this equation as
\begin{equation}
\beta(g^2)  \frac{\p}{\p g^2} \varrho (g^2) =  \epsilon (\varrho + \delta \varrho) - \beta(g^2)  \frac{\p}{\p g^2} (\delta \varrho)  + 2 \gamma_G(g^2) (\varrho + \delta \varrho) \;.
\end{equation}
As $\varrho$ is finite, we can even further simplify this into
\begin{equation}\label{diff2}
\beta(g^2)  \frac{\p}{\p g^2} \varrho (g^2) =   2 \gamma_G(g^2) \varrho + \epsilon \delta \varrho - \beta(g^2)  \frac{\p}{\p g^2} (\delta \varrho)  + 2 \gamma_G(g^2)  \delta \varrho \;.
\end{equation}
In an analogous fashion, we can find that differential equations for $\delta \zeta$, $\delta \omega$ and $\delta \chi$. If we derive \eqref{notatie} w.r.t.~$\mu$, we find the following set of coupled differential equations
\begin{align}\label{diff1}
&  \beta(g^2)  \frac{\p}{\p g^2} \frac{\zeta (g^2)}{2}  =  \frac{\epsilon}{2}  \delta \zeta  - \frac{1}{2}\beta(g^2)  \frac{\p}{\p g^2} (\delta \zeta)  +  \gamma_{\tau \tau}(g^2) (\zeta + \delta \zeta) \;,\nonumber\\
& \beta(g^2)  \frac{\p}{\p g^2} \alpha (g^2)  =  \epsilon \delta \alpha  - \beta(g^2)  \frac{\p}{\p g^2} (\delta \alpha)  + 2  \gamma_{QQ}(g^2) (\alpha + \delta \alpha) + \Gamma_{21}(g^2) (\chi + \delta \chi) \;, \nonumber\\
& \beta(g^2)  \frac{\p}{\p g^2} \chi (g^2) =  \epsilon  \delta \chi  - \beta(g^2)  \frac{\p}{\p g^2} (\delta \chi)  +   \gamma_{QQ}(g^2) (\chi+ \delta \chi) + \gamma_{\tau \tau}(g^2) (\chi + \delta \chi) \nonumber\\
& + \Gamma_{21} (g^2) (\zeta + \delta \zeta) \;.
\end{align}

\subsubsection{Determination of the LCO parameters $\delta \zeta$, $\delta \alpha$, $\delta \chi$ and $\delta \varrho$ \label{section4.2}}
In order to determine the LCO parameters $\delta \zeta$, $\delta \alpha$, $\delta \chi$ and $\delta \varrho$ at one loop, we need to calculate the one loop divergence of the energy functional $W(Q, \tau, G, \overline G)$. The details of these calculations can be found in appendix \ref{detailsvery}. From section \ref{diagrammatical}, we know that at one loop, $\delta \chi$ should be zero. This observation shall serve as a check on our computations.\\
\\
In the appendix, equation \eqref{deltas}, we have found
\begin{eqnarray}\label{deltasx}
\delta \zeta &=& -\frac{1}{\epsilon} \frac{3}{16 \pi^2} (N^2-1) \;,\nonumber\\
\delta \alpha &=& -\frac{1}{\epsilon} \frac{1}{4 \pi^2} (N^2-1)^2 \;, \nonumber\\
\delta \chi &=& 0 \;,\nonumber\\
\delta \varrho &=& \frac{1}{\epsilon} \frac{1}{4 \pi^2} (N^2-1)^2 \;.
\end{eqnarray}
Already a first check on these results is the value of $\delta \zeta$, which has been already calculated up to three loops, see \cite{Verschelde:2001ia,Browne:2003uv}. Comparison with these articles learns that the one loop value indeed coincides. Secondly, we also see that indeed $\delta \chi = 0$ at one loop, which nicely confirms our diagrammatical power counting argument.

\subsubsection{Solving the differential equations for $\zeta$, $\alpha$, $\chi$ and $\varrho$}
In this section, we shall try to solve the differential equations \eqref{diff2} and \eqref{diff1}, when possible. For these calculations, it is useful to keep in mind the $\beta$ function, here given up to two loops
\begin{eqnarray}
\beta(g^2) &=& -\epsilon g^2 - 2\left(  \beta_0 g^4 + \beta_1 g^6 + O(g^8)  \right)\;,
\end{eqnarray}
with
\begin{eqnarray}
\beta_0 &=&  \frac{11}{3}  \left( \frac{N}{16 \pi^2} \right)\;,\qquad \beta_1 ~=~  \frac{34}{3}                \left( \frac{N}{16 \pi^2}\right)^2\;,
\end{eqnarray}
in order to keep track of the orders.\\
\\
We start with \eqref{diff2},
\begin{equation}\label{diffa}
\beta(g^2)  \frac{\p}{\p g^2} \varrho (g^2) =   2 \gamma_G(g^2) \varrho + \epsilon \delta \varrho - \beta(g^2)  \frac{\p}{\p g^2} (\delta \varrho)  + 2 \gamma_G(g^2)  \delta \varrho \nonumber\\
\end{equation}
In order to solve this differential equation, we need to parameterize $\varrho$ as follows:
\begin{equation}
\varrho = \frac{\varrho_0}{g^2} + \varrho_1 + \varrho_2 g^2 + \mathcal O(g^4)\;.
\end{equation}
We also need the explicit value of the anomalous dimension $\gamma_G$. We have from the definition \eqref{anoG} that
\begin{equation}
 \gamma_G(g^2)= \mu \frac{\p}{\p \mu} \ln Z_G \;,
\end{equation}
and thus we need the value of $Z_G$. From the renormalization factors \eqref{Zfac} and \eqref{Z3}, we find that
\begin{equation}
 \gamma_G(g^2)= - \mu \frac{\p}{\p \mu} \ln Z_\varphi =  - \mu \frac{\p}{\p \mu} \ln (Z_g^{-1} Z_A^{-1/2}) \;.
\end{equation}
In \cite{Gracey:2002yt}, the factors $Z_g$ and $Z_A$ have been calculated up to three loops,
\begin{eqnarray}
Z_A &=& 1+ \frac{13}{6} \frac{1}{\epsilon} \frac{N g^2}{16\pi^2} + \left(\frac{-13}{8} \frac{1}{\epsilon^2} + \frac{59}{16} \frac{1}{\epsilon} \right) \left( \frac{N g^2}{16\pi^2} \right)^2  + \ldots \nonumber\\
Z_g &=& 1- \frac{11}{6} \frac{1}{\epsilon} \frac{N g^2}{16\pi^2} + \left(\frac{121}{24} \frac{1}{\epsilon^2} - \frac{17}{6} \frac{1}{\epsilon} \right) \left( \frac{N g^2}{16\pi^2} \right)^2 +\ldots \;.
\end{eqnarray}
So one can calculate $\gamma_G(g^2)$ up to three loops if necessary. Here only the first loop shall be useful for our calculations, i.e.
\begin{equation}
\gamma_G(g^2) = \frac{3}{4} \frac{N g^2}{16 \pi^2} + \ldots \;,
\end{equation}
as $\delta \varrho$, see equation \eqref{deltas} is only known up to lowest order. With this information, we can solve the differential equation \eqref{diffa} up to lowest order, by matching the corresponding orders in $g^2$
\begin{equation}
\varrho = \frac{24}{53} \frac{ (N^2 - 1)^2}{N g^2} + \varrho_1 + \ldots \;.
\end{equation}
Unfortunately, we cannot solve the differential equation for $\varrho_1$ as we would require the two loop value of $\delta \varrho$, which is on our to do list. Therefore, we leave this value as a still to be determined parameter. \\
\\
Let us now turn to the set of differential equations \eqref{diff1}. We can do a similar analysis as above for the first differential equation, namely
\begin{equation}
\beta(g^2)  \frac{\p}{\p g^2} \frac{\zeta (g^2)}{2}  =  \frac{\epsilon}{2}  \delta \zeta  - \frac{1}{2}\beta(g^2)  \frac{\p}{\p g^2} (\delta \zeta)  +  \gamma_{\tau \tau}(g^2) (\zeta + \delta \zeta)\;.
\end{equation}
We shall again parameterize $\zeta$ as follows:
\begin{equation}\label{opl1}
\zeta = \frac{\zeta_0}{g^2} + \zeta_1 + \zeta_2 g^2 + \mathcal O(g^4)\;.
\end{equation}
In fact, we can even solve this differential equation to two loops. From \cite{Verschelde:2001ia,Browne:2003uv,Dudal:2009tq}, we know that
\begin{multline}
\delta \zeta = \frac{N^2 - 1}{16 \pi^2} \Biggl[ -\frac{3}{\epsilon} + \left( \frac{35}{2} \frac{1}{\epsilon^2} -  \frac{139}{6} \frac{1}{\epsilon} \right)\left(  \frac{g^2 N}{16\pi^2} \right)   \\ +\left(-\frac{665}{6} \frac{1}{\epsilon^3} + \frac{6629}{36} \frac{1}{\epsilon^2} - \left(  \frac{71551}{432} + \frac{231}{16} \zeta(3) \right)  \frac{1}{\epsilon} \right)\left(  \frac{g^2 N}{16\pi^2} \right)^2 \Biggr]\;,
\end{multline}
and
\begin{multline}
 Z_{\tau\tau} = 1- \frac{35}{6} \frac{1}{\epsilon} \left( \frac{g^2 N }{16 \pi^2}\right)  + \left[ \frac{2765}{72}  \frac{1}{\epsilon^2}-\frac{449}{48}  \frac{1}{\epsilon}   \right] \left( \frac{g^2 N }{16 \pi^2}\right)^2 \\+ \left[ -\frac{113365}{432}\frac{1}{\epsilon^3} +  \frac{41579}{576}  \frac{1}{\epsilon^2}+\left(-\frac{75607}{2592}-\frac{3}{16}\zeta(3)\right)\frac{1}{\epsilon}  \right] \left( \frac{g^2 N }{16 \pi^2}\right)^3\;,
\end{multline}
so that from \eqref{gammas}
\begin{eqnarray}\label{ress1}
\gamma_{\tau\tau}(g^2) =  \frac{35}{6}\left(  \frac{g^2 N}{16\pi^2} \right) +  \frac{449}{24}\left(  \frac{g^2 N}{16\pi^2} \right)^2 +  \left( \frac{94363}{864} + \frac{9}{16} \zeta(3) \right)\left(  \frac{g^2 N}{16\pi^2} \right)^3\;.
\end{eqnarray}
By solving the differential equation for $\zeta$, we can determine $\zeta$ to one loop order. In principle, we can even go one loop further with the known results. However, as we shall only determine the effective potential to one loop order, we do not need this next loop result. We find,
\begin{eqnarray}\label{weljui}
\zeta &=& \frac{N^2 -1}{16 \pi^2} \left[\frac{9}{13} \frac{16 \pi^2}{ g^2 N} + \frac{161}{52} \right]\;,
\end{eqnarray}
see also \cite{Dudal:2009tq}. \\
\\
The second and third differential equation of \eqref{diff1} are coupled. They can be simplified and decoupled as $\delta \chi = 0$:
\begin{eqnarray}
\beta(g^2)  \frac{\p}{\p g^2} \alpha (g^2)  &=&  2  \gamma_{QQ}(g^2) \alpha + \epsilon  \delta \alpha  - \beta(g^2)  \frac{\p}{\p g^2} (\delta \alpha)  + 2  \gamma_{QQ}(g^2) \delta \alpha + \Gamma_{21}(g^2) \chi  \;,\nonumber\\
\beta(g^2)  \frac{\p}{\p g^2} \chi (g^2) &=&      \gamma_{QQ}(g^2) \chi + \gamma_{\tau \tau}(g^2) \chi  + \Gamma_{21} (g^2) (\zeta + \delta \zeta) \;.
\end{eqnarray}
Fortunately, we know that $\Gamma_{21} = 0$ at lowest order, from the diagrammatical argument in section \ref{diagrammatical}. Therefore, we can set $\Gamma_{21} = 0 + O(g^4)$. When parameterizing as usual
\begin{align}
\alpha &= \frac{\alpha_0}{g^2} + \alpha_1 + \alpha_2 g^2 + \mathcal O(g^4) \;, &  \chi &= \frac{\chi_0}{g^2} + \chi_1 + \chi_2 g^2 + \mathcal O(g^4) \;,
\end{align}
we find for the solution of the differential equations
\begin{eqnarray}\label{opl2}
\alpha_0 &=& -\frac{24  (N^2-1)^2}{35 N} \;,\nonumber\\
\chi_0 &=& 0 \;.
\end{eqnarray}

\subsubsection{Hubbard-Stratonovich transformations}
In this section, we shall get rid of the unwanted quadratic source dependence by the introduction of multiple Hubbard-Stratonovich fields.
We can then rewrite the relevant part of the action in terms of finite fields and sources:
\begin{eqnarray*}
 && \int \d^4 x  \Bigl[ \underbrace{ Z_{QQ} Z_\varphi }_{c} Q \overline \varphi^a_i \varphi^a_i   + \underbrace{ \frac{1}{2} Z_A Z_{\tau\tau}}_{b} \tau A_{\mu }^{a}A_{\mu }^{a}   + \underbrace{ \frac{1}{2} Z_A Z_{\tau Q }}_{a} Q A_{\mu }^{a}A_{\mu }^{a}  - \underbrace{ \frac{1}{2} Z_{\zeta \zeta} Z_{\tau\tau}^2 \zeta }_{\zeta'} \mu^{-\epsilon} \tau ^{2} \nonumber\\
 && -\underbrace{ Z_{QQ}^2 Z_{\alpha\alpha} \alpha }_{\alpha'} \mu^{-\epsilon} Q Q    - \underbrace{ Z_{QQ} Z_{\chi \chi} Z_{\tau\tau}  \chi}_{\chi'}  \mu^{-\epsilon} Q \tau\Bigr]\nonumber\\
 &&+ \int \d^4 x \Bigl[  Z_{ G} Z_\varphi \frac{1}{2} \overline G \overline \varphi^a_i \overline \varphi^a_i + Z_G Z_\varphi \frac{1}{2} G \varphi^a_i \varphi^a_i  + Z_\varrho Z_G^2 \varrho G \overline G\Bigr]\;.
\end{eqnarray*}
We shall now perform the following Hubbard-Stratonovich transformations by multiplying expression \eqref{enfunc} with the following unities,
\begin{eqnarray}\label{unities}
1 &=& \int [\d \sigma_1] \e^{-\frac{1}{4 \zeta'} \int\d^d x \left( \frac{\sigma_1 }{g}  + b \mu^{\epsilon/2} A^2 - 2 \zeta' \mu^{-\epsilon/2} \tau  - \chi' \mu^{-\epsilon/2} Q  \right)^2 }\;, \nonumber\\
1 &=& \int [\d \sigma_2] \e^{- \frac{1}{4 \zeta' [4 \alpha' \zeta' - \chi^{\prime 2}  ] } \int\d^d x \left( \frac{\sigma_2 }{g}  +  (b \chi' - 2 a \zeta') \mu^{\epsilon/2} A^2 - 2 c \zeta' \mu^{\epsilon/2} \overline \varphi \varphi + (4 \alpha' \zeta' - \chi^{\prime 2}) \mu^{-\epsilon/2} Q    \right)^2 } \;, \nonumber\\
1 &=& \int [\d \sigma_3] \e^{-\frac{1}{ 4 Z_\varrho Z_G^2  \varrho } \int\d^d x \left( \frac{\sigma_3 }{g}  + \frac{1}{2} \mu^{\epsilon/2} Z_{ G} Z_\varphi \overline \varphi \overline \varphi  + \frac{1}{2} \mu^{\epsilon/2} Z_{ G} Z_\varphi  \varphi \varphi + Z_{ G}^2 Z_\varrho \varrho \mu^{-\epsilon/2} \overline G + Z_G^2 Z_\varrho \varrho \mu^{-\epsilon/2} G   \right)^2 } \;, \nonumber\\
1 &=& \int [\d \sigma_4] \e^{-\frac{1}{ 4 Z_\varrho Z_G^2  \varrho } \int\d^d x \left( \frac{ \sigma_4 }{g}  + \frac{\ii}{2} \mu^{\epsilon/2} Z_{ G} Z_\varphi \overline \varphi \overline \varphi  -\frac{\ii}{2} \mu^{\epsilon/2} Z_{ G} Z_\varphi  \varphi \varphi - \ii Z_{ G}^2 Z_\varrho \varrho \mu^{-\epsilon/2} \overline G + \ii Z_G^2  Z_\varrho \varrho \mu^{-\epsilon/2} G   \right)^2 }  \;, \nonumber\\
\end{eqnarray}
whereby we have introduced four new fields, $\sigma_1$,$\sigma_2$, $\sigma_3$ and $\sigma_4$. By doing this HS transformation, we shall remove the quadratic sources and rewrite the functional energy as
\begin{multline}\label{enfunc2}
\e^{- W(Q, \tau, G, \overline G)} = \int [\d A_\mu][\d c][\d\overline c][\d b] [\d \sigma_1][\d \sigma_2][\d \sigma_3] [\d \sigma_4] [\d \varphi] [\d\overline \varphi] [\d \omega] [\d \overline \omega]\\
\times \e^{\left[ - \int\d^d x \left( \mathcal L(\phi, \sigma_1, \ldots,\sigma_4)   - \mu^{-\epsilon/2} \frac{\sigma_1}{g} \frac{  2 \zeta' \tau + \chi' Q }{2 \zeta'}    + \mu^{-\epsilon/2} \frac{\sigma_2}{g}  \frac{ Q}{2 \zeta'  }  +\frac{1}{2} \mu^{-\epsilon/2} \frac{\sigma_3 -\ii  \sigma_4}{g} \overline G+ \frac{1}{2} \frac{\sigma_3 +\ii \sigma_4}{g} \mu^{-\epsilon/2}  G  \right)  \right]}\;,
\end{multline}
with $\phi = ( A_\mu, c,\overline c, b,\varphi,\overline \varphi,\omega ,\overline \omega )$ and
\begin{align}
&\int \d^d x \mathcal L(\phi, \sigma_1,\ldots ,\sigma_4)  =S_\GZ +  \int \d^d x \Biggl(\frac{1}{4 \zeta^{\prime } } \frac{\sigma_1^2}{g^2} +    \frac{b}{2 \zeta'   } \frac{\sigma_1 }{g} \mu^{\epsilon/2} A^2   + \frac{b^2  }{4 \zeta' } \mu^{\epsilon} (A_\mu^a A_\mu^a)^2   \nonumber\\
& + \frac{1}{4 \zeta' [4 \alpha' \zeta' - \chi^{\prime 2}  ] } \frac{\sigma_2^2}{g^2}  + \frac{b \chi'  - 2 a \zeta'}{2 \zeta' [4 \alpha' \zeta' - \chi^{\prime 2}  ] }  \mu^{\epsilon/2} \frac{\sigma_2 }{g}   A^2 -    \frac{c }{  4 \alpha' \zeta' - \chi^{\prime 2}   }   \mu^{\epsilon/2}  \frac{\sigma_2}{g} \overline \varphi \varphi  \nonumber \\
& + \frac{(b \chi' - 2 a \zeta')^2}{4 \zeta' [4 \alpha' \zeta' - \chi^{\prime 2}  ] }  \mu^{\epsilon} (A_\mu^a A_\mu^a)^2   +  \frac{c^2 \zeta' }{ [4 \alpha' \zeta' - \chi^{\prime 2}  ] }  \mu^{\epsilon} (\overline \varphi^a_i \varphi^a_i)^2  -  \frac{c (b \chi' - 2 a \zeta') }{ 4 \alpha' \zeta'  - \chi^{\prime 2}   }  \mu^{\epsilon}  A_\mu^a A_\mu^a  \overline\varphi^b_i \varphi^b_i  \nonumber\\
&   + \frac{1}{ 4 Z_\varrho Z_G^2  \varrho }\left( \frac{\sigma_3^2}{g^2} + \frac{\sigma_4^2}{g^2} \right)  + \mu^{\epsilon/2}\frac{Z_\varphi}{ 4 Z_\varrho  Z_{G} \varrho } \frac{\sigma_3 }{g} \left( \overline \varphi \overline \varphi + \varphi  \varphi \right)   +\mu^{\epsilon/2}\frac{Z_\varphi}{ 4 Z_\varrho  Z_{ G} \varrho } \frac{\ii \sigma_4 }{g}\left( \overline \varphi \overline \varphi - \varphi  \varphi \right) \nonumber\\
&+ \mu^\epsilon \frac{Z_\varphi^2}{4 Z_\varrho \varrho} \overline \varphi^a_i \overline \varphi^a_i \varphi^b_j \varphi^b_j \Biggr) \;.
\end{align}

\noindent As the HS transformation does not put everything in the right form, we propose the following \textit{extra} transformation
\begin{eqnarray}
\sigma_1 \frac{\chi'}{2 \zeta'} - \frac{\sigma_2}{2 \zeta'} &=& \sigma_2' \;.
\end{eqnarray}
So \eqref{enfunc2} becomes
\begin{multline}
\e^{- W(Q, \tau, G, \overline G)} = \int [\d A_\mu][\d c][\d\overline c][\d b] [\d \sigma_1][\d \sigma_2][\d \sigma_3] [\d \sigma_4] [\d \varphi] [\d\overline \varphi] [\d \omega] [\d \overline \omega] \\
\times \e^{\left[ - \int\d^d x \left( \mathcal L(\phi, \sigma_1, \ldots,\sigma_4)   - \mu^{-\epsilon/2} \frac{\sigma_1}{g}  \tau     - \mu^{-\epsilon/2} \frac{\sigma_2'}{g}   Q   +\frac{1}{2} \mu^{-\epsilon/2} \frac{\sigma_3 -\ii  \sigma_4}{g} \overline G+ \frac{1}{2} \frac{\sigma_3 +\ii \sigma_4}{g} \mu^{-\epsilon/2}  G  \right)  \right]}\;,
\end{multline}
whereby
\begin{align}\label{lagrangian}
&\int \d^d x \mathcal L(\phi, \sigma_1,\ldots ,\sigma_4)  =S_\GZ +  \int \d^d x \Biggl(  \frac{ \alpha' }{4 \alpha' \zeta' - \chi^{\prime 2}}  \frac{\sigma_1^2}{g^2} + \frac{ \zeta' }{ 4 \alpha' \zeta' - \chi^{\prime 2}}  \frac{\sigma_2^2}{g^2}   -  \frac{ \chi' }{4 \alpha' \zeta' - \chi^{\prime 2}}  \frac{\sigma_1 \sigma_2}{g^2}  \nonumber\\
 & +   \frac{2 b \alpha' - a \chi'}{ 4 \alpha' \zeta' - \chi^{\prime 2}  } \frac{\sigma_1 }{g} \mu^{\epsilon/2} A^2  -    \frac{b \chi'  - 2 a \zeta'}{ [4 \alpha' \zeta' - \chi^{\prime 2}  ] }  \mu^{\epsilon/2} \frac{\sigma_2 }{g}   A^2  -    \frac{c \chi' }{  4 \alpha' \zeta' - \chi^{\prime 2}   }   \mu^{\epsilon/2}  \frac{\sigma_1}{g} \overline \varphi \varphi \nonumber\\
&  +    \frac{ 2 c \zeta' }{  4 \alpha' \zeta' - \chi^{\prime 2}   }   \mu^{\epsilon/2}  \frac{\sigma_2}{g} \overline \varphi \varphi  + \frac{b^2  }{4 \zeta' } \mu^{\epsilon} (A_\mu^a A_\mu^a)^2  + \frac{(b \chi' - 2 a \zeta')^2}{4 \zeta' [4 \alpha' \zeta' - \chi^{\prime 2}  ] }  \mu^{\epsilon} (A_\mu^a A_\mu^a)^2    \nonumber\\
&  +  \frac{c^2 \zeta' }{ [4 \alpha' \zeta' - \chi^{\prime 2}  ] }  \mu^{\epsilon} (\overline \varphi^a_i \varphi^a_i)^2  -  \frac{c (b \chi' - 2 a \zeta') }{ 4 \alpha' \zeta'  - \chi^{\prime 2}   }   \mu^{\epsilon}  A_\mu^a A_\mu^a  \overline\varphi^b_i \varphi^b_i  + \frac{1}{ 4 Z_\varrho Z_G^2  \varrho }\left( \frac{\sigma_3^2}{g^2}+ \frac{\sigma_4^2}{g^2} \right) \nonumber\\
 & + \mu^{\epsilon/2}\frac{Z_\varphi}{ 4 Z_\varrho  Z_{G} \varrho } \frac{\sigma_3 }{g} \left( \overline \varphi \overline \varphi + \varphi  \varphi \right)   +\mu^{\epsilon/2}\frac{Z_\varphi}{ 4 Z_\varrho  Z_{ G} \varrho } \frac{\ii \sigma_4 }{g}\left( \overline \varphi \overline \varphi - \varphi  \varphi \right) + \mu^\epsilon \frac{Z_\varphi^2}{4 Z_\varrho \varrho} \overline \varphi^a_i \overline \varphi^a_i \varphi^b_j \varphi^b_j \Biggr)\;.
\end{align}
Now acting with $\left . \frac{\delta }{\delta Q}\right|_{Q,\tau = 0}$ and $\left. \frac{\delta}{\delta \tau} \right|_{Q,\tau =0}$ on the equivalent energy functional before and after the HS transformation gives us the following two relations,
\begin{eqnarray}
Z_{QQ} Z_\varphi\Braket{\overline \varphi^a_i \varphi^a_i  } + \frac{1}{2} Z_A Z_{\tau W } \Braket {A_{\mu }^{a}A_{\mu }^{a}}  &=& - \mu^{-\epsilon/2}
\frac{\Braket{\sigma_2}}{g}  \;, \nonumber\\
\frac{1}{2} Z_A Z_{\tau \tau } \Braket {A_{\mu }^{a}A_{\mu }^{a}}  &=& - \mu^{-\epsilon/2}
\frac{\Braket{\sigma_1}}{g} \;,
\end{eqnarray}
while acting with $\left . \frac{\delta }{\delta G}\right|_{G,\overline G = 0}$ and $\left. \frac{\delta}{\delta \overline G} \right|_{G,\overline G=0}$
\begin{eqnarray}
Z_G Z_\varphi \Braket{\varphi \varphi} &=& \mu^{-\epsilon/2} \frac{\Braket{\sigma_3 + \ii \sigma_4}}{g} \;, \nonumber\\
Z_{ G} Z_\varphi \Braket{\overline \varphi \overline \varphi} &=&  \mu^{-\epsilon/2}\frac{\Braket{\sigma_3 -\ii \sigma_4}}{g} \;,
\end{eqnarray}
or equivalently
\begin{eqnarray}
Z_G Z_\varphi  \frac{1}{2}\Braket{\varphi \varphi + \overline \varphi \overline \varphi} &=& \mu^{-\epsilon/2} \frac{\Braket{\sigma_3}}{g} \;, \nonumber\\
Z_G Z_\varphi  \frac{\ii}{2}\Braket{ \overline \varphi \overline \varphi- \varphi \varphi } &=& \mu^{-\epsilon/2} \frac{\Braket{\sigma_4}}{g}\;.
\end{eqnarray}

\subsubsection{The effective action}
If we put the parameters
\begin{eqnarray}
\frac{m^2}{2} &=&   \frac{1}{4\alpha_0 \zeta_0 - 2 \chi_0^2} \left( 2 \alpha_0  g \sigma_1 - \chi_0 g \sigma_2\right) \;, \nonumber\\
M^2 &=&   \frac{1}{2\alpha_0 \zeta_0 -  \chi_0^2} \left(  \chi_0  g \sigma_1 - \zeta_0 g \sigma_2\right) \;, \nonumber\\
\rho &=& -\frac{53 N}{ 48 (N^2-1)^2   }  (\sigma_3 + \ii \sigma_4) g  \;, \nonumber\\
\rho^\dagger &=& -\frac{53 N}{ 48 (N^2-1)^2   }  (\sigma_3 - \ii \sigma_4)g \;,
\end{eqnarray}
with $\alpha_0, \zeta_0, \chi_0$ given in equations \eqref{weljui}-\eqref{opl2}, then the quadratical part of the Lagrangian \eqref{lagrangian} is given by
\begin{align}
&\int \d^d x \mathcal L(\phi, \sigma_1,\ldots ,\sigma_4)  =S_\GZ^{\quadr} +  \int \d^d x \Biggl(   \frac{ \alpha' }{4 \alpha' \zeta' - \chi^{\prime 2}}  \frac{\sigma_1^2}{g^2} + \frac{ \zeta' }{ 4 \alpha' \zeta' - \chi^{\prime 2}}  \frac{\sigma_2^2}{g^2}   -  \frac{ \chi' }{4 \alpha' \zeta' - \chi^{\prime 2}}  \frac{\sigma_1 \sigma_2}{g^2} \nonumber\\
 & + \frac{1}{ 4 Z_\varrho Z_G^2  \varrho }\left( \frac{\sigma_3^2}{g^2} + \frac{\sigma_4^2}{g^2} \right)
  +  \frac{ m^2}{2} \mu^{\epsilon/2} A^2  -    M^2 \mu^{\epsilon/2}  \overline \varphi \varphi     + \mu^{\epsilon/2} \frac{\rho}{2} \overline \varphi \overline \varphi +  \mu^{\epsilon/2} \frac{\rho^\dagger}{2} \varphi  \varphi  \Biggr)\;.
\end{align}
We have left out the higher order terms as we shall only calculate the one loop effective potential. \\
\\
We have collected all the details in the appendix \ref{detailsvery}, but the final result for the effective potential is given by
\begin{align}\label{finaleffective}
\Gamma^{(1)} &=  \frac{(N^2 -1)^2}{16 \pi^2} \Bigl[(M^2 - \sqrt{\rho \rho^\dagger})^2 \ln\frac{M^2 - \sqrt{\rho \rho^\dagger}}{\overline \mu^2}  + (M^2 + \sqrt{\rho \rho^\dagger})^2 \ln\frac{M^2 + \sqrt{\rho \rho^\dagger}}{\overline \mu^2} \nonumber\\
& - 2 (M^2 + \rho \rho^\dagger)\Bigr] + \frac{3(N^2 -1)}{64\pi^2} \Bigl[  - \frac{5}{6} (m^4 -2 \lambda^4) + y_1^2 \ln \frac{(-y_1)}{\overline \mu} + y_2^2 \ln \frac{(-y_2)}{\overline \mu} + y_3^2 \ln \frac{(-y_3)}{\overline \mu} \nonumber\\
&- y_4^2 \ln \frac{(-y_4)}{\overline \mu} - y_5^2 \ln \frac{(-y_5)}{\overline \mu} \Bigr]  -2 (N^2 - 1) \frac{\lambda^4}{N g^2}  + \frac{3}{2} \frac{\lambda^4}{32 \pi^2} (N^2-1) \nonumber\\
&+ \frac{1}{2} \frac{48 (N^2-1)^2}{53 N} \left( 1   - N g^2 \frac{53}{24} \frac{\varrho_1}{(N^2 - 1)^2} \right) \frac{\rho \rho^\dagger}{g^2} \nonumber\\
&  +  \frac{9}{13} \frac{N^2-1}{N}\frac{m^4}{2g^2}- \frac{24}{35}\frac{(N^2-1)^2}{N}\frac{M^4}{g^2}  - \frac{161}{52} \frac{N^2-1}{16 \pi^2}\frac{ m^4}{2 }- M^4 \alpha_1  +M^2 m^2 \chi_1\;.
\end{align}
whereby $y_1$, $y_2$ and $y_3$ are the solutions of the equation $y^3+(m^2 +2 M^2) y^2 +\bigl(\lambda^4+ M^4- \rho \rho^\dagger +2 M^2 m^2 \bigr) y + M^2 \lambda^4 + 1/2 ( \rho + \rho^\dagger) \lambda^4 + M^4 m^2  - m^2 \rho \rho^\dagger  =0$ and $y_4$ and $y_5$ of the equation $ y^2 + 2 M^2 y +M^4 -\rho \rho^\dagger =0$.

\subsubsection{Minimizing the effective potential to prove that the condensates are non-vanishing}
To simplify the calculations, let us set $\rho= \rho^\dagger = 0$, which is related to not considering the condensates $\braket{\overline \varphi \overline \varphi}$ and $\Braket{\varphi \varphi}$. For a moment, we are only considering $\braket{\varphi \overline \varphi}$, which also already has the desired influence on the propagators, see the next section. In this case, the effective action becomes:
\begin{align}\label{mineffpot}
\Gamma^{(1)} &=  \frac{(N^2 -1)^2}{16 \pi^2}\Bigl[2 M^4 \ln\frac{M^2}{\overline \mu^2}   - 2 M^2 \Bigr] + \frac{3(N^2 -1)}{64\pi^2} \Bigl[  - \frac{5}{6} (m^4 -2 \lambda^4)+ M^4 \ln \frac{(M^2)}{\overline \mu} \nonumber\\
&+ y_2^2 \ln \frac{(-y_2)}{\overline \mu} + y_3^2 \ln \frac{(-y_3)}{\overline \mu} - 2M^4 \ln \frac{M^2}{\overline \mu}  \Bigr]  -2 (N^2 - 1) \frac{\lambda^4}{N g^2}  + \frac{3}{2} \frac{\lambda^4}{32 \pi^2} (N^2-1) \nonumber\\
&  +  \frac{9}{13} \frac{N^2-1}{N}\frac{m^4}{2g^2}- \frac{24}{35}\frac{(N^2-1)^2}{N}\frac{M^4}{g^2}  - \frac{161}{52} \frac{N^2-1}{16 \pi^2}\frac{ m^4}{2 }- M^4 \alpha_1  +M^2 m^2 \chi_1\;.
\end{align}
whereby $y_2$ and $y_3$ are are given by $\frac{1}{2} \left(-m^2-M^2\pm\sqrt{m^4-2 M^2 m^2 + M^4 - 4 \lambda^4}\right)$. \\
\\
In order to find the minimum, we should derive this action w.r.t.~$m^2$ and $M^2$ and put the equations equal to zero. In addition, we should also impose the horizon condition \eqref{gapgamma}. Therefore, we have the following three conditions,
\begin{align}
\frac{\p \Gamma}{\p M^2} &=0 \;, &  \frac{\p \Gamma}{\p m^2} &=0 \;, &  \frac{\p \Gamma}{\p \lambda^4} &=0 \;,
\end{align}
which have to be solved for $M^2$, $m^2$ and $\lambda^4$. Unfortunately, it is impossible to solve these equations exactly due to the two unknown parameters $\alpha_1$ and $\chi_1$. However, we would like to know if the condensate $\braket{\overline \varphi \varphi}$ is present or not. For this, we need to find uncover if $M^2 = 0$ can be a solution of the above expression. In the appendix \ref{detailsvery}, we shall strongly argue that this is not the case, and thus that $M^2 \not=0$. Therefore, this is a strong indication that the condensate $\braket{\overline \varphi \varphi}$ is indeed present, thereby suggesting the dynamical transformation of GZ into a refined GZ.

\subsection{The gluon and the ghost propagator}
\subsubsection{The gluon propagator}
The gluon propagator shall still be infrared suppressed and non-zero at zero momentum. Indeed, starting from the very refined action \eqref{CGZ}, the quadratic action is given by
\begin{eqnarray}
S_{\quadr} &=& \frac{1}{4} (\p_\mu A_\nu - \p_\nu A_\mu)^2 + b \p_\mu A_\mu + \overline c \p^2 c + \overline \varphi \p^2 \varphi - \overline \omega \p^2 \omega - \gamma^2 g f^{abc} A_\mu^b (\varphi^{bc}_\mu + \overline \varphi^{bc}_\mu )\nonumber\\
&&  + \gamma^4 d (N^2 - 1) - M^2 \overline \varphi \varphi + \frac{m^2}{2} A_\mu A_\mu - \frac{\rho}{2} \overline \varphi \overline \varphi -\frac{\rho^{\dagger}}{2}\varphi  \varphi \;,
\end{eqnarray}
whereby we have replaced the source $\tau$ with $m^2$, $Q$ with $-M^2$, $\overline G^{ij}$ with $-\delta^{ij} \rho$ and $G^{ij}$ with $-\delta^{ij} \rho^{\dagger}$ and set all other sources equal to zero. From this, we can easily deduce the gluon propagator
\begin{multline} \label{gluonpropvra}
 \Braket{ A^a_{\mu}(p) A^b_{\nu}(-p)}  =  \left[\delta_{\mu\nu} - \frac{p_{\mu}p_{\nu}}{p^2} \right]\delta^{ab} \\\underbrace{ \frac{2 \left(M^2+p^2\right)^2-2 \rho  \rho^{\dagger}  }{ 2 M^4 p^2+2 p^6+2 M^2 \left(2 p^4+\lambda^4 \right)-\lambda^4  (\rho +\rho^{\dagger} )+2 m^2 \left(\left(M^2+p^2\right)^2-\rho  \rho^{\dagger} \right)+2 p^2 (\lambda^4 -\rho  \rho^{\dagger} ) }}_{\mathcal{D}(p^2)}\;,
\end{multline}
with $\lambda^4 = 2 g^2 N \gamma^4$. If we assume that $\rho = \rho^\dagger$, we then find the following gluon propagator:
\begin{eqnarray}
D(p^2) = \frac{M^2+p^2+\rho }{p^4 + M^2 p^2 +p^2 (\rho + m^2) +m^2 \left(M^2+\rho \right) +  \lambda ^4} \;,
\end{eqnarray}
which has exactly the same form as the refined gluon propagator \eqref{gluonprop2}. However, for the moment we cannot say whether $\rho = \rho^\dagger$ is the case or not. Notice that $\rho$, $\rho^\dagger$ as well as $M^2$ all separately make sure that $D(0) \not= 0$.

\subsubsection{The ghost propagator}
The one loop ghost propagator is given by
\begin{eqnarray} \label{ghostpropagator1}
\mathcal{G}^{ab}(k^2) &=&  \delta^{ab} \mathcal{G}(k^2) ~=~ \delta^{ab}\left( \frac{1}{k^2} +
\frac{1}{k^2} \left[g^2 \frac{N}{N^2 - 1} \int \frac{\d^4
q}{(2\pi)^4} \frac{(k-q)_{\mu} k_{\nu}}{(k-q)^2}
 \Braket{A^a_{\mu}A^a_{\nu}}\right] \frac{1}{k^2} \right) + \mathcal{O}(g^4) \nonumber\\
&=& \delta^{ab} \frac{1}{k^2} (1+ \sigma(k^2)) +
\mathcal{O}(g^4)\;,
\end{eqnarray}
with
\begin{align*}
&\sigma(k^2)=\frac{N}{N^2 - 1} \frac{g^2}{k^2}\int \frac{\d^4 q}{(2\pi)^4} \frac{(k-q)_{\mu} k_{\nu}}{(k-q)^2} \Braket{A^a_{\mu}A^a_{\nu}} \nonumber\\
&= Ng^2 \frac{k_{\mu} k_{\nu}}{k^2} \int \frac{\d^d q}{(2\pi)^d} \frac{1}{(k-q)^2} \left[ \delta_{\mu\nu} - \frac{q_{\mu}q_{\nu}}{q^2} \right]\nonumber\\
 &\times\frac{2 \left(M^2+q^2\right)^2-2 \rho  \rho^{\dagger}  }{ 2 M^4 q^2+2 q^6+2 M^2 \left(2 q^4+\lambda^4 \right)-\lambda^4  (\rho +\rho^{\dagger} )+2 m^2 \left(\left(M^2+q^2\right)^2-\rho  \rho^{\dagger} \right)+2 q^2 (\lambda^4 -\rho  \rho^{\dagger} ) } \;.
\end{align*}
As we are interested in the infrared behavior of this propagator, we expand the previous expression for small $k^2$
\begin{multline}\label{sigmaex}
\sigma (k^2\approx 0) =   Ng^2 \frac{d-1}{d}\int \frac{\d^d q}{(2\pi)^d}  \frac{1}{q^2} \\
\times \frac{2 \left(M^2+q^2\right)^2-2 \rho  \rho^{\dagger}  }{ 2 M^4 q^2+2 q^6+2 M^2 \left(2 q^4+\lambda^4 \right)-\lambda^4  (\rho +\rho^{\dagger} )+2 m^2 \left(\left(M^2+q^2\right)^2-\rho  \rho^{\dagger} \right)+2 q^2 (\lambda^4 -\rho  \rho^{\dagger} ) } \\ + O(k^2) \;.
\end{multline}
Let us now have a look at the gap equation. For this we can start from the (one-loop) effective action which can be written as (see the appendix \ref{detailsvery})
\begin{equation*}
\Gamma_\gamma^{(1)} = -d(N^{2}-1)\gamma^{4} +\frac{(N^{2}-1)}{2}\left( d-1\right) \int \frac{\d^{d}q}{\left( 2\pi \right) ^{d}} \ln A  + \ldots\;,
\end{equation*}
with \begin{equation*}
 A= \frac{2 M^4 q^2+2 q^6+2 M^2 \left(2 q^4+\lambda^4 \right)-\lambda^4  (\rho +\rho^{\dagger} )+2 m^2 \left(\left(M^2+q^2\right)^2-\rho  \rho^{\dagger} \right)+2 q^2 (\lambda^4 -\rho  \rho^{\dagger} )}{2 \left(M^2+q^2\right)^2-2 \rho  \rho^{\dagger}  } \;,
\end{equation*}
and the $\ldots$ indicating parts independent from $\lambda$. Setting $\lambda^4 = 2 g^2 N \gamma^4$, we rewrite the previous expression,
\begin{eqnarray*}
\mathcal{E}^{(1)} &=&  \frac{\Gamma_\gamma^{(1)}}{N^2 - 1} \frac{2 g^2 N}{d} ~=~ - \lambda^4  + g^2 N \frac{ d-1}{d} \int \frac{\d^{d}q}{\left(2\pi \right) ^{d}} \ln A + \ldots \;.
\end{eqnarray*}
The gap equation is given by $\frac{\p \mathcal{E}^{(1)}  }{\p \lambda^2} = 0$,
\begin{multline}
1 =    g^2 N \frac{d-1}{d} \int \frac{\d^{d}q}{\left(
2\pi \right) ^{d}}  \\
 \frac{ 2 M^2+2 q^2 - \rho -\rho^{\dagger} }{2 M^4 q^2+2 q^6+2 M^2 \left(2 q^4+\lambda^4 \right)-\lambda^4  (\rho +\rho^{\dagger} )+2 m^2 \left(\left(M^2+q^2\right)^2-\rho  \rho^{\dagger} \right)+2 q^2 (\lambda^4 -\rho  \rho^{\dagger} ) } ,
\end{multline}
where we have excluded the solution $\lambda =0$. With the help of this gap equation, we can rewrite equation \eqref{sigmaex},
\begin{multline}\label{sigmaex1}
\sigma (k^2\approx 0) = 1 +   Ng^2 \frac{d-1}{d}\int \frac{\d^d q}{(2\pi)^d}  \\
\times \frac{ 2 M^4/ q^2 + 2 M^2 - 2 \rho \rho^\dagger / q^2 + \rho + \rho^\dagger  }{ 2 M^4 q^2+2 q^6+2 M^2 \left(2 q^4+\lambda^4 \right)-\lambda^4  (\rho +\rho^{\dagger} )+2 m^2 \left(\left(M^2+q^2\right)^2-\rho  \rho^{\dagger} \right)+2 q^2 (\lambda^4 -\rho  \rho^{\dagger} ) } \\
+ O(k^2) \;.
\end{multline}
The integral in the above expression is finite. We can rewrite the integral as ($d = 4$)
\begin{multline*}
 I= Ng^2 \frac{3}{32 \pi^2}\int_0^\infty \d q \\
  \frac{ q ( M^4 -  (r^2 + s^2) )  +  q^3(  M^2  + r)  }{  M^4 q^2+  q^6+ M^2 \left(2 q^4+\lambda^4 \right)- r \lambda^4  +  m^2 \left(\left(M^2+q^2\right)^2-( r^2 + s^2)  \right)+ q^2 (\lambda^4 -(r^2 + s^2) ) }\;,
\end{multline*}
with $I = \sigma (k^2\approx 0)  - 1$, whereby we have parameterized
\begin{align}
\rho &= r + \ii s \;,  & \rho^\dagger &= r - \ii s \;.
\end{align}
We further write
\begin{multline}
 I =    \frac{3 Ng^2}{64 \pi^2} \int_0^\infty \d x \left(  M^4 -  (r^2 + s^2)   +  x (  M^2  + r)  \right) / \left(  x^3 + x^2 (2M^2 + m^2 ) \right. \\
  \left.  + x (M^4 + 2 m^2 M^2 +  \lambda^4 - (r^2 + s^2)) + \lambda^4 (M^2 - r) + m^2 (M^4 - (r^2 + s^2)) \right)\;.
\end{multline}
\\
\textbf{Solution of cubic equation}\\
\\
The next step would be to solve the cubic equation in the numerator of the equation above,
\begin{multline}
 x^3 + x^2 \underbrace{(2M^2 + m^2 )}_a + x \underbrace{(M^4 + 2 m^2 M^2 + \lambda^4 - (r^2 + s^2))}_b  \\
 + \underbrace{\lambda^4 (M^2 - r) + m^2 (M^4 - (r^2 + s^2)) }_c = 0 \;.
\end{multline}
In general, these are given by
\begin{eqnarray}
x_1 &=& \frac{-1}{3} \left( a  + \sqrt[3]{ \frac{m + \sqrt{n}}{2} } + \sqrt[3]{ \frac{m - \sqrt{n}}{2} } \right) \;, \nonumber\\
x_2 &=& \frac{-1}{3} \left( a  + \frac{-1 + \ii \sqrt{3}}{2} \sqrt[3]{ \frac{m + \sqrt{n}}{2} } + \frac{-1 - \ii \sqrt{3}}{2} \sqrt[3]{ \frac{m - \sqrt{n}}{2} } \right) \;, \nonumber\\
x_3 &=& \frac{-1}{3} \left( a  + \frac{-1 - \ii \sqrt{3}}{2} \sqrt[3]{ \frac{m + \sqrt{n}}{2} } + \frac{-1 + \ii \sqrt{3}}{2} \sqrt[3]{ \frac{m - \sqrt{n}}{2} } \right)\;,
\end{eqnarray}
with
\begin{eqnarray}
m &=& 2 \left(m^2-M^2\right) \left(\left(m^2-M^2\right)^2-9 \left(r^2+s^2\right)\right)-9 \left(m^2-M^2+3 r\right) \lambda ^4 \;,\nonumber\\
n&=& \left[2 \left(m^2-M^2\right) \left(\left(m^2-M^2\right)^2-9 \left(r^2+s^2\right)\right)-9 \left(m^2-M^2+3 r\right) \lambda ^4\right]^2 \nonumber\\
 &&-4 \left[\left(m^2-M^2\right)^2+3 \left(r^2+s^2-\lambda ^4\right)\right]^3\;.
\end{eqnarray}
Of course, it is possible that two (or three) solutions coincide. This can be checked by calculating the discriminant
\begin{equation}
\Delta = - 4 a^3 c + a^2 b^2 - 4 b^3 + 18 abc - 27 c^2 \;.
\end{equation}
If $\Delta =0$,  then the equation has three distinct real roots and at least two are equal. \\
\\
\textbf{Case 1: $x_1 \not = x_2 \not = x_3$}\\
\\
If $x_1 \not = x_2 \not = x_3$, we can rewrite the integral in $I$,
\begin{multline}
I=Ng^2 \frac{3}{64 \pi^2}\Biggl[  \int_0^\infty \d x  \frac{ M^4 -  (r^2 + s^2) + (  M^2  + r)x_1  }{(x_1-x_2)(x_1-x_3)} \frac{1}{x-x_1} \\
 + \int_0^\infty \d x  \frac{ M^4 -  (r^2 + s^2) + (  M^2  + r)x_2  }{(x_2-x_3)(x_2-x_1)(x-x_2)}+ \int_0^\infty \d x  \frac{ M^4 -  (r^2 + s^2) + (  M^2  + r)x_3  }{(x_3-x_1)(x_3-x_1)(x-x_3)} \Biggr]\;.
\end{multline}
These integrals are now easy to solve they all are of the type $\int \d x \frac{1}{x} = \ln x$.
\begin{multline*}
I=Ng^2 \frac{3}{64 \pi^2}\Biggl[ \underbrace{ \frac{ M^4 -  (r^2 + s^2) + (  M^2  + r)x_1  }{(x_1-x_2)(x_1-x_3)} }_{u_1} \left. \ln (x - x_1 )  \right|^{\infty}_0 \\
+   \underbrace{\frac{ M^4 -  (r^2 + s^2) + (  M^2  + r)x_2  }{(x_2-x_3)(x_2-x_1)} }_{v_1} \left. \ln (x - x_2 )  \right|^{\infty}_0  + \underbrace{ \frac{ M^4 -  (r^2 + s^2) + (  M^2  + r)x_3  }{(x_3-x_1)(x_3-x_1)}}_{w_1}  \left. \ln (x - x_3 )  \right|^{\infty}_0 \Biggr]\;.
\end{multline*}
One could expect there is a problem at infinity, in contrast with what we have concluded before. However, as $u_1 + v_1 + w_1 = 0$, the infinities cancel. We obtain,
\begin{multline}
I=Ng^2 \frac{3}{64 \pi^2}\Biggl[  \frac{ M^4 -  (r^2 + s^2) + (  M^2  + r)x_1  }{(x_1-x_2)(x_1-x_3)}   \ln ( - x_1 ) \\
 +   \frac{ M^4 -  (r^2 + s^2) + (M^2 + r)x_2  }{(x_2-x_3)(x_2-x_1)}\ln ( - x_2 ) +  \frac{ M^4 -  (r^2 + s^2) + (  M^2  + r)x_3  }{(x_3-x_1)(x_3-x_1)} \ln (- x_3 )  \Biggr]\;.
\end{multline}
\\
\textbf{Case 2: $x_1  = x_2 \not = x_3$}\\
\\
In this case, we can rewrite the integral in $I$ as
\begin{eqnarray}
I&=& Ng^2 \frac{3}{64 \pi^2}\Biggl[  \int_0^\infty \d x \underbrace{ \frac{ M^4 -  (r^2 + s^2) + (  M^2  + r)x_1  }{ (x_1-x_3)^2}}_{u_2} \frac{1}{x-x1}  \nonumber\\
 && - \int_0^\infty \d x  \underbrace{\frac{ M^4 -  (r^2 + s^2) + (  M^2  + r)x_1  }{ (x_1-x_3)^2}}_{v_2} \frac{1}{x-x_3} \nonumber\\
 && + \int_0^\infty \d x  \underbrace{\frac{ M^4 -  (r^2 + s^2) + (  M^2  + r)x_3  }{x_1-x_3}}_{w_2} \frac{1}{(x-x_3)^2} \Biggr]\;.
\end{eqnarray}
One can check that $u_2 + v_2 =0$, so we can execute the integrations,
\begin{multline}
I=Ng^2 \frac{3}{64 \pi^2}\Biggl[   \frac{ M^4 -  (r^2 + s^2) + (  M^2  + r)x_1  }{ (x_1-x_3)^2}  \ln (-x1)  \\
- \frac{ M^4 -  (r^2 + s^2) + (  M^2  + r)x_1  }{ (x_1-x_3)^2} \ln (-x_3)  -  \frac{ M^4 -  (r^2 + s^2) + (  M^2  + r)x_3  }{x_1-x_3}  \frac{1}{x_3^2} \Biggr]\;.
\end{multline}
\\
\textbf{Case 3: $x_1  = x_2  = x_3$}\\
\\
Finally, in this case we can write
\begin{equation}
I=Ng^2 \frac{3}{64 \pi^2}\Biggl[    ( M^2+ r )\int_0^\infty \d x   \frac{1}{ (x-x_1)^2 } + (M^4 -  (r^2 + s^2) + (  M^2  + r)x_1)\int_0^\infty \d x    \frac{1}{(x-x_1)^3} \Biggr]\;,
\end{equation}
so after integration
\begin{equation}
I=Ng^2 \frac{3}{64 \pi^2}\Biggl[   - \frac{ M^2+ r } { x_1 } + \frac{M^4 -  (r^2 + s^2) + (  M^2  + r)x_1}{ 2 x_1^2} \Biggr]\;.
\end{equation}
Now we can make some conclusions. Looking at the different cases, it looks almost certain that $I \not = 0$, as very specific values of the condensates would be needed to take care of this. Therefore, we have strong indications that the ghost propagator is non-enhanced.

\subsection{Conclusions for the further refining of the action}
In this section, we have investigated the possibility that more condensates are present than investigated so far. This has led us to the calculation of the one loop effective action. Unfortunately, due to the existence of two yet unknown higher loop parameters in the one loop effective action, we are unable to provide an estimate for the different condensates. Nevertheless, we have provided strong indications that some condensates are non-zero and shall lower the effective action. \\
\\
Besides this, we have also shown that in this further refined framework, the gluon propagator is non zero at zero momentum, and the ghost propagator is highly likely to be non-enhanced.\\
\\
A future research goal will be to compute the divergences of the vacuum diagram in Figure \ref{1loopd2} and the similar one for the mixing. Once this task will be executed, all information is available to actually work out the one loop effective potential and to investigate its structure and the associated formation of the RGZ condensates.

\chapter{The quest for physical operators, part I\label{chappart1}}

\section{Introduction}
So far, we have refined the Gribov-Zwanziger action so that the ghost and gluon propagator are in agreement with the lattice data. We recall that the gluon propagator has complex poles, as one can see from expression \eqref{gluonprop} and \eqref{gluonprop2}. This shows us that gluons cannot be considered as a part of the physical spectrum, which is clearly due to the restriction to the horizon $\Omega$. In this way, gluons can be seen as confined by the Gribov horizon. The natural question is then where are the physical particles in the (Refined) GZ action? If the Refined GZ does come close to the reality, which is quenched QCD here, we should be able to identify the particle content of quenched QCD, namely glueballs.\\
\\
Glueballs are entirely composed of gluons, and therefore the gauge field itself is a crucial ingredient. For standard hadronic particles on the other hand, also matter fields are indispensable. Hence, glueballs have been widely investigated, experimentally, on the lattice and in various theoretical models \cite{Mathieu:2008me}.\\
\\
So far, there is no clear experimental evidence for the existence of glueballs. If glueballs are observable particles, they would strongly mix with other states containing quarks.  Due to this feature, a clear observation of a glueball state turns out to be rather difficult. However, there are already many indications for the existence of glueballs, and the debate is currently ongoing.  It is worth mentioning here that several experiments are actually running and other ones are planned to start in the future: $
\overline{\mathrm{P}}$ANDA \cite{Bettoni:2005ut} , BES III \cite{Chanowitz:2006wf} and GlueX \cite{Carman:2005ps} to name only a few. Glueballs might also play an important role in the quark gluon plasma, a case that will be studied at e.g. the heavy ion collision experiment ALICE at CERN \cite{Alessandro:2006yt}.\\
\\
As no clear experimental data is yet available, the output of theoretical models ought to be compared with lattice data. In lattice gauge theories, there is no doubt about the existence of glueballs,  although lattice calculations are still limited as they cannot determine the decay channels of glueballs. In contrast with possible experimental data, lattice calculations can however also consider pure gauge theory. A consensus on the lowest lying scalar glueball mass in the pure gauge gauge theory has already been reached : $M_{0^{++}} \sim 1.6 $ GeV$^2$ for SU(3) \cite{Morningstar:1999rf,McNeile:2002en,Chen:2005mg,Teper:1998kw,Meyer:2008tr,West:1997sz}.\\
\\
Many theoretical models have been investigated and compared with the lattice data. An extensive recent overview is given in \cite{Mathieu:2008me}. Historically, the first model to describe glueballs is called the MIT bag model \cite{Jaffe:1975fd}. In this model, gluons are placed in a bag and confined by a boundary condition and a constant energy density $B$. This model, however, is rather phenomenological in nature. Other phenomenological models assume the gluons to have an effective mass \cite{Cornwall:1981zr,Bernard:1981pg}, which can be used to compose effective (potential) theories in which the masses of the different glueballs are calculated \cite{Cornwall:1982zn,Barnes:1981ac,Szczepaniak:1995cw,Kaidalov:1999yd}.\\
\\
A more direct way to deal with glueballs is by identifying suitable gauge invariant operators, which carry the correct quantum numbers to create/annihilate particular glueball states, and then calculating the corresponding correlators to get information on the  mass. In particular, this route is followed in the widely used QCD sumrule approach \cite{Novikov:1979va,Narison:2008nj}. For example, the operator relevant for the lightest scalar glueball is $F^2(x)\equiv F_{\mu\nu}^2(x)$, hence the study of the correlator
$\Braket{F^2(x) F^2(y)}$. One takes into account perturbative as well as non-perturbative contributions, which are associated with condensates and instantons \cite{Novikov:1979va,Shuryak:1982dp}. Also in the AdS/QCD approach, glueball (correlators) have been investigated based on the assumption that there is an approximate
dual gravity description \cite{Brower:2000rp,Forkel:2007ru}.\\
\\
In the light of such correlator studies, it would be interesting to investigate the correlator $\Braket{F^2(x) F^2(y)}$ within the Gribov-Zwanziger framework. For this, we need to study the renormalization of the operator $F^2_{\mu\nu}$ within the GZ action. As we show in this chapter, this is far from trivial.

\section{Ordinary Yang-Mills action with inclusion of the operator $F_{\mu\nu}^2$}
Before investigating the operator $F^2$ in the GZ framework, let us first, as an exercise, investigate $F^2$ in the much more simple standard Yang-Mills theory. Here, we shall encounter many nice features which are useful for the investigation of $F^2$ in the GZ framework. This section is based on \cite{Dudal:2008tg}.

\subsection{Renormalization of the YM action with inclusion of the operator $F_{\mu\nu}^2$}
\subsubsection{Introduction}
To study the correlator $\Braket{F^2(x) F^2(y)}$, we need to add the operator $F_{\mu\nu}^2$ to the ordinary Yang-Mills action by coupling it to a source $q(x)$. Indeed, the action we start from reads,
\begin{equation}\label{notrenorm}
\Sigma_{\mathrm{n.r.}} = \underbrace{\int \d^4 x \frac{1}{4}F_{\mu\nu}^2}_{ S_{\YM}} + \underbrace{\int \d^4 x\,\left(b^{a}\partial_\mu A_\mu^{a}+\overline{c}^{a}\partial _{\mu }
D_{\mu}^{ab}c^b \right)}_{S_{\gf}}  + \int \d^4 x \frac{q}{4}F_{\mu\nu}^2\,,
\end{equation}
whereby $S_\gf$ is the Landau gauge fixing part. This action is of course BRST invariant,
\begin{eqnarray}
s  \Sigma_{\mathrm{n.r.}} &=& 0\,,
\end{eqnarray}
whereby the BRST transformations of the fields are given by by equation \eqref{BRST}, and $s q = 0$. We recall that $s$ is nilpotent, $s^2 = 0$. In this fashion, the correlator is given by
\begin{eqnarray}
\left[{\frac{\delta}{\delta q(y)}  \frac{\delta}{\delta q(x)}} Z^c\right]_{q=0} &=& \Braket{F^2(x) F^2(y)}\,,
\end{eqnarray}
with $Z^c$ the generator of connected Green functions. However, it will turn out that the action \eqref{notrenorm} is not renormalizable. Indeed, as the operator $F_{\mu\nu}^2$ has mass dimension 4, it could mix with other operators of the same dimension. The question arises which kind of extra operators we need to consider.

\subsubsection{Three classes of operators\label{sectieB}}
In general, we can distinguish between 3 different classes of dimension 4 operators. Firstly, the class $C_1$ contains all the truly gauge invariant operators. These are the BRST closed but not exact operators like $F_{\mu\nu}^2$. These are constructed from the field strength $F_{\mu\nu}^a$ and the covariant derivative $D_{\mu}^{ab}$. Secondly, the class $C_2$ consists of BRST exact operators, e.g.~$s(\overline{c}^a \p_\mu A_\mu^a)$. The third class $C_3$ contains operators which will vanish upon using the equations
of motion, e.g.~$A_\mu^a \frac{\delta S}{\delta A_\mu^a}$, with $S=S_\YM+S_\gf$.\\
\\
Now, one can intuitively easily understand that these 3 different classes will mix in a certain way \cite{Collins:1984xc,Collins:1994ee}. Firstly, bare operators from
the class $C_2$ cannot receive contributions from gauge invariant operators ($C_1$). Indeed, taking the matrix element of a bare BRST exact operator from $C_2$ between physical states will give a vanishing result, if there would be a renormalized gauge invariant contribution from $C_1$ in its expansion, there would be a
nonvanishing contribution, clearly a contradiction. Secondly, as a $C_3$ operator will vanish upon using the equations of motion, while a $C_1$- and a $C_2$ operator in general do not, a $C_3$ operator cannot receive corrections from the $C_1$ and/or $C_2$ class.\\
\\
Thus, the mixing matrix will have an upper triangular form,
\begin{eqnarray}\label{upper}
\left(
\begin{array}{c} \mathcal F_0 \\ \mathcal E_0 \\ \mathcal H_0 \end{array} \right) &= & \left( \begin{array}{ccc} Z_{\mathcal
F\mathcal F}& Z_{\mathcal F\mathcal E}  & Z_{\mathcal F \mathcal H}
\\ 0  &Z_{\mathcal E\mathcal E} & Z_{\mathcal E \mathcal H}   \\ 0 &
0& Z_{\mathcal H \mathcal H}          \end{array} \right) \left(
\begin{array}{c} \mathcal F \\ \mathcal E \\ \mathcal H \end{array}
\right)\,,
\end{eqnarray}
whereby $\mathcal F$, $\mathcal E$, $\mathcal H$ are operators belonging,  respectively,  to the $C_1$, $C_2$ and $C_3$ class.\\
\\
We shall however not use these observations, and only rely on a formal algebraic analysis. All constraints on e.g.~the mixing matrix should be encoded in the Ward identities.\\
\\
For further use, let us elaborate a bit more on the equation of motion like terms, using a scalar field for notational simplicity. A term $\sim \frac{\delta S}{\delta\varphi}$ shall give rise to contact terms when taking expectation values. Using partial path integration, one finds
\begin{eqnarray}\label{s}
\Braket{\varphi(x_1)\varphi(x_2)\ldots\varphi(x_{n+1}) \frac{\delta S}{\delta \varphi(y)}}&=&\int [\d\phi]\varphi(x_1)\varphi(x_2)\ldots\varphi(x_{n+1}) \frac{\delta S}{\delta \varphi(y)}\e^{-S}\nonumber\\
&&\hspace{-4cm}=-\int [\d\Phi]\varphi(x_1)\varphi(x_2)\ldots\varphi(x_{n+1})\frac{\delta}{\delta \varphi(y)}\e^{-S}\nonumber\\
&&\hspace{-4cm}=\int[\d\Phi] \frac{\delta}{\delta \varphi(y)} \left[\varphi(x_1)\varphi(x_2)\ldots\varphi(x_{n+1})\right] \e^{-S}\nonumber\\
&&\hspace{-4cm}= \sum_{k=1}^{n+1}\delta(x_k-y)\braket{\varphi(x_1)\varphi(x_2)\ldots\varphi(x_{k-1})\varphi(x_{k+1})\ldots\varphi(x_{n+1})}\,.
\end{eqnarray}
We used the symbolic notation $ \int[\d \phi]$ for the integration over all the present fields. Introducing the $Z$-factors for the fields $\varphi$, one also learns that $\varphi(y)\frac{\delta S}{\delta \varphi(y)}$ does not need any renormalization factor, and thus that it is \emph{finite} when introduced into correlators\footnote{The implied limit $x_{n+1}\to y$ might seem problematic due to the appearance of a $\delta(0)$ in the last term of the r.h.s. of \eqref{s}.  However, $\delta(0)=0$ in dimensional regularization.}. Moreover, if $x_k\neq y$, $k=1,\ldots,n$, the l.h.s.~of \eqref{s} will vanish as the r.h.s.~does. On the other hand, it is easily recognized from \eqref{s} that the integrated operator $\int \d^4y \varphi(y)\frac{\delta S}{\delta\varphi(y)}$ is nothing more than a counting operator when inserted into a correlator, i.e.
\begin{equation}\label{count}
\Braket{\varphi(x_1)\varphi(x_2)\ldots\varphi(x_n)\int\d^4y\varphi(y)\frac{\delta S}{\delta \varphi(y)}}\\=n\Braket{\varphi(x_1)\varphi(x_2)\ldots\varphi(x_n)}\,.
\end{equation}

\subsubsection{The starting action}
We can now propose a more complete starting action than \eqref{notrenorm}. Besides the gauge invariant operator $F_{\mu\nu}^2$ belonging to the first class $C_1$, we also introduce the BRST closed operator $s (\overline \p c  A)\equiv s(\p_\mu \overline{c}^a  A_\mu^a) $, coupled to a new dimensionless source $\eta$. As we want this new source to only enter the cohomological trivial part of the action, we shall introduce a  BRST doublet $(\lambda, \eta)$,
\begin{equation}\label{lambdaeta}
s \eta = \lambda  \,,
\end{equation}
and add the following term to the action \eqref{notrenorm},
\begin{equation}
\int \d^4 x s ( \eta \overline c^a \p_\mu A_\mu^a) = \int \d^4 x (\lambda \p_\mu \overline c^a A_\mu^a + \eta ( \p_\mu b^a A_\mu^a + \p_\mu \overline c^a D^{ab}_{\mu} c^{b})  )\,.
\end{equation}
The BRST doublet-structure is highly useful in order to construct the most general invariant counterterm. Hence, the classical starting action is given by
\begin{multline}\label{klassiek}
S_{\mathrm{cl}}=S_{\YM} + \int \d^4x\,\left( b^{a}\partial_\mu A_\mu^{a}+\overline{c}^{a}\partial _{\mu } D_{\mu}^{ab}c^b \right) + \int \d^4 x q \underbrace{ \frac{1} {4} F_{\mu\nu}^2}_{\mathcal F} + \int \d^4 x \lambda \p_\mu \overline c^a A_\mu^a \\
  +\int \d^4 x \eta \underbrace{\left(  \p_\mu b^a A_\mu^a + \p_\mu\overline c^a D_\mu^{ab} c^b\right)}_{\mathcal E}\,.
\end{multline}
Later, we shall also introduce the equation of motion terms from class $C_3$. Notice that in principle, also $s(\overline{c}^a\p_\mu A_\mu^a)$ is another independent $d=4$ BRST exact operator which could play a role. It shall however turn out that the renormalization analysis closes without this operator, therefore we decided to immediately discard it.\\
\\
We can now proceed with the study of this action, using the formalism of algebraic renormalization as explained in chapter \ref{algebraic}. As usual, we first introduce $S_{\mathrm{ext}}$,
\begin{eqnarray}
S_{\mathrm{ext}}&=&\int \d^4x\left( -K_{\mu }^{a}\left( D_{\mu }c\right) ^{a}+\frac{1}{2}gL^{a}f^{abc}c^{b}c^{c}\right) \,,
\end{eqnarray}
Now, the enlarged action is given by
\begin{multline} \label{startactie}
\Sigma = S_{\YM} + \int \d^4x\,\left( b^{a}\partial_\mu A_\mu^{a}+\overline{c}^{a}\partial _{\mu } D_{\mu}^{ab}c^b \right)  + \int \d^4x \left( -K_{\mu }^{a}\left( D_{\mu }c\right) ^{a}+\frac{1}{2}gL^{a}f^{abc}c^{b}c^{c}\right) \\
+  \int \d^4 x q  \frac{1} {4} F_{\mu\nu}^2  + \int \d^4 x \lambda \p_\mu \overline c^a A_\mu^a +\int \d^4 x \eta \left(  \p_\mu b^a A_\mu^a + \p_\mu\overline c^a D_\mu^{ab} c^b \right)\,,
\end{multline}
and it will reduce itself to equation \eqref{klassiek}, once the sources $K_\mu^a$ and $L^a$ are set to zero at the end. Likewise, also $\lambda$ can be set to zero at that point.

\subsubsection{The Ward identities}
A second step in the process of algebraic renormalization is to determine all the Ward identities obeyed by the action \eqref{startactie}, which we have summarized here:
\begin{itemize}
\item The Slavnov-Taylor identity:
\begin{equation}
\mathcal{S}(\Sigma ) =\int \d^4x\left( \frac{\delta \Sigma}{\delta K_{\mu }^{a}}\frac{\delta \Sigma }{\delta A_{\mu}^{a}}+\frac{\delta \Sigma }{\delta L^{a}}\frac{\delta \Sigma }{\delta c^{a}}+b^{a}\frac{\delta \Sigma}{\delta \overline{c}^{a}} + \lambda \frac{\delta \Sigma }{\delta \eta} \right) = 0 \,.
\end{equation}
\item The Landau gauge condition:
\begin{eqnarray}
\frac{\delta \Sigma}{\delta b^{a}}&=&\p_\mu A_\mu^a -\p_\mu(\eta
A_\mu^a)\,.
\end{eqnarray}
\item The modified antighost equation:
\begin{eqnarray}
\frac{\delta \Sigma}{\delta \overline c^{a}}+\p_\mu\frac{\delta \Sigma}{\delta K_{\mu}^a}-\p_\mu\left( \eta\frac{\delta \Sigma}{\delta K_\mu^a } \right)&=& \p_\mu (\lambda A_\mu^a) \,.
\end{eqnarray}
\item The ghost Ward identity:
\begin{equation}
\int \d^4x\left( \frac{\delta }{\delta c^{a}}+gf^{abc}\left(\overline{c}^{b}\frac{\delta }{\delta b^{c}} \right)\right)\Sigma = g\int \d^4xf^{abc}\left(K_{\mu}^{b}A_{\mu }^{c}-L^{b}c^{c}\right)\,.
\end{equation}
The term $\Delta _{\mathrm{cl}}^{a}$, being linear in the quantum fields $A_{\mu }^{a}$, $c^{a}$, is a classical breaking.
\item The extra integrated Ward identity:
\begin{eqnarray}
\int \d^4 x\left( \frac{\delta \Sigma}{\delta \lambda} - \eta \frac{\delta \Sigma}{\delta \lambda} + \overline c^a \frac{\delta \Sigma}{\delta b^a} \right)&=& 0\,.
\end{eqnarray}
\end{itemize}
Apart from some small adaptations, the first 5 symmetries are similar to the ones in the ordinary Yang-Mills action, see p.\pageref{chap1wardidentities}. Moreover, we also find an extra Ward identity w.r.t.~the new doublet $(\lambda, \eta)$. This last identity will enable us to take into account in a purely algebraic way the effects related to the composite operators coupled to the sources $(\lambda, \eta)$. We underline here that this is the power of the algebraic formalism: by a well chosen set of sources to introduce the relevant operators, one can hope to find additional Ward identities  which, in turn, will constrain the theory at the quantum level, including the characterization of the most general counterterm. As such, a good choice of sources can considerably simplify the renormalization analysis.

\subsubsection{The counterterm}
When we turn to the quantum  level, we can use these symmetries to characterize the most general allowed invariant counterterm $\Sigma^{c}$. Following the algebraic renormalization procedure, $\Sigma^{c}$ is an integrated local polynomial in the fields and sources with dimension bounded by four, and with vanishing ghost number. The previous, nonanomalous, Ward identities imply the following constraints on $\Sigma^c$:
\begin{itemize}
\item The linearized Slavnov-Taylor identity:
\begin{equation}\label{5ward1}
\mathcal{B}_{\Sigma }\Sigma ^{c}=0\,, \qquad \mathcal{B}_{\Sigma}^{2} = 0\,,
\end{equation}
\begin{equation}
\mathcal{B}_{\Sigma} = \int \d^4x\left( \frac{\delta \Sigma}{\delta K_{\mu }^{a}}\frac{\delta }{\delta A_{\mu}^{a}}+\frac{\delta \Sigma }{\delta A_{\mu }^{a}}\frac{\delta}{\delta K_{\mu}^{a}}+\frac{\delta \Sigma }{\delta L^{a}}\frac{\delta }{\delta c^{a}}  +\frac{\delta\Sigma }{\delta c^{a}}\frac{\delta }{\delta L^{a}}+b^{a}\frac{\delta }{\delta \overline{c}^{a}} + \lambda \frac{\delta }{\delta \eta} \right)\,.
\end{equation}
\item The Landau gauge condition:
\begin{eqnarray}
\frac{\delta \Sigma^c}{\delta b^{a}}&=&0 \,.
\end{eqnarray}
\item The modified antighost equation:
\begin{eqnarray}
\frac{\delta \Sigma^c}{\delta \overline c^{a}}+\p_\mu\frac{\delta \Sigma^c}{\delta K_{\mu}^a}-\p_\mu \left( \eta \frac{\delta \Sigma}{\delta K_\mu^a } \right)&=&0 \,.
\end{eqnarray}
\item The ghost Ward identity:
\begin{eqnarray}
\int \d^4x\left( \frac{\delta }{\delta c^{a}}+gf^{abc}\left(\overline{c}^{b}\frac{\delta }{\delta b^{c}} \right)\right)\Sigma^c &=& 0 \,.
\end{eqnarray}
\normalsize
\item The extra integrated Ward identity:
\begin{eqnarray}\label{ward2}
\int \d^4 x\left( \frac{\delta \Sigma^c}{\delta \lambda} - \eta\frac{\delta \Sigma^c}{\delta \lambda}  \right)&=& 0 \,.
\end{eqnarray}
\end{itemize}
To construct the most general counterterm, Table \ref{groottable}, listing the dimension and ghost number of the various fields and sources, is useful.

\begin{table}[H]
\begin{center}
        \subtable{
        \begin{tabular}{|c|c|c|c|c|}
        \hline
        & $A_{\mu }^{a}$ & $c^{a}$ & $\overline{c}^{a}$ & $b^{a}$  \\
        \hline
        \hline
        \textrm{dimension} & $1$ & $0$ &$2$ & $2$  \\
        \hline
        $\mathrm{ghost\, number}$ & $0$ & $1$ & $-1$ & $0$   \\
        \hline
        \end{tabular}}
        \subtable{
    \begin{tabular}{|c|c|c|c|c|c|}
        \hline
        &$K_{\mu }^{a}$&$L^{a}$& $q$& $\eta$ & $\lambda$  \\
        \hline
        \hline
        \textrm{dimension} & $3$ & $4$ & $0$ & $0$ & 0    \\
        \hline
        $\mathrm{ghost\, number}$ & $-1$ & $-2$ & 0 & 0 & 1  \\
        \hline
        \end{tabular}}
        \caption{Quantum numbers of the fields and sources} \label{groottable}
\end{center}
\end{table}

\noindent There is however one subtlety concerning counterterms quadratic (or higher) in the sources. Only looking at the dimensionality, the ghost number and the constraints on the counterterm, it is a priori not forbidden to consider terms of the form $(q^2 \ldots)$, $(\eta^2 \ldots)$, $(q \eta \ldots)$, $(q^3 \ldots)$, etc., i.e.~terms of quadratic and higher order in the sources. If these terms are allowed, an infinite tower of counterterms would be generated and it would be impossible to prove the renormalizability of the
action as new divergences are being generated, which cannot be absorbed in terms already present in the starting action. However, we can give a simple argument why we may omit this class of terms. Assume that we would also introduce the following term of order $q^2$ in the action,
\begin{eqnarray}\label{term}
&\sim&  \int \d^4 x q^2 \frac{F_{\mu\nu}^2}{4}\,.
\end{eqnarray}
Subsequently, when calculating the correlator, this term would give rise to an extra contact term contribution,
\begin{equation}\label{argument}
\left[\frac{\delta }{\delta q(z)}\frac{\delta }{\delta q(y)}\int [\d \phi] \e^{- S} \right]_{q=0} = \underbrace{\Braket{\frac{F^{2}(z)}{4}
\frac{F^{2}(y)}{4} }}_{\mathrm{term\ due\ to\ part\ in\ }q} + \underbrace{\delta(y-z) \Braket{ \frac{F^2(y)}{2}}}_{\mathrm{term\ due\ to\
part\ in\ }q^2} \,.
\end{equation}
As eventually we are only interested in the correlator for $z \not = y$, we can thus neglect the term \eqref{term}. In fact, when looking at the case $z =y$, we should also couple a source to the novel composite operator $F^4$, which is not our current interest. We can repeat this kind of argument for all other terms of higher order in the sources.\\
\\
There is one exception to the previous remark: we cannot neglect higher order terms of  the type $(K q \ldots)$ and $(K \eta \ldots)$ due to the modified antighost equation,
 \begin{eqnarray}\label{5exception}
\frac{\delta \Sigma^c}{\delta \overline c^{a}}+\p_\mu\frac{\delta
\Sigma^c}{\delta K_{\mu}^a}-\p_\mu \left(\eta \frac{\delta \Sigma}{\delta
K_\mu^a }\right)&=&0 \,.
\end{eqnarray}
The second term of this equation  differentiates the counterterm w.r.t.~the source $K^a_\mu$, while the first term w.r.t. the field $\overline c^a$. Therefore, for the construction of the counterterm fulfilling all the constraints, we still need to include terms of order $K q$ and $K \eta$, as when deriving w.r.t~$K_\mu^a$, these terms will become of first order in the sources, just as the term $\propto \frac{\delta \Sigma^c}{\delta \overline c^a}$. However,  at the end, after having completely characterized the counterterm, we can ignore this class of terms again.\\
\\
We are now ready to construct the counterterm. Firstly, making use of general results on the cohomology of gauge theories, see chapter \ref{algebraic}, the most general integrated polynomial of dimension 4 in the fields and sources, with vanishing ghost number and  which takes into account the previous remarks on the  terms quadratic in the sources, can be written as
\begin{align*}
&\Sigma^c = a_0  \int \d^4 x\frac{1}{4}F_{\mu\nu}^2 + b_{0}  \int \d^4 x\frac{q}{4}F_{\mu\nu}^2 + \mathcal{B}_\Sigma \int \d^4 x \biggl\{ a_1 (K_{\mu}^{a}  +\partial _{\mu} \overline{c}^{a})A_{\mu}^{a} + a_{2} L^{a}c^{a} + a_3 b^a\overline c^a  \\
& + a_4 gf^{abc}\overline c^a\overline c^b c^c  \biggr\} + \mathcal{B}_\Sigma \int \d^4 x \biggl\{ b_{1}q (K_{\mu}^{a}+\partial _{\mu} \overline{c}^{a})A_{\mu }^{a} +  b_{2}q \overline{c}^{a} \partial _{\mu} A_{\mu}^{a} + b_3 q b^a\overline c^a  + b_4 q gf^{abc}\overline c^a\overline c^b c^c  \biggr\} \\
& + \mathcal{B}_\Sigma \int \d^4 x \biggl\{c_{1} \eta K_{\mu}^{a}A_{\mu }^{a} +c_{2}\eta\partial _{\mu} \overline{c}^{a}A_{\mu }^{a} + c_{3}\eta \overline{c}^{a} \partial _{\mu} A_{\mu}^{a}  +  c_4 \eta b^a\overline c^a
+ c_5 \eta gf^{abc}\overline c^a\overline c^b c^c   \biggr\} \\
&+ \mathcal{B}_\Sigma \int \d^4x\, \biggl\{d_1 \lambda \overline c^a \overline c^a \biggr\} \,.
\end{align*}
Secondly, we can simplify this counterterm by imposing all the constraints \eqref{5ward1}-\eqref{ward2}. After a certain amount of algebra, we eventually obtain
\begin{align}
&\Sigma^c= a_{0}  \int \d^4 x\frac{1}{4}F_{\mu\nu}^2 +  b_0 \int \d^4 x\frac{q}{4}F_{\mu\nu}^2  + a_{1}\int \d^4x\Biggl( A_{\mu}^{a}\frac{\delta S_{\YM}}{\delta A_{\mu }^{a}} +  A_{\mu}^{a}\frac{\delta \widehat{S}_{\YM}}{\delta A_{\mu }^{a}}+K_{\mu }^{a}\partial _{\mu }c^{a} \nonumber\\
&+ \p_\mu \overline{c}^a \p_\mu c^a - \eta \p_\mu \overline{c}^a \p_\mu c^a  \Biggr) + b_{1} \int \d^4x\, q \Biggl( A_{\mu}^{a}\frac{\delta S_{\YM}}{\delta A_{\mu }^{a}}  + K_{\mu }^{a}\partial _{\mu }c^{a} +\p_\mu \overline{c}^a \p_\mu c^a \Biggr)\,,
\end{align}
with
\begin{equation}\label{exra}
    \widehat{S}_{\YM}=\frac{1}{4}\int \d^4x qF_{\mu\nu}^2 \,.
\end{equation}
Now that we have constructed the most general counterterm obeying all the Ward identities, we can neglect, as previously described, the term in $K q$. Therefore, the final counterterm becomes,
\begin{align}
 \Sigma^c&= a_{0}  \int \d^4 x\frac{1}{4}F_{\mu\nu}^2 +  b_0 \int \d^4 x\frac{q}{4}F_{\mu\nu}^2  + a_{1}\int \d^4x\Biggl( A_{\mu}^{a}\frac{\delta S_{\YM}}{\delta A_{\mu }^{a}} +  A_{\mu}^{a}\frac{\delta \widehat{S}_{\YM}}{\delta A_{\mu }^{a}}  +K_{\mu }^{a}\partial _{\mu }c^{a}  \nonumber\\
& + \p_\mu \overline{c}^a \p_\mu c^a- \eta \p_\mu \overline{c}^a \p_\mu c^a  \Biggr)  + b_{1} \int \d^4x\, q \Biggl( A_{\mu}^{a}\frac{\delta S_{\YM}}{\delta A_{\mu }^{a}}  + \p_\mu \overline{c}^a \p_\mu c^a
\Biggr)\,.
\end{align}

\subsubsection{Introducing the equations of motion}
We still have to introduce the equations of motion as described in section \ref{sectieB}, as these can enter the operator $\mathcal F$. So far, we have found an action $\Sigma$ with corresponding counterterm $\Sigma^c$. Let us perform the linear shift on the gluon  field $A^a_\mu$,
\begin{eqnarray}
A^a_\mu \rightarrow A^a_\mu + J A^a_\mu\,,
\end{eqnarray}
with $J(x)$ a novel local source. This way of introducing the relevant gluon equation of motion operator shall turn out to be very efficient, as it allows us to uncover the finiteness of this kind of operator. Indeed, this shift basically corresponds to a redefinition of the gluon field, and has to be consistently done in the starting action and counterterm. Performing the shift in the action gives rise to the following shifted action $\Sigma'$,
\begin{eqnarray}\label{eindactie}
 \Sigma' &=& S_{\mathrm{\YM}} + \int \d^4x\,\left( b^{a}\partial_\mu A_\mu^{a}+\overline{c}^{a}\partial _{\mu } D_{\mu}^{ab}c^b \right)  +\int \d^4x \left( -K_{\mu }^{a}\left( D_{\mu }c\right) ^{a}+\frac{1}{2}gL^{a}f^{abc}c^{b}c^{c}\right) \nonumber\\
&&+  \int \d^4 x q  \frac{1} {4} F_{\mu\nu}^2  + \int \d^4 x \lambda \p_\mu \overline c^a A_\mu^a  +\int \d^4 x \eta\left(  \p_\mu b^a  A_\mu^a + \p_\mu\overline c^a D_\mu^{ab} c^b \right)\nonumber\\
&&+  \int \d^4 x J  \underbrace{ A_\mu^a\frac{\delta S_{\YM}}{\delta A_\mu^a}}_{\mathcal H}  +  \int \d^4 x  J  \left\{  - \p_\mu b^a A_\mu^a  + g f_{akb} A_\mu^k c^b \p_\mu \overline c^a  \right\} \,,
\end{eqnarray}
where we see the relevant gluon equation of motion term, $\mathcal H$, emerging. Again, we have neglected higher order terms in the sources, as the argument \eqref{argument} still holds. Analogously, we find a shifted counterterm,
\begin{multline*}
\Sigma'^c= a_{0}  \int \d^4 x\frac{1}{4}F_{\mu\nu}^2 +  b_0 \int \d^4 x\frac{q}{4}F_{\mu\nu}^2 + a_{1}\int \d^4x\Biggl( A_{\mu}^{a}\frac{\delta S_{\YM}}{\delta A_{\mu }^{a}} +  A_{\mu}^{a}\frac{\delta \widehat{S}_{\YM}}{\delta A_{\mu }^{a}}  +K_{\mu }^{a}\partial _{\mu }c^{a} \\
+ \p_\mu \overline{c}^a \p_\mu c^a - \eta \p_\mu \overline{c}^a \p_\mu c^a  \Biggr)  + b_{1} \int \d^4x\, q \Biggl( A_{\mu}^{a}\frac{\delta S_{\YM}}{\delta A_{\mu }^{a}}  + \p_\mu \overline{c}^a \p_\mu c^a \Biggr)+  a_{0}\int \d^4x\Biggl( J A_{\mu}^{a}\frac{\delta S_{\YM}}{\delta A_{\mu }^{a}} \Biggr)\nonumber\\
 + a_1 \int \d^4 x J \left( 2 A_\mu^a \p_\mu \p_\nu A_\nu^a - 2 A_\mu^a \p^2 A_\mu^a + 9 g f_{abc} A_\mu^a A_\nu^b \p_\mu A_\nu^c  + 4 g^2 f_{abc} f_{cde} A_\mu^a A_\nu^b A_\mu^d A_\nu^e\right)\,,
\end{multline*}
where one can neglect again the higher order terms in the sources.\\
\\
One could also introduce the other similar equation of motion terms, by introducing linear shifts for the $b^a$, $c^a$, $\overline c^a$ fields. However, the corresponding equation of motion operators will not mix with $F_{\mu\nu}^2$ and are therefore unnecessary to establish the renormalizability of the action \eqref{eindactie}.

\subsubsection{Stability and the renormalization (mixing) matrix}
Finally, it remains to discuss the stability of the classical action, i.e.~to check whether $\Sigma'^c$ can be reabsorbed in the classical action $\Sigma' $ by means of a multiplicative renormalization of the coupling constant $g$, the fields $\left\{ \phi =A,c,\overline{c},b\right\} $ and the sources $\left\{ \Phi = L,K, q, \eta, \lambda, J \right\} $, namely
\begin{equation}
\Sigma' (g,\phi ,\Phi )+h \Sigma'^c= \Sigma (g_{0},\phi
_{0},\Phi_{0})+{\cal O}(h^{2})\,,
\end{equation}
with $h$ the infinitesimal perturbation parameter. The bare fields, sources and parameters are defined as
\begin{align}
 K _{0\mu}^{a}&~=~Z_{K }K _{\mu }^{a}\,,                                  & A_{0\mu }^{a} &=Z_{A}^{1/2}A_{\mu }^{a}\,,                                  &  g_{0}~&=~Z_{g}g\,, \nonumber \\
 L_{0}^{a}~&=~Z_{L}L^{a}\,,                                               & c_{0}^{a} &=Z_{c}^{1/2}c^{a} \,,  \nonumber \\
 q_0 ~&=~Z_{q }q\,,                                                       & \overline{c}_{0}^{a} &=Z_{\overline{c}}^{1/2}\overline{c}^{a}\,,   &    \nonumber \\
 \eta_{0}~&=~Z_{\eta}\eta\,,                                      & b_{0}^{a} &=Z_{b}^{1/2}b^{a}\,,   \nonumber\\
 J_{0}~&=~Z_{J}J\,,   \nonumber\\
\lambda_0 &= Z_\lambda \lambda\,.
\end{align}
We also propose the following mixing matrix,
\begin{equation}
 \left(
  \begin{array}{c}
    q_0 \\
    \eta_0 \\
    J_0
  \end{array}
\right)=\left(
          \begin{array}{ccc}
            Z_{q      q} & Z_{q      \eta}  & Z_{q      J} \\
            Z_{\eta q} & Z_{\eta \eta}  & Z_{\eta J} \\
            Z_{J      q} & Z_{J      \eta}  & Z_{J      J}
          \end{array}
        \right)
\left(
  \begin{array}{c}
    q \\
    \eta \\
    J
  \end{array}
\right)\,,
\end{equation}
which will represent the mixing of the operators $\mathcal F$, $\mathcal E$ and $\mathcal H$. If we try to absorb the counterterm into the original action, we ultimately find,
\begin{align}\label{5Z1}
Z_{g} &=1-h \frac{a_0}{2}\;,   & Z_{A}^{1/2} &=1+h \left( \frac{a_0}{2}+a_{1}\right) \;,
\end{align}
and
\begin{align}\label{5Z2}
Z_{\overline{c}}^{1/2} &= Z_{c}^{1/2} = Z_A^{-1/4} Z_g^{-1/2} = 1-h \frac{a_{1}}{2}\,,&  Z_{b}&=Z_{A}^{-1}\,,&  Z_{K }&=Z_{c}^{1/2}\,,&  Z_{L} &=Z_{A}^{1/2}\,,
\end{align}
results which are known from the renormalization of the original Yang-Mills action in the Landau gauge \eqref{Za} and \eqref{Zalles}.\\
\\
In addition, we also find the following mixing matrix
\begin{eqnarray}\label{mixingmatrix}
\left(
          \begin{array}{ccc}
            Z_{q      q} & Z_{q      \eta}  & Z_{q      J} \\
            Z_{\eta q} & Z_{\eta \eta}  & Z_{\eta J} \\
            Z_{J      q} & Z_{J      \eta}  & Z_{J      J}
          \end{array}
        \right) &=& \left(
          \begin{array}{ccc}
            1 + h (b_0 - a_0) & 0  & 0 \\
            h b_1 & 1  & 0 \\
            h b_1 & 0  & 1
          \end{array}
        \right)\,,
\end{eqnarray}
and for completeness, the $Z$-factor of $\lambda$ reads,
\begin{eqnarray}\label{5lambda}
Z_{\lambda} &=& Z_{c}^{-1/2} Z_A^{-1/2} \,,
\end{eqnarray}
as the counterterm does not contain the source $\lambda$.\\
\\
Once having this mixing matrix at our disposal, we can of course pass to the corresponding bare operators. For this, we shall need the inverse of the mixing matrix \eqref{mixingmatrix},
\begin{equation}
 \left(
  \begin{array}{c}
    q \\
    \eta \\
    J
  \end{array}
\right)=\left(
          \begin{array}{ccc}
            \frac{1}{Z_{q    q}} & 0 & 0 \\
            -\frac{Z_{Jq}}{ Z_{q    q}} & 1 & 0 \\
            -\frac{Z_{Jq}}{ Z_{q    q}} & 0  & 1
          \end{array}
        \right)
\left(
  \begin{array}{c}
    q_0 \\
    \eta_0 \\
    J_0
  \end{array}
\right)\,.
\end{equation}
Now we can determine the corresponding mixing matrix for the operators, since insertions of an operator correspond to derivatives w.r.t. to the appropriate source of the generating functional $Z^c(q,\eta,J)$. In particular,
\begin{eqnarray}
    \mathcal{F}_0 &=& \frac{\delta Z^c(q, \eta, J)}{\delta q_0} \nonumber\\
    &=& \frac{\delta q}{\delta q_0}\frac{\delta Z^c(q, \eta, J) }{\delta q}+\frac{\delta \eta}{\delta q_0}\frac{\delta Z^c(q, \eta, J)}{\delta \eta } + \frac{\delta J}{\delta q_0}\frac{\delta Z^c(q, \eta, J)}{\delta J } \nonumber\\
    \Rightarrow  \mathcal{F}_0 &=&\frac{1}{Z_{q    q}}\mathcal{F}-\frac{Z_{Jq}}{ Z_{q    q}} \mathcal{G}  -\frac{Z_{Jq}}{ Z_{q    q}} \mathcal
    H\,,
\end{eqnarray}
and similarly for $\mathcal{G}_0$ and $\mathcal H_0$. In summary, we find
\begin{eqnarray}\label{operatormatrix}
 \left(
  \begin{array}{c}
    \mathcal F_0 \\
    \mathcal E_0 \\
    \mathcal H_0
  \end{array}
\right) &= & \left(
          \begin{array}{ccc}
            Z_{qq}^{-1}& -Z_{Jq}Z_{qq}^{-1}  &-Z_{Jq}Z_{qq}^{-1} \\
            0  &1 & 0   \\
           0 & 0& 1          \end{array}
        \right)
        \left(
\begin{array}{c}
    \mathcal F \\
    \mathcal E \\
    \mathcal H
  \end{array}
\right)\,.
\end{eqnarray}
From this matrix we can make several interesting observations. Firstly, we see that the operator $\frac{1}{4}F_{\mu\nu}^2 ~(= \mathcal F)$ indeed required the presence of the BRST exact operator $\mathcal E$ and  of the gluon equation of motion operator $\mathcal H$ as these operators are ``hidden'' in the bare operator $\mathcal F$. Secondly, we do retrieve an upper triangular matrix, in agreement with the earlier description in \eqref{upper}. Moreover, we also find that the BRST exact operator $\mathcal E$ does not mix with $\mathcal H$, a mixing which is in principle allowed, but has a $Z$-factor equal to $1$. This can be nicely understood: the integrated BRST exact operator is in fact proportional to a sum of two (integrated) equations of motion terms,
\begin{eqnarray}\label{countcon5}
\int \d^4 x ( \p_\mu b^a A_\mu^a + \p_\mu \overline c^a D_\mu^{ab} c^{b}) &=& - \int \d^4 x ( b^a \p_\mu A_\mu^a +  \overline c^a \p_\mu D_\mu^{ab} c^{b})\nonumber\\
 &=&-\int \d^4 x\left(b^a\frac{\delta S}{\delta b^a}+\overline c^a\frac{\delta S}{\delta \overline c^a}\right)   \,,
\end{eqnarray}
and therefore it does  not mix with other operators, just like $\mathcal H$.

\subsection{The operator mixing matrix to all orders\label{5sectmixingtoallorders}}
In this section, we shall demonstrate that we can determine the mixing matrix \eqref{operatormatrix} exactly, i.e.  to all orders of perturbation theory. For this purpose, we shall follow the lines of \cite{Brown:1979pq}, suitably adapted to the gauge theory under study. We start with the following most general $(n + 2m + r)$-point function defined as,
\begin{align}\label{definitie}
\mathcal G^{n + 2m + r}(x_1, \ldots, x_n, y_1, &\ldots, y_m,\hat y_1,\ldots, \hat y_m, z_1, \ldots, z_r ) \nonumber\\
&= \Braket{A(x_1)\ldots A(x_n) c(y_1)\ldots c(y_m ) \overline c(\hat y_1) \overline c (\hat y_m) b(z_1)\ldots b(z_r)  }\nonumber\\
&= \int [\d \phi] A(x_1)\ldots A(x_n) c(y_1)\ldots c(y_m ) \overline c(\hat y_1) \overline c (\hat y_m)  b(z_1)\ldots b(z_r) \e^{- S} \,,
\end{align}
with the action $S$ given by
\begin{eqnarray}
S &=& S_{\YM} + S_{\gf}\;.
\end{eqnarray}
We have immediately assumed that there is an equal amount of ghost and antighost fields present as in any other case, the Green function \eqref{definitie} would be zero, due to ghost number symmetry. Subsequently, from the definition \eqref{definitie}, we can immediately write down the connection between the renormalized Green function and the bare Green function,
\begin{eqnarray}\label{connectie}
\mathcal G^{n + 2m + r} = Z_A^{-n/2} Z_c^{-m}Z_b^{-r/2} \mathcal G_0^{n + 2m + r}\,.
\end{eqnarray}
From the previous equation, we shall be able to fix all the matrix elements of expression \eqref{operatormatrix}, based on the knowledge that
\begin{eqnarray}
\frac{\d \mathcal G^{n + 2m + r} }{\d g^2}\;,
\end{eqnarray}
is finite.\\
\\
We start by applying the chain rule when deriving the right hand side of equation \eqref{connectie} w.r.t.~$g^2$. We find,
\begin{eqnarray}\label{finalgoal2}
\frac{\p \mathcal G^{n'} }{\p g^2} &=& \frac{\p g_0^2 }{ \p g^2} \frac{\p \mathcal G_0^{n'} }{\p g_0^2} Z_A^{-n/2} Z_c^{-m}Z_b^{-r/2} + \frac{\p  Z_A^{-n/2} }{ \p g^2}  Z_c^{-m}Z_b^{-r/2}    \mathcal G_0^{n'} \nonumber\\
&&+    Z_A^{-n/2} \frac{\p  Z_c^{-m}}{ \p g^2} Z_b^{-r/2} \mathcal G_0^{n'} +   Z_A^{-n/2} Z_c^{-m} \frac{\p  Z_b^{-r/2} }{ \p g^2}     \mathcal G_0^{n'}\,,\nonumber\\
\end{eqnarray}
where we have replaced $(n + 2m + r)$ with $n'$ as a shorthand. Next, we have to calculate all the derivatives w.r.t.~$g^2$.
\begin{itemize}
\item \textbf{Calculation of $ \frac{\p g_0^2 }{ \p g^2}$}\\
In dimensional regularization, with $d=4-\varepsilon$, one can write down
\begin{eqnarray} \label{glabel5}
g_0^2 &=& \mu^\varepsilon Z_g^2 g^2\,.
\end{eqnarray}
Hence, if we derive this equation w.r.t.~$g^2$,
\begin{eqnarray}\label{ggnul}
\frac{\p g_0^2 }{ \p g^2} &=& \mu^\varepsilon \frac{\p Z_g^2}{ \p g^2} g^2 +  \mu^\varepsilon Z_g^2 ~=~ g_0^2 \left(  \frac{\p \ln Z_g^2}{ \p g^2}  + \frac{1}{g^2}\right) \,.
\end{eqnarray}
From the previous equation, we still have to determine $ \frac{\p \ln Z_g^2}{ \p g^2} $, which can be extracted from equation \eqref{glabel5}. Deriving this equation w.r.t.~$\mu$ gives,
\begin{equation}
\mu \frac{\p g_0^2}{\p \mu} = \varepsilon \mu^\varepsilon Z_g^2 g^2 + \mu^\varepsilon  \frac{\p Z_g^2}{\p g^2} \mu\frac{\p g^2}{\p \mu} g^2 + \mu^\varepsilon Z_g^2 \mu \frac{\p g^2}{\p \mu} = 0\,,
\end{equation}
were we have applied the chain rule again. We can rewrite this equation making use of the definition of the $\beta$-function
\begin{eqnarray}
\mu\frac{\p g^2}{\p \mu} &=& - \varepsilon g^2 + \beta(g^2)\,,
\end{eqnarray}
where we have immediately extracted the part in $\varepsilon$, and
we obtain,
\begin{equation}
   \frac{\p \ln Z_g^2}{\p g^2} =  \frac{1}{g^2} \left( \frac{-\varepsilon  g^2 }{\mu\frac{\p g^2}{\p \mu}}-1 \right) ~=~ \frac{1}{g^2} \left( \frac{ -\beta(g^2)}{  - \varepsilon g^2 + \beta(g^2) }   \right)\,.
\end{equation}
If we insert this result into expression \eqref{ggnul}, we ultimately find
\begin{eqnarray}
\frac{\p g_0^2 }{ \p g^2} &=& \frac{- \varepsilon g_0^2}{ - \epsilon g^2 + \beta(g^2)}\,.
\end{eqnarray}

\item \textbf{Calculation of $ \frac{\p Z_A^{-n/2} }{ \p g^2} $}\\
The next derivative w.r.t.~$g^2$ can  be calculated in a  similar way. We start by applying the chain rule,
\begin{equation}\label{one}
\frac{\p  Z_A^{-n/2} }{ \p g^2} = - n \frac{ Z_A^{-n/2}
}{Z_A^{1/2}} \frac{\p Z_A^{1/2}}{ \p g^2} ~=~ - n Z_A^{-n/2} \frac{
\p \ln Z_A^{1/2}}{ \p g^2}\,.
\end{equation}
Next, we derive $ \frac{ \p \ln Z_A^{1/2}}{ \p g^2}$ from the definition of the gluon anomalous dimension,
\begin{equation}\label{two}
\gamma_A =\mu \frac{\p \ln Z_A^{1/2}}{\p \mu} = \mu \frac{\p
g^2 }{\p \mu}  \frac{\p \ln Z_A^{1/2}}{\p g^2} = \left(  -
\varepsilon g^2 + \beta(g^2) \right) \frac{\p \ln Z_A^{1/2}}{\p
g^2}\,.
\end{equation}
From expression \eqref{one} and \eqref{two}, it now follows
\begin{eqnarray}
\frac{\p  Z_A^{-n/2} }{ \p g^2} &=& -n  Z_A^{-n/2} \frac{ \gamma_A}{
- \varepsilon g^2 + \beta(g^2)}\,.
\end{eqnarray}

\item \textbf{Calculation of $ \frac{\p Z_c^{-m} }{ \p g^2} $}\\
Completely analogously, we find with the help of the anomalous dimension of the ghost field,
\begin{eqnarray}
 \mu \frac{\p Z_c^{1/2}}{\p \mu} &=& \gamma_c Z_c^{1/2}\,,
\end{eqnarray}
\begin{eqnarray}
\frac{\p  Z_c^{-m} }{ \p g^2} &=& -2m  Z_A^{-n/2} \frac{ \gamma_A}{
- \varepsilon g^2 + \beta(g^2)}\,.
\end{eqnarray}

\item \textbf{Calculation of $ \frac{\p Z_b^{-r/2} }{ \p g^2} $}\\
Finally, from
\begin{eqnarray}
 \mu \frac{\p Z_b^{1/2}}{\p \mu} &=& \gamma_b Z_c^{1/2}\,,
\end{eqnarray}
we deduce
\begin{eqnarray}
\frac{\p Z_b^{-r/2}  }{ \p g^2} &=& -r Z_A^{-n/2} \frac{ \gamma_A}{ - \varepsilon g^2 + \beta(g^2)}\,.
\end{eqnarray}
\end{itemize}
Taking all the previous results into account, we can rewrite expression \eqref{finalgoal2},
\begin{equation}\label{finalgoal3}
\frac{\d \mathcal G^{n'} }{\d g^2} = \frac{Z_A^{-n/2}Z_c^{-m} Z_b^{-r/2} }{ - \varepsilon g^2 + \beta} \left[-\varepsilon g^2_0 \frac{\p}{\p g_0^2} - n \gamma_A -  2m \gamma_c -  r \gamma_b \right] \mathcal G_0^{n'}\,.
\end{equation}
The right hand side of \eqref{finalgoal3} still contains bare quantities, which we have to rewrite in terms of renormalized quantities. Notice also that we would like to get rid of the field numbers $n$, $m$ and $r$ as the mixing matrix will evidently will be independent of these numbers as they are arbitrary.\\
\\
We shall now alter the right hand side of equation \eqref{finalgoal3} by calculating $\frac{\p}{\p g_0^2} \mathcal G_0^{n'}$ and by removing the fields numbers. After a little bit of algebra, we obtain,
\begin{equation*}
\frac{\p (\e^{-S})}{\p g_0^2} = -\int \d^4 y \Biggl( - \frac{1}{g_0^2}  \left[ \frac{F_0^2(y)}{4}\right]  + \frac{1}{2g_0^2} \left[ A_0(y)\frac{\delta S}{\delta A_0(y)}
\right]- \frac{1}{2g_0^2} \left[ b_0(y) \p A_0(y)\right] \Biggr) \e^{-S}\,.
\end{equation*}
Consequently, deriving the $n'$ point Green function $\mathcal G^{n'}_0$ w.r.t.~$g^2_0$ will result in several insertions of integrated operators in this Green function,
\begin{equation}
g_0^2 \frac{\d \mathcal G^{n'}_0}{ \d g_0} = \int \d^4 y \Biggl( \mathcal G^{n'}_0 \biggl\{ \frac{F_0^2(y)}{4} \biggr\} - \frac{1}{2} \mathcal G^{n'}_0 \biggl\{ A_0(y)\frac{\delta S}{\delta A_0(y)} \biggr\}  + \frac{1}{2} \mathcal G^{n'}_0 \biggl\{ b_0(y) \p A_0(y) \biggr\} \Biggr)\,,
\end{equation}
where we have introduced a shorthand notation, e.g.
\begin{equation}
\mathcal G^{n'}_0 \biggl\{ \frac{F_0^2(y)}{4} \biggr\} = \Braket{\frac{F_0^2(y)}{4} A(x_1)\ldots A(x_n) c(y_1)\ldots c(y_m )\overline c(\hat y_1)\ldots \overline c (\hat y_m) b(z_1)\ldots b(z_r)}\,.
\end{equation}
The field numbers can be rewritten by inserting the corresponding counting operator. If we start by inserting the counting operator for the $n$ gluon fields , we find
\begin{eqnarray}
 \int \d^4 y \mathcal G_0^{n'}\biggl\{ A_0(y) \frac{\delta S}{\delta A_0(y)} \biggr\}  &=&n \mathcal
 G_0^{n'}\,,
\end{eqnarray}
as derived in equation \eqref{count}. We can derive analogous relations for the other counting operators,
\begin{align}
 \int \d^4 y \mathcal G_0^{n'}\biggl\{  c_0(y) \frac{\delta S}{\delta  c_0(y)} \biggr\} &= m \mathcal G_0^{n'}\,, &  \int \d^4 y \mathcal G_0^{n'}\biggl\{ b_0(y) \frac{\delta S}{\delta b_0(y)} \biggr\}  &= r \mathcal  G_0^{n'}\,.
\end{align}
Taking all these results together, expression \eqref{finalgoal3} now becomes,
\begin{multline*}
\frac{\d \mathcal G^{n'} }{\d g^2} = \frac{Z_A^{-n/2}Z_c^{-m} Z_b^{-r/2}}{-\varepsilon g^2 + \beta(g^2)} \int \d^4 y \Biggl[ - \epsilon \Bigl(  \mathcal G_0^{n'}\left\{\mathcal F_0 (y)\right\}     - \frac{1}{2} \mathcal G_0^{n'}\left\{\mathcal H_0 (y)  \right\} + \frac{1}{2} \mathcal G_0^{n'}\left\{\mathcal I_0 (y)  \right\} \Bigr)  \\
 -\gamma_A(g^2) \mathcal G_0^{n'}\left\{ \mathcal H_0(y) \right\} - 2  \gamma_c(g^2) \mathcal G_0^{n'}\left\{ \mathcal K_0(y) \right\}  -\gamma_b(g^2) \mathcal G_0^{n'}\left\{ \mathcal I_0(y)  \right\}\Biggr]\,.
\end{multline*}
We have again introduced a notational shorthand for the equation of motion operators, with $I_0 = b_0 \frac{\delta S}{\delta b_0}$ and $\mbox{$K_0 = c_0 \frac{\delta S}{\delta c_0}$}$ and with $\mathcal F$ and $\mathcal H$ already defined before.\\
\\
In the last part of the manipulation of the $n'$-point Green function we reexpress all the operators again in terms of their renormalized counterparts, thereby writing all the divergences explicitly in terms of $\varepsilon$. Firstly, we can reabsorb the $Z$-factors into $\mathcal G_0^{n'}$ to find,
\begin{multline}\label{finalgoal4}
\frac{\d \mathcal G^{n'} }{\d g^2} = \frac{1}{-\varepsilon g^2 + \beta(g^2)}\int \d^4 y \Biggl[ - \epsilon \Bigl(  \mathcal G_0^{n'}\left\{\mathcal F_0 (y)\right\}
- \frac{1}{2} \mathcal G_0^{n'}\left\{\mathcal H_0 (y)  \right\} + \frac{1}{2} \mathcal G_0^{n'}\left\{\mathcal I_0 (y)  \right\} \Bigr)  \\
-\gamma_A(g^2) \mathcal G_0^{n'}\left\{ \mathcal H_0(y) \right\} - 2  \gamma_c(g^2) \mathcal G_0^{n'}\left\{ \mathcal K_0(y) \right\}  -\gamma_b(g^2) \mathcal G_0^{n'}\left\{ \mathcal I_0(y) \right\}\Biggr]\,.
\end{multline}
Secondly, we parameterize the mixing matrix \eqref{operatormatrix},
\begin{eqnarray}
 \left(
  \begin{array}{c}
    \mathcal F_0 \\
    \mathcal E_0 \\
    \mathcal H_0
  \end{array}
\right) &= & \left(
          \begin{array}{ccc}
            1 + \frac{a}{\varepsilon}& -\frac{b}{\varepsilon}  &-\frac{b}{\varepsilon} \\
            0  &1 & 0   \\
           0 & 0& 1          \end{array}
        \right)
        \left(
\begin{array}{c}
    \mathcal F \\
    \mathcal E \\
    \mathcal H
  \end{array}
\right)\,.
\end{eqnarray}
Here we have displayed the fact that the entries associated with $a(g^2, \varepsilon)$ and $b(g^2, \varepsilon)$, which represent a formal power series in $g^2$, must at least have a simple pole in $\varepsilon$. We recall that the integrated operator $\mathcal E_0$ is proportional to the sum of the two counting operators $\int\mathcal I_0$ and $\int \mathcal K_0$, see expression \eqref{countcon5}. Therefore $\int\mathcal I_0 = \int\mathcal I$ and $\int\mathcal K_0 =\int\mathcal K$. Inserting all this information into expression \eqref{finalgoal4} yields,
\begin{multline}\label{finalgoal5}
\frac{\p \mathcal G^{n'} }{\p g^2} =   \frac{1}{-\varepsilon g^2 + \beta(g^2) }\int \d^4 y \Biggl[    \mathcal G^{n'}\Bigl\{ (- \varepsilon -a) \mathcal F  (y) - b \mathcal I(y) - b \mathcal K(y) +b \mathcal H(y)  + \frac{\varepsilon}{2} \mathcal H (y)\\
  -\frac{\varepsilon}{2} \mathcal I (y)  -\gamma_A(g^2) \mathcal H(y) - 2  \gamma_c(g^2) \mathcal K(y)   -\gamma_b(g^2)  \mathcal I(y)\Bigr\} \Biggr]\,,
\end{multline}
giving us the final result from which we shall be able to fix the matrix elements of expression \eqref{operatormatrix}.\\
\\
As the left hand side of our final expression \eqref{finalgoal5} is finite, the right hand side is finite too\footnote{We consider the finiteness properties in the sense of formal power series in $g^2$.}. Therefore, the following coefficients must be finite,
{\allowdisplaybreaks
\begin{eqnarray}
\mathcal F &:& \frac{- \varepsilon -a }{-\varepsilon g^2 + \beta(g^2) } = \frac{1}{g^2} \frac{(1  +a/ \varepsilon) }{1 - \beta(g^2) /(\varepsilon g^2) }\,, \nonumber\\
\mathcal I &:& \frac{ -\varepsilon/2 - b -  \gamma_b(g^2)  }{-\varepsilon g^2 + \beta(g^2) } = \frac{1}{2 g^2} \frac{ 1 + 2(b +  \gamma_b(g^2))/\varepsilon  }{1  - \beta(g^2)/(\varepsilon g^2) }\,,   \nonumber\\
\mathcal H &:& \frac{ \varepsilon/2 + b - \gamma_A(g^2)  }{-\varepsilon g^2 + \beta(g^2) } = -\frac{1}{2 g^2} \frac{1  + 2( b - \gamma_A(g^2))/\varepsilon }{1 - \beta(g^2) /(\varepsilon g^2) }\,, \nonumber\\
\mathcal K &:& \frac{  -b -  2\gamma_c(g^2)  }{-\varepsilon g^2 +
\beta(g^2) }\,,
\end{eqnarray}}

\noindent seen as a power series in $g^2$. This can only be true if
\begin{align*}
a(g^2,\varepsilon) =& - \frac{\beta(g^2)}{g^2}\,, & b(g^2,\varepsilon)=& \gamma_A(g^2) -\frac{1}{2}\frac{ \beta(g^2)}{g^2} = -\gamma_b(g^2) -\frac{1}{2}\frac{
\beta(g^2)}{g^2} = -2\gamma_c(g^2)\,.
\end{align*}
In fact, this last equation reveals a connection between the anomalous dimension of $A$ and $b$, and between the anomalous dimension of $A$, $g$ and $c$, namely
\begin{align}\label{rel}
\gamma_A + \gamma_b &= 0\,, &           \gamma_A + 2 \gamma_c &= \frac{ \beta }{ 2 g^2}\,.
 \end{align}
These relations are well-known to hold in the Landau gauge, since $Z_A Z_b = 1$ and $Z_c Z^{1/2}_A Z_g = 1$, as derived from the algebraic renormalization analysis, which  leads to equations \eqref{5Z1} and \eqref{5Z2}.\\
\\
In summary, we have completely fixed the mixing matrix in term of the elementary renormalization group functions, and this to all orders of perturbation theory,
\begin{equation}
\underbrace{ \left(
  \begin{array}{c}
    \mathcal F_0 \\
    \mathcal E_0 \\
    \mathcal H_0
  \end{array}
\right)}_{X_0} =  \underbrace{\left(
          \begin{array}{ccc}
            1 -\frac{\beta(g^2) / g^2}{\varepsilon}& -\frac{2 \gamma_c(g^2)}{\varepsilon}  &-\frac{ 2 \gamma_c(g^2)}{\varepsilon} \\
            0  &1 & 0   \\
           0 & 0& 1          \end{array}
        \right)}_{Z}
       \underbrace{ \left(
\begin{array}{c}
    \mathcal F \\
    \mathcal E \\
    \mathcal H
  \end{array}
\right)}_{X}\,.
\end{equation}
In addition, as a check of this result, we have also uncovered two relations, \eqref{rel}, between anomalous dimensions which must hold for consistency. These correspond to \eqref{5Z1} and \eqref{5Z2}, which are well-known nonrenormalization theorems in the Landau gauge.

\subsection{Constructing a renormalization group invariant\label{sectconstructingaRGI}}
As a last step, we can now look for a renormalization group invariant operator by determining the anomalous dimension $\Gamma$ coming from the mixing matrix $Z$. We define the anomalous dimension matrix $\Gamma$ as
\begin{eqnarray}
\mu \frac{\p}{\p \mu} Z &=& Z\, \Gamma\,.
\end{eqnarray}
For the calculation of $\Gamma$, we require
\begin{eqnarray}
\mu \frac{\p}{\p \mu} \left(1 -\frac{\beta / g^2}{\varepsilon}\right) &=& \frac{1}{\epsilon} (- \varepsilon g^2 + \beta(g^2)) \frac{\p (\beta/g^2)}{\p g^2} \;, \nonumber\\
\mu \frac{\p}{\p \mu} \frac{ 2\gamma_c}{\varepsilon} &=&
\frac{1}\varepsilon{}(- \varepsilon g^2 + \beta(g^2)) \frac{(2\p \gamma_c)}{\p g^2}\,,
\end{eqnarray}
so we obtain,
\begin{eqnarray}\label{Gamma1}
 \Gamma &= & \left(
          \begin{array}{ccc}
            g^2 \frac{\p (\beta / g^2)}{\p g^2}& -2g^2 \frac{\p \gamma_c}{\p g^2}  &-2g^2\frac{ \p \gamma_c}{\p g^2} \\
            0  &0 & 0   \\
           0 & 0& 0          \end{array}
        \right)\,,
\end{eqnarray}
which is indeed finite, a nice consistency check. This matrix is then related to the anomalous dimension of the operators:
\begin{align}\label{Gamma2}
X_0 &= Z X &\Rightarrow 0 &= \mu \frac{\p Z}{ \p \mu}  X + Z \mu \frac{\p X}{\p \mu} &\Rightarrow \mu \frac{\p X}{\p \mu} &= - \Gamma X \,.
\end{align}
We now have all the ingredients at our disposal to determine a renormalization group invariant operator. We are looking for a linear combination of $\mathcal F$, $\mathcal E$ and $\mathcal H$ which does not run,
\begin{eqnarray}
\mu \frac{\p}{\p \mu} \left[ k \mathcal F + \ell  \mathcal G + m \mathcal H \right] &=& 0\,,
\end{eqnarray}
whereby $k$, $\ell$ and $m$ are to be understood as functions of $g^2$. Invoking the chain rule gives
\begin{align}
&\mu \frac{\p k}{\p \mu}  \mathcal F  - k  g^2 \frac{\p (\beta / g^2)}{\p g^2} \mathcal F  + 2k g^2 \frac{\p \gamma_c}{\p g^2} \mathcal E + 2k g^2 \frac{\p \gamma_c}{\p g^2} \mathcal H  + \mu \frac{\p \ell }{\p \mu}  \mathcal E  + \mu \frac{\p m }{\p \mu} \mathcal H  = 0\,.
\end{align}
This previous equation results in two differential equations,
\begin{eqnarray}\begin{cases}
\mu \frac{\p k}{\p \mu}   - k  g^2 \frac{\p (\beta / g^2)}{\p g^2}  =0\,, \nonumber\\
\mu \frac{\p \ell }{\p \mu} + k g^2 \frac{\p \gamma_c}{\p g^2} = 0\,, \nonumber\\
\ell = m\,, \end{cases}
\end{eqnarray}
which can be solved by,
\begin{eqnarray}\begin{cases}
k(g^2) = \frac{\beta(g^2)}{g^2}\,, \nonumber\\
\ell(g^2)  = m(g^2) = -2\gamma_c(g^2)\,.\end{cases}
\end{eqnarray}
In summary, we have determined a renormalization group invariant scalar operator $\mathcal{R}$ containing $\mathcal F$. Explicitly,
\begin{equation}\label{YMRGI}
\mathcal{R}=\frac{1}{4}\frac{\beta(g^2)}{g^2}F_{\mu\nu}^2 -2\gamma_c(g^2)\left(A_\mu^a\p_\mu b^a  + \p_\mu \overline c^a D_\mu^{ab} c^{b}\right)-2 \gamma_c(g^2) A_{\mu}^a\frac{\delta S}{\delta A_\mu^a}\,,
\end{equation}
without having calculated any loop diagram. Moreover, this invariant is equal to the trace anomaly $\Theta_\mu^\mu$, which is expected as $\Theta_\mu^\mu$ is also a $d=4$ renormalization group invariant \cite{Collins:1976yq}.

\subsection{Conclusion\label{secconclusion}}
In conclusion, we have thus found that the operator $F^2_{\mu\nu}$ mixes with other $d=4$ operators, i.e.~a BRST invariant operator $\mathcal E = s(\p_\mu \overline{c}^a  A_\mu^a) $ and an equation of motion term $\mathcal H = A_\mu^a\frac{\delta S_{\YM}}{\delta A_\mu^a}$. We have also construction a renormalization group invariant, $\mathcal R$, see equation \eqref{YMRGI}. Notice however, that the second and the third term of the renormalization group invariant \eqref{YMRGI} shall drop when calculating correlators. Firstly, the third term is always zero when working on-shell and secondly, due to the BRST invariance of the Yang-Mills action
\begin{eqnarray}
\Braket{ \mathcal{R}(x) \mathcal{R} (y)} &=&  \Braket{   \left[ \frac{ \beta} {g^2} F^2 + s(\ldots) \right](x)   \left[\frac{ \beta} {g^2} F^2 + s(\ldots) \right](y)} \nonumber\\
                                 &=& \Braket{ \left( \frac{ \beta} {g^2}\right)^2 F^2(x) F^2(y) + s(\ldots)} =  \left(\frac{ \beta} {g^2}\right)^2  \Braket{F^2(x) F^2 (y)}\;,
\end{eqnarray}
BRST exact terms always drop out. However, we have paved the way for more complicated actions as the GZ action since this framework is very solid. We shall see that the correlator is no longer trivial in this case due to the breaking of the BRST.

\section{The (Refined) GZ action with the inclusion of the operator $F_{\mu\nu}^2$}
We shall now report an analogous story for the (Refined) GZ action, which is based on \cite{Dudal:2009zh}. We recall that the GZ action $S_\GZ$ breaks the BRST, see e.g.~section \ref{sectbreakingBRST} in chapter \ref{scrutinizing}. However, to be able to repeat the story of the previous section, we need an action which is BRST invariant. Luckily, we can embed the GZ action, into a ``larger'' action, which is BRST invariant, namely $\Sigma_\GZ$ in expression \eqref{brstinvariant}, by introducing 6 new sources, $U_\mu^{ai}$, $V_\mu^{ai}$, $M_\mu^{ai}$, $N_\mu^{ai}$, $R_\mu^{ai}$ and $T_\mu^{ai}$ with BRST variations given in expression \eqref{BRST2}. Only in the end, we shall give these sources their physical values \eqref{physlimit} to return to the original GZ action. It is then natural that also here we shall encounter mixing of the 4 dimensional operator $F^2_{\mu\nu}$ with other $d=4$ operators.

\subsection{Renormalization of the GZ action with inclusion of the operator $F_{\mu\nu}^2$}
\subsubsection{The starting action}
With the mixing of the 4 dimensional operators in mind, we can propose an enlarged Gribov-Zwanziger action containing the glueball operator $F^2_{\mu\nu}$. This action will turn out to be renormalizable. For this, we can make two observations. Firstly, the limit, $\{\varphi, \overline \varphi, \omega,  \overline \omega, U, V, N,$ $M, T, R \} \to 0$, has to lead to our original Yang-Mills action $S_\cl$ with the addition of the glueball terms given by equation \eqref{klassiek}. Secondly, setting all the terms related to the glueball term $q F^2$ equal to zero, we should recover the Gribov-Zwanziger action $\Sigma_\GZ$ in equation \eqref{brstinvariant}. Therefore, we propose the following starting action:
\begin{eqnarray}\label{glueballaction}
 \Sigma_\glue &=& \Sigma_{\GZ} + \int \d^d x \ q F_{\mu\nu}^a F_{\mu\nu}^a + \int \d^d x s \Bigl( \eta  \left[  \p_\mu \overline c^a A_\mu^a + \p \overline \omega \p \varphi + g f_{akb} \p \overline \omega^a A^k \varphi^b + U^a D^{ab} \varphi^b \right. \nonumber\\
 &&\left. + V^a D^{ab}\overline \omega^b + UV  - T_\mu^{a i} g f_{abc} D^{bd}_\mu c^d \overline \omega^c_i \right]\Bigr) \nonumber\\
&=& \Sigma_{\mathrm{GZ}} + \int \d^d x \ q F_{\mu\nu}^a F_{\mu\nu}^a +  \int \d^d x \Bigl( \lambda \left[  \p_\mu \overline c^a A_\mu^a + \p \overline \omega \p \varphi + g f_{akb} \p \overline \omega^a A^k \varphi^b + U^a D^{ab} \varphi^b \right.  \nonumber\\
&&\left. + V^a D^{ab}\overline \omega^b + UV  - T_\mu^{a i} g f_{abc} D^{bd}_\mu c^d \overline \omega^c_i  \right] + \eta \Bigl[ \p_\mu b^a A_\mu^a + \p_\mu \overline c^a D^{ab}_{\mu} c^b + \p \overline \varphi \p \varphi - \p \overline \omega \p \omega   \nonumber\\
 &&+ g f_{akb} \p \overline \varphi^a A^k \varphi^b+ g f_{akb} \p \overline \omega^a D^{kd} c^d \varphi^b  -  g f_{akb} \p \overline \omega^a A^k \omega^b  +M_\mu^{ai} D_\mu^{ab} \varphi_i^b  \nonumber\\
  && + g U_\mu^{ai} f^{abc}  D_\mu^{ab}c^b \varphi_i^c - U_\mu^{ai}  D_\mu^{ab}\omega_i^b  + N_\mu^{ai} D_\mu^{ab} \overline \omega_i^b-gV_\mu^{ai} f^{abc} D_\mu^{bd}c^d \overline \omega_i^c + V_\mu^{ai}  D_\mu^{ab} \overline \varphi_i^b\nonumber\\
   &&+M_\mu^{ai} V_\mu^{ai}-U_\mu^{ai} N_\mu^{ai} - R_\mu^{ai} g f^{abc} D_\mu^{bd} c^d \overline \omega^c_i  - T_\mu^{ai} g f_{abc} D^{bd}_\mu c^d \overline \varphi^c_i\Bigr]\Bigr)\,,
\end{eqnarray}
whereby $\lambda$ and $\eta$ were already introduced in expression \eqref{lambdaeta}. Now upon taking the limit $\{\varphi, \overline \varphi, \omega, \overline \omega, U, V, N, M, T,R  \} \to 0$, we indeed recover the Yang-Mills action\footnote{The term proportional to the equations of motion will be introduced later.} \eqref{klassiek} and setting all sources equal to zero ($q$, $\eta$, $\lambda$) $\to 0$, we find back our original Gribov-Zwanziger action, see equation \eqref{brstinvariant}. Notice that in principle, we could have taken other possible starting actions which also enjoy these two correct limits. We could have tried to couple different sources to the different BRST exact terms instead of employing only one source $\eta$. However, this would not lead to a renormalizable action, while the action \eqref{glueballaction} does turn out to be renormalizable, as we shall prove.\\
\\
Again, we shall establish the renormalizability of \eqref{glueballaction} by using the algebraic renormalization formalism, see chapter \ref{algebraic}. \\
\\
The first step is to introduce two auxiliary terms necessary for the process of renormalization. As always, we add the additional external term $S_{\ext,1}$ to the action,
\begin{eqnarray}\label{Sext1}
S_{\ext,1}&=&\int \d^d x \left( -K_\mu^a  D_\mu^{ab}  c^b + \frac{1}{2} g L^a f^{abc} c^b c^c  \right) \,,
\end{eqnarray}
In addition, we also introduce the following external term,
\begin{equation}\label{Sext2}
S_{\ext,2} = \int \d^d x s(X_i A_\mu^a \p \overline \omega^a_i) ~=~ \int \d^d x Y_i  A_\mu^a \p \overline \omega^a_i - \int \d^d x \left( X_i D^{ab}_\mu c^b \p_\mu \overline \omega^a_i + X_i A^a_\mu \p_\mu \overline \varphi^a_i \right)\,,
\end{equation}
whereby $(X_i, Y_i)$ is a new doublet of sources, i.e.~$s X_i = Y_i$. This additional term is necessary in order to have a sufficient powerful set of Ward identities. Without this term, two Ward identities of the original Gribov-Zwanziger action would be broken which are absolutely indispensable for the proof of the renormalization of the action (see Ward identity 7. and 8. in the list below). Again, in the end, we shall set
\begin{align}
\left. X_i \right|_\phys &= 0\,, &  \left. Y_i \right|_\phys &= 0\,.
\end{align}
We shall thus continue the analysis with the following action
\begin{eqnarray}
\Sigma &=& \Sigma_\glue + S_{\ext,1} + S_{\ext,2}\,.
\end{eqnarray}

\subsubsection{The Ward identities}
The second step is to search for all the Ward identities obeyed by the classical action $\Sigma$. Doing so, we find the following list of identities:
\begin{enumerate}
\item  The Slavnov-Taylor idenitity:
\begin{equation}
\mathcal{S}(\Sigma ) = 0  \;,
\end{equation}
where
\begin{multline}
\mathcal{S}(\Sigma )=\int \d^dx \Bigl( \frac{\delta \Sigma}{\delta K_{\mu}^{a}}\frac{\delta \Sigma }{\delta A_{\mu}^{a}}+\frac{\delta \Sigma }{\delta L^{a}}\frac{\delta
\Sigma}{\delta c^{a}}+b^{a}\frac{\delta \Sigma}{\delta \overline{c}^{a}}+\overline{\varphi }_{i}^{a}\frac{\delta \Sigma }{\delta \overline{\omega }_{i}^{a}}+\omega _{i}^{a}\frac{\delta \Sigma }{\delta \varphi _{i}^{a}}\\
+M_{\mu }^{ai}\frac{\delta \Sigma}{\delta U_{\mu}^{ai}}+ N_{\mu }^{ai}\frac{\delta \Sigma }{\delta V_{\mu }^{ai}} + R_{\mu }^{ai}\frac{\delta \Sigma }{\delta T_{\mu }^{ai}}+ \lambda \frac{\delta \Sigma }{\delta \eta} +  Y_i \frac{\delta \Sigma }{\delta X_i} \Bigr) \,.
\end{multline}

\item The $U(f)$ invariance:
\begin{eqnarray}\label{wardid1}
U_{ij} \Sigma &=&0\,,
\end{eqnarray}
with
\begin{multline}\label{Uij}
U_{ij}=\int \d^dx\left( \varphi_{i}^{a}\frac{\delta }{\delta \varphi _{j}^{a}}-\overline{\varphi}_{j}^{a}\frac{\delta }{\delta \overline{\varphi}_{i}^{a}}+\omega _{i}^{a}\frac{\delta }{\delta \omega _{j}^{a}}-\overline{\omega }_{j}^{a}\frac{\delta }{\delta \overline{\omega }_{i}^{a}}-  M^{aj}_{\mu} \frac{\delta}{\delta M^{ai}_{\mu}} -U^{aj}_{\mu}\frac{\delta}{\delta U^{ai}_{\mu}}   \right. \\
\left.+ N^{ai}_{\mu}\frac{\delta}{\delta N^{aj}_{\mu}} +V^{ai}_{\mu}\frac{\delta}{\delta V^{aj}_{\mu}} +   R^{aj}_{\mu}\frac{\delta}{\delta R^{ai}_{\mu}} + T^{aj}_{\mu}\frac{\delta}{\delta T^{ai}_{\mu}} +Y^{i}\frac{\delta}{\delta Y^{j}} +X^{i}\frac{\delta}{\delta X^{j}}  \right)\,.
\end{multline}
Using $Q_{f}=U_{ii}$, we can associate an extra quantum number to the $i$-valued fields and sources. One can find all quantum numbers in TABLE \ref{tabel3} and TABLE \ref{tabel4}.
\item The Landau gauge condition:
\begin{eqnarray}
\frac{\delta \Sigma}{\delta b^{a}}&=&\p_\mu A_\mu^a -\p_\mu(\eta A_\mu^a)\,.
\end{eqnarray}
\item The modified antighost equation :
\begin{eqnarray}
\frac{\delta \Sigma}{\delta \overline c^{a}}+\p_\mu\frac{\delta \Sigma}{\delta K_{\mu}^a}-\p_\mu\left( \eta\frac{\delta \Sigma}{\delta K_\mu^a } \right)&=& \p (\lambda A) \,.
\end{eqnarray}
\item\label{lincon} Two linearly broken local constraints:
\begin{eqnarray}\label{lincon2}
&&\frac{\delta\Sigma}{\delta \overline\varphi^{ai}}+\p_\mu\frac{\delta\Sigma}{\delta M_\mu^{ai}} + g f_{dba}    T^{d i}_\mu \frac{\delta \Sigma }{\delta K_{\mu }^{b i}}=gf^{abc}A_\mu^b V_\mu^{ci}-\eta gf^{abc}A_{\mu }^{b}V_{\mu}^{ci}-\p_\mu ( X_i A^a_\mu) \,, \nonumber\\
&&\frac{\delta \Sigma}{\delta\omega^{ai}}+\partial_{\mu}\frac{\delta\Sigma}{\delta N_{\mu}^{ai}}-gf^{abc}\overline{\omega}^{bi}\frac{\delta\Sigma}{\delta b^{c}} =gf^{abc}A_{\mu}^{b}U_{\mu}^{ci}  - \eta gf^{abc}A_{\mu}^{b}U_{\mu}^{ci} \,.
\end{eqnarray}
\item  The exact $\mathcal{R}_{ij}$ invariance:
\begin{equation}\label{wardander}
\mathcal{R}_{ij}\Sigma =0\,,
\end{equation}
with
\begin{equation*}
\mathcal{R}_{ij}=\int \d^dx\left( \varphi _{i}^{a}\frac{\delta}{\delta\omega _{j}^{a}}-\overline{\omega }_{j}^{a}\frac{\delta }{\delta \overline{\varphi }_{i}^{a}}+V_{\mu }^{ai}\frac{\delta }{\delta N_{\mu}^{aj}}-U_{\mu }^{aj}\frac{\delta }{\delta M_{\mu }^{ai}}  + T^{a i}_\mu \frac{\delta }{\delta R_{\mu }^{aj}}  - X^i \frac{\delta}{\delta Y^j}\right) \,.
\end{equation*}
\item An extra integrated Ward identity:
\begin{equation}\label{broken1}
\int \d^d x\left( \frac{\delta }{\delta \lambda} - \eta \frac{\delta }{\delta \lambda} + \overline c^a \frac{\delta }{\delta b^a}  + U_\mu^{ai} \frac{\delta }{\delta M_\mu^{ai}}   + \overline \omega_i^a  \frac{\delta }{\delta \overline \varphi_i^a}   - X_i  \frac{\delta }{\delta Y_i}    \right) \Sigma = 0\,,
\end{equation}
which expresses in functional form the BRST exactness of the operator coupled to $\lambda$.
\item The integrated Ward Identity:
\begin{eqnarray}\label{broken2}
\int \d^d x \left( c^a \frac{\delta }{ \delta \omega^{a i} } + \overline{\omega}^{a i}  \frac{\delta }{ \delta \overline{c}^a }  + U^{ai}_\mu  \frac{\delta }{ \delta K^{a}_\mu} -\eta U^{ai}_\mu  \frac{\delta }{ \delta K^{a}_\mu} - \lambda \frac{\delta }{ \delta Y_i}  \right) \Sigma = 0\,.
\end{eqnarray}
\item The $X$-and $Y$-Ward identities:
\begin{eqnarray} \label{wardideinde}
\int \d^d x \left[ (1 - \eta) \frac{\delta}{ \delta X^i} - \lambda \frac{\delta }{ \delta Y^i} + \overline \omega^{a}_i \frac{\delta }{ \delta \overline c^a}   + \overline \varphi_i^a \frac{\delta }{ \delta b^a} \right] \Sigma &=& 0 \,,\nonumber\\
\int \d^d x \left[ (1 - \eta) \frac{\delta}{ \delta Y^i} + \overline \omega_i^a \frac{\delta }{ \delta b^a} \right] \Sigma &=& 0\,.
\end{eqnarray}
\end{enumerate}
\begin{table}[H]
  \centering
        \begin{tabular}{|c|c|c|c|c|c|c|c|c|}
        \hline
        & $A_{\mu }^{a}$ & $c^{a}$ & $\overline{c}^{a}$ & $b^{a}$ & $\varphi_{i}^{a} $ & $\overline{\varphi }_{i}^{a}$ &            $\omega _{i}^{a}$ & $\overline{\omega }_{i}^{a}$ \\
        \hline
        \hline
        \textrm{dimension} & $1$ & $0$ &$2$ & $2$ & $1$ & $1$ & $1$ & $1$ \\
        \hline
        $\mathrm{ghost\, number}$ & $0$ & $1$ & $-1$ & $0$ & $0$ & $0$ & $1$ & $-1$ \\
        \hline
        $Q_{f}\textrm{-charge}$ & $0$ & $0$ & $0$ & $0$ & $1$ & $-1$& $1$ & $-1$\\
        \hline
        \end{tabular}
        \caption{Quantum numbers of the fields.}\label{tabel3}
        \end{table}
        \begin{table}[H]
    \centering
    \begin{tabular}{|c|c|c|c|c|c|c|c|c|c|c|c|c|c|}
        \hline
        &$U_{\mu}^{ai}$&$M_{\mu }^{ai}$&$N_{\mu }^{ai}$&$V_{\mu }^{ai}$&$R_{\mu }^{ai}$& $T_{\mu }^{ai}$ & $K_{\mu }^{a}$&$L^{a}$& $q$&  $\eta$& $\lambda$ & $X^i$ & $Y^i$ \\
        \hline
        \hline
        \textrm{dimension} & $2$ & $2$ & $2$ &$2$  & 2&2& $3$ & $4$ & $0$ & $0$ & $0$ &1&1 \\
        \hline
        $\mathrm{ghost\, number}$ & $-1$& $0$ & $1$ & $0$ & $0$&$-1$ & $-1$ & $-2$ & 0 & 0 & 1 & 0 & 1 \\
        \hline
        $Q_{f}\textrm{-charge}$ & $-1$ & $-1$ & $1$ & $1$ & $1$ &$1$ &$0$ & $0$  & 0 & 0&0 & 1 & 1 \\
        \hline
        \end{tabular}
        \caption{Quantum numbers of the sources.}\label{tabel4}
\end{table}

\noindent Let us stress here that it is of paramount importance to have a good set of Ward identities to start from. For the construction of the action $\Sigma$, one should keep in mind the limits to the ordinary Gribov-Zwanziger case and to the Yang-Mills action with the inclusion of the glueball term. It is logical that an identity which plays a crucial role in one of the two limit cases, should not be broken by the action $\Sigma$, as $\Sigma$ can be seen as an enlargement of the two limit cases. This is the reason why we have introduced $S_{\ext,2}$. Without the auxiliary sources $X_i$ and $Y_i$, the extra integrated Ward identity \eqref{broken1} and the integrated Ward identity \eqref{broken2} are broken, and without these two identities one cannot prove the renormalizability of the action in an algebraic way.

\subsubsection{The counterterm}
Let us now determine the counterterm. From the previous Ward identities, it follows that the counterterm $\Sigma^c$ is constrained by:
\begin{enumerate}
\item  The linearized Slavnov-Taylor identity:
\begin{equation}\label{ST}
\mathcal{B}_{\Sigma }\Sigma^c=0\,,
\end{equation}
where $\mathcal{B}_{\Sigma }$ is the nilpotent linearized Slavnov-Taylor operator,
\begin{multline}
\mathcal{B}_{\Sigma} =\int \d^dx\left( \frac{\delta \Sigma}{\delta K_{\mu }^{a}}\frac{\delta }{\delta A_{\mu }^{a}}+\frac{\delta \Sigma }{\delta A_{\mu }^{a}}\frac{\delta }{\delta K_{\mu }^{a}}+\frac{\delta\Sigma }{\delta L^{a}}\frac{\delta }{\delta c^{a}}+\frac{\delta\Sigma }{\delta c^{a}}\frac{\delta }{\delta L^{a}}+b^{a}\frac{\delta }{\delta \overline{c}^{a}}+\overline{\varphi}_{i}^{a}\frac{\delta }{\delta \overline{\omega }_{i}^{a}}+\omega_{i}^{a}\frac{\delta }{\delta \varphi_{i}^{a}}\right.\nonumber\\
\left.+M_{\mu }^{ai}\frac{\delta }{\delta U_{\mu }^{ai}}+N_{\mu }^{ai}\frac{\delta }{\delta V_{\mu }^{ai}} +  R_{\mu }^{ai}\frac{\delta  }{\delta T_{\mu }^{ai}} + \lambda \frac{\delta }{\delta \eta} + Y^i \frac{\delta  }{\delta X_i} \right)\,,
\end{multline}
and
\begin{equation}
\mathcal{B}_{\Sigma }\mathcal{B}_{\Sigma }=0\,.
\end{equation}
\item The $U(f)$ invariance:
\begin{eqnarray}\label{cont1}
U_{ij} \Sigma^c &=&0 \,.
\end{eqnarray}
$U_{ij}$ is given in expression \eqref{Uij}.
\item The Landau gauge condition
\begin{eqnarray}
\frac{\delta \Sigma^c}{\delta b^{a}}&=&0\,.
\end{eqnarray}
\item The modified antighost equation:
\begin{eqnarray}
\frac{\delta \Sigma^c}{\delta \overline c^{a}}+\p_\mu\frac{\delta \Sigma^c}{\delta K_{\mu}^a}-\p_\mu\left( \eta\frac{\delta \Sigma^c}{\delta K_\mu^a } \right)&=&0 \,.
\end{eqnarray}
\item  The linearly broken local constraints:
\begin{eqnarray}
&&\frac{\delta \Sigma^c}{\delta \overline{\varphi }^{ai}}+\partial _{\mu }\frac{\delta \Sigma^c }{\delta M_{\mu }^{ai}}  + g f_{dba}    T^{d i}_\mu \frac{\delta \Sigma }{\delta K_{\mu }^{b i}}=0\,, \nonumber\\
&&\frac{\delta \Sigma^c}{\delta\omega^{ai}}+\partial_{\mu}\frac{\delta\Sigma^c}{\delta N_{\mu}^{ai}}-gf^{abc}\overline{\omega}^{bi}\frac{\delta\Sigma^c}{\delta b^{c}}=0 \,.
\end{eqnarray}
\item  The exact $\mathcal{R}_{ij}$ symmetry:
\begin{equation}\label{cont2}
\mathcal{R}_{ij}\Sigma^c =0\,.
\end{equation}
\item The extra integrated Ward identity:
\begin{equation}
\int \d^d x\left( \frac{\delta }{\delta \lambda} - \eta \frac{\delta }{\delta \lambda} + \overline c^a \frac{\delta }{\delta b^a}  + U_\mu^{ai} \frac{\delta }{\delta M_\mu^{ai}}   + \overline \omega_i^a  \frac{\delta }{\delta \overline \varphi_i^a}  - X_i  \frac{\delta }{\delta Y_i}    \right) \Sigma^c = 0\,.
\end{equation}
\item The integrated Ward Identity:
\begin{eqnarray}
\int \d^d x \left( c^a \frac{\delta }{ \delta \omega^{a i} } + \overline{\omega}^{a i}  \frac{\delta }{ \delta \overline{c}^a }  + U^{ai}_\mu  \frac{\delta }{ \delta K^{a}_\mu} -\eta U^{ai}_\mu  \frac{\delta }{ \delta K^{a}_\mu} - \lambda \frac{\delta }{ \delta Y_i}  \right) \Sigma^c = 0\,.
\end{eqnarray}
\item The $X$-and $Y$-Ward identities:
\begin{eqnarray}\label{cont3}
\int \d^d x \left[ (1 - \eta) \frac{\delta}{ \delta X^i} - \lambda \frac{\delta }{ \delta Y^i} + \overline \omega^{a}_i \frac{\delta }{ \delta \overline c^a}   + \overline \varphi_i^a \frac{\delta }{ \delta b^a} \right] \Sigma^c &=& 0 \,, \nonumber\\
\int \d^d x \left[ (1 - \eta) \frac{\delta}{ \delta Y^i} + \overline \omega_i^a \frac{\delta }{ \delta b^a} \right] \Sigma^c &=& 0\,.
\end{eqnarray}
\end{enumerate}
At this point, we are ready to determine the most general integrated local polynomial $\Sigma^c$ in the fields and external sources of dimension bounded by four and with zero ghost number, limited by the constraints \eqref{ST}--\eqref{cont3}. As usual, the counterterm can be parameterized as follows:
\begin{eqnarray}
\Sigma^c &=& \underbrace{\left(\mathcal{B}_{\Sigma}\textrm{ closed but not exact part}\right)}_{\Sigma^c_1} + \underbrace{\mathcal{B}_{\Sigma }\Delta^{-1}}_{\Sigma^c_2}\,,
\end{eqnarray}
whereby $\Sigma^c_1$ is a cohomologically non-trivial part while $\Sigma^c_2$ represents the cohomologically trivial part. $\Delta^{-1}$ is the most general local polynomial with dimension 4 and ghost number $-1$. We have proven on p.\pageref{doublettheorem} that all fields and sources belonging to a doublet can only enter the cohomologically trivial part. This is exactly the reason why we have opted to introduce the source $\eta$, which is coupled to the BRST exact term, as part of a doublet. In this way, the source $\eta$ can only enter the trivial part, and turns out to be useful to explicitly prove the upper triangular form of the mixing matrix in equation \eqref{upper}. One can now check that the closed but not exact part is given by
\begin{eqnarray}
\Sigma^c_1 &=&a_0 S_{\YM} + b_0 \widehat S_{\YM}\,,
\end{eqnarray}
whereby
\begin{eqnarray}
\widehat S_{\YM} &=& \int \d^d x q \frac{1}{4} F_{\mu\nu}^a F_{\mu\nu}^a \,,
\end{eqnarray}
and the trivial part is given by the following rather lengthy expression:
{\allowdisplaybreaks\begin{align}\label{counterruw}
&\Sigma^c_2 =    \mathcal{B}_\Sigma \int \d^d x    \biggl\{ \biggl[ a_{1}(K_{\mu}^{a}+\partial _{\mu} \overline{c}^{a})A_{\mu}^{a}+a_{2} L^{a}c^{a} +a_{3}U_{\mu i}^{a} \partial _{\mu }\varphi _{i}^{a} +a_{4} V_{\mu i}^{a} \partial _{\mu }\overline{\omega }_{i}^{a} +a_{5} \overline{\omega }_{i}^{a}\partial ^{2}\varphi _{i}^{a}  \nonumber\\
&+a_{6}  U_{\mu i}^{a}V_{\mu i}^{a}+a_{7} gf^{abc}U_{\mu i}^{a} \varphi _{i}^{b}A_{\mu }^{c}+a_{8} gf^{abc}V_{\mu i}^{a} \overline{\omega }_{i}^{b}A_{\mu }^{c}+a_{9} gf^{abc}\overline{\omega }_{i}^{a}A_{\mu }^{c} \partial _{\mu }\varphi _{i}^{b}  \nonumber\\
&+a_{10} gf^{abc}\overline{\omega }_{i}^{a}(\partial _{\mu }A_{\mu}^{c})\varphi _{i}^{b} + a_{11} X^i \overline \omega^a_i \p A^a_\mu  + a_{12} X^i \p\overline  \omega^a_i  A^a_\mu + a_{13} X^i \overline \varphi^a_i \overline c^a + a_{14}g f_{abc} X^i \omega^a_i \overline \omega^b_j \overline \omega^c_j  \nonumber\\
& +  a'_{14} g f_{abc} X^i \omega^a_j \overline \omega^b_i \overline \omega^c_j   + a_{15} X^i \overline \omega^a_i b^a + a_{16} X^i U^{i a}_\mu A^a_\mu + a_{17} g f_{abc} X^i \overline \omega^a_i \varphi^b_j \overline \varphi_j^c +  a'_{17} g f_{abc} X^i \overline \omega^a_j \varphi^b_i \overline \varphi_j^c \nonumber\\
& +  a_{17}^{\prime\prime} g f_{abc} X^i \overline \omega^a_j \varphi^b_j \overline \varphi_i^c + a_{18} g f_{abc} X^i \overline \omega^a_i \overline c^b c^c  +a_{19} X^i X^i \overline \varphi^a_j \overline \omega^a_j + a_{19}^\prime X^i X^j \overline \varphi^a_i \overline \omega^a_j + a_ {20}X^i Y^j \overline \omega^i_a \overline \omega^j_a \nonumber\\
 &+ a_{21}g f_{abc} Y^i \overline \omega^a_i \overline \omega^b_j \varphi^c_j + a'_{21}g f_{abc} Y^i \overline \omega^a_j \overline \omega^b_j \varphi^c_i  +  a_{22} Y^i \overline \omega_i^a \overline c^a +  a_{23} R_{\mu}^{ai} U_{\mu}^{ai} + a_{24} T_{\mu }^{ai} M_{\mu }^{ai} \nonumber\\
&+ a_{25} g f_{abc} R_{\mu }^{ai} \overline{\omega }_{i}^{b} A_{\mu}^{c}  + a_{26} g f_{abc} T_{\mu}^{ai} \overline{\varphi }_{i}^{b} A_{\mu}^{c} + a_{27} R_{\mu}^{ai} \p_\mu \overline{\omega }_{i}^{a}  +  a_{28} T_{\mu}^{ai} \p_\mu \overline{\varphi }_{i}^{a}  \biggr]\nonumber\\
& +q \biggl[ b_{1} (K_{\mu}^{a}+\partial _{\mu} \overline{c}^{a})A_{\mu}^{a} +  c_{1}  \overline{c}^{a} \partial _{\mu} A_{\mu}^{a}  +b_{2} L^{a}c^{a} +b_{3} U_{\mu i}^{a} \partial _{\mu }\varphi _{i}^{a}+ c_{3} \partial _{\mu }U_{\mu i}^{a}\varphi _{i}^{a} +b_{4} V_{\mu i}^{a} \partial _{\mu }\overline{\omega }_{i}^{a}    \nonumber\\
&+c_{4} \partial _{\mu } V_{\mu i}^{a} \overline{\omega }_{i}^{a}+b_{5} \overline{\omega }_{i}^{a}\partial ^{2}\varphi _{i}^{a}+  c_{5} \partial_{\mu}\overline{\omega }_{i}^{a}\partial_{\mu}\varphi _{i}^{a}  + d_{5} \partial ^{2} \overline{\omega }_{i}^{a}\varphi _{i}^{a}  +b_{6}  U_{\mu i}^{a}V_{\mu i}^{a}+b_{7} gf^{abc}U_{\mu i}^{a} \varphi _{i}^{b}A_{\mu }^{c} \nonumber\\
&+  b_{8} gf^{abc}V_{\mu i}^{a} \overline{\omega }_{i}^{b}A_{\mu }^{c} +b_{9} gf^{abc}\overline{\omega }_{i}^{a}A_{\mu }^{c} \partial _{\mu }\varphi _{i}^{b} +c_{9} gf^{abc}\overline{\omega }_{i}^{a}(\partial _{\mu }A_{\mu}^{c})\varphi _{i}^{b} + d_{9} gf^{abc}\partial _{\mu } \overline{\omega }_{i}^{a}A_{\mu}^{c}\varphi _{i}^{b} \nonumber\\
& +b_{10} X^i \overline \omega^a_i \p A^a_\mu + c_{10} X^i \p\overline  \omega^a_i  A^a_\mu  + d_{10} \p X^i \overline  \omega^a_i  A^a_\mu+ b_{11} X^i \overline \varphi^a_i \overline c^a + b_{12}g f_{abc} X^i \omega^a_i \overline \omega^b_j \overline \omega^c_j  \nonumber\\
& + b_{12}'g f_{abc} X^i \omega^a_j \overline \omega^b_i \overline \omega^c_j + b_{13} X^i \overline \omega^a_i b^a + b_{14} X^i U^{i a}_\mu A^a_\mu + b_{15} g f_{abc} X^i \overline \omega^a_i \varphi^b_j \overline \varphi_j^c +b_{15}' g f_{abc} X^i \overline \omega^a_j \varphi^b_i \overline \varphi_j^c \nonumber\\
&  + b_{15}'' g f_{abc} X^i \overline \omega^a_j \varphi^b_j \overline \varphi_i^c + b_{16} g f_{abc} X^i \overline \omega^a_i \overline c^b c^c +b_{17} X^i X^i \overline \varphi^a_j \overline \omega^a_j  + b_{17}^\prime X^i X^j \overline \varphi^a_i \overline \omega^a_j + b_ {18}X^i Y^j \overline \omega^i_a \overline \omega^j_a \nonumber\\
&+ b_{19}g f_{abc} Y^i \overline \omega^a_i \overline \omega^b_j \varphi_j^c + b_{19}^\prime g f_{abc} Y^i \overline \omega^a_j \overline \omega^b_j \varphi_i^c  + b_{20} Y^i \overline \omega_i^a \overline c^a +  b_{21} R_{\mu}^{ai} U_{\mu}^{ai}  + b_{22} T_{\mu }^{ai} M_{\mu }^{ai}\nonumber\\
& + b_{23} g f_{abc} R_{\mu }^{ai} \overline{\omega }_{i}^{b} A_{\mu}^{c} + b_{24} g f_{abc} T_{\mu}^{ai} \overline{\varphi }_{i}^{b} A_{\mu}^{c} + b_{25} R_{\mu}^{ai} \p_\mu \overline{\omega }_{i}^{a}  + c_{25} \p_\mu R_{\mu}^{ai}  \overline{\omega }_{i}^{a}+  b_{26} T_{\mu}^{ai} \p_\mu \overline{\varphi }_{i}^{a} \nonumber\\
& +  c_{26}  \p_\mu T_{\mu}^{ai} \overline{\varphi }_{i}^{a} \biggr] + \eta \biggl[ e_{1}  K_{\mu}^{a}A_{\mu}^{a}+ e_1'\partial _{\mu} \overline{c}^{a} A_{\mu}^{a} +  f_{1}  \overline{c}^{a} \partial _{\mu} A_{\mu}^{a}  +e_{2} L^{a}c^{a} +e_{3}  U_{\mu i}^{a} \partial _{\mu }\varphi _{i}^{a}+ f_{3} \partial _{\mu }U_{\mu i}^{a}\varphi _{i}^{a} \nonumber\\
&  +e_{4} V_{\mu i}^{a} \partial _{\mu }\overline{\omega }_{i}^{a} +f_{4}\partial _{\mu } V_{\mu i}^{a} \overline{\omega }_{i}^{a} +e_{5} \overline{\omega }_{i}^{a}\partial ^{2}\varphi _{i}^{a} +  f_{5} \partial_{\mu}\overline{\omega }_{i}^{a}\partial_{\mu}\varphi _{i}^{a}  + g_{5} \partial ^{2} \overline{\omega }_{i}^{a}\varphi _{i}^{a}  +e_{6}  U_{\mu i}^{a}V_{\mu i}^{a} \nonumber\\
&+e_{7} gf^{abc}U_{\mu i}^{a} \varphi _{i}^{b}A_{\mu }^{c} +e_8 gf^{abc}V_{\mu i}^{a} \overline{\omega }_{i}^{b}A_{\mu }^{c}+e_{9} gf^{abc}\overline{\omega }_{i}^{a}A_{\mu }^{c} \partial _{\mu }\varphi _{i}^{b} +f_{9} gf^{abc}\overline{\omega }_{i}^{a}(\partial _{\mu }A_{\mu}^{c})\varphi _{i}^{b} \nonumber\\
& + g_{9} gf^{abc}\partial _{\mu } \overline{\omega }_{i}^{a}A_{\mu}^{c}\varphi _{i}^{b} +e_{10} X^i \overline \omega^a_i \p A^a_\mu + f_{10} X^i \p\overline  \omega^a_i  A^a_\mu   + g_{10} \p X^i \overline  \omega^a_i  A^a_\mu  + e_{11} X^i \overline \varphi^a_i \overline c^a  \nonumber\\
& + e_{12}g f_{abc} X^i \omega^a_i \overline \omega^b_j \overline \omega^c_j +e_{12}'g f_{abc} X^i \omega^a_j \overline \omega^b_i \overline \omega^c_j   + e_{13} X^i \overline \omega^a_i b^a + e_{14} X^i U^{i a}_\mu A^a_\mu  + e_{15} g f_{abc} X^i \overline \omega^a_i \varphi_b^j \overline \varphi^j_c  \nonumber\\
& + e_{15}' g f_{abc} X^i \overline \omega^a_j \varphi^b_i \overline \varphi_j^c + e_{15}'' g f_{abc} X^i \overline \omega^a_j \varphi^b_j \overline \varphi_i^c  + e_{16} g f_{abc} X^i \overline \omega^a_i \overline c^b c^c +e_{17} X^i X^i \overline \varphi^a_j \overline \omega^a_j \nonumber\\
&+ e_{17}^\prime X^i X^j \overline \varphi^a_i \overline \omega^a_j + e_ {18}X^i Y^j \overline \omega^i_a \overline \omega^j_a + e_{19}g f_{abc} Y^i \overline \omega^a_i \overline \omega^b_j \varphi^c_j + e_{19}'g f_{abc} Y^i \overline \omega^a_j \overline \omega^b_j \varphi^c_i + e_{20} Y^i \overline \omega_i^a \overline c^a  \nonumber\\
&  + e_{21} R_{\mu}^{ai} U_{\mu}^{ai} + e_{22} T_{\mu }^{ai} M_{\mu }^{ai}+ e_{23} g f_{abc} R_{\mu }^{ai} \overline{\omega }_{i}^{b} A_{\mu}^{c} + e_{24} g f_{abc} T_{\mu}^{ai} \overline{\varphi }_{i}^{b} A_{\mu}^{c} + e_{25} R_{\mu}^{ai} \p_\mu \overline{\omega }_{i}^{a}   \nonumber\\
&+ f_{25} \p_\mu R_{\mu}^{ai}  \overline{\omega }_{i}^{a} +  e_{26} T_{\mu}^{ai} \p_\mu \overline{\varphi }_{i}^{a} +  f_{26}  \p_\mu T_{\mu}^{ai} \overline{\varphi }_{i}^{a} \biggr] + \lambda \biggl[  h_1 g f_{abc} X^i  \varphi^{aj} \overline \omega^b_i \overline \omega^c_j + h_1' g f_{abc} X^i  \varphi^{ai} \overline \omega^b_j \overline \omega^c_j \nonumber\\
& + h_2 X^i \overline c^a \overline \omega^a_i   + h_3 \overline \omega^a_i \overline \omega^b_j \varphi^a_i \varphi^b_j  + (\mbox{variants of }h_3) +  h_4 T_{\mu i}^{a} \partial _{\mu }\overline{\omega }_{i}^{a} +  h_5 T_{\mu i}^{a}U_{\mu i}^{a}  + h_6 gf^{abc}T_{\mu i}^{a} \overline{\omega }_{i}^{b}A_{\mu }^{c}\biggr]  \biggr\}\,.
\end{align}}

\noindent \normalsize The coefficients $a_i$, $a'_i$, etc.~are a priori free parameters.\\
\\
Here again we did not include terms of the form $(q^2 \ldots)$, $(\eta^2 \ldots)$, $(q \eta \ldots)$, $(q^3 \ldots)$, $(\lambda q^2 \ldots)$ etc., into the counterterm as also here the argument \eqref{argument} holds.  This argument \eqref{argument} can be repeated for all the terms which are \textit{zero} in the physical limit. Therefore, this argument is not only valid for the dimensionless sources $q$, $\eta$ and $\lambda$, but also for the massive sources $K_\mu$, $L_\mu$, $X_i$, $Y_i$. Though, some care needs to be taken, see the example of the modified ghost equation \eqref{5exception}. Therefore, we choose to keep all the possible combinations of higher order in the massive sources in the counterterm \eqref{counterruw} as there are only a finite number of combinations, while keeping in mind the higher order combinations of the dimensionless sources. Only after imposing all the constraints, we can then safely neglect the terms quadratic in the sources.\\
\\
With the previous remark in mind, we can now impose all the constraints \eqref{cont1}-\eqref{cont3} on the counterterm, which is a very cumbersome job. We ultimately find
{\allowdisplaybreaks\begin{align}
&\Sigma^c= a_{0}S_{YM} +  b_0 \widehat S_{YM} + a_{1}\int \d^dx\Bigl(  A_{\mu}^{a}\frac{ \delta S_{YM}}{\delta A_{\mu }^{a}}+ A_{\mu}^{a}\frac{\delta \widehat{S}_{YM}}{\delta A_{\mu }^{a}}  + \p_\mu \overline{c}^a \p_\mu c^a + K_{\mu }^{a}\partial _{\mu }c^{a}  + M_\mu^{a i} \p_\mu \varphi_\mu^{ai}\nonumber\\
& -  U_\mu^{a i} \p_\mu \omega_\mu^{ai}  + N_\mu^{a i} \p_\mu \overline{\omega}_\mu^{ai} +  V_\mu^{a i} \p_\mu \overline{\varphi}_\mu^{ai}  +  \p_\mu \overline{\varphi}^{a i} \p_\mu \varphi_\mu^{ai} +  \p_\mu \omega^{a i} \p_\mu \overline{\omega}_\mu^{ai} + V_\mu^{a i} M_\mu^{a i} - U_\mu^{a i}N_\mu^{a i}\nonumber\\
&  - g f_{abc} U_\mu^{ia} \varphi^{bi} \p_\mu c^c - g f_{abc} V_\mu^{ia} \overline{\omega}^{bi} \p_\mu c^c - g f_{abc} \p_{\mu} \overline{\omega}^a \varphi^{bi}  \p_\mu c^c  - g f_{abc} R_\mu^{ai} \p_\mu c^b \overline \omega_i^c - g f_{abc} T_\mu^{ai} \p_\mu c^b \overline \varphi_i^c \Bigr)  \nonumber\\
&+ b_{1}\int \d^dx q \Bigl(  A_{\mu}^{a}\frac{ \delta S_{YM}}{\delta A_{\mu }^{a}}  + \p_\mu \overline{c}^a \p_\mu c^a + K_{\mu }^{a}\partial _{\mu }c^{a}  + M_\mu^{a i} \p_\mu \varphi_\mu^{ai} -  U_\mu^{a i} \p_\mu \omega_\mu^{ai} +  N_\mu^{a i} \p_\mu \overline{\omega}_\mu^{ai} +  V_\mu^{a i} \p_\mu \overline{\varphi}_\mu^{ai} \nonumber\\
&+  \p_\mu \overline{\varphi}^{a i} \p_\mu \varphi_\mu^{ai} +  \p_\mu \omega^{a i} \p_\mu \overline{\omega}_\mu^{ai} + V_\mu^{a i} M_\mu^{a i} - U_\mu^{a i}N_\mu^{a i} - g f_{abc} U_\mu^{ia} \varphi^{bi} \p_\mu c^c - g f_{abc} V_\mu^{ia} \overline{\omega}^{bi} \p_\mu c^c \nonumber\\
&- g f_{abc} \p_{\mu} \overline{\omega}^a \varphi^{bi}  \p_\mu c^c  - g f_{abc} R_\mu^{ai} \p_\mu c^b \overline \omega_i^c - g f_{abc} T_\mu^{ai} \p_\mu c^b \overline \varphi_i^c \Bigr)+ a_{1}\int \d^dx \eta \Bigl(  \p_\mu \overline{c}^a \p_\mu c^a  + M_\mu^{a i} \p_\mu \varphi_\mu^{ai} \nonumber\\
& -  U_\mu^{a i} \p_\mu \omega_\mu^{ai} +  N_\mu^{a i} \p_\mu \overline{\omega}_\mu^{ai} +  V_\mu^{a i} \p_\mu \overline{\varphi}_\mu^{ai} +  \p_\mu \overline{\varphi}^{a i} \p_\mu \varphi_\mu^{ai} +  \p_\mu \omega^{a i} \p_\mu \overline{\omega}_\mu^{ai} + V_\mu^{a i} M_\mu^{a i} - U_\mu^{a i}N_\mu^{a i}\nonumber\\
& - g f_{abc} U_\mu^{ia} \varphi^{bi} \p_\mu c^c - g f_{abc} V_\mu^{ia} \overline{\omega}^{bi} \p_\mu c^c - g f_{abc} \p_{\mu} \overline{\omega}^a \varphi^{bi}  \p_\mu c^c - g f_{abc} R_\mu^{ai} \p_\mu c^b \overline \omega_i^c - g f_{abc} T_\mu^{ai} \p_\mu c^b \overline \varphi_i^c  \Bigr) \nonumber\\
& -  a_{1}\int \d^dx \lambda \Bigl( U_\mu^{a i} \p_\mu \varphi^{a i} +  V_\mu^{a i} \p_\mu \overline \omega^{a i} + \p_\mu \overline \omega^{ai} \p_\mu \varphi^{ai} + U_\mu^{a i} V_\mu^{a i}  - gf_{abc} T^a \overline \omega^b \p_\mu c^c  \Bigr) \nonumber\\
&- a_{1}\int \d^dx  \Bigl( X^i \p_\mu \overline \omega^{ai} \p_\mu c^a  \Bigr)\,.
\end{align}}

\noindent Only now, we can discard the term $\sim q K_{\mu }^{a}\partial_{\mu }c^{a}$ as it is of quadratic order in the sources. One could argue that we can also neglect terms of higher order in $U_\mu^{ai}$, $N_\mu^{ai}$ and $T_\mu^{ai}$. However, all these sources belong to a BRST doublet. Moreover, the corresponding partner sources, $M_\mu^{ai}, V_\mu^{ai}, R_\mu^{ai}$, acquire a nonzero value in the physical limit, and it would be impossible to write the BRST exact term in our starting action $\Sigma_\glue$ (see expression \eqref{glueballaction}) as an $s$-variation when neglecting these kind of terms. In summary, the expression
{\allowdisplaybreaks\begin{align}\label{countertermfinal}
&\Sigma^c= a_{0}S_{YM} +  b_0 \widehat S_{YM} + a_{1}\int \d^dx\Bigl(  A_{\mu}^{a}\frac{ \delta S_{YM}}{\delta A_{\mu }^{a}}+ A_{\mu}^{a}\frac{\delta \widehat{S}_{YM}}{\delta A_{\mu }^{a}}  + \p_\mu \overline{c}^a \p_\mu c^a + K_{\mu }^{a}\partial _{\mu }c^{a}  + M_\mu^{a i} \p_\mu \varphi_\mu^{ai}\nonumber\\
& -  U_\mu^{a i} \p_\mu \omega_\mu^{ai}  + N_\mu^{a i} \p_\mu \overline{\omega}_\mu^{ai} +  V_\mu^{a i} \p_\mu \overline{\varphi}_\mu^{ai}  +  \p_\mu \overline{\varphi}^{a i} \p_\mu \varphi_\mu^{ai} +  \p_\mu \omega^{a i} \p_\mu \overline{\omega}_\mu^{ai} + V_\mu^{a i} M_\mu^{a i} - U_\mu^{a i}N_\mu^{a i}\nonumber\\
&  - g f_{abc} U_\mu^{ia} \varphi^{bi} \p_\mu c^c - g f_{abc} V_\mu^{ia} \overline{\omega}^{bi} \p_\mu c^c - g f_{abc} \p_{\mu} \overline{\omega}^a \varphi^{bi}  \p_\mu c^c  - g f_{abc} R_\mu^{ai} \p_\mu c^b \overline \omega_i^c - g f_{abc} T_\mu^{ai} \p_\mu c^b \overline \varphi_i^c \Bigr)  \nonumber\\
&+ b_{1}\int \d^dx q \Bigl(  A_{\mu}^{a}\frac{ \delta S_{YM}}{\delta A_{\mu }^{a}}  + \p_\mu \overline{c}^a \p_\mu c^a   + M_\mu^{a i} \p_\mu \varphi_\mu^{ai} -  U_\mu^{a i} \p_\mu \omega_\mu^{ai} +  N_\mu^{a i} \p_\mu \overline{\omega}_\mu^{ai} +  V_\mu^{a i} \p_\mu \overline{\varphi}_\mu^{ai} \nonumber\\
&+  \p_\mu \overline{\varphi}^{a i} \p_\mu \varphi_\mu^{ai} +  \p_\mu \omega^{a i} \p_\mu \overline{\omega}_\mu^{ai} + V_\mu^{a i} M_\mu^{a i} - U_\mu^{a i}N_\mu^{a i} - g f_{abc} U_\mu^{ia} \varphi^{bi} \p_\mu c^c - g f_{abc} V_\mu^{ia} \overline{\omega}^{bi} \p_\mu c^c \nonumber\\
&- g f_{abc} \p_{\mu} \overline{\omega}^a \varphi^{bi}  \p_\mu c^c  - g f_{abc} R_\mu^{ai} \p_\mu c^b \overline \omega_i^c - g f_{abc} T_\mu^{ai} \p_\mu c^b \overline \varphi_i^c \Bigr)+ a_{1}\int \d^dx \eta \Bigl(  \p_\mu \overline{c}^a \p_\mu c^a  + M_\mu^{a i} \p_\mu \varphi_\mu^{ai} \nonumber\\
& -  U_\mu^{a i} \p_\mu \omega_\mu^{ai} +  N_\mu^{a i} \p_\mu \overline{\omega}_\mu^{ai} +  V_\mu^{a i} \p_\mu \overline{\varphi}_\mu^{ai} +  \p_\mu \overline{\varphi}^{a i} \p_\mu \varphi_\mu^{ai} +  \p_\mu \omega^{a i} \p_\mu \overline{\omega}_\mu^{ai} + V_\mu^{a i} M_\mu^{a i} - U_\mu^{a i}N_\mu^{a i}\nonumber\\
& - g f_{abc} U_\mu^{ia} \varphi^{bi} \p_\mu c^c - g f_{abc} V_\mu^{ia} \overline{\omega}^{bi} \p_\mu c^c - g f_{abc} \p_{\mu} \overline{\omega}^a \varphi^{bi}  \p_\mu c^c - g f_{abc} R_\mu^{ai} \p_\mu c^b \overline \omega_i^c - g f_{abc} T_\mu^{ai} \p_\mu c^b \overline \varphi_i^c  \Bigr) \nonumber\\
& -  a_{1}\int \d^dx \lambda \Bigl( U_\mu^{a i} \p_\mu \varphi^{a i} +  V_\mu^{a i} \p_\mu \overline \omega^{a i} + \p_\mu \overline \omega^{ai} \p_\mu \varphi^{ai} + U_\mu^{a i} V_\mu^{a i}  - gf_{abc} T^a \overline \omega^b \p_\mu c^c  \Bigr) \nonumber\\
&- a_{1}\int \d^dx  \Bigl( X^i \p_\mu \overline \omega^{ai} \p_\mu c^a  \Bigr)\,.
\end{align}}

\noindent gives the general counterterm compatible with all Ward identities. \\
\\
We still need to introduce the operators belonging to the class $C_3$, which are related to the equations of motion, see section \ref{sectieB}. Therefore, we again perform a linear shift on the gluon field $A^a_\mu$ in the action $\Sigma$
\begin{eqnarray}
A^a_\mu \rightarrow A^a_\mu + \alpha A^a_\mu\,,
\end{eqnarray}
whereby $\alpha$ is a dimensionless new source. As this shift corresponds to a redefinition of the gluon field it has to be consistently done in the starting action as well as in the counterterm. Later on, we shall see that introducing the relevant gluon equation of motion operator through this shift, will allow us to uncover the finiteness of this kind of operator. Performing the shift in the classical action yields the following shifted action $\Sigma'$
{\allowdisplaybreaks\begin{align}\label{eindactie5}
& \Sigma' = S_{\mathrm{YM}} + \int \d^d x\,\left( b^{a}\partial_\mu A_\mu^{a}+\overline{c}^{a}\partial _{\mu } D_{\mu}^{ab}c^b \right) +\int \d^dx \left( -K_{\mu }^{a}\left( D_{\mu }c\right) ^{a}+\frac{1}{2}gL^{a}f^{abc}c^{b}c^{c}\right)   \nonumber\\
&+\int \d^d x\left( \overline{\varphi }_i^a \partial_{\nu}  D_\nu^{ab} \varphi_i^b - \overline{\omega}_{i}^{a}\partial_\nu D_\nu^{ab} \omega_i^b  -g \partial_\nu \overline \omega_i^a f^{abm}  D_\nu^{bd} c^d \varphi_i^m \right)+ \int \d^dx \Bigl( -M_{\mu }^{ai} D_\mu^{ab} \varphi_i^b\nonumber\\
&  - g U_\mu^{ai} f^{abc}  D_\mu^{bd} c^d \varphi_i^c+ U_\mu^{ai}  D_\mu^{ab} \omega_i^b  - N_\mu^{ai} D_\mu^{ab} \overline \omega_i^b  - V_\mu^{ai} D_\mu^{ab} \overline{\varphi}_i^b + g V_\mu^{ai} f^{abc} D_\mu^{bd} c^d \overline \omega_i^c - M_{\mu }^{ai}V_{\mu }^{ai}\nonumber\\
&  +U_{\mu }^{ai}N_{\mu }^{ai}  + R_\mu^{ai} g f^{abc} D_\mu^{bd} c^d \overline \omega^c_i  + T_\mu^{ai} g f_{abc} D^{bd}_\mu c^d \overline \varphi^c_i\Bigr)+ \int \d^d x  q F_{\mu\nu}^a F_{\mu\nu}^a +   \int \d^d x \lambda \Bigl[  \p_\mu \overline c^a A_\mu^a\nonumber\\
& + \p \overline \omega \p \varphi + g f_{akb} \p \overline \omega^a A^k \varphi^b + U^a D^{ab} \varphi^b + V^a D^{ab}\overline \omega^b + UV - T_\mu^{a i} g f_{abc} D^{bd}_\mu c^d \overline \omega^c_i \Bigr]  \nonumber\\
&+\int \d^d x  \eta \Bigl[\p_\mu b^a A_\mu^a + \p_\mu \overline c^a D^{ab}_{\mu} c^{b} + \p \overline \varphi \p \varphi - \p \overline \omega \p \omega + g f_{akb} \p \overline \varphi^a A^k \varphi^b  + g f_{akb} \p \overline \omega^a D^{kd} c^d \varphi^b  \nonumber\\
&-  g f_{akb} \p \overline \omega^a A^k \omega^b+M_{\mu }^{ai}\left( D_{\mu }\varphi_{i}\right) ^{a}+gU_{\mu }^{ai}f^{abc}\left( D_{\mu }c\right)^{b}\varphi _{i}^{c}-U_{\mu }^{ai}\left( D_{\mu }\omega _{i}\right)^{a}  + N_{\mu }^{ai}\left( D_{\mu }\overline{\omega }_{i}\right)^{a}\nonumber\\
&-gV_{\mu}^{ai}f^{abc}\left(D_{\mu }c\right)^{b}\overline{\omega}_{i}^{c} + V_{\mu }^{ai}\left( D_{\mu }\overline{\varphi}_{i}\right)^{a}+M_{\mu }^{ai}V_{\mu }^{ai}-U_{\mu }^{ai}N_{\mu }^{ai}  - R_\mu^{ai} g f^{abc} D_\mu^{bd} c^d \overline \omega^c_i \nonumber\\
 &- T_\mu^{ai} g f_{abc} D^{bd}_\mu c^d \overline \varphi^c_i\Bigr] + \int \d^d x \left( Y_i  A_\mu^a \p \overline \omega^a_i - X_i D^{ab}_\mu c^b \p_\mu \overline \omega^a_i + X_i A^a_\mu \p_\mu \overline \varphi^a_i\right) \nonumber\\
&+  \int \d^d x \alpha   A_\mu^a\frac{\delta S_{YM}}{\delta A_\mu^a}  +  \int \d^d x  \alpha  \left[  - \p_\mu b^a A_\mu^a  + g f_{akb} A_\mu^k c^b \p_\mu \overline c^a  \right] + \int \d^d x \alpha \Bigl[ - g f_{akb} \p_\mu \overline \varphi^a_i A^k_\mu \varphi^b \nonumber\\
& +  g f_{akb} \p_\mu \overline \omega^a_i A^k_\mu \omega^b - g^2 f_{abm} f_{bkd} \p_\mu \overline \omega^a \varphi^m A^k_\mu c^d\Bigr]+ \int \d^d x \alpha \Bigl[ - g f_{akb} M^a_i A^k_\mu \varphi^b_i  + g f_{akb} U^a_i A^k_\mu \omega^b_i  \nonumber\\
& - g f_{akb} N^a_i A^k_\mu \overline \omega^b_i  - g f_{akb} V^a_i A^k_\mu \overline \varphi^b_i\Bigr] - \int \d^d x \alpha \Bigl[ g^2 f_{abc} f_{bkd} U^a_i \varphi^c A^k c^d + g^2 f_{abc} f_{bkd} V^a \overline \omega^c A^k c^d\nonumber\\
&  + g^2 f_{abc} f_{bkd} R^a \overline \omega^c A^k c^d -g^2 f_{abc} f_{bkd} T^a \overline \varphi^c A^k c^d\Bigr]\,.
\end{align}}

\noindent Notice that we have neglected again higher order terms in the sources $\sim( \alpha \eta \ldots)$, $\sim ( \alpha \lambda \ldots)$ and $\sim ( \alpha q \ldots)$ as the argument
\eqref{argument} is still valid. The corresponding counterterm $\Sigma^{\prime c}$ reads:
\begin{multline}
\Sigma^{\prime c} = \Sigma^c +  a_{0}\int \d^dx\Bigl( \alpha A_{\mu}^{a}\frac{\delta S_{YM}}{\delta A_{\mu }^{a}} \Bigr)+ a_1 \int \d^d x \alpha \left( 2 A_\mu^a \p_\mu \p_\nu A_\nu^a - 2 A_\mu^a \p^2 A_\mu^a \right.\\
 \left. + 9 g f_{abc} A_\mu^a A_\nu^b \p_\mu A_\nu^c + 4 g^2 f_{abc} f_{cde} A_\mu^a A_\nu^b A_\mu^d A_\nu^e\right)\,,
\end{multline}
whereby $\Sigma^c$ is given in expression \eqref{countertermfinal} and we have once more dropped higher order terms in the sources.\\
\\
The final step in the renormalization procedure is to reabsorb the counterterm $\Sigma^{\prime c}$ into the original action $\Sigma'$,
\begin{equation}
\Sigma (g,\omega ,\phi ,\Phi )+h \Sigma^c = \Sigma (g_{0},\omega_{0},\phi _{0},\Phi_{0})+O(h^{2})\,.
\end{equation}
We set $\phi =(A_{\mu }^{a}$, $c^{a}$, $\overline{c}^{a}$, $b^{a}$, $\varphi_{i}^{a}$, $\omega_{i}^{a}$, $\overline{\varphi}_{i}^{a}$, $\overline{\omega}_{i}^{a})$ and $\Phi=(K^{a\mu }$, $L^{a}$, $M_{\mu }^{ai}$, $N_{\mu }^{ai}$, $V_{\mu}^{ai}$, $U_{\mu }^{ai}$, $\lambda$) and we define
\begin{align}
g_{0}&=Z_{g}g\,, &  \phi _{0} &=Z_{\phi}^{1/2}\phi \,, & \Phi _{0} &=Z_{\Phi }\Phi \,,
\end{align}
while for the other sources we propose the following mixing matrix
\begin{equation}
 \left(
  \begin{array}{c}
    q_0 \\
    \eta_0 \\
    \alpha_0
  \end{array}
\right)=\left(
          \begin{array}{ccc}
            Z_{q      q} & Z_{q      \eta}  & Z_{q      \alpha} \\
            Z_{\eta q} & Z_{\eta \eta}  & Z_{\eta \alpha} \\
            Z_{\alpha      q} & Z_{\alpha      \eta}  & Z_{\alpha      \alpha}
          \end{array}
        \right)
\left(
  \begin{array}{c}
    q \\
    \eta \\
    \alpha
  \end{array}
\right)\,.
\end{equation}
If we try to absorb the counterterm into the original action, we obviously find back all the renormalization factors of the original GZ action, see expression \eqref{Z1}- \eqref{Z3}. In addition, we also find the following mixing matrix
\begin{eqnarray}\label{5mix}
\left(
          \begin{array}{ccc}
            Z_{q      q} & Z_{q      \eta}  & Z_{q      \alpha} \\
            Z_{\eta q} & Z_{\eta \eta}  & Z_{\eta \alpha} \\
            Z_{\alpha      q} & Z_{\alpha      \eta}  & Z_{\alpha      \alpha}
          \end{array}
        \right) &=& \left(
          \begin{array}{ccc}
            1 + h (b_0 - a_0) & 0  & 0 \\
            h b_1 & 1  & 0 \\
            h b_1 & 0  & 1
          \end{array}
        \right)\,,
\end{eqnarray}
while for the $Z$-factor of $\lambda$ we have
\begin{eqnarray}
Z_{\lambda} &=& Z_{c}^{-1/2} Z_A^{-1/2} = Z_g^{1/2} Z_A^{-1/4} \,.
\end{eqnarray}
Both results are remarkable the same as \eqref{mixingmatrix} and \eqref{5lambda}. So far, we have proven that the two limit cases are at least correct. Finally, we find the new results
\begin{eqnarray}
Z_{Y} &=& Z_g Z_A^{-1/2} \,, \nonumber\\
Z_{X} &=& Z_g^{1/2} Z_A^{-1/4}\,.
\end{eqnarray}
In summary, the action $\Sigma'$ is renormalizable. Moreover, we have only 4 arbitrary parameters, $a_0$, $a_1$, $b_0$, $b_1$, which is the same number as in the limit case $\{\varphi, \overline \varphi, \omega, \overline \omega, U, V, N, M \} \to 0$, i.e.~the Yang-Mills case with the introduction of the glueball operator $\sim F^2_{\mu\nu}$, see the previous section. This is also a remarkable fact.

\subsection{Renormalization of the RGZ action with inclusion of the operator $F_{\mu\nu}^2$}
In analogy with chapter \ref{refined} we shall add the two dimensional mass term $ \sim \left( \overline{\varphi}^a_i \varphi^a_{i} - \overline{\omega}^a_i \omega^a_i \right)$ to the action $\Sigma_\glue$ in equation \eqref{glueballaction},
\begin{eqnarray}
\Sigma_\Rglue &=& \Sigma_\glue + \Sigma_{\overline{\varphi} \varphi} + \Sigma_{\en}\,,
\end{eqnarray}
whereby we recall that
\begin{eqnarray}\label{defex}
\Sigma_{\overline{\varphi} \varphi} &=& \int \d^d x \left( s(-J \overline{\omega}^a_i \varphi^a_{i})\right) =\int \d^d x\left( -J\left( \overline{\varphi}^a_i \varphi^a_{i} - \overline{\omega}^a_i \omega^a_i \right) \right) \,, \nonumber\\
\Sigma_{\en} &=&  \int \d^d x \varsigma \Theta J \,,
\end{eqnarray}
see equations\footnote{We are not including the condensate $\braket{A^2}$ here.} \eqref{nact} and \eqref{4sigma_en}-\eqref{4sigma_en2}.\\
\\
Let us now investigate the renormalizability of action $\Sigma_\Rglue$. We can go through the same steps as in the previous section. Therefore, we again add the two external pieces, $S_{\ext,1}$ and $S_{\ext,2}$ as defined in equation \eqref{Sext1} and \eqref{Sext2}, to the action $\Sigma_\Rglue$
\begin{eqnarray}
\Sigma_\R &=& \Sigma_\Rglue + S_{\ext,1} + S_{\ext,2}\,.
\end{eqnarray}
Subsequently, one can easily  check that all Ward identities \eqref{wardid1} - \eqref{wardander} and \eqref{wardideinde} remain unchanged up to potential harmless linear breaking terms. Therefore, the constraints \eqref{cont1} - \eqref{cont2} and \eqref{cont3} remain valid. Unfortunately, the extra integrated Ward identity \eqref{broken1} and the integrated Ward identity \eqref{broken2} are broken due to the introduction of the mass term. However, the mass term we have added is not a new interaction as it is only quadratic in the fields. Therefore, it cannot introduce new divergences into the massless theory $\Sigma$, and it can only influence its own renormalization\footnote{We employ massless renormalization schemes.} as well as  potentially vacuum terms, i.e.~pure source terms. Also, next to the Ward identities \eqref{wardid1} - \eqref{wardander} and \eqref{wardideinde}, we have a new identity
\begin{eqnarray}
    \frac{\delta \Sigma_\R }{\delta \Theta} &=&  \varsigma  J\,,
\end{eqnarray}
which is translated to the following constraint  at the level of the counterterm,
\begin{eqnarray}
    \frac{\delta \Sigma^{c}_\R }{\delta \Theta} &=& 0\,.
\end{eqnarray}
 As a consequence, $\Sigma^{c}_\R$ is independent from the source $\Theta$. Therefore, it follows that the form of the counterterm $\Sigma^c_\R$ can be written as
\begin{eqnarray}
    \Sigma^{c}_\R &=& \Sigma^c + \Sigma^c_J\,,
\end{eqnarray}
whereby $\Sigma^c$ is the counterterm \eqref{countertermfinal} of $\Sigma$ and $\Sigma^c_J$ is depending on $J$. One can now easily check that $\Sigma^c_J = \kappa J^2$, with $\kappa$ a new parameter as this is the only possible combination with the source $J$, which does not break the constraints \eqref{ST} - \eqref{cont2} and \eqref{cont3}.\\
 \\
$\kappa$ is in fact a redundant parameter, as no divergences in $J^2$ will occur, as explained on page \pageref{redundant}. Therefore, the counterterm $\Sigma^c_\R$ is actually equal to $\Sigma^c$. Defining
\begin{align}
   J_{0} &= Z_{J} J \,,
\end{align}
we find
\begin{align}\label{5ZJ}
Z_J &= Z_{\varphi}^{-1} = Z_g Z_A^{1/2}\,,
\end{align}
and we have proven the renormalizability of the action $\Sigma_\Rglue$.

\subsection{The operator mixing matrix to all orders}
\subsubsection{Preliminaries}
We can write the final action $\Sigma'$ from equation \eqref{eindactie5} in a more condensed form as
\begin{eqnarray}\label{eindactieproper}
 \Sigma' &=& \Sigma_\GZ + S_{\ext,1} + S_{\ext,2} + \int \d^d x  \left( q \mathcal F + \eta \mathcal E + \alpha \mathcal H \right)  + \int \d^d x \lambda \mathcal N\,,
\end{eqnarray}
whereby we have defined the operators
\begin{eqnarray}
\mathcal F &=& \frac{1}{4} F^a_{\mu\nu} F^a_{\mu\nu}\,,  \nonumber\\
\mathcal E &=& s \mathcal N\,, \nonumber\\
\mathcal H &=& A_\mu^a\frac{ \Sigma_\GZ}{A_\mu^a}\,,
\end{eqnarray}
with
\begin{equation}
\mathcal N = \left[  \p_\mu \overline c^a A_\mu^a + \p \overline \omega \p \varphi + g f_{akb} \p \overline \omega^a A^k \varphi^b + U^a D^{ab} \varphi^b + V^a D^{ab}\overline \omega^b + UV - T_\mu^{a i} g f_{abc} D^{bd}_\mu c^d \overline \omega^c_i\right]\,.
\end{equation}
Let us return to the mixing matrix of the sources $q$, $\eta$ and $J$ and pass to the corresponding operators. We have found in expression \eqref{5mix} that
\begin{equation}
 \left(
  \begin{array}{c}
    q_0 \\
    \eta_0 \\
    J_0
  \end{array}
\right)=\left(
          \begin{array}{ccc}
            Z_{q      q} & 0  & 0 \\
            Z_{J q} & 1  & 0 \\
            Z_{J q} & 0  & 1
          \end{array}
        \right)
\left(
  \begin{array}{c}
    q \\
    \eta \\
    J
  \end{array}
\right)\,.
\end{equation}
As this matrix is exactly the same as in the Yang-Mills case \eqref{mixingmatrix}, we can immediately write down the corresponding mixing matrix for the operators themselves, see expression \eqref{operatormatrix},
\begin{eqnarray}\label{operatormatrix2}
 \left(
  \begin{array}{c}
    \mathcal F_0 \\
    \mathcal E_0 \\
    \mathcal H_0
  \end{array}
\right) &= & \left(
          \begin{array}{ccc}
            Z_{qq}^{-1}& -Z_{Jq}Z_{qq}^{-1}  &-Z_{Jq}Z_{qq}^{-1} \\
            0  &1 & 0   \\
           0 & 0& 1          \end{array}
        \right)
        \left(
\begin{array}{c}
    \mathcal F \\
    \mathcal E \\
    \mathcal H
  \end{array}
\right)\,.
\end{eqnarray}
We recall that we recover the expected upper triangular form. Also in the GZ case, $\mathcal E$ has a $Z$-factor equal to $1$, and we also find that the BRST exact operator $\mathcal E$ does not mix with $\mathcal H$, although this mixing would in principle be allowed. This can be understood as follows. The integrated BRST exact operator $\mathcal E$ is in fact proportional to a sum of four (integrated) equations of motion terms and two other terms,
\begin{multline}\label{countcon}
\int \d^4 x \Bigl[ \p_\mu b^a A_\mu^a + \p_\mu \overline c^a D^{ab}_{\mu} c^b + \p \overline \varphi \p \varphi - \p \overline \omega \p \omega + g f_{akb} \p \overline \varphi^a A^k \varphi^b + g f_{akb} \p \overline \omega^a D^{kd} c^d \varphi^b \\
-  g f_{akb} \p \overline \omega^a A^k \omega^b +M_\mu^{ai} D_\mu^{ab} \varphi_i^b + g U_\mu^{ai} f^{abc}  D_\mu^{ab}c^b \varphi_i^c - U_\mu^{ai}  D_\mu^{ab}\omega_i^b  + N_\mu^{ai} D_\mu^{ab} \overline \omega_i^b \\
\hspace{3.5cm}-gV_\mu^{ai} f^{abc} D_\mu^{bd}c^d \overline \omega_i^c + V_\mu^{ai}  D_\mu^{ab} \overline \varphi_i^b +M_\mu^{ai} V_\mu^{ai}-U_\mu^{ai} N_\mu^{ai}\Bigr] \\
= -\int \d^4 x\left( b^a\frac{\delta \Sigma_\GZ }{\delta b^a}+ \overline c^a \frac{\delta \Sigma_\GZ}{\delta \overline c^a} +  \overline \varphi^a\frac{\delta \Sigma_\GZ }{\delta \overline \varphi^a} +  \overline \omega^a\frac{\delta \Sigma_\GZ }{\delta \overline \omega^a} +  M^{ai}_\mu\frac{\delta \Sigma_\GZ }{\delta M^{ai}_\mu} +  U^{ai}_\mu\frac{\delta \Sigma_\GZ }{\delta U^{ai}_\mu} \right)   \,,
\end{multline}
and therefore, like $\mathcal H$, it does not mix with the other operators. Notice that we can rewrite the integrated BRST operator in two other forms:
\begin{equation}\label{countcon2}
\eqref{countcon} =  -\int \d^4 x\left( b^a\frac{\delta \Sigma_\GZ }{\delta b^a}+ \overline c^a \frac{\delta \Sigma_\GZ}{\delta \overline c^a} +   \varphi^a\frac{\delta \Sigma_\GZ }{\delta \varphi^a} + \omega^a\frac{\delta \Sigma_\GZ }{\delta \omega^a} +  N^{ai}_\mu\frac{\delta \Sigma_\GZ }{\delta N^{ai}_\mu} +  V^{ai}_\mu\frac{\delta \Sigma_\GZ }{\delta V^{ai}_\mu} \right)   \,,
\end{equation}
or
\begin{equation}\label{countcon3}
\eqref{countcon}=  -\int \d^4 x\left( b^a\frac{\delta \Sigma_\GZ }{\delta b^a}+ c^a \frac{\delta \Sigma_\GZ}{\delta c^a} +   \overline \varphi^a\frac{\delta \Sigma_\GZ }{\delta\overline \varphi^a} + \omega^a\frac{\delta \Sigma_\GZ }{\delta \omega^a} +  M^{ai}_\mu\frac{\delta \Sigma_\GZ }{\delta M^{ai}_\mu} +  N^{ai}_\mu\frac{\delta \Sigma_\GZ }{\delta N^{ai}_\mu} \right)   \,.
\end{equation}

\noindent \textbf{Remark}\\
We can also use the refined action $\Sigma_\RGZ$ instead of $\Sigma_\GZ$. We define $\Sigma_\RGZ$ as
\begin{eqnarray}\label{RGZSigma}
 \Sigma_\RGZ &=& \Sigma_\GZ + \Sigma_{\overline \varphi \varphi}+  \Sigma_{\en}\,,
\end{eqnarray}
whereby $\Sigma_{\overline \varphi \varphi}$ and $\Sigma_{\en}$
are defined in equation \eqref{defex}. Replacing $\Sigma_\GZ$ by $\Sigma_\RGZ$ does not alter equation \eqref{operatormatrix2}, but it does slightly modify expression \eqref{countcon},
\begin{multline}\label{countconR}
\int \d^4 x \mathcal E =  -\int \d^4 x\Bigl( b^a\frac{\delta \Sigma_\RGZ }{\delta b^a}+ \overline c^a \frac{\delta \Sigma_\RGZ}{\delta \overline c^a} +  \overline \varphi^a\frac{\delta \Sigma_\RGZ }{\delta \overline \varphi^a} +  \overline \omega^a\frac{\delta \Sigma_\RGZ }{\delta \overline \omega^a} +  M^{ai}_\mu\frac{\delta \Sigma_\RGZ }{\delta M^{ai}_\mu} \\
+  U^{ai}_\mu\frac{\delta \Sigma_\RGZ }{\delta U^{ai}_\mu} - J \frac{\delta \Sigma_\RGZ }{\delta J}  + \Theta \frac{\delta \Sigma_\RGZ }{\delta \Theta }\Bigr)   \,,
\end{multline}
and analogously for expression \eqref{countcon2} and \eqref{countcon3}.

\subsubsection{The physical limit}
In the next subsection, we shall work in the physical limit as our final intention is to examine $n$-point functions with the (Refined) Gribov-Zwanziger action itself. In the physical limit, $\mathcal E$ becomes:
\begin{multline}\label{mult}
\left.\mathcal E\right|_\phys =  \p_\mu b^a  A_\mu^a + \p_\mu  \overline c^a D_\mu^{ab} c^b +  \p_\mu \overline \varphi_i^a  D_\mu^{ab} \varphi^b_i  - \p_\mu \overline \omega_i^a  D_\mu^{ab} \omega_i^b  + g f^{abc} \p_\mu \overline \omega_i^a    D_\mu^{bd} c^d  \varphi_i^c  \\
+ \gamma ^{2} g  f^{abc}A_\mu^a \varphi_\mu^{bc} +  \gamma^2 g f^{abc} A_\mu^a \overline \varphi_\mu^{bc} + d \left(N^{2}-1\right)  \gamma^4 \,.
\end{multline}
From this point, we can omit the constant term $d \left(N^{2}-1\right) \gamma^4$ as it shall not play a role in the calculation of the glueball correlator. Later, we shall determine the renormalization group invariant $\mathcal R(x)$ which contains $F^2_{\mu\nu}(x)$. As $\mathcal E$ mixes with $F^2_{\mu\nu}(x)$, this renormalization group invariant shall also contain this constant term. However, a constant term can never contribute to the final glueball correlator $\braket{\mathcal R(x) \mathcal R (y)}$ as it can never help to produce connected diagrams between the two space time points $x$ and $y$. Therefore, we shall simplify the calculations by omitting this term already from this point.\\
\\
In the physical limit $\mathcal H$ is given by
\begin{eqnarray}\label{H}
\left. \mathcal H\right|_\phys &=& A_\mu^a \frac{\delta S_\GZ}{ \delta A_\mu^a} \,,
\end{eqnarray}
whereby $S_\GZ$ is the physical Gribov-Zwanziger action \eqref{SGZphys}. Naturally, the mixing matrix \eqref{operatormatrix2} stays valid.

\subsubsection{The operator mixing matrix to all orders}
It this section, we shall determine the mixing matrix \eqref{operatormatrix2} to all orders. This proof is very elegant as it does not require to calculate any loop diagrams, and it is purely based on algebraic manipulations. We shall extend the proof given in section \ref{5sectmixingtoallorders}, which was based on \cite{Brown:1979pq}. Moreover, as a byproduct, the proof shall also reveal some identities between the anomalous dimensions of the different fields, which can serve as a check on relations as in \eqref{Z2} and \eqref{Z3}. We shall directly work with the physical action $S_\GZ$. In the end, we shall also look at the Refined Gribov-Zwanziger action, $S_\RGZ$. \\
\\
We start again our analysis with the following generic $n$-points function
\begin{eqnarray}\label{definitie5}
\mathcal G^n (x_1, \ldots, x_n ) &=& \Braket{\phi_i (x_1)\ldots \phi_j (x_n)  }~=~  \int [\d \phi] \phi_i (x_1)\ldots \phi_j (x_n) \e^{- S_\GZ} \,,
\end{eqnarray}
whereby $\phi_i$, $i = 1\ldots 8$ stands for one of the eight fields $(A_{\mu }^{a}$, $c^{a}$, $\overline{c}^{a}$, $b^{a}$, $\varphi_{\mu}^{ab}$, $\omega_{\mu}^{ab}$,
$\overline{\varphi}_{\mu}^{ab}$, $\overline{\omega}_{i}^{a})$, i.e~$\phi_1 = A_\mu$, $\ldots$, $\phi_8 =\overline{\omega}^{ab}_{\mu}$. We shall immediately omit the vacuum term $ \gamma^4 (N^2-1) d$ in the action $S_\GZ$, as it is relevant only for the calculation of the vacuum energy and not for the calculation of $n$-point functions. The total number of fields is given by $n$,
\begin{eqnarray}
n &=& \sum_i^8 n_i\,,
\end{eqnarray}
with $n_i$ the number of fields $\phi_i$ present in the $n$-points function \eqref{definitie5}. We are therefore considering the path integral for a random combination of fields. Subsequently, from the definition \eqref{definitie5}, we can immediately write down the connection between the renormalized Green function and the bare Green function, which is, in a very condensed notation,
\begin{eqnarray}\label{connectie5}
\mathcal G^{n} = \prod_{i=1}^8 Z_{\phi_i}^{-n_i/2}  \mathcal G_0^{n}\,.
\end{eqnarray}
From the previous equation, we shall be able to fix all the matrix elements of expression \eqref{operatormatrix2}, based on the knowledge that $\frac{\d \mathcal G^{n} }{\d g^2}$ must be finite in a renormalized theory.\\
\\
We shall therefore calculate this quantity. The first step is to apply the chain rule:
\begin{equation}\label{steponeone}
\frac{\d \mathcal G^{n} }{\d g^2} = \sum_{j=1}^8 \left( \frac{\p  Z_{\phi_j}^{-n_j/2} }{ \p g^2} \prod_{i \not= j} Z_{\phi_i}^{-n_i/2} \right)     \mathcal G_0^{n} + \prod_{i=1}^8 Z_{\phi_i}^{-n_i/2} \left[ \frac{\p g_0^2 }{ \p g^2} \frac{\p  }{\p g_0^2} +  \frac{\p \gamma_0^2 }{ \p g^2} \frac{\p  }{\p \gamma_0^2} \right] \mathcal G_0^{n}\,.
\end{equation}
Next, we need to calculate the derivatives w.r.t.~$g^2$.
\begin{itemize}
\item Firstly, we need to find $\p g_0^2  / \p g^2$. We employ dimensional regularization, with $d=4-\varepsilon$. If we derive
\begin{eqnarray} \label{glabel}
g_0^2 &=& \mu^\varepsilon Z_g^2 g^2\,,
\end{eqnarray}
w.r.t.~$\mu$ and $g^2$, combine these two equations and employ the following definition of the $\beta$-funtion\footnote{We have immediately extracted the part in $\varepsilon$.}
\begin{eqnarray}
\mu\frac{\p g^2}{\p \mu} &=& - \varepsilon g^2 + \beta(g^2)\,,
\end{eqnarray}
we obtain
\begin{eqnarray}\label{eq1}
\frac{\p g_0^2 }{ \p g^2} &=& \frac{- \varepsilon g_0^2}{ - \varepsilon g^2 + \beta(g^2)}\,.
\end{eqnarray}
\item Secondly, we calculate $ \frac{\p \gamma_0^2 }{ \p g^2}$. We start from
\begin{eqnarray}
\gamma^2_0 &=& Z_{\gamma^2} \gamma^2 \;,
\end{eqnarray}
whereby $Z_{\gamma^2} = Z_V =Z_M$ due to the limit \eqref{physlimit}. Deriving this equation w.r.t.~$g^2$ yields
\begin{equation*}
\frac{\p \gamma^2_0}{\p g^2} =  \frac{\p Z_{\gamma^2}}{\p g^2}  \gamma^2 = \frac{\p \ln Z_{\gamma^2}}{\p g^2}  \gamma^2_0 =  \frac{1}{\mu} \frac{\p \mu}{\p g^2} \mu  \frac{\p \ln Z_{\gamma^2}}{\p \mu}  \gamma^2_0 =\frac{1}{- \varepsilon g^2 + \beta(g^2)} \delta_{\gamma^2} \gamma_0^2\,,
\end{equation*}
and we have defined the anomalous dimension of $\gamma^2$ as
\begin{eqnarray}
\delta_{\gamma^2} &=& \mu  \frac{\p \ln Z_{\gamma^2}}{\p \mu} \,.
\end{eqnarray}
\item Finally, we search for $\p  Z_{\phi_j}^{-n_j/2}/ \p g^2$. Applying the chain rule gives
\begin{eqnarray}\label{one5}
\frac{\p  Z_{\phi^j}^{-n_j/2} }{ \p g^2} &=& - n_j  \frac{ Z_{\phi^j}^{-n_j/2} }{Z_{\phi^j}^{1/2}} \frac{\p Z_{\phi^j}^{1/2}}{ \p g^2} ~=~ - n_j Z_{\phi^j}^{-n_j/2} \frac{
\p \ln Z_{\phi^j}^{1/2}}{ \p g^2}\,.
\end{eqnarray}
Next, we derive $ \frac{ \p \ln Z_{\phi^i}^{1/2}}{ \p g^2}$ from the definition of the anomalous dimension,
\begin{eqnarray}\label{two5}
\gamma_{\phi^j} &=&\mu \frac{\p \ln Z_{\phi^j}^{1/2}}{\p \mu}  ~=~ \mu \frac{\p g^2 }{\p \mu}  \frac{\p \ln Z_{\phi^j}^{1/2}}{\p g^2} ~=~ \left(  - \varepsilon g^2 + \beta(g^2) \right) \frac{\p \ln Z_{\phi^j}^{1/2}}{\p g^2}\,.
\end{eqnarray}
From expression \eqref{one5} and \eqref{two5}, it now follows
\begin{eqnarray}\label{eq2}
\frac{\p  Z_{\phi^j}^{-n_j/2} }{ \p g^2} &=& - n_j  Z_{\phi^j}^{-n_j/2} \frac{ \gamma_{\phi^j}}{ - \varepsilon g^2 + \beta(g^2)}\,.
\end{eqnarray}
\end{itemize}
Inserting equation \eqref{eq1} and \eqref{eq2} into expression \eqref{steponeone}, we find:
\begin{eqnarray}\label{stepone}
\frac{\d \mathcal G^{n} }{\d g^2} &=& \frac{ \prod_{i} Z_{\phi_i}^{-n_i/2}}{- \varepsilon g^2 + \beta(g^2)} \left( - \sum_{j=1}^8 n_j  \gamma_{\phi^j} - \varepsilon g_0^2  \frac{\p  }{\p g_0^2} + \delta_{\gamma^2} \gamma^2_0 \frac{\p }{\p \gamma^2_0 }  \right)     \mathcal G_0^{n} \,.
\end{eqnarray}
The right hand side still contains bare and therefore divergent quantities. We would like to rewrite all these quantities in terms of finite quantities so that we can use the finiteness of the left hand side to make observations on the right hand side. Also, we should rewrite in some manner the number $n_j$ as the mixing matrix \eqref{operatormatrix2} is obviously independent from these arbitrary numbers.\\
\\
Therefore, as a second step, we shall rewrite the right hand side of \eqref{stepone} in terms of a renormalized quantity. Firstly, we calculate $\frac{\p}{\p g_0^2} \mathcal G_0^{n}$. Using
\begin{multline*}
\frac{\p \e^{-S_\GZ}}{\p g_0^2} ~=~-\int \d^4 y \left( - \frac{1}{g_0^2}  \left( \frac{F_0^2(y)}{4}\right)  + \frac{1}{2g_0^2} \left( A_0(y)\frac{\delta S_\GZ}{\delta A_0(y)}- b_0(y) \frac{\delta S_\GZ}{\delta b_0(y) } \right. \right. \\
\left. \left. +  \overline \omega_0(y) \frac{\delta S_\GZ}{\delta \overline \omega_0 (y)} - \omega_0(y) \frac{\delta S_\GZ}{\delta \omega_0(y) } \right)  \right) \e^{-S_\GZ}\,,
\end{multline*}
we can write,
\begin{multline}\label{vereenv1}
g_0^2 \frac{\d \mathcal G^{n}_0}{ \d g_0} ~=~  \int \d^4 y \left( \mathcal G^n_0 \left\{ \frac{F_0^2(y)}{4} \right\} - \frac{1}{2}  \mathcal G^n_0 \left\{  A_0(y)\frac{\delta S_\GZ}{\delta A_0(y)} \right\} + \frac{1}{2}  \mathcal G^n_0 \left\{  b_0(y) \frac{\delta S_\GZ}{\delta b_0(y) } \right\}  \right. \\
 \left. - \frac{1}{2}  \mathcal G^n_0 \left\{  \overline \omega_0(y) \frac{\delta S_\GZ}{\delta \overline \omega_0 (y)} \right\}+\frac{1}{2}  \mathcal G^n_0 \left\{  \omega_0(y) \frac{\delta S_\GZ}{\delta \omega_0(y)}  \right\} \right)\,.
\end{multline}
We have introduced a shorthand notation for an insertion in the $n$-points function, e.g.
\begin{eqnarray}
\mathcal G^{n}_0 \biggl\{ \frac{F_0^2(y)}{4} \biggr\} &=& \Braket{ \frac{F_0^2(y)}{4} \phi^i(x_1)\ldots  \phi^j(z_n) }\,.
\end{eqnarray}
Secondly, we analogously find
\begin{equation} \label{vereenv3}
\gamma_0^2 \frac{\p }{\p \gamma^2_0 } \mathcal G_0^{n} ~=~   \int \d^4 y \left(  \mathcal G^n_0 \left\{  \gamma^2_0 g_0 f^{abc}A_{\mu, 0}^a \varphi_{\mu,0}^{bc} +  \gamma_0^2 g_0 f^{abc} A_{\mu,0}^a \overline \varphi_{\mu,0}^{bc}  \right\} \right)\,.
\end{equation}
Thirdly, we rewrite $n_j \mathcal G^n_0$ by inserting the corresponding counting operator\footnote{It is easily checked that $\int \d^4 y \phi^j_0 \frac{\delta }{\delta \phi^j_0}$ counts the number of $\phi_0^j$ insertions.} into the Green function,
\begin{eqnarray}\label{vereenv2}
n_j \mathcal  G_0^{n}&=&  \int \d^4 y \mathcal G_0^{n}\biggl\{ \phi^j_0(y) \frac{\delta S_\GZ}{\delta \phi^j_0(y) } \biggr\}  \,.
\end{eqnarray}
Inserting \eqref{vereenv1}, \eqref{vereenv3} and \eqref{vereenv2} into our main expression \eqref{stepone} results in
\begin{align} \label{steptwo}
\frac{\d \mathcal G^{n} }{\d g^2}& = \frac{1}{- \varepsilon g^2 + \beta(g^2)} \int \d^d y \Biggl[ - \sum_{j=1}^8 \gamma_{\phi^j} \mathcal G^{n}\biggl\{ \phi^j_0(y) \frac{\delta S_\GZ}{\delta \phi^j_0(y) } \biggr\}   - \varepsilon  \mathcal G^n \left\{ \frac{F_0^2(y)}{4} \right\}  \nonumber\\
 &+  \frac{\varepsilon}{2}  \mathcal G^n \left\{  A_0(y)\frac{\delta S_\GZ}{\delta A_0(y)} \right\}- \frac{\varepsilon}{2}  \mathcal G^n \left\{  b_0(y) \frac{\delta S_\GZ}{\delta b_0(y) } \right\} + \frac{\varepsilon}{2}  \mathcal G^n \left\{  \overline \omega_0(y) \frac{\delta S_\GZ}{\delta \overline \omega_0 (y)} \right\}     \nonumber\\
& -\frac{\varepsilon}{2}  \mathcal G^n \left\{  \omega_0(y) \frac{\delta S_\GZ}{\delta \omega_0(y)}  \right\}+  \delta_{\gamma^2}  \mathcal G^n \left\{  \gamma^2_0 g_0 f^{abc}A_{\mu, 0}^a \varphi_{\mu,0}^{bc} +  \gamma_0^2 g_0 f^{abc} A_{\mu,0}^a \overline \varphi_{\mu,0}^{bc}  \right\}  \Biggr]\,.
\end{align}
Notice that we have also absorbed the factor $\prod_{i}Z_{\phi_i}^{-n_i/2}$ into the Green functions, and therefore we can replace $\mathcal G_0^n$ again by  $\mathcal G^n$.
Finally, we need to rewrite all the inserted operators in the $n$-points function $\mathcal G^n$ in terms of their renormalized counterparts. For this we return to the mixing matrix
\eqref{operatormatrix2} and parameterize it as follows
\begin{eqnarray}
 \left(
  \begin{array}{c}
    \mathcal F_0 \\
    \mathcal E_0 \\
    \mathcal H_0
  \end{array}
\right) &= & \left(
          \begin{array}{ccc}
            1 + \frac{a}{\varepsilon}& -\frac{b}{\varepsilon}  &-\frac{b}{\varepsilon} \\
            0  &1 & 0   \\
           0 & 0& 1          \end{array}
        \right)
        \left(
\begin{array}{c}
    \mathcal F \\
    \mathcal E \\
    \mathcal H
  \end{array}
\right)\,.
\end{eqnarray}
Here we have displayed the fact that the entries associated with $a(g^2, \varepsilon)$ and $b(g^2, \varepsilon)$, which represent a formal power series in $g^2$, must at least have a simple pole in $\varepsilon$. Therefore, we can rewrite
\begin{eqnarray}
 - \varepsilon  \mathcal F_0(y) &=& \frac{F_0^2(y)}{4} ~=~  \left(  - \varepsilon  - a\right) \mathcal F(y) + b \left. \mathcal E(y) \right|_\phys + b  A(y)\frac{\delta S_\GZ}{\delta A(y)}\,,\nonumber\\
 \left. \mathcal H_0 \right|_\phys &=& A_0(y)\frac{\delta S_\GZ}{\delta A_0(y)} ~=~ A(y)\frac{\delta S_\GZ}{\delta A(y)}\,,
\end{eqnarray}
whereby we recall that we are working in the physical limit and we
have replaced $\left. \mathcal H \right|_\phys$ by the
expression \eqref{H}. Subsequently,
\begin{eqnarray}
 \gamma^2_0 g_0 f^{abc}A_{\mu, 0}^a \varphi_{\mu,0}^{bc} &=&  \gamma^2 g f^{abc}A_{\mu}^a \varphi_{\mu}^{bc}\,, \nonumber\\
 \gamma_0^2 g_0 f^{abc} A_{\mu,0}^a \overline \varphi_{\mu,0}^{bc} &=&   \gamma^2 g f^{abc} A_{\mu}^a \overline \varphi_{\mu}^{bc}\,,
\end{eqnarray}
as one can check with the $Z$-factors in \eqref{Z3}. Finally, all the other operators are equations of motion terms, which appear in expression \eqref{countcon}, \eqref{countcon2} and \eqref{countcon3} and therefore have the same $Z$-factor as the operator $\mathcal E$, i.e.~$Z=1$. Summarizing, expression \eqref{steptwo} becomes:
\begin{align}\label{steptwobla}
&\frac{\d \mathcal G^{n} }{\d g^2} = \frac{1}{- \varepsilon g^2 + \beta(g^2)} \int \d^d y \Biggl[ (-\varepsilon - a)\mathcal G^n\left\{ \mathcal F \right\} + \left(\frac{\varepsilon}{2} + b -\gamma_A \right)   \mathcal G^{n} \biggl\{A \frac{\delta S_\GZ}{\delta A } \biggr\} \nonumber    \\
 & + \left( -\frac{\varepsilon}{2} -\gamma_b -b \right) \mathcal G^n \left\{  b(y)\frac{\delta S_\GZ}{\delta b(y)} \right\} + \left(- \gamma_{\overline c}  -b \right) \mathcal G^n \left\{  \overline c(y) \frac{\delta S_\GZ}{\delta  \overline c(y)}  \right\} - \gamma_c \mathcal G^n \left\{  c(y) \frac{\delta S_\GZ}{\delta  c(y)}  \right\}  \nonumber  \\
 & +  \left( - \frac{\varepsilon}{2} - \gamma_{\overline \omega} \right)  \mathcal G^n \left\{  \overline \omega(y) \frac{\delta S_\GZ}{\delta \overline \omega (y)} \right\}  +\left( \frac{\varepsilon}{2}- \gamma_{ \omega}\right) \mathcal G^n \left\{  \omega(y) \frac{\delta S_\GZ}{\delta \omega(y)}  \right\} -  \gamma_\varphi   \mathcal G^n \left\{  \varphi(y) \frac{\delta S_\GZ}{\delta \varphi (y)}  \right\} \nonumber \\
 & -  \gamma_{\overline \varphi}   \mathcal G^n \left\{  \overline \varphi(y) \frac{\delta S_\GZ}{\delta \overline \varphi (y)}  \right\}+ b \mathcal G^n \left\{  \p_\mu \overline \varphi_i^a  D_\mu^{ab} \varphi^b_i  - \p_\mu \overline \omega_i^a  D_\mu^{ab} \omega_i^b  + g f^{abc} \p_\mu \overline \omega_i^a    D_\mu^{bd} c^d  \varphi_i^c  \right. \nonumber  \\
&\left. + \gamma ^{2} g  f^{abc}A_\mu^a \varphi_\mu^{bc} +  \gamma^2 g f^{abc} A_\mu^a \overline \varphi_\mu^{bc}   \right\}+  \delta_{\gamma^2}  \mathcal G^n \left\{  \gamma^2 g f^{abc}A_{\mu}^a \varphi_{\mu}^{bc} +  \gamma^2 g f^{abc} A_{\mu}^a \overline \varphi_{\mu}^{bc}  \right\}  \Biggr]\,,
\end{align}
where we have immediately taken the full expression of $\left.\mathcal E\right.|_\phys$ in equation \eqref{mult}.\\
\\
From expression \eqref{steptwobla}, we can determine $a(g^2, \varepsilon)$ and $b(g^2, \varepsilon)$. As $\frac{\d \mathcal G^{n} }{\d g^2}$ is a finite expression, we know that the right hand side of equation \eqref{steptwobla} must also be finite. Therefore, as all the Green functions are expressed in terms of finite quantities, we can choose a set of linearly independent terms and demand that their coefficients are finite:
\begin{subequations}
\begin{align}
\mathcal G^n\left\{ \mathcal F \right\} &: \frac{- \varepsilon -a }{-\varepsilon g^2 + \beta(g^2) } \,,  &  \mathcal G^{n} \left\{A \frac{\delta S_\GZ}{\delta A } \right\} &:\frac{ \varepsilon/2 + b - \gamma_A(g^2)  }{-\varepsilon g^2 + \beta(g^2) } \,, \label{a}\\
\mathcal G^n \left\{  b\p_\mu A_\mu \right\} &:\frac{ -\frac{\varepsilon}{2} -\gamma_b -b  }{-\varepsilon g^2 + \beta(g^2) }\,,   & \mathcal G^n \left\{  \overline{c}^{a}\partial _{\mu } D_\mu^{ab} c^b \right\} &: \frac{ - \gamma_{\overline c}  -b - \gamma_c }{-\varepsilon g^2 + \beta(g^2) }\,, \label{b}\\
\mathcal G^n\left\{ \overline \varphi_i^a  \p_\mu D_\mu^{ab}\varphi_i^b \right\} &:  \frac{ -\gamma_\varphi -\gamma_{\overline \varphi} -b  }{-\varepsilon g^2 + \beta(g^2) }\,,  &  \mathcal G^n\left\{ \overline \omega_i^a  \p_\mu D_\mu^{ab} \omega_i^b \right\} &:  \frac{ -\gamma_\omega -\gamma_{\overline \omega} -b  }{-\varepsilon g^2 + \beta(g^2) }\,, \label{c}\\
\mathcal G^n\left\{  -\gamma^2 g f^{abc} A_\mu^a \overline \varphi^{bc}  \right\} &:  \frac{  -\gamma_\varphi - \delta_{\gamma^2} - b  }{-\varepsilon g^2 + \beta(g^2) }\,,   & \mathcal G^n\left\{ -\gamma^2 g f^{abc} A_\mu^a \varphi^{bc}  \right\} &:  \frac{   -\gamma_{\overline \varphi} - \delta_{\gamma^2}- b }{-\varepsilon g^2 + \beta(g^2) }\,,  \label{e}
\end{align}
\vspace{-0.7cm}
\begin{align}
\mathcal G^n\left\{ -g f^{abc} \p_\nu \overline \omega_i^a  D_\nu^{bd} c^{d} \varphi_i^c \right\} :  \frac{ -\gamma_c -\gamma_{\overline \omega} - \gamma_\varphi + \frac{\varepsilon}{2} -b} {-\varepsilon g^2 + \beta(g^2) }\,.     \label{d}
\end{align}
\end{subequations}
We can rewrite the coefficients of $\mathcal G^n\left\{ \mathcal F \right\}$ and $\mathcal G^{n} \left\{A \frac{\delta S_\GZ}{\delta A } \right\}$ in \eqref{a} as
\begin{align}
\frac{- \varepsilon -a }{-\varepsilon g^2 + \beta(g^2) } &= \frac{1}{g^2} \frac{(1  +a/ \varepsilon) }{1 - \beta(g^2) /(\varepsilon g^2) }\,, &  \frac{ \varepsilon/2 + b - \gamma_A(g^2)  }{-\varepsilon g^2 + \beta(g^2) } &= -\frac{1}{2 g^2} \frac{1  + 2( b - \gamma_A(g^2))/\varepsilon }{1 - \beta(g^2) /(\varepsilon g^2) }\,.
\end{align}
Hence, in order to be finite, we must conclude that
\begin{align}\label{aenb}
a(g^2,\varepsilon) =& - \frac{\beta(g^2)}{g^2}\,, &
b(g^2,\varepsilon) =& \gamma_A(g^2) -\frac{1}{2}\frac{\beta(g^2)}{g^2}\,.
\end{align}
Notice that $a$ and $b$ depends on $g^2$, but not on $\varepsilon$. Therefore, the matrix elements of the first row of the parametrization \eqref{para} only display a simple pole in $\varepsilon$.\\
\\
Moreover, from the other equations we shall obtain relations between the anomalous dimensions of the fields and sources. Let us start with the coefficient of $\mathcal G^n \left\{  b\p_\mu A_\mu \right\}$ in equation \eqref{b}, yielding
\begin{eqnarray}
 \frac{ -\varepsilon/2 - b -  \gamma_b(g^2)  }{-\varepsilon g^2 + \beta(g^2) } &=& \frac{1}{2 g^2} \frac{ 1 + 2(b +  \gamma_b(g^2))/\varepsilon  }{1  - \beta(g^2)/(\varepsilon g^2) }\,,
\end{eqnarray}
which means that
\begin{eqnarray}
b(g^2,\varepsilon)&=&  -  \gamma_b(g^2)  -  \frac{1}{2}  \frac{\beta(g^2)}{g^2}\,.
\end{eqnarray}
Inserting the value of $b(g^2,\varepsilon)$ from expression \eqref{aenb} gives the following relation
\begin{eqnarray}\label{rel5}
\gamma_A + \gamma_b &=& 0\,.
\end{eqnarray}
This relation is a translation of the relation $Z_A^{1/2}Z_b^{1/2} = 1$ found in equation \eqref{Z2}. Indeed, deriving both sides w.r.t.~$\mu$ gives
\begin{equation}
 \frac{1}{Z_A^{1/2}Z_b^{1/2}  }\mu \frac{\p}{\p \mu}\left( Z_A^{1/2}Z_b^{1/2} \right) ~= ~ \gamma_A + \gamma_b  ~=~ 0\,.
\end{equation}
Analogously, for the coefficient of $\mathcal G^n \left\{  \overline{c}^{a}\partial _{\mu } D_\mu^{ab} c^b \right\}$, we find
\begin{eqnarray}\label{waardeb}
b(g^2,\varepsilon)&=& -\gamma_c - \gamma_{\overline c}\,,
\end{eqnarray}
yielding
\begin{eqnarray}
\gamma_A + \gamma_c + \gamma_{\overline c} &=& \frac{ \beta }{ 2 g^2}\,,
\end{eqnarray}
which is a translation of $Z_c^{1/2} Z_{\overline c}^{1/2} Z^{1/2}_A Z_g = 1$ as $\mu \frac{\d Z_g}{\d \mu} = - \frac{ \beta }{ 2 g^2}$. Next, the coefficients of \eqref{c} and \eqref{d} lead to
\begin{align}
\gamma_\varphi + \gamma_{\overline \varphi} + \gamma_A & = \frac{\beta}{2 g^2}\,, &  \gamma_\omega + \gamma_{\overline \omega} + \gamma_A & = \frac{\beta}{2 g^2}\,, & \gamma_c + \gamma_{\overline \omega} + \gamma_\varphi + \gamma_A &= \frac{\beta}{g^2}\,,
\end{align}
stemming from
\begin{align}
Z^{1/2}_\varphi Z^{1/2}_{\overline \varphi}Z_A^{1/2} Z_g & = 1\,, &  Z^{1/2}_\omega Z^{1/2}_{\overline \omega} Z^{1/2}_A Z_g & =1 \,, &  Z^{1/2}_c Z^{1/2}_{ \overline \omega} Z_\varphi^{1/2} Z_A^{1/2} Z_g & = 1 \,.
\end{align}
These relations originate from the relations derived in \eqref{Z2} and \eqref{Z3}. Finally, the coefficients in equation \eqref{e} are finite if
\begin{equation}
  -\gamma_{\overline \varphi} - \delta_{\gamma^2} ~=~-\gamma_{ \varphi} - \delta_{\gamma^2} ~=~ b ~= ~\gamma_A(g^2) -\frac{1}{2}\frac{\beta(g^2)}{g^2}\,,
\end{equation}
or equivalently
\begin{align}
Z^{1/2}_{\overline \varphi}Z_A^{1/2} Z_g Z_{\gamma^2} & =1\,, &  Z^{1/2}_{\varphi}Z_A^{1/2} Z_g Z_{\gamma^2} & =1\,,
\end{align}
which is also fulfilled as $Z_{\gamma^2} = Z_V = Z_g^{-1/2} Z_A^{-1/4}$. \\
\\
In summary, we have determined to all orders the mixing matrix \eqref{operatormatrix2}. For notational simplicity, we take the value \eqref{waardeb} for $b$ and we use the equality $\gamma_c = \gamma_{\overline c}$:
\begin{eqnarray}\label{para}
Z &= & \left(
          \begin{array}{ccc}
            1 -  \frac{\beta(g^2)}{\varepsilon g^2}& \frac{2\gamma_c}{\varepsilon }  & \frac{2\gamma_c}{\varepsilon }  \\
            0  &1 & 0   \\
           0 & 0& 1          \end{array}
        \right)\,.
\end{eqnarray}
We have encountered numerous checks which show the consistency of our results.\\
\\
\textbf{Remark}\\
This matrix is also valid for the refined action $S_\RGZ$. One can repeat the proof by replacing $S_\GZ$ with $S_\RGZ$ and by adding the following term in $M^2 = J$ to the game,
\begin{eqnarray}
 S_{\overline{\varphi} \varphi} &=& - M^2 \int \d^d x  \left( \overline \varphi^a_i \varphi^a_i - \overline \omega^a_i \omega^a_i \right)\,,
 \end{eqnarray}
see equation \eqref{defex}. In the end, expression \eqref{steptwobla} will collect an extra term
\begin{align}
&\frac{\d \mathcal G^{n} }{\d g^2} = \eqref{steptwobla}+  \frac{1}{- \varepsilon g^2 + \beta(g^2)} \int \d^d y \left[ \delta_{M^2}  \mathcal G^n \left\{ M^2 (\overline \varphi \varphi - \overline \omega \omega)   \right\} \right]\,,
\end{align}
where we have introduced the anomalous dimension of $M^2$,
\begin{eqnarray}
\delta_{M^2} &=& \mu  \frac{\p \ln Z_{M^2}}{\p \mu} \,.
\end{eqnarray}
This leads to the following extra coefficients
\begin{align}
 \mathcal G^n \left\{ -M^2  \overline \varphi^a_i \varphi^a_i   \right\}&: \frac{- \gamma_{\overline \varphi} -\gamma_{ \varphi} - \delta_{M^2}  }{-\varepsilon g^2 + \beta(g^2) }\,,  &   \mathcal G^n \left\{ M^2 \overline \omega^a_i \omega^a_i \right\}&: \frac{ - \gamma_{\overline \omega} -\gamma_{ \omega} - \delta_{M^2} }{-\varepsilon g^2 + \beta(g^2) }\,.
\end{align}
so that
\begin{align}
  \gamma_{\overline \varphi}+\gamma_{ \varphi}+ \delta_{M^2} &=0\,, &   \gamma_{\overline \omega} +\gamma_{ \omega} + \delta_{M^2} & =0\,,
\end{align}
or equivalently
\begin{align}
  Z^{1/2}_{\overline \varphi}Z^{1/2}_{ \varphi}Z_{M^2} &=1\,, &   Z^{1/2}_{\overline \omega}Z^{1/2}_{ \omega}Z^{1/2}_{M^2} & =1\,,
\end{align}
which is correct as $Z_J = Z_{M^2} = Z_g Z_A^{1/2}$, see equation \eqref{5ZJ}. All the other relations stay valid of course.

\subsection{Constructing a renormalization group invariant}
As the final step of our analysis, we shall try to determine a renormalization group invariant operator which contains $\mathcal F \equiv \frac{F^2_{\mu\nu}(x)}{4}$. This is
useful as we would want to obtain a renormalization group invariant estimate for the glueball mass, i.e.~the pole of the corresponding correlator. This analysis is completely
similar to the one presented in section \ref{sectconstructingaRGI}, due to the fact that the mixing matrix $Z$ is exactly the same.
Therefore, we can immediately conclude that (see expression \eqref{YMRGI})
\begin{eqnarray}\label{RGE}
\mathcal{R}&=& \frac{\beta(g^2)}{g^2} \mathcal F -2\gamma_c(g^2)  \mathcal E  -2\gamma_c(g^2) \mathcal H\;,
\end{eqnarray}
is a renormalization group invariant scalar operator containing $F_{\mu\nu}^2$, in the case of the Gribov-Zwanziger action $\Sigma_\GZ$ as well as in the case of the refined action $\Sigma_\RGZ$.

\subsection{Conclusion}
We have found a renormalization group invariant, the final goal would be that of evaluating the glueball correlator
\begin{multline} \label{mainpoint}
\Braket{\mathcal R(x) \mathcal R(y)}_\phys=\left< \left( \frac{\beta(g^2)}{g^2}  \mathcal F(x) -2 \gamma_c(g^2)  \mathcal E (x) - 2\gamma_c(g^2)  \mathcal H(x) \right) \right. \times \\ \left. \left( \frac{\beta(g^2)}{g^2}  \mathcal F(y) -2 \gamma_c(g^2)  \mathcal E(y)  - 2\gamma_c(g^2)  \mathcal H(y)\right)\right>_\phys \,,
\end{multline}
using the (Refined) Gribov-Zwanziger action.\\
\\
As usual the equation of motion terms like $\mathcal H$ will not play a role. Let us demonstrate this with a simple example,
\begin{eqnarray}
\Braket{ \mathcal F(x) \mathcal H(y)}_\phys &=& \Braket{  \mathcal F(x) A^a_\mu(y) \frac{\delta S_{\RGZ} }{ \delta A_\mu^a(y)} }~=~ \int [\d \Phi ] \mathcal F(x)  A_\mu^a (y) \frac{\delta S_{\RGZ} }{ \delta A_\mu^a (y)} \e^{-S_{\RGZ}}  \nonumber\\
&=&   - \int [\d \Phi ] \mathcal F(x) A_\mu^a (y) \frac{\delta  \e^{-S_{\RGZ}} }{ \delta A_\mu^a (y)} = \int [\d \Phi ]   \e^{-S_{\RGZ}}  \frac{\delta  \left( A_\mu^a (y) \mathcal F(x) \right) }{ \delta A_\mu^a (y)}  \nonumber\\
&=& \ldots \delta(x-y) + \delta(0) \Braket{\mathcal F(x)}\,,
\end{eqnarray}
which is zero as $x \not= y$ and $\delta(0) = 0$ in dimensional regularization. Therefore, expression \eqref{mainpoint} reduces to,
\begin{multline}
\Braket{\mathcal R(x) \mathcal R(y)}_\phys ~=~ \left( \frac{\beta(g^2)}{g^2} \right)^2 \Braket{  \mathcal F(x) \mathcal F(y) }+ \left(2\gamma_c(g^2) \right)^2 \Braket{  \mathcal E (x) \mathcal E (y)}_\phys \\ -2 \gamma_c(g^2)  \frac{\beta(g^2)}{g^2}  \left(  \Braket{ \mathcal F(x)  \mathcal E(y)}_\phys  +   \Braket{ \mathcal E(x) \mathcal F(y) }_\phys \right)\,,
\end{multline}
and only $\mathcal E$ is of importance. We recall that $\mathcal E$ is given by
\begin{multline}
\mathcal E =  \p_\mu b^a  A_\mu^a + \p_\mu  \overline c^a D_\mu^{ab} c^b +  \p_\mu \overline \varphi_i^a  D_\mu^{ab} \varphi^b_i  - \p_\mu \overline \omega_i^a  D_\mu^{ab} \omega_i^b  + g f^{abc} \p_\mu \overline \omega_i^a    D_\mu^{bd} c^d  \varphi_i^c  \\
+ \gamma ^{2} g  f^{abc}A_\mu^a \varphi_\mu^{bc} +  \gamma^2 g f^{abc} A_\mu^a \overline \varphi_\mu^{bc} + d \left(N^{2}-1\right)  \gamma^4 \,.
\end{multline}
From this point we can compare with the Yang-Mills case, see section \ref{secconclusion}.  For the standard YM action, we have seen that gauge invariant operators ${\cal F}$ only mix with BRST exact and equation of motion type terms. While the latter always yield trivial information at the level of correlators, the BRST exact pieces drop out due to the BRST invariance of the gauge invariant operator ${\cal F}$ and of the vacuum. The situation is quite different here in the (Refined) GZ framework. Namely, in the physical limit, $\mathcal E$ is no longer a BRST invariant operator. In addition, the BRST symmetry of the GZ action is softly broken, see section \ref{sectbreakingBRST} of chapter \ref{scrutinizing}.  Therefore, when turning to physical states, $\mathcal E$ will no longer be irrelevant, and will explicitly influence the value of the correlator. This is not the only observation we can make. ${\cal R}(x)$ is not the only renormalization group invariant of dimension 4. Indeed, also the operator $\mathcal {E}(x)$ does not run with the scale, as we directly infer from equations \eqref{Gamma1} and \eqref{Gamma2}. We can therefore imagine to study correlators of linear combinations of the operators ${\cal F}$ and $\cal E$, where the linear combination is chosen in such a way that the emerging pole structure would be real. We shall elaborate on this in the next chapter, but one can already notice that this is not a trivial issue in the Gribov-Zwanziger framework, basically due to the fact that the poles of the gluon propagator itself are already not necessarily real-valued. Finally, we observe that when the Gribov parameter $\gamma^2$ is formally set back to zero, we recover the correlators of the usual kind in Yang-Mills gauge theories, as the BRST symmetry gets restored, as well as the BRST exactness of the operator $\mathcal{E}$.

\section{Intermediate conclusion}
Now that we have studied the renormalization of $F^2_{\mu\nu}$ in the GZ action, and shown that this is far from trivial, we should study the spectral properties of the correlator $\Braket{\mathcal R(x) \mathcal R(y)}$. In practice, we have to investigate whether the correlator $\Braket{\mathcal R(x) \mathcal R(y)}$ can be cast in the form of a K\"all\'{e}n-Lehmann  representation, i.e.~a spectral representation  with a positive spectral function, and whose analytic continuation in the complex Euclidean $k^2$-plane exhibits a cut along the negative real axis only. Such a spectral representation would thus imply that, when moving to Minkowski space, the cuts are located along the positive real axis. Moreover, positivity of the spectral function then guarantees that a meaningful interpretation in terms of states of a physical spectrum can be attached to those  operators.  This is precisely what one would expect from a confining theory. This is a highly nontrivial task, given the complexity of the Gribov type propagator, see \eqref{gluonprop} as well as of the Gribov-Zwanziger action. In fact, the correlator $\Braket{ F^2(x) F^2(y)}$ was already studied in \cite{Zwanziger:1989mf} at one loop order. The results found in \cite{Zwanziger:1989mf} can be summarized as follows:
\begin{eqnarray}\label{physicalcutunphysicalcut}
G(k^2) &=& \int \d^{4}x\ e^{-ikx\ }\Braket{ F^{2}(x)F^{2}(0)} = G^{\rm phys}(k^2) + G^{\rm unphys}(k^2) \;.
\end{eqnarray}
The unphysical part, $G^{\rm unphys}(k^2)$, displays cuts along the imaginary axes beginning at the unphysical values $k^2=\pm 4i{ \gamma}^2$, whereas the physical part, $G^{\rm phys}(k^2)$, has a cut beginning at the physical threshold $k^2=-2{\gamma}^2$. Moreover, the spectral function of  $G^{\rm phys}(k^2)$ turns out to be positive, so that it  possesses a K\"all\'{e}n-Lehmann representation \cite{Zwanziger:1989mf}. As such, $G^{\rm phys}(k^2)$ is an acceptable correlation function for physical glueball excitations. What is also interesting in this expression is that a physical cut has emerged in the correlation function of a gauge invariant quantity, even if it has
been evaluated with a gluon propagator exhibiting only unphysical complex poles.\\
\\
In the paper \cite{Zwanziger:1989mf} however, the renormalization of $F^2_{\mu\nu}$ was not taken into account. As we have proven, due to the breaking of the BRST $\Braket{\mathcal R(x) \mathcal R(y)} \not= \Braket{\mathcal F(x) \mathcal F(y)}$. The hope was therefore that $\mathcal E$ would cancel the unphysical pole. However, it would seem like a deus ex machina if this would really be the case, as there is no residual freedom left in $\mathcal R$. Therefore, in the next chapter, we shall go a deeper into the cut structure of correlator to find out where exactly the physical cut in expression \eqref{physicalcutunphysicalcut} comes from.

\chapter{The quest for physical operators, part II\label{chappart2}}

\section{Introduction}
In this chapter, we shall look at correlators from a whole different perspective. In the last chapter, we started from a gauge invariant quantity, namely $F^2_{\mu\nu}$, and investigated its renormalization, to construct a RGI operator, while in this chapter we shall focus on the spectral representation of the correlators. If we want obtain a particle interpretation out of an action, we need to look for operators which correlation functions (1) exhibiting only real cuts and (2) having positive spectral functions \cite{Peskin}.  For the first requirement (1), we need to find an operator $\mathcal O$, so that the corresponding correlator $\Braket{\mathcal O(k) \mathcal O(-k)}$ can be cast into the a spectral representation, i.e.
\begin{align}\label{spectrepr}
 \Braket{ \mathcal O(k) \mathcal O(-k) }  =  \int_{\tau_{0}}^{\infty} \d\tau \; \rho({\tau}) \; \frac{1}{\tau+k^2} \;,
\end{align}
where the quantity $\tau_{0}>0$ stands for the threshold. If we now introduce the complex function
\begin{align} \label{6a4}
F(z) =    \int_{\tau_{0}}^{\infty} \d\tau \; \rho(\tau) \; \frac{1}{\tau+z} \;,
\end{align}
then from complex analysis, it follows that $F(z)$ is an analytic function in the cut complex plane, where the interval $(-\infty, -\tau_{0})$ has been excluded. Therefore, when moving from Euclidean to Minkowski space, i.e.~$k^2_{Eucl} \rightarrow - k^2_{Mink}$, expression \eqref{6a4} gives the spectral representation\footnote{This is called the K\"all\'{e}n-Lehmann represention, see \cite{Peskin}.} of a quantity exhibiting a discontinuity along the positive real axis, starting at the threshold $\tau_{0}$ and extending till $+\infty$.  In order to meet the second requirement (2), we need $\rho(\tau)$ to be positive. \\


\noindent To meet both requirements, we shall introduce $i$-particles: a pair of fields with complex conjugate masses which emerge in a natural way when dealing with a gluon propagator which behaves like \eqref{gluonprop}. As it will be discussed in details, we shall be able to show that local composite operators built up with pairs of $i$-particles display cuts along the negative real axis in the complex Euclidean $k^2$-plane, while giving rise to positive spectral functions. For the benefit of the reader, we shall first work in a toy model, before going to the complex GZ action. This chapter is mainly based on \cite{Baulieu:2009ha}. We emphasize that we shall only work with the GZ action in this chapter for simplicity, and not with the more complicated RGZ action.

\section{A scalar field theory toy model\label{sec:model1}}
In this section, we shall construct a toy model, before going over to the more complicated GZ action. The main feature of this toy model, is that it shall also exhibit a Gribov type of gluon propagator like \eqref{gluonprop}, as this is the key ingredient in constructing operators in the GZ framework which give rise to correlators which have real cuts and positive spectral functions.

\subsection{Constructing the toy model}
A simple way of constructing a field theory model exhibiting a confining Gribov type propagator is through a scalar field $\psi$ whose Euclidean action is nonlocal, being specified by
\begin{align}
S = \int \d^d x \; \frac{1}{2}  \psi \left( -\partial^2 + 2\frac{\theta^4}{-\partial^2} \right) \psi \;. \label{act}
\end{align}
Indeed, the resulting propagator is of the Gribov type as
\begin{align}
\Braket{ \psi(k) \psi(p)} = (2\pi)^d \delta(k+p) \frac{k^2}{k^4+2 \theta^4} \;. \label{gp}
\end{align}
The massive parameter $\theta$ is introduced by hand and is the analog of the Gribov parameter $\gamma$. Analogous to section \ref{localization} in chapter \ref{chapgribovtoGZ}, we can localize the non-local expression \eqref{act} by introducing a pair of bosonic complex conjugate fields $({\overline \varphi}, \varphi)$ and a pair of anticommuting fields $({\overline \omega}, \omega)$, so that
\begin{align}
S =  \int \d^d x  \left(  \frac{1}{2}  \psi ( -\partial^2 ) \psi +  {\overline \varphi} (-\partial^2) \varphi +  {\theta^2}\psi (\varphi-\overline{\varphi})
- {\overline \omega}(-\partial^2) \omega   \right) \;.  \label{lact}
\end{align}
Let us have a look at the propagators of the model. If we want to apply equation \eqref{gauss1}, we need to rewrite the complex conjugate bosonic fields $(\overline \varphi, \varphi)$ in terms of real fields, i.e.
\begin{align}
\varphi  & = \frac{U+\ii V}{\sqrt{2}} \;,  &
{\overline \varphi}  & = \frac{U- \ii V}{\sqrt{2}} \; , \label{v1}
\end{align}
and the action becomes
\begin{equation}
S = \int \d^4 x \left( \frac{1}{2} \psi (-\partial^2) \psi + \frac{1}{2} V (-\partial^2) V + \sqrt{2} \ii \theta^2 \psi V
+ \frac{1}{2} U (-\partial^2) U - {\overline \omega} (-\partial^2) \omega \; \right) \;. \label{act1}
\end{equation}
We can rewrite this expression in matrixform,
\begin{equation}
\exp[-S] = \exp[ -\frac{1}{2} \int \d^4 x
\begin{bmatrix}
 \psi(x) & V(x) &  U(x)
\end{bmatrix}
\underbrace{\begin{bmatrix} - \p^2   & \ii \theta^2 \sqrt{2} & 0\\
                 \ii \theta^2 \sqrt{2} & -\p^2 & 0 \\
                 0 & 0 & - \p^2
 \end{bmatrix}}_A
  \begin{bmatrix}
 \psi(x) \\ V(x)  \\ U(x)
\end{bmatrix}   - {\overline \omega} (-\partial^2) \omega \; ] \;.
\end{equation}
If we now want to apply formula \eqref{gauss1}, we have to calculate $A^{-1}$,
\begin{equation}
A^{-1} =\begin{bmatrix} \frac{- \p^2}{\p^4 + 2  \theta^4 }   & -\frac{\ii \theta^2 \sqrt{2}}{ \p^4 + 2  \theta^4 } & 0\\
                 -\frac{\ii \theta^2 \sqrt{2}}{ \p^4 + 2  \theta^4 } & \frac{- \p^2}{\p^4 + 2  \theta^4 } & 0 \\
                 0 & 0 &\frac{1}{ - \p^2}\;.
 \end{bmatrix}
\end{equation}
Now going to Fourierspace, we can find the propagators,
\begin{align}\label{nodig1}
\Braket{ \psi(p) \psi(k)} & = (2\pi)^d \delta(p+k)  \frac{p^2}{p^4+2\theta^4} \;, &
\Braket{ V(p) V(k)}       & = (2\pi)^d \delta(p+k)  \frac{p^2}{p^4+2\theta^4} \;, \nonumber \\
\Braket{ V(p) \psi(k) }   & = (2\pi)^d \delta(p+k)  \frac{-\ii\sqrt{2} \theta^2}{p^4+2\theta^4} \;, &
\Braket{ U(p) U(k)}       & = (2\pi)^d \delta(p+k)  \frac{1}{p^2} \;.
\end{align}
Having evaluated the propagators of the fields $(\psi, U, V)$, we can now check what  the propagators in terms of the fields $(\varphi, {\overline \varphi})$ are.
One finds
\begin{align}
\Braket{ \psi(p) \psi(k)}                    & = (2\pi)^d \delta(p+k)   \frac{\theta^2}{p^4 + 2 \theta^4} \;, &
\Braket{ \psi (p) {\overline \varphi} (k) }  & = (2\pi)^d \delta(p+k)   \frac{-\theta^2}{p^4 + 2 \theta^4} \;, \nonumber \\
\Braket{ \varphi(p) {\overline \varphi} (k)} & = (2\pi)^d \delta(p+k)   \frac{p^4 +\theta^4}{p^2(p^4 + 2 \theta^4)} \;, &
\Braket{ \varphi(p) \varphi (k)}             & = (2\pi)^d \delta(p+k) \frac{\theta^4}{p^2(p^4 + 2 \theta^4)} \;.  \label{np}
\end{align}

\subsection{Introducing the $i$-particles}\label{ip}
The $i$-particles are new variables introduced in such a way that \eqref{act1} can be cast in complete diagonal form:
\begin{align}
\psi & = \frac{1}{\sqrt{2}} (\lambda +\eta) \;, & V & = \frac{1}{\sqrt{2}} (\lambda -\eta) \;. \label{le}
\end{align}
Indeed, the action is now diagonal,
\begin{equation}
S = \int \d^d x \; \left( \; \frac{1}{2} \lambda (-\partial^2+ \ii \sqrt{2}\theta^2) \lambda + \frac{1}{2} \eta (-\partial^2-\ii \sqrt{2}\theta^2) \eta
+ \frac{1}{2} U (-\partial^2) U - {\overline \omega} (-\partial^2) \omega \; \right) \;. \label{actGZ2}
\end{equation}
From this expression one immediately sees that the fields $\lambda$ and $\eta$ correspond to the propagation of unphysical modes with complex masses $\pm \ii \sqrt{2} \theta^2$. These are the $i$-particles of the model, namely
\begin{align}\label{nodig2}
\Braket{ \lambda(k) \lambda(p)} & = (2\pi)^d \delta(p+k) \frac{1}{k^2+\ii \sqrt{2}\theta^2} \;, &
\Braket{ \eta(k) \eta(p) }      & = (2\pi)^d \delta(p+k) \frac{1}{k^2-\ii\sqrt{2}\theta^2} \;.
 \end{align}

\subsection{Proposal for a ``good'' operator}
Let us now try to construct an operator which exhibits real cuts and has a positive spectral density. The $i$-particles shall form the basis of these operators. The simplest example which one can consider at one loop is that of the dimension two composite operator consisting of
one $i$-particle of the type $\lambda$ and one $i$-particle of the type $\eta$, namely
\begin{align}
O_1(x) = \lambda(x) \eta(x) \;, \label{o1}
\end{align}
The correlation function $\Braket{ O_1(k) O_1(-k) } $ in $d$ Euclidean dimensions is given by
\begin{align} \label{o1c}
\Braket{ O_1(k) O_1(-k) } = \int \frac{\d^dp}{(2\pi)^d}\; \frac{1}{(k-p)^2-\ii \sqrt{2}\theta^2} \frac{1}{p^2+\ii\sqrt{2}\theta^2} \;.
\end{align}
By direct inspection of the action \eqref{actGZ2}, it follows that the  correlation function of three operators $O_1(x)$ vanishes
\begin{align}
\Braket{ O_1(x) O_1(y) O_1(z) } = 0 \;. \label{v}
\end{align}
Only correlation functions with an even number of operators $O_1$ are nonvanishing. We shall here evaluate this integral for $d=4$ and $d=2$, which can be done with the same technique. For the case $d=3$, we refer to \cite{Baulieu:2009ha}.

\subsubsection{The spectral representation in $d =2$}
As explained in the introduction, we need to find the spectral function of the two-point function $\Braket{ O_1(k) O_1(-k) }$. As a first step, we shall employ formula \eqref{combdenominator} from the appendix, so \eqref{o1c} becomes
\begin{align}
\Braket{ O_1(k) O_1(-k) } &= \int \frac{\d^dp}{(2\pi)^d}\; \int^{1}_{0} \d x \frac{1}{\left[x(k^2-2p\cdot k-2\ii\sqrt{2}\theta^2) + p^2 + \ii\sqrt{2}\theta^2\right]^2}
\;.
\end{align}
Next, we define $q\equiv p-kx$,
\begin{align}
\Braket{ O_1(k) O_1(-k) }&= \int \frac{\d^dq}{(2\pi)^d}\; \int^{1}_{0} \d x \frac{1}{\left[q^2 + (x-x^2)k^2 - (2x-1)\ii\sqrt{2}\theta^2 \right]^2}\;.
\end{align}
Using now the identity \eqref{loopint} with $n=2$ and $\Delta^2 = (x-x^2)k^2 - (2x-1)\ii\sqrt{2}\theta^2$, we obtain
\begin{align}
\Braket{ O_1(k) O_1(-k) } &=  \frac{\Gamma(2-\frac{d}{2})}{(4\pi)^{\frac{d}{2}}} \int^{1}_{0} \d x \left[(x-x^2)k^2 - (2x-1)\ii\sqrt{2}\theta^2 \right]^{\frac{d}{2}-2}
\;. \label{fey-par2}
\end{align}
We can now consider the special case of $d=2$. We have then
\begin{align}
\Braket{ O_1(k) O_1(-k) } &=  \frac{1}{4\pi} \int^{1}_{0} \d x \frac{1}{\left[(x-x^2)k^2 - (2x-1)\ii\sqrt{2}\theta^2 \right]}
\;. \label{fey-parD2}
\end{align}
Our goal is to bring this integral into a form where its analytic structure as a function of the external momentum $k^2$ becomes manifest. If we change the variable of integration as
\begin{align}
u\equiv \frac{(2x-1)}{2(x-x^2)}
\;, \label{fey-chgvar}
\end{align}
the integral takes the form
\begin{align}\label{fey-parD2-u}
\Braket{ O_1(k) O_1(-k) } &=  \frac{1}{4\pi} \int^{\infty}_{-\infty}  \frac{\d u}{\sqrt{u^2 + 1}}\frac{1}{k^2 - 2\ii\sqrt{2}\theta^2 u}
\;.
\end{align}
We are working in the Euclidean region, that is, $k^2$ is real and positive. The integral is then well defined in the upper-half complex $u$-plane, except for a branch cut due to the square root. This branch cut starts at  $u=\ii$ and extends along the imaginary axis to $\ii\infty$. We can then consider a contour of integration\footnote{Use is made of the fact that the integrand of \eqref{fey-parD2-u} falls off sufficiently fast at complex infinity.} which is deformed to surround this cut, i.e.
\begin{equation}
\Braket{ O_1(k) O_1(-k) } =  \frac{1}{4\pi}\left[ \int_{+\ii \infty}^{\ii}  \frac{\d u}{\sqrt{u^2 + 1}}\frac{1}{k^2 - 2\ii\sqrt{2}\theta^2 u} +   \int^{+\ii \infty}_{\ii}  \frac{\d u}{\sqrt{u^2 + 1}}\frac{1}{k^2 - 2\ii\sqrt{2}\theta^2 u} \right]
\;.
\end{equation}
whereby the integrals are evaluated on respectively the left and right side of the branch cut. Now replacing $u=\ii y$, we find $\sqrt{u^2 + 1} = \ii \sqrt{|-y^2 +1|}$ for the first integral and $\sqrt{u^2 + 1} = -\ii \sqrt{|-y^2 +1|}$ for the second integral, and thus
\begin{align}
\Braket{ O_1(k) O_1(-k) } &=  \frac{1}{2\pi} \int^{\infty}_{1}  \frac{\d y}{\sqrt{y^2 - 1}}\frac{1}{k^2 + 2\sqrt{2}\theta^2 y}\;.
\; \label{fey-parD2-y}
\end{align}
Observe that this has the exact structure of the K\"all\'{e}n-Lehmann representation. More explicitly, writing $\tau = 2\sqrt{2}\theta^2 y$ we have
\begin{align}
\Braket{ O_1(k) O_1(-k) } &=   \int^{\infty}_{2\sqrt{2}\theta^2} \d \tau\frac{\rho(\tau)}{k^2 + \tau}
\;, \label{fey-parD2-KL}
\end{align}
where
\begin{align}
\rho(\tau) = \frac{1}{2\pi\sqrt{\tau^2 - 8\theta^4}}\;,
 \label{fey-parD2-spectral}
\end{align}
is the spectral function. Interpreting this result as an expression of the physical spectrum of the theory, we see that it has a threshold at $2\sqrt{2}\theta^2$.\\
\\
Finally, we can also evaluate \eqref{fey-parD2-spectral} explicitly, resulting in
\begin{equation}
\Braket{ O_1(k) O_1(-k) } = \frac{1}{2\pi} \frac{1}{\sqrt{8 \theta^4 - k^4}} \arccos \frac{k^2}{2 \sqrt{2} \theta^2}\;,
\end{equation}
a function, which has indeed a branch cut from $-\infty$ until $-2 \sqrt{2} \theta^2$.

\begin{figure}[H]
  \begin{center}
  \subfigure[]{\includegraphics[width=7cm]{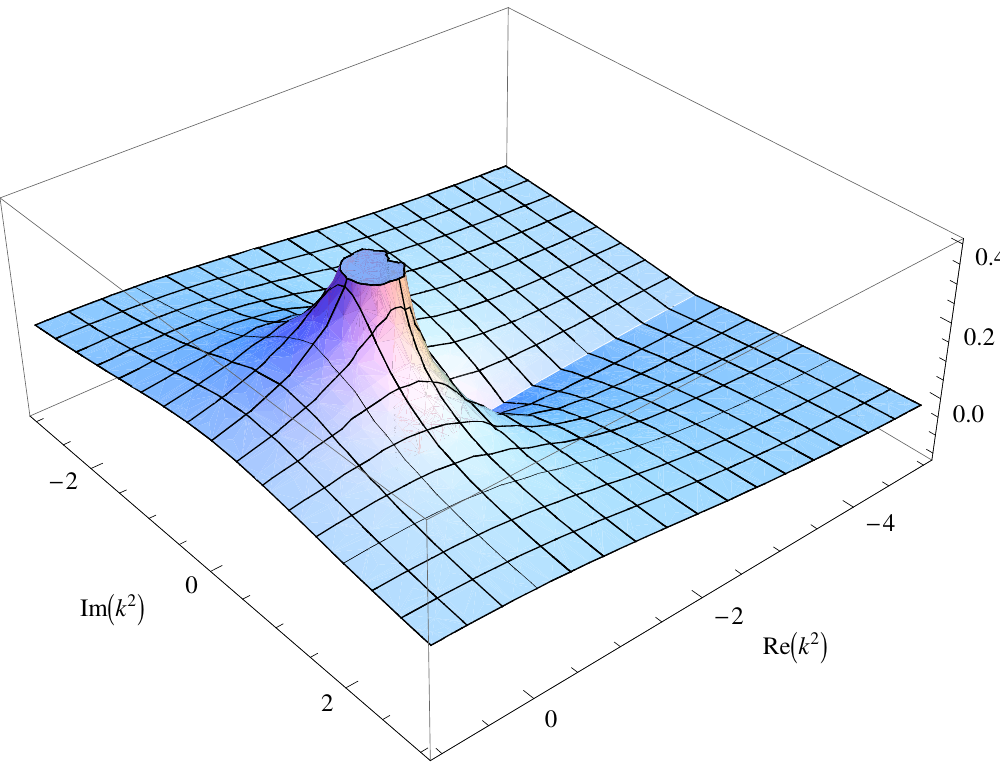} }
    \hspace{1cm}
  \subfigure[]{\includegraphics[width=7cm]{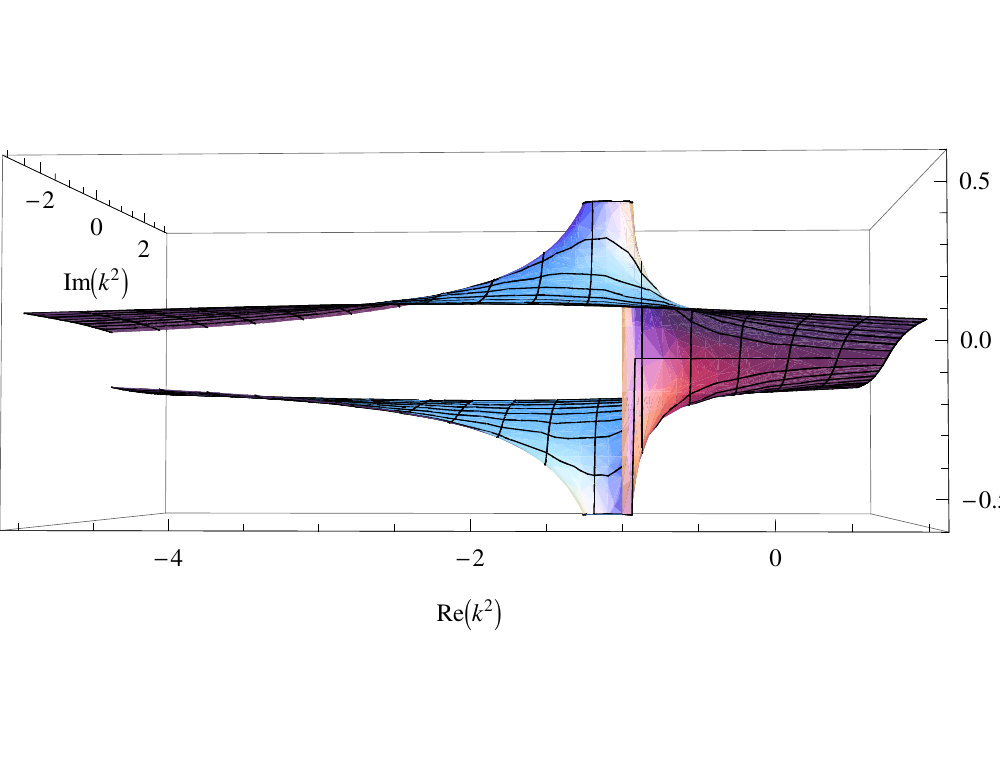}}
     \end{center}\label{6extrafig}
  \caption{Plot of (a) $\mathrm{Re}\left[\frac{1}{2\pi} \frac{1}{\sqrt{1 - k^4}} \arccos k^2\right]$ and (b) $\mathrm{Im}\left[\frac{1}{2\pi} \frac{1}{\sqrt{1 - k^4}} \arccos k^2\right]$.}
\end{figure}

\subsubsection{The spectral representation in $d =4$ \label{chap6sectspectrd4}}
\begin{figure}[H]
  \begin{center}
  \subfigure[]{\includegraphics[width=7cm]{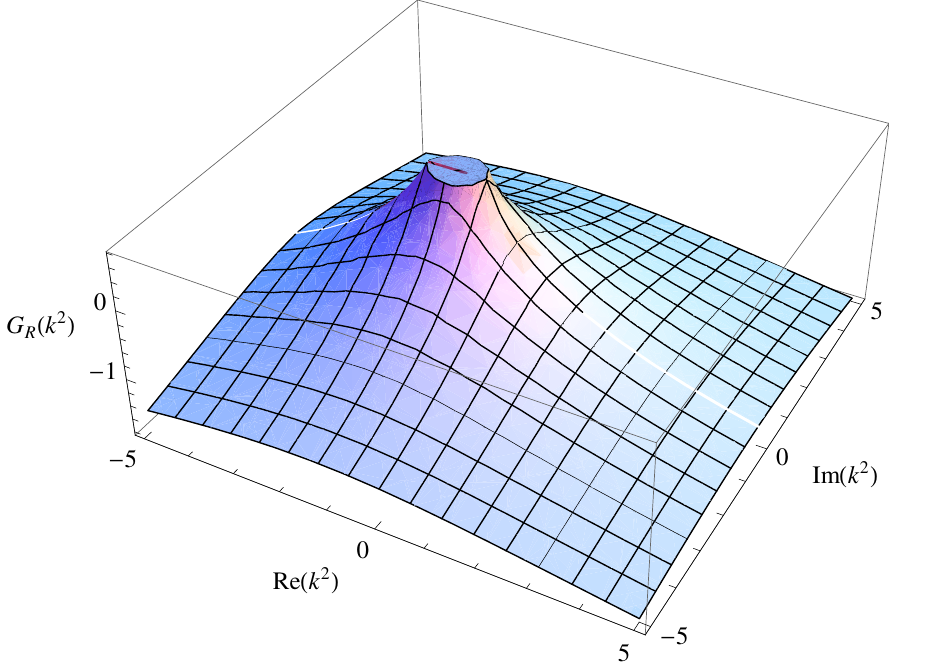} }
    \hspace{1cm}
  \subfigure[]{\includegraphics[width=7cm]{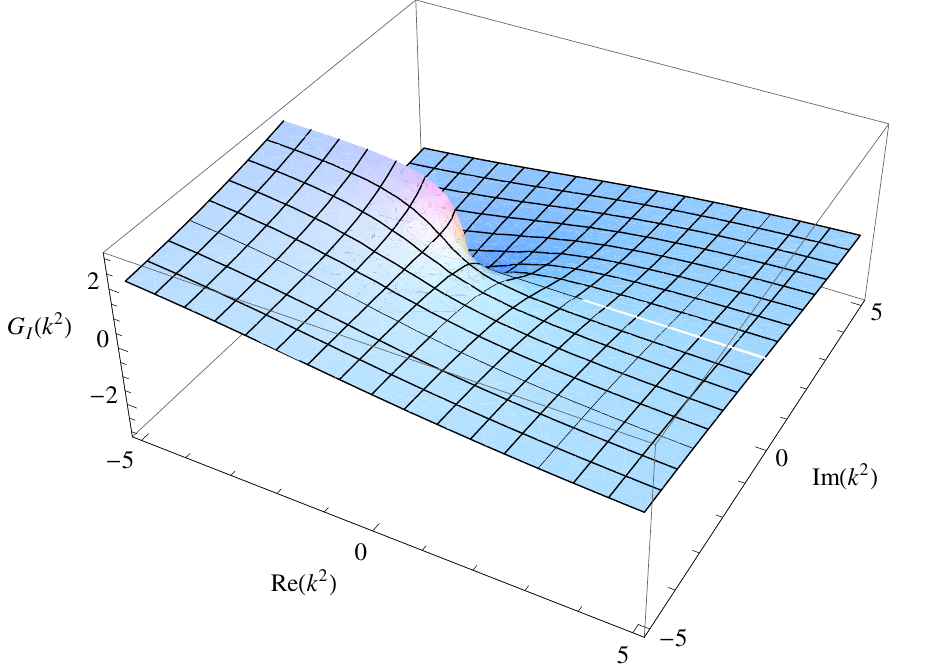}}
     \end{center}
  \caption{Plot of (a) $G_R(k^2)\equiv16\pi^2 \mathrm{Re}[F(k^2)-F(0)]$  and (b) $G_I(k^2)\equiv16\pi^2 \mathrm{Im}[F(k^2)-F(0)]$ with $F(k^2)-F(0)$ given in \eqref{int12}.}\label{6fig1}
\end{figure}

We can repeat this calculation for $d=4$. We define
\begin{equation}
F(k^2)=\Braket{ O_1(k) O_1(-k) }\,.
\end{equation}
We start from \eqref{fey-par2} and act on it with $\frac{\p}{\p k^2}$. This  regularizes the original integral for $F(k^2)$, which is ultraviolet divergent.
After setting $d=4 - \epsilon$, expanding for small $\epsilon$ and setting $\epsilon \to 0$ in the end, we find
\begin{equation}\label{int9}
\frac{\p}{\p k^2}F(k^2)= -\frac{1}{16\pi^2}\int_0^1 \d x\frac{x(1-x)}{x(1-x)k^2+x\ii-\ii/2}=-\frac{1}{16\pi^2}\int_0^1 \d x\frac{x(1-x)}{x(1-x)k^2-x\ii+\ii/2}\,,
\end{equation}
where we temporarily switched to units $2\sqrt{2}\theta^2=1$ for notational convenience. The substitution $s=\frac{2x-1}{2x(1-x)}$, or $x=\frac{-1+s+\sqrt{1+s^2}}{2s}$, brings us to
\begin{eqnarray}\label{int10}
\frac{\p}{\p k^2}F(k^2)&=& -\frac{1}{16\pi^2}\int_{-\infty}^{+\infty} \d s \frac{1}{k^2-\ii s} \frac{\d \left(\frac{-1+s+\sqrt{1+s^2}}{2s}\right)}{\d s}
\nonumber\\&=&\frac{1}{16\pi^2}\int_{-\infty}^{+\infty} \frac{\ii \d s}{(k^2- \ii s)^2}\frac{-1+s+\sqrt{1+s^2}}{2s}\,,
\end{eqnarray}
where we employed partial integration. There is no problem at $s=0$ if $k^2>0$, since $\lim_{s\to0}\frac{-1+s+\sqrt{1+s^2}}{2s}=\frac{1}{2}$. Similar remarks apply as in the $d=2$ case. There are no poles in the upper half $s$-plane for $k^2>0$, so we can deform the contour to be located around the cut for $s\in[\ii\infty,\ii]$. Setting $s=\ii\tau$, we compute \eqref{int10} as
\begin{align}\label{int10b}
\frac{\p}{\p k^2}F(k^2)&= \frac{1}{16\pi^2}\left[ \int_{\infty}^{1}\frac{-\d\tau}{(k^2+\tau)^2}\frac{-1+\ii\tau-\ii\sqrt{\tau^2-1}}{2\ii\tau}+ \int^{\infty}_{1}\frac{-\d\tau}{(k^2+\tau)^2}\frac{-1+\ii\tau+\ii\sqrt{\tau^2-1}}{2\ii\tau}\right]\nonumber\\
&=-\frac{1}{16\pi^2}\int^{\infty}_{1}\frac{\sqrt{\tau^2-1}}{\tau}\frac{\d\tau}{(k^2+\tau)^2}\,.
\end{align}
We can subsequently integrate this expression from $0$ to $k^2$, finding
\begin{eqnarray}\label{int11}
F(k^2)-F(0)=\frac{1}{16\pi^2}\int_1^{\infty}\frac{\sqrt{\tau^2-1}}{\tau}\left(\frac{1}{\tau+k^2}-\frac{1}{\tau}\right)\d\tau\,,
\end{eqnarray}
or, by restoring the units,
\begin{eqnarray}\label{int11b}
F(k^2)-F(0)=\frac{1}{16\pi^2}\int_{2\sqrt{2}\theta^2}^{\infty}\frac{\sqrt{\tau^2-8\theta^4}}{\tau}\left(\frac{1}{\tau+k^2}-\frac{1}{\tau}\right)\d\tau\,.
\end{eqnarray}
From this expression, we notice the importance of the subtraction of $F(0)$ to find a finite result, otherwise we would find a divergent spectral integral. The spectral density can be read off from \eqref{int11b},
\begin{equation}\label{int12b}
\rho(\tau)=\frac{1}{16\pi^2}\frac{\sqrt{\tau^2-8\theta^4}}{\tau}\,,
\end{equation}
which is clearly positive for $\tau\geq 2\sqrt{2}\theta^2$, thus showing that the correlation function has a well defined probabilistic interpretation also in $d=4$.\\
\\
Finally, an explicit integration of \eqref{int11} leads to
\begin{eqnarray}\label{int12}
F(k^2)-F(0)=\frac{1}{16\pi^2}\left(1-\frac{\pi}{2k^2}+\frac{\sqrt{1-k^4}}{k^2}\mathrm{arccos}(k^2)\right)\,.
\end{eqnarray}
In Figure \ref{6fig1}, we have displayed the (rescaled) real and imaginary part of $F(k^2)$. The cut for $z\in[-\infty,-1]$ is clearly visible.

\noindent \textbf{Remark} The same calculation can be repeated in the case of two particles with real mass $\mu$,  where the spectrum is found to begin at the threshold $4\mu^2 = (\mu + \mu)^2$. In the present case we have an analogous situation for the $i$-particles, even though they have complex masses: $2\sqrt{2}\theta^2 = \left( \sqrt{\ii \sqrt{2} \theta^2} + \sqrt{-\ii \sqrt{2} \theta^2} \right)^2$.

\subsection{Intermezzo: A closer look at the analytic continuation by means of the spectral representation}\label{details}
In this section, we shall give a closer look at analytic continuation of functions, and we shall demonstrate how careful one should be when working with complex functions.

\subsubsection{The case of complex masses}
Let us again start with the expression \eqref{o1c}, whereby for simplicity we shall work in $d=2$, and set $2\sqrt{2} \theta^2 = 1$
\begin{equation}\label{h1}
\Braket{ O_1(k) O_1(-k) }=\frac{1}{4\pi^2}\int \d^2p \frac{1}{\vec{p}^2-\ii/2}\frac{1}{\vec{p}^2+\vec{k}^2-2\vec{p}\cdot\vec{k}+\ii /2}\,.
\end{equation}
For any real external momentum vector $\vec{k}$, so that $k^2>0$, the integral \eqref{h1} certainly makes sense.\\
\\
Now let us recall that in the previous section, we have rewritten \eqref{h1} in terms of \eqref{fey-parD2}, namely
\begin{equation}\label{ha1}
\Braket{ O_1(k) O_1(-k) } = \frac{1}{4\pi}\int_0^1 \frac{\d x}{xk^2-x^2k^2+\ii x-\ii/2}\,,
\end{equation}
as the Feynman trick is certainly valid for $k^2>0$. Next, we have further manipulated this integral, whereby still assuming that $k^2 >0$. We have found that
\begin{equation}\label{h6}
  \Braket{ O_1(k) O_1(-k) }=\frac{1}{2\pi}\int_1^{\infty}\frac{\d\tau}{\tau+k^2}\frac{1}{\sqrt{\tau^2-1}}=\frac{1}{2\pi}\frac{\mathrm{arccos}(k^2)}{\sqrt{1-k^4}} = F_1(k^2)\,,
\end{equation}
which has branch cut along the negative axis. Here, we have called this result $F_1(k^2)$\\
\\
However, we could also continue with equation \eqref{ha1}. By rewriting it as
\begin{equation}\label{g2}
\Braket{ O_1(k) O_1(-k) } = \frac{1}{4\pi}\int_0^1 \frac{\d x}{xk^2-x^2k^2+\ii x-\ii/2}=\frac{1}{4\pi}\int_0^1 \d x\frac{1}{\sqrt{k^4-1}}\left(\frac{1}{x-x_+}-\frac{1}{x-x_-}\right)\;,
\end{equation}
with $x_{\pm}=\frac{\ii+k^2\mp\sqrt{k^4-1}}{2k^2}$ we can integrate this expression exactly by making use of \cite{'tHooft:1978xw}
\begin{equation}\label{g1}
    \int_0^1\frac{\d x}{ax+b}=\frac{1}{a}\ln\frac{a+b}{b}\,,
\end{equation}
valid for any complex number $a$ and $b$.  The ill-definedness of the integral in the l.h.s. of \eqref{g1} for $\frac{-b}{a}\in[0,1]$ corresponds exactly to the branch cut of the $\ln$ in the r.h.s. of \eqref{g1}. Applying this formula twice on \eqref{g2} yields
\begin{equation}
\Braket{ O_1(k) O_1(-k) } = \frac{1}{4\pi}\frac{1}{\sqrt{k^4-1}}\left[\ln\left( \ii\left(k^2+\sqrt{k^4-1}\right)\right)-\ln \left(\ii\left(k^2-\sqrt{k^4-1}\right)\right)\right] = F_2(k^2)\;.
\end{equation}
We call this result $F_2(k^2)$.\\
\\
Now we have obtained something very peculiar. On the positive $k^2$-axis, one verifies that $F_1(k^2)=F_2(k^2)$, which is necessary for consistency as both derivations we have done, are surely valid for $k^2>0$. However, the situation changes drastically in the complex $k^2$-plane. For $\mathrm{Re}(k^2)>0$, it stills holds that $F_1(k^2)=F_2(k^2)$, but the functions differ for $\mathrm{Re}(k^2)\leq 0$. Two questions arise:
\begin{itemize}
\item How can we explain this difference?
\item More importantly, how can be know which function is the correct analytic continuation of \eqref{h1}?
\end{itemize}

\begin{figure}[h]
  \begin{center}
\includegraphics[width=7cm]{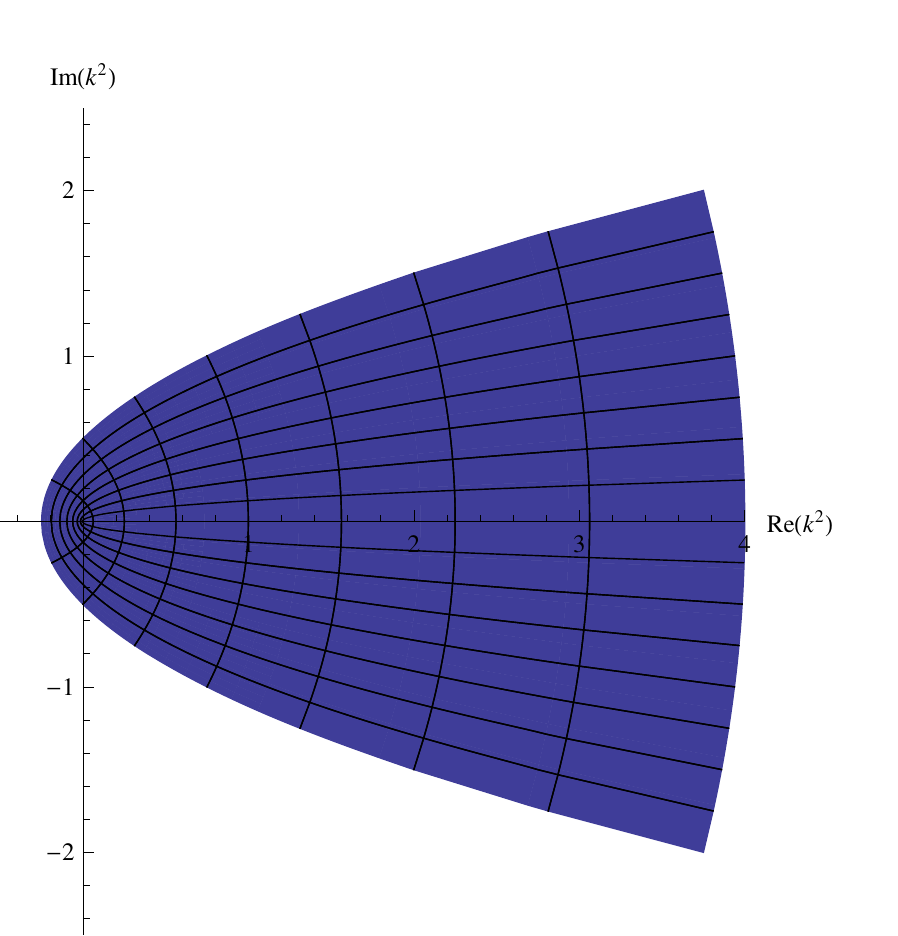}
     \end{center}
  \caption{Parabolic region in the complex $k^2$ plane where the original momentum integral $O(k^2)$ exists. There is a small region in the left half plane where $F(k^2)$ is well-defined, even without invoking analytical continuation. A priori, expression \eqref{h2} could only have a branch cut on the negative axis starting from $k^2 < -1/4$ and on the imaginary axis for $|\mathrm{Im}(k^2)|\geq 1/2$.}\label{cone}
\end{figure}

\noindent To answer the first question, this difference can be traced back to the branch cut of $F_2(k^2)$, which is given by the complete imaginary axis, while $F_1(k^2)$ only has a branch cut on a part negative real axis. Therefore, we can understand that for $\mathrm{Re}(k^2)>0$, $F_1(k^2)$ has to be equal to $F_2(k^2)$. Indeed, two complex functions which are analytic in an open region and equal for a converging series of points inside this open region, are always equal\footnote{This is a theorem of complex analysis.}. As here, $F_1 = F_2$ on the positive $k^2$ axis, they have to be equal $\forall\ \mathrm{Re} (k^2)>0$ as this plane forms an open region. However, this open region is limited to the right side of the complex plane, due to the branch cut of $F_2(k^2)$ smeared out over the entire imaginary axis. Therefore, points on the left side of the complex plane, i.e.~$\mathrm{Re} (k^2)>0$ do not necessarily have to lead to coinciding values of $F_1$ and $F_2$.\\
\\
The second question is a bit more involved to answer. We shall argue that only $F_1(k^2)$ obtained via the spectral representation gives a decent analytic continuation of the original momentum integral \eqref{h1}.  Let us thus choose an external momentum vector $\vec{k}$ which can be complex, or
\begin{equation}
\vec{k}=\mathrm{Re}(\vec{k}) + \ii \mathrm{Im}(\vec{k})\equiv\vec{k}_R + \ii \vec{k}_I\,.
\end{equation}
Let us take start with the most general $\vec k$, namely $\vec{k}=(u+\ii v,u'+\ii v')$. From expression \eqref{h1}, we observe that we have the $O(2)$ rotational symmetry for $\vec k$, as we can always rotate $\vec{k}_R$ and $\vec{k}_I$ simultaneously. Therefore, we can already choose without loss of generality $\vec{k}=(u+\ii v,u')$, since we can always rotate our $\vec{k}_R$ and $\vec{k}_I$ simultaneously so that $\vec k_I$ coincides with the real axis.  The corresponding integral for $\vec{k}=(u+\ii v,u')$ now reads
\begin{multline*}
    \Braket{ O_1(k) O_1(-k) } \\
    =\frac{1}{4\pi^2}\int \d p_x \d p_y \frac{1}{p_x^2+p_y^2-\ii/2}\frac{1}{p_x^2+p_y^2+u^2-v^2+2\ii uv+u'^2-2p_x(u+\ii v)-2p_yu'+\ii/2}\,,
\end{multline*}
but the simple shift $p_y\to p_y+u$, brings us to $\vec{k}=(u+\ii v,0)$. Therefore, from the beginning, we can set $\vec{k}=(u+\ii v,0)$ without loss of generality.\\
\\
Now that we have restricted ourselves to complex momenta of the type $\vec{k}=(u+\ii v,0)$, which allows to reach any value of
\begin{equation}\label{h3}
k^2=\vec{k}^2=u^2-v^2+2\ii uv\;.
\end{equation}
we can rewrite \eqref{h1} as
\begin{multline}\label{h2}
    \Braket{ O_1(k) O_1(-k) }=\frac{1}{4\pi^2}\int \d p_x \d p_y \frac{1}{p_x^2+p_y^2-\ii /2}\\ \times \frac{1}{p_x^2+p_y^2+u^2-v^2+2\ii uv-2p_x(u+\ii v)+\ii/2}= F(k^2)\,.
\end{multline}
whereby we have renamed this expression $F(k^2)$. Although we already chose $k^2$ to be the argument of $F(k^2)$, this is not a priori clear when $\vec{k}$ is not real. However, we  notice that the integral in the r.h.s.~of \eqref{h1}, if it exists at least\footnote{We shall shortly see that such $\vec{k}$ do exist.}, will define an analytic function of $k\equiv u+\ii v$, which can be easily checked by means of the Cauchy-Riemann equations. Since we still have the $O(2)$ rotational symmetry, as we can rotate $\vec{k}_R$ and $\vec{k}_I$ simultaneously, which thereby defines a real rotation of the complex vector $\vec{k}=k\vec{e}_x$, it appears that \eqref{h1} must be a function of $\vec{k}^2$.\\
\\
To proceed, let us check when the integral \eqref{h2} is well-defined. This is only the case, when the integrand of \eqref{h2} is free of poles whereby we notice that $\vec{p}\in\mathbb{R}^2$. Such poles can only appear when\footnote{We can exclude here the $v=0$ case as this corresponds to the anyhow well-defined case of real external momentum.} the imaginary part in the denominator is equal to zero while simultaneously the real part is equal to zero
\begin{eqnarray}\label{h4}
    p_x=u+\frac{1}{4v}\,,\nonumber\\
    p_x^2 + p_y^2 + u^2 - v^2 - 2 p_x u = 0\;.
\end{eqnarray}
We can combine both equations into
\begin{equation}\label{h5}
    \frac{1}{16v^2}-v^2+p_y^2=0\,.
\end{equation}
Hence, if we take $|v|<1/2$, with $u$ arbitrary, the equation \eqref{h5} never can be solved for any $p_y^2>0$ and thus the integral \eqref{h2} is will exist as such. It is interesting to notice that this does \emph{not} mean that the Feynman trick is applicable for any such vector $\vec{k}$. Nevertheless, the integral has a well-defined value. As such, it must coincide with other ways to compute or define it.\\
\\
We can now discriminate between $F_1(k^2)$ and $F_2(k^2)$ by taking a test value in the left half complex $k^2$ plane, such that the original integral \eqref{h2} exists. Its value should then coincide with either $F_1(k^2)$ or $F_2(k^2)$.\\
\\
Firstly, we can immediately motivate that $F_2(k^2)$ cannot be correct in the whole complex plane. Looking at \eqref{h3}, it is clear that $k^2\in \ii \mathbb{R}$ if $u=\pm v$, in which case we have $k^2=\pm2\ii v^2$. But as we have shown, for $|v|<1/2$, the original integral is well-defined, meaning that there cannot be a cut on the \emph{whole} imaginary axis, whereas $F_2(k^2)$ has a cut for any $k^2\in \ii\mathbb{R}$. This shows that $F_2(k^2)$ cannot be the analytic continuation of the original momentum integral \eqref{h1}.\\
\\
Secondly, it is also now clear that that $F(k^2)$ coincides with $F_1(k^2)$ in the region $F(k^2)$ is well defined. In Figure \ref{cone}, the shaded parabolic region corresponds to the values of $k^2\in\mathbb{C}$  for which the integral \eqref{h2} is certainly well-defined and analytical. As this is an open region, and $F_1(k^2)$ is also analytical in this region, $F_1(k^2)$ and $F(k^2)$ have to coincide.

\begin{figure}[h]
  \begin{center}
    \subfigure[]{\includegraphics[width=4.8cm]{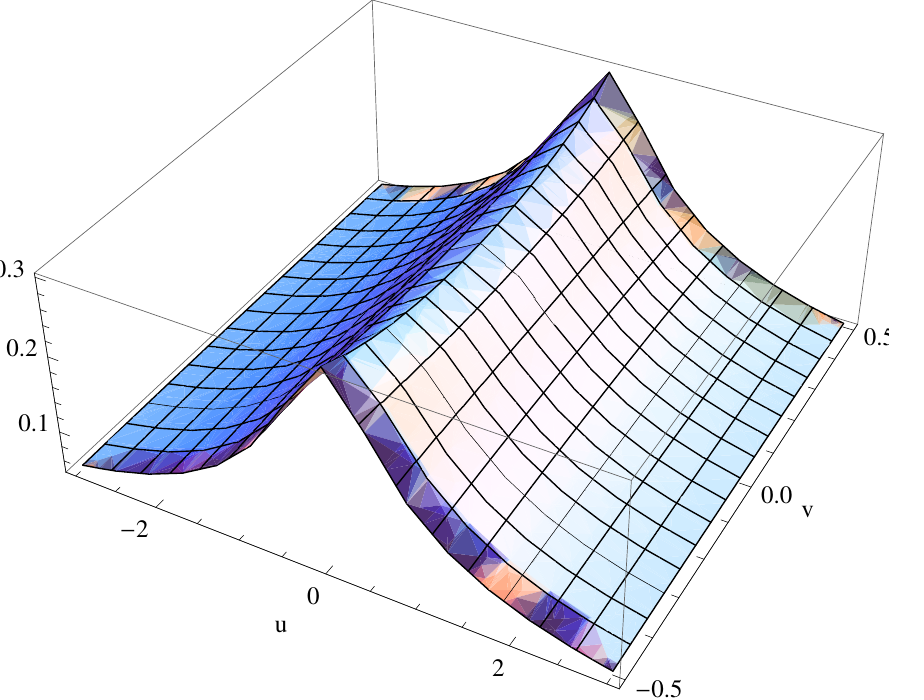} \label{fig3a}}
    \hspace{0.5cm}
    \subfigure[]{\includegraphics[width=4.8cm]{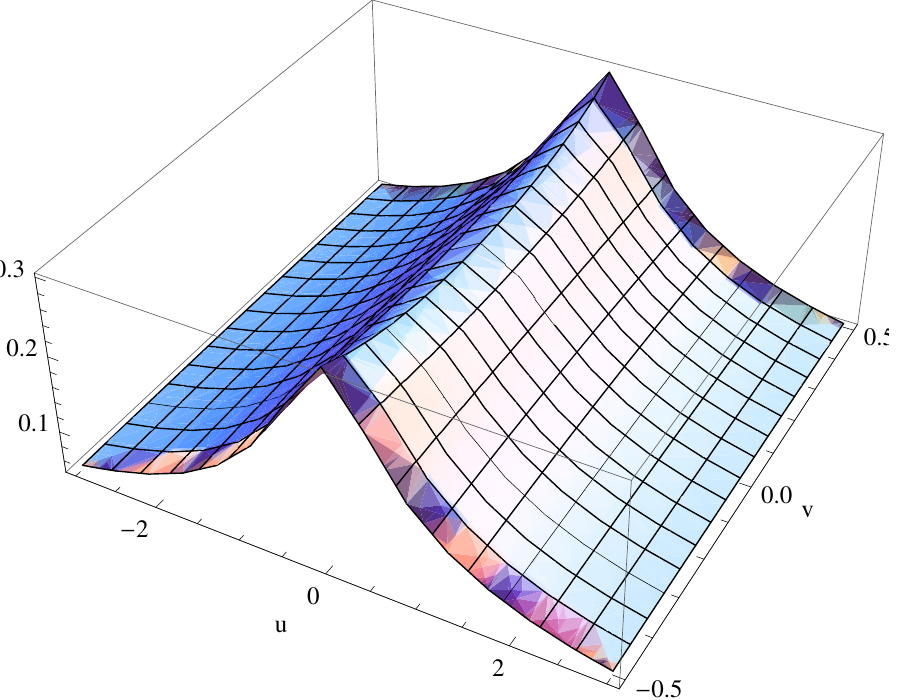}\label{fig3b}}
   \hspace{0.5cm}
    \subfigure[]{\includegraphics[width=4.8cm]{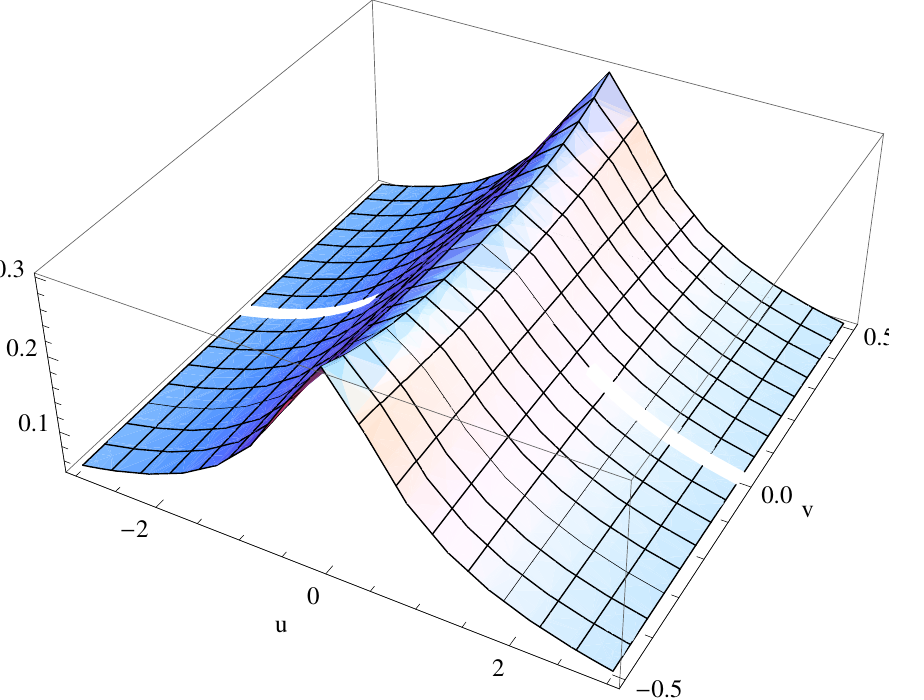}\label{fig3c}}
     \end{center}
  \caption{$\mathrm{Re}(O)$, $\mathrm{Re}(\hat{O})$ and $\mathrm{Re}(F_1)$. }
\end{figure}

\begin{figure}[h]
  \begin{center}
    \subfigure[]{\includegraphics[width=4.8cm]{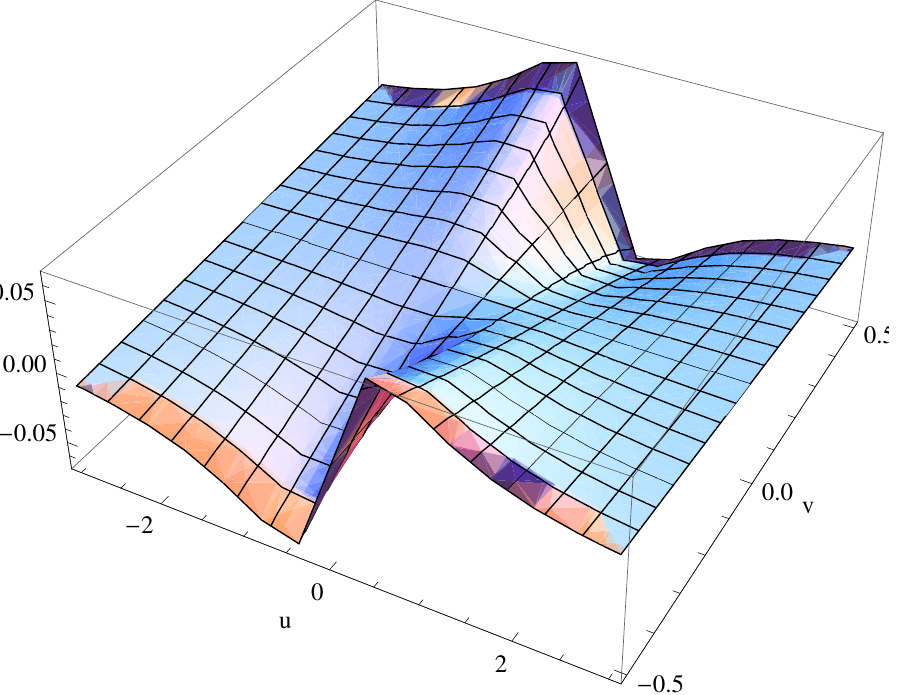} \label{fig4a}}
    \hspace{0.5cm}
    \subfigure[]{\includegraphics[width=4.8cm]{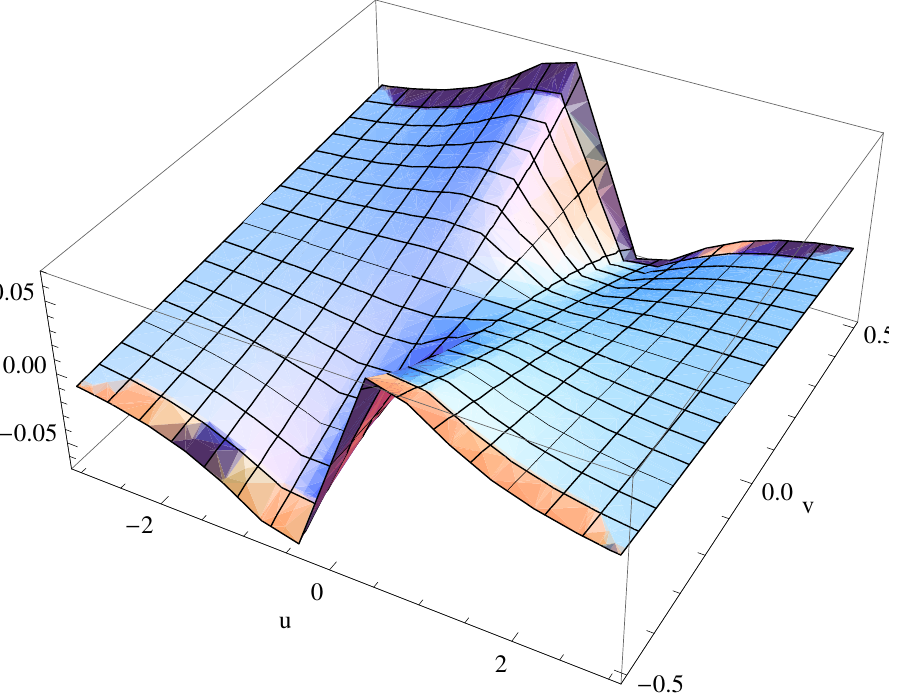}\label{fig4b}}
    \hspace{0.5cm}
    \subfigure[]{\includegraphics[width=4.8cm]{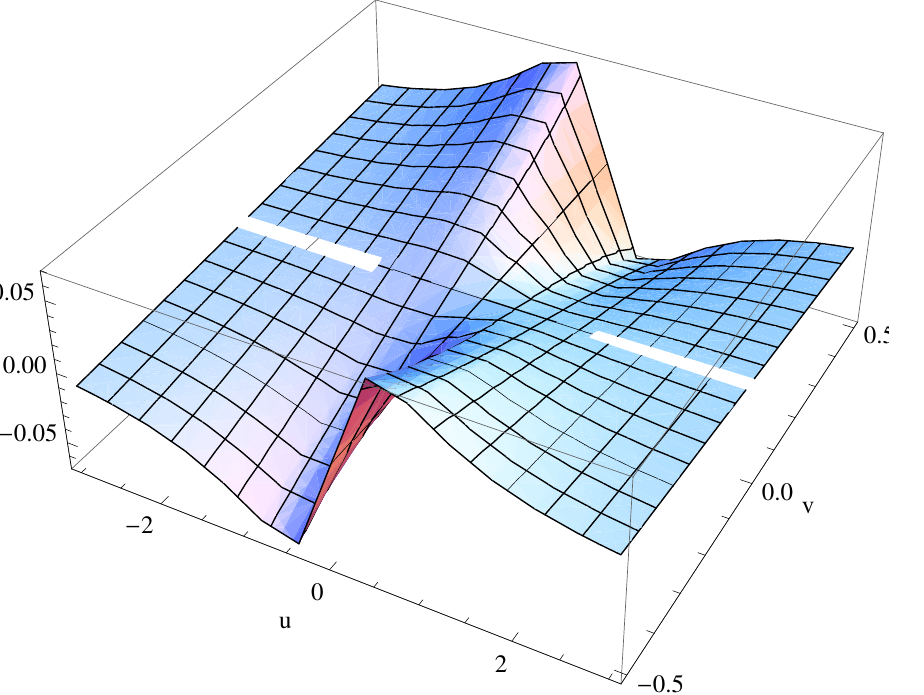}\label{fig4c}}
     \end{center}
  \caption{$\mathrm{Im}(O)$, $\mathrm{Im}(\hat{O})$ and $\mathrm{Im}(F_1)$. }
\end{figure}

\noindent We have provided a nontrivial verification of the correctness of the analytically continued function $F_1(k^2)$. In fact, much more can be learned from our analysis. Let us take expression \eqref{h3} into reconsideration,
\begin{equation}\label{h3again}
k^2=u^2-v^2+2\ii uv\,.
\end{equation}
The results obtained do, of course, not mean that there \emph{must} be a cut for $k^2\in[-\infty,-1/4]$ or that there is a ``cut region'' outside of the displayed parabola. But we have clearly demonstrated that $F_1(k^2)$ is a good analytical continuation.\\
\\
To close this section, we mention that we could also have started with the alternative definition
\begin{equation}\label{h14}
    \hat{O}(k^2)=\frac{1}{4\pi^2}\int d^2p \frac{1}{\vec{p}^2+\vec{k}^2-2\vec{p}\cdot\vec{k}-\ii/2}\frac{1}{\vec{p}^2+\ii/2}\,,
\end{equation}
which corresponds to switching the momenta running in the two legs. For real external momenta $\vec{k}$, we obviously have $O(k^2)=\hat{O}(k^2)$, but for complex $\vec{k}$, we can no longer perform a translation on the real integration momentum $\vec{p}$ to prove this. \\
\\
However, again setting $\vec{k}=(u+\ii v,0)$, it is easily checked that $|v|<1/2$ is sufficient to also guarantee that $\hat{O}(k^2)$ is well-defined, while from Figure~\ref{fig3a}, \ref{fig3b}, \ref{fig4a}, \ref{fig4b}, it is clear that  $O(k^2)=\hat{O}(k^2)$ for $u$ arbitrary, $|v|<1/2$. For completeness, we have also shown
$F_1(k^2)$ in Figure.~\ref{fig3c}, \ref{fig4c}, which illustrates that indeed $O(k^2)=\hat{O}(k^2)=F_1(k^2)$ over the parabolic $k^2$-region shown in FIG.~2. To avoid confusion, we point out that we have plotted $O(u,v)\equiv O(k^2)$, with $k^2=u^2-v^2+2iuv$, and analogously for the other functions.

\noindent Although we focused exclusively on the $d=2$ case in this section, similar conclusions can be drawn for $d=3$ or $d=4$.

\subsubsection{The case of two real masses}\label{detailsb}
\begin{figure}[H]
  \begin{center}
\includegraphics[width=7cm]{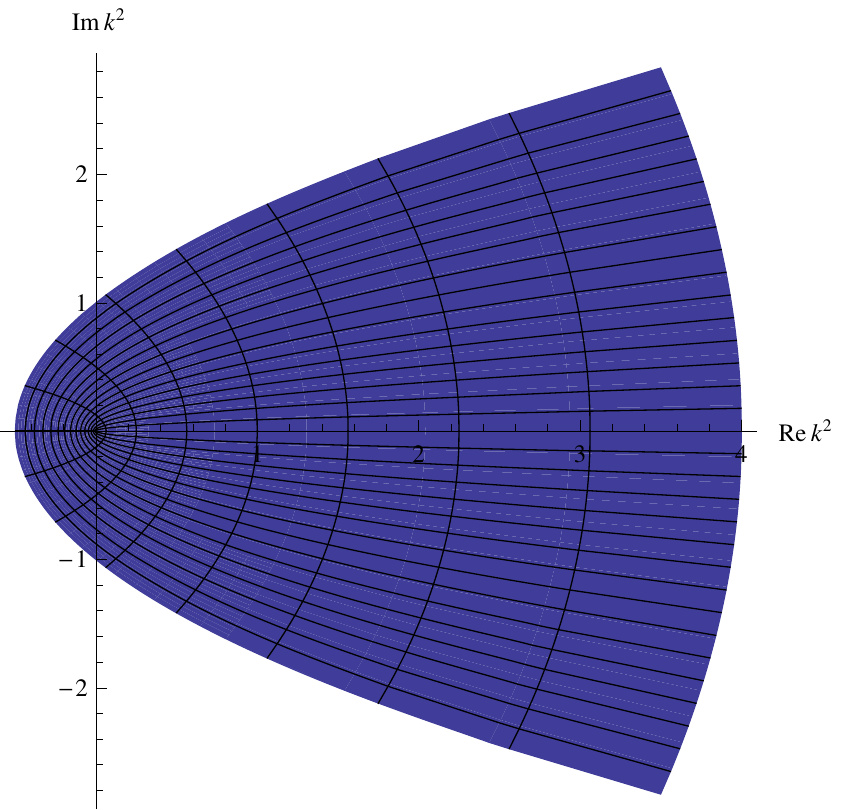}\label{cone2}
     \end{center}
  \caption{Parabolic region in the complex $k^2$ plane  where the momentum integral $O_2(k^2)$ exists with $m_2^2 = 1/2$.}
\end{figure}

In order to corroborate the previous nontrivial conclusions about a relatively simple Feynman integral with complex masses, we find it instructive to also include a similar analysis of the probably more familiar case of two real and positive masses, being $m_1$ and $m_2$. This has been investigated in great detail in e.g.~\cite{Itzykson:1980rh}, albeit in Minkowski space. The analog of \eqref{h1} is given by the following correlation function,
\begin{equation}\label{h10}
    O_2(k^2)=\frac{1}{4\pi^2}\int d^2p \frac{1}{\vec{p}^2+ m_1^2}\frac{1}{\vec{p}^2+\vec{k}^2-2\vec{p}\cdot\vec{k}+ m_2^2 }\,.
\end{equation}
Translating the results of \cite{Itzykson:1980rh} to Euclidean space, it was shown that this function of $k^2$ has a positive spectral density in combination with a branch cut on the negative real $k^2$-axis starting from $-(m_1+ m_2)^2$ until $-\infty$.\\
\\
Let us now also investigate the region where this integral actually exists. We can again restrict ourselves to complex momenta of the type $\vec{k} = (u + iv,0)$ without loss of generality. Inserting this in equation \eqref{h10}, we obtain,
\begin{equation}\label{h2a}
    O_2(k^2)=\frac{1}{4\pi^2}\int \d p_x \d p_y \frac{1}{p_x^2+p_y^2+ m_1^2}\frac{1}{p_x^2+p_y^2+u^2-v^2+2iuv-2p_x(u+\ii v)+m_2^2}\,.
\end{equation}
Let us check when poles can emerge. The first denominator is always positive, however, the second denominator can have poles when the imaginary part vanishes,
\begin{eqnarray}\label{im}
2 \ii u v - 2 \ii p_x v = 0 \Leftrightarrow  v = 0 \quad\text{   or  }\quad u = p_x\;,
\end{eqnarray}
simultaneously with the real part being equal to zero,
\begin{eqnarray}
p_x^2 + p_y^2 + u^2 - v^2 + 2 p_x u + m_2^2 = 0\;.
\end{eqnarray}
Inserting the first solution of equation \eqref{im},\textit{i.e.}~$v=0$, in the equation above, we find
\begin{eqnarray}
(p_x + u)^2 + p_y^2 = - m_2^2\;,
\end{eqnarray}
which can never be fulfilled. The second solution, \textit{i.e.}~$u=p_x$, results in
\begin{eqnarray}
 p_y^2  &=& v^2 - m_2^2\;.
\end{eqnarray}
Therefore, no poles shall occur for $|v| < m_2$, while for $|v|\geq m_2$, the integral \eqref{h10} becomes ill-defined.
This is important as it shows us that also in the well-studied case of two real masses, the original integral \eqref{h10} is also only well-defined in a certain region of the complex plane, here displayed in Figure \ref{cone2}. Consequently, one also needs to perform an analytic continuation outside this region, just as in the case of pure complex masses.

\subsection{Higher order: a Two loop example}\label{higherloop}
\begin{figure}[H]
  \begin{center}
\includegraphics[width=5cm]{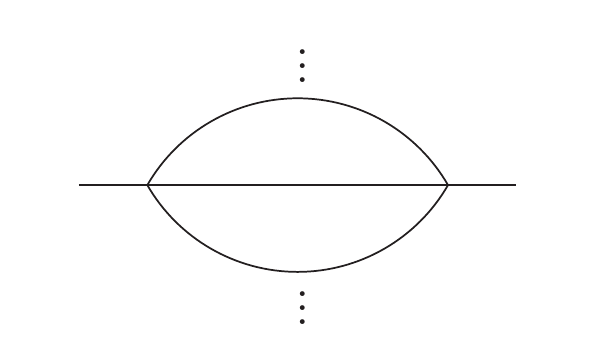}
     \end{center}
     \vspace{-1cm}
  \caption{The water melon diagrams.}\label{watermelon}
\end{figure}

In this easy toy model, we can also quite easily investigate higher loop diagrams. As there are no interactions, only one type of diagrams can appear in higher order, namely diagrams of the water melon type, see Figure~\ref{watermelon}. This shall allow us to construct a rather interesting iterative procedure valid for an $n$-loop integral. Unfortunately, this kind of construction is not applicably to the more complicated GZ action, which does have interactions.

\noindent In order to obtain an example of a  two loop correlation function, we consider the local operator
\begin{align}
O_2(x) = \lambda(x) \eta(x) U(x) \;. \label{o2}
\end{align}
For the correlation function $\Braket{O_2(k) O_2(-k) }$  we find with the help of \eqref{nodig1} and \eqref{nodig2},
\begin{align}\label{blub}
\Braket{O_2(k) O_2(-k) } = \int {\frac{\d^4p}{(2\pi)^4}}  {\frac{\d^4q}{(2\pi)^4}} \frac{1}{q^2} \frac{1}{p^2+\ii\sqrt{2}\theta^2} \frac{1}{(k-q-p)^2-\ii\sqrt{2}\theta^2}  \;,
\end{align}
where $k^2$ is the Euclidean external momentum. In order to evaluate it, we rewrite it as
\begin{align}\label{6rewrite}
\Braket{O_2(k) O_2(-k) } = \int {\frac{\d^4q}{(2\pi)^4}} \frac{1}{q^2} \left( \int {\frac{\d^4p}{(2\pi)^2}}   \frac{1}{p^2+\ii\sqrt{2}\theta^2} \frac{1}{(k-q-p)^2-\ii\sqrt{2}\theta^2} \right) \;.
\end{align}
Now we can use the results from section \ref{chap6sectspectrd4}, see also expression \eqref{int11b}, namely\footnote{From now on, we shall not bother about potential subtractions to make it well-defined. These can always be obtained by taking a suitable number of derivatives w.r.t.~the external momentum $k^2$.}
\begin{align}\label{6bla}
\Braket{ O_1(k) O_1(-k) } = \int \frac{\d^4 p}{(2\pi)^4}\; \frac{1}{(k-p)^2-\ii \sqrt{2}\theta^2} \frac{1}{p^2+\ii\sqrt{2}\theta^2} = \int_{2 \sqrt{2} \theta^2}^{\infty} \d \tau \; \rho({\tau}) \frac{1}{\tau+ k^2} \;,
\end{align}
with $\rho(\tau) = \frac{1}{16\pi^2} \frac{\sqrt{\tau^2 - 8 \theta^4}}{\tau}$, so we find for expression \eqref{6rewrite}
\begin{align}\label{6rewrite2}
\Braket{O_2(k) O_2(-k) } = \int_{\tau_{0}}^{\infty} \d\tau \; \rho({\tau}) \left( \int {\frac{\d^4q}{(2\pi)^4}} \frac{1}{q^2} \frac{1}{\tau+ (k-q)^2}\right) \,.
\end{align}
whereby $\tau_0 = 2 \sqrt{2} \theta^2$. The $q$-integral becomes now straightforward. It corresponds to the one loop integral in which one particle is massless, and the other one has a real and positive mass $\tau$. Such integrals also have a spectral representation with positive spectral density, as we have calculated in appendix \ref{sigma}\ref{appendixspectral}, namely
\begin{equation}
\int {\frac{\d^4 p}{(2\pi)^4}} \frac{1}{p^2} \frac{1}{(k-p)^2 + \tau} =  \int_{m^2}^{\infty}  \d s\; \rho_{1}(s) \frac{1}{s+k^2}  \,,
\end{equation}
with $\rho_1(s) = 1- \frac{\tau}{s} $. Applying this formula on \eqref{6rewrite2} yields,
\begin{align}\label{j2}
\Braket{O_2(k) O_2(-k) }  = \int_{\tau_{0}}^{\infty} \d\tau \; \rho({\tau}) \int_{\tau}^{\infty}  \d s \;\rho_{1}(s) \frac{1}{s+k^2}\,.
\end{align}
This is not yet in the form of a spectral representation. By switching the order of integration, we can however reexpress the double integral as
\begin{equation}\label{dd1}
 \Braket{O_2(k) O_2(-k) } = \int_{\tau_0}^\infty \d s \frac{1}{s+k^2}\underbrace{\int_{\tau_0}^{s}\d\tau\rho_1(s)\rho(\tau)}_{\rho_2(s)}\;.
\end{equation}
This means that the spectral density is given by
\begin{multline}\label{dd2}
\rho_2(s) =     \int_{\tau_0}^{s}\d\tau\rho(\tau)\rho_1(s) = \int_{2 \sqrt{2} \theta^2}^{s} \d\tau  \frac{1}{16\pi^2} \frac{\sqrt{\tau^2 - 8 \theta^4}}{\tau} \left( 1- \frac{\tau}{s} \right)
     = \frac{1}{(16 \pi^2)^2} \Biggl(\frac{1}{2} \sqrt{s^2- 8 \theta ^4} \\+ 2  \sqrt{2}\theta ^2 \left(\arctan \left[\frac{2 \sqrt{2} \theta ^2}{\sqrt{s^2-8 \theta ^4}}\right] - \frac{\pi}{2}  \right) +  \frac{4 \theta^4}{ s} \left(  \ \ln \frac{s+\sqrt{s^2-8 \theta ^4}}{2 \sqrt{2} \theta^2} \right) \Biggr)\,.
\end{multline}
In conclusion, (1) the two loop correlation function \eqref{blub} can also be reexpressed in a spectral form, and moreover (2) the spectral density is positive. Indeed, for $s\in[\tau_0,\infty]$, we have $\rho(\tau)\geq 0$ due to the positivity of $\rho(\tau)$ over the interval $[\tau_0,\infty]$. We shall thus have that $\rho_2(s)\geq 0$ itself if $\rho_1(s)\geq 0$ for $s\in[\tau_0,\infty]$. The latter turns out to be true, as $\rho_1(s)$ is defined for all $\tau$ with $\tau\geq \tau_0$, where it is positive. Also one can check this from the explicit expression in the r.h.s.~of \eqref{dd2}.\\
\\
With the same techniques, it is also possible to prove this at three loops, and beyond three loops. However, as this is not applicable to the GZ action, we refer to \cite{Baulieu:2009ha} for the details.

\section{The Gribov-Zwanziger action\label{secttheGZaction}}
Let us discuss here how $i$-particles can arise in the Gribov-Zwanziger action. In what follows, we shall limit ourselves to evaluate correlation functions of suitable composite operators at one loop order only.

\subsection{Introducing the $i$-particles}
To introduce the $i$-particles in the Gribov-Zwanziger action, it suffices thus to consider the quadratic part of the GZ action, see expression \eqref{GZstart}, namely
\begin{align}
S_{\GZ}^{\rm quad} = \int \d^4x\; \left( \frac{1}{2} A^a_{\mu} (-\partial^2\delta_{\mu\nu}-\p_\mu\p_\nu) A^a_{\nu} + {\overline \varphi}^{ab}_{\mu}\partial^2\varphi^{ab}_{\mu} -\gamma^2\,g\,f^{abc}A_\mu^{a}(\varphi_\mu^{bc}+{\overline\varphi}_\mu^{bc}) +b^a \p_\mu A_\mu^a\right) \;,
\end{align}
We proceed by decomposing the fields $(\varphi_\mu^{ab}, {\overline\varphi}_\mu^{ab})$ in symmetric and anti-symmetric components in color space,
\begin{align}
\varphi^{ab}_{\mu} & = \varphi^{[ab]}_{\mu} + \varphi^{(ab)}_{\mu}  \;,  &
\varphi^{[ab]}_{\mu}  & = \frac{1}{2} \left( \varphi^{ab}_{\mu} - \varphi^{ba}_{\mu} \right) \;, &
\varphi^{(ab)}_{\mu}  & = \frac{1}{2} \left( \varphi^{ab}_{\mu} + \varphi^{ba}_{\mu} \right) \;,\nonumber\\
{\overline \varphi}^{ab}_{\mu} & = {\overline \varphi}^{[ab]}_{\mu} + {\overline \varphi}^{(ab)}_{\mu}  \;, &
{\overline \varphi}^{[ab]}_{\mu}  & = \frac{1}{2} \left( {\overline \varphi}^{ab}_{\mu} - {\overline \varphi}^{ba}_{\mu} \right) \;, &
{\overline \varphi}^{(ab)}_{\mu}  & = \frac{1}{2} \left( {\overline \varphi}^{ab}_{\mu} + {\overline \varphi}^{ba}_{\mu} \right) \;.
\end{align}
Thus
\begin{multline}
S_{\GZ}^{\rm quad} = \int \d^4x\; \left( \frac{1}{2} A^a_{\mu} (-\partial^2\delta_{\mu\nu}-\p_\mu\p_\nu) A^a_{\nu} + {\overline \varphi}^{[ab]}_{\mu}\partial^2\varphi^{[ab]}_{\mu} + {\overline \varphi}^{(ab)}_{\mu}\partial^2\varphi^{(ab)}_{\mu}\right.\\
\left.-\gamma^2\,g\,f^{abc}A_\mu^{a}(\varphi_\mu^{[bc]}+{\overline\varphi}_\mu^{[bc]}) +b^a \p_\mu A_\mu^a\right) \;.
\end{multline}
A first  step towards diagonalization of this expression is achieved by setting
\begin{align}\label{vu}
\varphi^{[ab]}_{\mu} = \frac{1}{\sqrt{2}} \left( U^{[ab]}_{\mu} + \ii V^{[ab]}_{\mu} \right) \;, \nonumber \\
{\overline \varphi}^{[ab]}_{\mu} = \frac{1}{\sqrt{2}} \left( U^{[ab]}_{\mu} - \ii V^{[ab]}_{\mu} \right)  \;,
\end{align}
so that
\begin{align}
S_{\GZ}^{\rm quad} &= \int \d^4x\; \left( \frac{1}{2} A^a_{\mu}(-\partial^2\delta_{\mu\nu}-\p_\mu\p_\nu) A^a_{\nu} + \frac{1}{2} U^{[ab]}_{\mu}\partial^2 U^{[ab]}_{\mu} - \sqrt{2} g\gamma^2 f^{abc}A^a_{\mu} V^{[bc]}_{\mu}+b^a \p_\mu A_\mu^a\right.\nonumber\\
&+ \left.  \frac{1}{2} V^{[ab]}_{\mu}\partial^2V^{[ab]}_{\mu} +  {\overline \varphi}^{(ab)}_{\mu}\partial^2\varphi^{(ab)}_{\mu} \right) \;. \label{quad2}
\end{align}
From expression \eqref{quad2} one sees that the gauge field $A^a_{\mu}$ mixes with the adjoint projection of $U^{[ab]}_{\mu}$, obtained by employing the following decomposition
\begin{align}
U^{[ab]}_{\mu}= \frac{1}{N} f^{abp}f^{pmn} U^{[mn]}_{\mu} + \left( U^{[ab]}_{\mu} - \frac{1}{N} f^{abp}f^{pmn} U^{[mn]}_{\mu} \right) = f^{abp} U^{p}_{\mu} + S^{[ab]}_{\mu} \;, \label{dec}
\end{align}
where
\begin{align}
U^{p}_{\mu}= \frac{1}{N} f^{pmn} U^{[mn]}_\mu \;, \label{adjp}
\end{align}
stands for the adjoint projection of $U^{[ab]}_{\mu}$ in color space, and
\begin{align}
S^{[ab]}_{\mu} = U^{[ab]}_{\mu} - \frac{1}{N} f^{abp}f^{pmn} U^{[mn]}_{\mu} \;, \label{indep}
\end{align}
denote the remaining independent components of $U^{[ab]}_{\mu}$ which are orthogonal to the tensor $f^{abc}$. In fact, making use of
\begin{align}
f^{abc} f^{dbc} = N \delta^{ad} \;, \label{ff1}
\end{align}
it is easily checked that
\begin{align}
f^{abc} S^{[ab]}_{\mu}=0 \;. \label{ort}
\end{align}
Therefore, expression \eqref{quad2} becomes
\begin{align}
  S_{\GZ}^{\rm quad}=  \int \d^4x  & \left( \frac{1}{2} A^a_{\mu} (-\partial^2\delta_{\mu\nu}-\p_\mu\p_\nu) A^a_{\nu} + \frac{N}{2} U^{a}_{\mu}\partial^2 U^{a}_{\mu} -\sqrt{2} g \gamma^2 N A^a_\mu V^a_\mu  +b^a\p_\mu A_\mu^a\right) \nonumber \\
+ \int \d^4x & \left( \frac{1}{2} S^{[ab]}_{\mu} \partial^2 S^{[ab]}_{\mu} +  \frac{1}{2} V^{[ab]}_{\mu}\partial^2 V^{[ab]}_{\mu} +  {\overline \varphi}^{(ab)}_{\mu}\partial^2\varphi^{(ab)}_{\mu} \right) \;. \label{quad3}
\end{align}
Finally setting,
\begin{align}
A^{a}_{\mu} & = \frac{1}{\sqrt{2}} \left( \lambda^{a}_{\mu}+ \eta^{a}_{\mu} \right) \;, &
U^{a}_{\mu} & = \frac{-\ii}{\sqrt{2N}} \left( \lambda^{a}_{\mu}- \eta^{a}_{\mu} \right) \;, \label{fv}
\end{align}
$S_{\GZ}^{\rm quad}$ becomes
\begin{align}
  S_{\GZ}^{\rm quad} =& \int \d^4x  \left( \frac{1}{2} {\lambda}^{a}_{\mu} \left(-\partial^2+ \ii\sqrt{2N}g\gamma^2 \right) {\lambda}^{a}_{\mu} + \frac{1}{2} {\eta}^{a}_{\mu}\left(-\partial^2- \ii\sqrt{2N}g\gamma^2\right){\eta}^{a}_{\mu}  \right)\nonumber \\
&+ \int \d^4x  \left( \frac{1}{2} S^{[ab]}_{\mu} (-\partial^2) S^{[ab]}_{\mu} +  \frac{1}{2} V^{[ab]}_{\mu}(-\partial^2)V^{[ab]}_{\mu} +  {\overline \varphi}^{(ab)}_{\mu}\partial^2\varphi^{(ab)}_{\mu} \right) \;. \label{quad4} \nonumber\\
 &+\int \d^4x\left(\frac{1}{4} \left(\p_\mu\lambda_\mu^a\right)^2+\frac{1}{4}\left(\p_\mu\eta_\mu^a\right)^2 +\frac{1}{2}\left(\p_\mu\lambda_\mu^a\right) \left(\p_\nu\eta_\nu^a\right) +\frac{b^a}{\sqrt{2}}\p_\mu\lambda_\mu^a+\frac{b^a}{\sqrt{2}}\p_\mu\eta_\mu^a\right)\;.
\end{align}
The fields $\lambda^{a}_{\mu}, \eta^{a}_{\mu}$ shall describe the $i$-particles of the Gribov-Zwanziger action.\\
\\
We shall now calculate the (relevant) propagators of this action. Using the fields definitions, we find
\begin{eqnarray}\label{r5f}
  \lambda_\mu^a&=&\frac{1}{\sqrt{2}}A_\mu^a+\frac{\ii}{2\sqrt{N}}f^{abc}\left(\varphi_\mu^{bc}+\overline \varphi_\mu^{bc}\right)\;,   \nonumber\\
\eta_\mu^a&=&\frac{1}{\sqrt{2}}A_\mu^a-\frac{\ii}{2\sqrt{N}}f^{abc}\left(\varphi_\mu^{bc}+\overline \varphi_\mu^{bc}\right)\;,
\end{eqnarray}
so with the propagators of the original GZ action, see expression \eqref{summarypropGZ}, we can calculate
\begin{eqnarray}
    \braket{\lambda_\mu^a \lambda_\nu^b}&=& \Braket{\left[ \frac{1}{\sqrt{2}}A_\mu^a + \frac{\ii}{2\sqrt{N}} f^{ak\ell}\left(\varphi_\mu^{k \ell}-\overline \varphi_\mu^{k \ell}\right)\right] \left[ \frac{1}{\sqrt{2}}A_\nu^b+ \frac{\ii}{2\sqrt{N}}f^{bpq}\left(\varphi_\mu^{pq}-\overline \varphi_\mu^{pq}\right)\right] }\nonumber\\
    &=& \frac{1}{2} \Braket{A_\mu^a A_\nu^b} + \frac{\ii}{\sqrt{2N}} f^{bpq} \Braket{A_\mu^a \varphi^{pq}_\nu} + \frac{\ii}{\sqrt{2N}} f^{ak \ell} \Braket{A_\nu^b \varphi^{k\ell}_\mu}  \nonumber\\
    && -\frac{1}{2N} f^{ak\ell}f^{bpq} \left( \Braket{\varphi^{k\ell}_\mu \varphi^{pq}_\nu}  - \Braket{\varphi^{k\ell}_\mu \overline \varphi^{pq}_\nu} \right) \nonumber\\
    &=& \delta^{ab} \left( \frac{1}{p^2 + \ii \lambda^2}  P_{\mu\nu}  + \frac{1}{2} \frac{1}{p^2} L_{\mu\nu} \right)\;,
\end{eqnarray}
whereby
\begin{align}
P_{\mu\nu}(p) &= \delta_{\mu\nu} - \frac{p_{\mu} p_\nu}{p^2}\;, &  L_{\mu\nu}(p) &=  \frac{p_{\mu} p_\nu}{p^2} \;.
\end{align}
In an analogous fashion, we also find
\begin{eqnarray}
    \braket{\eta_\mu^a \eta_\nu^b}&=& \delta^{ab} \left( \frac{1}{p^2 - \ii \lambda^2}  P_{\mu\nu}  + \frac{1}{2} \frac{1}{p^2} L_{\mu\nu} \right) \;,\nonumber\\
     \braket{\lambda_\mu^a \eta_\nu^b}&=& -\delta^{ab}  \frac{1}{2} \frac{1}{p^2} L_{\mu\nu}\;.
\end{eqnarray}
Although, the propagators are neither completely color diagonal nor transverse, we shall still show that we can find good operators using these $i$-fields.

\subsection{Proposal for a  ``good'' operator}
Let us give here two examples of one loop correlation functions of composite operators constructed from the $i$-fields. To this order, one can introduce the $i$-field strengths defined by
\begin{align}\label{fs}
\lambda^{a}_{\mu\nu} & = \partial_{\mu}  \lambda^{a}_{\nu} -  \partial_{\nu}  \lambda^{a}_{\mu} \;, \nonumber \\
\eta^{a}_{\mu\nu} & = \partial_{\mu}  \eta^{a}_{\nu} -  \partial_{\nu}  \eta^{a}_{\mu} \;.
\end{align}
As simplest examples we investigate the following composite operators at leading order:
\begin{align}\label{goodoperators}
O^{(1)}_{\lambda\eta}(x) &=  \left( \lambda^{a}_{\mu\nu}(x) \eta^{a}_{\mu\nu}(x) \right) \;, \nonumber  \\
O^{(2)}_{\lambda\eta}(x) &=  \varepsilon_{\mu\nu\rho\sigma} \left( \lambda^{a}_{\mu\nu}(x) \eta^{a}_{\rho\sigma}(x) \right) \;.
\end{align}
After some calculation, see appendix \ref{sigma}\ref{appcorr}, we find:
\begin{align}
\braket{ O^{(1)}_{\lambda\eta}(k) O^{(1)}_{\lambda\eta}(-k)} & = 4(N^2-1) \int \frac{\d^dp}{(2\pi)^d} \frac{p^2(p-k)^2+(d-2)(p^2-p\,k)^2}{(p^2-\ii \lambda^2)((p-k)^2+\ii \lambda^2)}  \;,\\
\braket{ O^{(2)}_{\lambda\eta}(k) O^{(2)}_{\lambda\eta}(-k)}  &= 32(N^2-1) \int \frac{\d^dp}{(2\pi)^d} \frac{(k^2 p^2-(k\,p)^2)}{(p^2-\ii \lambda^2)((p-k)^2+\ii \lambda^2)} \;.
\end{align}

\subsubsection{The spectral representation in $d = 2$ }
We shall provide some details concerning the derivation for the case $d=2$. Let us begin with the analysis of
\begin{eqnarray}
\braket{ O^{(1)}_{\lambda\eta}(k) O^{(1)}_{\lambda\eta}(-k)} = 4 (N^2-1) F(k^2)\label{O1}\;,
\end{eqnarray}
with
\begin{eqnarray}
F(k^2)= \int\frac{\d^dp}{(2\pi)^d} \frac{p^2(p-k)^2+(d-2)(p^2-p\,k)^2}{(p^2+\ii \lambda^2)((p-k)^2-\ii \lambda^2)}\;.  \label{O2}
\end{eqnarray}
We first derive a Feynman parametrization of \eqref{O2}. Proceeding in the usual way one finds
\begin{eqnarray}
F(k^2)= \int_0^1 \d x \int\frac{\d^d q}{(2\pi)^d} \frac{N(q,k,x)}{(q^2+\Delta^2)^2}\;,  \label{O3}
\end{eqnarray}
where we used the substitution $q=p-kx$, and whereby
\begin{equation}
\Delta^2=x(1-x)k^2-(2x-1)\ii \lambda^2\;.
\end{equation}
We shall temporarily work in units $2\lambda^2=1$. We still have to identify the numerator $N(q,k,x)$. Keeping in mind that terms odd in $q_\mu$ will vanish upon integration, and that we may replace $q_\mu q_\nu\to q^2\frac{\delta_{\mu\nu}}{d}$ within the $q$-integral, we are brought to
\begin{eqnarray}
F(k^2)= (d-1)\int_0^1 \d x \int\frac{\d^d q}{(2\pi)^d} \frac{x^2(1-x)^2 k^4 + \frac{2}{d}\left[1-(d+2)x(1-x)\right]k^2q^2 + q^4}{(q^2+\Delta^2)^2} \;,
\end{eqnarray}
after a bit of algebra. Subsequently, from \eqref{loopint} it follows that
\begin{equation}\label{O6}
    \int \frac{\d^dq}{(2\pi)^d}\frac{q^2}{(q^2+\Delta)^n}=\frac{1}{(4\pi)^{d/2}}\frac{d}{2}\frac{\Gamma(n-d/2-1)}{\Gamma(n)}(\Delta^2)^{d/2-n+1}\;,
\end{equation}
and consequently also
\begin{equation}
    \int \frac{\d^dq}{(2\pi)^d}\frac{q^4}{(q^2+\Delta)^n}=\frac{1}{(4\pi)^{d/2}}\frac{d(d+2)}{4}\frac{\Gamma(n-d/2-2)}{\Gamma(n)}(\Delta^2)^{d/2-n+2}\;.
\end{equation}
To obtain a finite result, we prefer to look at
\begin{multline}
\frac{\p^2 F(k^2)}{(\p k^2)^2}= \frac{1}{4\pi}\int_0^1 \d x \left[\left(12x^4-24x^3+14x^2-2x\right)\frac{1}{\Delta^2}+(8x^6-24x^5+25x^4 \right.\\
\left. -10x^3+x^2) \frac{k^2}{(\Delta^2)^2} +2x^4(1-x)^4\frac{k^4}{(\Delta^2)^3}\right]\;,
\end{multline}
where we set $d=2$. We consequently find
\begin{eqnarray}
\frac{\p^2 F(k^2)}{(\p k^2)^2}&=&\frac{1}{4\pi}\int_0^1 \d x\left[\frac{-12x^2+12x-2}{k^2-\ii s}+k^2\frac{8x^2-8x+1}{(k^2-is)^2}+2k^4\frac{x(1-x)}{(k^2-\ii s)^3}\right]\;,
\end{eqnarray}
where we reintroduced $s=\frac{2x-1}{2x(1-x)}$, hence $x=\frac{-1+s+\sqrt{1+s^2}}{2s}$, which gives rise to
\begin{eqnarray}
\frac{\p^2 F(k^2)}{(\p k^2)^2}&=& \frac{1}{4\pi}\int_{-\infty}^{+\infty}\frac{\d s}{2(1+s^2+\sqrt{1+s^2})}\left[-\frac{2}{s^2}(3+s^2-3\sqrt{1+s^2})\frac{1}{k^2-\ii s}\right.\nonumber\\
&&\left.+\frac{k^2}{s^2}(4+s^2-4\sqrt{1+s^2})\frac{1}{(k^2- \ii s)^2}+\frac{k^4}{1+\sqrt{1+s^2}}\frac{1}{(k^2-\ii s)^3}\right]\;.
\end{eqnarray}
We first rewrite everything in terms of $k^2-\ii s$ as follows
\begin{eqnarray}\label{O11}
\frac{\p^2 F(k^2)}{(\p k^2)^2}&=& \frac{1}{4\pi}\int_{-\infty}^{+\infty}\frac{\d s}{2(1+s^2+\sqrt{1+s^2})}\left[-\frac{2}{s^2}(3+s^2-3\sqrt{1+s^2})\frac{1}{k^2-\ii s}\right.\nonumber\\
&&\left.+\frac{1}{s^2}(4+s^2-4\sqrt{1+s^2})\frac{k^2- \ii s+ \ii s}{(k^2-\ii s)^2}+\frac{(k^2- \ii s+ \ii s)^2}{1+\sqrt{1+s^2}}\frac{1}{(k^2-\ii s)^3}\right]\nonumber\\
&=&\frac{1}{4\pi}\int_{-\infty}^{+\infty}\frac{\d s}{2(1+s^2+\sqrt{1+s^2})}\left[-\frac{1}{s^2}\left(3+s^2-3\sqrt{1+s^2}\right)\frac{1}{k^2- \ii s}\right.\nonumber\\
&&\left.+  \ii s\frac{-1+\sqrt{1+s^2}}{1+\sqrt{1+s^2}}\frac{1}{(k^2-  \ii s)^2}-\frac{s^2}{1+\sqrt{1+s^2}}\frac{1}{(k^2- \ii s)^3}\right]\;.
\end{eqnarray}
Using two consecutive partial integrations, we can show that
\begin{eqnarray}
&&\int_{-\infty}^{+\infty}\frac{\d s}{2(1+s^2+\sqrt{1+s^2})}\left[-\frac{1}{s^2}(3+s^2-3\sqrt{1+s^2})\frac{1}{k^2- \ii s}\right]\nonumber\\
&=&-\int_{-\infty}^{+\infty}\d s\left(\frac{-1+\sqrt{1+s^2}}{s^2}+\ln\left(1+\sqrt{1+s^2}\right)\right)\frac{1}{(k^2-  \ii s)^3}\;.
\end{eqnarray}
Similarly, partial integration  leads to
\begin{eqnarray}
&&\int_{-\infty}^{+\infty}\frac{\d s}{2(1+s^2+\sqrt{1+s^2})}\left[is\frac{-1+\sqrt{1+s^2}}{1+\sqrt{1+s^2}}\frac{1}{(k^2- \ii s)^2}\right]\nonumber\\
&=&\int_{-\infty}^{+\infty}\d s \left[\frac{2}{1+\sqrt{1+s^2}}+\ln\left(1+\sqrt{1+s^2}\right)\right]\frac{1}{(k^2- \ii s)^3}\;.
\end{eqnarray}
Hence, we can rewrite \eqref{O11} as
\begin{eqnarray}
\frac{\p^2 F(k^2)}{(\p k^2)^2}&=& \frac{1}{4\pi}\int_{-\infty}^{+\infty}\d s\left[\frac{1}{2\sqrt{1+s^2}}\right]\frac{1}{(k^2- \ii s)^3}\;.
\end{eqnarray}
after simplification. We observe that there are no poles in the upper half $s$-plane for $k^2>0$, so we can fold our contour around the cut for $s\in[\ii \infty,\ii ]$. With $s=\ii \tau$, we can write
\begin{eqnarray}
\frac{\p^2 F(k^2)}{(\p k^2)^2}&=& \frac{1}{4\pi}\int_{+\infty}^1\ii  \d\tau\left[\frac{1}{-2\ii \sqrt{\tau^2-1}}\right]\frac{1}{(k^2+\tau)^3}+\frac{1}{4\pi}\int_1^{+\infty}\ii \d\tau\left[\frac{1}{2\ii \sqrt{\tau^2-1}}\right]\frac{1}{(k^2+\tau)^3}\nonumber\\
&=& \frac{1}{4\pi}\int_1^{+\infty}\d \tau\frac{1}{\sqrt{\tau^2-1}}\frac{1}{(k^2+\tau)^3}\;.
\end{eqnarray}
We can now return to the original function $F(k^2)$. A first integration from 0 to $k^2$ gives
\begin{equation*}
\frac{\p F(k^2)}{\p k^2}-\left[\frac{\p F(k^2)}{\p k^2}\right]_{k^2=0} = \frac{1}{4\pi}\int_1^{+\infty}\d\tau\frac{1}{-2\sqrt{\tau^2-1}}\frac{1}{(k^2+\tau)^2}-\frac{1}{4\pi}\int_1^{+\infty}\d\tau\frac{1}{-2\sqrt{\tau^2-1}}\frac{1}{\tau^2}\;,
\end{equation*}
so that we get,
\begin{eqnarray}\label{O17}
&& F(k^2)-k^2\left[\frac{\p F(k^2)}{\p k^2}\right]_{k^2=0}-F(0)\nonumber\\&=& \frac{1}{4\pi}\int_1^{+\infty}\d\tau\frac{1}{2\sqrt{\tau^2-1}}\frac{1}{k^2+\tau}+\frac{k^2}{4\pi}\int_1^{+\infty}\d\tau\frac{1}{2\sqrt{\tau^2-1}}\frac{1}{\tau^2}- \frac{1}{4\pi}\int_1^{+\infty}\d\tau\frac{1}{2\sqrt{\tau^2-1}}\frac{1}{\tau}\;. \nonumber\\
\end{eqnarray}
We thus find
\begin{equation}\label{O18}
\rho(\tau)=\frac{1}{8\pi}\frac{1}{\sqrt{\tau^2-1}}\;.
\end{equation}
We conclude that, upon restoring units, we \emph{formally} have
\begin{equation}\label{O19}
\braket{ O^{(1)}_{\lambda\eta}(k) O^{(1)}_{\lambda\eta}(-k)}  =\int_{2\lambda^2}^{+\infty}\frac{2 (N^2-1) \lambda^4}{\pi\sqrt{\tau^2-4\lambda^4}}\frac{\d\tau}{\tau+k^2}\;.
\end{equation}
We clearly notice that the spectral density $\rho(\tau)$ is positive for $\tau\geq 2\lambda^2$. The result as written in \eqref{O19} is indeed only formally correct, since the l.h.s.~of \eqref{O19} is divergent, directly seen upon inspection of its definition \eqref{O3}. Nevertheless, the spectral representation \eqref{O19} appearing in the r.h.s.~defines a finite function. The apparent contradiction is easily resolved by realizing that one should in fact refer to \eqref{O17}, which gives the correctly subtracted result. \\
\\
Let us now turn to the analysis of
\begin{eqnarray}
\braket{ O^{(2)}_{\lambda\eta}(k) O^{(2)}_{\lambda\eta}(-k)} = 32 (N^2-1) G(k^2)\;,
\end{eqnarray}
with
\begin{eqnarray}
G(k^2)= \int\frac{\d^dp}{(2\pi)^d} \frac{k^2 p^2 - (k p)^2}{(p^2+\ii \lambda^2)((p-k)^2-\ii \lambda^2)}\;.
\end{eqnarray}
Invoking the Feynman trick this time yields
\begin{eqnarray}
G(k^2)= \int_0^1 \d x \int\frac{\d^d q}{(2\pi)^d} \frac{M(k,q)}{(q^2+\Delta^2)^2}\;,
\end{eqnarray}
with
\begin{eqnarray}
M(k,q) &=& k^2 q^2 \left( 1- \frac{1}{d}\right)\;.
\end{eqnarray}
As usual, we shall work in units $2\lambda^2=1$. By using \eqref{O6}, we obtain the finite function
\begin{eqnarray}
\frac{\p^2 G(k^2)}{(\p k^2)^2}&=& \frac{1}{8\pi}\int_0^1 \d x \left[\left(x^2 (1-x)^2\right)\frac{k^2}{(\Delta^2)^2} -\left(2 (1-x) x \right)\frac{1}{\Delta^2}\right]\;.
\end{eqnarray}
We substitute $x=\frac{-1+s+\sqrt{1+s^2}}{2s}$, so we obtain
\begin{eqnarray}
\frac{\p^2 G(k^2)}{(\p k^2)^2}&=& \frac{1}{8\pi}\int_{-\infty}^{+\infty}\frac{\d s}{2(1+s^2+\sqrt{1+s^2})}\left[ \frac{k^2}{ (k^2 + \ii s)^2} -\frac{2}{ k^2 + \ii s} \right]\;.
\end{eqnarray}
Rewriting in terms of $k^2-\ii s$ gives
\begin{eqnarray}\label{P4}
\frac{\p^2 G(k^2)}{(\p k^2)^2}&=& \frac{1}{8\pi}\int_{-\infty}^{+\infty}\frac{\d s}{2(1+s^2+\sqrt{1+s^2})}\left[ \frac{-\ii s}{ (k^2 + \ii s)^2} -\frac{1}{ k^2 + \ii s} \right]\;.
\end{eqnarray}
Subsequently, using partial integration gives us the following identity
\begin{eqnarray*}
\int_{-\infty}^{+\infty}\frac{\d s}{2(1+s^2+\sqrt{1+s^2})}\left[ \frac{-\ii s}{ (k^2 + \ii s)^2}\right]&=&-\int_{-\infty}^{+\infty}\d s\left[ \ln \left( \sqrt{s^2 +1 } + 1 \right) \right]\frac{1}{(k^2-\ii s)^3}\;.
\end{eqnarray*}
Analogically,
\begin{equation*}
\int_{-\infty}^{+\infty}\frac{\d s}{2(1+s^2+\sqrt{1+s^2})}\left[\frac{-1}{(k^2-\ii s)^2}\right]=\int_{-\infty}^{+\infty}\d s \left[\sqrt{s^2 +1} -  \ln \left( \sqrt{s^2 +1 } + 1 \right)\right]\frac{1}{(k^2-\ii s)^3}\;.
\end{equation*}
Therefore, equation \eqref{P4} becomes
\begin{eqnarray}
\frac{\p^2 G(k^2)}{(\p k^2)^2}&=& \frac{1}{8\pi}\int_{-\infty}^{+\infty}\d s\frac{\sqrt{1+s^2}}{(k^2-\ii s)^3}\,.
\end{eqnarray}
As before, we can fold our contour around the cut for $s\in[\ii \infty,\ii]$ and by setting $s=\ii \tau$, we find
\begin{eqnarray}
\frac{\p^2 G(k^2)}{(\p k^2)^2}&=& \frac{1}{4\pi}\int_{1}^{+\infty}\d\tau \frac{\sqrt{\tau^2-1}}{(k^2+\tau)^3}\;,
\end{eqnarray}
and thus
\begin{eqnarray*}
\frac{\p G(k^2)}{\p k^2}-\left[\frac{\p G(k^2)}{\p k^2}\right]_{k^2=0}&=& \frac{1}{4\pi}\int_1^{+\infty}\d\tau\frac{\sqrt{\tau^2-1}}{-2}\frac{1}{(k^2+\tau)^2}-\frac{1}{4\pi}\int_1^{+\infty}\d\tau\frac{\sqrt{\tau^2-1}}{-2}\frac{1}{\tau^2}\;.
\end{eqnarray*}
Integrating a second time gives,
\begin{eqnarray*}
&& G(k^2)-G^2\left[\frac{\p G(k^2)}{\p k^2}\right]_{k^2=0}-G(0)\nonumber\\&=& \frac{1}{8\pi}\int_1^{+\infty}\d\tau \sqrt{\tau^2-1}\frac{1}{k^2+\tau}+\frac{k^2}{8\pi}\int_1^{+\infty}\d\tau\sqrt{\tau^2-1}\frac{1}{\tau^2}- \frac{1}{8\pi}\int_1^{+\infty}\d\tau \sqrt{\tau^2-1} \frac{1}{\tau}\;.
\end{eqnarray*}
The spectral density can be read off,
\begin{equation}
\rho(\tau)=\frac{1}{8\pi}\sqrt{\tau^2-1} \;,
\end{equation}
whereby $\rho(\tau)\geq0$ for $\tau\geq 1$.  We reintroduce the units, and we conclude that
\begin{equation}\label{P19}
\braket{ O^{(2)}_{\lambda\eta}(k) O^{(2)}_{\lambda\eta}(-k)}  =\frac{4(N^2-1)}{\pi}  \int_{2\lambda^2}^{+\infty}\d\tau\frac{\sqrt{\tau^2-4\lambda^4}}{\tau+k^2}\;,
\end{equation}
which is again a formal result due to the divergent nature of both l.h.s.~and r.h.s.~.

\subsubsection{The spectral representation in $d = 4$ }
In four dimensions, we can do a similar analysis to obtain \cite{Baulieu:2009ha}
\begin{equation}\label{GZres2}
  \Braket{ O^{(1)}_{\lambda\eta}(k) O^{(1)}_{\lambda\eta}(-k) }  = 12N \int_{2\gamma^2}^\infty \d\tau \frac{1}{\tau+k^2} \frac{ \sqrt{\tau^2-4 \gamma^4}(2\gamma^4+\tau^2)}{32   \pi ^2 \tau}\;,
\end{equation}
and
\begin{equation}
\Braket{ O^{(2)}_{\lambda\eta}(k) O^{(2)}_{\lambda\eta}(-k) }  = 96N \int_{2\gamma^2}^\infty \d\tau \frac{1}{\tau+k^2} \frac{\left(\tau^2-4\gamma^4\right)^{3/2}}{64\pi^2 \tau}\;,
\end{equation}
which is again a formal result. We see that the spectral densities are positive again, and thus at least at leading order, the operators  $ O^{(1)}_{\lambda\eta}$ and $ O^{(2)}_{\lambda\eta}$ appear to be physical.

\section{The relation between $F^2$ and the operator $O^{(1)}_{\lambda\eta}(x)$}
Now that we have found an operator with a good spectral representation, it is useful to see how the operator $F^2_{\mu\nu}$, which we have investigated in great detail in the previous chapter, is related to this operator. Let us introduce the following definitions
\begin{eqnarray}\label{6sim}
\varphi_{\mu\nu}^a &=&\frac{1}{N} f_{abc}  ( \p_\mu \varphi^{[bc]}_\nu - \p_\nu \varphi^{[bc]}_\mu) \;,\nonumber\\
\overline \varphi_{\mu\nu}^a &=& \frac{1}{N}f_{abc}  ( \p_\mu \overline\varphi^{[bc]}_\nu - \p_\nu \overline \varphi^{[bc]}_\mu)\;,
\end{eqnarray}
Using the relations \eqref{r5f}, which expresses the $i$-fields in terms of the original variables, we find for the $i$-fields strengths, see equation \eqref{fs},
\begin{eqnarray}\label{lambdaf}
\lambda_{\mu\nu}^a &=&  \frac{1}{\sqrt{2}}   f_{\mu\nu}^a   - \frac{\ii \sqrt{N} }{2 }   \left(   \varphi^{a}_{\mu\nu}  + \overline\varphi^{a}_{\mu\nu} \right) \;,\nonumber\\
\eta_{\mu\nu}^a &=&  \frac{1}{\sqrt{2}}   f_{\mu\nu}^a   +\frac{\ii \sqrt{N}  }{2 }   \left(   \varphi^{a}_{\mu\nu}  + \overline\varphi^{a}_{\mu\nu} \right)\;,
\end{eqnarray}
whereby with $ f_{\mu\nu}^a$ we denote only the abelian part of the field strength
\begin{equation}
f^a_{\mu\nu}= \partial_\mu A^a_\nu  - \partial_\nu A^a_\mu  \;.
\end{equation}
Therefore, we can write $f_{\mu\nu}^2$ in terms of the $i$-particles variables,
\begin{equation}\label{eerstef}
\frac{1}{2} f_{\mu\nu}^2 = \frac{1}{2} \underbrace{\lambda^a_{\mu\nu} \eta^a_{\mu\nu}}_{O^{(1)}_{\lambda\eta}} + \frac{1}{4} \eta^a_{\mu\nu} \eta^a_{\mu\nu}  +\frac{1}{4} \lambda^a_{\mu\nu} \lambda^a_{\mu\nu}\;.
\end{equation}
From this relation, we can make some interesting observations. In the end of the previous section, we have mentioned that the correlator $\Braket{F^{2}(x)F^{2}(y)}$ displays unphysical cuts. From the  previous equation, we can now see that these unphysical cuts are in fact stemming from the last two terms of equation \eqref{eerstef}, namely $\frac{1}{4} \eta^a_{\mu\nu} \eta^a_{\mu\nu}  +\frac{1}{4} \lambda^a_{\mu\nu} \lambda^a_{\mu\nu}$. We have thus uncovered at lowest order what the structure of the correlator $\Braket{F^{2}(x)F^{2}(y)}$ is.\\
\\
The main question is, if we want to continue with the operator $\lambda^a_{\mu\nu} \eta^a_{\mu\nu}$, whether this operator is \textit{renormalizable}. For this, we should first try to find the non-abelian generalization of the operator  $\lambda^{a}_{\mu\nu} \eta^{a}_{\mu\nu}$. However, already at lowest order, we can see that this operator breaks many crucial Ward identities of the GZ action which are needed for renormalization. Indeed, from equation \eqref{lambdaf}, we also find that rewriting $O^{(1)}_{\lambda\eta}(x)$ in terms of old variables gives
\begin{equation}
O^{(1)}_{\lambda\eta}(x) =   \lambda^{a}_{\mu\nu}(x) \eta^{a}_{\mu\nu}(x)  = \frac{1}{2} f_{\mu\nu}^2 - \frac{N}{4} \left(   \varphi^{a}_{\mu\nu}  + \overline\varphi^{a}_{\mu\nu} \right)^2\;.
\end{equation}
E.g.~due to the second term, this operator already breaks the most important symmetry, i.e.~the BRST symmetry. This breaking is not soft\footnote{This was the case for the GZ action, see \eqref{breaking}, and therefore, the breaking does not spoil renormalizability.}, i.e.~proportional to the mass parameter $\gamma^2$, but this is a hard breaking and therefore, impossible to restore. Next to this breaking, one can also check that other Ward identities are broken, see p.\pageref{pagewardidentities}. It looks therefore highly unlikely that the operator $O^{(1)}_{\lambda\eta}$ is renormalizable.

\section{Conclusion}
In this chapter, we have pursued the investigation of the analyticity properties of correlation functions evaluated with a confining propagator of the Gribov type. We started with a simple toy model to characterize examples of composite operators, whose correlations functions display cuts only on the negative real axis, while possessing a positive spectral function. For this, we have introduced $i$-particles, which seem rather natural objects when dealing with a Gribov type propagator. We could then do a similar analysis for the more complicated GZ action and we have found two operators, namely $O^{(1)}_{\lambda\eta}$ and $O^{(2)}_{\lambda\eta}$ which are given in equation \eqref{goodoperators} and which have the desired analytical properties. Naturally, the next step is to investigate the renormalization of these good operators $O^{(1)}$ and $O^{(2)}$. Unfortunately, focusing on the easiest operator $O^{(1)}$, we already see that this operator breaks the BRST symmetry in a hard way. Therefore, we would have to look for other operators, whose correlations functions also display cuts on the negative real axis, while possessing a positive spectral function, but which do not break the BRST symmetry $s$ in a hard way.\\
\\
Let us give an example of this kind of operator: if we consider the following operator
\begin{equation}
O_\new = \lambda^a_{\mu\nu} \eta^a_{\mu\nu}  -  \frac{N}{2} U^a_{\mu\nu} U^a_{\mu\nu} +  N    \omega_{\mu\nu}^a  \overline \omega_{\mu\nu}^a\;,
\end{equation}
whereby $\omega_{\mu\nu}^a$ and $\overline \omega_{\mu\nu}^a$ are similarly defined as in equation \eqref{6sim}
\begin{eqnarray}
\omega_{\mu\nu}^a &=& \frac{1}{N} f_{abc}  ( \p_\mu \omega^{[bc]}_\nu - \p_\nu \omega^{[bc]}_\mu) \;,\nonumber\\
\overline \omega_{\mu\nu}^a &=& \frac{1}{N} f_{abc}  ( \p_\mu \overline  \omega^{[bc]}_\nu - \p_\nu \overline  \omega^{[bc]}_\mu)\;,
\end{eqnarray}
we see that we have an $s$ invariant operator. Indeed, rewriting $O_\new$ in the old variables, we find
\begin{equation}\label{Onew}
    O_{\new}= \frac{1}{2}f_{\mu\nu}^2+ N  s( \varphi_{\mu\nu}^a \overline \omega_{\mu\nu}^a)\;,
\end{equation}
which is indeed $s$ invariant. However, the problem with the current operator is that we can easily see that this operator shall have a negative spectral density. The part in the ghost sector, $\sim \braket{\omega_{\mu\nu}^a  \overline \omega_{\mu\nu}^a(x) \omega_{\mu\nu}^a  \overline \omega_{\mu\nu}^a(y)}$, will induce a cut along the whole negative axis, with negative discontinuity, due to the ghost character. The corresponding cut should start at $k^2=0$, as the GZ ghosts are massless. The part $\sim \Braket{ U^a_{\mu\nu} U^a_{\mu\nu}(x)  U^a_{\mu\nu} U^a_{\mu\nu}(y) }$ cannot cancel this cut, as it has only $\frac{N}{2}$ degrees of freedom, while the ghost sector still has $N$ degrees of freedom. Therefore, in the GZ action, this operator does not look like a physical operator.\\
\\
Let us therefore consider the Refined Gribov-Zwanziger action \eqref{nact}. So far, we have not investigated the consequences of adding another massive parameter stemming from the condensate $\braket{(\overline{\varphi}\varphi-\overline{\omega}\omega)}$ into the game. However, many results will change significant when considering the RGZ. For example, considering the operator $O_{\new}$ again, we expect that the cut of the fields strength correlator $\braket{\omega_{\mu\nu}^a  \overline \omega_{\mu\nu}^a(x) \omega_{\mu\nu}^a  \overline \omega_{\mu\nu}^a(y)}$ shall start at $-(M+M)^2$, with a negative spectral density of course. However, due to this shift, it becomes possible now that the branch cut has shifted enough to have an overlap with the branch cut of the correlator stemming from the $\lambda^a_{\mu\nu} \eta^a_{\mu\nu}$ part. This opens new perspectives for further research.\\
\\
In fact, we can even generalize the form of the operator \eqref{Onew}. We expect the following form of operator to have a descent spectral density and to be renormalizable
\begin{equation}
O = F_\mu\nu^2 + s(\ldots) + \gamma^2 \times (d=2\text{ operators})\;,
\end{equation}
so that in the limit $\gamma \to 0$, we recover the usual BRST cohomology. The extra terms proportional to $\gamma^2$ usually do not spoil the renormalizability and are therefore in principle allowed. The first step would be to investigate the abelian operator. If this operator would turn out to have good spectral properties, we would at least be able to write down an extension of the operator to the quantum level. Next, one can try to investigate its renormalization and, if possible, its higher order spectral properties. This is currently under investigation.

\chapter{Extracting glueball masses from the $i$ particle approach\label{last}}
\section{The process}
As a final part of this thesis, it remains to discuss \cite{Dudal:2010cd}. In this work, a rough estimate for the glueball masses was extracted for the scalar ($0^{++}$), the pseudoscalar ($0^{-+}$) and a tensorial glueball ($2^{++}$), but in the RGZ framework, using numbers from the fit described in section \ref{orlandof} of chapter \ref{refined}. Let us summarize here the followed way. Firstly, the spectral representations were calculated, not only for the scalar glueball, i.e.
\begin{equation}
F_{\mu\nu} F_{\mu\nu} \;,
\end{equation}
see also the previous chapter, but also for the pseudoscalar glueball, i.e.
\begin{equation}
\frac{1}{2} \epsilon^{\mu \nu \alpha \beta} F_{\mu \nu} F_{\alpha \beta}\;,
\end{equation}
and the tensorial glueball $2^{++}$, i.e.
\begin{eqnarray}
&&\Theta_{\mu \nu} = \p^4 \theta_{\mu \nu} - \p^2 \p_\mu \p_\alpha \theta_{\alpha \nu} - \p^2 \p_\nu \p_\alpha \theta_{\alpha \mu} + \p^2 \left( \delta_{\mu\nu} - \frac{2}{3} P_{\mu\nu} \right) \p_\alpha \p_\beta \theta_{\alpha \beta} \;, \nonumber\\
&& \text{with }\theta_{\mu \nu} = F_{\alpha \mu} F_{\alpha \nu} - \frac{\delta_{\mu \nu}}{4} F_{\alpha \beta}^2\;,
\end{eqnarray}
which is symmetric, traceless\footnote{We are considering here only the classical level, as at the quantum level, the energy momentum tensor is no longer traceless.} and divergence-free. Normally, one would expect $\theta_{\mu \nu}$ to be the candidate for the tensorial glueball as it is the energy momentum tensor: it is  a rank two tensor which is symmetric, traceless and divergence-free. However, this is only true within the normal Yang-Mills action as here in the GZ action, there is a mass scale present. Therefore, the
energy-momentum tensor is not given by $\theta_{\mu \nu}$, as $\theta_{\mu \nu}$ has a non-vanishing trace now. Therefore, instead, the authors of \cite{Dudal:2010cd} have proposed the rank two tensor $\Theta_{\mu \nu}$, given in the expression above, which has only 5 remaining degrees of freedom as desired. Moreover, in the limit $\gamma \to 0$, this tensor reduces to $\p^4 \theta_{\mu \nu}$, i.e.~a derivative of the energy momentum tensor\footnote{Extra derivatives in the operators play no role for the corresponding correlation functions.}. Therefore, $\Theta_{\mu \nu}$ seems like a reasonable extension of the known energy momentum tensor in Yang-Mills theory $\theta_{\mu \nu}$. \\
\\
The glueball correlation functions were split in a physical part (the $i$ particles) and an unphysical part, which was consequently left out. \\
\\
We recall that the spectral representation is given by
\begin{equation}\label{specf}
F(k^2) = \int_{\tau_0}^\infty \frac{\rho(t)}{ t + k^2} \d t \;.
\end{equation}
However, as shown in the previous chapter, see e.g.~expression \eqref{int11}, one always needs to subtract the UV divergent parts. In general, this can be done by first deriving $F(k^2)$ w.r.t.~$k^2$ until $\frac{\p^r F(k^2)}{(\p k^2)^r}$ is finite, with $r$ a natural number. Next, we integrate back each time from $T$ to $k^2$ so we obtain a finite spectral density:
\begin{multline}\label{poly}
F^{\sub}(k^2) =  F(k^2) - F(T) - \frac{\p F(T)}{\p k^2} (k^2 - T) - \frac{\p^2 F(T)}{(\p k^2)^2} \frac{(k^2 - T)^2}{2!} - \ldots - \frac{\p^{r-1} F(T)}{(\p k^2)^{r-1}} \frac{(k^2 - T)^{r-1}}{(r-1)!} \\= (-1)^{r} (k^2 - T)^r \int_{\tau_0}^\infty \underbrace{ \frac{ \rho(t)}{ t (t + T)^r }}_{\rho(t)'} \frac{\d t}{t + k^2}\;.
\end{multline}
$T \geq 0$ indicates the momentum substraction scale, which was always taken equal to zero in the previous chapter. Here, the scale $T$ is a parameter and the idea is that the physics should not depend on the substraction scale.\\
\\
$F^{\sub}(k^2)$ can be calculated in a similar fashion as in the previous chapter. However, as we are working in the RGZ framework, it becomes a bit more complicated to calculate the spectral densities. However, by using the results of \cite{Dudal:2010wn} it was found that \cite{Dudal:2010cd}
\begin{eqnarray}\label{spectf}
F^{\sub}_{0^{++}}(k^2) &=& - (k^2 - T)^3 \int_{\tau_0}^\infty  c_{0^{++}} \frac{1}{t (t+T)^3} \sqrt{ t^2 - 8 \theta^4 - 4 \mu^2 t}\nonumber\\
 && \times \left( \frac{1}{2} t^2 + 2 \theta^4 - 2 t \mu^2 + 3 \mu^4\right) \frac{1}{ t + k^2} \d t \;,\nonumber\\
F^{\sub}_{0^{-+}}(k^2) &=& - (k^2 - T)^3 \int_{\tau_0}^\infty  c_{0^{-+}} \frac{1}{t (t+T)^3} \sqrt{ t^2 - 8 \theta^4 - 4 \mu^2 t}\nonumber\\
 && \times \left( 2 \theta^4 +  t \mu^2 - \frac{1}{4} t^2 \right) \frac{1}{ t + k^2} \d t \;, \nonumber\\
 F^{\sub}_{2^{++}}(k^2) &=& - (k^2 - T)^7 \int_{\tau_0}^\infty  c_{2^{++}} \frac{1}{t (t+T)^7} \sqrt{ t^2 - 8 \theta^4 - 4 \mu^2 t} \;, \nonumber\\
 && \times \left( 16t^2 \theta^8 - 4 \theta^4 \mu^2 t^3 + 16 t^4 \theta^4 + 9 \mu^4 t^4 - \frac{9}{2} \mu^4 t^5 + \frac{3}{2} t^6 \right) \frac{1}{ t + k^2} \d t \;. \nonumber\\
\end{eqnarray}
The parameters $c$ are positive constants irrelevant for the calculations, and $\mu$ and $\theta$ are given by
\begin{eqnarray}
\mu^2 &=& Re \left[ \frac{1}{2} \left(m^2 +M^2 + \sqrt{m^4- 2 M^2 m^2+M^4-4 \lambda ^4}\right) \right] \;,\nonumber\\
\sqrt{2} \theta^2 &=& Im \left[ \frac{1}{2} \left(m^2 +M^2 + \sqrt{m^4- 2 M^2 m^2+M^4-4 \lambda ^4}\right) \right] \;, \nonumber\\
\end{eqnarray}
corresponding to the real and imaginary part of the poles of the refined gluon propagator \eqref{gluonprop2}.\\
\\
Secondly, let us have a closer look at the general form of the spectral representation \eqref{specf}. In fact, in is better to set $t=1/s$ in the spectral density, so we find
\begin{equation}
F(k^2) = \int_{0}^{1/\tau_0} \frac{\rho(1/s)}{ 1 + s k^2} \d s \;,
\end{equation}
so we can expand for small $k^2$:
\begin{equation}
F(k^2) = \sum_{n=0}^\infty \nu_n (-1)^n (k^2)^n \;,
\end{equation}
with
\begin{equation}
\nu_n = \int_{0}^{1/\tau_0}  s^n  \rho(1/s) \d s \;,
\end{equation}
called the moments. Even more appropriate is to look at the following function
\begin{equation}\label{formss}
f(z) = \frac{1}{z} F(-1/z) =  \int_{0}^{1/\tau_0} \frac{\rho(1/s)}{ z-s} \d s \;,
\end{equation}
which can be expanded for large negative $z$ (small positive $k^2$):
\begin{equation}\label{moments}
f(z) = \sum_{n=0}^\infty \nu_n  \left(\frac{1}{z}\right)^{n+1} \;.
\end{equation}
This function can be approximated by a Pad\'{e} approximant. As for large $z$, the function behaves like $1/z$, we should use the $[N-1,N]$ Pad\'{e} approximant:
\begin{equation}
f(z) = \frac{P_{N-1}(z)}{Q_N(z)} \;,
\end{equation}
whereby $P_{N-1}(z)$ and $Q_N(z)$ are polynomials in $z$ of order $N-1$ respectively $N$. This Pad\'{e} approximation can be completely fixed in terms of the moments $\nu_n$. Let us take the first two moments of the Taylor expansion and set it equal to the $[0,1]$ Pad\'{e} approximant
\begin{eqnarray}\label{glue4}
    f(z)=\frac{\nu_0}{z}+\frac{\nu_1}{z^2}+\mathcal{O}(z^{-3}) = \frac{a_0}{b_0 + b_1 z} +\mathcal{O}(z^{-3})\;,
\end{eqnarray}
then follows that the $[0,1]$ Pad\'{e} approximant is given by
\begin{eqnarray}\label{glue5}
    f(z) \approx \frac{-\frac{\nu^{2}_0}{\nu_1}}{1-\frac{\nu_0}{\nu_1}z}\;.
\end{eqnarray}
One can see that this function displays a pole at
\begin{equation}\label{glue6}
    z^*=\frac{\nu_1}{\nu_0}\Rightarrow k^2=-\frac{\nu_0}{\nu_1} \;,
\end{equation}
meaning that the mass estimate itself is given by
\begin{equation}\label{glue7}
    m=\sqrt{\frac{\nu_0}{\nu_1}} \;.
\end{equation}
A general property of the Pad\'{e} approximants is that the $Q_N$ are othogonal polynomials over $[0, 1/\tau_0]$ with weight $\rho(1/s)$, see \cite{Yn,shohat}. As the weight is positive, it follows that the poles\footnote{This is a general property of orthogonal polynomials with positive weight.} of $Q_N$, namely $z^*$, will be real, all different and lying in the interval $]0, 1/\tau_0[$. Therefore, the mass estimate will always be larger then $\tau_0$. Let us now look at this result in relation with the previous chapter. There, we found that a positive spectral density $\rho(t)$ gives a branch cut for $F(k^2)$ on the negative $k^2$ axis starting from $-\tau$ to $-\infty$. Here, by doing a Padé approximation, we replaced in a sense the branch cut with a number of poles, depending on the order of the polynomial $Q_N$. In fact, in the limit $N \to \infty$, the whole branch cut would be covered with poles starting from the point $-\tau_0$. \\
\\
So far, we have built the reasoning of the Padé approximation on the infinite expression $F(k^2)$ instead of on the subtracted expression $F^\sub(k^2)$. However, the difference between both expressions shall be a polynomial in $k^2$. Indeed, to find $F^\sub(k^2)$, we derive $F(k^2)$ $r$ times w.r.t.~$k^2$. In this way, we kill the moments $\nu_0, \ldots, \nu_{r-1}$. After this operation, the expression $\frac{\p^r F(k^2)}{(\p k^2)^r}$ is finite, meaning that the infinities are hidden in the first $r$ moments, while all the succeeding moments are finite. From expression \eqref{poly} it is then clear that the difference between $F^\sub(k^2)$ for a certain substraction scale $T$ and $F(k^2)$ is only a polynomial in $k^2$, and therefore, the pole structure does not change. We can thus equally work with $F^\sub(k^2)$ instead of $F(k^2)$.\\
\\
Let us thus rewrite \eqref{spectf} in the form \eqref{formss},
\begin{align}
f_{0^{++}}(k^2) &=  \int_{0}^{1/\tau_0}  c_{0^{++}} \frac{s^3}{ (1+sT)^3} \sqrt{ \frac{1}{s^2} - 8 \theta^4 - \frac{4 \mu^2}{ s}} \left( \frac{1}{2s^2} + 2 \theta^4 - \frac{2}{s}  \mu^2 + 3 \mu^4\right) \frac{1}{ z-s} \d s \;,\nonumber\\
f_{0^{-+}}(k^2) &= \int_{0}^{1/\tau_0}  c_{0^{-+}} \frac{s^3}{ (1+sT)^3}  \sqrt{ \frac{1}{s^2} - 8 \theta^4 - \frac{4 \mu^2}{s} } \left( 2 \theta^4 +  \frac{ \mu^2}{s} - \frac{1}{4s^2}  \right) \frac{1}{ z-s} \d s \;, \nonumber\\
f_{2^{++}}(k^2) &= \int_{0}^{1/\tau_0} c_{2^{++}} \frac{s^7}{ (1+sT)^7} \sqrt{ \frac{1}{s^2} - 8 \theta^4 - \frac{4 \mu^2}{ s}}\nonumber\\
 & \times \left( \frac{16 \theta^8}{s^2} - \frac{4 \theta^4 \mu^2}{s^3}  + \frac{16 \theta^4 }{s^4}+  \frac{9 \mu^4}{ s^4} - \frac{9 \mu^4}{2 s^5}  + \frac{3}{2s^6} \right) \frac{1}{ z -s} \d s \;,
\end{align}
whereby we omitted $(-1)^r(k^2 -T)^r$, as it has no influence on the pole structure.\\
\\
Now as a third step, we need to determine the moments of the previous expression as described in \eqref{moments}. The question remains, how many moments are necessary for a reliable determination of the poles. The idea is to restrict to the first two moments as they are related to the IR region of the glueball operator. Indeed, for the one loop expression of the correlation, we can always write
\begin{equation}\label{uit1}
    \frac{F^{\text{sub}}(k^2)}{(-1)^r(k^2-T)^r}=\int \d q \frac{r(k,q)}{q} \;,
\end{equation}
whereby $\lim_{q \to \infty} r(k,q) < \infty$ as we are working with $F^{\sub}$ which is a UV finite expression. For small incoming momentum $k^2$, we can expand $r(k,q)$,
\begin{equation}
r(k,q) = r(0,q) + k^2 \frac{\p r (0,q) }{\p k ^2} + \ldots\;.
\end{equation}
From this expression we can deduce that certainly the first term $r(0,q)$ will have the most important contribution at small $q^2$. Also the higher order derivatives of $r(0,q)$ w.r.t.~$k^2$, will dominate for small $q^2$, as deriving w.r.t.~$k^2$ can only bring down more powers of $q^2$. In conclusion, for low incoming momentum $k^2$, only the low momentum information of the other quantities (propagators, vertices, etc) encoded in $r(k,q)$ will be relevant. Therefore, as we believe that our gluon propagator is a typical IR valid object, the first two moments of $F^\sub(k^2)$ will already encode the low momentum information we have on the glueball propagator. Moreover, the first two moments give already a good approximation of the starting function $F^\sub(k^2)$. We are thus basing ourselves on IR data, which we believe to be trustworthy, and that we are then extrapolating to the mass estimates using the Padé approximant. \\
\\
The reader shall have noticed that the foregoing glueball estimates were obtained using the tree level glueball correlation functions. One could wonder why tree level information, i.e.~obtained in a perturbative way, would provide any basis for an extrapolation to a nontrivial bound state mass. Without going into details, the main reason behind this is as follows. A very good approximation to the fully nonperturbative gluon propagator was used, as quasi-exactly known from the lattice simulations. Evidently, this means that a specific amount of nontrivial physics input is already present, under the form of this "exact" gluon propagator. The positivity violation of it is indeed interpreted as a confinement signal. The fact that the gluons cannot appear as separated physical observables is hinting towards the formation of a bound state. Two (or more) of such unphysical gluons have thus been put together in a composite operator like $F^2$, and it turns out that already at tree level, a physical signal is created, under the form of the physical branch cut. It is essentially this branch which is then used to get at least some, hopefully meaningful, information on the possible pole structure of the investigated glueball correlation functions. It is a priori by no means trivial that combining two unphysical gluons together would have created a physical branch cut. As already explained, one can argue that in the deep infrared, there are reasons to believe in the glueball correlation function, and this is what has been extrapolated using the Padé approximation theory.\\

\section{The results}
Let us show the results of this calculation. The masses are given by expression \eqref{glue7} and depend on the substraction scale $T$. In the following figure, one can see the result for the scalar $0^{++}$, the pseudoscalar $0^{-+}$ and the $2^{++}$ glueball masses.

\begin{figure}[H]
  \centering
     \includegraphics[width=8cm]{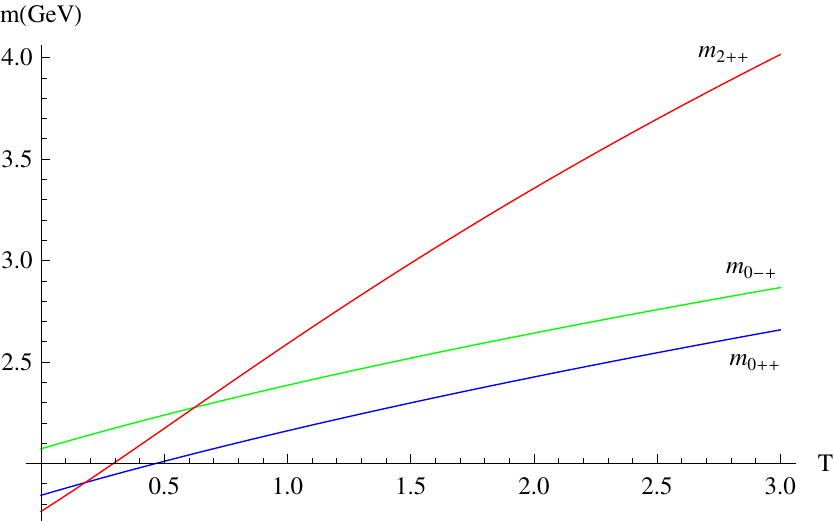}
  \caption{The masses of the $0^{++}$, the $0^{-+}$ and the $2^{++}$ in function of the subtraction scale $T$. }  \label{remarkfig}
\end{figure}

\noindent As a final step, it remain to determine the masses with the least dependence on the scale $T$. As one can see from the figure, there are no minima for the masses w.r.t.~$T$. However, when taking the relative masses, $\frac{m_{2^{++}}}{m_{0^{++}}}$ and $\frac{m_{2^{++}}}{m_{0^{-+}}}$, there are reflection points w.r.t.~$T$ given by $T \approx 0.34$ respectively $T \approx 0.35$, two values which are very close. Now setting $T = 0.34$, the masses become:
\begin{equation}
 m_{0++}\approx 1.98 \text{ GeV}\,,\qquad m_{0-+}\approx 2.21 \text{ GeV}\,,\qquad m_{2'++} \approx 2.17 \text{ GeV} \;.
\end{equation}
Comparing this with the lattice values of \cite{Mathieu:2008me,Chen:2005mg,Crede:2008vw,Narison:1996fm}, namely
\begin{equation}
 m_{0++}^\lat\approx 1.73 \text{ GeV}\,,\qquad m_{0-+}\approx 2.59 \text{ GeV}\,,\qquad m_{2'++} \approx 2.40 \text{ GeV} \;,
\end{equation}
one can see that the values obtained here are within $20\%$ range of the lattice results. It is worth mentioning that instanton contributions were omitted here. These are known to be relevant for the scalar and pseudoscalar channel, giving an attractive, resp.~repulsive contribution around 200-300 MeV, see e.g.~\cite{Mathieu:2008me,Schafer:1994fd}  and references therein.

\chapter{Conclusion and outlook\label{conclusions}}
The Gribov-Zwanziger action in the Landau gauge can be seen as an improved Faddeev-Popov action, where the latter approach to gauge fixing suffers from Gribov copies. Indeed, the GZ action implements a restriction of the region of integration to the so called Gribov region $\Omega$,
\begin{eqnarray}
\Omega &\equiv &\{ A^a_{\mu}, \, \p_{\mu} A^a_{\mu}=0, \, \mathcal{M}^{ab}  >0  \} \,,
\end{eqnarray}
whereby $\mathcal{M}^{ab}$ is the Faddeev-Popov operator, see equation \eqref{mab2}. One can prove that all gauge orbits intersect with this region, which is of paramount importance for having a good gauge fixing. Unfortunately, it has been proven that this region is not free of Gribov copies itself, but it is currently the best tool on the market. In this thesis, we have reviewed the current status of the Gribov-Zwanziger framework. \\
\\
The main results of this research can be divided into two parts.

\section*{From propagators...}
The first part concerns the infrared behavior of the gluon and the ghost propagator. The latest lattice data have shown that in 3d and 4d in the Landau gauge, the ghost propagator is not enhanced, in contrast to the general credence. Also, the gluon propagator is infrared suppressed and non-vanishing at zero momentum. For a long time, it was believed that the gluon propagator did vanish at zero momentum. Unfortunately, these recent lattice data are in contrast with the predictions of the Gribov-Zwanziger action. Using this framework, the gluon propagator does vanish at zero momentum and the ghost propagator is infrared enhanced and thus something was missing in the GZ action. Therefore, in this thesis, we have tried to restore the agreement with the recent lattice data by refining the GZ action, by including $d=2$ condensates. Doing so, we have found a gluon propagator which does not vanish at zero momentum and a ghost propagator which is no longer infrared enhanced. We have also fitted our form of the gluon propagator with the lattice results, and we have found a quite remarkable agreement.\\
\\
In 2d, something peculiar is going on. The lattice data show that the ghost propagator does display enhancement and the gluon propagator vanishes at zero momentum. We have shown that in 2d, it is impossible to refine the GZ action due to the typical infrared problems of 2d gauge theories. Therefore we are lead to the original GZ results, which indeed qualitatively agree with the lattice results.\\
\\
In conclusion, we have constructed a model which agrees with the lattice data when comparing quantities like propagators. We have obtained satisfying results, but some critical remarks should be made:
\begin{itemize}
\item We have not really succeeded in computing  good dynamical values for the condensates $\braket{A^2}$ and $\braket{\overline{\varphi}\varphi-\overline{\omega}\omega}$. This is due to the difficulties of the calculations when having more than one mass parameter into the game. However, the effective potential, calculated in the end of chapter \ref{refined} is better suited to discuss these condensates. However, still two parameters are unknown and need to be calculated before being able to extract a value for the condensates. Although, we should notice that even this does not allow to include the full-nonperturbative value of these condensates\footnote{We are still not able to take into account e.g.~topological effects,\ldots. }.
\item As higher loop calculations are very difficult using the GZ action, no one has really succeeded in obtaining a gluon or a ghost propagator analytically which completely agrees with the lattice data for a momentum regime 0-1.5 GeV without any input from the lattice. We can only predict qualitative results, which could then be fitted to the lattice. This itself is already a non trivial point.
\item One can also wonder whether the very deep infrared behavior of the propagators really matters? One could argue that only the mid-momentum regime 0.1 GeV -0.8 GeV is important, as this regime agrees with the size of particles, namely 0.8 fm - 6.4 fm. Is has already been argued that whether the gluon propagator reaches zero/not zero at zero momentum has absolutely no influence on the properties of the particles \cite{Blank:2010pa}. Perhaps all the commotion around the zero momentum behavior of the propagators is not that important after all. Nevertheless, a whole confinement story was built in the past on the IR singularity of the ghost, in which case the zero momentum regime was of utmost importance.
\end{itemize}

\section*{...to glueballs}
So far, we have thus concentrated on the behavior of propagators in the Landau gauge. However, these are unphysical quantities and at a certain point, one should start looking for the physical degrees of freedom in pure QCD, i.e.~glueballs. For the (R)GZ framework, the start of this investigation was done in chapters \ref{chappart1}, \ref{chappart2} and \ref{last}. This comprehends the second part of the results.\\
\\
As a first attempt, we have investigated in chapter \ref{chappart1} the renormalization of $F^2_{\mu\nu}$ using the (R)GZ action, as this operator is usually associated with the scalar glueball. This renormalization was far from trivial, as this operator mixes with other $d=4$ operators, namely
\begin{multline}
\mathcal E =  \p_\mu b^a  A_\mu^a + \p_\mu  \overline c^a D_\mu^{ab} c^b +  \p_\mu \overline \varphi_i^a  D_\mu^{ab} \varphi^b_i  - \p_\mu \overline \omega_i^a  D_\mu^{ab} \omega_i^b  + g f^{abc} \p_\mu \overline \omega_i^a    D_\mu^{bd} c^d  \varphi_i^c  \\
+ \gamma ^{2} g  f^{abc}A_\mu^a \varphi_\mu^{bc} +  \gamma^2 g f^{abc} A_\mu^a \overline \varphi_\mu^{bc} + d \left(N^{2}-1\right)  \gamma^4 \,.
\end{multline}
Moreover, due to the breaking of the BRST in the GZ action, this mixing has serious consequences\footnote{In ordinary Yang-Mills theory, this mixing is irrelevant as the Yang-Mills action is invariant under the BRST symmetry.} on the correlator $\Braket{F^2(x) F^2(y)}$. In fact, instead of considering $\Braket{F^2(x) F^2(y)}$, one should really consider the following correlator, $\Braket{\mathcal R(x) \mathcal R(y)}$, whereby $\mathcal R$ is a renormalization group invariant,
\begin{eqnarray}
\mathcal{R}= \frac{\beta(g^2)}{g^2} \mathcal F -2\gamma_c(g^2)  \mathcal E \;.
\end{eqnarray}
We emphasize that this analysis has been done for the GZ action as well as for the RGZ action, but the same operator appears for both actions. The problem with this operator, is that it does not have a proper K\"all\'{e}n-Lehmann representation,
\begin{align}
 \Braket{ \mathcal O(k) \mathcal O(-k) }  =  \int_{\tau_{0}}^{\infty} \d\tau \; \rho({\tau}) \; \frac{1}{\tau+k^2} \;,
\end{align}
with $\rho({\tau})\geq 0$. The correlator $\Braket{F^2(x) F^2(y)}$ and most likely also $\Braket{\mathcal R(x) \mathcal R (y)}$ has branch cuts at the imaginary axis. This seems to indicate that something is still missing.\\
\\
Therefore, in chapter \ref{chappart2} we have looked for an operator at lowest order, which does have physical cuts and can be written in a K\"all\'{e}n-Lehmann representation with positive spectral density. For this, we needed to introduce a new concept: $i$-particles.  In fact, this is nothing more than rewriting the fields of the GZ action into a new notation so the quadratic part of the GZ action becomes (almost) diagonal. We have found that the lowest order correlator corresponding to the following operators can be written into a proper spectral representation with positive spectral density
\begin{align}
O^{(1)}_{\lambda\eta}(x) &=  \left( \lambda^{a}_{\mu\nu}(x) \eta^{a}_{\mu\nu}(x) \right) \;, \nonumber  \\
O^{(2)}_{\lambda\eta}(x) &=  \varepsilon_{\mu\nu\rho\sigma} \left( \lambda^{a}_{\mu\nu}(x) \eta^{a}_{\rho\sigma}(x) \right) \;.
\end{align}
whereby the $\lambda^{a}_{\mu\nu}(x)$ and $\eta^{a}_{\mu\nu}(x)$ are given in terms of the original GZ fields in equation \eqref{r5f}. Although at lowest order, these operators could be interpreted as being ``physical'' in a sense, these operators do not really look renormalizable. Here we emphasize that so far, we only did the analysis within the GZ framework.\\
\\
In conclusion, so far, it seems that we have to choose: or we have a renormalizable operator, or we have an operator with a good spectral density. The question remains how we can unite both properties.\\
\\
Finally, in the last chapter, chapter \ref{last}, we have demonstrated that estimates for the masses obtained with the $i$ particles, thereby using the fits from the RGZ propagator with the lattice data (see \ref{orlandof} of chapter \ref{refined}), are in fact no so far of the lattice data. This gives some good hope for the RGZ model. However, we still need to explain the discrepancy between the $i$-particles and the glueball operators, i.e.~how to get rid in a calculable fashion, of the unphysical pieces in the correlation functions.

\section*{Outlook}
Let us end by saying that this research is far from finished. At the end of the chapter \ref{chappart2}, we have already given a sneak preview on how we could continue with this research. As we have only considered the GZ action in chapter \ref{chappart2}, the possibility remains that in the RGZ model, we can combine both properties of renormalizability and a good spectral representation. Of course, even if one would succeed in this task, one has not solved confinement. If one would really want to find a particle, one should be able to do some kind of all order resummation which would lead to a pole instead of a branch cut. However, this does not seem feasible for the moment with the current techniques. Therefore, it would already be a nice first step to find a renormalizable operator which has a  K\"all\'{e}n-Lehmann representation with positive spectral density, and continue the research from there on. \\
\\
Let us also point out that the described attempt was the first trying to convert unphysical degrees of freedom (gluons and ghosts) to physical degrees of freedom (glueballs), directly starting from the elementary propagators in the Landau gauge. Evidently, this is not an easy task, hence the results are of a rather humble nature. Let us however point out that the kind of difficulties encountered will be common to all approaches, and should not be seen as disfavouring the GZ formalism focused on in this thesis.\\
\\
In particular, focusing on the Landau gauge and its propagators, whatever approach one is using, at the end of the day, the form of the gluon/ghost propagator will, and should be, very similar to that of the lattice estimate, in particular there will be an infrared suppressed gluon propagator, with a positivity violation. One could even assume that $D(0)=0$. It is clear that no normal, physical particle corresponds to these gluon degrees of freedom.\\
\\
We would also like to point out that the BRST symmetry, broken or not, would appear to be only of a minor help in the discussion of positivity and unitarity in the confined phase. The physical spectrum should correspond to the classically gauge invariant operators. If the usual BRST is there, this is a trivial fact. But from our present research, it should be observed that using the highly nontrivial Landau gauge gluon propagator, which form is clear from the lattice, would seem to give rise to a rather complicated cut structure of the correlation functions of the gauge invariant composite operators. As a consequence, it will unavoidably be a hard nut to crack to understand in a qualitative and quantitative fashion how the  ``strange'' gluon (which is definitely not of a massless nature) will conspire with the massless ghost, quadratically or worse divergent in the infrared, to give gauge invariant correlation functions only exhibiting a physical cut and poles. The usual BRST analysis does not seem of much help here, as the BRST itself is not what is guaranteeing the positivity, not even in the perturbative case. It serves as a powerful tool to select a subspace, whereafter the positivity of the remaining physical degrees of freedom (the transverse gluon polarizations) is indeed observed, but the latter irrespective of the BRST symmetry transformation itself.\\
\\
Also other attempts to extract glueball mass estimates suffer from drawbacks. For example, using potential models to construct the bound state spectrum, one needs to make crucial assumptions about the interaction potential. Frequently, confinement is built in using a linear potential. At other instances, even assumptions about the nature of the gluons (massive/massless?) need to be made. When using the Bethe-Salpeter equations for bound states, one has to make assumptions about the invoked kernel, of how to perform resummations, $\ldots$. Aspects of gauge invariance are also not always transparent. Using the celebrated sum rules, one also needs to assume the existence of the bound state, and unknown nonperturbative physics is parametrized under the form of nonvanishing quark and gluon condensates. With lattice simulations, one can find nice estimates devoid of any gauge fixing ambiguities, but in this case, the precise physics behind the glueballs is not always clear. As lattice gauge theory is also purely Euclidean in nature, not much can be learnt about the analytic structure of the correlation functions.\\
\\
All these problems are merely a reflection of the intrinsic difficulty of constructing the bound state spectrum in a strongly interacting theory with confinement, as is QCD.

\appendix
\chapter{Formulae}
\section{Gaussian integrals}
\subsection{Gaussian integral for scalar variables}
\begin{eqnarray}\label{gauss1}
	I(A, J) &=& \int [\d \varphi] \exp \left[ - \frac{1}{2} \int \d^d x \d^d y\  \varphi(x) A(x,y) \varphi(y) +   \int \d^d x \ \varphi (x) J(x) \right] \nonumber \\
	&=& C (\det A)^{-1/2} \exp \frac{1}{2} \int \d^dx \d^dy\ J(x) A^{-1}(x,y) J(y),
\end{eqnarray}
with $C$ an infinite constant, which, in practice, can always be omitted.

\subsection{Gaussian integral for complex conjugated scalar variables}
\begin{eqnarray}\label{gauss8}
	I(A, J) &=& \int [\d \varphi] [\d \overline \varphi] \exp \left[ -  \int \d^d x \d^d y\  \overline \varphi(x) A(x,y) \varphi(y) +   \int \d^d x \left( \overline \varphi(x)  J_{\overline \varphi}(x) + \varphi (x) J_\varphi(x) \right)\right] \nonumber \\
	&=& C (\det A)^{-1} \exp  \int \d^d x \d^d y\ J_\varphi(x) A^{-1}(x,y) J_{\overline \varphi}(y)\;,
\end{eqnarray}
again with $C$ an infinite constant.

\subsection{Gaussian integral for Grassmann variables}
\begin{eqnarray}\label{ghostapp}
	I(A,\eta,\bar{\eta}) &=& \int [\d\theta][\d\bar{\theta}] \exp\left[ \int \d^d x \d^d y \ \bar{\theta}(x) A(x,y) \theta(y) + \int \d^dx \ (\bar{\eta}(x) \theta(x) + \bar{\theta}(x) \eta(x)) \right]  \nonumber \\
	&=&C \det A \exp  -\int \d^d x \d^d y \   \bar{\eta}(x) A^{-1}(x,y) \eta(y),
\end{eqnarray}
again with $C$ an infinite constant.

\section{Loop integrals}
To combine propagator denominators, we can use the following Feynman trick
\begin{eqnarray}\label{combdenominator}
\frac{1}{AB} &=& \int_0^1 \d x \frac{1}{[xA + (1-x)B]^2} \;,
\end{eqnarray}
which can be safely applied when no poles appear in the right hand side of \eqref{combdenominator}. This means that one is in a dangerous zone when $\frac{B}{B-A}$ is real and
$0 < \frac{B}{B-A} < 1$.\\
\\
To evaluate a $d$ dimensional loop integral, one can use
\begin{equation}\label{loopint}
\int \frac{\d^dq}{(2\pi)^d}\;\frac{1}{\left[q^2 + \Delta^2\right]^n} = \frac{1}{(4\pi)^{\frac{d}{2}}} \frac{\Gamma(n-\frac{d}{2})}{\Gamma(n)}(\Delta^2)^{\frac{d}{2}-n}\;,
\end{equation}
which is valid for $q^2>0$. \\
\\
The structure constants of the $SU(N)$ group have the following property,
\begin{eqnarray}\label{liestructure}
f^{abc}f^{dbc} &=& N \delta^{ad} \;.
\end{eqnarray}



\chapter{Some loose ends\label{sigma}}
\section{$D_\mu (A) \omega = 0$ is a gauge invariant equation.\label{appgaugeinv}}
To prove that $D_\mu (A) \omega = 0$ is a gauge invariant equation, we can write from expression \eqref{covariantderivativeadjoint},
\begin{equation}
D_\mu \omega = \p_\mu \omega - \ii g A_\mu \omega + \ii g \omega A_\mu \;.
\end{equation}
Now performing a $SU(N)$ transformation, we know that $D_\mu \omega$ is in the adjoint representation by definition,
\begin{eqnarray}
D_\mu \omega  = 0 \to U D_\mu \omega U^\dagger = 0 \;,
\end{eqnarray}
so working out this equation we find,
\begin{eqnarray}
U D_\mu \omega U^\dagger &=& U \p_\mu \Omega U^\dagger - \ii g U A_\mu U^\dagger U \omega U^\dagger + \ii g U \omega U^\dagger U A_\mu U^\dagger \nonumber\\
&=& \p_\mu \omega' - (\p_\mu U) \omega U^\dagger  - U  \omega (\p_\mu U^\dagger ) - \ii g (A_\mu' + \frac{\ii}{g} \p_\mu U U^\dagger ) \omega ' + \ii g \omega '(A_\mu' + \frac{\ii}{g} \p_\mu U U^\dagger ) \nonumber\\
&=& D_\mu' \omega' - \p_\mu U \omega  U^\dagger - U \omega \p_\mu U^\dagger + \p_\mu U U^\dagger (U \omega U^\dagger) - (U \omega U^\dagger) \p_\mu U U^\dagger \nonumber\\
&=&  D_\mu' \omega'\;,
\end{eqnarray}
whereby we made use of equation \eqref{notinf} and the simple formula
\begin{eqnarray}
U U^\dagger = 1 &\Rightarrow& \p_\mu U U^\dagger +  U \p_\mu U^\dagger = 0 \;.
\end{eqnarray}
We have thus indeed proven that $D_\mu (A) \omega = 0$ is a gauge invariant equation.

\section{$\sigma$ decreases with increasing $k^2$.}
We shall prove that the following function
\begin{eqnarray*}
f(k,A) &=& \frac{k_\mu  k_\nu}{k^2} \int\frac{ \d^d q}{(2 \pi)^2} f(q^2)  \frac{ 1 }{(k-q)^2} P_{\mu\nu} = \int\frac{ \d^4 q}{(2 \pi)^2} f(q^2)  \frac{ 1 }{(k-q)^2}  \left( 1 - \frac{k_\mu k_\nu}{k^2} \frac{q_\mu q_\nu}{q^2} \right)\;,
\end{eqnarray*}
decreases with increasing $k^2$. Let us prove this in 2 dimensions for simplicity. We assume $k = (k_x, k_y)$ to be oriented along the $x$ axis. Using polar coordinates, we obtain
\begin{eqnarray}
f(k,A) &=&  \int_0^{\infty}\frac{ \d q}{(2 \pi)^2} q f(q^2)  \int_{0}^{2\pi} \d \theta \frac{ 1 - \cos^2 \theta }{k^2 + q^2 - k q \cos \theta }  \nonumber\\
&=&  \int_0^{\infty}\frac{ \d q}{(2 \pi)^2} q f(q^2)  \left( \theta(q^2 - k^2) \frac{\pi}{k^2} + \theta(k^2 - q^2) \frac{\pi}{k^2} \right) \;,
\end{eqnarray}
whereby we have used the result \eqref{contourint} from section \ref{chap42dform} in chapter \ref{refined}. Now deriving $f(k,A)$ w.r.t.~$k^2$ and using the property of the $\theta$ function: $\frac{\p}{\p x} \theta(x-y) = \delta(x-y)$, we find
\begin{eqnarray}
\frac{\p}{\p k^2} f(k,A) &=& -\int_0^{\infty}\frac{ \d q}{(2 \pi)^2} q f(q^2)  \theta(k^2 - q^2) \frac{\pi}{k^4} = - \theta(k)\frac{\pi}{k^4} \int_0^{k}\frac{ \d q}{(2 \pi)^2} q f(q^2) \;,
\end{eqnarray}
and thus $f(k,A)$ is a decreasing function for increasing $k^2$.

\section{Determinant of $K_{\mu\nu}$}
We calculate the determinant of
\begin{eqnarray}
K_{\mu \nu}^{ab} (k) &=&\delta^{ab} \left( \underbrace{\beta\frac{1}{V} \frac{2}{d} \frac{N g^2}{N^2 - 1}}_{\lambda} \delta_{\mu\nu} \frac{1}{k^2} + \delta_{\mu\nu} k^2 + \left(\frac{1}{\alpha} - 1 \right)k_\mu k_\nu \right)\;.
\end{eqnarray}
We can write
\begin{eqnarray}\label{B4}
\left(\det  K_{\mu \nu}^{ab} (k) \right)^{-1/2} &=& \e^{-\frac{1}{2} \ln \det  K_{\mu \nu}^{ab}} = \e^{-\frac{1}{2}\Tr \ln   K_{\mu \nu}^{ab}}\;.
\end{eqnarray}
Therefore, we need to determine
\begin{eqnarray}
\Tr \ln   K_{\mu \nu}^{ab} 
&=& (N^2 - 1) \Tr \ln \left( \delta_{\mu\kappa} \left(\frac{\lambda}{k^2} + k^2 \right) \left( \delta_{\kappa\nu} +  \frac{1}{\frac{\lambda}{k^2} + k^2}\left(\frac{1}{\alpha} - 1 \right)k_\kappa k_\nu \right) \right) \nonumber\\
&=& (N^2 - 1) \left[ \Tr \ln \left( \delta_{\mu\nu} \left(\frac{\lambda}{k^2} + k^2 \right)\right) + \Tr \ln \left( \delta_{\mu\nu} +  \frac{k^2}{\lambda + k^4}\left(\frac{1}{\alpha} - 1 \right)k_\mu k_\nu \right)   \right] \nonumber\\
&=& (N^2 - 1)\Biggl[ d \sum_k \ln \frac{k ^4 + \lambda}{k ^2}\nonumber\\
 && + \Tr \left( \frac{k^2}{\lambda + k^4}\left(\frac{1}{\alpha} - 1 \right)k_\mu k_\nu + \left( \frac{k^2}{\lambda + k^4}\left(\frac{1}{\alpha} - 1\right) \right)^2 k_\mu k_\kappa k_\kappa k_\nu\right)  \Biggr]\;,
\end{eqnarray}
whereby we used $\ln (1 + x) = x - \frac{x^2}{2} + \ldots$. We can now take the trace of the diagonal elements of the second term, and again use $x - \frac{x^2}{2} + \ldots = \ln (1 + x)$. We obtain,
\begin{eqnarray}
\Tr \ln   K_{\mu \nu}^{ab} &=& (N^2 - 1)\left[ d \sum_k \ln \frac{k ^4 + \lambda}{k ^2}  + \sum_k \ln \left( 1 +  \frac{k^2}{\lambda + k^4}\left(\frac{1}{\alpha} - 1 \right) k^2 \right)\right] \nonumber\\
&=& (N^2 - 1)\left[ d \sum_k \ln \frac{k ^4 + \lambda}{k ^2} - \sum_k \ln \frac{k^4 + \lambda}{k^2} + \sum_k \ln \left( \frac{\lambda}{k^2} + \frac{k^2}{\alpha}\right) \right]\;.
\end{eqnarray}
By working out the last term, we see that it is proportional to $\alpha$,
\begin{eqnarray}
\sum_k \ln \left( \frac{\lambda}{k^2} + \frac{k^2}{\alpha}\right) &=&  \sum_k \ln \left( \frac{k^4}{\alpha} + \lambda \right)  - \sum_k \ln k^2 \nonumber\\
&=& V \int \frac{\d^d k}{(2\pi)^d} \ln \left( \frac{k^2}{\sqrt{\alpha}} + \ii \sqrt \lambda \right)+ V \int \frac{\d^d k}{(2\pi)^d} \ln \left( \frac{k^2}{\sqrt{\alpha}} - \ii \sqrt \lambda \right) \nonumber\\
&\sim& \alpha^{d/4} \;,
\end{eqnarray}
whereby $\int \d^q q \ln q^2$ is zero in dimensional regularization. Therefore, in the limit $\alpha \to 0$, becomes zero.  In conclusion, we find
\begin{align}\label{B8}
\left(\det  K_{\mu \nu}^{ab} (k) \right)^{-1/2}  &=\exp  \left[(N^2 - 1) \frac{(d-1)}{2}V \int \frac{\d^d k}{(2\pi)^d} \ln \frac{k^4 + \lambda}{k^2} \right] \nonumber\\
&=\exp  \left[(N^2 - 1) \frac{(d-1)}{2}V  \int \frac{\d^d k}{(2\pi)^d} \ln \left( k^2 + \frac{1}{V} \frac{2}{d} \frac{\beta N g^2}{N^2 - 1}\frac{1}{k^2} \right) \right]\;.
\end{align}

\section{Addendum chapter 4: The one loop effective potential\label{appendix3}}
\subsection{The 4d case}
We explain in  detail how we obtained equation \eqref{wnull}. Starting from\footnote{Notice that $W^{(0)}(J)$ is in fact $Z_c(J)$ as defined in equation \eqref{exp}.}
\begin{eqnarray}
\e^{-W^{(0)}(J)} &=& \int [\d \Phi] \e^{-S_{\RGZ}^0} \;,
\end{eqnarray}
whereby $S_\RGZ^0$ is the quadratic part of the action $S_\RGZ$, we can integrate out over all fields so we obtain,
\begin{eqnarray}
 \int [\d \Phi] \e^{-S_{\RGZ}^0} &=& \e^{d(N^2-1) \gamma^4}  \int [\d A] \e^{-\frac{1}{2} \int \frac{\d^d p}{(2\pi)^d} A_\mu^a \Delta_{\mu\nu}^{ab}  A_\nu^b} \;,
\end{eqnarray}
whereby
\begin{eqnarray}
\Delta^{ab}_{\mu\nu} &=&\left[ \left(p^2 + m^2 + \frac{2 g^2 N \gamma^4}{p^2 + M^2} \right) \delta_{\mu\nu}  + p_{\mu} p_{\nu} \left(\frac{1}{\alpha} - 1\right) \right] \delta^{ab} \;.
\end{eqnarray}
Using equation \eqref{gauss1}, we find
\begin{eqnarray}\label{2qh}
 \int [\d \Phi] \e^{-S_{\RGZ}^0} &=& \e^{d(N^2-1) \gamma^4}  \left(\det \Delta_{\mu\nu}^{ab} \right)^{-1/2} =  \e^{d(N^2-1) \gamma^4} \e^{-\frac{1}{2} \Tr \ln \Delta^{ab}_{\mu\nu}} \;,
\end{eqnarray}
similar as in equation \eqref{B4}. We can now perform the same steps as in \eqref{B4}-\eqref{B8} finding,
\begin{eqnarray*}
\frac{1}{2} \Tr \ln \Delta^{ab}_{\mu\nu} &=&(N^2 - 1) \frac{(d-1)}{2}\Tr \ln \left( p^2 + m^2 + \frac{\lambda^4}{p^2 + M^2} \right) \;. \nonumber\\
&=&   (N^2 - 1) \frac{(d-1)}{2}\Tr \left[ \ln\left[ ( p^2 + m^2) (p^2 + M^2) + \lambda^4 \right] -  \ln (p^2 + M^2) \right]\;,
\end{eqnarray*}
whereby we have used the notational shorthand \eqref{lambda4}. The second part is a standard integral \cite{Peskin} and evaluated as:
\begin{eqnarray}\label{standard}
\Tr \ln ( p^2 + M^2) &=& \frac{-\Gamma(-d/2)}{(4\pi)^{d/2}} \frac{1}{(M^2)^{-d/2}}\;,
\end{eqnarray}
with $\Gamma$ the Euler Gamma-function. Using dimensional regularization, $d= 4- \epsilon$ we obtain,
\begin{eqnarray}\label{f1}
 -\frac{N^2 - 1}{2} (d-1) \Tr \ln ( p^2 + M^2) &=& -3 \frac{N^2 - 1}{64 \pi^2} M^4 \left( -\frac{5}{6} - \frac{2}{\epsilon} + \ln \frac{M^2}{\overline{\mu}^2}
 \right)\;.
\end{eqnarray}
We recall that we work in the $\MSbar$ scheme. Next, we try to convert the first part into the standard form,
\begin{align}\label{f2}
\frac{N^2 - 1}{2}& (d-1)  \Tr \ln \bigl(( p^2 + m^2) (p^2 + M^2) + \lambda^4  \bigr) \nonumber\\
=& \frac{N^2 - 1}{2} (d-1) \Tr \ln \left( p^2 + m_1^2 \right) + \Tr \ln \left( p^2 + m_2^2 \right) \nonumber\\
=&\frac{N^2 - 1}{2} (d-1)\left[ \frac{-\Gamma(-d/2)}{(4\pi)^{d/2}} \frac{1}{(m_1^2)^{-d/2}} + \frac{-\Gamma(-d/2)}{(4\pi)^{d/2}} \frac{1}{(m_2^2)^{-d/2}} \right]\nonumber\\
=&  3 \frac{N^2 - 1}{64 \pi^2} \left( m_1^4 \left( -\frac{5}{6} -
\frac{2}{\epsilon} + \ln \frac{m_1^2}{\overline{\mu}^2} \right)  +
m_2^4 \left( -\frac{5}{6} - \frac{2}{\epsilon} + \ln
\frac{m_2^2}{\overline{\mu}^2} \right) \right)+O(\epsilon) \;,
\end{align}
where $m_1$ and $m_2$ are given by
\begin{align}\label{mm2}
m_1^2 &= \frac{  (m^2 + J) - \sqrt{(J + m^2)^2  - 4( m^2 J + \lambda^4)  }}{2} \;, \nonumber\\
 m_2^2 &= \frac{ (m^2 + J) + \sqrt{(J + m^2)^2  - 4( m^2 J + \lambda^4)  }}{2} \;.
\end{align}
We still have to calculate the first term of \eqref{2qh}. From equation \eqref{Zgamma}, we can calculate that\footnote{For the explicit loop calculations of the $Z$-factors, we refer to \cite{Gracey:2002yt}.}
\begin{eqnarray}
\gamma_0^4 &=& Z_{\gamma^2}^2 \gamma^4\;,  \qquad \text{with} \qquad Z_{\gamma^2}^2 = 1+ \frac{3}{2} \frac{g^2N}{16 \pi^2} \frac{1}{\epsilon}\;,
\end{eqnarray}
so we find
\begin{eqnarray}\label{f3}
-d(N^2 - 1)\gamma^4_0 &=& -4 (N^2 - 1)\gamma^4 - 4 \frac{3}{2} (N^2 - 1)\frac{g^2 N}{16\pi^2}\frac{1}{\epsilon} \gamma^4 + \frac{3}{2} \frac{g^2N}{16 \pi^2}\gamma^4 (N^2-1) \;.
\end{eqnarray}
From equation \eqref{f1}, \eqref{f2} and \eqref{f3} we see that the infinities cancel out nicely, so that the functional energy reads,
\begin{eqnarray*}
W^{(0)}(J) &=& - \frac{4 (N^2 - 1)}{2 g^2 N} \lambda^4  + \frac{3(N^2 - 1)}{64 \pi^2} \left( \frac{8}{3} \lambda^4 + m_1^4 \ln \frac{m_1^2}{\overline{\mu}^2}
+ m_2^4 \ln \frac{m_2^2}{\overline{\mu}^2} - J^2 \ln \frac{J}{\overline{\mu}^2}\right) \;,
\end{eqnarray*}
which is exactly expression \eqref{wnull}.

\subsection{The 3d case\label{appendix4}}
The 3d case can be calculated in a similar fashion as the 4d case. As expression \eqref{2qh} is still general for all $d$, we have that
\begin{eqnarray}
W^{(0)}(J) &=& - d(N^2-1) \gamma^4 + \frac{1}{2} \Tr \ln \Delta^{ab}_{\mu\nu}\;,
\end{eqnarray}
whereby
\begin{eqnarray*}
\frac{1}{2} \Tr \ln \Delta^{ab}_{\mu\nu} &=&  (N^2 - 1) \frac{(d-1)}{2}\Tr \left[ \ln\left[ ( p^2 + m^2) (p^2 + M^2) + \lambda^4 \right] -  \ln (p^2 + M^2) \right]\nonumber\\
&=& (N^2 - 1) \frac{(d-1)}{2}\Tr \left[ \ln\left[  p^2 + m_1^2 \right] + \ln \left[  p^2 + m_1^2 \right] -  \ln (p^2 + M^2) \right]\;,
\end{eqnarray*}
whereby $m_1$ and $m_2$ are given by expression \eqref{mm2}. By employing the standard formula \eqref{standard}, we find
\begin{eqnarray}
\frac{1}{2} \Tr \ln \Delta^{ab}_{\mu\nu} &=& \frac{N^2-1}{6\pi}\left(-m_1^3-m_2^3+J^{3/2}\right)\;,
\end{eqnarray}
and thus
\begin{equation}\label{Wnull3d}
  W^{(0)}(J)=-3(N^2-1)\frac{\lambda^4}{2g^2N}+\frac{N^2-1}{6\pi}\left(-m_1^3-m_2^3+J^{3/2}\right) \;.
\end{equation}

\section{Spectral density for one real mass and one vanishing mass}\label{appendixspectral}
We start from
\begin{eqnarray}
F(k^2) &=& \int \frac{\d^d p}{(2\pi)^d} \frac{1}{(k-p)^2 + m^2} \frac{1}{p^2}\,.
\end{eqnarray}
For $k^2>0$, we can employ the Feynman trick and differentiating w.r.t.~$k^2$, yields for $d=4$
\begin{eqnarray}
\frac{\p F(k^2)}{\p k^2} &=& - \frac{1}{16\pi^2}  \int_0^1 \d x  \frac{x (1-x)}{x (1-x) k^2  + x m^2} \nonumber\\
&=& - \frac{1}{16\pi^2}  \int_0^1 \d x  \frac{1}{ k^2  +  \frac{m^2}{1-x}}\,.
\end{eqnarray}
We perform a transformation of variables, by setting $s = \frac{m^2}{1-x}$,
\begin{eqnarray}
\frac{\p F(k^2)}{\p k^2} &=&  - \frac{m^2}{16\pi^2}  \int_{m^2}^{+\infty} \d s  \frac{1}{s^2} \frac{1}{ k^2  +   s  } \nonumber\\
&=& + \frac{1}{16\pi^2}  \int_{m^2}^{+\infty} \d s  \frac{\d }{ \d s} \left( \frac{m^2}{s}\right) \frac{1}{ k^2  +   s  }\,.
\end{eqnarray}
After doing a partial integration, we obtain
\begin{eqnarray}
\frac{\p F(k^2)}{\p k^2} &=&  \frac{1}{16\pi^2} \left[  \left. \frac{1}{ k^2  +   s  } \frac{m^2}{s} \right|_{s = m^2}^{+\infty} -  \int_{m^2}^{+\infty} \d s \frac{m^2}{s}  \frac{d }{ \d s} \left( \frac{1}{ k^2  +   s  }\right) \right] \nonumber\\
&=& \frac{-1}{16\pi^2} \left[  \frac{1}{ k^2  +   m^2 }   +  \int_{m^2}^{+\infty} \d s \frac{m^2}{s}  \frac{-1}{( k^2  +   s )^2 }\right] \nonumber\\
&=&  \frac{-1}{16\pi^2}  \frac{\p }{\p k^2}   \left[  \ln ( k^2  +   m^2 )  +    \int_{m^2}^{+\infty} \d s \frac{m^2}{s}\frac{1}{ k^2  +   s  }  \right]\,,
\end{eqnarray}
so that
\begin{eqnarray}
 F(k^2) - F(0) &=&    \frac{1}{16\pi^2}  \left[ - \ln ( k^2  +   m^2 ) + \ln m^2  -   \int_{m^2}^{+\infty} \d s \frac{m^2}{s}\frac{1}{ k^2  +   s  }   + \int_{m^2}^{+\infty} \d s \frac{m^2}{s}\frac{1}{ s  }  \right]  \nonumber\\
 &=&    \frac{1}{16\pi^2}  \left[  \int_{m^2}^{+\infty} \d s \frac{1}{k^2 +s}-  \int_{m^2}^{+\infty} \d s \frac{1}{s} -   \int_{m^2}^{+\infty} \d s \frac{m^2}{s}\frac{1}{ k^2  +   s  }   + \int_{m^2}^{+\infty} \d s \frac{m^2}{s}\frac{1}{ s  }  \right] \nonumber\\
 &=&    \frac{1}{16\pi^2}   \int_{m^2}^{+\infty} \d s  \left[ \frac{1}{k^2 +s}- \frac{1}{s}\right]\left( 1-   \frac{m^2}{s}\right)\,.
\end{eqnarray}
We conclude that the spectral density is given by
\begin{eqnarray}\label{appspec}
\rho_1(s) &=& 1-   \frac{m^2}{s}\,,
\end{eqnarray}
which is indeed positive for $s\geq m^2$.

\section{Calculating correlators\label{appcorr}}
\subsection{The correlator $\braket{O^{(1)}_{\lambda\eta}(x) O^{(1)}_{\lambda\eta}(y)}$}

We shall now try to calculate $\braket{ O^{(1)}_{\lambda\eta}(k) O^{(1)}_{\lambda\eta}(-k)}$. Already at lowest order, this asks for some algebra. We depart from
\begin{eqnarray} \label{eq6}
\braket{O^{(1)}_{\lambda\eta}(x) O^{(1)}_{\lambda\eta}(y)}&=&  \braket{ \lambda_{\mu\nu}^a(x) \eta_{\mu\nu}^a(x) \lambda_{\alpha\beta}^b(y) \eta_{\alpha\beta}^b(y)}\;,
\end{eqnarray}
or
\begin{eqnarray}
\braket{O^{(1)}_{\lambda\eta}(x) O^{(1)}_{\lambda\eta}(y)}&=& 4\braket{\p_\mu \lambda_\nu^a \p_\mu \eta_\nu^a \p_\alpha \lambda_\beta^b\p_\alpha \eta_\beta^b }-4\braket{\p_\mu \lambda_\nu^a \p_\mu \eta_\nu^a \p_\alpha \lambda_\beta^b\p_\beta
\eta_\alpha^b }\nonumber\\
&-&4\braket{\p_\mu \lambda_\nu^a \p_\nu \eta_\mu^a \p_\alpha \lambda_\beta^b\p_\alpha \eta_\beta^b }+4\braket{\p_\mu \lambda_\nu^a \p_\nu \eta_\mu^a \p_\alpha \lambda_\beta^b\p_\beta \eta_\alpha^b}\;,
\end{eqnarray}
at lowest order. It is understood that $\{\mu,\nu,a\}$ refers to $x$ and $\{\alpha,\beta,b\}$ to $y$. We shall now pass to Fourier space, in which case we find
\begin{eqnarray*}
\braket{O^{(1)}_{\lambda\eta}(x) O^{(1)}_{\lambda\eta}(y)}&=&4\int \d k \d \ell  \d p \d q \e^{\ii py}\e^{\ii qy}\e^{\ii kx}\e^{\ii \ell  x}k_\mu \ell_\mu p_\alpha q_\alpha \braket{\lambda_\nu^a(k) \eta_\nu^a(\ell ) \lambda_\beta^b(p) \eta_\beta^b(q)}\nonumber\\
&-&4\int \d k \d \ell  \d p \d q \e^{\ii py}\e^{\ii qy}\e^{\ii kx}\e^{\ii\ell  x}k_\mu \ell _\mu p_\alpha q_\beta \braket{\lambda_\nu^a(k) \eta_\nu^a(\ell ) \lambda_\beta^b(p) \eta_\alpha^b(q)}\nonumber\\
&-&4\int \d k \d\ell  \d p \d q \e^{\ii py}\e^{\ii qy}\e^{\ii kx}\e^{\ii\ell  x}k_\mu \ell _\nu p_\alpha q_\alpha \braket{\lambda_\nu^a(k) \eta_\mu^a(\ell ) \lambda_\beta^b(p) \eta_\beta^b(q)}\nonumber\\
&+&4\int \d k\d\ell  \d p \d q \e^{\ii py}\e^{\ii qy}\e^{\ii kx}\e^{\ii\ell  x}k_\mu \ell _\nu p_\alpha q_\beta \braket{\lambda_\nu^a(k) \eta_\mu^a(\ell ) \lambda_\beta^b(p) \eta_\alpha^b(q)}\;,
\end{eqnarray*}
so that we can now perform the contractions using Wick's theorem. We did not write down any factors of $(2\pi)^d$. Each time we contract 2 fields, we will pick up a propagator, and we have appropriately implemented momentum conservation, i.e.
\begin{equation}\label{1}
    \braket{\lambda_\mu^a(p) \lambda_\nu^b(k) \ldots}\to     D_{\mu\nu}\delta^{ab}\delta(p+k)\;,
\end{equation}
whereby we use the symbols $D_{\mu\nu}$, $D_{\mu\nu}^\dagger$ and $D'_{\mu\nu}$  for the $\braket{\lambda_\mu \lambda_\nu}$, $\braket{\eta_\mu \eta_\nu}$ and $\braket{\lambda_\mu \eta_\nu}$ propagator
Doing so, we can write down for example
\begin{multline}
\braket{\lambda_\nu^a(k) \eta_\nu^a(\ell ) \lambda_\beta^b(p) \eta_\beta^b(q)}=\delta^{aa}\delta^{bb}D'_{\nu\nu}(k)\delta(k+{}\ell )D'_{\beta\beta}(p)\delta(p+q)\\
+\delta^{ab}\delta^{ab}D_{\nu\beta}(k)\delta(k+p)D_{\nu\beta}^\dagger(q)\delta(\ell +q)+\delta^{ab}\delta^{ab}D_{\nu\beta}'(k)\delta(k+q)D_{\nu\beta}'(p)\delta(\ell +p)\nonumber\;.
\end{multline}
In each case, we may drop the first term, as this gives rise to a disconnected contribution. The other terms are given by
\begin{align*}
\braket{\lambda_\nu^a(k) \eta_\nu^a(\ell ) \lambda_\beta^b(p) \eta_\alpha^b(q)} =& \delta^{ab}\delta^{ab}D_{\nu\beta}(k)\delta(k+p)D_{\nu\alpha}^\dagger(q)\delta(\ell +q)\\
&+\delta^{ab}\delta^{ab}D_{\nu\alpha}'(k)\delta(k+q)D_{\nu\beta}'(p)\delta(\ell +p)\;,\\
\braket{\lambda_\nu^a(k) \eta_\mu^a(\ell ) \lambda_\beta^b(p) \eta_\beta^b(q)} =& \delta^{ab}\delta^{ab}D_{\nu\beta}(k)\delta(k+p)D_{\mu\beta}^\dagger(q)\delta(\ell +q)\\
&+\delta^{ab}\delta^{ab}D_{\nu\beta}'(k)\delta(k+q)D_{\mu\beta}'(p)\delta(\ell +p)\;,\\
\braket{\lambda_\nu^a(k) \eta_\mu^a(\ell ) \lambda_\beta^b(p) \eta_\alpha^b(q)} =& \delta^{ab}\delta^{ab}D_{\nu\beta}(k)\delta(k+p)D_{\mu\alpha}^\dagger(q)\delta(\ell +q)\\
&+\delta^{ab}\delta^{ab}D_{\nu\alpha}'(k)\delta(k+q)D_{\mu\beta}'(p)\delta(\ell +p)\;.
\end{align*}
Collecting all the terms, we find
\begin{multline*}
\braket{O^{(1)}_{\lambda\eta}(x) O^{(1)}_{\lambda\eta}(y)} = 4(N^2-1) \int \d k \d q \e^{\ii(k-q)(x-y)}\left[ k_\mu q_\mu k_\alpha q_\alpha \left(D_{\nu\beta}(k)D_{\nu\beta}^\dagger(q) + D'_{\nu\beta}(k)D'_{\nu\beta}(q)  \right) \right.\\
 - 2 k_\mu q_\mu k_\alpha q_\beta \left(D_{\nu\beta}(k)D_{\nu\alpha}^\dagger(q) + D'_{\nu\beta}(k)D'_{\nu\alpha}(q)  \right)
\left.  + k_\mu q_\nu k_\alpha q_\beta \left(D_{\nu\beta}(k)D_{\mu\alpha}^\dagger(q) + D'_{\nu\beta}(k)D'_{\mu\alpha}(q)  \right) \right]\;,
\end{multline*}
after some suitable renaming of Lorentz indices and momenta which allowed to recombine several terms. By the substitution
\begin{equation}
p= k-q\;,
\end{equation}
we can rewrite \eqref{eq6} as a Fourier transform
\begin{eqnarray}
\braket{O^{(1)}_{\lambda\eta}(x) O^{(1)}_{\lambda\eta}(y)} &=&\int\frac{\d^dp}{(2\pi)^d}\e^{ip(x-y)}
\braket{ O^{(1)}_{\lambda\eta}(p) O^{(1)}_{\lambda\eta}(-p)}\;,
\end{eqnarray}
whereby
\begin{multline*}
\braket{ O^{(1)}_{\lambda\eta}(p) O^{(1)}_{\lambda\eta}(-p)}= 4(N^2-1)\int \frac{\d^dk}{(2\pi)^d} \Bigl[ k_\mu (k-p)_\mu k_\alpha (k-p)_\alpha \left(D_{\nu\beta}(k)D_{\nu\beta}^\dagger(k-p) \right.\\
\left. + D'_{\nu\beta}(k)D'_{\nu\beta}(k-p)  \right)  - 2 k_\mu (k-p)_\mu k_\alpha (k-p)_\beta \left(D_{\nu\beta}(k)D_{\nu\alpha}^\dagger(k-p) \right.\\
 \left. + D'_{\nu\beta}(k)D'_{\nu\alpha}(k-p)  \right) + k_\mu (k-p)_\nu k_\alpha (k-p)_\beta \left(D_{\nu\beta}(k)D_{\mu\alpha}^\dagger(k-p) + D'_{\nu\beta}(k)D'_{\mu\alpha}(k-p)  \right) \Bigr]\;.
\end{multline*}
We can simplify this expression to find
\begin{equation}
\braket{ O^{(1)}_{\lambda\eta}(p) O^{(1)}_{\lambda\eta}(-p)}= 4(N^2-1)\int \frac{\d^dk}{(2\pi)^d} \Biggl[  \frac{(d-1) k^4+\left(p^2-2 (d-1) k\cdot p\right) k^2+(d-2) (k\cdot p)^2  }{(k^2 - \ii \lambda^2)((p-k)^2 + \ii \lambda^2)}\Biggr]\;.
\end{equation}

\subsection{The correlator $\braket{O^{(2)}_{\lambda\eta}(x) O^{(2)}_{\lambda\eta}(y)}$}
Let us now calculate $\braket{ O^{(2)}_{\lambda\eta}(k) O^{(2)}_{\lambda\eta}(-k)}$. We depart from
\begin{eqnarray}
\braket{O^{(2)}_{\lambda\eta}(x) O^{(2)}_{\lambda\eta}(y)}&=& \varepsilon_{\mu\nu\rho\sigma}  \varepsilon_{\alpha\beta\gamma\delta}  \braket{ \lambda_{\mu\nu}^a(x) \eta_{\rho\sigma}^a(x) \lambda_{\alpha\beta}^b(y) \eta_{\gamma\delta}^b(y)}\;,
\end{eqnarray}
or
\begin{eqnarray}
\braket{O^{(2)}_{\lambda\eta}(x) O^{(2)}_{\lambda\eta}(y)}&=& 4 \varepsilon_{\mu\nu\rho\sigma}  \varepsilon_{\alpha\beta\gamma\delta} \bigl[ \braket{\p_\mu \lambda_\nu^a \p_\rho \eta_\sigma^a \p_\alpha \lambda_\beta^b\p_\gamma \eta_\delta^b }-\braket{\p_\mu \lambda_\nu^a \p_\rho \eta_\sigma^a \p_\alpha \lambda_\beta^b\p_\delta
\eta_\gamma^b }\nonumber\\
&-&\braket{\p_\mu \lambda_\nu^a \p_\sigma \eta_\rho^a \p_\alpha \lambda_\beta^b\p_\gamma \eta_\delta^b }+\braket{\p_\mu \lambda_\nu^a \p_\sigma \eta_\rho^a \p_\alpha \lambda_\beta^b\p_\delta \eta_\gamma^b  } \bigr]\;,
\end{eqnarray}
at lowest order. It is understood that $\{\mu,\nu,\rho, \sigma,a\}$ refers to $x$ and $\{\alpha,\beta,\gamma, \sigma, b\}$ to $y$. We shall now pass to Fourier space, in which case we find
\begin{eqnarray*}
\braket{O^{(1)}_{\lambda\eta}(x) O^{(1)}_{\lambda\eta}(y)}&=&  \varepsilon_{\mu\nu\rho\sigma}  \varepsilon_{\alpha\beta\gamma\delta} \Bigl[4\int \d k \d \ell  \d p \d q \e^{\ii py}\e^{\ii qy}\e^{\ii kx}\e^{\ii \ell  x}k_\mu \ell_\rho p_\alpha q_\gamma \braket{\lambda_\nu^a(k) \eta_\sigma^a(\ell ) \lambda_\beta^b(p) \eta_\delta^b(q)}\nonumber\\
&-&4\int \d k \d \ell  \d p \d q \e^{\ii py}\e^{\ii qy}\e^{\ii kx}\e^{\ii\ell  x}k_\mu \ell _\rho p_\alpha q_\delta \braket{\lambda_\nu^a(k) \eta_\sigma^a(\ell ) \lambda_\beta^b(p) \eta_\gamma^b(q)}\nonumber\\
&-&4\int \d k \d\ell  \d p \d q \e^{\ii py}\e^{\ii qy}\e^{\ii kx}\e^{\ii\ell  x}k_\mu \ell _\sigma p_\alpha q_\gamma \braket{\lambda_\nu^a(k) \eta_\rho^a(\rho ) \lambda_\beta^b(p) \eta_\delta^b(q)}\nonumber\\
&+&4\int \d k\d\ell  \d p \d q \e^{\ii py}\e^{\ii qy}\e^{\ii kx}\e^{\ii\ell  x}k_\mu \ell_\sigma p_\alpha q_\delta \braket{\lambda_\nu^a(k) \eta_\rho^a(\ell ) \lambda_\beta^b(p) \eta_\gamma^b(q)}\Bigr]\;,
\end{eqnarray*}
so that we can now perform the contractions again,
\begin{align*}
\braket{\lambda_\nu^a(k) \eta_\sigma^a(\ell ) \lambda_\beta^b(p) \eta_\delta^b(q)}=& \delta^{aa} (D_{\nu\beta}(k)\delta(k+p)D_{\sigma\delta}^\dagger(q)\delta(\ell +q)
+D_{\nu\delta}'(k)\delta(k+q)D_{\sigma\beta}'(p)\delta(\ell +p))\nonumber\\
\braket{\lambda_\nu^a(k) \eta_\sigma^a(\ell ) \lambda_\beta^b(p) \eta_\gamma^b(q)}=& \delta^{aa}(D_{\nu\beta}(k)\delta(k+p)D_{\sigma\gamma}^\dagger(q)\delta(\ell +q)
+D_{\nu\gamma}'(k)\delta(k+q)D_{\sigma\beta}'(p)\delta(\ell +p)) \\
\braket{\lambda_\nu^a(k) \eta_\rho^a(\rho ) \lambda_\beta^b(p) \eta_\delta^b(q)}  =& \delta^{aa}(D_{\nu\beta}(k)\delta(k+p)D_{\rho\delta}^\dagger(q)\delta(\ell +q) +D_{\nu\delta}'(k)\delta(k+q)D_{\rho\beta}'(p)\delta(\ell +p))\\
\braket{\lambda_\nu^a(k) \eta_\rho^a(\ell ) \lambda_\beta^b(p) \eta_\gamma^b(q)}  =& \delta^{aa}(D_{\nu\beta}(k)\delta(k+p)D_{\rho\gamma}^\dagger(q)\delta(\ell +q) +D_{\nu\gamma}'(k)\delta(k+q)D_{\rho\beta}'(p)\delta(\ell +p))\;.
\end{align*}
Collecting all the terms, we find
\begin{multline*}
\braket{O^{(1)}_{\lambda\eta}(x) O^{(1)}_{\lambda\eta}(y)} = 4(N^2-1)\varepsilon_{\mu\nu\rho\sigma}  \varepsilon_{\alpha\beta\gamma\delta}  \int \d k \d q \e^{\ii(k-q)(x-y)}\Bigl[ k_\mu q_\rho k_\alpha q_\gamma D_{\nu\beta}(k)D_{\sigma\delta}^\dagger(q)\\
 + k_\mu q_\rho q_\alpha k_\gamma  D_{\nu\delta}'(k)D_{\sigma\beta}'(q)  -  k_\mu q_\rho k_\alpha q_\delta  D_{\nu\beta}(k)D_{\sigma\gamma}^\dagger(q) - k_\mu q_\rho q_\alpha k_\delta D'_{\nu\gamma}(k)D'_{\sigma\beta}(q)  \\
  -  k_\mu q_\sigma k_\alpha q_\gamma  D_{\nu\beta}(k)D_{\rho\delta}^\dagger(q) - k_\mu q_\sigma q_\alpha k_\gamma D'_{\nu\delta}(k)D'_{\rho\beta}(q)  + k_\mu q_\sigma k_\alpha q_\delta D_{\nu\beta}(k)D_{\rho\gamma}^\dagger(q) \\
  + k_\mu q_\sigma q_\alpha k_\delta D'_{\nu\gamma}(k)D'_{\rho\beta}(q)   \Bigr]\;.
\end{multline*}
By the substitution $p= k-q$ we find
\begin{eqnarray}
\braket{O^{(2)}_{\lambda\eta}(x) O^{(2)}_{\lambda\eta}(y)} &=&\int\frac{\d^dp}{(2\pi)^d}\e^{ip(x-y)}
\braket{ O^{(2)}_{\lambda\eta}(p) O^{(1)}_{\lambda\eta}(-p)}\;,
\end{eqnarray}
whereby
\begin{align*}
&\braket{ O^{(1)}_{\lambda\eta}(p)  O^{(1)}_{\lambda\eta}(-p)}= 4(N^2-1)\varepsilon_{\mu\nu\rho\sigma}  \varepsilon_{\alpha\beta\gamma\delta}  \int \frac{\d^d k}{(2\pi)^d} \\
\Bigl[& k_\mu (k-p)_\rho k_\alpha (k-p)_\gamma D_{\nu\beta}(k)D_{\sigma\delta}^\dagger(k-p) + k_\mu (k-p)_\rho (k-p)_\alpha k_\gamma  D_{\nu\delta}'(k)D_{\sigma\beta}'(k-p) \\
& -  k_\mu (k-p)_\rho k_\alpha (k-p)_\delta  D_{\nu\beta}(k)D_{\sigma\gamma}^\dagger(k-p) - k_\mu (k-p)_\rho (k-p)_\alpha k_\delta D'_{\nu\gamma}(k)D'_{\sigma\beta}(k-p)  \\
 & -  k_\mu (k-p)_\sigma k_\alpha (k-p)_\gamma  D_{\nu\beta}(k)D_{\rho\delta}^\dagger(k-p) - k_\mu (k-p)_\sigma (k-p)_\alpha k_\gamma D'_{\nu\delta}(k)D'_{\rho\beta}(k-p) \\
 &  + k_\mu (k-p)_\sigma k_\alpha (k-p)_\delta D_{\nu\beta}(k)D_{\rho\gamma}^\dagger(k-p)   + k_\mu (k-p)_\sigma (k-p)_\alpha k_\delta D'_{\nu\gamma}(k)D'_{\rho\beta}(k-p)   \Bigr] \nonumber\\
 &= 16(N^2-1)\varepsilon_{\mu\nu\rho\sigma}  \varepsilon_{\alpha\beta\gamma\delta}  \int \frac{\d^d k}{(2\pi)^d} \Bigl[ k_\mu (k-p)_\rho k_\alpha (k-p)_\gamma D_{\nu\beta}(k)D_{\sigma\delta}^\dagger(k-p) \\
 &+ k_\mu (k-p)_\rho (k-p)_\alpha k_\gamma  D_{\nu\delta}'(k)D_{\sigma\beta}'(k-p) \Bigr]\;.
\end{align*}
We can simplify this expression to find
\begin{equation}
\braket{ O^{(2)}_{\lambda\eta}(p) O^{(2)}_{\lambda\eta}(-p)}= 32(N^2-1)\int \frac{\d^d k}{(2\pi)^d} \Biggl[  \frac{ k^2 p^2 - (kp)^2}{(k^2 - \ii \lambda^2)((p-k)^2 + \ii \lambda^2)}\Biggr]\;.
\end{equation}

\chapter{Alternative proof of the renormalizability of the GZ action\label{appendixalternative}}
In this appendix, we shall write down an alternative proof of the renormalization of the Gribov-Zwanziger action, useful for chapter \ref{scrutinizing}. This proof shall be very similar to section \ref{algebraicrenormGZ} of chapter \ref{chapgribovtoGZ}.

\section{The starting action and the BRST}
Looking at  expression \eqref{brstinvariant}, we may also try to also start with another possible BRST invariant action,
\begin{eqnarray}\label{B1}
\Sigma_\GZ^{(2)} &=& S_{\YM} + S_{\gf} + S_0 + S_\s^{(2)} + S_{\mathrm{ext}} \,,
\end{eqnarray}
whereby we have immediately added $S_{\mathrm{ext}}$ still given by \eqref{ext} and with
\begin{eqnarray}\label{s2}
S_\s^{(2)} &=& s\int \d^d x \left( -U_\mu^{ai} \p_\mu \varphi_i^a - U_\mu^{\prime ai} g f_{akb} A^k_\mu \varphi_i^b - V_\mu^{ai} \p_{\mu} \overline \omega_i^{a} - V_\mu^{\prime ai} g f_{akb} A^k_\mu \overline \omega_i^b      - U_\mu^{\prime ai} V_\mu^{ \prime ai} \right. \nonumber\\
  && \left.+ T_\mu^{a i} g f_{abc} D^{bd}_\mu c^d \overline \omega^c_i \right)\nonumber\\
&=& \int \d^d x \left( -M_\mu^{ai}  \p_\mu  \varphi_i^a   + U_\mu^{ai}  \p_\mu \omega_i^a  - M^{\prime a i}_{\mu} g f^{akb} A_\mu^k \varphi^{b}_i    - gf^{abc} U_\mu^{\prime ai}   D^{bd}_\mu c^d  \varphi_i^c \right. \nonumber\\
&& + U_\mu^{\prime ai}  g f_{akb} A_\mu^{k} \omega_i^b -N_\mu^{ai}  \p_\mu \overline \omega_i^a   + V_\mu^{ai}  \p_\mu \overline \varphi_i^a  - N^{\prime a i}_{\mu} g f^{akb} A_\mu^k \overline \omega^{b i}  - gf^{abc} V_\mu^{\prime ai}   D^{bd}_\mu c^d  \overline \omega_i^c  \nonumber \\
&& \left. + V_\mu^{\prime ai}  g f_{akb} A_\mu^{k} \overline \varphi_i^b + R_\mu^{ai} g f^{abc} D_\mu^{bd} c^d \overline \omega^c_i  + T_\mu^{ai} g f_{abc} D^{bd}_\mu c^d \overline \varphi^c_i\right) \,,
\end{eqnarray}
whereby in contrast with section \ref{algebraicrenormGZ} of chapter \ref{chapgribovtoGZ} we have to treated $g f_{akb} A^k_\mu \varphi^{bc}_\nu$ and $g f_{akb} A^k_\mu \overline \varphi^{bc}_\nu$ as the relevant composite operators. We have now introduced 5 doublets, ($U_\mu^{ai}$, $M_\mu^{ai}$), ($U_\mu^{\prime ai}$, $M_\mu^{\prime ai}$), ($V_\mu^{ai}$, $N_\mu^{ai}$), ($V_\mu^{\prime ai}$, $N_\mu^{\prime ai}$) and ($T_\mu^{ai}$, $R_\mu^{ai}$) with the following BRST transformations,
\begin{align}
sU_{\mu }^{ai} &= M_{\mu }^{ai}\,, & sM_{\mu }^{ai}&=0\,,  \nonumber \\
sU_{\mu }^{\prime ai} &= M_{\mu }^{\prime ai}\,, & sM_{  \mu }^{\prime ai}&=0\,,  \nonumber \\
sV_{\mu }^{ai} &= N_{\mu }^{ai}\,, & sN_{\mu }^{ai}&=0\,,\nonumber \\
sV_{\mu }^{\prime ai} &= N_{\mu }^{\prime ai}\,, & sN_{\mu }^{\prime ai}&=0\,,\nonumber \\
sT_{\mu }^{ai} &= R_{\mu }^{ai}\,, & sR_{\mu }^{ai}&=0\;.
\end{align}
In order to go back from $S_\s^{(2)}$ to $S_\s$ of equation \eqref{previous}, we just need to set $U = U'$, $V= V'$, $N = N'$ and $M= M'$. Eventually, it appears natural to give
the primed sources the same physical value of their corresponding unprimed counterparts, see equation \eqref{physlimit}.

\section{The Ward identities}
Just as in section \ref{algebraicrenormGZ} of chapter \ref{chapgribovtoGZ}, we enlist all the Ward identities obeyed by $\Sigma_\GZ^{(2)}$, which of course look very similar.

\begin{enumerate}
\item The Slavnov-Taylor identity is now given by
\begin{equation}
\mathcal{S}(\Sigma^{(2)}_\GZ )=0\;,
\end{equation}
with
\begin{multline*}
\mathcal{S}(\Sigma^{(2)}_\GZ ) =\int \d^4x\Bigl( \frac{\delta \Sigma^{(2)}_\GZ }{\delta K_{\mu }^{a}}\frac{\delta \Sigma^{(2)}_\GZ }{\delta A_{\mu }^{a}}+\frac{\delta \Sigma^{(2)}_\GZ }{\delta L^{a}}\frac{\delta \Sigma^{(2)}_\GZ }{\delta c^{a}} +b^{a}\frac{\delta \Sigma^{(2)}_\GZ }{\delta \overline{c}^{a}}+\overline{\varphi }_{i}^{a}\frac{\delta \Sigma^{(2)}_\GZ }{\delta \overline{\omega }_{i}^{a}}+\omega _{i}^{a}\frac{\delta \Sigma^{(2)}_\GZ }{\delta \varphi _{i}^{a}} \nonumber\\
  + R_{\mu }^{ai}\frac{\delta \Sigma^{(2)}_\GZ }{\delta T_{\mu }^{ai}}+M_{\mu }^{ai}\frac{\delta \Sigma^{(2)}_\GZ }{\delta U_{\mu}^{ai}}+N_{\mu }^{ai}\frac{\delta \Sigma^{(2)}_\GZ }{\delta V_{\mu }^{ai}} +M_{\mu }^{\prime ai}\frac{\delta \Sigma^{(2)}_\GZ
}{\delta U_{\mu}^{\prime ai}}+N_{\mu }^{\prime ai}\frac{\delta \Sigma^{(2)}_\GZ }{\delta V_{\mu }^{\prime ai}} \Bigr) \;.
\end{multline*}

\item The $U(f)$ invariance is easily adapted
\begin{equation}
U_{ij} \Sigma^{(2)}_\GZ =0\;,
\end{equation}
\begin{multline}
U_{ij}=\int \d^dx\Bigl( \varphi_{i}^{a}\frac{\delta }{\delta \varphi _{j}^{a}}-\overline{\varphi}_{j}^{a}\frac{\delta }{\delta \overline{\varphi}_{i}^{a}}+\omega _{i}^{a}\frac{\delta }{\delta \omega _{j}^{a}} -\overline{\omega }_{j}^{a}\frac{\delta }{\delta \overline{\omega }_{i}^{a}}-  M^{aj}_{\mu} \frac{\delta}{\delta M^{ai}_{\mu}} -  M^{\prime aj}_{\mu} \frac{\delta}{\delta M^{\prime ai}_{\mu}} -U^{aj}_{\mu}\frac{\delta}{\delta U^{ai}_{\mu}}    \\
- U^{\prime aj}_{\mu}\frac{\delta}{\delta U^{\prime ai}_{\mu}}  + N^{ai}_{\mu}\frac{\delta}{\delta N^{aj}_{\mu}}+ N^{\prime ai}_{\mu}\frac{\delta}{\delta N^{\prime aj}_{\mu}} +V^{ai}_{\mu}\frac{\delta}{\delta V^{aj}_{\mu}}  +V^{\prime ai}_{\mu}\frac{\delta}{\delta V^{\prime aj}_{\mu}}
 +  R^{aj}_{\mu}\frac{\delta}{\delta R^{ai}_{\mu}} + T^{aj}_{\mu}\frac{\delta}{\delta T^{ai}_{\mu}} \Bigr)  \;. \nonumber
\end{multline}
We have again that the $i$-valued fields and sources turn out to possess an additional quantum number. All the quantum number are still the same as in Table \ref{2tabel1} and Table \ref{2tabel2}, whereby we keep in mind that the quantum numbers of the primed sources are obviously the same as those of the unprimed ones.

\item The Landau gauge condition does not change,
\begin{eqnarray}
\frac{\delta \Sigma^{(2)}_\GZ }{\delta b^{a}}&=&\partial_\mu A_\mu^{a}\;.
\end{eqnarray}

\item The same goes for the antighost equation,
\begin{eqnarray}
\frac{\delta \Sigma^{(2)}_\GZ }{\delta \overline{c}^{a}}+\partial _{\mu}\frac{\delta \Sigma^{(2)}_\GZ }{\delta K_{\mu }^{a}}&=&0\;.
\end{eqnarray}

\item The linearly broken local constraints now become
\begin{align}
\frac{\delta \Sigma^{(2)}_\GZ }{\delta \overline{\varphi }^{a}_i}+\partial _{\mu }\frac{\delta \Sigma^{(2)}_\GZ }{\delta M_{\mu }^{ai}}  +\partial _{\mu }\frac{\delta \Sigma^{(2)}_\GZ }{\delta M_{\mu }^{ \prime ai}} + g f_{dba}    T^{d i}_\mu \frac{\delta \Sigma^{(2)}_\GZ }{\delta K_{\mu }^{b i}} &=gf^{abc}A_{\mu }^{b}V_{\mu}^{\prime ci} \;, \nonumber\\
\frac{\delta \Sigma^{(2)}_\GZ }{\delta \omega^{a}_i}+\partial _{\mu}\frac{\delta \Sigma^{(2)}_\GZ }{\delta N_{\mu}^{ai}} +\partial _{\mu}\frac{\delta \Sigma^{(2)}_\GZ }{\delta N_{\mu}^{\prime ai}} -gf^{abc}\overline{\omega }^{bi}\frac{\delta \Sigma^{(2)}_\GZ }{\delta b^{c}}&=gf^{abc}A_{\mu }^{b}U_{\mu }^{\prime ci} \;.
\end{align}
We also find some extra linearly broken constraints
\begin{align}
\frac{\delta \Sigma^{(2)}_\GZ }{\delta M_{\mu}^{ai}} &= \p_\mu \varphi^{ai}    \;,&
\frac{\delta \Sigma^{(2)}_\GZ }{\delta N_{\mu}^{ai}} &= \p_\mu \overline \omega^{ai}  \;, &
\frac{\delta \Sigma^{(2)}_\GZ }{\delta U_{\mu}^{ai}} &= \p_\mu \omega^{ai}   \;,&
\frac{\delta \Sigma^{(2)}_\GZ }{\delta V_{\mu}^{ai}} &= \p_\mu  \overline \varphi^{ai}   \;.
\end{align}

\item The exact $\mathcal{R}_{ij}$ symmetry can be adapted to
\begin{equation}
\mathcal{R}_{ij}\Sigma^{(2)}_\GZ =0\;,
\end{equation}
with
\begin{equation*}
\mathcal{R}_{ij} = \int \d^4x\Bigl( \varphi _{i}^{a}\frac{\delta}{\delta\omega _{j}^{a}}-\overline{\omega }_{j}^{a}\frac{\delta }{\delta \overline{\varphi }_{i}^{a}}+V_{\mu }^{ai}\frac{\delta }{\delta N_{\mu}^{ai}}  +V_{\mu }^{\prime ai}\frac{\delta }{\delta N_{\mu}^{\prime aj}}-U_{\mu }^{aj}\frac{\delta }{\delta M_{\mu }^{ai}} -U_{\mu }^{\prime ai}\frac{\delta }{\delta M_{\mu }^{\prime ai}} + T^{a i}_\mu \frac{\delta }{\delta R_{\mu }^{aj}}  \Bigr) \;.
\end{equation*}

\item The integrated Ward identity is now linearly broken as follows
\begin{equation}
\int \d^4 x \left( c^a \frac{ \delta \Sigma^{(2)}_\GZ }{ \delta \omega^{ a }_i} + \overline \omega^{a}_i \frac{ \delta \Sigma^{(2)}_\GZ }{ \delta  \overline c^a } + U^{\prime a i}_\mu \frac{ \delta \Sigma^{(2)}_\GZ }{ \delta  K^a_\mu }  \right)= U^{ai}_\mu \p_\mu c^a - U^{\prime ai}_\mu \p_\mu c^a \;.
\end{equation}
\end{enumerate}

\section{The counterterm}
We again translate all the identities into contraints for the counterterm $\Sigma^{(2)c}_\GZ$
\begin{enumerate}
\item The linearized Slavnov-Taylor identity:
\begin{equation}
\mathcal{B}^{(2)}\Sigma_\GZ^{(2)c}=0\;,
\end{equation}
with $\mathcal{B}^{(2)}$ the nilpotent linearized Slavnov-Taylor operator,
\begin{multline}
\mathcal{B}^{(2)} =\int \d^{4}x\Bigl( \frac{\delta \Sigma^{(2)}_\GZ}{\delta K_{\mu }^{a}}\frac{\delta }{\delta A_{\mu }^{a}}+\frac{\delta \Sigma^{(2)}_\GZ }{\delta A_{\mu }^{a}}\frac{\delta }{\delta K_{\mu }^{a}}+\frac{\delta \Sigma^{(2)}_\GZ }{\delta L^{a}}\frac{\delta }{\delta c^{a}}+\frac{\delta\Sigma^{(2)}_\GZ }{\delta c^{a}}\frac{\delta }{\delta L^{a}}+b^{a}\frac{\delta }{\delta \overline{c}^{a}}\\
+\overline{\varphi}_{i}^{a}\frac{\delta }{\delta \overline{\omega }_{i}^{a}}+\omega_{i}^{a}\frac{\delta }{\delta \varphi_{i}^{a}}+M_{\mu }^{ai}\frac{\delta }{\delta U_{\mu }^{ai}}
 + N_{\mu }^{ai}\frac{\delta }{\delta V_{\mu }^{ai}}  +M_{\mu }^{\prime ai}\frac{\delta }{\delta U_{\mu }^{\prime  ai}} + N_{\mu }^{\prime ai}\frac{\delta }{\delta V_{\mu }^{\prime ai}}   + R_{\mu }^{ai}\frac{\delta }{\delta T_{\mu }^{ai}}  \Bigr) \nonumber\,.
\end{multline}

\item The $U(f)$ invariance
\begin{eqnarray}
U_{ij} \Sigma^{(2)c}_\GZ &=&0 \;.
\end{eqnarray}

\item The Landau gauge condition
\begin{eqnarray}
\frac{\delta \Sigma^{(2)c}_\GZ}{\delta b^{a}}&=&0\,.
\end{eqnarray}

\item The antighost equation
\begin{eqnarray}
\frac{\delta \Sigma^{(2)c}_\GZ}{\delta \overline c^{a}}+\p_\mu\frac{\delta \Sigma^{(2)c}_\GZ}{\delta K_{\mu}^a} &=&0 \,.
\end{eqnarray}

\item The linearly broken local constraints
\begin{eqnarray}
\left( \frac{\delta  }{\delta \overline{\varphi }^{a}_i}+\partial _{\mu }\frac{\delta}{\delta M_{\mu }^{ai}} +  \partial _{\mu }\frac{\delta  }{\delta M_{\mu }^{ \prime ai}}+ g f_{abc}    T^{b i}_\mu \frac{\delta }{\delta K_{\mu }^{c i}}\right) \Sigma^{(2)c}_\GZ  &=&0 \;, \nonumber\\
\left( \frac{\delta }{\delta \omega^{a}_i}+\partial _{\mu}\frac{\delta }{\delta N_{\mu}^{ai}} +  +\partial _{\mu }\frac{\delta  }{\delta N_{\mu }^{ \prime ai}} -gf^{abc}\overline{\omega }^{b}_i\frac{\delta }{\delta b^{c}} \right)  \Sigma^{(2)c}_\GZ  &=&0 \;,
\end{eqnarray}
and
\begin{align}\label{bigc}
\frac{\delta \Sigma^{(2)c}_\GZ}{\delta M_{\mu}^{ai}} &= 0 \;,&
\frac{\delta \Sigma^{(2)c}_\GZ}{\delta N_{\mu}^{ai}} &= 0\;,&
\frac{\delta \Sigma^{(2)c}_\GZ}{\delta U_{\mu}^{ai}} &= 0\;,&
\frac{\delta \Sigma^{(2)c}_\GZ}{\delta V_{\mu}^{ai}} &= 0\;.
\end{align}
\item The exact $\mathcal{R}_{ij}$ symmetry
\begin{equation}
\mathcal{R}_{ij}\Sigma^{(2)c}_\GZ=0\;.
\end{equation}

\item Finally, the integrated Ward identity becomes
\begin{equation}
\int \d^4 x \left( c^a \frac{ \delta \Sigma^{(2)c}_\GZ}{ \delta \omega^{ a}_i} + \overline \omega^{a}_i \frac{ \delta \Sigma^{(2)c}_\GZ}{ \delta  \overline c^a } + U^{\prime a i}_\mu \frac{ \delta \Sigma^{(2)c}_\GZ}{ \delta  K^a_\mu }  \right) = 0 \;.
\end{equation}
\end{enumerate}
Now we can write down the most general counterterm $\Sigma^{(2)c}_\GZ$ of $d=4$, which obeys the linearized Slavnov-Taylor identity, has ghost number zero, and vanishing $Q_f$ number,
\begin{multline*}
\Sigma^{(2)c}_\GZ= a_0 S_{\YM} + \mathcal{B}^{(2)} \int \d^d x \biggl\{ \biggl[ a_{1} K_{\mu}^{a} A_{\mu}^{a} + a_2 \partial _{\mu} \overline{c}^{a} A_{\mu}^{a}+a_3 L^{a}c^{a}
 +a_4 U_{\mu}^{ai}\,\partial _{\mu }\varphi_i^a +a_5 V_{\mu}^{ai}\,\partial _{\mu }\overline{\omega }_{i}^{a} \\
  +a_6\,\overline{\omega }_{i}^{a}\partial ^{2}\varphi _{i}^{a} +a_7 U_{\mu}^{ai}V_{\mu }^{ai} +a_8 gf^{abc}U_{\mu }^{ai}\,\varphi _{i}^{b} A_{\mu }^{c}+a_9 gf^{abc}V_{\mu }^{ai}\,\overline{\omega }_{i}^{b}A_{\mu }^{c}+a_{10} gf^{abc}\overline{\omega }_{i}^{a}A_{\mu }^{c}\,\partial _{\mu }\varphi _{i}^{b} \\
 +a_{11}\,gf^{abc}\overline{\omega }_{i}^{a}(\partial _{\mu }A_{\mu}^{c})\varphi _{i}^{b}
+ b_1 R_{\mu}^{ai} U_{\mu}^{ai}+b_2 T_{\mu}^{ai} M_{\mu }^{ai}   + b_3 g f_{abc} R_{\mu}^{ai} \overline{\omega }_{i}^{b} A_{\mu}^{c}
+ b_4 g f_{abc} T_{\mu}^{ai} \overline{\varphi }_{i}^{b} A_{\mu}^{c} \\
+ b_5 R_{\mu}^{ai} \p \overline{\omega }_{i}^{a} + b_6 T_{\mu}^{ai} \p \overline{\varphi }_{i}^{a} +a_4^\prime U_{\mu }^{\prime  ai}\,\partial _{\mu }\varphi _{i}^{a}
+a_5^\prime  \,V_{\mu}^{\prime ai}\,\partial _{\mu }\overline{\omega }_{i}^{a}  +a_6^\prime \overline{\omega }_{i}^{\prime a}\partial ^{2}\varphi _{i}^{a}
+a_7^\prime  \,U_{\mu}^{\prime  ai}V_{\mu}^{\prime ai} \nonumber\\
+a_8^\prime \,gf^{abc}U_{\mu}^{\prime ai}\,\varphi _{i}^{b}A_{\mu }^{c}+a_9^\prime \,gf^{abc}V_{\mu}^{\prime ai}\,\overline{\omega }_{i}^{b}A_{\mu }^{c}
+a_{10}^\prime \,gf^{abc}\overline{\omega }_{i}^{a}A_{\mu }^{c}\,\partial _{\mu }\varphi _{i}^{b} +a_{11}^\prime \,gf^{abc}\overline{\omega }_{i}^{a}(\partial _{\mu }A_{\mu}^{c})\varphi _{i}^{b}  \biggr] \biggr\} \;.
\end{multline*}
Notice that the part in $a$ and $b$ parameters is exactly the same as in section \ref{algebraicrenormGZ} of chapter \ref{chapgribovtoGZ}, see equation \eqref{counterterm}. We shall now impose all the constraints induced by the Ward identities. We keep in mind that the argument concerning the broken ghost Ward identity still holds. Also, the 4 constraints \eqref{bigc} invoke the counterterm to be independent of the sources $U'$, $V'$, $M'$ and $N'$. Ultimately, we find
\begin{multline}
\Sigma^{(2)c}_\GZ= a_{0}S_{YM}  + a_{1}\int \d^dx\Biggl(  A_{\mu}^{a}\frac{ \delta S_{YM}}{\delta A_{\mu }^{a}}  + \p_\mu \overline{c}^a \p_\mu c^a + K_{\mu }^{a}\partial _{\mu }c^{a}  + M_\mu^{\prime a i} \p_\mu \varphi^{a}_i -  U_\mu^{\prime a i} \p_\mu \omega^{a}_i + N_\mu^{\prime a i} \p_\mu \overline{\omega}^{a}_i \nonumber\\
+  V_\mu^{\prime a i} \p_\mu \overline{\varphi}_\mu^{ai}  +  \p_\mu \overline{\varphi}^{a}_i \p_\mu \varphi_i^{a} +  \p_\mu \omega^{a}_i \p_\mu \overline{\omega}^{a}_i + V_\mu^{\prime a i} M_\mu^{\prime a i} - U_\mu^{\prime a i}N_\mu^{\prime a i} - g f_{abc} U_\mu^{\prime ia} \varphi^{b}_i \p_\mu c^c \nonumber\\
- g f_{abc} V_\mu^{\prime ia} \overline{\omega}^{b}_i \p_\mu c^c - g f_{abc} \p_{\mu} \overline{\omega}^a \varphi^{b}_i  \p_\mu c^c  - g f_{abc} R^{ai}_\mu \p_\mu c^b \overline \omega^c + g f_{abc} T^{ai}_\mu \p_\mu c^b \overline \varphi^c \Biggr) \;.
\end{multline}
Notice the close similarity between this counterterm and the one in expression \eqref{final}.

\section{The renormalization factors}
The last step is to find all the renormalization factors. Due to the close similarity with the output of section \ref{algebraicrenormGZ} of chapter \ref{chapgribovtoGZ}, many $Z$ factors will be the same. One can indeed check that equations \eqref{Z1} and \eqref{Z2} still hold, and also the $Z$-factors of $Z_{\varphi}^{1/2}$, $Z_{\overline \varphi}^{1/2} $, $Z_\omega^{1/2}$, $Z_{\overline \omega}^{1/2} $, $Z_T$ and $Z_R$  do not change. Only the renormalization of the sources $U$, $V$, $M$, $N$ is different as they mix with respectively $U'$, $V'$, $M'$, $N'$. Indeed, we find that
\begin{align}\label{zmatrix1}
 \left[
  \begin{array}{c}
    M_0 \\
    M'_0 \\
  \end{array}
\right]&=\left[
          \begin{array}{cc}
              Z_g^{1/2} Z_A^{1/4} & -a_1 \\
            0 & Z_g^{-1/2} Z_A^{-1/4}  \\
          \end{array}
        \right]
\left[
  \begin{array}{c}
    M\\
    M'
  \end{array}
\right]\,,  \nonumber\\
 \left[
  \begin{array}{c}
    U_0 \\
    U'_0 \\
  \end{array}
\right]&=\left[
          \begin{array}{cc}
               Z_A^{1/2} & -a_1 \\
            0 & Z_g^{-1}   \\
          \end{array}
        \right]
\left[
  \begin{array}{c}
    U\\
    U'
  \end{array}
\right]\,,  \nonumber\\
 \left[
  \begin{array}{c}
    N_0 \\
    N'_0 \\
  \end{array}
\right]&=\left[
          \begin{array}{cc}
              Z_g^{1}  & -a_1 \\
            0 &  Z_A^{-1/2}  \\
          \end{array}
        \right]
\left[
  \begin{array}{c}
    N\\
    N'
  \end{array}
\right]\,,  \nonumber\\
 \left[
  \begin{array}{c}
    V_0 \\
    V'_0 \\
  \end{array}
\right]&=\left[
          \begin{array}{cc}
               Z_g^{1/2}Z_A^{1/4} & -a_1 \\
            0 & Z_g^{-1/2}Z_A^{-1/4}   \\
          \end{array}
        \right]
\left[
  \begin{array}{c}
    V\\
    V'
  \end{array}
\right]\,,
\end{align}
which again proves the renormalizability of the Gribov-Zwanziger action. The consequences of this mixing are explained in section \ref{secthorizon} of chapter \ref{scrutinizing}.

\chapter{The effective action and the gap equations}
Let us first explain the idea behind the method before going into detailed calculations. With $m=0$, the tree level propagator \eqref{gluonprop2} yields:
\begin{eqnarray}\label{wishedform}
\mathcal{D}(p^2) &=& \frac{p^2 + M^2}{p^4 + M^2 p^2 + 2 g^2 N \gamma^4  }\;.
\end{eqnarray}
Expanding the mass $M^2$ as a series in $g^2$, gives
\begin{eqnarray}
M^2 &=& M^2_0 + g^2 M_1^2 + g^4 M_2^2 + \ldots \;.
\end{eqnarray}
We only need to consider $M_0$, which is of order unity, as we are considering the tree level propagator. We know that at the end of our calculations we have to set our sources equal to zero, or $J= M^2 =0$. If we work at lowest order, this means we have to set $M_0 =0$ (and the gluon propagator will not display the desired behavior). However,  going one order higher gives:
\begin{eqnarray}
 M_0^2 + g^2 M_1^2 &=& 0     \;.
\end{eqnarray}
The last equation might imply that $M_0^2$ is no longer equal to zero, and consequently, the tree level gluon propagator will attain the desired form. Let us elaborate further on this aspect.

\section{One loop effective potential}
\begin{figure}[H]
   \centering
       \includegraphics[width=8cm]{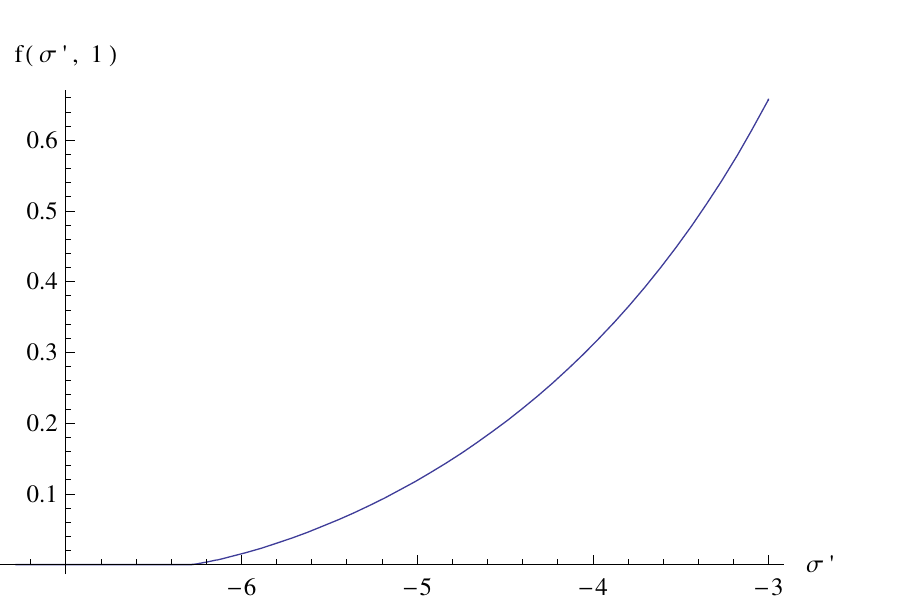}
   \caption{A plot of $f(\widehat{\sigma}' , 1)$ in terms of $\widehat{\sigma}' = \frac{2}{3} 64 \pi^2 \frac{ \widehat{\sigma}}{N^2 - 1}$ }
  \label{plot1}
\end{figure}

To implement the above-mentioned ideas, we shall first calculate the one loop energy functional. We start with the action $S_{\RGZ}'$, whereby setting $m = 0$ is equivalent with putting $\tau = 0$. We replace the mass $M^2$ again with the source $J$. Similar as has been done in the appendix \ref{sigma}\ref{appendix3}, we obtain
\begin{multline}\label{xo}
W^{(0)}(J) = - \frac{4 (N^2 - 1)}{2 g^2 N} \lambda^4 +  \frac{d (N^2 -1)}{ g^2 N}  \varsigma \ \lambda^2 M^2 +  \frac{3(N^2 - 1)}{64 \pi^2} \left( \frac{8}{3} \lambda^4 + m_1^4 \ln \frac{m_1^2}{\overline{\mu}^2}\right. \\
\left. + m_2^4 \ln \frac{m_2^2}{\overline{\mu}^2} - J^2 \ln \frac{J}{\overline{\mu}^2}\right) \;,
\end{multline}
whereby $m_1$ and $m_2$ are now given by
\begin{align}\label{notationalshorthandmnul}
m_1^2 &= \frac{J - \sqrt{J^2 - 4\lambda^4  }}{2} \;, & m_2^2 &= \frac{J + \sqrt{J^2 - 4\lambda^4  }}{2} \;.
\end{align}
As we have determined the energy functional $W^{(0)}(J)$, we can now calculate the one loop effective action via the Legendre transform of $W(J)$. If we define
\begin{align}\label{klassiekveld}
\sigma(x) &~=~ \frac{\delta  W(J)}{\delta J(x)}\;, & \sigma_{\mathrm{cl}} & ~=~  \frac{d (N^2 -1)}{ g^2 N}  \varsigma \ \lambda^2\;,
\end{align}
then
\begin{eqnarray}
\widehat{\sigma}(x) &=& \sigma(x)  - \sigma_{\mathrm{cl}} =  - \frac{\int [\d \Phi] \left(\overline{\varphi}\varphi-\overline{\omega}\omega \right) \e^{-S'_\RGZ}}{\int [ \d \Phi]  \e^{-S'_\RGZ}} \;,
\end{eqnarray}
represents the expectation value of the local composite operator, $-\left(\overline{\varphi}\varphi-\overline{\omega}\omega\right)$. The effective action is given by
\begin{eqnarray}
\Gamma(\sigma) &=& W(J) - \int \d^4 x \ J(x)\sigma(x)  \;,
\end{eqnarray}
or equivalently, as we prefer to work in the variable $\widehat{\sigma}$,
\begin{eqnarray}\label{gamma}
\Gamma(\widehat{\sigma}) &=& W(J) - \int \d^4 x \ J(x) \left( \widehat{\sigma}(x) + \sigma_{\mathrm{cl}} \right)  \;.
\end{eqnarray}
Calculating $\Gamma(\widehat{\sigma})$  by explicitly doing the inversion is a rather cumbersome task. In  most cases one can perform a Hubbard-Stratonovich transformation to eliminate the term $J \left(\overline{\varphi}\varphi-\overline{\omega}\omega \right)$ from the action and introduce a new field $\sigma'$ which couples linearly to $J$. This greatly simplifies the calculation. However,  in this case, it seems impossible to do such a transformation as a required term in $J^2$ is missing. Hence, there is no other option than to actually perform the inversion. In order to calculate this inversion, we shall limit ourself to constant $J$ and $\widehat{\sigma}$ as we are mainly interested in the (space time) independent vacuum expectation value of the operator $-\left(\overline{\varphi}\varphi-\overline{\omega}\omega \right)$ coupled to the source $J$. This vacuum expectation value is given by
\begin{eqnarray}
\left. \widehat{\sigma} \right|_{J=0}  &=&- \frac{\int  [\d \Phi]\left( \overline{\varphi}\varphi-\overline{\omega}\omega
 \right) \e^{-S}}{\int [\d \Phi] \e^{-S_\GZ}} \;,
\end{eqnarray}
where $S_\GZ$ represents the ordinary Gribov-Zwanziger action \eqref{GZstart}. As we already have calculated $W(J)$ up to one loop is it straightforward to verify that
\begin{equation}
\widehat{\sigma}=\frac{\p}{\p J} W_{0}(J) - \sigma_{\mathrm{cl}} = \frac{1}{2} \frac{3(N^2 - 1)}{64 \pi^2} J \left( 2 \ln \frac{t}{4} + \left( \sqrt{1-t} + \frac{1}{\sqrt{1-t}} \right) \ln \frac{1+ \sqrt{1-t}}{1- \sqrt{1-t}}\right) \;,
\end{equation}
whereby we shortened the notation by putting $t= 4\lambda^4 / J^2 $. From the previous expression we find for the condensate
\begin{eqnarray}\label{perturbative1}
\left. \widehat{\sigma} \right|_{J=0} &=& -    \frac{3(N^2 - 1)}{64 \pi} \lambda \;,
\end{eqnarray}
which is obviously exactly the same result as \eqref{condpert}.\\
\\
We are now ready to compute the effective action up to one loop along the lines of \cite{Yokojima:1995hy}. The energy functional can be written as a series in the coupling constant $g^2$,
\begin{eqnarray}
  W(J) &=& W_0(J) + g^2 W_1 (J) + \ldots =  \sum_{i=0}^{\infty} (g^2)^i W_{i}(J).
\end{eqnarray}
As a consequence, looking at the definition \eqref{klassiekveld}, we can write
\begin{eqnarray} \label{phi}
\widehat{\sigma} &=& \widehat{\sigma}_{0}(J) + g^2 \widehat{\sigma}_{1}(J)  + \ldots = \sum_{i=0}^{\infty} (g^2)^i \widehat{\sigma}_{i}(J),
\end{eqnarray}
where $\widehat{\sigma}_{i}(J)$ corresponds to the $i$th order in $g^2$ (regarding $J$ as of order unity). This is called the original series. The inverted series is defined as
\begin{eqnarray} \label{J}
J &=& J_{0}(\widehat{\sigma}) + g^2 J_{1}(\widehat{\sigma})  + \ldots = \sum_{j=0}^{\infty} (g^2)^j J_{j}(\widehat{\sigma}),
\end{eqnarray}
with $J_{j}(\widehat{\sigma})$  the $j$th order coefficient.
Substituting \eqref{J} into \eqref{phi} gives,
\begin{eqnarray}
\widehat{\sigma} &=& \sum_{i=0}^{\infty} (g^2)^i \widehat{\sigma}_{i}\left[\sum_{j=0}^{\infty} (g^2)^j J_{j}(\widehat{\sigma})\right] \nonumber\\
&=&  \widehat{\sigma}_{0}(J_{0}(\widehat{\sigma})) + g^2 \left( \widehat{\sigma}_{0}'(J_{0}(\widehat{\sigma})) \cdot J_{1}(\widehat{\sigma}) + \widehat{\sigma}_{1}(J_{0}(\widehat{\sigma})) \right) + \ldots \;.
\end{eqnarray}
By regarding $\widehat{\sigma}$ as of the order unity and by comparing both sides of the last equation, one finds
\begin{eqnarray}\label{laagsteorde}
\widehat{\sigma} &=& \widehat{\sigma}_{0}\left(J_{0}(\widehat{\sigma})\right) \;,  \\
 J_{1}(\widehat{\sigma}) &=& - \frac{\widehat{\sigma}_{1}\left( J_{0}(\widehat{\sigma})\right)}{ \widehat{\sigma}_{0}^\prime\left( J_{0}(\widehat{\sigma})\right)} \;.\\
&\vdots& \nonumber
\end{eqnarray}
For the moment, as we are working at lowest order, we only need equation \eqref{laagsteorde}. We can invert this equation,  so we find for $J_{0}(\widehat{\sigma})$:
\begin{eqnarray}
J_{0}(\widehat{\sigma}) &=& \widehat{\sigma}_{0}^{-1}(\widehat{\sigma})\;,
\end{eqnarray}
meaning that we have to solve
\begin{multline*}
\widehat{\sigma} \equiv \widehat{\sigma}_0 (J_0, \lambda)= \\
 \frac{1}{2} \frac{3(N^2 - 1)}{64 \pi^2} J_0 \left( 2 \ln \frac{t(\lambda,J_0)}{4} + \left( \sqrt{1-t(\lambda,J_0)} + \frac{1}{\sqrt{1-t(\lambda,J_0)}} \right) \ln \frac{1+\sqrt{1-t(\lambda,J_0)}}{1- \sqrt{1-t(\lambda,J_0)}}\right) \;,
\end{multline*}
for $J_0$,  so we can write
\begin{eqnarray}
J_0 &=& f(\widehat{\sigma}, \lambda)\;.
\end{eqnarray}
We immediately suspect that this inversion will not give rise to an analytical expression. Once we have found $f(\widehat{\sigma}, \lambda)$, we substitute this expression into the effective action,
\begin{eqnarray}\label{2effectieveactie2}
\Gamma(\widehat{\sigma}, \lambda) &=& W(f(\widehat{\sigma}, \lambda), \lambda) -   f(\widehat{\sigma}, \lambda) \widehat{\sigma}  \;.
\end{eqnarray}
At this point, as we have found an expression for the one loop effective action, we can implement two equations to fix $\widehat{\sigma}$ and
$\lambda$. Firstly, the minimization condition reads
\begin{eqnarray}
\frac{\p}{\p \widehat{\sigma}}\Gamma(\widehat{\sigma}, \lambda) &=& 0 \;,
\end{eqnarray}
and secondly, the horizon condition \eqref{gapgamma} can be translated as
\begin{eqnarray}\label{gap2}
\frac{\p}{\p \lambda}\Gamma(\widehat{\sigma}, \lambda) &=& 0 \;.
\end{eqnarray}
We start with the first gap equation.  Replacing
$\Gamma$ by equation \eqref{gamma} leads to
\begin{align}
\frac{\p}{\p \widehat{\sigma}}\Gamma(\widehat{\sigma}, \lambda) &= 0 &\Rightarrow&& \frac{\p W}{\p J} \frac{\p J}{\p \widehat{\sigma}} -
\frac{\p J}{\p \widehat{\sigma}} \widehat{\sigma} - \frac{\p J}{\p \widehat{\sigma}}\sigma_{\mathrm{cl}} - J  &= 0  &\Rightarrow&&  J &=0 &\Rightarrow&& f(\widehat{\sigma}, \lambda) &=0 \;.
\end{align}
Since there are only 2 explicit scales, $\lambda$ and $\widehat{\sigma}$, present, the first gap equation can be used to express e.g.~$\widehat{\sigma}$ in terms of $\lambda$. For the sake of a numerical
computation, we can therefore momentarily set $\lambda=1$. From Figure \ref{plot1} one can obtain an estimate $\widehat{\sigma}'$ of
$f(\widehat{\sigma}', 1) = 0$, with $\widehat{\sigma}' = \frac{2}{3} 64 \pi^2 \frac{ \widehat{\sigma}}{N^2 -1}\;$. Doing so, we find $\widehat{\sigma}' \approx - 6.28$, so that
\begin{eqnarray} \label{oplossing1}
\widehat{\sigma} &\approx & - 6.28  \times \left(  \frac{3(N^2 - 1)}{128 \pi^2} \right) \lambda \;,
\end{eqnarray}
which of course corresponds to the already obtained perturbative solution \eqref{perturbative1}.

\begin{figure}[H]
   \centering
       \includegraphics[width=8cm]{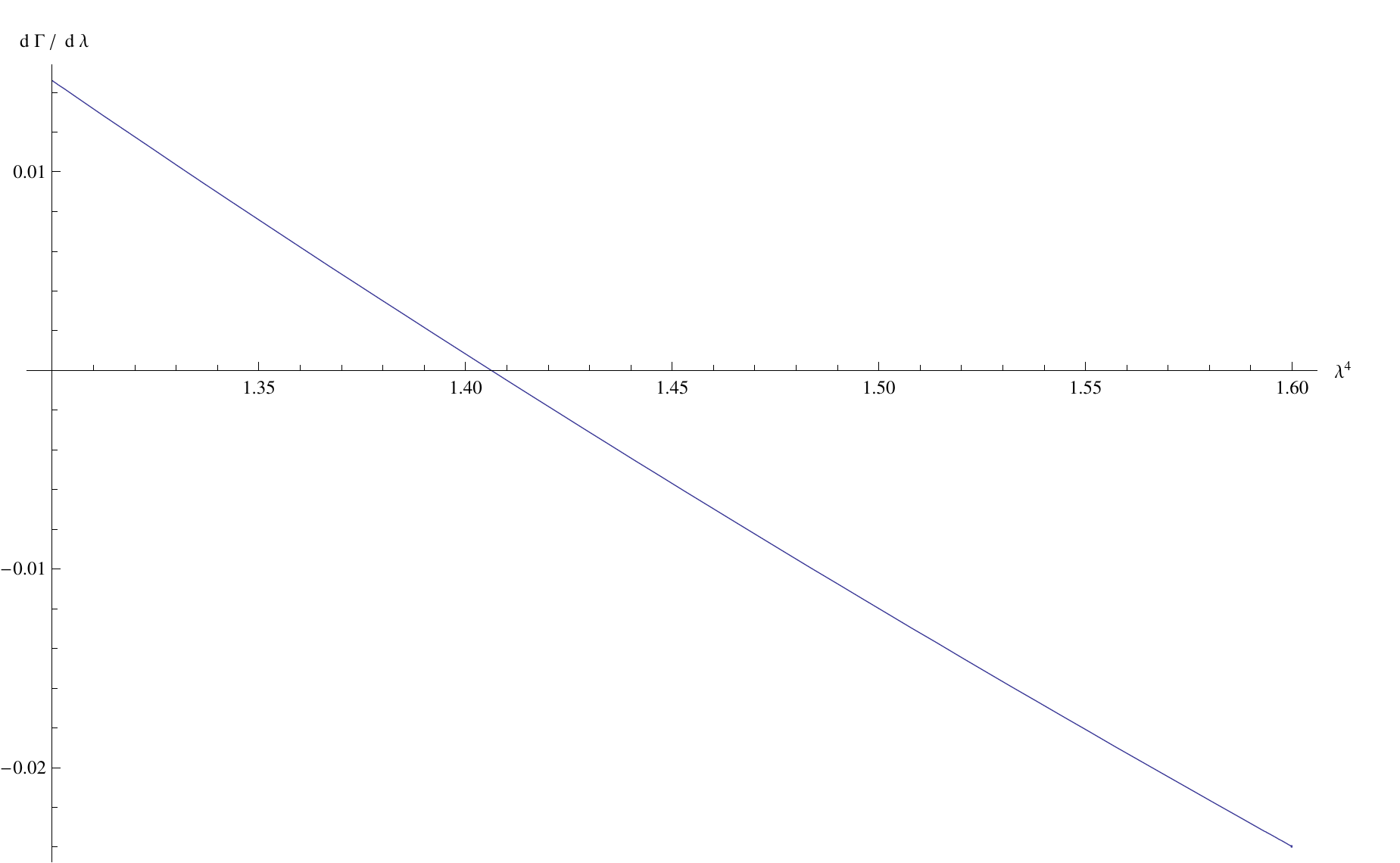}
   \caption{The horizon function $\frac{\p \Gamma}{\p \lambda}$ for $N = 3$.}
   \label{plot2}
\end{figure}

\noindent The second gap equation \eqref{gap2} must then consequently also give us back the perturbative solution. To check this, we first calculate the perturbative result for $\lambda$ by taking the limit $J \rightarrow 0$ in expression \eqref{xo}
\begin{eqnarray}\label{energy}
\Gamma_0 &=&  - \frac{2 (N^2 - 1)}{ g^2 N} \lambda^4 + \frac{3(N^2 - 1)}{64 \pi^2} \left( \frac{8}{3} \lambda^4 - 2\lambda^4 \ln \frac{\lambda^2}{\overline{\mu}^2}
\right) \;.
\end{eqnarray}
Next, we take the partial derivative with respect to $\lambda$ which
gives,
\begin{eqnarray}
\frac{ \p \Gamma_0}{\p \lambda} &=&  4 \lambda^3 \left( - \frac{2 (N^2 - 1)}{ g^2 N} + \frac{3(N^2 - 1)}{64 \pi^2} \left( \frac{5}{3}  - 2 \ln \frac{\lambda^2}{\overline{\mu}^2} \right)
\right) \;.
\end{eqnarray}
The natural choice for the renormalization constant is to set $\overline{\mu} = \lambda$ to kill the logarithms. Imposing the gap equation $\frac{ \p \Gamma_0}{\p \lambda} = 0$ gives us,
\begin{eqnarray}\label{gkwadraat}
\frac{g^2 N}{16 \pi^2} &=& \frac{8}{5} \;.
\end{eqnarray}
We remark that we have neglected the solution $\gamma =0$, as for $\gamma = 0$ the Gribov-Zwanziger reduces to the Faddeev-Popov action see p.\pageref{sectbreakingBRST}. From
\begin{eqnarray}\label{gkwadraat2}
g^2(\overline{\mu}^2)&=& \frac{1}{\beta_0\ln\frac{\overline{\mu}^2}{\lms^2} }\;,\;\;\;\;\;\textrm{with }\;\;\;\;\beta_0=\frac{11}{3}\frac{N}{16\pi^2}\;,
\end{eqnarray}
and expression \eqref{gkwadraat} we find an estimate for $\lambda$:
\begin{eqnarray}\label{perturbative}
\lambda^4 &=& \e^{44/15} \;,
\end{eqnarray}
where we have worked in units $\lms = 1$. This perturbative solution is also in compliance with \cite{Dudal:2005na}. Now, we return to the effective action \eqref{2effectieveactie2}. We first take the partial derivative with respect to $\lambda$, afterwards we set $N=3$, we explicitly replace $g^2$  by expression \eqref{gkwadraat2} and we use the minimizing condition \eqref{oplossing1}. Numerically, we find the following value for $\lambda^4$:
\begin{eqnarray}
\lambda^4 &=& 1.41 \;,
\end{eqnarray}
as one can read off from Figure \ref{plot2}. This is exactly the perturbative result \eqref{perturbative}.
\noindent If we calculate the vacuum energy with this value for $\lambda$, we find from \eqref{energy},
\begin{eqnarray}\label{positiveenergy}
E_{\mathrm{vac}} &=& \frac{3}{64}\frac{N^2-1}{\pi^2} \e^{44/15}\;.
\end{eqnarray}
We notice that the vacuum energy is positive.

\section{Conclusion}
We can conclude at this point, that in the framework we have used, we recover only the perturbative solution. Unfortunately, at lowest order, one finds $J_0 = 0$ as explained in the beginning of this section, so we were unable to find a dynamical value for $M^2$ at first order. However, if we would be able to go one order higher, with $J_0 + g^2 J_1 = 0$, we might find $J_0 \not=0$ and consequently the gluon propagator at tree level would attain the desired form \eqref{wishedform}. In addition, we might even discover a nonperturbative solution. Unfortunately, this is not as straightforward as  at leading order. The main difficulty resides in the evaluation of two loop vacuum bubbles for the effective potential with three different mass scales.  Whilst the master integrals are known, \cite{vanderBij:1983bw,Ford:1992pn,Davydychev:1992mt}, the main complication is that the propagator of \eqref{gluonprop} with $m^2$~$=$~$0$ needs to be split into standard form but this introduces the masses of \eqref{notationalshorthandmnul} which are either complex or negative. In either scenario the master two loop vacuum bubble is known for distinct positive masses and involves several dilogarithm functions. Therefore in our case for even the simplest of mass choices the resulting dilogarithms will be complex as well as being a complicated function of $m_1^2$, $m_2^2$ and $\lambda$. Moreover, this is prior to computing the full effective potential itself by adding all the relevant combinations of master integrals together. Therefore, it seems to us that whilst such a computation could be completed in principle, currently the resulting huge expression could not possibly lend itself to a tractable analysis similar to the relatively simple one we have carried out at one loop.

\chapter{Propagators of the Refined GZ action\label{propagatorsRGZ}}
The propagators of the RGZ action can be calculated in a similar fashion as in section \ref{propGZ1} of chapter \ref{scrutinizing}. We obtain the following propagators
\begin{eqnarray}
\Braket{\widetilde{ \overline \omega}^{ab}_\mu(k) \widetilde \omega^{cd}_\nu(p)} &=& \delta^{ac}\delta^{bd} \delta^{\mu\nu} \frac{-1}{p^2 + M^2} \delta(p+k) (2\pi)^4 \;,\nonumber\\
\Braket{ \widetilde{ \overline c}^a(k) \widetilde c^b(p) } &=& \delta^{ab} \frac{1}{p^2} \delta(p+k) (2\pi)^4 \;,\nonumber\\
\Braket{ \widetilde A_\mu^a (p)  \widetilde A_\nu^b (k)}&=&   \frac{ p^2 + M^2}{p^4 + M^2 p^2 + \lambda^4}  P_{\mu \nu} \delta^{ab} \delta(k+ p) (2\pi)^4 \;,\nonumber\\
\Braket{\widetilde A_\mu^a(p) \widetilde b^b(k)} &=& - \ii \frac{p_\mu}{ p^2} \delta^{ab} \delta(p+k) (2\pi)^4 \;,  \nonumber\\
\Braket{b^a(p) b^b(k)} &=&   \delta^{ab} \frac{ \lambda^4}{p^2 (p^2 + M^2)}\delta(p+k)(2\pi)^4  \;,\nonumber\\
\Braket{ \widetilde A^a_{\mu}(p) \widetilde{ \varphi}^{bc}_{\nu}(k)} &=& \Braket{ \widetilde A^a_{\mu}(p) \widetilde{\overline \varphi}^{bc}_{\nu}(k)} ~=~ f^{abc}  \frac{- g \gamma^2}{p^4 + M^2 p^2  + \lambda^4}  P_{\mu \nu}(p) (2\pi)^4 \delta(p+k) \;, \nonumber\\
 \Braket{ \widetilde b^a(p) \widetilde{ \varphi}^{bc}_{\nu}(k)} &=&  \Braket{ \widetilde b^a(p) \widetilde{ \overline \varphi}^{bc}_{\nu}(k)} ~=~  f^{abc} \ii p_\nu \frac{-  g \gamma^2}{ p^2 (p^2 + M^2)}  (2\pi)^4 \delta(p+k) \;, \nonumber\\
  \Braket{\widetilde{ \varphi}^{ab}_{\mu}(p) \widetilde{ \overline \varphi }^{cd}_{\nu}(k)} &=& \Bigl(  f^{abr} f^{cdr} P_{\mu \nu} \frac{ g^2 \gamma^4}{ (M^2 + p^2)( p^4 + M^2 p^2 + 2 g^2 N \gamma^4 )}  \nonumber\\
  &&\hspace{5cm} +  \frac{-1}{ p^2 + M^2} \delta^{ac} \delta^{bd} \delta_{\mu \nu}  \Bigr)    (2\pi)^4 \delta(p+k) \;, \nonumber\\
   \Braket{\widetilde{ \varphi}^{ab}_{\mu}(p) \widetilde{ \varphi }^{cd}_{\nu}(k)} &=&  \Braket{\widetilde{\overline \varphi}^{ab}_{\mu}(p) \widetilde{\overline \varphi }^{cd}_{\nu}(k)} ~=~ f^{abr} f^{cdr} P_{\mu \nu} \frac{ g^2 \gamma^4}{ (M^2 + p^2)( p^4 + M^2 p^2 + 2 g^2 N \gamma^4 )} \nonumber\\
   && \hspace{6cm}\times   (2\pi)^4 \delta(p+k) \;,.
\end{eqnarray}

\chapter{Renormalization of the further refined action\label{renvery}}

\section{The starting action}
Let us repeat the starting action \eqref{CGZ},
\begin{eqnarray}
\Sigma_\CGZ &=& \Sigma_\GZ'  + \Sigma_{A^2} + S_{\varphi \overline \varphi}  + S_{\overline \omega \omega} + S_{\overline{\varphi} \overline \varphi, \overline \omega \overline \varphi }  + S_{\varphi \varphi, \omega \varphi } + S_\vac \;,
\end{eqnarray}
whereby $\Sigma_\GZ'$ is given by equation \eqref{enlarged}, $\Sigma_{A^2}$ by \eqref{sigmaakwadraat} and
\begin{eqnarray}
S_{\varphi \overline \varphi } & =&  \int \d^4 x s(  P \overline \varphi^a_i \varphi^a_i ) ~=~  \int \d^4 x \left[  Q  \overline \varphi^a_i \varphi^a_i- P  \overline \varphi^a_i \omega^a_i\right]  \;, \nonumber\\
S_{\overline \omega \omega } & =&  \int \d^4 x s(  V \overline \omega^a_i \omega^a_i) ~=~  \int \d^4 x \left[ W\overline \omega^a_i \omega^a_i - V \overline \varphi^a_i \omega^a_i \right]  \;, \nonumber\\
S_{\overline{\varphi} \overline \varphi, \overline \omega \overline \varphi } & =&  \frac{1}{2} \int \d^4 x s(  \overline G^{ij} \overline \omega^a_i \overline \varphi^a_j) ~=~  \int \d^4 x \left[  \overline H^{ij} \overline \omega^a_i \overline \varphi^a_j + \frac{1}{2} \overline G^{ij} \overline \varphi^a_i \overline \varphi^a_j \right]  \;, \nonumber\\
S_{\varphi \varphi, \omega \varphi } &=& \frac{1}{2} \int \d^4 x s(  H^{ij} \varphi^a_i  \varphi^a_j) ~=~   \int \d^4 x  \left[ \frac{1}{2} G^{ij} \varphi^a_i \varphi^a_j - H^{ij} \omega^a_i \varphi^a_j \right] \;, \nonumber\\
S_\vac &=&  \int \d^4 x \left[ \kappa (G^{ij} \overline G^{ij} - 2 H^{ij} \overline H^{ij}) +  \lambda (G^{ii} \overline G^{jj} - 2 H^{ii} \overline H^{jj}) \right] \nonumber\\
 && - \int \d^4 x \left[ \alpha   (Q Q +  Q W) + \beta ( QW + W W)  + \chi Q \tau + \delta W \tau \right] \;.
\end{eqnarray}

\section{The Ward identities}
With the help of section \ref{algebraicrenormGZ} of chapter \ref{algebraic}, we can easily summarize all Ward identities obeyed by the action $\Sigma_\CGZ$
\begin{enumerate}
\item  The Slavnov-Taylor identity reads
\begin{equation}\label{qqf1}
\mathcal{S}(\Sigma_\CGZ )=0\;,
\end{equation}
with
\begin{eqnarray*}
\mathcal{S}(\Sigma_\CGZ ) &=&\int \d^{4}x\left( \frac{\delta \Sigma_\CGZ}{\delta K_{\mu }^{a}}\frac{\delta \Sigma_\CGZ }{\delta A_{\mu}^{a}}+\frac{\delta \Sigma_\CGZ }{\delta L^{a}}\frac{\delta \Sigma_\CGZ}{\delta c^{a}} + b^{a}\frac{\delta \Sigma_\CGZ}{\delta \overline{c}^{a}}+\overline{\varphi }_{i}^{a}\frac{\delta \Sigma_\CGZ }{\delta \overline{\omega }_{i}^{a}} \right. \\
&&+\left.\omega _{i}^{a}\frac{\delta \Sigma_\CGZ }{\delta \varphi _{i}^{a}} +M_{\mu }^{ai}\frac{\delta \Sigma_\CGZ}{\delta U_{\mu}^{ai}} +N_{\mu }^{ai}\frac{\delta \Sigma_\CGZ }{\delta V_{\mu }^{ai}} + R_{\mu }^{ai}\frac{\delta \Sigma_\CGZ }{\delta T_{\mu }^{ai}}  +  Q \frac{\delta \Sigma_\CGZ }{\delta P} \right.\nonumber\\
&&\left.+  W \frac{\delta \Sigma_\CGZ }{\delta V} + \tau \frac{\delta \Sigma_\CGZ }{\delta \eta} + 2  \overline H^{ij} \frac{\delta \Sigma_\CGZ }{\delta \overline G^{ij}} +  G^{ij} \frac{\delta \Sigma_\CGZ }{\delta H^{ij}} \right) \;.
\end{eqnarray*}

\item For the $U(f)$ invariance we now have
\begin{eqnarray}
U_{ij} \Sigma_\CGZ &=&0\;,
\end{eqnarray}
whereby
\begin{multline*}
U_{ij}=\int \d^{4}x\left( \varphi_{i}^{a}\frac{\delta }{\delta \varphi _{j}^{a}}-\overline{\varphi
}_{j}^{a}\frac{\delta }{\delta \overline{\varphi}_{i}^{a}}+\omega _{i}^{a}\frac{\delta }{\delta \omega _{j}^{a}}-\overline{\omega }_{j}^{a}\frac{\delta }{\delta \overline{\omega }_{i}^{a}}
-  M^{aj}_{\mu} \frac{\delta}{\delta M^{ai}_{\mu}} -U^{aj}_{\mu}\frac{\delta}{\delta U^{ai}_{\mu}} +N^{ai}_{\mu}\frac{\delta}{\delta N^{aj}_{\mu}}  \right.\\
\left. +V^{ai}_{\mu}\frac{\delta}{\delta V^{aj}_{\mu}}+ R^{aj}_{\mu}\frac{\delta}{\delta R^{ai}_{\mu}} + T^{aj}_{\mu}\frac{\delta}{\delta T^{ai}_{\mu}} +2 \overline G^{ki} \frac{\delta}{\delta \overline G^{kj}}  -2 G^{kj} \frac{\delta}{\delta G^{ki}} + 2 \overline H ^{ki} \frac{\delta}{\delta \overline H^{kj}} -2  H^{kj} \frac{\delta}{\delta H^{ki}  } \right) \;.
\end{multline*}
By means of the diagonal operator $Q_{f}=U_{ii}$, the single $i$-valued fields and sources turn out to possess an additional quantum number.

\item  The Landau gauge condition and the antighost equation are given by
\begin{eqnarray}
\frac{\delta \Sigma_\CGZ }{\delta b^{a}}&=&\partial_\mu A_\mu^{a}\;,\\
\frac{\delta \Sigma_\CGZ }{\delta \overline{c}^{a}}+\partial _{\mu
}\frac{\delta \Sigma_\CGZ }{\delta K_{\mu }^{a}}&=&0\;.
\end{eqnarray}

\item  The linearly broken local constraints yield
\begin{eqnarray}
\frac{\delta \Sigma_\CGZ }{\delta \overline{\varphi }^{a}_i}+\partial _{\mu }\frac{\delta \Sigma_\CGZ }{\delta M_{\mu }^{ai}} + g f_{dba}    T^{d i}_\mu \frac{\delta \Sigma_\CGZ }{\delta K_{\mu }^{b i}} &=&gf^{abc}A_{\mu }^{b}V_{\mu}^{ci} + \ldots \;, \nonumber\\
\frac{\delta \Sigma_\CGZ }{\delta \omega ^{a}_i}+\partial _{\mu}\frac{\delta \Sigma_\CGZ }{\delta N_{\mu}^{ai}}-gf^{abc}\overline{\omega }^{b}_i\frac{\delta \Sigma_\CGZ }{\delta b^{c}}&=&gf^{abc}A_{\mu }^{b}U_{\mu }^{ci} + \ldots\;.
\end{eqnarray}
whereby the $\ldots$ are extra linear breaking terms.

\item  The exact $\mathcal{R}_{ij}$ symmetry is \textbf{broken}

\item The integrated Ward Identity is \textbf{broken}

\item There is also a new identity:
\begin{eqnarray}\label{qqf8}
\frac{\delta \Sigma_\CGZ }{\delta P} &=& \frac{\delta \Sigma_\CGZ }{\delta V}\;.
\end{eqnarray}

\end{enumerate}
\begin{table}[H]
        \begin{tabular}{|c|c|c|c|c|c|c|c|c||c|c|c|c|}
        \hline
        & $A_{\mu }^{a}$ & $c^{a}$ & $\overline{c}^{a}$ & $b^{a}$ & $\varphi_{i}^{a} $ & $\overline{\varphi }_{i}^{a}$ & $\omega _{i}^{a}$ & $\overline{\omega }_{i}^{a}$ &$U_{\mu}^{ai}$&$M_{\mu }^{ai}$&$N_{\mu }^{ai}$&$V_{\mu }^{ai}$  \\
        \hline
        \hline
        \textrm{dimension} & $1$ & $0$ &$2$ & $2$ & $1$ & $1$ & $1$ & $1$ &$2$ & $2$ & $2$ &$2$  \\
        \hline
        $\mathrm{ghost\; number}$ & $0$ & $1$ & $-1$ & $0$ & $0$ & $0$ & $1$ & $-1$ & $-1$& $0$ & $1$ & $0$ \\
        \hline
        $Q_{f}\textrm{-charge}$ & $0$ & $0$ & $0$ & $0$ & $1$ & $-1$& $1$ & $-1$&  $-1$& $-1$&$1$  & $1$ \\
        \hline
        \end{tabular}
        \end{table}
        \begin{table}[H]
        \begin{tabular}{|c|c|c|c|c|c|c|c|c|c|c|c|c|c|c|}
        \hline
        &$R_{\mu }^{ai}$&$T_{\mu }^{ai}$&$K_{\mu }^{a}$&$L^{a}$ & $Q$  & $P$ & $W$ & $V$  & $\tau$ & $\eta$ & $G^{ij}$ &$\overline G^{ij}$ &$H^{ij}$&$\overline H^{ij}$ \\
        \hline
        \hline
         \textrm{dimension} &$2$ & $2$& $3$ & $4$ & $2$ & 2  &2&2 & 2&2 &2&2&2&2\\
        \hline
        $\mathrm{ghost\; number}$ &$0$ & $-1$ & $-1$ & $-2$ & 0 &-1 &0&-1 &0&-1&0&0&-1&1\\
        \hline
        $Q_{f}\textrm{-charge}$ &1 & 1 & $0$  & $0$ &0 &0 &0&0 &0&0&-2&2&-2&2\\
        \hline
        \end{tabular}
        \caption{Quantum numbers of the fields and sources.}
        \end{table}

\section{The counterterm}
These identities \eqref{qqf1}-\eqref{qqf8} can be translated into constraints on the counterterm according to the QAP. Unfortunately, many identities are broken due to the introduction of these $d=2$ operators. However, we are using mass independent renormalization schemes and therefore, the new massive sources ($P$, $Q$, $V$, $W$, $G^{ij}$, $\overline G^{ij}$, $H^{ij}$, $\overline H^{ij}$) cannot influence the counterterm of the original GZ action \eqref{final}. Therefore, the counterterm is given by
\begin{equation}
\Sigma^c_\CGZ = \Sigma^c_\GZ + \Sigma^c_A + \Sigma^c_{P - H}\;,
\end{equation}
with $\Sigma^c_\GZ$ given by equation \eqref{final}, and $\Sigma^c_A$ given by
\begin{equation}
\Sigma^c_A = \int \d^{4}x\left( \frac{a_{2}}{2}\tau A_{\mu }^{a}A_{\mu }^{a}+ \frac{a_{3}}{2}\zeta \tau ^{2}+\left( a_{2}-a_{1}\right) \eta A_{\mu}^{a}\partial _{\mu }c^{a}\right)\;,
\end{equation}
as already determined in \eqref{countertermAGZ}. $\Sigma^c_{P \ldots H}$ is dependent of all the sources  ($P$, $Q$, $V$, $W$, $G^{ij}$, $\overline G^{ij}$, $H^{ij}$, $\overline H^{ij}$), is of dimension 4, ghost number $-1$ and $Q_f = 0$ and obeys the remaining Ward identities. Due to the linearly broken constraints we find
\begin{align}
\frac{ \p \Sigma^c_{P - H}}{ \p \varphi} &= 0\;, & \frac{ \p \Sigma^c_{P - H}}{ \p \overline \varphi} &= 0 \;, & \frac{ \p \Sigma^c_{P - H}}{ \p \omega} &= 0 \;, &\frac{ \p \Sigma^c_{P - H}}{ \p \overline \omega } &= 0 &\;.
\end{align}
Therefore,
\begin{multline}
\Sigma^c_{P -H}=  \mathcal B_\Sigma \bigl( b_1 P A_\mu^a A_\mu^a + b_2 V A_\mu^a A_\mu^a + b_3 Q P + b_4 Q V + b_5 W P + b_6 W V + b_7 P \tau + b_8 V \tau \\+ b_9 Q \eta + b_{10} W \eta + c_1 H^{ij} \overline G^{ij} +  c_2 H^{ii}  \overline  G^{jj}  \bigr)\;,
\end{multline}
whereby $b_1$, $\ldots$, $c_2$ are arbitrary constants. By invoking the new identity
\begin{eqnarray}
\frac{\delta\Sigma^c_{P -H} }{\delta P} &=& \frac{\delta \Sigma^c_{P -H} }{\delta V}\;,
\end{eqnarray}
we can write
\begin{multline}
\Sigma^c_{P -H}=  b_1 [ (Q+W) A_\mu^a A_\mu^a +2 (P+V) \p_\mu c^a A_\mu^a] + b_3 Q Q + b_4 Q W + b_6 W W  + b_7 Q \tau + b_8 W \tau \\
+  c_1 (G^{ij} \overline G^{ij} - 2 H^{ij} \overline H^{ij}) + c_{2}   (G^{ii} \overline G^{jj} - 2 H^{ii} \overline H^{jj})\;.
\end{multline}
Let us notice that due to the $U(f)$ constraint, the term in $c_2$ is only present when
\begin{equation}
G^{ij} \overline G^{qq} + 2 H^{pp} \overline H^{ij} ~=~ G^{qq} \overline G^{ij} + 2 H^{ij} \overline H^{qq}\;,
\end{equation}
which is indeed the case due to hermiticity.

\section{The renormalization factors}
Let us now try to reabsorb this counterterm into the starting action \eqref{CGZ}. We shall split this analysis into three parts, according to
\begin{equation}
\Sigma^c_A + \Sigma^c_{P - H} = \Sigma^c_I + \Sigma^c_{II} + \Sigma^c_{III}\;,
\end{equation}
whereby
\begin{eqnarray}
\Sigma^c_I &=&  c_1  (G^{ij} \overline G^{ij} - 2 H^{ij} \overline H^{ij}) + c_{2} (G^{ii} \overline G^{jj} - 2 H^{ii} \overline H^{jj})\;, \nonumber\\
\Sigma^c_{II} &=& b_1 [ (Q+W) A_\mu^a A_\mu^a +2 (P+V) \p_\mu c^a A_\mu^a] +   \frac{a_{2}}{2}\tau A_{\mu }^{a}A_{\mu }^{a} + \left( a_{2}-a_{1}\right) \eta A_{\mu }^{a}\partial _{\mu }c^{a} \;,\nonumber\\
\Sigma^c_{III} &=&  b_3 Q Q + b_4 Q W + b_6 WW  + b_7 Q \tau + b_8 W \tau + \frac{a_{3}}{2}\zeta \tau ^{2}\;,
\end{eqnarray}
are the three parts which we shall try to absorb separately.\\
\\
Firstly, we start with the vacuum counterterm connected to the arbitrary parameters $c_1$ and $c_2$. If we redefine $c_1$ and $c_2$, we can write
\begin{equation}
\Sigma^c_I =   c_1 \kappa (G^{ij} \overline G^{ij} - 2 H^{ij} \overline H^{ij}) + c_{2}  \lambda (G^{ii} \overline G^{jj} - 2 H^{ii} \overline H^{jj})\;,
\end{equation}
and if we define
\begin{align}
\overline H^{ij}_0 &= Z_{\overline H }  \overline H^{ij} &  H^{ij}_0 &= Z_{ H }  H^{ij} & \overline G^{ij}_0 &= Z_{\overline G }  \overline G^{ij} & G^{ij}_0 &= Z_{G }  \overline G^{ij} & \kappa_0 &= Z_\kappa \kappa & \lambda_0 &= Z_\lambda \lambda\;,
\end{align}
we find for the renormalization factors of the new sources and the LCO parameters $\kappa$ and $\lambda$:
\begin{eqnarray}\label{Zfac}
Z_{\overline H} &=& Z_{\overline \varphi}^{-1/2} Z_{\overline \omega}^{-1/2} \;,\nonumber\\
Z_{\overline G } &=& Z_{\overline \varphi}^{-1} \;,\nonumber\\
Z_{ H } &=& Z_{ \varphi}^{-1/2} Z_{ \omega}^{-1/2}\;, \nonumber\\
Z_{G } &=& Z_{ \varphi}^{-1}\;,\nonumber\\
Z_{\kappa} &=& (1+ c_1) Z_{\overline G}^{-1} Z_{G}^{-1} = (1+ c_1) Z_{\overline H}^{-1} Z_{H}^{-1} \;,\nonumber\\
Z_{\lambda} &=& (1+ c_2) Z_{\overline G}^{-1} Z_{G }^{-1} = (1+ c_2) Z_{\overline H }^{-1} Z_{H}^{-1}\;,
\end{eqnarray}
and thus the part $\Sigma^c_I$ can absorbed in the starting action.\\
\\
Secondly, let us focus on $\Sigma^c_{II}$
\begin{equation}
\Sigma^c_{II} = b_1 [ (Q+W) A_\mu^a A_\mu^a +2 (P+V) \p_\mu c^a A_\mu^a] +   \frac{a_{2}}{2}\tau A_{\mu }^{a}A_{\mu }^{a} + \left( a_{2}-a_{1}\right) \eta A_{\mu }^{a}\partial _{\mu }c^{a}\;.
\end{equation}
We propose the following mixing matrix:
\begin{eqnarray}
 \left(
  \begin{array}{c}
     Q_0 \\
     W_0 \\
     \tau_0
  \end{array}
\right) &= & \left(
          \begin{array}{ccc}
            Z_{QQ}&Z_{QW} &Z_{Q\tau} \\
            Z_{WQ}  &Z_{WW} & Z_{W\tau}   \\
           Z_{\tau Q} & Z_{\tau W}& Z_{\tau\tau}          \end{array}
        \right)
        \left(
\begin{array}{c}
    Q\\
    W \\
    \tau
  \end{array}
\right)\,.
\end{eqnarray}
\begin{itemize}
\item From
\begin{equation}
Q_0 \varphi^a_{i,0} \overline \varphi^a_{i,0} =  [Z_{QQ}Q + Z_{QW} W + Z_{Q\tau} \tau] Z_{\varphi} \varphi^a_{i} \overline \varphi^a_{i} ~=~ Q \varphi^a_{i} \overline \varphi^a_{i}\;,
\end{equation}
we find that $Z_{QQ} = Z_{\overline \varphi}^{-1}$, while $Z_{QW} = Z_{Q\tau}=0$.
\item From
\begin{equation}
W_0 \overline \omega^a_{i,0} \omega^a_{i,0} = [Z_{WQ}Q + Z_{WW} W + Z_{W\tau} \tau] Z_{\varphi}\overline \omega^a_{i} \omega^a_{i} =W \varphi^a_{i} \overline \varphi^a_{i}\;,
\end{equation}
we find that $Z_{WW} = Z_{\overline \varphi}^{-1}$, while $Z_{WQ} = Z_{W\tau}=0$.

\item Finally, from
\begin{multline}
\frac{1}{2}\tau_0 A^a_{\mu,0} A^a_{\mu,0} = \frac{1}{2} [Z_{\tau Q} Q + Z_{\tau W} W + Z_{ \tau \tau} \tau] Z_A  A^a_{\mu} A^a_{\mu}  \\=  \frac{1}{2} \left( 1 +  a_2 \right) \tau A_{\mu}^{a}A_{\mu }^{a} + b_1 Q  A_{\mu }^{a}A_{\mu }^{a} + b_1 W  A_{\mu }^{a}A_{\mu }^{a} \;,
\end{multline}
we obtain $Z_{\tau \tau} = Z_\tau= \left( 1 + a_2 \right)Z_A^{-1}$, and $Z_{\tau Q} = Z_{\tau W} = 2 b_1$.
\end{itemize}
In summary, we find the following matrix
\begin{eqnarray}\label{mixing}
 \left(
  \begin{array}{c}
     Q_0 \\
     W_0 \\
     \tau_0
  \end{array}
\right) &= & \left(
          \begin{array}{ccc}
            Z_{ \varphi}^{-1}& 0 &0 \\
            0  & Z_{\varphi}^{-1} & 0   \\
           Z_{\tau W} & Z_{\tau W}& Z_{\tau\tau}          \end{array}
        \right)
        \left(
\begin{array}{c}
    Q\\
    W \\
    \tau
  \end{array}
\right)\,.
\end{eqnarray}
Now that we have the mixing matrix at our disposal, we can pass to the corresponding bare operators by taking the inverse of this matrix,
\begin{eqnarray}
 \left(
  \begin{array}{c}
     Q \\
     W \\
     \tau
  \end{array}
\right) &= & \left(
          \begin{array}{ccc}
            Z_{\varphi}& 0 &0 \\
            0  & Z_{ \varphi} & 0   \\
           - \frac{ Z_{\tau W} Z_{\varphi} }{ Z_{\tau\tau}} & - \frac{ Z_{\tau W} Z_{ \varphi} }{ Z_{\tau\tau}} & \frac{1}{Z_{\tau\tau}}          \end{array}
        \right)
        \left(
\begin{array}{c}
    Q_0\\
    W_0 \\
    \tau_0
  \end{array}
\right)\,.
\end{eqnarray}
Subsequently, we can derive the corresponding mixing matrix for the operators, since insertions of an operator correspond to derivatives w.r.t.~to the appropriate source of the generating functional $Z^c(Q,W,\tau)$. In particular,
\begin{eqnarray}
\frac{1}{2}      A^2_0 &=& \left. \frac{\delta Z^c(Q,W,\tau)}{\delta \tau_0} \right|_{\tau_0 = 0} \nonumber\\
    &=& \frac{\delta Q}{\delta \tau_0}\frac{\delta Z^c(Q,W,\tau) }{\delta Q}+\frac{\delta W}{\delta \tau_0}\frac{\delta Z^c(Q,W,\tau)}{\delta W } + \frac{\delta \tau}{\delta \tau_0}\frac{\delta Z^c(Q,W,\tau)}{\delta \tau} \nonumber\\
    \Rightarrow  A^2_0 &=&   \frac{1}{Z_{\tau\tau}}   A^2 \;,
\end{eqnarray}
and similarly for $ \overline \varphi^a_{i,0} \varphi^a_{i,0}$ and $\overline \omega^a_{i,0} \omega^a_{i,0} $. We thus need to take the transpose of the previous matrix,
\begin{eqnarray}
 \left(
  \begin{array}{c}
      \overline \varphi^a_{i,0} \varphi^a_{i,0} \\
     \overline \omega^a_{i,0} \omega^a_{i,0}  \\
     A^2_0
  \end{array}
\right) &= & \left(
          \begin{array}{ccc}
            Z_{ \varphi}& 0                     & - \frac{ Z_{\tau W} Z_{ \varphi} }{ Z_{\tau\tau}} \\
            0                    & Z_{ \varphi} & - \frac{ Z_{\tau W} Z_{ \varphi} }{ Z_{\tau\tau}}   \\
            0                     &  0                   &   \frac{1}{Z_{\tau\tau}}          \end{array}
        \right)
        \left(
\begin{array}{c}
     \overline \varphi^a_{i} \varphi^a_{i} \\
     \overline \omega^a_{i} \omega^a_{i} \\
    A^2
  \end{array}
\right)\,.
\end{eqnarray}
We can make some observations from this matrix. Firstly, we find that $A^2_0$ does not contain the operators $ \overline \varphi^a_{i} \varphi^a_{i} $ and $ \overline \omega^a_{i} \omega^a_{i}$. This is already a first check on our results as without these two latter operators the GZ action including $A^2$ is renormalizable as we have shown already in section \ref{condensateAkwadraat} of chapter \ref{refined}. Secondly, we observe that
\begin{eqnarray}
\overline \varphi^a_{i,0} \varphi^a_{i,0} - \overline \omega^a_{i,0} \omega^a_{i,0}  &=& Z_\varphi (  \overline \varphi^a_{i} \varphi^a_{i} - \overline \omega^a_{i} \omega^a_{i})\;,
\end{eqnarray}
meaning that the mixing with $A^2$ disappears again when recombining the two operators in a certain way. In fact, this is the operator $(  \overline \varphi^a_{i} \varphi^a_{i} - \overline \omega^a_{i} \omega^a_{i})$ which we have investigated using the RGZ action  and no mixing with $A^2$ appears for this operator.  \\
\\
We can do a completely analogous reasoning for the part in $\p_\mu c^a A^a_\mu$. We first set $V + P = X$. We propose
\begin{eqnarray}
 \left(
  \begin{array}{c}
     X_0 \\
     \eta_0
  \end{array}
\right) &= & \left(
          \begin{array}{ccc}
            Z_{XX}& Z_{X\eta} \\
           Z_{\eta X} &  Z_{\eta\eta}          \end{array}
        \right)
        \left(
\begin{array}{c}
    X\\
    \eta
  \end{array}
\right)\,.
\end{eqnarray}
\begin{itemize}
\item From
\begin{eqnarray*}
- (X_0)  [\overline \varphi^a_{i,0}  \omega^a_{i,0} ]  &=&  - [ Z_{XX} X + Z_{X \eta}  \eta]  Z_{\overline \varphi}^{1/2} Z_{\omega}^{1/2} \overline \varphi^a_{i} \omega^a_{i} ~=~ - X [\overline \varphi^a_{i}  \omega^a_{i} ]
\end{eqnarray*}
we find that $Z_{XX} = Z_{\overline \varphi}^{-1/2} Z_{\omega}^{-1/2} $, while $Z_{X\eta} =0$.

\item Also, from
\begin{equation*}
\eta_0 A^a_{\mu,0} \p_{\mu} c^a_0 =  [Z_{\eta X} X  + Z_{ \eta \eta} \eta]  Z_A^{1/2} Z_c^{1/2}  A^a_{\mu} \p_{\mu} c^a =  \left( 1 +  a_2-a_1 \right) \eta A^a_{\mu} \p_{\mu} c^a +  2 b_1 X A^a_{\mu} \p_{\mu} c^a \;,
\end{equation*}
we obtain $Z_{\eta \eta} = Z_\eta= \left( 1 +  a_2-a_1 \right) Z_A^{-1/2} Z_c^{-1/2} $, and $Z_{\eta X} = 2 b_1$.
\end{itemize}
Therefore, we find that
\begin{eqnarray}
 \left(
  \begin{array}{c}
      \overline \varphi^a_{i,0} \omega^a_{i,0} \\
     A_{\mu,0} \p_\mu c_0
  \end{array}
\right) &= & \left(
          \begin{array}{ccc}
           Z_A^{1/2} Z_c^{1/2}          & -2 b_1                       \\
           0                     &  Z_{\eta}^{-1}
                     \end{array}
        \right)
        \left(
\begin{array}{c}
     \overline \varphi^a_{i} \omega^a_{i} \\
     A_{\mu} \p_\mu c
  \end{array}
\right)\,.
\end{eqnarray}
Again, we find $ A_{\mu,0} \p_\mu c_0$ does not contain $\overline \varphi^a_{i,0} \omega^a_{i,0} $, which is necessary as the GZ action with the inclusion of $A^2$ is renormalizable. We also see that, when setting $V = - P$, $X =0$, the mixing with $A^2$ disappears again.\\
\\
Thirdly, the vacuumterm $\Sigma_{III}^c$ has the following form
\begin{eqnarray}
 b_3 Q Q + b_4 Q W + b_6 W W  + b_7 Q \tau + b_8 W \tau + \frac{a_{3}}{2}\zeta \tau ^{2}\;,
\end{eqnarray}
we know that setting $Q = -W$ has to return the vacuumterm from the RGZ action $\sim a_4 Q \tau + \frac{a_{3}}{2}\zeta \tau ^{2}$. Therefore, we may set
\begin{eqnarray}
b_3 -b_4 + b_6 = 0 \;.
\end{eqnarray}
In this case, the vacuumterm reduces to
\begin{eqnarray}
- c_1 \alpha (Q Q +  Q W) - c_2 \beta( QW + W W)  - c_3 \chi Q \tau - c_4  \delta W \tau  + \frac{a_{3}}{2}\zeta \tau ^{2}\;,
\end{eqnarray}
where we have extracted $\alpha$, $\beta$, $\chi$ and $\delta$ and some minus signs for convenience. If we allow mixing between the different parameters,
\begin{eqnarray}
 \left(
  \begin{array}{c}
      \alpha_0 \\
      \beta_0\\
      \chi_0\\
      \delta_0 \\
      \zeta_0
  \end{array}
\right) &= & \left(
          \begin{array}{ccccc}
           Z_{\alpha\alpha}          & Z_{\alpha\beta}  &  Z_{\alpha\chi}  & Z_{\alpha\delta}  & Z_{\alpha \zeta}         \\
           Z_{\beta\alpha}           & Z_{\beta\beta}   &  Z_{\beta\chi}   & Z_{\beta\delta}   & Z_{\beta \zeta}          \\
           Z_{\chi\alpha}          & Z_{\chi\beta}  &  Z_{\chi\chi}  & Z_{\chi\delta}  & Z_{\chi \zeta}              \\
           Z_{\delta\alpha}          & Z_{\delta\beta}  &  Z_{\delta\chi}  & Z_{\delta\delta}  & Z_{\delta \zeta}               \\
           Z_{\zeta\alpha}           & Z_{\zeta\beta}   &  Z_{\zeta\chi}   & Z_{\zeta\delta}   & Z_{\zeta \zeta}
                     \end{array}
        \right)
        \left(
\begin{array}{c}
 \alpha \\
      \beta \\
      \chi \\
      \delta \\
      \zeta
  \end{array}
\right)\,.
\end{eqnarray}
when absorbing the counterterm, we find for the mixing matrix
\begin{equation} \left(
          \begin{array}{ccccc}
           Z_{\alpha\alpha}          & Z_{\alpha\beta}  &  Z_{\alpha\chi}  & Z_{\alpha\delta}  & Z_{\alpha \zeta}         \\
           Z_{\beta\alpha}           & Z_{\beta\beta}   &  Z_{\beta\chi}   & Z_{\beta\delta}   & Z_{\beta \zeta}          \\
           Z_{\chi\alpha}          & Z_{\chi\beta}  &  Z_{\chi\chi}  & Z_{\chi\delta}  & Z_{\chi \zeta}              \\
           Z_{\delta\alpha}          & Z_{\delta\beta}  &  Z_{\delta\chi}  & Z_{\delta\delta}  & Z_{\delta \zeta}               \\
           Z_{\zeta\alpha}           & Z_{\zeta\beta}   &  Z_{\zeta\chi}   & Z_{\zeta\delta}   & Z_{\zeta \zeta}
                     \end{array}
        \right) =
\left(
          \begin{array}{ccccc}
           \frac{1+c_1}{Z_{QQ}^2}        & 0 &  - \frac{Z_{\chi \chi} Z_{\tau W}}{Z_{QQ}} & 0   & \frac{Z_{\tau W}^2 Z_{\zeta \zeta}}{2Z_{QQ}^2}     \\
           0          &\frac{1+c_2}{Z_{QQ}^2}  &  0  & - \frac{Z_{\delta \delta} Z_{\tau W}}{Z_{QQ}}  & \frac{Z_{\tau W}^2 Z_{\zeta \zeta}}{2Z_{QQ}^2}    \\
           0          & 0 &  \frac{1+c_3}{Z_{QQ}Z_{\tau\tau}}  & 0    &  -\frac{ Z_{\tau W} Z_{\zeta \zeta}}{Z_{QQ} }   \\
           0          & 0 &  0  &  \frac{1+c_4}{Z_{QQ}Z_{\tau\tau}}   &  -\frac{ Z_{\tau W} Z_{\zeta \zeta}}{Z_{QQ} }     \\
           0&0&0&0& \frac{1-a_3}{Z_{\tau\tau}^2}
                     \end{array}
        \right)\;.
\end{equation}
In summary, we have proven the action to be renormalizable.

\chapter{Details of the calculation of the effective action for the further refined GZ action\label{detailsvery}}
\section{Determination of the LCO parameters $\delta \zeta$, $\delta \alpha$, $\delta \chi$ and $\delta \varrho$ \label{appsection4.2}}
We shall start from expression \eqref{startxx}, determine the quadratic part, and integrate out all the fields. The quadratic action is given by
\begin{multline}\label{start2}
\Sigma_\CGZ^{\quadr} = \int \d^d x  \left[ A_\mu^a \delta^{ab} \left( - \delta_{\mu\nu} \p^2 + \left( 1 - \frac{1}{\alpha}\right) \p_\mu \p_\nu \right)A_\nu^b  + \overline \varphi \p^2 \varphi - \gamma^2 g f_{abc} A_\mu^a (\varphi^{bc}_\mu + \overline \varphi^{bc}_\mu) \right]\nonumber\\
\\+ \int \d^4 x  \left[ Q  \overline \varphi^a_i \varphi^a_i   + \frac{1}{2} \tau A_{\mu }^{a}A_{\mu }^{a}  - \frac{1}{2}\zeta \tau ^{2}  - \alpha   Q Q    - \chi Q \tau\right]  + \int \d^4 x \left[ \frac{1}{2} \overline G \overline \varphi^a_i \overline \varphi^a_i + \frac{1}{2} G \varphi^a_i \varphi^a_i  + \varrho G \overline G \right]\;,
\end{multline}
whereby we have immediately integrated out the ghost fields, $c, \overline c, \omega, \overline \omega$, as they are only appear trivially. We have also already integrated out the $b$-field whereby $\alpha$ is formally equal to zero.\\
\\
As a first step, we integrate out the $\varphi$ ad $\overline \varphi$ fields. For this, we shall split $\varphi$, $\overline \varphi$, $G$ and $\overline G$ into real components:
\begin{align}
\overline \varphi^a_i &= U^a_i + \ii V^a_i \;,   & \varphi^a_i &= U^a_i - \ii V^a_i  \;,\nonumber\\
\overline G &= X + \ii Y  \;,  & G &= X- \ii Y \;,
\end{align}
so the part depending on $\varphi$ and $\overline \varphi$ in expression \eqref{start2} becomes
\begin{eqnarray*}
&&\int \d^d x \left( U^a_i\p^2 U^a_i + V^a_i \p^2 V^a_i  - 2 \gamma^2 g f_{abc} A_\mu^a U^{bc}_\mu + Q U^2 + Q V^2 + X U^2 - X V^2 \right. \nonumber\\
 && \left.- 2 Y U^a_iV^a_i + \varrho X^2 + \varrho Y^2 \right) \nonumber\\
&=& \int \d^d x \left( \frac{1}{2} \begin{bmatrix}
     U_\mu^{ab} &      V_\mu^{ab}
  \end{bmatrix}   \begin{bmatrix}
    2 (\p^2 + Q + X) &      -2 Y \\
    -2Y &  2 (\p^2 + Q - X)
  \end{bmatrix} \begin{bmatrix}
     U_\mu^{ab} \\     V_\mu^{ab}
  \end{bmatrix}   - 2 \gamma^2 g f_{abc} A_\mu^a U^{bc}_\mu \right)\;.
\end{eqnarray*}
Therefore, applying Gaussian integration, we find for the integration over $\varphi$ and $\overline \varphi$
\begin{multline} \label{uitinto}
\int [\d \varphi] [\d \overline \varphi] \exp [- \Sigma_\CGZ^{\quadr} ]\\ = \exp \left[ \frac{1}{2} \lambda^4 A_\mu^k \left( \frac{\p^2 + Q - X}{\p^4 + 2 Q \p^2 + Q^2 - X^2 - Y^2}\right) A_\mu^k + \ldots\right] (\det P_{\mu\nu}^{ab})^{-1/2}\;,
\end{multline}
whereby we recall that $\lambda$ is defined as $\lambda^4 =  2 \gamma^4 g^2 N$. $P$ is given by
\begin{equation}
P_{\mu\nu}^{ab,cd} =\delta_{\mu\nu} \delta^{ab} \delta^{cd}\begin{bmatrix}
    2 (\p^2 + Q + X) &      -2 Y \\
    -2Y &  2 (\p^2 + Q - X)
  \end{bmatrix}\;,
\end{equation}
and the $\ldots$ stand for the other terms in $\Sigma_\CGZ^{\quadr}$, see \eqref{start2}, i.e.~terms purely in $A$ and the vacuum terms. The second step is to integrate out the gluon field $A_\mu^a$. Combining the expression \eqref{uitinto} with the terms purely in $A$ from the quadratic action, we obtain,
\begin{multline}
\int [\d A] \e^{ \Bigl[- \frac{1}{2}  A_\mu^a \delta^{ab} \left( - \delta_{\mu\nu} \p^2 + \left( 1 - \frac{1}{\alpha}\right) \p_\mu \p_\nu  -  \lambda^4 \left( \frac{\p^2 + Q - X}{\p^4 + 2 Q \p^2 + Q^2 - X^2 - Y^2}\right) + \tau \delta_{\mu\nu}  \right)A_\nu^b \Bigr]}\\
= \left[ \det \left( - \delta_{\mu\nu} \p^2 + \left( 1 - \frac{1}{\alpha}\right) \p_\mu \p_\nu  -  \lambda^4 \left( \frac{\p^2 + Q - X}{\p^4 + 2 Q \p^2 + Q^2 - X^2 - Y^2}\right) + \tau \delta_{\mu\nu}  \right) \right]^{-1/2}\;.
\end{multline}
Therefore, the total effective action at one loop is given by
\begin{multline}\label{totaleffaction}
\e^{-W(Q, \tau, G, \overline G)}= (\det P)^{-1/2} \Bigl[ \det \Bigl( - \delta_{\mu\nu} \p^2 + \left( 1 - \frac{1}{\alpha}\right) \p_\mu \p_\nu \\ -  \lambda^4 \delta_{\mu\nu}\left( \frac{\p^2 + Q - X}{\p^4 + 2 Q \p^2 + Q^2 - X^2 - Y^2}\right) + \tau \delta_{\mu\nu}  \Bigr) \Bigr]^{-1/2} \e^{\left[ - \int \d^4 x \left[ - \frac{1}{2}\zeta \tau ^{2}  - \alpha   Q Q    - \chi Q \tau \frac{1}{2}  + \varrho G \overline G \right]\right] }\;.
\end{multline}
In order to find $\delta \zeta$, $\delta \alpha$, $\delta \chi$ and $\delta \varrho$ at one loop, we need to find the first order infinities of the previous expression. These shall be present in the two determinants which we need to evaluate.\\
\\
Let us start with the first determinant of $P$. In general, we can write
\begin{equation}
(\det P_{\mu\nu}^{ab})^{-1/2} = \e^{-\frac{1}{2} \Tr \ln P_{\mu\nu}^{ab,cd}} = \e^{-\frac{1}{2} d (N^2-1)^2 \Tr \ln P}\;.
\end{equation}
As we are taking the trace, we know that $\Tr \ln P = \Tr \ln P'$ with $P'$ the diagonalization of $P$. Therefore, after diagonalization, we find
\begin{multline}
(\det P_{\mu\nu}^{ab,cd})^{-1/2} \\= \exp \left[ -\frac{1}{2} d (N^2-1)^2 \Tr  \left( \ln ( -\p^2 - Q + \sqrt{X^2 + Y^2}) + \ln ( -\p^2 - Q - \sqrt{X^2 + Y^2}) \right)   \right]\;.
\end{multline}
Employing the standard formula, \cite{Peskin}
\begin{equation}\label{standard2}
\Tr \ln (-\p^2 + M^2) = - \frac{\Gamma(-d/2) }{(4\pi)^{d/2}} \frac{1}{ (M^2)^{-d/2}}\;,
\end{equation}
we obtain the following infinity
\begin{eqnarray}\label{res1}
(\det P)^{-1/2} &=& \exp \left[\frac{1}{\epsilon}\frac{ (N^2 - 1)^2}{4 \pi^2} \left[ Q^2 + X^2 + Y^2\right] + c_1\right]\;,
\end{eqnarray}
whereby $c_1$ is a constant term. \\
\\
The second determinant requires a bit more effort to calculate. Let us call the corresponding matrix $K$. We thus calculate
\begin{equation}
(\det K_{\mu\nu}^{ab})^{-1/2} = \e^{-\frac{1}{2} (N^2-1) \Tr \ln K_{\mu\nu} }\;,
\end{equation}
Therefore, we need to determine
\begin{multline}\label{terms}
\Tr \ln K_{\mu\nu} = \Tr \ln \left(   \delta_{\mu\nu}\left( -\p^2  -  \lambda^4 \left( \frac{\p^2 + Q - X}{\p^4 + 2 Q \p^2 + Q^2 - X^2 - Y^2}\right) + \tau \right)\right) \\ + \Tr \ln \left( \delta_{\mu\nu} + \frac{1}{\left( -\p^2  -  \lambda^4 \left( \frac{\p^2 + Q - X}{\p^4 + 2 Q \p^2 + Q^2 - X^2 - Y^2}\right) + \tau  \right)}  \left( 1 - \frac{1}{\alpha}\right) \p_\mu \p_\nu  \right)\;.
\end{multline}
For the first term, we can easily take the trace over the Lorentz indices, while for the second term, we need to use $\ln (1 + x) = x - \frac{x^2}{2} + \ldots$, then take the trace of the diagonal elements of the second term, and again employ $x - \frac{x^2}{2} + \ldots = \ln (1 + x)$. After these operations, we obtain
\begin{multline*}
\Tr \ln K_{\mu\nu} = d \Tr \ln \left(  \left( -\p^2  -  \lambda^4 \left( \frac{\p^2 + Q - X}{\p^4 + 2 Q \p^2 + Q^2 - X^2 - Y^2}\right) + \tau \right)\right) \\ + \Tr \ln \left( 1 + \frac{1}{\left( -\p^2  -  \lambda^4 \left( \frac{\p^2 + Q - X}{\p^4 + 2 Q \p^2 + Q^2 - X^2 - Y^2}\right) + \tau  \right)}  \left( 1 - \frac{1}{\alpha}\right) \p^2  \right)\;,
\end{multline*}
which can be written as
\begin{multline*}
\Tr \ln K_{\mu\nu} = (d-1) \Tr \ln \left(  \left( -\p^2  -  \lambda^4 \left( \frac{\p^2 + Q - X}{\p^4 + 2 Q \p^2 + Q^2 - X^2 - Y^2}\right) + \tau \right)\right) \\ + \Tr \ln \left( \left( -\p^2  -  \lambda^4 \left( \frac{\p^2 + Q - X}{\p^4 + 2 Q \p^2 + Q^2 - X^2 - Y^2}\right) + \tau  \right) +   \left( 1 - \frac{1}{\alpha}\right) \p^2  \right)\;.
\end{multline*}
The first term of this expression can be written as\footnote{We shall replace $-\p^2$ by $p^2$ from now on and work in momentum space.}
\begin{align}
& (d-1) \Bigl[ \Tr \ln \bigl(  p^6+(\tau -2 Q) p^4+\left(\lambda ^4+Q^2-X^2-Y^2-2 Q \tau \right) p^2-Q \lambda ^4+X \lambda ^4+Q^2 \tau \nonumber\\
 &-X^2 \tau -Y^2 \tau  \bigr)  -\Tr \ln \left(  p^4-2 Q p^2+Q^2-X^2-Y^2 \right) \Bigr]\nonumber\\
=& (d-1) \left( \Tr \ln (p^2 - x_1) + \Tr \ln (p^2 - x_2) + \Tr \ln (p^2 - x_3) - \Tr \ln (p^2 - x_4) - \Tr \ln (p^2 - x_5)  \right)\;,
\end{align}
whereby $x_1$, $x_2$ and $x_3$ are the solutions of the equation $x^3+(\tau -2 Q) x^2+\bigl(\lambda ^4+Q^2-X^2-Y^2-2 Q \tau \bigr) x -Q \lambda ^4+X \lambda ^4+Q^2 \tau -X^2 \tau -Y^2 \tau =0$
and $x_4$ and $x_5$ of the equation $ x^2 -2 Q x +Q^2-X^2-Y^2 =0$. After determining $x_1, \ldots, x_5$, we can apply the standard formula \eqref{standard2} again, so we ultimately find for the first term
\begin{equation}
- \frac{3}{16 \pi^2} \frac{1}{\epsilon} \left( \tau^2 - 2 \lambda^4 \right) + c_2\;,
\end{equation}
with $c_2$ a constant, which is not our current interest. For the second term of \eqref{terms}, we can perform an analogous analysis, whereby we find that this term is proportional to $\alpha$ and therefore does not contribute to the determinant. Therefore, the second determinant ultimately gives:
\begin{equation}\label{res2}
(\det K_{\mu\nu}^{ab})^{-1/2} = \exp\left[  (N^2-1)  \frac{3}{32 \pi^2} \frac{1}{\epsilon} \left( \tau^2 - 2 \lambda^4 \right) + c_2 \right] \;.
\end{equation}
We can now combine both results \eqref{res1} and \eqref{res2} to find
\begin{equation}
W(Q, \tau, G, \overline G) = - \frac{(N^2 -1)}{4 \pi^2} \frac{1}{\epsilon} \left( \frac{3}{8}\tau^2 + (N^2-1)( Q^2 + G \overline G) - \frac{3}{4} \lambda^4 \right) + c \;,
\end{equation}
with $c$ a constant term. Therefore, at one loop we obtain
\begin{eqnarray}\label{deltas}
\delta \zeta &=& -\frac{1}{\epsilon} \frac{3}{16 \pi^2} (N^2-1) \;,\nonumber\\
\delta \alpha &=& -\frac{1}{\epsilon} \frac{1}{4 \pi^2} (N^2-1)^2 \;,\nonumber\\
\delta \chi &=& 0 \;, \nonumber\\
\delta \varrho &=& \frac{1}{\epsilon} \frac{1}{4 \pi^2} (N^2-1)^2\;.
\end{eqnarray}

\section{Calculation of the effective action}
We can now proceed in a very similar fashion as in section \ref{appsection4.2}. We find a few parts for the effective potential. A first part, $\Gamma^{(1)}_a$, is the equivalent of $(\det P)^{-1/2}$ in expression \eqref{totaleffaction}
\begin{multline}
\Gamma^{(1)}_a = (N^2 -1)^2 \Biggl[ -\frac{1}{\epsilon} \frac{1}{4\pi^2} (M^4 + \rho \rho^\dagger) + \frac{1}{16 \pi^2} \left( (M^2 - \sqrt{\rho \rho^\dagger})^2 \ln\frac{M^2 - \sqrt{\rho \rho^\dagger}}{\overline \mu^2} \right. \\
\left. + (M^2 + \sqrt{\rho \rho^\dagger})^2 \ln \frac{M^2 + \sqrt{\rho \rho^\dagger}}{\overline \mu^2} - 2 (M^2 + \rho \rho^\dagger) \right) \Biggr]\;.
\end{multline}
The second part, the equivalent of $(\det K)^{-1/2}$ is given by
\begin{multline}
\Gamma^{(1)}_b = \frac{3(N^2 -1)}{64\pi^2} \Biggl[ -\frac{2}{\epsilon}  (m^4 -2 \lambda^4) - \frac{5}{6} (m^4 -2 \lambda^4) + y_1^2 \ln \frac{(-y_1)}{\overline \mu} + y_2^2 \ln \frac{(-y_2)}{\overline \mu} + y_3^2 \ln \frac{(-y_3)}{\overline \mu} \\- y_4^2 \ln \frac{(-y_4)}{\overline \mu} - y_5^2 \ln \frac{(-y_5)}{\overline \mu} \Biggr]\;,
\end{multline}
whereby $y_1$, $y_2$ and $y_3$ are the solutions of the equation $y^3+(m^2 +2 M^2) y^2 +\bigl(\lambda^4+ M^4- \rho \rho^\dagger +2 M^2 m^2 \bigr) y + M^2 \lambda^4 + 1/2 ( \rho + \rho^\dagger) \lambda^4 + M^4 m^2  - m^2 \rho \rho^\dagger  =0$
and $y_4$ and $y_5$ of the equation $ y^2 + 2 M^2 y +M^4 -\rho \rho^\dagger =0$. \\
The third part is the constant term of the GZ action,
\begin{equation}
\Gamma^{(1)}_c = -d \gamma^4_0 (N^2 - 1)\;.
\end{equation}
From equation \eqref{Zgamma}, we can calculate that\footnote{For the explicit loop calculations of the $Z$-factors, we refer to \cite{Gracey:2002yt}.}
\begin{eqnarray}
\gamma_0^4 &=& Z_{\gamma^2}^2 \gamma^4\;,  \qquad \text{with} \qquad Z_{\gamma^2}^2 = 1+ \frac{3}{2} \frac{g^2N}{16 \pi^2} \frac{1}{\epsilon}\;,
\end{eqnarray}
so we find
\begin{eqnarray}
\Gamma^{(1)}_c&=& -d(N^2 - 1)\gamma^4_0 = -4 (N^2 - 1)\gamma^4 - 4 \frac{3}{2} (N^2 - 1)\frac{g^2 N}{16\pi^2}\frac{1}{\epsilon} \gamma^4 + \frac{3}{2} \frac{g^2N}{16 \pi^2}\gamma^4 (N^2-1) \nonumber\\
&=& -2 (N^2 - 1) \frac{\lambda^4}{N g^2} - 6 (N^2 - 1)\frac{\lambda^4}{32\pi^2}\frac{1}{\epsilon}  + \frac{3}{2} \frac{\lambda^4}{32 \pi^2} (N^2-1)  \;.
\end{eqnarray}
The fourth part requires some calculation. We firstly find
\begin{equation}
 \frac{1}{ 4 Z_\varrho Z_G^2  \varrho }\left( \frac{\sigma_3^2}{g^2} + \frac{\sigma_4^2}{g^2} \right) = \frac{1}{2} \frac{48 (N^2-1)^2}{53 N} \left( 1 - \frac{53}{6} \frac{1}{\epsilon} \frac{N g^2}{16 \pi^2}  - N g^2 \frac{53}{24} \frac{\varrho_1}{(N^2 - 1)^2} \right) \frac{\rho \rho^\dagger}{g^2}\;,
\end{equation}
and secondly
\begin{multline}
\frac{ \alpha' }{4 \alpha' \zeta' - \chi^{\prime 2}}  \frac{\sigma_1^2}{g^2} + \frac{ \zeta' }{ 4 \alpha' \zeta' - \chi^{\prime 2}}  \frac{\sigma_2^2}{g^2}   -  \frac{ \chi' }{4 \alpha' \zeta' - \chi^{\prime 2}}  \frac{\sigma_1 \sigma_2}{g^2} \\= \frac{\zeta_0 m^4}{2 g^2}+ \frac{ \alpha_0 M^4}{g^2} + \frac{1}{\epsilon}\left( \frac{13 N \zeta_0 m^4}{96 \pi ^2}+\frac{M^4 (N^2-1)^2}{4 \pi ^2} \right)  -\frac{\zeta_1 m^4}{2}- M^4 \alpha_1 + M^2 m^2 \chi_1\;,
\end{multline}
so that
\begin{multline}
\Gamma^{(1)}_d = \frac{1}{2} \frac{48 (N^2-1)^2}{53 N} \left( 1 - \frac{53}{6} \frac{1}{\epsilon} \frac{N g^2}{16 \pi^2}  - N g^2 \frac{53}{24} \frac{\varrho_1}{(N^2 - 1)^2} \right) \frac{\rho \rho^\dagger}{g^2} + \frac{\zeta_0 m^4}{2 g^2}+\frac{ \alpha_0 M^4}{g^2}\\
+ \frac{1}{\epsilon}\left( \frac{13 N \zeta_0 m^4}{96 \pi ^2}+\frac{M^4 (N^2-1)^2}{4 \pi ^2} \right)  -\frac{\zeta_1 m^4}{2}- M^4 \alpha_1 + M^2 m^2 \chi_1\;.
\end{multline}
As a check on our results, we see that all the infinities cancel, so we find
\begin{align}
\Gamma^{(1)} &=  \frac{(N^2 -1)^2}{16 \pi^2} \Bigl[(M^2 - \sqrt{\rho \rho^\dagger})^2 \ln\frac{M^2 - \sqrt{\rho \rho^\dagger}}{\overline \mu^2}  + (M^2 + \sqrt{\rho \rho^\dagger})^2 \ln\frac{M^2 + \sqrt{\rho \rho^\dagger}}{\overline \mu^2} \nonumber\\
& - 2 (M^2 + \rho \rho^\dagger)\Bigr] + \frac{3(N^2 -1)}{64\pi^2} \Bigl[  - \frac{5}{6} (m^4 -2 \lambda^4) + y_1^2 \ln \frac{(-y_1)}{\overline \mu} + y_2^2 \ln \frac{(-y_2)}{\overline \mu} + y_3^2 \ln \frac{(-y_3)}{\overline \mu} \nonumber\\
&- y_4^2 \ln \frac{(-y_4)}{\overline \mu} - y_5^2 \ln \frac{(-y_5)}{\overline \mu} \Bigr]  -2 (N^2 - 1) \frac{\lambda^4}{N g^2}  + \frac{3}{2} \frac{\lambda^4}{32 \pi^2} (N^2-1) \nonumber\\
&+ \frac{1}{2} \frac{48 (N^2-1)^2}{53 N} \left( 1   - N g^2 \frac{53}{24} \frac{\varrho_1}{(N^2 - 1)^2} \right) \frac{\rho \rho^\dagger}{g^2} \nonumber\\
&  +  \frac{9}{13} \frac{N^2-1}{N}\frac{m^4}{2g^2}- \frac{24}{35}\frac{(N^2-1)^2}{N}\frac{M^4}{g^2}  - \frac{161}{52} \frac{N^2-1}{16 \pi^2}\frac{ m^4}{2 }- M^4 \alpha_1  +M^2 m^2 \chi_1\;.
\end{align}

\section{The minimum of the effective action}
We shall start from expression \eqref{mineffpot} and derive w.r.t.~$M^2$, $m^2$ and $\lambda^4$. As we would like to know if $M^2 = 0$ can be a minimum of the potential, we further set $M^2 = 0$. We then obtain the following equations
\begin{align}
&\frac{3 \left(\ln \left(m^2-\sqrt{m^4-4 \lambda^4}\right)-\ln \left(m^2+\sqrt{m^4-4 \lambda^4}\right)\right) \lambda^4}{4 \pi ^2 \sqrt{m^4-4 \lambda^4}}+m^2 \chi_1 -\frac{8}{\pi ^2}  = 0 \;, \nonumber\\
& \frac{1}{ \sqrt{m^4-4 \lambda^4}} \Biggl[ 11 \sqrt{m^4 - 4 \lambda^4} (24 \ln 2 -17 ) m^2\nonumber\\
 & +39 \left(-m^4+\sqrt{m^4-4 \lambda^4} m^2+2 \lambda^4\right) \ln \left(\frac{1}{8} \left(m^2-\sqrt{m^4-4 \lambda^4}\right)\right)\nonumber\\
&  +39 \left(m^4+\sqrt{m^4-4 \lambda^4} m^2-2 \lambda^4\right) \ln \left(\frac{1}{8} \left(m^2+\sqrt{m^4-4 \lambda^4}\right)\right) \Biggr] = 0 \;, \nonumber\\
&\frac{ \lambda^2 }{ \sqrt{m^4-4 \lambda^4}} \Biggl[ 9 \left(\sqrt{m^4-4 \lambda^4}-m^2\right)\ln \left(\frac{1}{8} \left(m^2-\sqrt{m^4-4 \lambda^4}\right)\right)+9 \left(m^2+\sqrt{m^4-4 \lambda^4}\right)\nonumber\\
& \times  \ln \left(\frac{1}{8} \left(m^2+\sqrt{m^4-4 \lambda^4}\right)\right)+\sqrt{m^4-4 \lambda^4} (-15+176 \ln 2 ) \Biggr] = 0
\end{align}
whereby we have chosen to set\footnote{We work in units $\lms = 1$.} $\overline \mu = 2 \lms $ and $N = 3$. Now looking at the equation, we see that the second and third equation can be solved exactly for $m^2$ and $\lambda$. There are even multiple solutions possible. We take the solution which has the lowest value for the effective action with $M^2 = 0$. However, for this solution to be a solution of the first equation, these values should be very specific and the chance that they will also satisfy the first equation is practically non-existent, with a certain value of $\chi_1$. Moreover, at a different scale $\overline \mu$, the three equations will look slightly different. However, $\chi_1$ is a number and stays the same. Therefore, it would be necessary at all different scales that these three equations can be solved exactly for only two parameters. This is practically impossible, leading to the conclusion that almost certainly, $M^2 \not= 0$. Said otherwise, it is apparent that the GZ theory dynamically converts itself into a RGZ-like theory. We plan to work this out into future detail in ongoing research.

\providecommand{\href}[2]{#2}\begingroup\raggedright

\endgroup


\begin{thebibliography}{100}

\bibitem{Faddeev:1967fc}
L.~D. Faddeev and V.~N. Popov, ``{Feynman diagrams for the Yang-Mills field},''
\href{http://dx.doi.org/10.1016/0370-2693(67)90067-6}{{\em Phys. Lett.} {\bf
  B25} (1967)  29--30}.

\bibitem{Gribov:1977wm}
V.~N. Gribov, ``{Quantization of non-Abelian gauge theories},''
\href{http://dx.doi.org/10.1016/0550-3213(78)90175-X}{{\em Nucl. Phys.} {\bf
  B139} (1978)  1}.

\bibitem{Zwanziger:1989mf}
D.~Zwanziger, ``{Local and renormalizable action from the Gribov horizon},''
\href{http://dx.doi.org/10.1016/0550-3213(89)90122-3}{{\em Nucl. Phys.} {\bf
  B323} (1989)  513--544}.

\bibitem{Cucchieri:2007md}
A.~Cucchieri and T.~Mendes, ``{What's up with IR gluon and ghost propagators in
  Landau gauge? A puzzling answer from huge lattices},'' {\em PoS} {\bf
  LAT2007} (2007)  297,
\href{http://arxiv.org/abs/0710.0412}{{\tt arXiv:0710.0412 [hep-lat]}}.

\bibitem{Peskin}
M.~E. Peskin and D.~V. Schroeder, ``{An Introduction to quantum field
  theory},''. Reading, USA: Addison-Wesley (1995) 842 p.

\bibitem{Maas:2007uv}
A.~Maas, ``{Two- and three-point Green's functions in two-dimensional
  Landau-gauge Yang-Mills theory},''
  \href{http://dx.doi.org/10.1103/PhysRevD.75.116004}{{\em Phys. Rev.} {\bf
  D75} (2007)  116004},
\href{http://arxiv.org/abs/0704.0722}{{\tt arXiv:0704.0722 [hep-lat]}}.

\bibitem{Cucchieri:2003di}
A.~Cucchieri, T.~Mendes, and A.~R. Taurines, ``{SU(2) Landau gluon propagator
  on a 140$^3$ lattice},''
  \href{http://dx.doi.org/10.1103/PhysRevD.67.091502}{{\em Phys. Rev.} {\bf
  D67} (2003)  091502},
\href{http://arxiv.org/abs/hep-lat/0302022}{{\tt arXiv:hep-lat/0302022}}.

\bibitem{Bogolubsky:2007ud}
I.~L. Bogolubsky, E.~M. Ilgenfritz, M.~Muller-Preussker, and A.~Sternbeck,
  ``{The Landau gauge gluon and ghost propagators in 4D SU(3) gluodynamics in
  large lattice volumes},'' {\em PoS} {\bf LAT2007} (2007)  290,
\href{http://arxiv.org/abs/0710.1968}{{\tt arXiv:0710.1968 [hep-lat]}}.

\bibitem{Piguet1995}
O.~Piguet and S.~P. Sorella, ``{Algebraic renormalization: Perturbative
  renormalization, symmetries and anomalies},''
{\em Lect. Notes Phys.} {\bf M28} (1995)  1--134.

\bibitem{Zinn-Justin2002}
J.~Zinn-Justin, ``{Quantum field theory and critical phenomena},''
{\em Int. Ser. Monogr. Phys.} {\bf 113} (2002)  1--1054.

\bibitem{Ryder}
L.~H. Ryder, ``{Quantum field theory},''. Cambridge, Uk: Univ. Pr. ( 1985)
  443p.

\bibitem{Zee:2003mt}
A.~Zee, ``{Quantum field theory in a nutshell},''. Princeton, UK: Princeton
  Univ. Pr. (2003) 518 p.

\bibitem{Casalbuoni}
R.~Casalbuoni, ``{Advanced Quantum Field},''.

\bibitem{Lehmann:1954rq}
H.~Lehmann, K.~Symanzik, and W.~Zimmermann, ``{On the formulation of quantized
  field theories},''
\href{http://dx.doi.org/10.1007/BF02731765}{{\em Nuovo Cim.} {\bf 1} (1955)
  205--225}.

\bibitem{Lowenstein1971}
J.~H. Lowenstein, ``{Normal product quantization of currents in Lagrangian
  field theory},''
\href{http://dx.doi.org/10.1103/PhysRevD.4.2281}{{\em Phys. Rev.} {\bf D4}
  (1971)  2281--2290}.

\bibitem{Lam1972}
Y.-M.~P. Lam, ``{Perturbation Lagrangian theory for scalar fields: Ward-
  Takahasi identity and current algebra},''
\href{http://dx.doi.org/10.1103/PhysRevD.6.2145}{{\em Phys. Rev.} {\bf D6}
  (1972)  2145--2161}.

\bibitem{Clark1976}
T.~E. Clark and J.~H. Lowenstein, ``{Generalization of Zimmermann's
  Normal-Product Identity},''
\href{http://dx.doi.org/10.1016/0550-3213(76)90457-0}{{\em Nucl. Phys.} {\bf
  B113} (1976)  109}.

\bibitem{Lowenstein:1971jk}
J.~H. Lowenstein, ``{Differential vertex operations in Lagrangian field
  theory},''
\href{http://dx.doi.org/10.1007/BF01907030}{{\em Commun. Math. Phys.} {\bf 24}
  (1971)  1--21}.

\bibitem{Lam:1973qa}
Y.-M.~P. Lam, ``{Equivalence theorem on Bogolyubov-Parasiuk-Hepp-Zimmermann
  renormalized Lagrangian field theories},''
\href{http://dx.doi.org/10.1103/PhysRevD.7.2943}{{\em Phys. Rev.} {\bf D7}
  (1973)  2943--2949}.

\bibitem{Haussling}
R.~Häussling, ``{Quantum Action Principle and Zimmermann identities},''. Lect.
  Notes from Conference held at Hesselberg Academy, {Theory of Renormalization
  and Regularization}.

\bibitem{'tHooft:2005cq}
G.~'t~Hooft, (Ed.~), ``{50 Years of Yang-Mills Theory},''.

\bibitem{Nakishi:1966di}
N.~Nakishi, ``{Covariant quantization of the electromagnetic field in the
  Landau gauge},''
{\em Prog. Theor. Phys.} {\bf 35} (1966)  1111.

\bibitem{Becchi:1974xu}
C.~Becchi, A.~Rouet, and R.~Stora, ``{The Abelian Higgs-Kibble Model. Unitarity
  of the S Operator},''
\href{http://dx.doi.org/10.1016/0370-2693(74)90058-6}{{\em Phys. Lett.} {\bf
  B52} (1974)  344}.

\bibitem{Becchi:1996yh}
C.~Becchi, ``{Introduction to BRS symmetry},''
\href{http://arxiv.org/abs/hep-th/9607181}{{\tt arXiv:hep-th/9607181}}.

\bibitem{Kraus:1997bi}
E.~Kraus, ``{Renormalization of the electroweak standard model to all
  orders},'' \href{http://dx.doi.org/10.1006/aphy.1997.5746}{{\em Annals Phys.}
  {\bf 262} (1998)  155--259},
\href{http://arxiv.org/abs/hep-th/9709154}{{\tt arXiv:hep-th/9709154}}.

\bibitem{Zwanziger:1992qr}
D.~Zwanziger, ``{Renormalizability of the critical limit of lattice gauge
  theory by BRS invariance},''
\href{http://dx.doi.org/10.1016/0550-3213(93)90506-K}{{\em Nucl. Phys.} {\bf
  B399} (1993)  477--513}.

\bibitem{DeWitt:1964yg}
B.~S. DeWitt, ``{Theory of radiative corrections for non-abelian gauge
  fields},''
\href{http://dx.doi.org/10.1103/PhysRevLett.12.742}{{\em Phys. Rev. Lett.} {\bf
  12} (1964)  742--746}.

\bibitem{DeWitt:1967yk}
B.~S. DeWitt, ``{Quantum Theory of Gravity. 1. The Canonical Theory},''
\href{http://dx.doi.org/10.1103/PhysRev.160.1113}{{\em Phys. Rev.} {\bf 160}
  (1967)  1113--1148}.

\bibitem{Weinberg:1996kr}
S.~Weinberg, ``{The quantum theory of fields. Vol. 2: Modern applications},''.
  Cambridge, UK: Univ. Pr. (1996) 489 p.

\bibitem{Masujima}
M.~Masujima, ``{Path Integral Quantization and Stochastic Quantization},'' {\em
  Springer Tracts in Modern Physics} {\bf 165}  .

\bibitem{thooft2007}
G.~'t~Hooft, {\em {The Conceptual Basis of Quantum Field Theory}}.
\newblock Elsevier, 2007.

\bibitem{Riley}
K.~Riley, M.~Hobson, and S.~Bence, {\em {Mathematical Methods for Physics and
  Engineering}}.
\newblock University of Cambridge, 1998.

\bibitem{bach}
V.~Bach, ``{Quantization of Nonabelian Gauge Fields and BRST Formalism},''.

\bibitem{Pokorski}
S.~Pokorski, {\em {Gauge Field Theories}}.
\newblock {University of Cambridge}, 2000.

\bibitem{'tHooft:1971fh}
G.~'t~Hooft, ``{Renormalization of Massless Yang-Mills Fields},''
\href{http://dx.doi.org/10.1016/0550-3213(71)90395-6}{{\em Nucl. Phys.} {\bf
  B33} (1971)  173--199}.

\bibitem{DeWitt:1967ub}
B.~S. DeWitt, ``{Quantum theory of gravity. II. The manifestly covariant
  theory},''
\href{http://dx.doi.org/10.1103/PhysRev.162.1195}{{\em Phys. Rev.} {\bf 162}
  (1967)  1195--1239}.

\bibitem{Kleinert:2006}
H.~Kleinert, ``{Particles and Quantum Fields },''. Lecture notes, 2006.

\bibitem{Leibbrandt:1987qv}
G.~Leibbrandt, ``{Introduction to Noncovariant Gauges},''
\href{http://dx.doi.org/10.1103/RevModPhys.59.1067}{{\em Rev. Mod. Phys.} {\bf
  59} (1987)  1067}.

\bibitem{Shinohara:2001cw}
T.~Shinohara, T.~Imai, and K.-I. Kondo, ``{The most general and renormalizable
  maximal Abelian gauge},''
  \href{http://dx.doi.org/10.1142/S0217751X03016008}{{\em Int. J. Mod. Phys.}
  {\bf A18} (2003)  5733--5756},
\href{http://arxiv.org/abs/hep-th/0105268}{{\tt arXiv:hep-th/0105268}}.

\bibitem{Klaus}
K.~Lichtenegger, {\em {Aspects of Confinement in a Functional Approach to
  Coulomb Gauge QCD}}.
\newblock PhD thesis, {Karl-Franzens Universit\"at Graz}, 2010.

\bibitem{Slavnov:1989jh}
A.~A. Slavnov, ``{Physical unitarity in the BRST approach},''
\href{http://dx.doi.org/10.1016/0370-2693(89)91521-9}{{\em Phys. Lett.} {\bf
  B217} (1989)  91--94}.

\bibitem{Frolov:1989az}
S.~A. Frolov and A.~A. Slavnov, ``{Construction of the effective action for
  general gauge theories via unitarity},''
\href{http://dx.doi.org/10.1016/0550-3213(90)90562-R}{{\em Nucl. Phys.} {\bf
  B347} (1990)  333--346}.

\bibitem{Henneaux:1992ig}
M.~Henneaux and C.~Teitelboim, ``{Quantization of gauge systems},''. Princeton,
  USA: Univ. Pr. (1992) 520 p.

\bibitem{Dudal:2007ch}
D.~Dudal, N.~Vandersickel, and H.~Verschelde, ``{Unitarity analysis of a
  non-Abelian gauge invariant action with a mass},''
  \href{http://dx.doi.org/10.1103/PhysRevD.76.025006}{{\em Phys. Rev.} {\bf
  D76} (2007)  025006},
\href{http://arxiv.org/abs/0705.0871}{{\tt arXiv:0705.0871 [hep-th]}}.

\bibitem{Baulieu:1983tg}
L.~Baulieu, ``{Perturbative Gauge Theories},''
  \href{http://dx.doi.org/10.1016/0370-1573(85)90091-2}{{\em Phys.Rept.} {\bf
  129} (1985)  1}.

\bibitem{Nakanishi:1990qm}
N.~Nakanishi and I.~Ojima, ``{Covariant operator formalism of gauge theories
  and quantum gravity},''
{\em World Sci. Lect. Notes Phys.} {\bf 27} (1990)  1--434.

\bibitem{Sobreiro:2004us}
R.~F. Sobreiro, S.~P. Sorella, D.~Dudal, and H.~Verschelde, ``{Gribov horizon
  in the presence of dynamical mass generation in Euclidean Yang-Mills theories
  in the Landau gauge},''
  \href{http://dx.doi.org/10.1016/j.physletb.2004.03.084}{{\em Phys. Lett.}
  {\bf B590} (2004)  265--272},
\href{http://arxiv.org/abs/hep-th/0403135}{{\tt arXiv:hep-th/0403135}}.

\bibitem{Henyey:1978qd}
F.~S. Henyey, ``{Gribov ambiguity without topological charge},''
\href{http://dx.doi.org/10.1103/PhysRevD.20.1460}{{\em Phys. Rev.} {\bf D20}
  (1979)  1460}.

\bibitem{vanBaal:1991zw}
P.~van Baal, ``{More (thoughts on) Gribov copies},''
\href{http://dx.doi.org/10.1016/0550-3213(92)90386-P}{{\em Nucl. Phys.} {\bf
  B369} (1992)  259--275}.

\bibitem{Semenov1986}
M.~Semenov-Tyan-Shanskii and V.~Franke, ``{A variational principle for the
  Lorentz condition and restriction of the domain ofpath integration in
  non-abelian gauge theory},'' {\em Zap. Nauch. Sem. Leningrad. Otdeleniya
  Matematicheskogo Instituta im V. A. Steklov, AN SSSR} {\bf 120} (1982)  p
  159. (English translation: New York: Plenum Press 1986).

\bibitem{Maskawa:1978rf}
T.~Maskawa and H.~Nakajima, ``{How dense is the Coulomb gauge fixing
  degeneracies? Geometrical formulation of the Coulomb gauge},''
\href{http://dx.doi.org/10.1143/PTP.60.1526}{{\em Prog. Theor. Phys.} {\bf 60}
  (1978)  1526}.

\bibitem{Zwanziger1982a}
D.~Zwanziger, ``{Nonperturbative modification of the Faddeev-Popov formula and
  banishment of the naive vacuum},''
\href{http://dx.doi.org/10.1016/0550-3213(82)90260-7}{{\em Nucl. Phys.} {\bf
  B209} (1982)  336}.

\bibitem{Capri:2005dy}
M.~A.~L. Capri {\em et al.}, ``{A study of the gauge invariant, nonlocal mass
  operator $\Tr \int \d^4 x F_{\mu\nu}\frac{1}{D^2} F_{\mu\nu}$ in Yang-Mills
  theories},'' \href{http://dx.doi.org/10.1103/PhysRevD.72.105016}{{\em Phys.
  Rev.} {\bf D72} (2005)  105016},
\href{http://arxiv.org/abs/hep-th/0510240}{{\tt arXiv:hep-th/0510240}}.

\bibitem{Dell'Antonio:1991xt}
G.~Dell'Antonio and D.~Zwanziger, ``{Every gauge orbit passes inside the Gribov
  horizon},''
\href{http://dx.doi.org/10.1007/BF02099494}{{\em Commun. Math. Phys.} {\bf 138}
  (1991)  291--299}.

\bibitem{Zwanziger:2003cf}
D.~Zwanziger, ``{Non-perturbative Faddeev-Popov formula and infrared limit of
  QCD},'' \href{http://dx.doi.org/10.1103/PhysRevD.69.016002}{{\em Phys. Rev.}
  {\bf D69} (2004)  016002},
\href{http://arxiv.org/abs/hep-ph/0303028}{{\tt arXiv:hep-ph/0303028}}.

\bibitem{Dell'Antonio:1989jn}
G.~Dell'Antonio and D.~Zwanziger, ``{Ellipsoidal bound on the Gribov horizon
  contradicts the perturbative renormalization group},''
\href{http://dx.doi.org/10.1016/0550-3213(89)90135-1}{{\em Nucl. Phys.} {\bf
  B326} (1989)  333--350}.

\bibitem{Cucchieri:1997dx}
A.~Cucchieri, ``{Gribov copies in the minimal Landau gauge: The influence on
  gluon and ghost propagators},''
  \href{http://dx.doi.org/10.1016/S0550-3213(97)00629-9}{{\em Nucl. Phys.} {\bf
  B508} (1997)  353--370},
\href{http://arxiv.org/abs/hep-lat/9705005}{{\tt arXiv:hep-lat/9705005}}.

\bibitem{Zwanziger:1993dh}
D.~Zwanziger, ``{Fundamental modular region, Boltzmann factor and area law in
  lattice gauge theory},''
\href{http://dx.doi.org/10.1016/0550-3213(94)90396-4}{{\em Nucl. Phys.} {\bf
  B412} (1994)  657--730}.

\bibitem{Singer:1978dk}
I.~M. Singer, ``{Some Remarks on the Gribov Ambiguity},''
\href{http://dx.doi.org/10.1007/BF01609471}{{\em Commun. Math. Phys.} {\bf 60}
  (1978)  7--12}.

\bibitem{Bassetto:1983rq}
A.~Bassetto, I.~Lazzizzera, and R.~Soldati, ``{Absence of Gribov Copies in the
  space-like planar gauge},''
\href{http://dx.doi.org/10.1016/0370-2693(83)91115-2}{{\em Phys. Lett.} {\bf
  B131} (1983)  177}.

\bibitem{Ghiotti:2005ih}
M.~Ghiotti, A.~C. Kalloniatis, and A.~G. Williams, ``{Landau gauge Jacobian and
  BRST symmetry},''
  \href{http://dx.doi.org/10.1016/j.physletb.2005.09.015}{{\em Phys. Lett.}
  {\bf B628} (2005)  176--182},
\href{http://arxiv.org/abs/hep-th/0509053}{{\tt arXiv:hep-th/0509053}}.

\bibitem{Slavnov:2008xz}
A.~A. Slavnov, ``{A Lorentz invariant formulation of the Yang-Mills theory with
  gauge invariant ghost field Lagrangian},''
  \href{http://dx.doi.org/10.1088/1126-6708/2008/08/047}{{\em JHEP} {\bf 08}
  (2008)  047},
\href{http://arxiv.org/abs/0807.1795}{{\tt arXiv:0807.1795 [hep-th]}}.

\bibitem{Quadri:2010vt}
A.~Quadri and A.~Slavnov, ``{Renormalization of the Yang-Mills theory in the
  ambiguity-free gauge},''
  \href{http://dx.doi.org/10.1007/JHEP07(2010)087}{{\em JHEP} {\bf 1007} (2010)
   087}, \href{http://arxiv.org/abs/1002.2490}{{\tt arXiv:1002.2490 [hep-th]}}.

\bibitem{Zwanziger:1981kg}
D.~Zwanziger, ``{Covariant Quantization of Gauge Fields Without Gribov
  Ambiguity},''
\href{http://dx.doi.org/10.1016/0550-3213(81)90202-9}{{\em Nucl. Phys.} {\bf
  B192} (1981)  259}.

\bibitem{Parisi:1980ys}
G.~Parisi and Y.-s. Wu, ``{Perturbation Theory Without Gauge Fixing},''
{\em Sci. Sin.} {\bf 24} (1981)  483.

\bibitem{Hirschfeld:1978yq}
P.~Hirschfeld, ``{Strong evidence that Gribov copying does not affect gauge
  theory functional integral},''
\href{http://dx.doi.org/10.1016/0550-3213(79)90052-X}{{\em Nucl. Phys.} {\bf
  B157} (1979)  37}.

\bibitem{Friedberg:1995ty}
R.~Friedberg, T.~D. Lee, Y.~Pang, and H.~C. Ren, ``{A soluble gauge model with
  Gribov type copies},''
\href{http://dx.doi.org/10.1006/aphy.1996.0032}{{\em Ann. Phys.} {\bf 246}
  (1996)  381--445}.

\bibitem{Greensite:2004ke}
J.~Greensite, S.~Olejnik, and D.~Zwanziger, ``{Coulomb energy, remnant
  symmetry, and the phases of non- Abelian gauge theories},''
  \href{http://dx.doi.org/10.1103/PhysRevD.69.074506}{{\em Phys. Rev.} {\bf
  D69} (2004)  074506},
\href{http://arxiv.org/abs/hep-lat/0401003}{{\tt arXiv:hep-lat/0401003}}.

\bibitem{private}
{Private communication with D. Zwanziger.}

\bibitem{Ezawa:1982bf}
Z.~F. Ezawa and A.~Iwazaki, ``{Abelian Dominance and Quark Confinement in
  Yang-Mills Theories},''
\href{http://dx.doi.org/10.1103/PhysRevD.25.2681}{{\em Phys. Rev.} {\bf D25}
  (1982)  2681}.

\bibitem{Cucchieri:1997fy}
A.~Cucchieri, ``{Infrared behavior of the gluon propagator in lattice Landau
  gauge},'' \href{http://dx.doi.org/10.1016/S0370-2693(98)00046-X}{{\em Phys.
  Lett.} {\bf B422} (1998)  233--237},
\href{http://arxiv.org/abs/hep-lat/9709015}{{\tt arXiv:hep-lat/9709015}}.

\bibitem{Cucchieri:1999sz}
A.~Cucchieri, ``{Infrared behavior of the gluon propagator in lattice Landau
  gauge: The three-dimensional case},''
  \href{http://dx.doi.org/10.1103/PhysRevD.60.034508}{{\em Phys. Rev.} {\bf
  D60} (1999)  034508},
\href{http://arxiv.org/abs/hep-lat/9902023}{{\tt arXiv:hep-lat/9902023}}.

\bibitem{Oliveira:2004gy}
O.~Oliveira and P.~J. Silva, ``{The infrared Landau gauge gluon propagator from
  lattice QCD},'' \href{http://dx.doi.org/10.1063/1.1920970}{{\em AIP Conf.
  Proc.} {\bf 756} (2005)  290--292},
\href{http://arxiv.org/abs/hep-lat/0410048}{{\tt arXiv:hep-lat/0410048}}.

\bibitem{Bloch:2003sk}
J.~C.~R. Bloch, A.~Cucchieri, K.~Langfeld, and T.~Mendes, ``{Propagators and
  running coupling from SU(2) lattice gauge theory},''
  \href{http://dx.doi.org/10.1016/j.nuclphysb.2004.03.021}{{\em Nucl. Phys.}
  {\bf B687} (2004)  76--100},
\href{http://arxiv.org/abs/hep-lat/0312036}{{\tt arXiv:hep-lat/0312036}}.

\bibitem{Furui:2003jr}
S.~Furui and H.~Nakajima, ``{Infrared features of the Landau gauge QCD},''
  \href{http://dx.doi.org/10.1103/PhysRevD.69.074505}{{\em Phys. Rev.} {\bf
  D69} (2004)  074505},
\href{http://arxiv.org/abs/hep-lat/0305010}{{\tt arXiv:hep-lat/0305010}}.

\bibitem{Cucchieri:2010xr}
A.~Cucchieri and T.~Mendes, ``{Numerical test of the Gribov-Zwanziger scenario
  in Landau gauge},''
\href{http://arxiv.org/abs/1001.2584}{{\tt arXiv:1001.2584 [hep-lat]}}.

\bibitem{Zwanziger1989}
D.~Zwanziger, ``{Action from the Gribov horizon},''
\href{http://dx.doi.org/10.1016/0550-3213(89)90263-0}{{\em Nucl. Phys.} {\bf
  B321} (1989)  591}.

\bibitem{munster}
A.~M\"unster, {\em Statistical Thermodynamics, Volume 1}.
\newblock Springer-Verlag Academic Press, 1969.

\bibitem{private2}
{Private communication with D. Zwanziger, A. Cucchieri.}

\bibitem{Ford:2009ar}
F.~R. Ford and J.~A. Gracey, ``{Two loop MSbar Gribov mass gap equation with
  massive quarks},''
  \href{http://dx.doi.org/10.1088/1751-8113/42/32/325402}{{\em J. Phys.} {\bf
  A42} (2009)  325402},
\href{http://arxiv.org/abs/0906.3222}{{\tt arXiv:0906.3222 [hep-th]}}.

\bibitem{Gracey:2005cx}
J.~A. Gracey, ``{Two loop correction to the Gribov mass gap equation in Landau
  gauge QCD},'' \href{http://dx.doi.org/10.1016/j.physletb.2005.10.046}{{\em
  Phys. Lett.} {\bf B632} (2006)  282--286},
\href{http://arxiv.org/abs/hep-ph/0510151}{{\tt arXiv:hep-ph/0510151}}.

\bibitem{Dudal:2010fq}
D.~Dudal, S.~P. Sorella, and N.~Vandersickel, ``{More on the renormalization of
  the horizon function of the Gribov-Zwanziger action and the Kugo-Ojima Green
  function(s)},'' \href{http://dx.doi.org/10.1140/epjc/s10052-010-1304-8}{{\em
  Eur. Phys. J.} {\bf C68} (2010)  283--298},
\href{http://arxiv.org/abs/1001.3103}{{\tt arXiv:1001.3103 [hep-th]}}.

\bibitem{Capri:2010hb}
M.~Capri, A.~Gomez, M.~Guimaraes, V.~Lemes, S.~Sorella, {\em et al.}, ``{A
  remark on the BRST symmetry in the Gribov-Zwanziger theory},''
  \href{http://dx.doi.org/10.1103/PhysRevD.82.105019}{{\em Phys.Rev.} {\bf D82}
  (2010)  105019}, \href{http://arxiv.org/abs/1009.4135}{{\tt arXiv:1009.4135
  [hep-th]}}.

\bibitem{Dudal:2008sp}
D.~Dudal, J.~A. Gracey, S.~P. Sorella, N.~Vandersickel, and H.~Verschelde, ``{A
  refinement of the Gribov-Zwanziger approach in the Landau gauge: infrared
  propagators in harmony with the lattice results},''
  \href{http://dx.doi.org/10.1103/PhysRevD.78.065047}{{\em Phys. Rev.} {\bf
  D78} (2008)  065047},
\href{http://arxiv.org/abs/0806.4348}{{\tt arXiv:0806.4348 [hep-th]}}.

\bibitem{Gomez:2009tj}
A.~J. Gomez, M.~S. Guimaraes, R.~F. Sobreiro, and S.~P. Sorella, ``{Equivalence
  between Zwanziger's horizon function and Gribov's no-pole ghost form
  factor},'' \href{http://dx.doi.org/10.1016/j.physletb.2009.12.001}{{\em Phys.
  Lett.} {\bf B683} (2010)  217--221},
\href{http://arxiv.org/abs/0910.3596}{{\tt arXiv:0910.3596 [hep-th]}}.

\bibitem{Maggiore:1993wq}
N.~Maggiore and M.~Schaden, ``{Landau gauge within the Gribov horizon},''
  \href{http://dx.doi.org/10.1103/PhysRevD.50.6616}{{\em Phys. Rev.} {\bf D50}
  (1994)  6616--6625},
\href{http://arxiv.org/abs/hep-th/9310111}{{\tt arXiv:hep-th/9310111}}.

\bibitem{Dudal:2005na}
D.~Dudal, R.~F. Sobreiro, S.~P. Sorella, and H.~Verschelde, ``{The Gribov
  parameter and the dimension two gluon condensate in Euclidean Yang-Mills
  theories in the Landau gauge},''
  \href{http://dx.doi.org/10.1103/PhysRevD.72.014016}{{\em Phys. Rev.} {\bf
  D72} (2005)  014016},
\href{http://arxiv.org/abs/hep-th/0502183}{{\tt arXiv:hep-th/0502183}}.

\bibitem{Dudal:2010hj}
D.~Dudal and N.~Vandersickel, ``{On the reanimation of a local BRST invariance
  in the (Refined) Gribov-Zwanziger formalism},''
\href{http://arxiv.org/abs/1010.3927}{{\tt arXiv:1010.3927 [hep-th]}}.

\bibitem{Sorella:2009vt}
S.~P. Sorella, ``{Gribov horizon and BRST symmetry: a few remarks},''
  \href{http://dx.doi.org/10.1103/PhysRevD.80.025013}{{\em Phys. Rev.} {\bf
  D80} (2009)  025013},
\href{http://arxiv.org/abs/0905.1010}{{\tt arXiv:0905.1010 [hep-th]}}.

\bibitem{Kondo:2009qz}
K.-I. Kondo, ``{The nilpotent 'BRST' symmetry for the Gribov-Zwanziger
  theory},''
\href{http://arxiv.org/abs/0905.1899}{{\tt arXiv:0905.1899 [hep-th]}}.

\bibitem{Kondo:2009wk}
K.-I. Kondo, ``{Decoupling and scaling solutions in Yang-Mills theory with the
  Gribov horizon},''
\href{http://arxiv.org/abs/0909.4866}{{\tt arXiv:0909.4866 [hep-th]}}.

\bibitem{Kugo:1979gm}
T.~Kugo and I.~Ojima, ``{Local Covariant Operator Formalism of Nonabelian Gauge
  Theories and Quark Confinement Problem},''
{\em Prog. Theor. Phys. Suppl.} {\bf 66} (1979)  1.

\bibitem{Kugo:1995km}
T.~Kugo, ``{The universal renormalization factors Z(1) / Z(3) and color
  confinement condition in non-Abelian gauge theory},''
\href{http://arxiv.org/abs/hep-th/9511033}{{\tt arXiv:hep-th/9511033}}.

\bibitem{Kondo:2009ug}
K.-I. Kondo, ``{Kugo-Ojima color confinement criterion and Gribov- Zwanziger
  horizon condition},''
  \href{http://dx.doi.org/10.1016/j.physletb.2009.06.026}{{\em Phys. Lett.}
  {\bf B678} (2009)  322--330},
\href{http://arxiv.org/abs/0904.4897}{{\tt arXiv:0904.4897 [hep-th]}}.

\bibitem{Aguilar:2009nf}
A.~C. Aguilar, D.~Binosi, J.~Papavassiliou, and J.~Rodriguez-Quintero,
  ``{Non-perturbative comparison of QCD effective charges},''
  \href{http://dx.doi.org/10.1103/PhysRevD.80.085018}{{\em Phys. Rev.} {\bf
  D80} (2009)  085018},
\href{http://arxiv.org/abs/0906.2633}{{\tt arXiv:0906.2633 [hep-ph]}}.

\bibitem{Boucaud:2009sd}
P.~Boucaud {\em et al.}, ``{Gribov's horizon and the ghost dressing
  function},'' \href{http://dx.doi.org/10.1103/PhysRevD.80.094501}{{\em Phys.
  Rev.} {\bf D80} (2009)  094501},
\href{http://arxiv.org/abs/0909.2615}{{\tt arXiv:0909.2615 [hep-ph]}}.

\bibitem{Fischer:2008uz}
C.~S. Fischer, A.~Maas, and J.~M. Pawlowski, ``{On the infrared behavior of
  Landau gauge Yang-Mills theory},''
  \href{http://dx.doi.org/10.1016/j.aop.2009.07.009}{{\em Annals Phys.} {\bf
  324} (2009)  2408--2437},
\href{http://arxiv.org/abs/0810.1987}{{\tt arXiv:0810.1987 [hep-ph]}}.

\bibitem{Dudal:2009xh}
D.~Dudal, S.~P. Sorella, N.~Vandersickel, and H.~Verschelde, ``{Gribov no-pole
  condition, Zwanziger horizon function, Kugo-Ojima confinement criterion,
  boundary conditions, BRST breaking and all that},''
  \href{http://dx.doi.org/10.1103/PhysRevD.79.121701}{{\em Phys. Rev.} {\bf
  D79} (2009)  121701},
\href{http://arxiv.org/abs/0904.0641}{{\tt arXiv:0904.0641 [hep-th]}}.

\bibitem{Cucchieri:2004mf}
A.~Cucchieri, T.~Mendes, and A.~R. Taurines, ``{Positivity violation for the
  lattice Landau gluon propagator},''
  \href{http://dx.doi.org/10.1103/PhysRevD.71.051902}{{\em Phys. Rev.} {\bf
  D71} (2005)  051902},
\href{http://arxiv.org/abs/hep-lat/0406020}{{\tt arXiv:hep-lat/0406020}}.

\bibitem{Sternbeck:2004xr}
A.~Sternbeck, E.~M. Ilgenfritz, M.~Muller-Preussker, and A.~Schiller, ``{The
  gluon and ghost propagator and the influence of Gribov copies},''
  \href{http://dx.doi.org/10.1016/j.nuclphysbps.2004.11.154}{{\em Nucl. Phys.
  Proc. Suppl.} {\bf 140} (2005)  653--655},
\href{http://arxiv.org/abs/hep-lat/0409125}{{\tt arXiv:hep-lat/0409125}}.

\bibitem{Alkofer:2000wg}
R.~Alkofer and L.~von Smekal, ``{The infrared behavior of QCD Green's
  functions: Confinement, dynamical symmetry breaking, and hadrons as
  relativistic bound states},''
  \href{http://dx.doi.org/10.1016/S0370-1573(01)00010-2}{{\em Phys. Rept.} {\bf
  353} (2001)  281},
\href{http://arxiv.org/abs/hep-ph/0007355}{{\tt arXiv:hep-ph/0007355}}.

\bibitem{Lerche:2002ep}
C.~Lerche and L.~von Smekal, ``{On the infrared exponent for gluon and ghost
  propagation in Landau gauge QCD},''
  \href{http://dx.doi.org/10.1103/PhysRevD.65.125006}{{\em Phys. Rev.} {\bf
  D65} (2002)  125006},
\href{http://arxiv.org/abs/hep-ph/0202194}{{\tt arXiv:hep-ph/0202194}}.

\bibitem{Pawlowski:2003hq}
J.~M. Pawlowski, D.~F. Litim, S.~Nedelko, and L.~von Smekal, ``{Infrared
  behaviour and fixed points in Landau gauge QCD},''
  \href{http://dx.doi.org/10.1103/PhysRevLett.93.152002}{{\em Phys. Rev. Lett.}
  {\bf 93} (2004)  152002},
\href{http://arxiv.org/abs/hep-th/0312324}{{\tt arXiv:hep-th/0312324}}.

\bibitem{Alkofer:2003jj}
R.~Alkofer, W.~Detmold, C.~S. Fischer, and P.~Maris, ``{Analytic properties of
  the Landau gauge gluon and quark propagators},''
  \href{http://dx.doi.org/10.1103/PhysRevD.70.014014}{{\em Phys. Rev.} {\bf
  D70} (2004)  014014},
\href{http://arxiv.org/abs/hep-ph/0309077}{{\tt arXiv:hep-ph/0309077}}.

\bibitem{Cucchieri:2007rg}
A.~Cucchieri and T.~Mendes, ``{Constraints on the IR behavior of the gluon
  propagator in Yang-Mills theories},''
  \href{http://dx.doi.org/10.1103/PhysRevLett.100.241601}{{\em Phys. Rev.
  Lett.} {\bf 100} (2008)  241601},
\href{http://arxiv.org/abs/0712.3517}{{\tt arXiv:0712.3517 [hep-lat]}}.

\bibitem{Cucchieri:2008fc}
A.~Cucchieri and T.~Mendes, ``{Constraints on the IR behavior of the ghost
  propagator in Yang-Mills theories},''
  \href{http://dx.doi.org/10.1103/PhysRevD.78.094503}{{\em Phys. Rev.} {\bf
  D78} (2008)  094503},
\href{http://arxiv.org/abs/0804.2371}{{\tt arXiv:0804.2371 [hep-lat]}}.

\bibitem{Sternbeck:2007ug}
A.~Sternbeck, L.~von Smekal, D.~B. Leinweber, and A.~G. Williams, ``{Comparing
  SU(2) to SU(3) gluodynamics on large lattices},'' {\em PoS} {\bf LAT2007}
  (2007)  340,
\href{http://arxiv.org/abs/0710.1982}{{\tt arXiv:0710.1982 [hep-lat]}}.

\bibitem{Bogolubsky:2009dc}
I.~L. Bogolubsky, E.~M. Ilgenfritz, M.~Muller-Preussker, and A.~Sternbeck,
  ``{Lattice gluodynamics computation of Landau gauge Green's functions in the
  deep infrared},''
  \href{http://dx.doi.org/10.1016/j.physletb.2009.04.076}{{\em Phys. Lett.}
  {\bf B676} (2009)  69--73},
\href{http://arxiv.org/abs/0901.0736}{{\tt arXiv:0901.0736 [hep-lat]}}.

\bibitem{Bornyakov:2008yx}
V.~G. Bornyakov, V.~K. Mitrjushkin, and M.~Muller-Preussker, ``{Infrared
  behavior and Gribov ambiguity in SU(2) lattice gauge theory},''
  \href{http://dx.doi.org/10.1103/PhysRevD.79.074504}{{\em Phys. Rev.} {\bf
  D79} (2009)  074504},
\href{http://arxiv.org/abs/0812.2761}{{\tt arXiv:0812.2761 [hep-lat]}}.

\bibitem{Gong:2008td}
M.~Gong, Y.~Chen, G.~Meng, and C.~Liu, ``{Lattice Gluon Propagator in the
  Landau Gauge: A Study Using Anisotropic Lattices},''
  \href{http://dx.doi.org/10.1142/S021773230903031X}{{\em Mod. Phys. Lett.}
  {\bf A24} (2009)  1925--1935},
\href{http://arxiv.org/abs/0811.4635}{{\tt arXiv:0811.4635 [hep-lat]}}.

\bibitem{Dudal:2007cw}
D.~Dudal, S.~P. Sorella, N.~Vandersickel, and H.~Verschelde, ``{New features of
  the gluon and ghost propagator in the infrared region from the
  Gribov-Zwanziger approach},''
  \href{http://dx.doi.org/10.1103/PhysRevD.77.071501}{{\em Phys. Rev.} {\bf
  D77} (2008)  071501},
\href{http://arxiv.org/abs/0711.4496}{{\tt arXiv:0711.4496 [hep-th]}}.

\bibitem{Dudal:2008rm}
D.~Dudal, J.~A. Gracey, S.~P. Sorella, N.~Vandersickel, and H.~Verschelde,
  ``{The Landau gauge gluon and ghost propagator in the refined
  Gribov-Zwanziger framework in 3 dimensions},''
  \href{http://dx.doi.org/10.1103/PhysRevD.78.125012}{{\em Phys. Rev.} {\bf
  D78} (2008)  125012},
\href{http://arxiv.org/abs/0808.0893}{{\tt arXiv:0808.0893 [hep-th]}}.

\bibitem{Dudal:2008xd}
D.~Dudal, S.~P. Sorella, N.~Vandersickel, and H.~Verschelde, ``{The effects of
  Gribov copies in 2D gauge theories},''
  \href{http://dx.doi.org/10.1016/j.physletb.2009.08.055}{{\em Phys. Lett.}
  {\bf B680} (2009)  377--383},
\href{http://arxiv.org/abs/0808.3379}{{\tt arXiv:0808.3379 [hep-th]}}.

\bibitem{Dudal:2010tf}
D.~Dudal, O.~Oliveira, and N.~Vandersickel, ``{Indirect lattice evidence for
  the Refined Gribov-Zwanziger formalism and the gluon condensate
  $\braket{A^2}$ in the Landau gauge},''
  \href{http://dx.doi.org/10.1103/PhysRevD.81.074505}{{\em Phys. Rev.} {\bf
  D81} (2010)  074505},
\href{http://arxiv.org/abs/1002.2374}{{\tt arXiv:1002.2374 [hep-lat]}}.

\bibitem{Binosi:2009qm}
D.~Binosi and J.~Papavassiliou, ``{Pinch Technique: Theory and Applications},''
  \href{http://dx.doi.org/10.1016/j.physrep.2009.05.001}{{\em Phys. Rept.} {\bf
  479} (2009)  1--152},
\href{http://arxiv.org/abs/0909.2536}{{\tt arXiv:0909.2536 [hep-ph]}}.

\bibitem{Boucaud:2008ky}
P.~Boucaud {\em et al.}, ``{On the IR behaviour of the Landau-gauge ghost
  propagator},'' \href{http://dx.doi.org/10.1088/1126-6708/2008/06/099}{{\em
  JHEP} {\bf 06} (2008)  099},
\href{http://arxiv.org/abs/0803.2161}{{\tt arXiv:0803.2161 [hep-ph]}}.

\bibitem{Shifman:1978bx}
M.~A. Shifman, A.~I. Vainshtein, and V.~I. Zakharov, ``{QCD and Resonance
  Physics. Sum Rules},''
\href{http://dx.doi.org/10.1016/0550-3213(79)90022-1}{{\em Nucl. Phys.} {\bf
  B147} (1979)  385--447}.

\bibitem{Greensite:1985vq}
J.~Greensite and M.~B. Halpern, ``{Variational computation of glueball masses
  in continuum QCD},''
{\em Nucl. Phys.} {\bf B271} (1986)  379.

\bibitem{Lavelle:1988eg}
M.~J. Lavelle and M.~Schaden, ``{Propagators and condensates in QCD},''
\href{http://dx.doi.org/10.1016/0370-2693(88)90433-9}{{\em Phys. Lett.} {\bf
  B208} (1988)  297}.

\bibitem{Gubarev:2000eu}
F.~V. Gubarev, L.~Stodolsky, and V.~I. Zakharov, ``{On the significance of the
  quantity $A^2$},'' \href{http://dx.doi.org/10.1103/PhysRevLett.86.2220}{{\em
  Phys. Rev. Lett.} {\bf 86} (2001)  2220--2222},
\href{http://arxiv.org/abs/hep-ph/0010057}{{\tt arXiv:hep-ph/0010057}}.

\bibitem{Gubarev:2000nz}
F.~V. Gubarev and V.~I. Zakharov, ``{On the emerging phenomenology of
  $<(A^a_\mu)^2_\mini>$},''
  \href{http://dx.doi.org/10.1016/S0370-2693(01)00085-5}{{\em Phys. Lett.} {\bf
  B501} (2001)  28--36},
\href{http://arxiv.org/abs/hep-ph/0010096}{{\tt arXiv:hep-ph/0010096}}.

\bibitem{Narison:2005hb}
S.~Narison, ``{The SVZ-expansion and beyond},''
  \href{http://dx.doi.org/10.1016/j.nuclphysbps.2006.11.070}{{\em Nucl. Phys.
  Proc. Suppl.} {\bf 164} (2007)  225--231},
\href{http://arxiv.org/abs/hep-ph/0508259}{{\tt arXiv:hep-ph/0508259}}.

\bibitem{Grunberg:1997ud}
G.~Grunberg, ``{Power corrections and Landau singularity},''
\href{http://arxiv.org/abs/hep-ph/9705290}{{\tt arXiv:hep-ph/9705290}}.

\bibitem{Akhoury:1997ys}
R.~Akhoury and V.~I. Zakharov, ``{Renormalons and $(1/Q)^2$ corrections},''
\href{http://arxiv.org/abs/hep-ph/9705318}{{\tt arXiv:hep-ph/9705318}}.

\bibitem{Akhoury:1997by}
R.~Akhoury and V.~I. Zakharov, ``{On non perturbative corrections to the
  potential for heavy quarks},''
  \href{http://dx.doi.org/10.1016/S0370-2693(98)00932-0}{{\em Phys. Lett.} {\bf
  B438} (1998)  165--172},
\href{http://arxiv.org/abs/hep-ph/9710487}{{\tt arXiv:hep-ph/9710487}}.

\bibitem{Gubarev:1998ew}
F.~V. Gubarev, M.~I. Polikarpov, and V.~I. Zakharov, ``{Small-size strings in
  Abelian Higgs model},''
\href{http://arxiv.org/abs/hep-th/9812030}{{\tt arXiv:hep-th/9812030}}.

\bibitem{Chetyrkin:1998yr}
K.~G. Chetyrkin, S.~Narison, and V.~I. Zakharov, ``{Short-distance tachyonic
  gluon mass and $1/Q^2$ corrections},''
  \href{http://dx.doi.org/10.1016/S0550-3213(99)00167-4}{{\em Nucl. Phys.} {\bf
  B550} (1999)  353--374},
\href{http://arxiv.org/abs/hep-ph/9811275}{{\tt arXiv:hep-ph/9811275}}.

\bibitem{Zakharov:1999jj}
V.~I. Zakharov, ``{Gluon condensate and beyond},''
  \href{http://dx.doi.org/10.1142/S0217751X9900230X}{{\em Int. J. Mod. Phys.}
  {\bf A14} (1999)  4865--4880},
\href{http://arxiv.org/abs/hep-ph/9906264}{{\tt arXiv:hep-ph/9906264}}.

\bibitem{Burgio:1997hc}
G.~Burgio, F.~Di~Renzo, G.~Marchesini, and E.~Onofri,
  ``{Lambda$^2$-contribution to the condensate in lattice gauge theory},''
  \href{http://dx.doi.org/10.1016/S0370-2693(98)00057-4}{{\em Phys. Lett.} {\bf
  B422} (1998)  219--226},
\href{http://arxiv.org/abs/hep-ph/9706209}{{\tt arXiv:hep-ph/9706209}}.

\bibitem{Bali:1999ai}
G.~S. Bali, ``{Are there short distance non-perturbative contributions to the
  QCD static potential?},''
  \href{http://dx.doi.org/10.1016/S0370-2693(99)00757-1}{{\em Phys. Lett.} {\bf
  B460} (1999)  170},
\href{http://arxiv.org/abs/hep-ph/9905387}{{\tt arXiv:hep-ph/9905387}}.

\bibitem{Chernodub:2000bk}
M.~N. Chernodub, F.~V. Gubarev, M.~I. Polikarpov, and V.~I. Zakharov,
  ``{Confinement and short distance physics},''
  \href{http://dx.doi.org/10.1016/S0370-2693(00)00093-9}{{\em Phys. Lett.} {\bf
  B475} (2000)  303--310},
\href{http://arxiv.org/abs/hep-ph/0003006}{{\tt arXiv:hep-ph/0003006}}.

\bibitem{Boucaud:2000ey}
P.~Boucaud {\em et al.}, ``{Lattice calculation of $1/p^2$ corrections to
  alpha(s) and of Lambda(QCD) in the MOM~ scheme},'' {\em JHEP} {\bf 04} (2000)
   006,
\href{http://arxiv.org/abs/hep-ph/0003020}{{\tt arXiv:hep-ph/0003020}}.

\bibitem{Lavelle:1995ty}
M.~Lavelle and D.~McMullan, ``{Constituent quarks from QCD},''
  \href{http://dx.doi.org/10.1016/S0370-1573(96)00019-1}{{\em Phys. Rept.} {\bf
  279} (1997)  1--65},
\href{http://arxiv.org/abs/hep-ph/9509344}{{\tt arXiv:hep-ph/9509344}}.

\bibitem{Boucaud:2001st}
P.~Boucaud {\em et al.}, ``{Testing Landau gauge OPE on the lattice with a
  condensate},'' \href{http://dx.doi.org/10.1103/PhysRevD.63.114003}{{\em Phys.
  Rev.} {\bf D63} (2001)  114003},
\href{http://arxiv.org/abs/hep-ph/0101302}{{\tt arXiv:hep-ph/0101302}}.

\bibitem{Boucaud:2002nc}
P.~Boucaud {\em et al.}, ``{Instantons and condensate},''
  \href{http://dx.doi.org/10.1103/PhysRevD.66.034504}{{\em Phys. Rev.} {\bf
  D66} (2002)  034504},
\href{http://arxiv.org/abs/hep-ph/0203119}{{\tt arXiv:hep-ph/0203119}}.

\bibitem{Boucaud:2008gn}
P.~Boucaud {\em et al.}, ``{Ghost-gluon running coupling, power corrections and
  the determination of $\Lambda_{\bar {\rm MS}}$},''
  \href{http://dx.doi.org/10.1103/PhysRevD.79.014508}{{\em Phys. Rev.} {\bf
  D79} (2009)  014508},
\href{http://arxiv.org/abs/0811.2059}{{\tt arXiv:0811.2059 [hep-ph]}}.

\bibitem{Verschelde:2001ia}
H.~Verschelde, K.~Knecht, K.~Van~Acoleyen, and M.~Vanderkelen, ``{The
  non-perturbative groundstate of QCD and the local composite operator
  $A_\mu^2$},'' \href{http://dx.doi.org/10.1016/S0370-2693(01)00929-7}{{\em
  Phys. Lett.} {\bf B516} (2001)  307--313},
\href{http://arxiv.org/abs/hep-th/0105018}{{\tt arXiv:hep-th/0105018}}.

\bibitem{Dudal:2002pq}
D.~Dudal, H.~Verschelde, and S.~P. Sorella, ``{The anomalous dimension of the
  composite operator $A^2$ in the Landau gauge},''
  \href{http://dx.doi.org/10.1016/S0370-2693(03)00043-1}{{\em Phys. Lett.} {\bf
  B555} (2003)  126--131},
\href{http://arxiv.org/abs/hep-th/0212182}{{\tt arXiv:hep-th/0212182}}.

\bibitem{Dudal:2003vv}
D.~Dudal, H.~Verschelde, R.~E. Browne, and J.~A. Gracey, ``{A determination of
  and the non-perturbative vacuum energy of Yang-Mills theory in the Landau
  gauge},'' \href{http://dx.doi.org/10.1016/S0370-2693(03)00541-0}{{\em Phys.
  Lett.} {\bf B562} (2003)  87--96},
\href{http://arxiv.org/abs/hep-th/0302128}{{\tt arXiv:hep-th/0302128}}.

\bibitem{Browne:2003uv}
R.~E. Browne and J.~A. Gracey, ``{Two loop effective potential for in the
  Landau gauge in quantum chromodynamics},'' {\em JHEP} {\bf 11} (2003)  029,
\href{http://arxiv.org/abs/hep-th/0306200}{{\tt arXiv:hep-th/0306200}}.

\bibitem{Vercauteren:2007gx}
D.~Vercauteren and H.~Verschelde, ``{Resolving the instability of the Savvidy
  vacuum by dynamical gluon mass},''
  \href{http://dx.doi.org/10.1016/j.physletb.2008.01.013}{{\em Phys. Lett.}
  {\bf B660} (2008)  432--438},
\href{http://arxiv.org/abs/0712.0570}{{\tt arXiv:0712.0570 [hep-th]}}.

\bibitem{Chernodub:2008kf}
M.~N. Chernodub and E.~M. Ilgenfritz, ``{Electric-magnetic asymmetry of the
  $A^2$ condensate and the phases of Yang-Mills theory},''
  \href{http://dx.doi.org/10.1103/PhysRevD.78.034036}{{\em Phys. Rev.} {\bf
  D78} (2008)  034036},
\href{http://arxiv.org/abs/0805.3714}{{\tt arXiv:0805.3714 [hep-lat]}}.

\bibitem{Dudal:2009tq}
D.~Dudal, J.~A. Gracey, N.~Vandersickel, D.~Vercauteren, and H.~Verschelde,
  ``{The asymmetry of the dimension 2 gluon condensate: the zero temperature
  case},'' \href{http://dx.doi.org/10.1103/PhysRevD.80.065017}{{\em Phys. Rev.}
  {\bf D80} (2009)  065017},
\href{http://arxiv.org/abs/0907.0380}{{\tt arXiv:0907.0380 [hep-th]}}.

\bibitem{Kondo:2001nq}
K.-I. Kondo, ``{Vacuum condensate of mass dimension 2 as the origin of mass gap
  and quark confinement},''
  \href{http://dx.doi.org/10.1016/S0370-2693(01)00817-6}{{\em Phys. Lett.} {\bf
  B514} (2001)  335--345},
\href{http://arxiv.org/abs/hep-th/0105299}{{\tt arXiv:hep-th/0105299}}.

\bibitem{Li:2004te}
X.-d. Li and C.~M. Shakin, ``{Description of gluon propagation in the presence
  of an $A^2$ condensate},''
  \href{http://dx.doi.org/10.1103/PhysRevD.71.074007}{{\em Phys. Rev.} {\bf
  D71} (2005)  074007},
\href{http://arxiv.org/abs/hep-ph/0410404}{{\tt arXiv:hep-ph/0410404}}.

\bibitem{Gubarev:2005it}
F.~V. Gubarev and S.~M. Morozov, ``{ condensate, Bianchi identities and
  chromomagnetic fields degeneracy in SU(2) YM theory},''
  \href{http://dx.doi.org/10.1103/PhysRevD.71.114514}{{\em Phys. Rev.} {\bf
  D71} (2005)  114514},
\href{http://arxiv.org/abs/hep-lat/0503023}{{\tt arXiv:hep-lat/0503023}}.

\bibitem{Kekez:2005ie}
D.~Kekez and D.~Klabucar, ``{eta and eta' mesons and dimension 2 gluon
  condensate },'' \href{http://dx.doi.org/10.1103/PhysRevD.73.036002}{{\em
  Phys. Rev.} {\bf D73} (2006)  036002},
\href{http://arxiv.org/abs/hep-ph/0512064}{{\tt arXiv:hep-ph/0512064}}.

\bibitem{Andreev:2006vy}
O.~Andreev, ``{$(1/q)^2$ corrections and gauge / string duality},''
  \href{http://dx.doi.org/10.1103/PhysRevD.73.107901}{{\em Phys. Rev.} {\bf
  D73} (2006)  107901},
\href{http://arxiv.org/abs/hep-th/0603170}{{\tt arXiv:hep-th/0603170}}.

\bibitem{RuizArriola:2004en}
E.~Ruiz~Arriola, P.~O. Bowman, and W.~Broniowski, ``{Landau-gauge condensates
  from the quark propagator on the lattice},''
  \href{http://dx.doi.org/10.1103/PhysRevD.70.097505}{{\em Phys. Rev.} {\bf
  D70} (2004)  097505},
\href{http://arxiv.org/abs/hep-ph/0408309}{{\tt arXiv:hep-ph/0408309}}.

\bibitem{RuizArriola:2006gq}
E.~Ruiz~Arriola and W.~Broniowski, ``{Dimension-two gluon condensate from
  large-N(c) Regge models},''
  \href{http://dx.doi.org/10.1103/PhysRevD.73.097502}{{\em Phys. Rev.} {\bf
  D73} (2006)  097502},
\href{http://arxiv.org/abs/hep-ph/0603263}{{\tt arXiv:hep-ph/0603263}}.

\bibitem{Megias:2009mp}
E.~Megias, E.~Ruiz~Arriola, and L.~L. Salcedo, ``{Trace Anomaly, Thermal Power
  Corrections and Dimension Two condensates in the deconfined phase},''
  \href{http://dx.doi.org/10.1103/PhysRevD.80.056005}{{\em Phys. Rev.} {\bf
  D80} (2009)  056005},
\href{http://arxiv.org/abs/0903.1060}{{\tt arXiv:0903.1060 [hep-ph]}}.

\bibitem{Vercauteren:2010rk}
D.~Vercauteren and H.~Verschelde, ``{The Asymmetry of the dimension 2 gluon
  condensate: The Finite temperature case},''
  \href{http://dx.doi.org/10.1103/PhysRevD.82.085026}{{\em Phys.Rev.} {\bf D82}
  (2010)  085026}, \href{http://arxiv.org/abs/1007.2789}{{\tt arXiv:1007.2789
  [hep-th]}}.

\bibitem{Bowman:2007du}
P.~O. Bowman {\em et al.}, ``{Scaling behavior and positivity violation of the
  gluon propagator in full QCD},''
  \href{http://dx.doi.org/10.1103/PhysRevD.76.094505}{{\em Phys. Rev.} {\bf
  D76} (2007)  094505},
\href{http://arxiv.org/abs/hep-lat/0703022}{{\tt arXiv:hep-lat/0703022}}.

\bibitem{Jackiw:1995nf}
R.~Jackiw and S.-Y. Pi, ``{Threshhold Singularities and the Magnetic Mass in
  Hot {QCD}},'' \href{http://dx.doi.org/10.1016/0370-2693(95)01509-4}{{\em
  Phys. Lett.} {\bf B368} (1996)  131--136},
\href{http://arxiv.org/abs/hep-th/9511051}{{\tt arXiv:hep-th/9511051}}.

\bibitem{Stevenson:1981vj}
P.~M. Stevenson, ``{Optimized Perturbation Theory},''
\href{http://dx.doi.org/10.1103/PhysRevD.23.2916}{{\em Phys. Rev.} {\bf D23}
  (1981)  2916}.

\bibitem{Gracey:2006dr}
J.~A. Gracey, ``{One loop gluon form factor and freezing of alpha(s) in the
  Gribov-Zwanziger QCD Lagrangian},''
  \href{http://dx.doi.org/10.1007/JHEP02(2010)078}{{\em JHEP} {\bf 05} (2006)
  052},
\href{http://arxiv.org/abs/hep-ph/0605077}{{\tt arXiv:hep-ph/0605077}}.

\bibitem{Silva:2006bs}
P.~J. Silva and O.~Oliveira, ``{Exploring the infrared Landau gauge propagators
  using large asymmetric lattices},'' {\em PoS} {\bf LAT2006} (2006)  075,
\href{http://arxiv.org/abs/hep-lat/0609069}{{\tt arXiv:hep-lat/0609069}}.

\bibitem{Cucchieri:2006tf}
A.~Cucchieri, A.~Maas, and T.~Mendes, ``{Exploratory study of three-point
  Green's functions in Landau-gauge Yang-Mills theory},''
  \href{http://dx.doi.org/10.1103/PhysRevD.74.014503}{{\em Phys. Rev.} {\bf
  D74} (2006)  014503},
\href{http://arxiv.org/abs/hep-lat/0605011}{{\tt arXiv:hep-lat/0605011}}.

\bibitem{Cucchieri:2008qm}
A.~Cucchieri, A.~Maas, and T.~Mendes, ``{Three-point vertices in Landau-gauge
  Yang-Mills theory},''
  \href{http://dx.doi.org/10.1103/PhysRevD.77.094510}{{\em Phys. Rev.} {\bf
  D77} (2008)  094510},
\href{http://arxiv.org/abs/0803.1798}{{\tt arXiv:0803.1798 [hep-lat]}}.

\bibitem{Zwanziger:2001kw}
D.~Zwanziger, ``{Non-perturbative Landau gauge and infrared critical exponents
  in QCD},'' \href{http://dx.doi.org/10.1103/PhysRevD.65.094039}{{\em Phys.
  Rev.} {\bf D65} (2002)  094039},
\href{http://arxiv.org/abs/hep-th/0109224}{{\tt arXiv:hep-th/0109224}}.

\bibitem{Gracey:2007bj}
J.~A. Gracey, ``{Recent results for Yang-Mills theory restricted to the Gribov
  region},''
\href{http://arxiv.org/abs/0711.3622}{{\tt arXiv:0711.3622 [hep-th]}}.

\bibitem{Furui:2004cx}
S.~Furui and H.~Nakajima, ``{What the Gribov copy tells about confinement and
  the theory of dynamical chiral symmetry breaking},''
\href{http://dx.doi.org/10.1103/PhysRevD.70.094504}{{\em Phys. Rev.} {\bf D70}
  (2004)  094504}.

\bibitem{Ilgenfritz:2006gp}
E.~M. Ilgenfritz, M.~Muller-Preussker, A.~Sternbeck, and A.~Schiller,
  ``{Gauge-variant propagators and the running coupling from lattice QCD},''
\href{http://arxiv.org/abs/hep-lat/0601027}{{\tt arXiv:hep-lat/0601027}}.

\bibitem{Jackiw:1980kv}
R.~Jackiw and S.~Templeton, ``{How Superrenormalizable Interactions Cure their
  Infrared Divergences},''
\href{http://dx.doi.org/10.1103/PhysRevD.23.2291}{{\em Phys. Rev.} {\bf D23}
  (1981)  2291}.

\bibitem{Lucini:2002wg}
B.~Lucini and M.~Teper, ``{SU(N) gauge theories in 2+1 dimensions: Further
  results},'' \href{http://dx.doi.org/10.1103/PhysRevD.66.097502}{{\em Phys.
  Rev.} {\bf D66} (2002)  097502},
\href{http://arxiv.org/abs/hep-lat/0206027}{{\tt arXiv:hep-lat/0206027}}.

\bibitem{Bassetto:1999ah}
A.~Bassetto, ``{Two-dimensional Yang-Mills theory: Perturbative and instanton
  contributions, and its relation to QCD in higher dimensions},''
  \href{http://dx.doi.org/10.1016/S0920-5632(00)00767-2}{{\em Nucl. Phys. Proc.
  Suppl.} {\bf 88} (2000)  184--193},
\href{http://arxiv.org/abs/hep-th/9911208}{{\tt arXiv:hep-th/9911208}}.

\bibitem{Sobreiro:2005ec}
R.~F. Sobreiro and S.~P. Sorella, ``{Introduction to the Gribov ambiguities in
  Euclidean Yang- Mills theories},''
\href{http://arxiv.org/abs/hep-th/0504095}{{\tt arXiv:hep-th/0504095}}.

\bibitem{wolfram}
http://mathworld.wolfram.com.

\bibitem{Bali:1992ru}
G.~S. Bali and K.~Schilling, ``{Running coupling and the Lambda parameter from
  SU(3) lattice simulations},''
  \href{http://dx.doi.org/10.1103/PhysRevD.47.661}{{\em Phys. Rev.} {\bf D47}
  (1993)  661--672},
\href{http://arxiv.org/abs/hep-lat/9208028}{{\tt arXiv:hep-lat/9208028}}.

\bibitem{Gracey:2010cg}
J.~Gracey, ``{Alternative refined Gribov-Zwanziger Lagrangian},''
  \href{http://dx.doi.org/10.1103/PhysRevD.82.085032}{{\em Phys.Rev.} {\bf D82}
  (2010)  085032}, \href{http://arxiv.org/abs/1009.3889}{{\tt arXiv:1009.3889
  [hep-th]}}.

\bibitem{Verschelde:1995jj}
H.~Verschelde, ``{Perturbative calculation of nonperturbative effects in
  quantum field theory},''
\href{http://dx.doi.org/10.1016/0370-2693(95)00338-L}{{\em Phys. Lett.} {\bf
  B351} (1995)  242--248}.

\bibitem{Gracey:2002yt}
J.~A. Gracey, ``{Three loop MS-bar renormalization of the Curci-Ferrari model
  and the dimension two BRST invariant composite operator in QCD},''
  \href{http://dx.doi.org/10.1016/S0370-2693(02)03077-0}{{\em Phys. Lett.} {\bf
  B552} (2003)  101--110},
\href{http://arxiv.org/abs/hep-th/0211144}{{\tt arXiv:hep-th/0211144}}.

\bibitem{Mathieu:2008me}
V.~Mathieu, N.~Kochelev, and V.~Vento, ``{The Physics of Glueballs},''
  \href{http://dx.doi.org/10.1142/S0218301309012124}{{\em Int. J. Mod. Phys.}
  {\bf E18} (2009)  1--49},
\href{http://arxiv.org/abs/0810.4453}{{\tt arXiv:0810.4453 [hep-ph]}}.

\bibitem{Bettoni:2005ut}
D.~Bettoni, ``{Physics with the PANDA detector at GSI},''
\href{http://dx.doi.org/10.1088/1742-6596/9/1/059}{{\em J. Phys. Conf. Ser.}
  {\bf 9} (2005)  309--314}.

\bibitem{Chanowitz:2006wf}
M.~S. Chanowitz, ``{Hunting the scalar glueball: Prospects for BES III},''
  \href{http://dx.doi.org/10.1142/S0217751X06034719}{{\em Int. J. Mod. Phys.}
  {\bf A21} (2006)  5535--5542},
\href{http://arxiv.org/abs/hep-ph/0609217}{{\tt arXiv:hep-ph/0609217}}.

\bibitem{Carman:2005ps}
D.~S. Carman, ``{GlueX: The search for gluonic excitations at Jefferson
  Laboratory},'' \href{http://dx.doi.org/10.1063/1.2176480}{{\em AIP Conf.
  Proc.} {\bf 814} (2006)  173--182},
\href{http://arxiv.org/abs/hep-ex/0511030}{{\tt arXiv:hep-ex/0511030}}.

\bibitem{Alessandro:2006yt}
{\bf ALICE} Collaboration, G.~Alessandro, (Ed.~) {\em et al.}, ``{ALICE:
  Physics performance report, volume II},''
\href{http://dx.doi.org/10.1088/0954-3899/32/10/001}{{\em J. Phys.} {\bf G32}
  (2006)  1295--2040}.

\bibitem{Morningstar:1999rf}
C.~J. Morningstar and M.~J. Peardon, ``{The glueball spectrum from an
  anisotropic lattice study},''
  \href{http://dx.doi.org/10.1103/PhysRevD.60.034509}{{\em Phys. Rev.} {\bf
  D60} (1999)  034509},
\href{http://arxiv.org/abs/hep-lat/9901004}{{\tt arXiv:hep-lat/9901004}}.

\bibitem{McNeile:2002en}
C.~McNeile, ``{Lattice predictions for hybrids and glueballs},''
  \href{http://dx.doi.org/10.1016/S0375-9474(02)01233-2}{{\em Nucl. Phys.} {\bf
  A711} (2002)  303--308},
\href{http://arxiv.org/abs/hep-lat/0207001}{{\tt arXiv:hep-lat/0207001}}.

\bibitem{Chen:2005mg}
Y.~Chen {\em et al.}, ``{Glueball spectrum and matrix elements on anisotropic
  lattices},'' \href{http://dx.doi.org/10.1103/PhysRevD.73.014516}{{\em Phys.
  Rev.} {\bf D73} (2006)  014516},
\href{http://arxiv.org/abs/hep-lat/0510074}{{\tt arXiv:hep-lat/0510074}}.

\bibitem{Teper:1998kw}
M.~J. Teper, ``{Glueball masses and other physical properties of SU(N) gauge
  theories in D = 3+1: A review of lattice results for theorists},''
\href{http://arxiv.org/abs/hep-th/9812187}{{\tt arXiv:hep-th/9812187}}.

\bibitem{Meyer:2008tr}
H.~B. Meyer, ``{Glueball matrix elements: a lattice calculation and
  applications},'' \href{http://dx.doi.org/10.1088/1126-6708/2009/01/071}{{\em
  JHEP} {\bf 01} (2009)  071},
\href{http://arxiv.org/abs/0808.3151}{{\tt arXiv:0808.3151 [hep-lat]}}.

\bibitem{West:1997sz}
G.~B. West, ``{Glueballs and nonperturbative QCD},''
{\em Nucl. Phys. Proc. Suppl.} {\bf 54A} (1997)  353--361.

\bibitem{Jaffe:1975fd}
R.~L. Jaffe and K.~Johnson, ``{Unconventional States of Confined Quarks and
  Gluons},''
\href{http://dx.doi.org/10.1016/0370-2693(76)90423-8}{{\em Phys. Lett.} {\bf
  B60} (1976)  201}.

\bibitem{Cornwall:1981zr}
J.~M. Cornwall, ``{Dynamical Mass Generation in Continuum QCD},''
\href{http://dx.doi.org/10.1103/PhysRevD.26.1453}{{\em Phys. Rev.} {\bf D26}
  (1982)  1453}.

\bibitem{Bernard:1981pg}
C.~W. Bernard, ``{Monte Carlo Evaluation of the Effective Gluon Mass},''
\href{http://dx.doi.org/10.1016/0370-2693(82)91228-X}{{\em Phys. Lett.} {\bf
  B108} (1982)  431}.

\bibitem{Cornwall:1982zn}
J.~M. Cornwall and A.~Soni, ``{Glueballs as Bound States of Massive Gluons},''
\href{http://dx.doi.org/10.1016/0370-2693(83)90481-1}{{\em Phys. Lett.} {\bf
  B120} (1983)  431}.

\bibitem{Barnes:1981ac}
T.~Barnes, ``{A transverse gluonium potential model with Breit-Fermi hyperfine
  effects},''
\href{http://dx.doi.org/10.1007/BF01549736}{{\em Zeit. Phys.} {\bf C10} (1981)
  275}.

\bibitem{Szczepaniak:1995cw}
A.~Szczepaniak, E.~S. Swanson, C.-R. Ji, and S.~R. Cotanch, ``{Glueball
  Spectroscopy in a Relativistic Many-Body Approach to Hadron Structure},''
  \href{http://dx.doi.org/10.1103/PhysRevLett.76.2011}{{\em Phys. Rev. Lett.}
  {\bf 76} (1996)  2011--2014},
\href{http://arxiv.org/abs/hep-ph/9511422}{{\tt arXiv:hep-ph/9511422}}.

\bibitem{Kaidalov:1999yd}
A.~B. Kaidalov and Y.~A. Simonov, ``{Glueball spectrum and the pomeron in the
  Wilson loop approach},'' \href{http://dx.doi.org/10.1134/1.1307465}{{\em
  Phys. Atom. Nucl.} {\bf 63} (2000)  1428--1447},
\href{http://arxiv.org/abs/hep-ph/9911291}{{\tt arXiv:hep-ph/9911291}}.

\bibitem{Novikov:1979va}
V.~A. Novikov, M.~A. Shifman, A.~I. Vainshtein, and V.~I. Zakharov, ``{In a
  Search for Scalar Gluonium},''
\href{http://dx.doi.org/10.1016/0550-3213(80)90306-5}{{\em Nucl. Phys.} {\bf
  B165} (1980)  67}.

\bibitem{Narison:2008nj}
S.~Narison, ``{Light scalar mesons in QCD},''
  \href{http://dx.doi.org/10.1016/j.nuclphysbps.2008.12.069}{{\em Nucl. Phys.
  Proc. Suppl.} {\bf 186} (2009)  306--311},
\href{http://arxiv.org/abs/0811.0563}{{\tt arXiv:0811.0563 [hep-ph]}}.

\bibitem{Shuryak:1982dp}
E.~V. Shuryak, ``{The Role of Instantons in Quantum Chromodynamics. 2. Hadronic
  Structure},''
\href{http://dx.doi.org/10.1016/0550-3213(82)90479-5}{{\em Nucl. Phys.} {\bf
  B203} (1982)  116}.

\bibitem{Brower:2000rp}
R.~C. Brower, S.~D. Mathur, and C.-I. Tan, ``{Glueball Spectrum for QCD from
  AdS Supergravity Duality},''
  \href{http://dx.doi.org/10.1016/S0550-3213(00)00435-1}{{\em Nucl. Phys.} {\bf
  B587} (2000)  249--276},
\href{http://arxiv.org/abs/hep-th/0003115}{{\tt arXiv:hep-th/0003115}}.

\bibitem{Forkel:2007ru}
H.~Forkel, ``{Holographic glueball structure},''
  \href{http://dx.doi.org/10.1103/PhysRevD.78.025001}{{\em Phys. Rev.} {\bf
  D78} (2008)  025001},
\href{http://arxiv.org/abs/0711.1179}{{\tt arXiv:0711.1179 [hep-ph]}}.

\bibitem{Dudal:2008tg}
D.~Dudal, S.~P. Sorella, N.~Vandersickel, and H.~Verschelde, ``{A purely
  algebraic construction of a gauge and renormalization group invariant scalar
  glueball operator},''
  \href{http://dx.doi.org/10.1140/epjc/s10052-009-1139-3}{{\em Eur. Phys. J.}
  {\bf C64} (2009)  147--159},
\href{http://arxiv.org/abs/0812.2401}{{\tt arXiv:0812.2401 [hep-th]}}.

\bibitem{Collins:1984xc}
J.~C. Collins, ``{Renormalization. An introduction to renormalization, the
  renormalization group, and the Operator Product Expansion},''. Cambridge, Uk:
  Univ. Pr. ( 1984) 380p.

\bibitem{Collins:1994ee}
J.~C. Collins and R.~J. Scalise, ``{The Renormalization of composite operators
  in Yang-Mills theories using general covariant gauge},''
  \href{http://dx.doi.org/10.1103/PhysRevD.50.4117}{{\em Phys. Rev.} {\bf D50}
  (1994)  4117--4136},
\href{http://arxiv.org/abs/hep-ph/9403231}{{\tt arXiv:hep-ph/9403231}}.

\bibitem{Brown:1979pq}
L.~S. Brown, ``{Dimensional regularization of composite operators in scalar
  field theory},''
\href{http://dx.doi.org/10.1016/0003-4916(80)90377-2}{{\em Ann. Phys.} {\bf
  126} (1980)  135}.

\bibitem{Collins:1976yq}
J.~C. Collins, A.~Duncan, and S.~D. Joglekar, ``{Trace and Dilatation Anomalies
  in Gauge Theories},''
\href{http://dx.doi.org/10.1103/PhysRevD.16.438}{{\em Phys. Rev.} {\bf D16}
  (1977)  438--449}.

\bibitem{Dudal:2009zh}
D.~Dudal, S.~P. Sorella, N.~Vandersickel, and H.~Verschelde, ``{A
  renormalization group invariant scalar glueball operator in the (Refined)
  Gribov-Zwanziger framework},''
  \href{http://dx.doi.org/10.1088/1126-6708/2009/08/110}{{\em JHEP} {\bf 08}
  (2009)  110},
\href{http://arxiv.org/abs/0906.4257}{{\tt arXiv:0906.4257 [hep-th]}}.

\bibitem{Baulieu:2009ha}
L.~Baulieu, D.~Dudal, M.~Guimaraes, M.~Huber, S.~Sorella, {\em et al.},
  ``{Gribov horizon and i-particles: About a toy model and the construction of
  physical operators},''
  \href{http://dx.doi.org/10.1103/PhysRevD.82.025021}{{\em Phys.Rev.} {\bf D82}
  (2010)  025021}, \href{http://arxiv.org/abs/0912.5153}{{\tt arXiv:0912.5153
  [hep-th]}}.

\bibitem{'tHooft:1978xw}
G.~'t~Hooft and M.~J.~G. Veltman, ``{Scalar One Loop Integrals},''
\href{http://dx.doi.org/10.1016/0550-3213(79)90605-9}{{\em Nucl. Phys.} {\bf
  B153} (1979)  365--401}.

\bibitem{Itzykson:1980rh}
C.~Itzykson and J.~B. Zuber, ``{Quantum field theory},''. New York, Usa:
  Mcgraw-hill (1980) 705 P.(International Series In Pure and Applied Physics).

\bibitem{Dudal:2010cd}
D.~Dudal, M.~Guimaraes, and S.~Sorella, ``{Glueball masses from an infrared
  moment problem and nonperturbative Landau gauge},''
  \href{http://arxiv.org/abs/1010.3638}{{\tt arXiv:1010.3638 [hep-th]}}.

\bibitem{Dudal:2010wn}
D.~Dudal and M.~S. Guimaraes, ``{On the computation of the spectral density of
  two-point functions: complex masses, cut rules and beyond},''
  \href{http://dx.doi.org/10.1103/PhysRevD.83.045013}{{\em Phys.Rev.} {\bf D83}
  (2011)  045013}, \href{http://arxiv.org/abs/1012.1440}{{\tt arXiv:1012.1440
  [hep-th]}}.

\bibitem{Yn}
F.~Yndur\'{a}in, {\em Pad\'{e} Approximants}, ch.~The moment problem and
  applications.
\newblock The Institute of Physics, London and Bristol.

\bibitem{shohat}
J.~A. Shohat and J.~D. Tamarkin, {\em The Problem of Moments}.
\newblock the American Mathematical Society, 1970.

\bibitem{Crede:2008vw}
V.~Crede and C.~A. Meyer, ``{The Experimental Status of Glueballs},''
  \href{http://dx.doi.org/10.1016/j.ppnp.2009.03.001}{{\em Prog. Part. Nucl.
  Phys.} {\bf 63} (2009)  74--116},
\href{http://arxiv.org/abs/0812.0600}{{\tt arXiv:0812.0600 [hep-ex]}}.

\bibitem{Narison:1996fm}
S.~Narison, ``{Masses, decays and mixings of gluonia in QCD},''
  \href{http://dx.doi.org/10.1016/S0550-3213(97)00562-2}{{\em Nucl. Phys.} {\bf
  B509} (1998)  312--356},
\href{http://arxiv.org/abs/hep-ph/9612457}{{\tt arXiv:hep-ph/9612457}}.

\bibitem{Schafer:1994fd}
T.~Schafer and E.~V. Shuryak, ``{Glueballs and instantons},''
  \href{http://dx.doi.org/10.1103/PhysRevLett.75.1707}{{\em Phys.Rev.Lett.}
  {\bf 75} (1995)  1707--1710}, \href{http://arxiv.org/abs/hep-ph/9410372}{{\tt
  arXiv:hep-ph/9410372 [hep-ph]}}.

\bibitem{Blank:2010pa}
M.~Blank, A.~Krassnigg, and A.~Maas, ``{Rho-meson, Bethe-Salpeter equation, and
  the far infrared},'' \href{http://dx.doi.org/10.1103/PhysRevD.83.034020}{{\em
  Phys.Rev.} {\bf D83} (2011)  034020},
  \href{http://arxiv.org/abs/1007.3901}{{\tt arXiv:1007.3901 [hep-ph]}}.

\bibitem{Yokojima:1995hy}
S.~Yokojima, ``{Effective action of a local composite operator},''
\href{http://dx.doi.org/10.1103/PhysRevD.51.2996}{{\em Phys. Rev.} {\bf D51}
  (1995)  2996--3008}.

\bibitem{vanderBij:1983bw}
J.~van~der Bij and M.~J.~G. Veltman, ``{Two Loop Large Higgs Mass Correction to
  the rho Parameter},''
\href{http://dx.doi.org/10.1016/0550-3213(84)90284-0}{{\em Nucl. Phys.} {\bf
  B231} (1984)  205}.

\bibitem{Ford:1992pn}
C.~Ford, I.~Jack, and D.~R.~T. Jones, ``{The Standard Model Effective Potential
  at Two Loops},'' \href{http://dx.doi.org/10.1016/0550-3213(92)90165-8}{{\em
  Nucl. Phys.} {\bf B387} (1992)  373--390},
\href{http://arxiv.org/abs/hep-ph/0111190}{{\tt arXiv:hep-ph/0111190}}.

\bibitem{Davydychev:1992mt}
A.~I. Davydychev and J.~B. Tausk, ``{Two loop selfenergy diagrams with
  different masses and the momentum expansion},''
\href{http://dx.doi.org/10.1016/0550-3213(93)90338-P}{{\em Nucl. Phys.} {\bf
  B397} (1993)  123--142}.

\end{thebibliography}
\end{document}